\documentclass[twoside,12pt]{article}

\usepackage{amsmath}
\usepackage{amssymb}
\usepackage{physics}
\usepackage[utf8x]{inputenc}
\usepackage[inline]{enumitem}

\usepackage{tocloft}
\tocloftpagestyle{empty}

\usepackage[T1]{fontenc}
\usepackage{lmodern}
\usepackage{microtype}
\usepackage{bbm}

\usepackage[dvipsnames]{xcolor}
\definecolor{darkblue}{rgb}{0.0, 0.0, 0.55}
\usepackage[colorlinks=true,allcolors=darkblue]{hyperref}

\usepackage{epsfig}

\usepackage[capitalise]{cleveref}

\usepackage{etoolbox}
\usepackage{xstring}

\DeclareRobustCommand{\refcite}[2][]{%
\IfSubStr{#2}{,}{refs.}{ref.}~\ifstrempty{#1}{\cite{#2}}{\cite[#1]{#2}}%
}
\DeclareRobustCommand{\Refcite}[2][]{%
\IfSubStr{#2}{,}{Refs.}{Ref.}~\ifstrempty{#1}{\cite{#2}}{\cite[#1]{#2}}%
}

\usepackage{cite}

\let\ocite\cite
\renewcommand{\cite}{\unskip~\ocite}

\hypersetup{linktoc=all}

\numberwithin{equation}{section}

\newcommand{\abstractindent}{\noindent}
\newcommand{\address}{\normalsize\itshape}
\newcommand{\datecommand}{\date{}}
\newcommand{\titlestyle}{\vspace{1cm} \LARGE}
\newcommand{\authorstyle}{\large}

\usepackage{geometry}
\usepackage{mathtools}
\usepackage{siunitx}
\usepackage{xspace}
\usepackage[normalem]{ulem}

\newcommand{\overbar}[1]{\mkern 3mu\overline{\mkern-3mu#1\mkern-1.5mu}\mkern 1.5mu}

\newcommand{\SE}{\ensuremath{S_\text{E}}}
\newcommand{\ST}{\ensuremath{S_\text{T}}}
\newcommand{\SV}{\ensuremath{S_\text{V}}}
\newcommand{\tv}{\ensuremath{\phi_t}}
\newcommand{\fv}{\ensuremath{\phi_f}}

\newcommand{\bounce}{\ensuremath{\bar{\phi}}}
\DeclareMathOperator{\detprime}{{\det}^\prime}
\DeclareMathOperator{\detplus}{{\det}^{+}}
\newcommand{\Tau}{\mathcal{T}}
\newcommand{\code}{\textsf}
\newcommand{\vol}{\mathcal{V}}
\newcommand{\Vext}{\vol_t^{\text{ext}}}
\newcommand{\Vphys}{\vol_{\text{phys}}}
\newcommand{\Ekin}{E_{\text{kin}}}
\newcommand{\rhokin}{\rho_{\text{kin}}}
\newcommand{\rhokinavg}{\hspace{-0.2ex}\overbar{\hspace{0.2ex} \rho \hspace{0.2ex}}_{\hspace{-0.2ex} \text{kin}}} 

\newcommand{\rhotot}{\rho_{\text{tot}}}
\newcommand{\FED}{\mathcal{F}} 
\newcommand{\Uf}{\overbar{U}_{\!\!f}} 
\newcommand{\Ulong}{\overbar{U}_{\!\parallel}} 
\newcommand{\Utrans}{\overbar{U}_{\!\!\perp}} 

\newcommand{\Treh}{T_{\text{reh}}}
\newcommand{\thb}{\bar{\theta}}
\newcommand{\Ttr}{T_*} 
\newcommand{\lenscale}{L_*} 
\newcommand{\bubsep}{R_{\text{sep}}} 
\newcommand{\bubrad}{\bar{R}} 
\newcommand{\bubradweight}{R_\rho} 
\newcommand{\ptot}{p_{\text{tot}}}
\newcommand{\pdr}{p_{\text{dr}}}
\newcommand{\pfr}{p_{\text{fr}}}
\newcommand{\pfrlo}{p_{\text{fr}}^{\text{LO}}}

\newcommand{\redshift}{\mathcal{R}}

\newcommand{\dint}{\textrm{d}}

\newcommand{\MSbar}{\ensuremath{\overline{\text{MS}}}\xspace}
\newcommand{\DRbar}{\ensuremath{\overline{\text{DR}}}\xspace}
\newcommand{\gev}{\,\text{GeV}}

\newcommand{\half}{\frac{1}{2}}
\newcommand{\recip}[1]{\frac{1}{#1}}

\newcommand{\myunit}[1]{\ensuremath{\, \mathrm{#1}}}

\newcommand{\vecb}[1]{\boldsymbol{#1}}

\newcommand{\be}{\begin{equation}}
\newcommand{\ee}{\end{equation}}
\newcommand{\bea}{\begin{eqnarray}}
\newcommand{\eea}{\end{eqnarray}}

\newcommand{\field}{\vecb{\phi}}
\newcommand{\Trans}[2]{\mathcal{T}_{#1 \rightarrow #2}}

\def\comments{true}

\ifx\comments\undefined
       \newcommand{\comment}[1]{}
\else
       \newcommand{\comment}[1]{#1}
\fi

\begin{document}

\title{\titlestyle Cosmological phase transitions: from perturbative particle physics to gravitational waves}
\author{\authorstyle Peter Athron$^{1,2}$, Csaba Bal\'azs$^2$, Andrew Fowlie$^3$,\\ 
\authorstyle Lachlan Morris$^2$, Lei Wu$^1$\\
\\
\address $^1$Department of Physics and Institute of Theoretical Physics, \\ 
\address Nanjing Normal University, Nanjing, Jiangsu 210023, China\\
\address $^2$School of Physics and Astronomy, Monash University,\\
\address Melbourne, Victoria 3800, Australia\\
\address $^3$Department of Physics, School of Mathematics and Physics,\\ 
\address Xi'an Jiaotong-Liverpool University, Suzhou, Jiangsu 215123, China}
\datecommand
\clearpage\maketitle
\thispagestyle{empty}

\begin{abstract}
\abstractindent Gravitational waves (GWs) were recently detected for the first time.
This revolutionary discovery opens a new way of learning about particle physics through GWs from first-order phase transitions (FOPTs) in the early Universe.
FOPTs could occur when new fundamental symmetries are spontaneously broken down to the Standard Model and are a vital ingredient in solutions of the matter anti-matter asymmetry problem.
The purpose of our work is to review the path from a particle physics model to GWs, which contains many specialized parts, so here we provide a timely review of all the required steps, including:~%
\begin{enumerate*}[label=(\roman*),font=\itshape]%
  \item building a finite-temperature effective potential in a particle physics model and checking for FOPTs;
  \item computing transition rates;
  \item analyzing the dynamics of bubbles of true vacuum expanding in a thermal plasma;
  \item characterizing a transition using thermal parameters; and, finally, 
  \item making predictions for GW spectra using the latest simulations and theoretical results and considering the detectability of predicted spectra at future GW detectors.
\end{enumerate*}
For each step we emphasize the subtleties, advantages and drawbacks of different methods, discuss open questions and review the state-of-art approaches available in the literature. 
This provides everything a particle physicist needs to begin exploring GW phenomenology.
\end{abstract}

\newpage
\tableofcontents
\newpage
\pagenumbering{arabic}
\setcounter{page}{1}

\section{Introduction}

In 2016 the first direct observation of gravitational waves (GWs) was reported
by the LIGO Collaboration~\cite{LIGOScientific:2016aoc}, leading to the award of the 2017 Nobel Prize for physics.
The discovery began a new era of multi-messenger astronomy~\cite{Meszaros:2019xej}, as the 15 datasets \cite{LIGO-data} released so far provide a new way of learning about astro-physics when GWs are used alongside regular electromagnetic radiation, neutrinos and cosmic rays. The prospect of increasingly sensitive GW observatories in the future, furthermore, heralds a new era in particle physics. 

Fundamental theories, such as Grand Unified Theories (GUTs; see e.g., \refcite{Langacker:1980js}), predict that symmetries were broken in the early Universe, potentially through violent first-order phase transitions (FOPTs; see \refcite{Mazumdar:2018dfl} for a review of cosmological phase transitions), and the existence of cosmic strings, domain walls and magnetic monopoles. These particle physics predictions could finally be tested through their impact on the GW spectrum at future GW observatories. For example, in the early Universe, GWs may originate from fundamental physics, including quantum fluctuations during inflation and (p)re-heating or topological defects during symmetry breaking, including cosmic strings, domain walls and vacuum bubbles. On the other hand, in the late Universe GWs may originate from galaxy mergers or binary mergers of e.g., black holes and neutron stars. See e.g., \refcite{Cai:2017cbj, Bian:2021ini} for reviews on sources of GWs.

We focus on GW signals from FOPTs in this review, as FOPTs appear naturally in extensions of the Standard Model of particle physics (SM) and are a crucial ingredient in explanations of the abundance of matter rather than anti-matter in the Universe, as they help to satisfy one of Sakharov's conditions for baryogenesis. Although the SM predicts that there was an electroweak phase transition (EWPT), FOPTs must be connected to new physics as lattice simulations show that the SM predicts that the EWPT was a smooth crossover transition. As such, any evidence that the EWPT was first order or evidence of any FOPT would indicate new physics beyond the SM.  

\begin{figure}[t]
\centering
\includegraphics[width=0.29\linewidth]{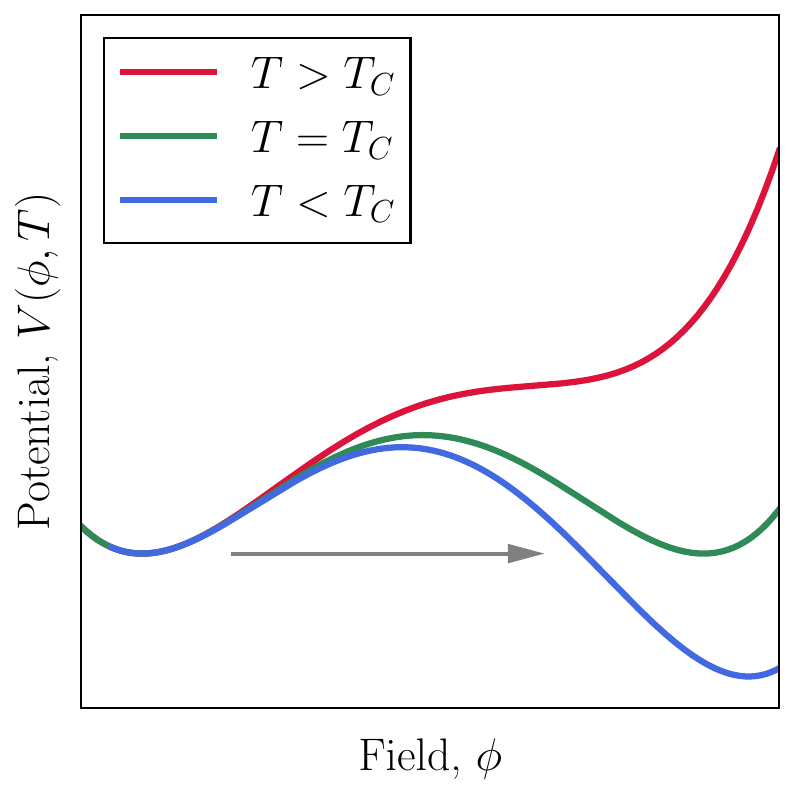}%
\raisebox{2.75cm}{\hspace{1.25mm}\Large$\Rightarrow$\hspace{.5mm}}
\raisebox{-0.mm}{\includegraphics[width=0.2875\linewidth]{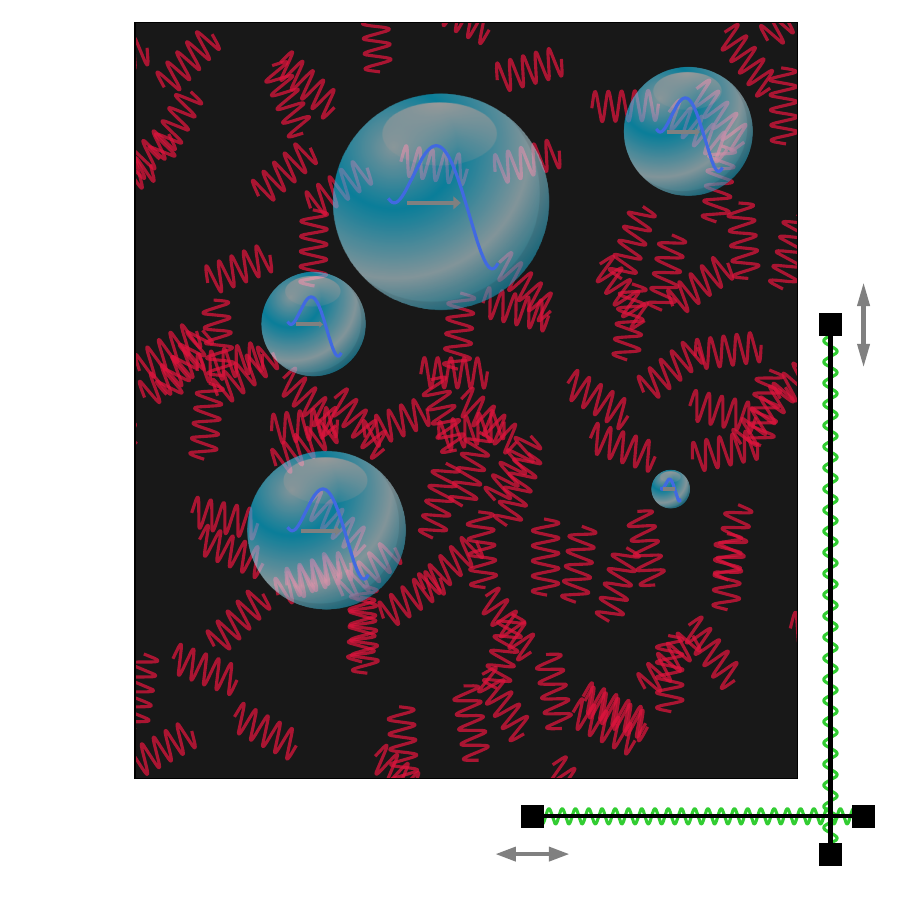}}%
\raisebox{2.75cm}{\hspace{0.5mm}\Large$\Rightarrow$\hspace{.5mm}}
\includegraphics[width=0.29\linewidth]{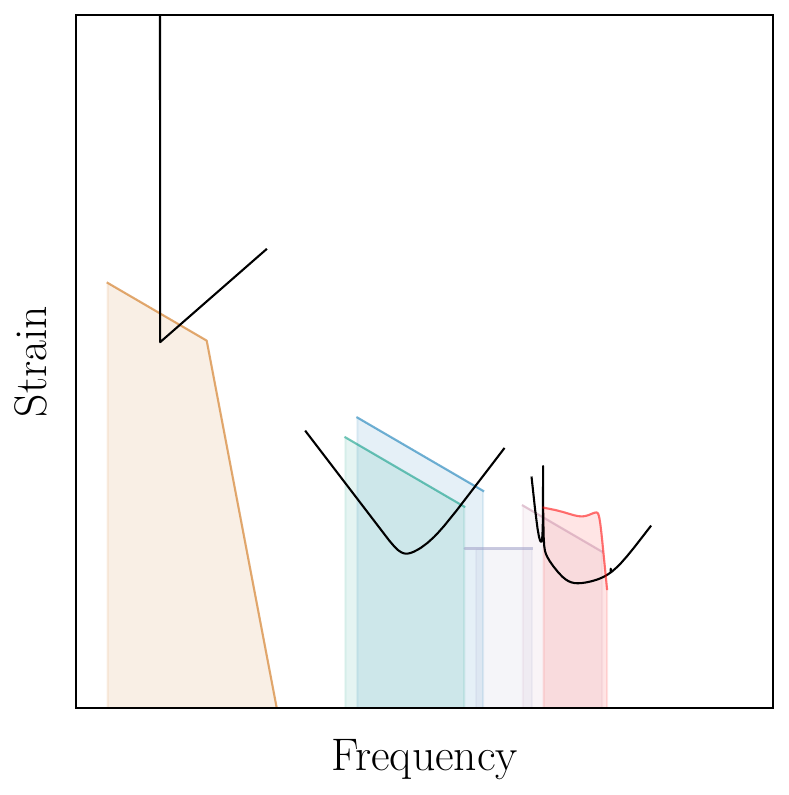}
\caption{As the Universe cools, the potential develops a new minima away from the origin (left). A first-order phase transition occurs through bubbles, which appear spontaneously and expand in the thermal plasma (center). The GWs from bubble collisions and the plasma may be measured using a laser interferometer, resulting in a stochastic GW background spectrum (right).}
\label{fig:bubbles}
\end{figure}

The GWs from a FOPT would be a message from the physics of the early Universe. As the Universe cooled and expanded, we expect that the shape of the scalar potential, which plays the role of the free-energy, changed. At a particular temperature, a new vacuum could appear separated from the existing one by a barrier, as in the first panel in \cref{fig:bubbles}. The transition to the new vacuum through the barrier would proceed through bubbles which expand,
generating GWs through friction with the thermal
plasma and collisions~(see e.g., \refcite{Binetruy:2012ze,Caprini:2015zlo, Caprini:2019egz,Hindmarsh:2020hop} for reviews), until they dominate the Universe. The
GWs signals may be detected in the near future using laser interferometers. We divide analysis of this phenomenon into five stages:
\begin{enumerate}
\item The behavior of the effective potential as the Universe cools
\item The decay of a false vacuum and spontaneous creation of bubbles of the true vacuum
\item The growth of bubbles of the true vacuum and completion of a FOPT
\item Thermal parameters associated with a FOPT
\item The production and detection of GWs from those thermal parameters
\end{enumerate}
We expect each aspect to be important in the era of GWs. In this review we cover what is known about them, focusing on a particle physics perspective and covering perturbative methods familiar and accessible to particle physicists. Although powerful, we mention lattice methods only briefly as they are computationally demanding and beyond most phenomenological studies. We begin in \cref{sec:effective_potential} by reviewing the effective potential at zero- and finite-temperature. This tells us how the ground state of the Universe might change as the Universe cools. We continue in \cref{sec:transition_rates} by reviewing the physics of tunneling and thermal fluctuations over a barrier separating different ground states. These processes lead to phase transitions that proceed through bubbles; we review the nucleation and growth of bubbles until a transition completes in \cref{sec:transitionAnalysis} and their characteristic thermal parameters in \cref{sec:thermal_parameters}. We consider the creation and detection of GWs from the bubbles in \cref{Section:GWs-Sources}, before closing in \cref{sec:summary} with a discussion of open issues.

First-order cosmological phase transitions can also be part of a
successful electroweak baryogenesis mechanism, and this is one of the
strongest motivations for expecting strong first-order phase
transitions to be part of our cosmological phase history.  As a result
many of the calculations of the phase transitions represent a
significant overlap between the calculations required for testing
GWs and electroweak baryogenesis. Therefore some
parts of this review, specifically \cref{sec:effective_potential,sec:transition_rates}, can also be useful for
electroweak baryogenesis, but beyond that we will not provide any further
pedagogy on that topic and recommend that interested readers refer to
the large number of pedagogical texts already available for electroweak
baryogenesis
\cite{Cohen:1993nk,Trodden:1998ym,Cline:2006ts,Morrissey:2012db,White:2016nbo}.

Whilst this is a review, there a few novel arguments and derivations in places. These include new derivations of the JMAK equation in \cref{sec:altJMAK} and the connection between the effective potential and the convex hull of the potential in \cref{App:Convex}. Lastly, in \cref{sec:from_fopt_to_gws} we slightly extend previous works~\cite{Kosowsky:1991ua} to provide a simple justification of the form of the GW spectrum on dimensional grounds alone.

\section{Effective potential}\label{sec:effective_potential}

We wish to explore transitions between different vacua in a model of
particle physics, ultimately leading to GW signals. In this setting
the vacua are minima of the effective potential, which was essentially introduced in
\refcite{Heisenberg:1936nmg,Schwinger:1951nm}, first applied to
spontaneous symmetry breaking in
\refcite{Jona-Lasinio:1964zvf,Goldstone:1962es}, and the formalism
was presented clearly in the manner it is understood and used today in
\refcite{Coleman:1973jx,Jackiw:1974cv}.  The latter two references
provide very clear explanations and proofs and are highly recommended
as a starting point for the effective potential.  Further pedagogical
descriptions can be found in
\refcite{Brandenberger:1984cz,Sher:1988mj,Quiros:1999jp}.
Our first step is thus to
construct the effective potential in models of particle physics.

In general, the effective potential is a real  scalar function of one or more
scalar fields.\footnote{While in principle the effective
  potential is real, when calculated perturbatively the effective
  potential can have imaginary parts, as we discuss in
  \cref{sec:imaginary_contributions}.} 
  More precisely, it is a
  function of the classical fields that are independent of
  spacetime. 
  It can be calculated perturbatively, and
receives both zero-temperature and finite-temperature quantum corrections.
   The effective potential is then constructed as an object
  that includes quantum corrections in such a way that it can be
  minimized to give the vacua of the theory at any order in perturbation theory.
  Thus the effective potential allows an extension of the simple
  geometric picture of the tree-level potential --- in which the shape of the potential shows whether a symmetry is spontaneously broken --- to higher
  orders.
We review
the contributions to it in the following subsections.

\subsection{Tree-level potentials}
\label{sec:V0}
The zeroth-order contribution to the effective potential is the
tree-level scalar potential, $V_0(\{\phi_i\})$.  This is just the
classical scalar potential of the Lagrangian, i.e.\ the negative sum of
non-derivative Lagrangian interactions involving only scalar fields and represents the free energy density of ``constant'' scalar
fields $\phi_i$. The minima therefore correspond to the
vacua at this tree-level approximation.  In $d = 4$ dimensions, requiring renormalizable
interactions restricts the potential to no higher powers of
fields than quartic, meaning that the most general renormalizable
tree-level scalar potential for a model involving a set of real scalar
fields $\{\phi_i\}$ where $i=1 \ldots N$ may be written as
\begin{equation}
V(\{\phi_i\}) = \Lambda^4 + t_i \phi_i + \frac{1}{2} m^2_{ij}\phi_i\phi_j + \frac{1}{6} \kappa_{ijk}\phi_i\phi_j\phi_k + \frac{1}{24}\lambda_{ijkl}\phi_i\phi_j\phi_k\phi_l .
\label{eq:general_potential}
\end{equation}
The constant $\Lambda^4$ has no impact when exploring the shape of
the potential and the associated minima.\footnote{While only the
  energy difference matters in quantum field theory (QFT), the energy density in the
  zero-temperature ground state acts as a cosmological constant,
  affecting the expansion of the Universe. This is discussed in
  \cref{sec:cosmicExpansion}.}  In concrete models the Lagrangian must
respect the underlying symmetries and this will further restrict the
form of this scalar potential. Linear coefficients vanish when defining fields relative to the
minimum. More generally, there is a freedom
to shift the fields such that it only results in a redefinition of
the parameters.  This can be used to eliminate one parameter, such that
the linear $t_i$ coefficients are rarely present\footnote{Though in
  some cases the authors may choose to eliminate another term instead.
  See e.g.\ Section~2.1 of \refcite{Espinosa:2011ax} for a discussion of
  this in the scalar singlet extension of the Standard Model.} even
when they are not forbidden by the symmetries.

When investigating the structure of the tree-level potential, the
fields can be treated classically, using a simple geometric 
interpretation of the shape of the potential.  Stationary points are at field
locations where single derivatives of the tree-level potential, with
respect to each field $\phi_i$, vanish. That is, where
\begin{equation}
\label{eq:tree-level-stat-points}  
\frac{\partial V(\{\phi_i \})}{\partial \phi_i} = 0, \;\;\; \forall \, \phi_i.  
\end{equation}
The nature of the stationary points can be determined by constructing the
Hessian from second derivatives, and evaluating at the field values
$\phi_i=\phi_i^s$ of a stationary point $s$,
\begin{equation}
\left. \frac{\partial^2 V(\{\phi_i \})}{\partial\phi_i\partial\phi_j}\right|_{\phi_i = \phi_i^s} ,
\end{equation}  
and applying the second derivative test.  In this way the local minima
of the potential may be identified and provide a tree-level estimate
of the vacua.

\begin{figure}[t]
    \centering
    \includegraphics[width=0.7\linewidth]{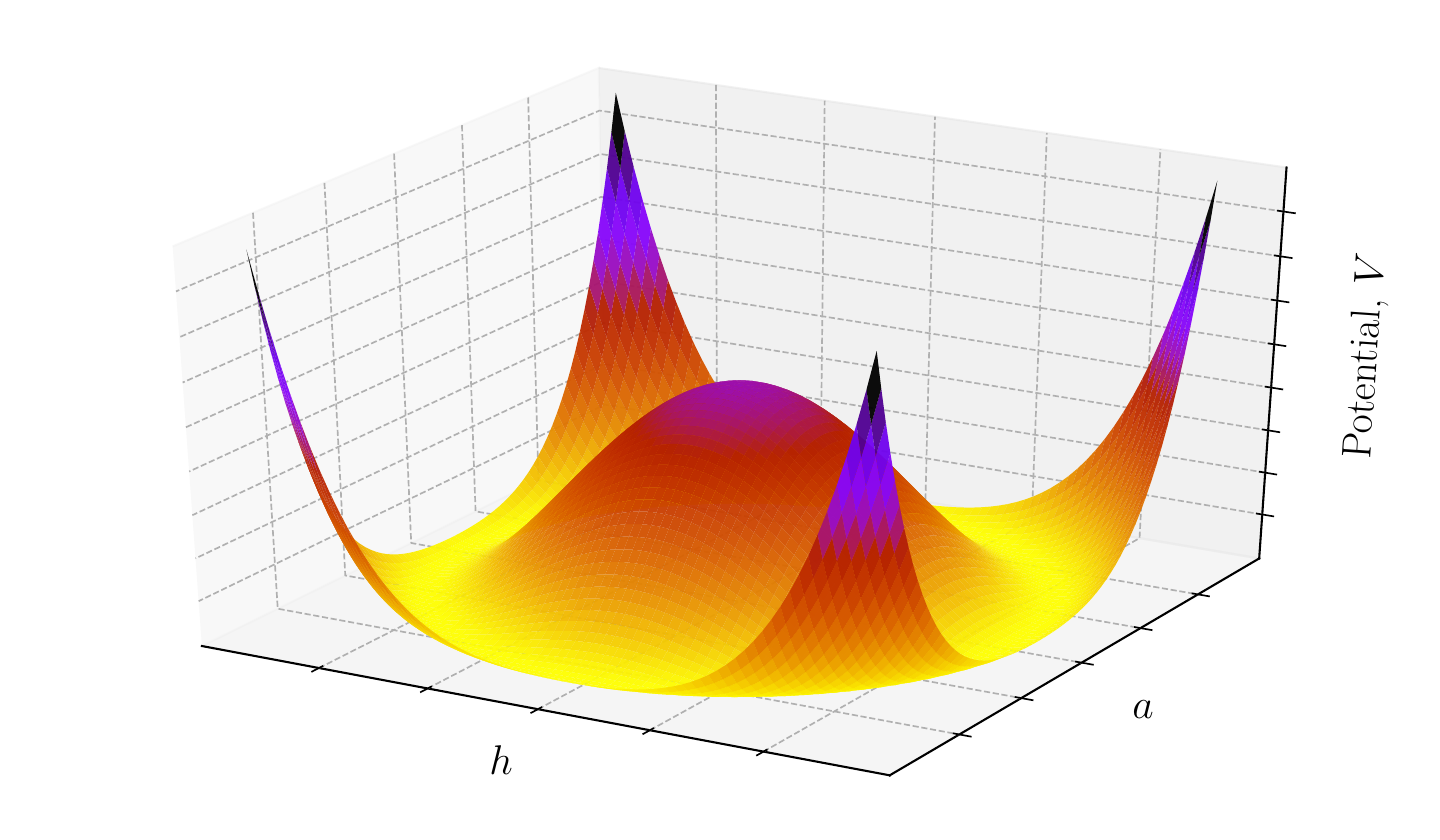}
    \caption{A two-dimensional Higgs potential with a circle of degenerate minima. A particular vacuum is chosen, breaking the symmetry.}
    \label{fig:higgs}
\end{figure}

For example, unless the electroweak symmetry is broken radiatively, at
zero temperature the tree-level potential should have a stable minima
located at a non-zero value of the Higgs field.  In the Higgs
mechanism the origin is destabilized at zero temperature, reaching a
local maximum along the direction of a real,
neutral scalar field (or fields) that plays the role of the Higgs, and
gives rise to the well known Higgs potential shown in
\cref{fig:higgs}. 

The location of the minima represent vacua and are referred to as the
vacuum expectation values (VEVs) of the field, $\langle \phi_i
\rangle$.   To describe physics in a vacuum where
one or more scalar fields have non-zero values, we can shift the
definition of the field to be about the VEV,
\begin{equation}
  \phi_i^\prime \coloneqq \phi_i - \langle \phi_i\rangle.
\label{eq:shift_field}
\end{equation}
The masses\footnote{Here we refer to the scheme-dependent tree-level
  masses.  If the effective potential is only evaluated at tree level
  this may be sufficient. However, if one is calculating the effective
  potential at higher orders as we discuss in the following subsections, then
  at least for the Higgs potential in realistic models, one should
  make sure the parameters appearing in the tree-level potential are
  extracted from (or are used to calculate) the measured Higgs pole mass at
  the same order for consistency.} can be obtained either by inserting
\cref{eq:shift_field} into the Lagrangian and carefully expanding
terms about the minima, or by evaluating double derivatives at the
minima,
\begin{equation}\label{eq:tree_level_mass}
  m_{\phi_i\phi_j}^2  = \left.\frac{\partial^2 V(\{\phi_i \})}{\partial\phi_i\partial\phi_j}\right|_{\phi_i=\langle \phi_i\rangle}  = \left\langle\frac{\partial^2 V(\{\phi_i \})}{\partial\phi_i\partial\phi_j}\right\rangle_0 ,
\end{equation}
where the Lagrangian parameters and VEVs satisfy the constraints
\begin{equation}
\left\langle\frac{\partial V(\{\phi_i \})}{\partial \phi_i}\right\rangle_0 = 0 \;\;\; \forall \, \phi_i,
\end{equation}
 and we have adopted the convention that any quantity sandwiched
between angled brackets with a zero subscript, $\langle \cdot \rangle_0$, is to be evaluated
at a zero-temperature vacuum.

Since we are ultimately interested in phase transitions between vacua, we are particularly concerned here by the number of vacua and their locations. As a simple example, let us consider $\phi^4$ theory. In this model we have a tree-level potential
\begin{equation}
\label{eq:phi_four}  
V_0(\phi)=\frac{1}{2} m^2 \phi^2+\frac{\lambda}{4!}\phi^4
\end{equation} for a real scalar
field $\phi$, with a quadratic field coefficient $m^2$ and a
dimensionless quartic coupling $\lambda$. Since the potential should
be bounded from below, we require $\lambda \ge 0$. If $m^2 > 0$, the
potential contains a single minimum at the origin, and the mass of the
scalar is just $m_s^2 = m^2$. If, on the other hand, $m^2 < 0$, the minimum
lies at $\langle \phi \rangle_0 = \sqrt{-6 m^2 / \lambda}$ and the mass
is given by $m_s^2 = \frac13\lambda \langle\phi^2\rangle_0$.  We may
also define a field-dependent mass,
\begin{equation}
m^2_s (\phi) \coloneqq m^2 + \frac{1}{2}\lambda \phi^2 ,
\label{Eq:field_dep_mass}
\end{equation}
where $m^2 = -\frac{\lambda}{6}\langle \phi^2 \rangle_0$ is still
satisfied. This is the tree-level mass in \cref{eq:tree_level_mass}
evaluated at an arbitrary field coordinate and will be useful for
studying the shape of the potential at the loop level.

\begin{figure}[t]
    \centering
    \includegraphics[width=0.6\linewidth]{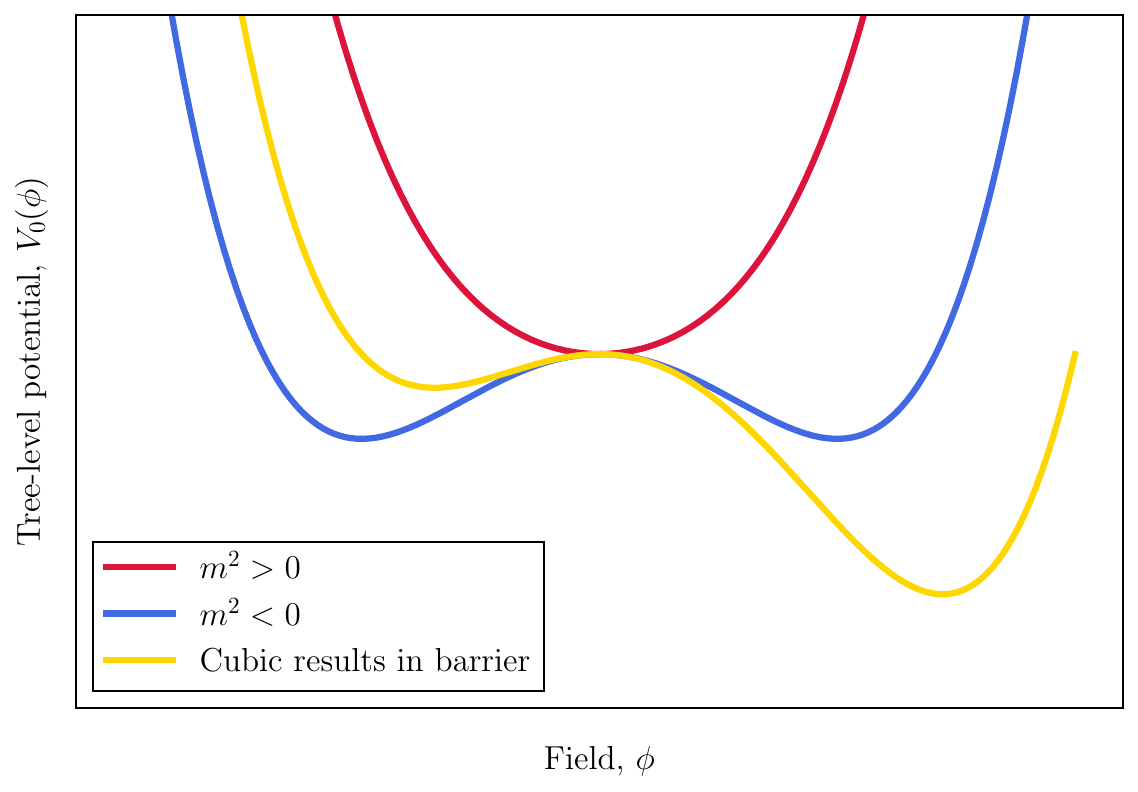}
    \caption{Possible shapes of the tree-level potential. The red line
      shows the shape of the potential given in \cref{eq:phi_four}
      when the parameter $m^2 > 0$, where in this case the minimum is
      located at the origin. The blue line shows the shape of the same
      potential when $m^2 < 0$, this is the well-known Mexican hat
      shaped potential of the Higgs mechanism with degenerate minima
      at non-zero values of the field $\phi$.  Finally the yellow line
      shows a potential where there are two non-degenerate minima at
      different field locations and a barrier separating them, which
      is a shape that can appear already at tree-level when there is a
      cubic term in the potential.  }
    \label{fig:phi_four}
\end{figure}

Although the location of the vacuum
changes to non-zero values when $m^2<0$, there remains only one physically distinct
vacuum.  Even at tree level, however, we can write potentials with
more than one physically distinct vacuum that are separated by a
barrier. This may be achieved by adding a cubic term to the model,
\begin{equation}
  V_0(\phi)=\frac{1}{2} m^2 \phi^2 + \kappa \phi^3 + \frac{\lambda}{4!}\phi^4.
  \label{Eq:V0cubic}
\end{equation}
These possibilities are shown in \cref{fig:phi_four}. In general, though, we could be misled by considering only the tree-level potential, as the structure of the potential at loop level could be qualitatively different from that at tree level. Besides shifting tree-level minima, loop corrections could result in new minima and barriers between them. This is especially true when symmetries forbid tree-level barriers, as is the case in the SM. We must therefore  consider the impact of loop corrections on the tree-level potential. The resulting potential is called the effective potential. As we shall see, the beauty of the effective potential is that we can in principle obtain the true vacua by the same minimization procedure as in the simple tree-level case, without any approximation. In practice, however, the effective potential can only be calculated perturbatively to a given loop order, as we now describe in the following subsection.

\subsection{Zero-temperature radiative corrections at higher orders}
\label{sec:ZeroTempDeltaV}
In QFT, the tree-level effective potential in
\cref{eq:general_potential} will receive higher-order radiative
corrections.  These corrections may be
calculated diagrammatically in three ways: either directly from the original action
\cite{Coleman:1973jx}, through the background field method, or using
tadpole diagrams~\cite{Lee:1974fj}. Alternatively, the corrections can also be
computed using functional methods
\cite{Jackiw:1974cv,Dolan:1974gu,Coleman:1974jh,Schnitzer:1974ue}.
Computing the effective potential directly from the action involves
calculating an infinite set of diagrams with varying numbers of
external legs.  The functional method is computationally simpler
allowing one to calculate the effective potential from the vacuum
diagrams.

The direct calculation is an infinite sum of all connected one-particle
irreducible (1PI) diagrams with classical fields as external legs, at
zero external momentum. An intuitive explanation of this is given in
section 2.2.1 of \refcite{Sher:1988mj} which is roughly as follows.
The scalar potential should be the sum of all contributions to the
free energy density from non-derivative (i.e.\ non-kinetic) scalar
interactions.  At the loop level this should then include the sum of
all loop contributions to these scalar interactions. These loop
contributions will be the 1PI diagrams with any number of scalar
fields as external legs and the momentum set to zero to remove kinetic
contributions.

More formally, the effective potential of a scalar field may be obtained
as follows.  We begin from the generating functional representing vacuum
to vacuum amplitudes,
\begin{equation}
  Z[J]\coloneqq \int {\cal D}\phi \exp[i\int d^4 x({\cal L} + J \phi)], \;\;\;\;  Z[J] = \exp[iW[J]].
  \label{Eq:Zdef}
\end{equation}
The expectation value
\begin{equation}
\frac{\delta W[J]}{\delta J} = \frac{1}{Z[J]} \int {\cal D}\phi \exp[i\int d^4 x({\cal L} + J \phi)] \phi = \langle\Omega|\phi(x)|\Omega\rangle_J \equiv \overline{\phi}(x)
  \label{Eq:classical_field_def}
\end{equation}
is commonly referred to as the classical field, since we have taken a weighted average of the quantum fluctuations. We may now define the effective action through the Legendre transform of $W[J]$ with $J(x)$ and $\overline{\phi}(x)$ as conjugate variables,
\begin{equation}
  \Gamma[\overline{\phi}] \coloneqq W[J] - \int d^4x J(x) \overline{\phi}(x).
  \label{Eq:eff_action_def}
\end{equation}
When the external source is set to zero the functional derivative of the effective action vanishes, i.e.\
\begin{equation}\label{eq:vanish_gamma}
\left.\frac{\delta \Gamma[\overline{\phi}]}{\delta \overline{\phi}(x)}\right|_{J=0} = 0.
\end{equation}
Thus, \cref{eq:vanish_gamma} plays the role of an equation of motion, motivating why we called $\Gamma$ the effective action. When expanded in powers of scalar fields the effective action can be written as
  \begin{equation}
    \Gamma[\overline{\phi}] = \sum_{n=0}^{\infty}\frac{1}{n!}\int d^4x_1 \ldots d^4x_n\Gamma^{[n]}(x_1, \ldots x_n)\overline{\phi}_1(x_1) \ldots \overline{\phi}_n(x_n) ,
    \label{eq:eff_action_expansion}
  \end{equation} 
where the $\Gamma^{[n]}$ are the 1PI Green's functions. This reflects the fact that the effective action encodes all quantum effects because at tree level it generates loop amplitudes.

We now want to find a solution to \cref{eq:vanish_gamma} for a translationally invariant vacuum where
\begin{equation}
  \overline{\phi}(x) = \phi_{\textrm{cl}} \;\; \forall x. 
\label{eq:phicl}
\end{equation}  
The effective action may then be written as
\begin{equation}
  \Gamma[\phi_\textrm{cl}] = - \int d^4x V_{\textrm{eff}} (\phi_{\textrm{cl}}),
\label{Eq:Gamma_to_V}
\end{equation}
defining the effective potential. This is analogous to the classical relationship between action and potential for static field configurations. If we then do a Fourier
transformation of the effective action from position into momentum space, we can
extract the effective potential in the following form,
\begin{equation}
  V_{\textrm{eff}} = - \sum_{n=0}^\infty\frac{\phi_\textrm{cl}^n}{n!}\Gamma^{(n)}(p_i=0) .
  \label{Eq:eff_pot__def}
\end{equation}
The vanishing external momenta comes from the translational
invariance of $\phi_{\textrm{cl}}$, which introduces a Dirac delta
$\delta^{(4)}(p)$ during the Fourier transform. \Cref{Eq:eff_pot__def} reflects the fact that the effective potential encodes quantum effects at tree level, though at zero external momentum. Moreover, by \cref{eq:vanish_gamma},
\begin{equation}\label{eq:v_eff_turning_point}
\frac{\partial V_{\textrm{eff}}}{\partial \phi_\textrm{cl}} = 0.
\end{equation}
Thus $V_{\textrm{eff}}$ may be interpreted as a potential that is
minimized by the VEV of the scalar field.  Thus,
we have arrived at an equation that expresses the description in the
second paragraph of this section prior to introducing the effective
action formalism and matches our intuitive expectation.  More detailed
and pedagogical expositions of this can be found in
e.g.\ \refcite{Peskin:1995ev,Quiros:1999jp}.

 When we further expand \cref{Eq:eff_pot__def} perturbatively, this forms an infinite sum
over both the number of external legs and the loop order.  In
perturbative calculations the loop expansion is then truncated at some
point and the calculations are approximations that are only carried
out up to fixed loop order, or some fixed orders in the couplings.
At each loop order the vacua can be found by finding the minimum of
the effective potential extending the tree-level expression in
\cref{eq:tree-level-stat-points} to all loop orders.  The double
derivatives of the effective potential also correspond to the masses
of the particles when the momentum in the self-energy is set to
zero. This extends the tree-level relation
\cref{eq:tree_level_mass} to higher loop orders, though in these
cases it is an approximation since it neglects the momentum.
Nonetheless, this {\it effective potential approximation} is
widely used due to its convenience over computing self energies. In
fact, it is possible to compute all scalar vertices in this
approximation using the effective potential and these have been
presented in analytical form for a general model in
\refcite{Camargo-Molina:2016moz}.

\subsubsection{Brief review of the one-loop \MSbar calculation in the Landau gauge}
\label{sec:EffPot1LMSbarLandau}
One-loop corrections to the effective potential were first calculated
in \refcite{Coleman:1973jx,Jackiw:1974cv}.  The general form of the
one-loop corrections are well known, and have been previously
presented pedagogically in e.g.\ \refcite{Sher:1988mj,Quiros:1999jp}.
The results depend on both gauge and the renormalization scheme, with
the Landau gauge ($\xi=0$ in general $R_\xi$ gauges) and the $\MSbar$
scheme being the most common choices for these.

As a demonstration of the direct approach to calculating one-loop
corrections to the potential, we briefly review the calculation of
them at zero temperature in the $\phi^4$ model shown at tree-level in
\cref{eq:phi_four}. More details can be found in e.g.\ Section~1.2 of
\refcite{Quiros:1999jp}, which we broadly follow here.  As is quite
common in the literature, here and elsewhere in this review, we will
write the scalar fields in the effective potential as $\phi$ rather
than $\phi_\textrm{cl}$ to avoid notational clutter.

\begin{figure}[t]
    \centering
    \includegraphics[width=0.6\linewidth]{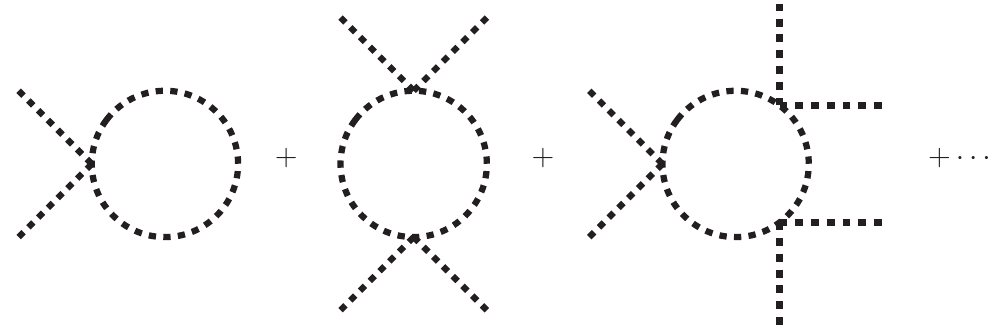}
    \caption{1PI diagrams of scalar fields contributing to the one-loop effective potential in the simplest $\phi^4$ model.}
    \label{fig:1LScalar}
\end{figure}

One-loop mass corrections are obtained by summing over all
1PI diagrams with a single loop and zero
external momenta. In $\phi^4$ we only have the scalar contributions,
the diagrams for which are shown in \cref{fig:1LScalar}. Using the Feynman rules we can obtain a general expression for the $n$-th 1PI
diagram, which has $n$ vertices, $n$ propagators and $2n$ external
legs.  After also applying Feynman rules and symmetry factors this leads to one-loop corrections of the form
\begin{align}
    \Delta V_1(\phi)&=i\sum_{n=1}^{\infty}\int\frac{d^4k}{(2\pi)^4}
    \frac{1}{2n} \left[\frac{i}{(k^2-m^2_s+i\epsilon)}\right]^n\left[\frac{-i\lambda}{2}\right]^n \phi^{2n}. 
    \label{Eq:1L_Sc_sum_ex_legs}
\end{align}
Noting that the infinite sum over $n$ is just the Taylor series of a
logarithmic function, and performing the usual Wick rotation to express
it as an integral over the Euclidean momentum $k_E = (-ik^0, k^1, k^2,
k^3)$, we get
\begin{equation}
    \label{Eq:1LScalar}
    \Delta V_1(\phi)=\frac{1}{2}\int\frac{d^4k_E}{(2\pi)^4}\log\left[1 +\frac12 \frac{\lambda \phi^2}{k_E^2 + m^2} \right].
\end{equation}
In the ultraviolet the integrand grows as $d^4 k_E / k_E^2$ and the integral diverges.

We deal with the ultraviolet divergence in \cref{Eq:1LScalar} by using dimensional regularization (DR; see e.g., \cite{Leibbrandt:1975dj}) to isolate the divergent part
in the loop integrals, leaving us with an integral over $d$
dimensions,
\begin{equation}
    \label{Eq:V_1L_DR_scalar}
    \Delta V_1(\phi)=\frac{1}{2} \mu^{4-d}
    \int\frac{d^d k_E}{(2\pi)^d}\log\left[
    1 +\frac12 \frac{\lambda \phi^2 }{k_E^2 + m^2}\right].
\end{equation}
The dimensional parameter $\mu$ is introduced to ensure the
dimension of the one-loop effective potential remains four.  The logarithm appearing in \cref{Eq:1LScalar,Eq:V_1L_DR_scalar} may be re-expressed as $\log[(k_E^2 + m^2 + 1/2\lambda \phi^2) / (k_E^2 + m^2) ]$ to show that the field dependence only appears in the numerator, and dropping the field independent part we are left with
\begin{equation}
    \label{Eq:1loopdr_phidep}
    \Delta V_1(\phi)=\frac{1}{2} \mu^{4 - d}
    \int\frac{d^d k_E}{(2\pi)^d}\log\left[
    k_E^2+m^2_s (\phi)\right],
\end{equation}
where $m^2_s (\phi)$ is the field-dependent scalar mass we introduced in \cref{Eq:field_dep_mass}.

After integrating we subtract the divergent terms with the modified
minimal subtraction scheme (\MSbar{}) to obtain the one-loop
corrections to the effective potential in the \MSbar{} scheme,
\begin{equation}
    \Delta V^{\MSbar}_1(\phi)= \frac{1}{4(4 \pi)^2} \, m^4_s(\phi) \left[
    \log\left(\frac{m^2_s (\phi)}{\mu^2}\right) -\frac{3}{2}\right].
\label{Eq:1loopms}
\end{equation}
 We note that the dimensionful parameter, $\mu$, introduced to maintain the dimension in \cref{Eq:V_1L_DR_scalar}, remains in our final expression. This loop correction to the scalar potential changes the shape and minima of the potential.

The result depends only on this field-dependent mass, with no
dependence on vertices beyond this.  This is true quite generally at
one loop, even when fermions and vector bosons can be in the loop. It
is a consequence of the effective potential diagrams at one loop
containing only vertices to the external classical field.  It can also
be understood from the functional method approach where the one-loop
topologies of the vacuum diagrams do not admit any vertices. This
makes it quite easy to obtain general expressions for the one-loop
corrections to the effective potential that depend only on masses of
the new particle.

In general the one-loop corrections to the effective potential at zero
temperature include contributions from scalars, fermions and gauge
bosons. For example in the \MSbar\ scheme and the Landau gauge, the effective potential may be decomposed as
\begin{equation} V^{\MSbar{},\text{Landau}}_{1,\, T=0} =  V^{\MSbar{}, \, \textrm{scalar}}_{1, \, T=0} + V^{\MSbar{}, \, \textrm{fermion}}_{1, \, T=0} + V^{\MSbar{}, \, \textrm{gauge boson}}_{1, \, T=0}.
\label{Eq:Veff1LMSbarLandau}
\end{equation}
For the scalar contribution in \cref{Eq:Veff1LMSbarLandau}, we can easily generalize our result in
\cref{Eq:1loopms} to the case where we have a potential that is a
general polynomial function of $n$ scalar fields.  
Terms in the
potential with $m$-fields will contribute a vertex factor multiplied by
the $m-2$ classical external fields that must be associated with
that vertex. These terms will be handled just like the $\lambda
\phi^2/2$ term in our $\phi^4$ example in
\cref{Eq:1L_Sc_sum_ex_legs,Eq:1LScalar} such that they
contribute all the field-dependent scalar mass terms and the general result is simply
\begin{equation}
    V^{\MSbar{}, \, \textrm{scalar}}_{1, \, T=0} =\frac{1}{4(4\pi)^2} \sum_s m^4_s (\phi)\left[
    \log\left(\frac{m^2_s (\phi)}{\mu^2}\right) -\frac{3}{2}\right].
\label{Eq:Veff1Lscalar}
\end{equation}
The field dependent
masses, $m_s^2(\phi)$, may be found by taking double derivatives of the potential and
diagonalizing as described in the previous section.

Similar results can be obtained for two-component Weyl fermions and vector bosons, giving
\begin{align}
 V^{\MSbar{}, \, \textrm{fermion}}_{1, \, T=0}  = - \frac{1}{2(4\pi)^2} &\sum_f m^4_f (\phi)\left[
    \log\left(\frac{m^2_f(\phi)}{\mu^2}\right) -\frac{3}{2}\right],
\label{Eq:Veff1Lfermion}\\
 V^{\MSbar{}, \, \textrm{gauge boson}}_{1, \, T=0}  =  \frac{1}{4(4\pi)^2} &\sum_v m^4_v (\phi)\left[
    \log\left(\frac{m^2_v (\phi)}{\mu^2}\right) -\frac{5}{6}\right].
\label{Eq:Veff1Lgauge}
\end{align}
As in the scalar case, the fermions or vector bosons in the loop only
have vertices that connect to the external classical field, leading to
these very simple general expressions that depend only on the field-dependent masses.

These results may be written more compactly as
\begin{equation} V^{\MSbar{},\text{Landau}}_{1,\, T=0} =  \frac{1}{4(4\pi)^2} \sum_i (-1)^{2s_i} (1 + 2 s_i) m^4_i (\phi)\left[
  \log\left(\frac{m^2_i (\phi)}{\mu^2}\right) -k_i\right].
  \label{Eq:xi0_Eff_Pot_final}
\end{equation}
where $i$ runs over all real scalar, Weyl fermion and vector boson degrees
of freedom, $s_i$ is the spin of field $i$, and $k_i = 3/2$ for scalars and
fermions and $5/6$ for gauge bosons.

As stated earlier, this result is both gauge and scheme dependent.  For
example, in the \DRbar scheme\footnote{More precisely
  we refer to the widely used variant of the Dimensional Reduction
  renormalization scheme where the epsilon scalars are decoupled
  \cite{Jack:1994rk}, see e.g.\ \refcite{Martin:2001vx} for a
  discussion about the differences to this result in the
  \DRbar scheme \cite{Siegel:1979wq,Capper:1979ns}.}
the result is given by \cref{Eq:xi0_Eff_Pot_final} but with all
$k_i=3/2$.  In fact taking the $k_i$ as scheme-dependent constants,
\cref{Eq:xi0_Eff_Pot_final} holds for any regularization scheme based
on dimensional continuation \cite{Martin:2001vx}.


A common alternative to the \MSbar and \DRbar schemes is to use on-shell-like schemes, where renormalization conditions are defined such that the single and double derivatives of the one-loop effective potential vanish. This ensures that in the effective potential approximation for the masses,\footnote{This is equivalent to a pole mass calculation with momentum in the self-energy set to zero.} the tree-level masses can be interpreted as the physical masses. This was originally used in \refcite{Anderson:1991zb}. This can be achieved by adding an additional counterterm potential to the \MSbar or \DRbar effective potential, where the counterterms are fixed according to these renormalization conditions. Following \refcite{Basler:2018cwe} we can write a counterterm potential for a general potential with parameters $\{p_i\}$ and a set of fields  $\{\phi_k\}$ for which we consider the possibility of VEVs $\{v_k\}$ with non-zero values as
\begin{align}
  V_\text{CT} = \sum_i \frac{\partial V_0}{\partial p_i}\delta p_i + \sum_k \delta T_k(\phi_k + v_k).
\end{align}
Thus we introduce counter terms $\delta T_k$ for the tadpoles $T_k$ and $\delta p_i$ for the parameters.  The counterterms are then fixed by the renormalization conditions, although their form depends on the specific model.  At the one-loop level imposing such renormalization conditions at $T=0$ in the tree-level minimum means applying the conditions,
\begin{align}
  \left. \frac{\partial (V_\text{CW} + V_\text{CT})}{\partial \phi_i}\right|_{(v_i = v_i^\text{tree})} &= 0 \\
  \left. \frac{\partial^2 (V_\text{CW} + V_\text{CT})}{\partial \phi_i\partial \phi_j}\right|_{(v_i = v_i^\text{tree})} &= 0
\end{align}
For a detailed presentation of an on-shell like renormalization scheme in a SM extension, see e.g.\ Appendix A of \refcite{Xie:2020wzn} for the scalar singlet model. 

\subsubsection{Gauge dependence of the effective potential}
\label{sec:DeltaVT0GaugeDep}
  As stated above, the effective potential is gauge dependent.  This gauge dependence is not
surprising since the effective potential is not directly a physical quantity.  However, in
\cref{sec:gauge_indep_approaches} we discuss how gauge dependence has nonetheless
presented challenges in phase transition calculations, and we discuss the
development of methods aimed at tackling these challenges. The
  one-loop effective potential has been calculated in the linear
  $R_\xi$ gauge \cite{Fujikawa:1972fe,Yao:1973am},\footnote{Note that in \refcite{Fujikawa:1972fe} the
    $\xi$ parameter is written such that the Landau gauge is $\xi
    \rightarrow \infty$ while the unitary gauge is in the limit $\xi
    \rightarrow 0$. However, the standard way of writing the $R_\xi$
    gauge in the literature is the opposite, such that the Landau
    gauge is obtained with $\xi=0$ and the unitary gauge is obtained
    with $\xi \rightarrow \infty$.  We will adopt the latter
    convention in this review.}
  where a gauge fixing term of the form
  \begin{align}
    {\cal L}_{\text{g.f}} = - \frac{1}{2\xi_a}(\partial^\mu A_\mu^a  +i \xi\phi_i g_a t^a_{ij}\overline{\phi}_j)^2
    \label{Eq:RxiBgFgaugefix}
  \end{align}  
 is added to the Lagrangian. For illustration we have written this for a set of real gauge bosons $\{A_\mu^a\}$ for a group with generators $t^a_{ij}$ under which a set of real scalar fields $\{\phi_j\}$ transform.
The effective potential in this general gauge may be written as \cite{Patel:2011th}
\begin{equation}
\begin{split}
    V^{R_\xi}_{1,\, T=0} ={}& \frac{1}{4(4\pi)^2}\Bigg[\sum_\phi n_\phi m_{\phi}^4(\{\phi_j\}, \xi) \left(\ln\left( \frac{m_{\phi}^2(\{\phi_j\},\xi)}{Q^2}\right) - k_s\right) \\
+ & \sum_V n_V m_V^4(\{\phi_j\})\left(\ln\left(\frac{m_V^2(\{\phi_j\})}{Q^2}\right) - k_V\right)\\
- & \sum_V (\xi m_V^2(\{\phi_j\}))^2 \left(\ln\left(\frac{\xi m_V^2(\{\phi_j\})}{Q^2}\right) - k_V \right)\\
- & \sum_f n_f m_f^4(\{\phi_j\}) \left(\ln\left(\frac{m_f(\{\phi_j\})^2}{Q^2}\right) - k_f
\right) \Bigg] ,
    \label{Eq_1L_EP_Rxi}
\end{split}
\end{equation} 
where $n_\phi=1, n_V = 3, n_f = 2$ are the degrees of freedom for
scalar, massive vector boson and fermion fields and the sums over
$\phi$, $V$ and $f$ run over all real scalar, Weyl fermion and massive
vector fields.\footnote{Massless fields give vanishing contributions;
  however, if a massless vector boson needs to be included it would
  have $n_V=2$ and be included in the sum over vector states.}  Note
that one could in principle account for states with degenerate masses
in $n_\phi, n_V, n_f$, but we do not do this so e.g.\ one
must also explicitly sum over the different colors for quarks etc.  For
clarity we have explicitly written the masses as functions of the
classical scalar fields, and, where relevant, of the gauge parameter.
We have adapted the expression given in \refcite{Patel:2011th} to
include fermions and to use the renormalization scheme dependent
$k_{s,f,V}$ factors that are specified below
\cref{Eq:xi0_Eff_Pot_final} for the cases of $\MSbar$ and
$\overline{\text{DR}}$.  A pedagogical derivation of the SM
effective potential in the $R_\xi$ gauge may also be found in
Appendix B of \refcite{DiLuzio:2014bua}.

This is still not the most general form of the effective potential
however, as many other gauge choices beyond the standard $R_\xi$ gauge
are possible, and are warranted when considering the shape of the
effective potential.  In particular, the choice of gauge fixing term in
\cref{Eq:RxiBgFgaugefix} has some disadvantages. If the $\phi_i$
in that equation are fixed to the values of the field in the minimum
of the potential, as is the standard implementation in most
calculations of observables, then when the scalar fields in the rest
of the Lagrangian are allowed to fluctuate away from their values in
the minimum, the gauge fixing criteria would not cancel the kinetic
terms that introduce mixing between the scalar Goldstones and the
associated vector gauge bosons, which one would then have to take into
account.  Instead for effective potential studies it is usually
allowed to vary and is sometimes referred to as the background field
$R_\xi$ gauge to distinguish it from the standard implementation of
the $R_\xi$ gauge.  It is really this background field $R_\xi$ gauge
that we have used to derive \cref{Eq_1L_EP_Rxi}.  However, this
has another disadvantage: the gauge itself is varying when the field
fluctuates
\cite{Arnold:1992fb,Laine:1994bf,Laine:1994zq,Andreassen:2013hpa,Andreassen:2014eha}. As an alternative, a simpler gauge fixing term of the form
 \begin{align}
    {\cal L}_{\text{g.f}} = - \frac{1}{2\xi_a}(\partial^\mu A_\mu^a )^2
    \label{Eq:Fermigaugefix}
  \end{align}
 is often employed in the literature
 \cite{Arnold:1992fb,Kastening:1993zn,Laine:1994bf,Laine:1994zq,Andreassen:2013hpa,DiLuzio:2014bua,Andreassen:2014eha,Papaefstathiou:2020iag}
  and variously referred to as the Lorenz, Fermi or
 the covariant gauge. In these gauges, the ability to cancel the mixing
 between Goldstones and the gauge bosons is sacrificed in order to
 avoid having the gauge changing with the field, and was considered a
 safer approach due to concerns that the latter issue may be
 pathological \cite{Laine:1994zq}.  While this may have been too
 conservative, the background field $R_\xi$ gauge is not
 renormalization group invariant \cite{Martin:2018emo}. On the other
 hand the covariant gauge suffers from $\xi$-dependent infrared
 divergences that mean the perturbativity may be restricted to very
 low values of $\xi$ in the broken phase \cite{Laine:1994zq}, while
 the background field $R_\xi$ gauge does not appear to suffer from
 these \cite{Kripfganz:1995jx}. The gauge dependence in this covariant
 gauge and $R_\xi$ gauge were recently studied and compared in
 \refcite{Athron:2022jyi}.

A much more general set of gauges which includes both of these as
special cases was considered in \refcite{Martin:2018emo}, where
readers can find useful commentary on the advantages and disadvantages
of different special cases.  There the effective potential is
presented at two-loop in a gauge which generalizes the background
field $R_\xi$ gauge in such a way that it is renormalization group
invariant \cite{Martin:2018emo}, while the two-loop effective
potential loop functions it is expressed in terms of are actually applicable
to any of the gauges they consider. 

\subsubsection{Fixed loop order calculations beyond one loop}
The Standard Model of particle physics (SM) has been the
minimal description of known data for a long time, and understanding
the structure of its electroweak symmetry-breaking Higgs potential or,
e.g.\ whether it can have a successful baryogenesis mechanism, has been of
great interest and significance.  Therefore, the SM
effective potential has been subjected to a great deal of scrutiny
over the years and in particular it has been calculated at quite high
orders in perturbation theory.  The two-loop effective potential for
the SM itself in the \MSbar\ scheme was first presented in the Landau
gauge in \refcite{Ford:1992pn}. This was augmented with leading-order three-loop
corrections involving the strong coupling, $g_s$, and top Yukawa
coupling, $y_t$ \cite{Martin:2013gka}, with the complete three-loop
potential derived (as a special case) in
\refcite{Martin:2017lqn}. Additional leading-order four-loop
contributions from QCD were obtained and presented in
\refcite{Martin:2015eia}. Two-loop calculations of the effective potential in a three-dimensional
effective field theory (see also the discussion in
\cref{sec:3defftheory}) that capture the relevant features of the SM
in the high-temperature limit have also been presented.  These were
performed the Fermi gauges in \refcite{Laine:1994zq} and the
(background) $R_\xi$ gauges in \refcite{Kripfganz:1995jx}.

For general renormalizable theories the two-loop \MSbar effective
potential in the Landau gauge was presented in
\refcite{Martin:2001vx}, almost ten years after it was first
presented in the SM.  Recently in \refcite{Martin:2018emo} this was
extended to a general gauge, as already mentioned in
\cref{sec:DeltaVT0GaugeDep}.  The loop order in the Landau gauge was
also improved upon with the three-loop effective potential in the
Landau for a general renormalizable theory
\cite{Martin:2017lqn}. 

\subsection{Finite-temperature corrections}
\label{sec:VT}
The thermal effects play an important role in the study of the
cosmological phase transitions.  For example, the shape of the
potential can be substantially changed such that the Mexican hat
potential with a non-zero minimum at zero temperature is washed out at
high temperature, leaving a global minimum at the origin, with the
implication being that electroweak symmetry (and/or other symmetries)
may be restored at high temperature
\cite{Kirzhnits:1972iw,Kirzhnits:1976ts,Weinberg:1974hy,Dolan:1973qd}.

To explore this we must rely on finite-temperature field theory.
While QFT describes how the fundamental
particles interact through fundamental forces at zero temperature,
finite-temperature field theory gives the behavior when the
background temperature cannot be neglected. QFT is an excellent tool
for predictions in particle physics experiments such as the LEP,
Tevatron and LHC particle colliders, while thermodynamics describes
the properties of large ensemble systems, through parameters such as
the temperature, entropy, pressure, volume and the
chemical potential,
and can best be understood through statistical mechanics as aggregate
properties arising from the averaging over the effects of microscopic
interactions of the basic constituents.  Finite-temperature field
theory \cite{Matsubara:1955ws,Kubo:1957mj,Martin:1959jp,Schwinger:1960qe, Keldysh:1964ud,Bernard:1974bq} is
the marriage between these two approaches where the quantum processes
of interest do not take place in a cold vacuum but instead within a
background of many interactions that can be described as a heat bath
of temperature $T$.  While QFT remains the fundamental description of
nature, it is not tractable to describe all interactions that create a
finite-temperature background in a fine-grained manner using
it. Instead the zero-temperature quantum field calculations are still
used in the calculations of the processes of interest, but with the
additional interaction to the surrounding thermal heat bath treated
thermodynamically, i.e.\ in a coarse-grained manner that just uses
aggregate properties of the system.

There are many good pedagogical finite-temperature field theory reviews
\cite{Brandenberger:1984cz, Landsman:1986uw, Quiros:1994dr, Landshoff:1998ku, Quiros:1999jp, Zinn-Justin:2000ecv, Lombardo:2000rs, Das:2000ft, Laine:2016hma} and classic textbooks
\cite{Kapusta:1989tk,Bellac:2011kqa,Kapusta:2006pm}.  A detailed
pedagogical introduction or review for finite-temperature field theory
is well beyond the scope of this review, and for which we refer the
reader to the aforementioned texts.  Instead here we will briefly
summarize key points that lead to the finite-temperature corrections
to the effective potential before presenting the well-known results.
These corrections were first looked at in
\refcite{Kirzhnits:1972iw,Kirzhnits:1976ts,Weinberg:1974hy,Dolan:1973qd},
but we will mostly follow the presentation given in the pedagogical
\refcite{Brandenberger:1984cz,Quiros:1999jp} where more detailed
steps can be found.

The thermodynamic state of a system is governed by the canonical partition function, ${\cal Z} = \Tr e^{-\beta H}$.  From this we can calculate the quantities that characterize the thermodynamic state of the system, such as the entropy, average pressure and energy.
The partition function can be represented by a path integral and the effective action can be determined from it in finite-temperature field theory in a manner very reminiscent of the relations in \cref{sec:ZeroTempDeltaV}, with
\begin{align}
  {\cal Z} = \exp{i {\cal W}} \;\; \text{and} \;\;  \overline{\phi}(x) = \frac{\delta {\cal W}[J]}{\delta J(x)} 
\end{align}
for a temperature-dependent effective action given by
\begin{align}
  \Gamma^{T}[\overline{\phi}] &= {\cal W}[J] - \int d^4x \overline{\phi}(x) J(x),
\end{align}
which satisfies
\begin{equation}\label{eq:vanish_gamma_finite_t}
\left.\frac{\delta \Gamma^T[\overline{\phi}]}{\delta \overline{\phi}(x)}\right|_{J=0} = 0.
\end{equation}
Once again for a translation-invariant vacuum where, $\overline{\phi}(x) = \phi_\text{cl}$ for all $x$, we get
\begin{align}
  \Gamma^T[\phi_\text{cl}] = - \int d^4 x V^T_\text{eff}(\phi_\text{cl}) ,
\end{align}  
where $V^T_\text{eff}$ is the finite-temperature generalization of
 the effective potential. In analogy to the relationship between free energy and the partition function in statistical mechanics, $e^{-\beta F} = \cal Z$, $V^T_\text{eff}$ plays the role of the free energy density. As expected by the second law of thermodynamics and the principle of minimum free energy, by \cref{eq:vanish_gamma_finite_t}, at equilibrium, the system lies in the state that minimizes the free energy. In other words, the location of the vacua may be found as stationary
 points of $V^T_\text{eff}$ and temperature-dependent masses may be
 approximated as double derivatives of $V^T_\text{eff}$ in the
 effective potential approximation where we neglect the momenta in the
 self-energy.

There are several different, but equivalent ways of evaluating the
path integral in finite-temperature QFT. Among them,
imaginary-time (or Matsubara) \cite{Matsubara:1955ws} and
real-time formalisms
\cite{Schwinger:1960qe,Keldysh:1964ud,Bakshi:1962dv,Bakshi:1963bn,Matsumoto:1982ry,Matsumoto:1984au,Landsman:1986uw}
are often used in the calculations~\cite{Das:2000ft}. In the former,
the dynamical time is traded for the equilibrium temperature. This
method is not applicable when the system is far away from
equilibrium. In the latter, there is a time variable that describes
the dynamics of the system and a temperature variable that represents
the equilibrium temperature, but this approach is much more
complicated than the former formalism. The results we present for the
effective potential will be independent of the method used, but
intermediate steps we show will depend on this, for which we adopt the
imaginary-time formalism as it is widely used and simpler.  In this
approach the zero-temperature calculations in
\cref{sec:ZeroTempDeltaV} are modified by replacing the momentum in
the propagators with $p_\mu =(i \omega_n, \vec{k})$, where $\vec{k}$
is the usual three-momentum and $\omega_n$ are the Matsubara frequencies
with
\begin{align}
  \omega_n = \begin{cases}
   2\pi n / \beta &  \text{for bosons} \\
   (2n+1) \pi / \beta &  \text{for fermions}
  \end{cases}
\end{align}
where $\beta=T^{-1}$. In calculations we replace integrals over $p^0$
with sums over Matsubara modes,
\begin{align}
  \int \frac{dp^0}{2\pi} \rightarrow \frac{1}{\beta} \sum_n \, ,
\end{align}
and impose conservation of Matsubara frequency and three-momentum at each vertex. With these adaptions, evaluating scalar loops as in \cref{fig:1LScalar} at finite temperature in the imaginary-time formalism gives one-loop corrections to the effective potential of the form
\begin{equation}
    \label{Eq:1L_Scalar_T}
    V^S_{1}(\phi)=\frac{1}{2\beta} \sum_{n=-\infty}^{\infty}
    \int \frac{d^3 k}{(2\pi)^3}\log(\omega_n^2+\omega^2),
\end{equation}
with 
\begin{equation}
    \label{Eq:omega_shifted}
    \omega^2= |\vec{k}|^{2}+m^2_s (\phi),
\end{equation}
where $\omega_n = 2\pi n / \beta$
are the bosonic Matsubara frequencies and $\phi$ is the classical field though once again we omit the subscript in this part to avoid notational clutter. The sum over Matsubara frequencies, $\omega_{n}$, is ultraviolet-divergent, but the divergent part is independent of the classical field $\phi$ contained in $\omega$, so the finite part contains all the $\phi$ dependence.  To handle the infinite sum and extract this crucial information, we use a trick that was first presented in \refcite{Dolan:1973qd} to rewrite \cref{Eq:1L_Scalar_T}.  The trick allows us to replace the infinite sum as
\begin{align}
 v(\omega):=\sum_{n=-\infty}^{\infty} \log(\omega_n^2+\omega^2) \rightarrow 2\beta\left[\frac{w}{2}+\frac{1}{\beta}\log\left(
    1-e^{-\beta\omega}\right) \right] ,
\end{align}
where the replacement is dropping terms independent of $\phi$. 

This trick is described in detail in \cref{sec:FTVStrick}, but the essence of it is that a derivative with respect to $\omega$ is taken, the derivative can then be summed correctly, before finally integrating with respect to $\omega$.  Since we  differentiate and then integrate with respect to $\omega$ the final result must have the same dependence on $\omega$, and therefore on the classical field.  However, the divergent terms appear in the $\omega$-independent integration constant which we drop.

Following this, the one-loop thermal correction of the effective potential in \cref{Eq:1L_Scalar_T} can be rewritten as
\begin{equation}
    \label{Eq:VS1_finT}
    V^S_{1}(\phi)=\int \frac{d^3 k}{(2\pi)^3}\left[\frac{\omega}{2}+\frac{1}{\beta}
    \log\left(1-e^{-\beta\omega} \right) \right],
\end{equation}
where again we note that divergent terms independent of the classical field $\phi$ have been dropped.  
Here the first term in \cref{Eq:VS1_finT} is the temperature-independent part, and as one may anticipate this can be shown to match the zero-temperature result obtained in the previous section.  This is divergent and can be handled by the usual process of renormalization to obtain the scheme-dependent result in \cref{Eq:1loopms}. On the other hand, the second term, i.e.\ the finite-temperature part, is exponentially small for $p \gg T$ and thus is free from ultraviolet divergence. Since the exponential only depends on the magnitude of the momentum $k$, the angular parts of the integral can be performed. As a result the finite-temperature contributions to the effective potential from a real scalar may be written in terms of the bosonic thermal function $J_{B}(y^2)$,
\begin{equation}
V^S_{1T}(\phi) =  T\int\frac{d^3 k}{(2\pi)^3}
    \log\left(1-e^{-\omega/T} \right)=\frac{T^4}{2\pi^2}
    J_B\left(\frac{m^2_s(\phi)}{T^2}\right),
    \label{Eq:V1BT}
\end{equation}  
where we now write this in terms of $T$ instead of $\beta$ to make the temperature dependence explicit and we have introduced the thermal function
\begin{equation}
    \label{Eq:jb}
    J_B(y^2)=\int_0^{\infty} dk\
    k^2\log\left[1-e^{-\sqrt{k^2+y^2}}\right].
\end{equation}
This function can also be expanded in the high-temperature limit ($y^2 \ll 1$), giving
\begin{equation}
\begin{split}
    \label{eq:jb_high_t_exp}
    J^{HT}_B(y^2)  ={}&  -\frac{\pi^4}{45}+
    \frac{\pi^2}{12}y^2-\frac{\pi}{6} y^{3}-\frac{1}{32}
    y^4\log\frac{y^2}{a_B}\\
    & 
    -2\pi^{7/2}\sum_{\ell=1}^{\infty}(-1)^{\ell}\frac{\zeta(2\ell+1)}
    {(\ell+1)!}\Gamma\!\left(\ell+\frac{1}{2}\right) \! \left(\frac{y^2}{4\pi^2}
    \right)^{\!\!\ell+2},
\end{split}
\end{equation}
where $a_B=16\pi^2\exp(3/2-2\gamma_E)$, $\gamma_E$ is the Euler-Mascheroni constant and $\zeta$ is the Riemann $\zeta$-function. Note that the second term in this expansion introduces terms of the form $T^2\phi^2$ to the potential, showing that the temperature can induce a mass correction, which can play a significant role in the symmetry restoration of  spontaneously broken symmetries. As suggested at the start of this section this means the cosmological phase transition could exist at high temperature in the early Universe. 

On the other hand, in the low-temperature limit ($y^2 \gg 1$),  where we perform an expansion of the $J_B$ function (\cref{Eq:jb}) in $\exp(-y)$ and $1/y$ \cite{Laine:2016hma},
\begin{align}
    \label{eq:jb_low_t_exp}
    J^{LT}_B(y^2) & =  -\sqrt{2} \, \Gamma(3/2) \, y^{\frac{3}{2}} e^{-y} \left[1+{\cal O}(1/y)+{\cal O}(e^{-y}) \right], \\ 
    & =-y^2 K_2(y)  \left[1+{\cal O}(e^{-y}) \right].
\end{align}
In the above equation, the higher order power-suppressed terms in the first line actually form an asymptotic series. Nonetheless there is also a convergent expansion in terms of modified Bessel functions of the second kind, where we include the leading term in the second line \cite{Laine:2016hma}.  If further terms from that expansion are kept one gets \cite{Laine:2016hma,Curtin:2016urg}
\begin{align}
    \label{eq:jb_low_t_exp_bessel}
    J^{LT}_B(y^2) & \approx  - \sum^{m}_{n=1} \frac{1}{n^2}y^2 K_2(yn) ,
\end{align}
where $m$ is the order the series is truncated at and gives a good approximation with only $m=2,3$ \cite{Curtin:2016urg}.  As $y \gg 1$, the Boltzmann factor, $\exp(-y)$, exponentially suppresses the thermal effects in \cref{eq:jb_low_t_exp}. Therefore the zero-temperature corrections discussed earlier give a good approximation in this limit.

One-loop finite-temperature corrections of the effective potential from fermion loops can also be derived using similar methods.  In this case, applying the imaginary-time formalism to the relevant one-loop diagrams gives
\begin{equation}
    \label{Eq:V1L_fer_T}
    V^F_{1}(\phi)=-\frac{n_F}{2\beta} \sum_{n=-\infty}^{\infty}
    \int \frac{d^3 k}{(2\pi)^3}\log(\omega_n^2+\omega^2),
\end{equation}
with
\begin{equation}
    \label{Eq:OmegaFerm}
    \omega^2=\vec{k}^{\,2}+m_f^2,
\end{equation}
where $n_F=2$ ($4$) for Weyl (Dirac) fermions and $\omega_n=(2n+1)\pi/\beta$ are the fermionic Matsubara frequencies. 

Again we have a sum over Matsubara frequencies that is divergent, but the part that depends on $\omega$, and therefore the classical field $\phi$, is finite. We use essentially the same trick as we did for the scalar parts to separate the finite part that depends on $\omega$ from the divergence.  As shown in \cref{sec:FTVStrick}, this leads to the replacement
\begin{align} v(\omega)=\sum_{n=-\infty}^{\infty} \log(\omega_n^2+\omega^2)
  \rightarrow \beta \omega + 2 \log[1 + \exp{-\beta\omega}] .
\end{align}

Making this replacement for $v(\omega)$ in the one-loop thermal correction of the effective potential in \cref{Eq:V1L_fer_T}, we can obtain
\begin{equation}
    \label{Eq:V1LF}
    V^{F}_{1}(\phi)= -n_F \int \frac{d^3
    k}{(2\pi)^3}\left[\frac{\omega}{2}+\frac{1}{\beta}
    \log\left(1+e^{-\beta\omega} \right) \right],
\end{equation}
where again we stress that this has been obtained after dropping
divergent parts from the infinite sum over the Matsubara frequencies, but
this does not alter the field dependence of the result. As with the
scalar case, the first term is temperature independent, and can be re-expressed to agree with the zero-temperature result. The second term is the temperature-dependent contribution to the effective potential, which can be expressed through the fermionic thermal function $J_F(y^2)$ as
\begin{equation}
  \label{Eq:V1FT_JF}
  V^{F}_{1T}(\phi)=
    - n_FT\int\frac{d^3 k}{(2\pi)^3}
    \log\left(1+e^{-\frac{\omega}{T}} \right)=
    -\frac{n_F T^4}{2\pi^2}
    J_F\left(\frac{m_f^2(\phi)}{T^2}\right) ,
\end{equation}
where again we write $T$ explicitly here and introduce a fermionic thermal function $J_{F}(m_f^2(\phi) / T^2)$ given by
\begin{equation}
    \label{Eq:jf}
    J_F(y^2)=\int_0^{\infty} dk\
    k^2\log\left[1+e^{-\sqrt{k^2+y^2}}\right].
\end{equation}
In the high-temperature limit ($y^2 \ll 1$), $J_{F}(y^2)$ can be expanded in $y^2$ as
\begin{equation}
\begin{split}
    \label{Eq:jf_HT_exp}
    J^{HT}_F(y^2)  ={}&  \frac{7\pi^4}{360}-\frac{\pi^2}{24}y^2-\frac{1}{32}y^4\log\frac{y^2}{a_F} \\
    & -\frac{\pi^{7/2}}{4}\sum_{\ell=1}^{\infty}(-1)^{\ell}
    \frac{\zeta(2\ell+1)}{(\ell+1)!}
    \left(1-2^{-2\ell-1}\right)
    \Gamma\left(\ell+\frac{1}{2}\right)\left(\frac{y^2}{\pi^2}
    \right)^{\ell+2},
\end{split}
\end{equation}
where $a_F=\pi^2\exp(3/2-2\gamma_E)$. In the low-temperature limit, similarly to the bosonic case, the fermionic thermal function $J_{F}(y^2)$ is approximately given by
\begin{align}
    \label{Eq:jf_LT_exp}
    J^{LT}_F(y^2) & \approx  - \sum^{m}_{n=1} \frac{(-1)^n}{n^2}y^2 K_2(yn),
\end{align}
where again $m$ is the order the series is truncated at and $m=2,3$ gives a good approximation \cite{Curtin:2016urg}.

Note that comparing \cref{Eq:jf,Eq:jb}, the two thermal functions are very similar.  The only difference is a sign that appears from the difference between Fermi-statistics and Bose-Einstein statistics. This already suggests that the result for the gauge bosons may be very similar to that of the scalars and indeed when we compute this in a similar manner to the calculations for the scalar and fermion contributions we obtain
\begin{equation}
   \label{Eq:gb_1loopT}
   V^B_{1T}(\phi)= 
   \frac{n_VT^4}{2\pi^2}
   J_B\left(\frac{M^2_{V}(\phi)}{T^2} \right) ,
\end{equation}
where the common factor outside the brackets, $n_V = 3$ for a massive gauge boson, is the degrees of freedom of the vector state. The second term involving the $J_{B}$ function is exactly the same as for scalars in \cref{Eq:jb}, except this extra factor from the degrees of freedom,  while the first term gives the usual zero-temperature contribution to the effective potential discussed in \cref{sec:ZeroTempDeltaV}.

Finally, putting the bosonic and fermionic contributions together, we can obtain the full one-loop thermal corrections to the effective potential in the Landau gauge,
\begin{align}
  \label{Eq:VT1L_Landau}
    V^{\text{Landau}}_{1T}=\frac{T^4}{2 \pi^2}\left[\sum_{i} n_\phi J_{B} \left( \frac{m_{\phi_i}^2}{T^2} \right)  +  \sum_{j}n_VJ_B \left( \frac{m_{V_j}^2}{T^2} \right)  -\sum_{l} n_f J_{F} \left( \frac{m_{f_l}^2}{T^2} \right) \right] ,
\end{align}
where, as in the zero-temperature case, $n_\phi=1, n_V = 3, n_f = 2$ are
the degrees of freedom for scalar, massive vector boson and Weyl
fermion fields, while for massless vector bosons $n_V=2$ should be
used, and the sum is over the set of real scalars $\{\phi_i\}$, gauge
bosons $\{V_j\}$ and Weyl fermions $\{f_l\}$.  For simplicity we do
not write the explicit field dependence of the masses here.

Similar to the discussion in \cref{sec:DeltaVT0GaugeDep}, the finite-temperature corrections to the effective potential also depend on the gauge, just like the zero-temperature corrections.  The finite-temperature effective potential in the $R_\xi$ gauge is \cite{Patel:2011th}  
\begin{align}
 \label{Eq:VT_Rxi} 
V^{R_{\xi}}_{1T}  &=  \frac{T^4}{2 \pi^2}\left[ \sum_i n_{\phi} J_B\left(\frac{m_{\phi_i}^2(\xi)}{T^2}\right)+ \sum_{j}n_{V}J_B\left(\frac{m_{V_j}^2}{T^2} \right)  - \frac13 \sum_{j}n_{V}J_B\left(\frac{\xi m_{V_j}^2}{T^2}\right)+ \sum_{l} n_f J_F\left(\frac{m_{f_l}^2}{T^2}\right)   \right] ,
\end{align}
and it has also been calculated in the covariant gauge
\cite{Arnold:1992fb,Kastening:1993zn,Laine:1994bf,Laine:1994zq,Andreassen:2013hpa,Andreassen:2014eha,Papaefstathiou:2020iag}.  Precision
calculations of the effective potential at finite temperature have
focused on the {\it three-dimensional effective field theory} or {\it
  dimensional reduction} approach which we will discuss in
\cref{sec:3defftheory}.

In calculations related to phase transitions, the effective potential
needs to be evaluated many times and therefore the computational cost
of such evaluations can be very important.  The $J_{B/F}$
functions can be rather expensive to evaluate, however fast evaluations of these
functions are available \cite{Fowlie:2018eiu}. Instead one may use the
analytic expression from the high- and low-temperature limits where
appropriate, or construct a function based on both limits
as is done in \refcite{Cline:1996mga}. Another approach used in public codes
\cite{Wainwright:2011kj,Athron:2020sbe} is to simply tabulate the
values of these functions since they only admit one argument.

\subsection{Imaginary contributions in the perturbative calculation}
\label{sec:imaginary_contributions}
One can easily see that the one-loop corrections to the effective
potential that we introduced in \cref{sec:EffPot1LMSbarLandau} gives
rise to imaginary contributions to the effective potential. Consider the
contributions from SM Goldstones in the loop in \cref{Eq:Veff1Lscalar}
where the Goldstones have \MSbar masses given by $m_G^2 = \mu^2 +
\lambda \phi^2$, where $\phi$ is the Higgs field and in the minimum
this vanishes because $\mu^2 = - \lambda v^2$.  If $\phi < v$ then
$m_G^2 < 0$ and the logarithm for this term in \cref{Eq:Veff1Lscalar}
will introduce an imaginary component to the effective potential.  The
Higgs itself will similarly introduce imaginary components when
$3\phi^2 < v^2$.  Imaginary components may also appear in the
finite-temperature potential when a field-dependent mass becomes
tachyonic, in this case coming from the negative mass squared entering
the $J_{B/F}$ functions of the finite-temperature potential.

This introduces two questions, one foundational and one practical:
what is the meaning of a complex-valued effective potential and how do
we handle the effective potential when it has non-zero imaginary
components?

To address these first note that the imaginary components enter due to
symmetry breaking which occurs when the potential is non-convex, i.e.\
when one can have $V''(\phi_{\text{cl}}) < 0$.  The fact that the effective
potential can even be non-convex is itself strange when one considers
its definition as the Legendre Transform of the generating functional
of the connected Green's functions
(\cref{Eq:eff_action_def}).  Since a Legendre Transform must
be either concave wherever it is defined, or convex wherever it is
defined, this should imply that the effective potential should a be
convex function\footnote{A concave potential is of course unbounded
  from below.} of the scalar field
\cite{Symanzik:1969ek,Iliopoulos:1974ur} as well as real
\cite{Fujimoto:1982tc,Weinberg:1987vp}. However, a convex potential
cannot allow spontaneous symmetry breaking \cite{Fujimoto:1982tc}, so
if the vacua we consider must always come from a convex function this
would undermine the whole topic of this review. However, the effective
potential obtained from an effective action defined by the Legendre
Transform in \cref{Eq:eff_action_def} does not directly correspond
to what we calculate perturbatively \cite{Dannenberg:1987fw} from the
1PI Green's functions.

\begin{figure}
\centering
\includegraphics[width=0.6\linewidth]{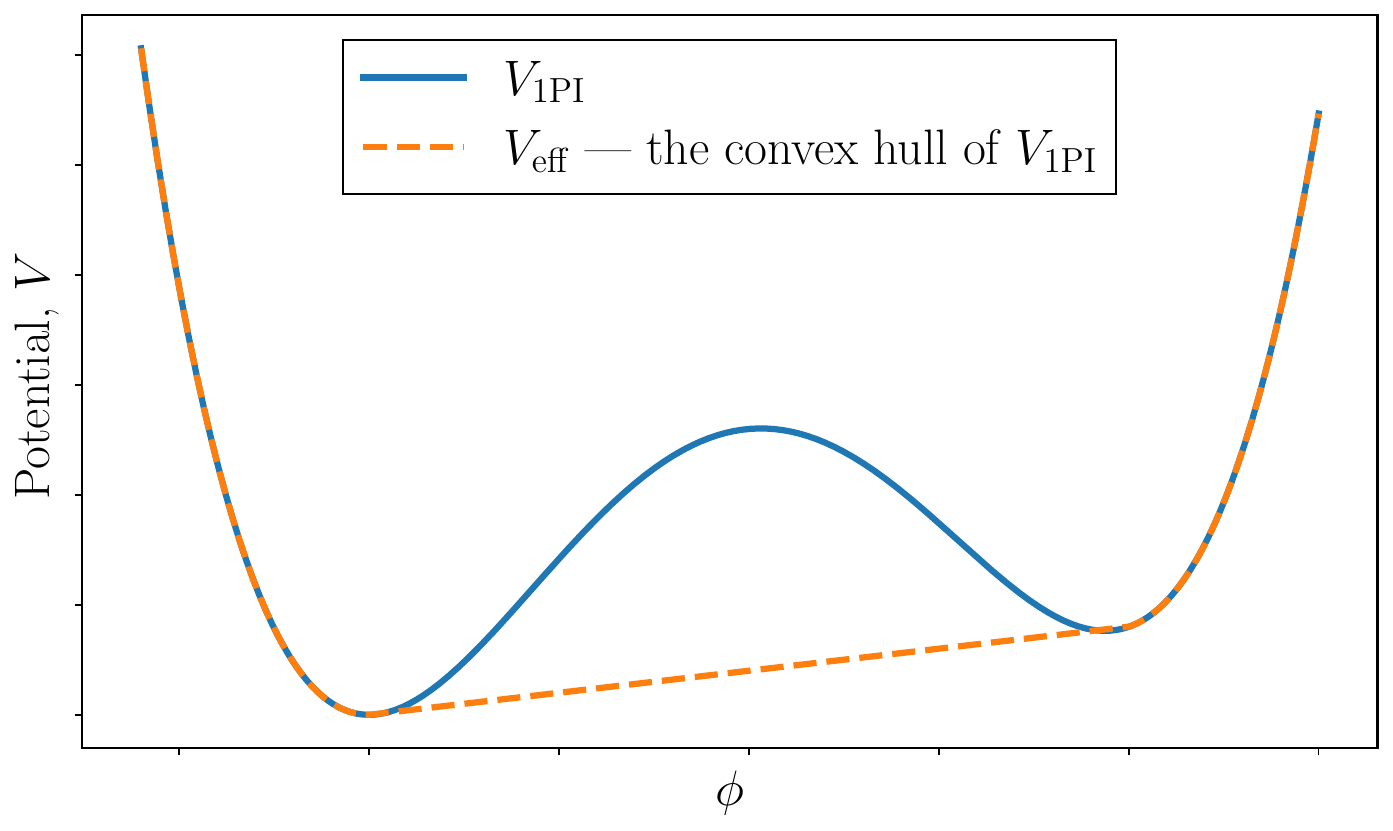}
\caption{An illustration showing how applying the Legendre Transform twice
  maps the original function to its convex envelope (or hull). In
  this example we show a non-convex function labeled $V_{\text{1P1}}$
  as a blue line, while the convex envelope of this function is shown by
  the dashed orange line and labeled $V_\text{eff}$.  This is effectively an update of Figs.\ 3 and 4 appearing in \refcite{Dannenberg:1987fw}, though for this illustration we have chosen $V_\text{1PI} = \phi^4 - 5.98 \phi^3 +12.94 \phi^2 -11.94\phi +4.98$, based on some mild modifications to the examples used for Fig.\ 1 in \refcite{Fujimoto:1982tc}.} 
\label{fig:VeffConvexHull}
\end{figure}

The connection between the two only follows when the classical field
$\overline{\phi}$ defined in \cref{Eq:classical_field_def} is a
single-valued functional of $J$ \cite{Fujimoto:1982tc}.  As pointed out in \refcite{Dannenberg:1987fw} when computing
the effective potential perturbatively we use the relation
\begin{align}
 \exp{W[J(x)] + {\cal O}(\hbar)} = N \int D\overline{\phi} \exp{\Gamma_\text{1PI}[\overline{\phi}] + \int d^4x J(x) \overline{\phi}(x)} ,
\end{align}  
which as $\hbar\rightarrow 0$ leads to
\begin{align}
  W[J] =\left[\Gamma_\text{1PI}[\overline{\phi}] + \int d^4x J(x)\overline{\phi}(x)\right]_{\delta \Gamma_{\text{1PI}}/ \delta\overline{\phi} = -J}. 
\end{align}
Therefore, $W[J]$ is in fact a Legendre Transform of
$\Gamma_\text{1PI}$, which is the generating functional of the 1PI
Green's functions. However, we defined the effective action in
\cref{Eq:eff_action_def} as the Legendre Transform of $W[J]$. For
a convex function, this is consistent, as the
Legendre Transform of the Legendre Transform returns the original function, 
\begin{equation}
W \equiv \text{LT}[\Gamma_\text{1PI}] \quad\Rightarrow\quad
\Gamma_\text{eff} \equiv \text{LT}[W] =  \Gamma_\text{1PI} .
\end{equation}
One can also
perform Legendre Transforms for non-convex functions where an
otherwise convex (concave) function has a localized bump in the
middle, like the one shown as a solid blue line in
\cref{fig:VeffConvexHull}. In this case the Legendre Transform of
the Legendre Transform does not return the original function, but
rather the convex (concave) envelope (or hull as its sometimes
described) of it,
\begin{equation}
W \equiv \text{LT}[\Gamma_\text{1PI}] \quad\Rightarrow\quad 
\Gamma_\text{eff} \equiv \text{LT}[W] = \left.\Gamma_\text{1PI}\right|_{\text{Convex hull}} .
\end{equation}
A convex hull is shown in \cref{fig:VeffConvexHull} by the
dashed red line. This convex hull must still satisfy \cref{eq:v_eff_turning_point} and so can be used to find the ground state. Being convex, however, it cannot tell us about the false vacua and so is not obviously useful for studying transitions.

Therefore, when $\overline{\phi}$ is
multivalued, $\Gamma_{\text{eff}}$ becomes the concave envelope of
$\Gamma_\text{1PI}$. Similarly it follows that we should then define $V_\text{1PI}$ as the
 negative of the spatially independent part of $\Gamma_\text{1PI}$ and
 distinguish between $V_\text{1PI}$ which we calculate perturbatively
 and can be non-convex and $V_\text{eff}$ which must be convex, and
 real, and is in fact the convex envelope of $V_\text{1PI}$.\footnote{It has also
 been suggested that $V_\text{1PI}$ is the analytic
 continuation of $V_\text{eff}$ into the region where $\phi$ is
 between the two minima \cite{Langer:1967ax,Langer:1969bc,Weinberg:1987vp}.}

Now $V_\text{eff}$ should be interpreted as the expectation value of
the energy density, $V_\text{eff} = \langle \Omega | H | \Omega \rangle$
where $|\Omega\rangle$ is a normalized state, $\langle \Omega|\Omega\rangle = 1$, that minimizes the energy, $\delta\langle
|\Omega | H | \Omega \rangle = 0$,  while satisfying
$\overline{\phi}(x) = \langle \Omega | \phi(x) | \Omega \rangle$. This state does not need to be localized and instead for a non-convex
potential it can be a superposition of multiple vacuum states located
at different field values. As a result it effectively interpolates
between the minima of the free energy density for the case where
$\overline{\phi}$ is concentrated around a single value, like the
orange dashed line in \cref{fig:VeffConvexHull}. This is further explained using purely classical arguments in \cref{App:Convex}.  When
we add an additional constraint that the state must
be localized, that is $|\Omega\rangle$ must be concentrated around field values $\phi(x) = \overline{\phi}$,\footnote{For a more precise definition of what is meant by a localized state, see Section\ II of \refcite{Weinberg:1987vp}.}   we obtain $V_\text{1PI}$. Physically we are interested in
a localized state in one of the minima some time
before the phase transition.  This  then transitions to another homogeneous
state that is localized around the other minimum. $V_\text{1PI}$  is the
correct object to consider in these two situations, and it does not
need to be convex or real. 

\Refcite{Weinberg:1987vp} then shows that the imaginary components
correspond to the decay of a spatially homogeneous quantum state which
can be unstable when the potential is non-convex.  This can be
understood because the additional restriction we have placed on
$|\Psi\rangle$, making it a localized state, does not commute with the
Hamiltonian.  Note that this decay associated with the imaginary
components of the effective potential should not be confused with the
decay of the false vacuum, nor with the decay of the scalar
fields. The former should be non-perturbative rather than arising
from imaginary components in the perturbative potential, and these
imaginary components may appear for $V_\text{1PI}$ potential around
the stable global minimum, while the latter can only happen when
decays into other fields are kinematically allowed, which would not be
the case if there is only one field.  Instead the decay from the
imaginary components of $V_\text{1PI}$ is simply, as stated above, the
decay of the unstable state representing a localized configuration of
the scalar fields.

\Refcite{Weinberg:1987vp} argues that when the times involved are
short relative to the decay rate of this state, the real part of the
effective potential can be used reliably to study its behavior. It is
further argued in the appendix of \refcite{Sher:1988mj} that when
studying the electroweak potential and phase transitions associated
with it, using the real part of the perturbative potential is
justified, but that this may not be the case with other applications.
The convexity of the potential and the work of
\Refcite{Weinberg:1987vp} also plays a role in a more recent work
\cite{Plascencia:2015pga} on gauge-independent calculations of the
false vacuum decay.  One may nonetheless still worry about these
imaginary parts, especially without a clear prescription for checking
if the decay rates from the imaginary parts are too large compared to
the time scales involved.  Therefore, it is worthwhile noting that a
resummation procedure that cures infrared divergences associated with
the Goldstone bosons when they become massless (discussed in
\cref{sec:goldstone_catastrophe}) can also remove the imaginary
components associated with those states \cite{Martin:2014bca}.  More
detailed discussion and additional commentary on these issues can be
found in the appendix of \refcite{Sher:1988mj} and in Section~4.6 of a
very recent review \cite{Devoto:2022qen}, as well as the original
references cited above.

\subsection{Zero-temperature resummations}
\subsubsection{Renormalization Group Improvement}
Perturbative calculations of the effective potential in the $\MSbar$
and $\DRbar$ schemes depend on the renormalization scale. An
appropriate choice for the renormalization scale is one which keeps
the logarithms appearing from higher order corrections small, thus
avoiding large logarithms which can appear as multiplicative factors
with each power of the expansion parameter, and thereby keeping the
perturbation series under control.

When there are multiple mass scales in the theory minimizing all large
logarithms with a single choice of renormalization scale may not be
possible.\footnote{ Note this is only a problem if the coefficient
  times the logarithm will be large, so if the perturbative expansion
  parameter is small, large logarithms may not be an issue.}  In such
cases the effective field theory method of matching and running may be
used. In these procedures an effective field theory containing only
light fields is constructed. The effects of the heavy fields are
incorporated through matching conditions performed at a scale close to
the heavy masses. After matching at this scale renormalization group
equations (RGEs) can be used to run the parameters of the effective field
theory down to the scale of the light masses. The large logarithms
are resummed using this procedure (see \refcite{Delamotte:2002vw} for a nice
pedagogical explanation of this resummation). In the context of the
one-loop effective potential light masses are normally suppressed by
from the $(\text{mass})^4$ factor, but this above procedure may still
be required for other calculations.

However when we are interested in the shape of the effective
potential, we must consider the field dependent masses, which means
that the mass scale is changing as we vary the field.  Therefore
exploring the shape of the effective potential at field values which
vary over orders of magnitude with a fixed renormalization scale can
lead to lead substantial uncertainties \cite{Kastening:1991gv}.  A
particularly extreme example of this is when people consider the
vacuum stability of the SM vacuum where a deeper minimum
exists at very large field values (see e.g.\ \cite{Coleman:1977py,Isidori:2001bm,Buttazzo:2013uya,DiLuzio:2014bua,Espinosa:2015qea,Andreassen:2017rzq,Chigusa:2017dux}). While clearly not as acute,
the scale dependence in studies of the electroweak phase transition (EWPT)
can also be very large (see e.g.~\refcite{Croon:2020cgk, Athron:2022jyi} for recent evaluations of
this uncertainty). To reduce such uncertainties one can use the
renormalization group improved effective potential where the RGEs are
used to run the couplings to the appropriate scale for different
values of $\phi$.  One-loop RGEs will resum the leading logarithms
while two-loop RGEs will resum the next-to-leading logarithms and so
on.  More details on renormalization
group improved effective potentials can be found in
\refcite{Sher:1988mj,Ford:1992mv}.

\subsubsection{Goldstone Boson catastrophe}\label{sec:goldstone_catastrophe}

The Landau gauge ($\xi=0$) is the most common and convenient choice
for calculating the effective potential and it is only in this gauge
that the effective potential is known to three-loop order
\cite{Martin:2017lqn}.  In fact it was only recently that the two-loop
effective potential became available in other gauges
\cite{Martin:2018emo}.  However, the Goldstones are massless in the
Landau gauge, which can lead to infra-red divergences
\cite{Ford:1992mv,Martin:2013gka}, an issue commonly referred to as
the ``Goldstone boson
catastrophe''~\cite{Martin:2013gka,Martin:2014bca,Elias-Miro:2014pca}.
The Goldstone boson catastrophe does not appear in the one-loop
corrections to the effective potential because the contributions are
of the form $m_G^4\ln m_G^4$ and $\lim_{m_G \rightarrow 0}m_G^4\ln
m_G^4=0$. However the Goldstone Boson catastrophe does appear directly
in the effective potential at the three-loop level and higher, and at
the two-loop level in the single derivatives of the effective
potential that appear in the electroweak symmetry breaking (EWSB)
conditions \cite{Martin:2013gka,Martin:2014bca,Elias-Miro:2014pca}.\footnote{
In Fermi gauges there is a related infrared divergence that appears at
one-loop in the EWSB conditions \cite{Nielsen:1975fs,
  Aitchison:1983ns,Loinaz:1997td,Espinosa:2016uaw}, methods for handling this are discussed in \refcite{Espinosa:2016uaw}.}

Furthermore when parameters in the tree-level effective potential are
extracted from the Higgs mass using the effective potential
approximation, the issue can appear already at the one-loop
level. While the Goldstone boson catastrophe here can be avoided
simply by calculating the Higgs mass using the full one-loop self
energy with the momentum set to the Higgs mass
\cite{Casas:1994us,Cline:1996mga}, the effective potential
approximation is widely used because it is a significantly simpler
procedure, and makes use of the elegant formalism of the effective
potential and the geometric picture mirroring the treatment of the
tree-level potential that it allows.  Indeed this may be especially
appealing when working on the effective potential itself.

\begin{figure}
  \centering
  \includegraphics[width=0.33\linewidth]{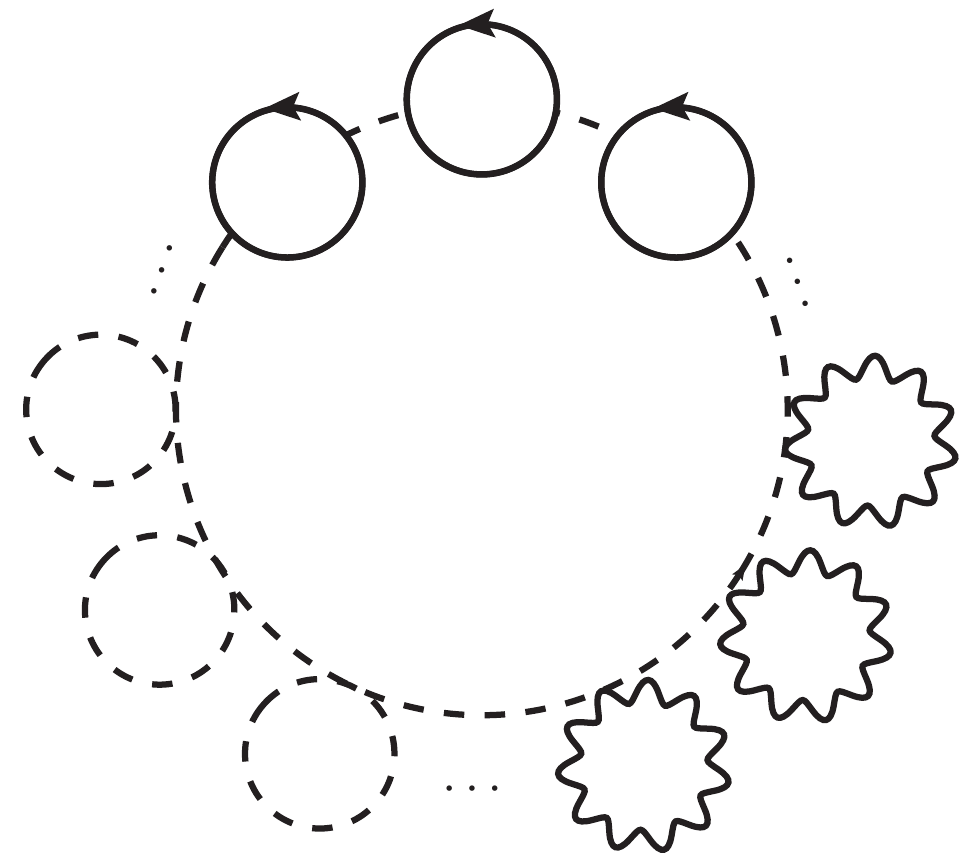}
  \caption{A ring of Goldstone propagators separated by loops that contribute to the Goldstone self-energy.}
  \label{fig:rings}
\end{figure}

The Goldstone boson catastrophe is solved
\cite{Martin:2014bca,Elias-Miro:2014pca} by resumming to all orders an
infinite series of infra-red divergent diagrams from Goldstones to
produce an infra-red finite result. The resummed diagrams give leading
contributions from Goldstones in the limit $m_G\rightarrow 0$.  At
one loop the contribution comes simply from the neutral and charged
Goldstone boson vacuum diagram, while at higher orders the
contributions appear from ring diagrams of the form shown in \cref{fig:rings}, i.e.\ a ring of Goldstone
propagators separated by subdiagrams that are the Goldstone self energies. So at the $\ell$ loop order there is a ring of $\ell - 1$
Goldstone propagators and $\ell -1$ self energies.  The self energies
should include all contributions with heavy masses or massless
contributions with large momentum (referred to as the hard part
of the self-energy in \refcite{Espinosa:2017aew}). 

For the leading behavior it is sufficient to evaluate the self
energies at zero momentum, simplifying the inclusion of the self
energies to just constant mass squared insertions $\Delta$, leading to a
summation of~\cite{Martin:2014bca}
\begin{align}
  V_\text{eff}^\text{resum} = \frac{3}{16\pi^2}\sum_{n=0}^{\infty}\frac{\Delta^n}{n!} \left(\frac{d}{d m^2_G}\right)^n f(m_G^2) = \frac{3}{16\pi^2}f(m_G^2 + \Delta) ,
\label{Eq:GoldstoneResummation}
\end{align}
where the loop function $f$ is
\begin{align}
  f(x) = \frac{x^2}{4}(\ln x/Q^2 -3/2) .
  \end{align}
The fixed-order effective potential $V_\text{eff}$  calculated at loop order $\ell$ may then be corrected to give \cite{Martin:2014bca}
\begin{align}
  \label{Eq:Veff_GSResum}
  \hat{V}_\text{eff} =  V_\text{eff} + V_\text{eff}^\text{resum} - \frac{3}{16\pi^2}\left[\sum_{n=0}^{\ell -1}\frac{\Delta^n}{n!} \left(\frac{d}{d m^2_G}\right)^n f(m_G^2)\right] .
\end{align}
Note that at one loop this reduces to
\begin{align}
  \label{Eq:V1L_GSResum}
  \hat{V}^{\text{1L}}_\text{eff} =  V^\text{1L}_\text{eff} + \frac{3}{16\pi^2}\left[f(m_G^2+\Delta) - f(m_G^2)\right] .
\end{align}

In the SM the form of the correction $\Delta$ applied is
given in Eq.~(2.3) in \refcite{Martin:2014bca} and Eq.~(16)
in \refcite{Elias-Miro:2014pca}, which is the Goldstone self-energy without
the contribution from the Goldstone itself, evaluated at zero
momentum. In extensions of the SM it should be computed
using the prescription stated above where $\Delta$ is defined as the
hard part of the self-energy.  An all orders proof of the Goldstone
resummation when $\Delta$ is defined this way is given in
\refcite{Espinosa:2017aew}. Further discussions of the Goldstone boson catastrophe and these solutions can also be found in \refcite{Pilaftsis:2015bbs,Marko:2016wtw,Pilaftsis:2017enx,Kumar:2016ltb,Braathen:2016cqe,Braathen:2017izn}.

\subsection{Finite-temperature resummations}

A significant challenge in calculating the effective potential at
finite temperature is posed by the fact that finite-temperature
perturbation theory breaks down at high temperatures~\cite{Dolan:1973qd,Weinberg:1974hy,Kirzhnits:1974as,Linde:1978px,Linde:1980ts,Gross:1980br}.
The infrared bosonic modes at high temperatures lead to a temperature
enhancement of the couplings, $g^2 \rightarrow g^2T/m $ which spoils
the perturbative expansion when $T \gg m$.  Infrared divergences also
appear directly in the effective potential at the 4-loop order
\cite{Linde:1980ts}.  Furthermore the observation of symmetry
restoration at high temperature suggests that high temperatures are
essential for understanding the cosmological phase history.

The two main methods for dealing with this are i) to resum the
leading infrared-divergent contributions in a process
known as {\it daisy resummation} and ii) to integrate out the temperature
dependence by matching to a three-dimensional effective field theory (3dEFT).

Daisy resummation can be applied as a simple modification to the
construction of the effective potential and we provide a pedagogical
description of it in \cref{sec:daisyresum}.  Constructing a three
dimensional effective field theory requires a more dramatic departure
from the procedure described in the previous sections.  However by
effectively integrating out the high-temperature dependence it
resolves the issue in a more fundamental way and as we will discuss
has additional benefits that go beyond this.  A thorough pedagogical
description of the 3dEFT approach is beyond
the scope of this review, since a proper treatment is much more
involved than for the daisy resummation.  However in
\cref{sec:3defftheory} we provide some brief details about the method
and provide references to recent literature where the details can be
found.

\subsubsection{Daisy resummation}
\label{sec:daisyresum}
Contributions from resummed `daisy' or `ring' diagrams (see
\cref{fig:rings}) that account for the leading infrared-divergent
contributions \cite{Dolan:1973qd,Weinberg:1974hy,Kirzhnits:1974as} may
be added to the effective
potential~\cite{Takahashi:1985vx,Fendley:1987ef,Pisarski:1988vd,Braaten:1989mz,Carrington:1991hz}. This
daisy resummation is quite similar to the resummation in solving the
Goldstone catastrophe discussed in \cref{sec:goldstone_catastrophe},
though this finite-temperature daisy resummation predates the
Goldstone resummation by decades and is an issue of greater severity.
If each `petal' in the daisy can be recognized as a
constant temperature-dependent mass squared insertion $\Delta_T$, then
the corrections may be resummed in a manner that is analogous to
\cref{Eq:GoldstoneResummation}.

In practice this daisy resummation is usually performed in one of two
approaches, commonly referred to as the Parwani method
\cite{Parwani:1991gq} and the Arnold-Espinosa method
\cite{Arnold:1992rz} where these approaches were used to improve the
two-loop potential with this resummation. In both methods, one
evaluates each petal in the high-temperature expansion (see
\cref{eq:jb_high_t_exp}), such that both the zero-temperature corrections and
non-leading terms from the high-temperature expansion of the
finite-temperature self-energy are neglected. The temperature-corrected masses obtained in this manner are referred to as the
`thermal' masses or Debye masses.

The temperature-dependent mass corrections, $\Delta_T$, are calculated
from the high-temperature expansion of the finite-temperature self
energies for diagrams with zero Matsubara modes in the limit of
vanishing momenta.  Similar to tree-level and zero-temperature
corrections the thermal corrections to the scalar masses can also be
easily obtained from double derivatives of the high-temperature
expansion of $V_{1T}$.  Only thermal mass corrections at the lowest
order in $m/T$ are included, which are of the form $\Delta_{T,i} = c_i
T^2$ where the coefficients $c_i$ depend on the dimensionless
couplings.  All scalar and vector bosons appearing in the one-loop
corrections to the potential contribute zero Matsubara modes in these
petals and thus receive corrections of this form. This includes the
photon mass which picks up a temperature-dependent mass at finite
temperature. However it does not include fermion masses, as the
fermions do not contribute the problematic zero Matsubara modes and
also do not get temperature corrections at this order as they are
protected by the chiral symmetry, nor does it include the ghosts (see
Appendix~A of Section~5 in \refcite{Laine:2016hma} for
this). Furthermore for the vector bosons it is only the longitudinal
mass that obtains such corrections, splitting the longitudinal and
transverse masses.  The transverse masses are protected by gauge
symmetry such that the dominant high-temperature correction varies
like $g^2mT$ rather than $g^2T^2$ for the longitudinal mode (see
e.g.\ equations (15) and (16) in \refcite{Espinosa:1992kf}).

In the Parwani method at one-loop order the resummation is simply
achieved by replacing the tree-level masses appearing in the $J_{B}$
functions with the thermal masses that include the corrections
$\Delta_{T,\tilde{B}}$ for each contributing boson $\tilde{B}$.  Thus, all the
  scalar and vector boson masses in $V_T$ include thermal corrections,
\begin{align}
  \label{Eq:VeffParwani}
  \text{Parwani:} \;\;\;\; V^\text{1L,Par}_\text{eff}(\{\phi_i\},T) = \left.[V_0(\{\phi_i\}) + V_{1,T=0}(\{\phi_i\}) + V_{1T}(\{\phi_i\},T)]\right|_{m_{\tilde{B}}^2 \rightarrow m_{\tilde{B}}^2 + \Delta_{T, \tilde{B}}} ,
\end{align}
where the set of masses $\{m_{\tilde{B}}\}$ should be all scalars and longitudinal vector boson masses appearing in the
one-loop corrections to the potential.  $V_0$ is the tree-level
potential discussed in \cref{sec:V0}, $V_{1,T=0}$ are the one-loop
zero-temperature corrections discussed in \cref{sec:ZeroTempDeltaV}
and $V_{1T}$ are the one-loop finite-temperature corrections given in
\cref{sec:VT}. We do not specify the gauge here, as one should be free
to choose which gauge to apply this procedure in and the details of
the resummation do not change.
 
The one-loop Parwani procedure given in \cref{Eq:VeffParwani} is very
similar to the one-loop Goldstone resummation of
\cref{Eq:V1L_GSResum}.  Furthermore, similar to the more general
expression for Goldstone resummation, \cref{Eq:Veff_GSResum}, which is
valid at higher orders, in the Parwani method one must take care to
consistently cancel the higher corrections that are already included
by the resummation procedure to avoid a double counting of these
contributions.  This procedure was demonstrated at the two-loop level for scalar
$\phi^4$ theory in \refcite{Parwani:1991gq}.

The Arnold-Espinosa procedure is an alternative prescription for
including the resummed corrections that was proposed in
\refcite{Arnold:1992rz} and which may be simpler to apply at higher
orders.  In this prescription, at the one-loop level one first performs a
high-temperature expansion on the finite-temperature potential $V_{1T}$
using \cref{eq:jb_high_t_exp}.\footnote{We only need to consider the
  high-temperature expansion of bosonic contributions since the
  fermions do not get a thermal masses at one loop.}  The masses in
this expression are then replaced with the thermal masses that were
also calculated in the high-temperature limit.  In general the
precise form of the corrections depends on the
gauge.  For simplicity we will first do this in the Landau gauge starting from
\cref{Eq:VT1L_Landau} and substituting in \cref{eq:jb_high_t_exp}.

Thus, we obtain the high-temperature contributions of the {\it
  contributing} bosonic modes,\footnote{A high-temperature expansion of
  the fermion contributions may also be carried out but since we only
  resum scalar and longitudinal gauge boson contributions here we
  will omit this.}
\begin{align}
  \label{Eq:VHTB}
  V^\text{HTB}_\text{1T,Landau}(\{\phi_i\}, T) &= -\frac{T^4 \pi^2}{90}N + \frac{T^2}{24}\sum_{\tilde{B}} m_{\tilde{B}}^2(\{\phi_i\}) - \frac{T}{12\pi}\sum_{\tilde{B}}m^3_{\tilde{B}}(\{\phi_i\}) + {\cal O}(m_{\tilde{B}}^4) ,
\end{align}
where the sums are over the set of all scalar and longitudinal bosons and  $N$ is the total number of these contributing bosonic degrees of freedom\footnote{If we also kept the transverse bosonic modes then it would instead be  $N=\sum_{\phi_i}n_\phi + \sum_{V_j}n_V = N_\phi n_{\phi} + N_V n_V$, i.e.\ the total number of bosonic degrees of freedom.}, which is just the number of real scalars plus the number of vector bosons since there is just one longitudinal mode per vector boson.  
We then replace the tree-level mass with the thermal mass,  
\begin{equation}
\begin{split}
  \label{Eq:VHTBresum}
  V^\text{resum}_\text{1T,Landau}(\{\phi_i\}, T) ={}& -\frac{T^4 \pi^2}{90}N + \frac{T^2}{24}\sum_{\tilde{B}}(m_{\tilde{B}}^2(\{\phi_i\}) + \Delta_{T, \tilde{B}}) \\
  &- \frac{T}{12\pi}\sum_{\tilde{B}}(m^2_{\tilde{B}}(\{\phi_i\}) + \Delta_{T, \tilde{B}})^{3/2} + {\cal O}(m_{\tilde{B}}^4) .
\end{split}
\end{equation}

The first term is field independent, the second
term includes the $T^2 m^2$ contribution that usually leads to
symmetry restoration at high temperature and an additional
field-independent contribution.  Finally, we have the third term that provides
field-dependent modifications to the potential from resumming the
high-temperature corrections.  At one loop if one is only interested in the
high-temperature limit then this could be directly used as a
replacement of $V_{1T}$. This is the basic procedure reviewed in
Section\ II of \refcite{Arnold:1992rz}.  However, to preserve the
one-loop precision at low temperatures one can improve upon this by
keeping the original $V_{1T}$ and then adding resummed corrections
on top. One must then explicitly subtract the contributions from the
zero-temperature masses that are already accounted for in $V_{1T}$ to
avoid double counting them, giving
\begin{align}
  \label{Eq:Vring}
  V^\text{Landau}_\text{daisy}(\{\phi_i\}, T)  &= V^\text{resum}_{1T}(\{\phi_i\}, T)  - V^\text{HTB}_{1T}(\{\phi_i\}, T)\\
&= -\frac{T}{12\pi} \sum_{\tilde{B}} \left[ 
    \left(m_{\tilde{B}}^2(\{\phi_i\}) + \Delta_{T,\tilde{B}}\right)^{3/2} -  \left(m_{\tilde{B}}^2(\{\phi_i\})\right)^{3/2}
    \right] + \frac{T^2}{24} \Delta_{T, {\tilde{B}}} ,
\end{align}
where the field-independent term at the end is usually neglected.
Therefore in the Arnold-Espinosa approach to daisy resummation the
effective potential is given by
\begin{equation}
\begin{split}
  \label{Eq:VeffAE}
  \text{Arnold-Espinosa:} \;\;\;\; V^\text{1L,A-E}_\text{eff}(\{\phi_i\}, T) ={}& V_0(\{\phi_i\}) + V_{1,T=0}(\{\phi_i\}) + V_{1T}(\{\phi_i\}, T)\\
  & + V_\text{daisy}(\{\phi_i\}, T) ,
\end{split}
\end{equation}
where the last term gives the daisy corrections and depends on the
gauge it is calculated in.

In the more general $R_\xi$ gauge one has explicit $\xi$ dependence in
$V_\text{daisy}$. We will again construct
the resummed daisy contribution from the high-temperature expansion of
the effective potential, but this time using \cref{Eq:VT_Rxi}. The gauge
dependence cancels at the lowest order in the $m/T$ expansion such
that the second term in \cref{Eq:VHTB} is unchanged, while the
first field-independent term will cancel.  Thus, the relevant terms
are
\begin{align}
 \label{Eq:VHTB_Rxi} 
 V^\text{HTB}_{\text{1T},R_\xi}(\{\phi_i\}, T) &\ni - \frac{T}{12}  \left[\sum_\phi (m^2_\phi(\xi,\{\phi_i\}))^{3/2}  + \sum_{\tilde{B}}(m^2_{\tilde{B}}(\{\phi_i\}))^{3/2} \right] ,
\end{align}
where the first term includes both physical scalars and Goldstone
bosons that have masses of the form $m_G^2 = m^2_{G,0} + \xi m_B^2$,
where $m^2_{G,0}$ is the $\xi=0$ mass and $m_B$ is the mass of the
gauge boson mass associated with that Goldstone boson. Note that we do
not include a contribution from the last term in \cref{Eq:VT_Rxi} as these
Fadeev-Popov ghost contributions do not get a thermal mass.

Adding thermal masses to these terms gives cubic contributions to the resummed potential,
\begin{align}
\label{Vresum_cube_Rxi}  
V_\text{resum} \ni -\frac{T}{12} \left [\sum_{\phi} ( m^2_\phi(\xi,\{\phi_i\}) +  \Delta_{T, {\phi}} )^{3/2}  + \sum_{\tilde{B}}(m_{\tilde{B}}^2\{\phi_i\} +  \Delta_{T, {\tilde{B}}}  )^{3/2} \right] .
\end{align}
Subtracting the two we are left with the daisy contributions in the $R_\xi$ gauge,
\begin{equation}
\begin{split}
  V^{R_\xi}_{\text{daisy}}(\{\phi_i\}, T)  ={}&
  -\frac{T}{12}\sum_\phi \left[( m^2_\phi(\xi,\{\phi_i\}) +  \Delta_{T, {\phi}} )^{3/2} -  ( m^2_\phi(\xi,\{\phi_i\}))^{3/2}\right] \\ 
&-\frac{T}{12}\sum_{\tilde{B}}\left[(m_{\tilde{B}}^2(\{\phi_i\}) + \Delta_{T, {\tilde{B}}} )^{3/2}  - (m_{\tilde{B}}^2(\{\phi_i\}))^{3/2}\right] .
\end{split}
\end{equation}
While the Arnold-Espinosa and Parwani procedures both resum leading
infrared divergent terms, they differ in the additional higher-order
corrections that are included, which are not complete in either
scheme.  Comparisons between the two approaches have been used as an
estimate of the size of the uncertainties from the missing corrections
(see e.g.\ \refcite{Cline:1996mga,Athron:2022jyi}). However this approach has also been discouraged in \refcite{Lofgren:2023sep}, arguing that when a correct power counting is implemented the schemes no longer differ by higher order and strongly advocating that power counting schemes should be used.

At the one-loop level both methods are relatively easy to implement
and fast to evaluate.  Therefore daisy resummation represents a
practical solution to the problem of the perturbative expansion being
spoiled by infrared divergent terms. One may also carefully check the
expansion parameter is now under control as is done in
\Refcite{Cline:1996mga}. Further improvement of the
perturbative expansion may be achieved by resumming subleading
infrared-divergent terms represented by the diagrams referred to as ``superdaisy'' diagrams
\cite{Dolan:1973qd}. Further two-loop corrections known as sunset and
lollipop diagrams may also contribute.

One may consider improving the precision by going beyond the leading-order
high-temperature expansion for the thermal masses. The full
thermally-corrected masses are found by iteratively solving the
equations
\cite{Brahm:1991nh,Brahm:1991rr,Espinosa:1992gq,Espinosa:1992kf,Quiros:1992ez}
\begin{align}
  M_{\tilde{B_i}}^2 = m_{\tilde{B_i}}^2 + \Sigma_T(M_{\tilde{B_i}}^2,T),
  \label{Eq:Full1LTheramalMass}
\end{align}
where $\Sigma_T$ is the full self-energy at finite temperature.  While
intuitively one may expect that including these masses in the
effective potential should improve the calculation it actually
miscounts two-loop daisy diagrams and gives spurious linear terms
\cite{Dine:1992vs,Dine:1992wr,Arnold:1992rz,Boyd:1992xn,Curtin:2016urg,Laine:2017hdk}. Instead
\Refcite{Dine:1992vs,Boyd:1993tz,Curtin:2016urg,Curtin:2022ovx} find
that a tadpole resummation approach the appropriate way to consistently
improve beyond the daisy resummation. In this approach the mass in
\cref{Eq:Full1LTheramalMass} is inserted into the first derivative of
the effective potential, and this is then integrated to obtain an
improved effective potential. They find this consistently resums
higher order corrections in addition to the leading terms in the
high-temperature expansion.

Further useful pedagogical discussion of daisy resummation may be
found in Section~3 of \Refcite{Senaha:2020mop}, and we also recommend
in particular the original papers \Refcite{Dolan:1973qd} and
\Refcite{Arnold:1992rz} for clear explanations.

\subsubsection{Three-dimensional effective field theory}\label{sec:3defftheory}

It can be demonstrated (see e.g.\ \refcite{Trodden:1998ym,Kajantie:1995dw}) that a four-dimensional theory at finite temperature is equivalent to a three-dimensional Euclidean theory that has an infinite number of fields that appear in the Matsubara decomposition of the original fields,
\begin{align}
  \Phi(x,\tau) = \sum_{n=-\infty}^\infty \varphi_n(x)\exp{i \omega_n \tau} ,
\end{align}
where $\tau = -it$ is the imaginary time. This is just a Fourier
expansion in the imaginary time coordinate. Integrating out the
non-zero Matsubara modes results in a zero-temperature
three-dimensional theory (3dEFT) with only the bosons
\cite{Ginsparg:1980ef,Appelquist:1981vg,Nadkarni:1982kb}.  The
behavior of the original four-dimensional theory at high temperatures
may be determined from this 3dEFT.

This means the matching and running method of effective field theories
can then be applied to this problem to resum the effects from high
temperature \cite{Farakos:1994kx,Braaten:1995cm,Kajantie:1995dw} that
come from the non-zero Matsubara modes. This approach has been named
Dimensional Reduction in the literature \cite{Nadkarni:1982kb},
through we mostly avoid it here due to the clash with other uses such
as the dimensional reduction regularization and methods for handling
data from a high dimensional space.  It is an alternative to the daisy
resummation approach in the previous subsection where we are resumming
these effects from the non-zero Matsubara modes by including thermal
corrections to the zero Matsubara modes. Thus it should be understood
that the 3dEFT approach resums the leading infrared-divergent
contributions captured by daisy resummation, while additionally
correctly resumming other temperature contributions. In particular
large logarithmic contributions that may also spoil the perturbation
theory can also be resummed with this method \cite{Farakos:1994kx}.
This ensures that the effective potential remains perturbative at high
temperatures and is a more complete solution than the daisy
resummation, even when that is improved with the tadpole resummation
method to go beyond the leading infrared-divergent terms.

More precisely the full four-dimensional theory can be matched to the
three-dimensional theory containing only the zero Matsubara modes by
computing $n$-point correlation functions in both theories and equating
the corresponding correlation functions at the matching scale.  The
matching scale is taken to be about $\pi T$ to make logarithms in the
matching small. \Refcite{Farakos:1994kx} finds that the choice $\mu
= 4\pi T e^{-\gamma}$, where $\gamma$ is the Euler–Mascheroni
constant, is the scale that removes all logarithmic contributions from
the matching. The temperature enters the three-dimensional theory
only through the couplings set by the matching conditions.  The
renormalization group equations are derived in the three-dimensional
theory and used to resum the large logarithms. The three-dimensional
theory is super-renormalizable, which has important consequences.  The
beta functions for the masses are exact at the two-loop level, while
the dimensionless couplings are renormalization group invariants
\cite{Farakos:1994kx}.

In comparison to the daisy resummation discussed in \cref{sec:daisyresum}, this procedure also resums large logarithmic
contributions and therefore appears to be a formally more precise
solution when the temperature is large compared to the field.  On the
other hand the 3dEFT approach is considerably more work
to implement, especially in models with many new states beyond the
SM. Perhaps for this reason daisy resummation has been
more commonly used in the study of the EWPT and
perturbative approaches to phase transitions in BSM physics.

Recently however the uncertainties in the effective field theory
approach have been compared with the daisy resummation approach in
\refcite{Croon:2020cgk}.  They find that the uncertainties in the
critical temperature, other thermal parameters and the gravitational
wave spectrum when calculated using the 3dEFT approach are much
smaller than when calculated using the daisy resummation.  Furthermore
very recently this 3dEFT approach
has been implemented in a public code \code{DRalgo}
\cite{Ekstedt:2022bff} which should make it substantially more
accessible.  It was also argued (in the context of an inert doublet
model) in \refcite{Laine:2017hdk} that to get a precision at or below
$10\%$ dimensional reduction is required in purely perturbative
studies.

Explicit details for performing the matching and running may be found
in original papers~\cite{Farakos:1994kx,Braaten:1995cm,Kajantie:1995dw}.
Very recent pedagogical discussion can be found in Appendix B of
\refcite{Croon:2020cgk} and an explicit demonstration of the
procedure for a toy scalar singlet field theory is given in
\refcite{Schicho:2021gca}, which they then generalize to the scalar
singlet extension of the SM.  Furthermore a general
procedure is clearly described in
\refcite{Ekstedt:2022bff}. Although we intend this review to be
pedagogical and show the reader how to construct the effective
potential, in this case the expressions are rather lengthy to
reproduce here and the treatments in these recent papers cover the
situation better than we could hope to here, so we instead refer the
reader to \refcite{Croon:2020cgk,Schicho:2021gca,Ekstedt:2022bff} for more
details.

The 3dEFT approach can be used
perturbatively to study phase transitions in BSM physics.  It can also
be used to remove the fermions from the theory to provide a setup
where the ultrasoft modes can be studied non-perturbatively in lattice
simulations.  While the lattice simulations clearly provide the gold
standard for analyzing phase transitions they are not really feasible
for detailed exploration of the many specific SM extensions, though a
few limited studies have been carried out
\cite{Kainulainen:2019kyp,Niemi:2020hto,Gould:2021dzl}.  For
sufficiently heavy new physics states the three dimensional effective
field theory can be the same as that obtained from the standard model
(but with different matching conditions), allowing existing lattice
results to be used \cite{Andersen:2017ika,Gould:2019qek}, though see
also recent updated results \cite{Gould:2022ran}. However using the
3dEFT in a perturbative approach is
also a reasonable compromise between time and rigor and this can also
be a first step towards a lattice treatment. This approach has been
used in the Minimal Supersymmetric SM
\cite{Losada:1996ju,Farrar:1996cp,Cline:1997bm,Bodeker:1996pc}, the
Two Higgs Doublet Model
\cite{Losada:1996ju,Andersen:1998br,Gorda:2018hvi}, the scalar singlet
model \cite{Brauner:2016fla,Schicho:2021gca,Niemi:2021qvp} and a real triplet
extension of the SM \cite{Niemi:2018asa}.

The 3dEFT constructed here has further benefits beyond improving the
convergence of the perturbation theory. As we discuss in
\cref{sec:gauge_indep_approaches} one method designed to avoid
problems from the gauge dependence of the effective potential is the
$\hbar$-expansion. While the 3dEFT
is still gauge dependent, performing the expansion in $\hbar$
consistently should be straightforward, while trying to combine daisy
resummation and the $\hbar$-expansion can be rather challenging.
Furthermore as discussed in \cref{sec:separation_of_scales} the
separation of scales used in the 3dEFT approach can help resolve some rather foundational issues in
the calculation of tunneling between minima.

\subsection{Gauge-independent approaches}
\label{sec:gauge_indep_approaches} 
In \cref{sec:DeltaVT0GaugeDep} we showed the form of the one-loop potential in the
$R_\xi$ gauge, with the manifest $\xi$ dependence. This $\xi$ dependence is
not just some artifice for shuffling around different contributions as
they appear in the various gauges it incorporates, but actually
represents a real dependence of the effective potential on the $R_\xi$
gauge parameter.  

This gauge dependence of the effective potential was first pointed out
in \refcite{Jackiw:1974cv} and a clear discussion and demonstration of
it can be found in \refcite{Patel:2011th}.  While it may be disturbing
that the effective potential is gauge dependent it is not itself an
observable, and observables correctly calculated from it must be gauge
invariant.  Nonetheless the gauge dependence of the effective
potential still presents a thorny problem, with gauge dependence
appearing in the critical temperature marking the point where phase
transitions become possible, at intermediate stages in the calculation
of observables and in widely used proxy indicators such as the order
parameter, $\gamma = v / T_c$.\footnote{This is used in checking if a generated baryon
asymmetry would be washed out by the sphaleron process inside the
bubbles of new phase. Frequently a criteria of  $\gamma \geq 1$ is employed but it is also better to make a more careful estimation of the right-hand side of that criteria \cite{Patel:2011th}.} In particular gauge dependence in the effective
potential can also propagate to calculations of the decay rate of the
vacuum (see \cref{sec:transition_rates}) if care is not taken.

There have been many works on understanding the gauge dependence of
the effective potential and developing gauge-independent techniques
for calculating the effective potential in the context of SM and
various new physics
scenarios~\cite{Dolan:1974gu,Kang:1974yj,Fischler:1974ue,Frere:1974ia,Fukuda:1975di,Nielsen:1975fs,Aitchison:1983ns,Kobes:1990dc,DelCima:1999gg,DelCima:1999dr,Alexander:2008hd,Patel:2011th,DiLuzio:2014bua,Nielsen:2014spa,Espinosa:2016uaw,Ekstedt:2018ftj,Ekstedt:2020abj}. This
was discussed specifically in the context of gravitational waves in
\refcite{Wainwright:2011qy,Croon:2020cgk}. The implications for
calculations of vacuum decay has also received a lot of attention
\cite{Metaxas:1995ab,Garny:2012cg, Plascencia:2015pga,
  Endo:2017gal,Endo:2017tsz,Lofgren:2020zzn,Arunasalam:2021zrs,Hirvonen:2021zej,Lofgren:2021ogg,Croon:2021vtc}. This
remains an active but somewhat contentious topic and we cannot do all
of the ideas and arguments justice here.

Of great importance to this topic are the Nielsen identities \cite{Nielsen:1975fs,Fukuda:1975di} which show that the value of the exact effective potential at extrema are gauge independent as they imply that
\begin{equation}
\frac{\partial V}{\partial \xi} \propto \frac{\partial V}{\partial \phi}.
\end{equation}

\begin{figure}
\centering
\includegraphics[width=0.9\linewidth]{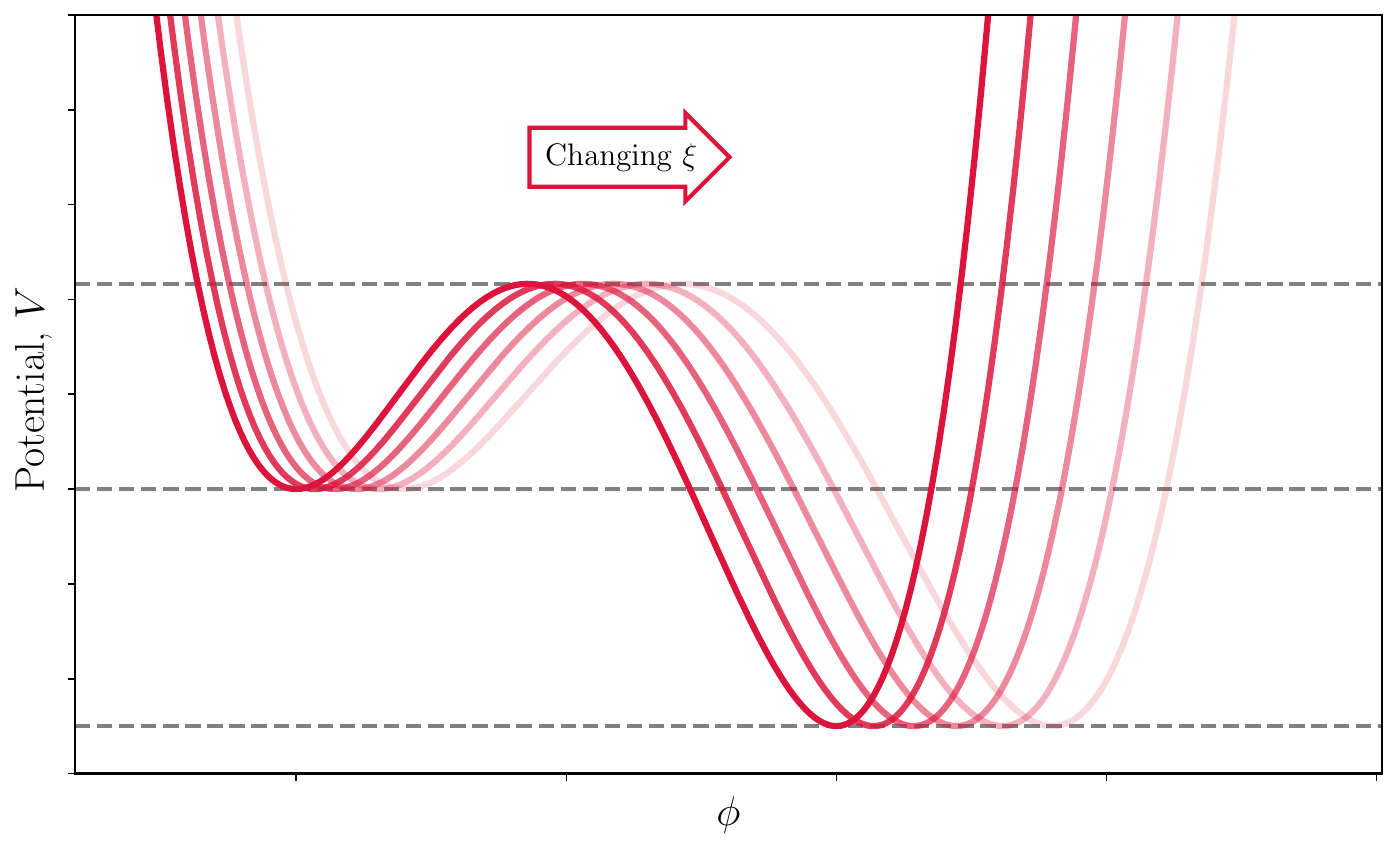}
\caption{Schematic illustration of Nielsen's identity based on Fig.~2 in \refcite{Patel:2011th}. Changing the gauge parameter, $\xi$, in the exact effective potential does not change the values of the potential at the extrema. The locations of the extrema change, however, as the potential compresses and expands.}
\label{fig:nielsen}
\end{figure}

\Refcite{Fukuda:1975di} showed this independence holds beyond the
standard (linear) $R_\xi$ gauges.  The locations of the extrema,
however, are gauge dependent. This is illustrated in
\cref{fig:nielsen}.

As explained in \refcite{Patel:2011th}, truncating the effective
potential at a fixed order in $\hbar$ impacts Nielsen's
identities. Carefully matching powers of $\hbar$,
\refcite{Patel:2011th} shows that the one-loop effective potential is
invariant at \emph{tree-level} extrema.

This so-called $\hbar$-expansion method has been used in the
literature in the 1990s \cite{Laine:1994zq,Farakos:1994xh}, but its
recent use mostly stems from \refcite{Patel:2011th}. As described
there, this can be applied to compute gauge-independent perturbative
estimates of the critical temperature and the order parameter.  It is
not easy to use this method together with the Daisy resummation
methods described in \cref{sec:daisyresum} without spoiling either the
gauge independence or the resummation, but it can be combined with the
3dEFT approach discussed in \cref{sec:3defftheory}
\cite{Laine:1994zq,Patel:2011th,Schicho:2022wty}.  The Nielsen
identities are further exploited in
\refcite{Lofgren:2020zzn,Arunasalam:2021zrs,Hirvonen:2021zej,Lofgren:2021ogg}
to develop a broader framework of gauge-invariant bubble nucleation
calculations.

\subsection{Tracing the structure of the effective potential}\label{sec:tracing_phases}

The effective potentials are complicated objects, as they are functions of several fields and the temperature, as well as the model's parameters, renormalization scale and gauge fixing parameters. There could be several minima that evolve as the temperature changes, potentially appearing and disappearing, and changing in depth. Typically, spontaneously broken symmetries are restored at high temperatures, as thermal corrections make the potential more convex, moving the minima to the origin~\cite{Kirzhnits:1972ut,Kirzhnits:1972iw,Weinberg:1974hy,Linde:1978px,Dolan:1973qd}. To study phase transitions systematically, we here describe how one can trace field values that minimize the effective potential as the temperature changes. We later in \cref{sec:transition_rates} describe how to analyze transitions between the minima in QFT at finite temperature. 

In simple cases and under simplifying assumptions, it may be possible to tackle the problem analytically by solving for the minima as functions of temperature. For example, in an approximation of the potential that neglects zero-temperature one-loop corrections and keeps only the leading terms in a high-temperature expansion of the thermal corrections~\cite{Profumo:2007wc,Barger:2008jx,Vaskonen:2016yiu,Chiang:2017nmu,Ghorbani:2020xqv}. Even then, though, the tadpole equations could be a set of coupled cubic equations that are quite challenging to solve~\cite{Maniatis:2006jd} unless the potential possesses symmetries that drastically simplify the problem. Thus tracing phases usually requires special numerical and analytic techniques to repeatedly find the minima of the potential. Even at tree-level, this is a daunting task. Analytic developments were made in \refcite{Maniatis:2012ex} and \code{Vevacious}~\cite{Camargo-Molina:2013qva}, following similar developments in materials science~\cite{Mehta:2011xs,Mehta:2012qr} and string theory~\cite{Mehta:2011wj}. By exploiting homotopy continuation, it is possible to ensure that all minima of the tree-level potential were found. However, even the zero-temperature loop corrections may shift them or introduce new minima, and the effects of finite-temperature corrections are even more dramatic. We thus sketch an algorithm developed in \code{CosmoTransitions}~\cite{Wainwright:2011kj} and \code{PhaseTracer}~\cite{Athron:2020sbe} for numerically finding and tracing minima of the finite-temperature effective potential.

We trace minima up from $T = 0$ and downwards from high temperature. We begin tracing by drawing a collection of starting points from a uniform distribution over the field values (see \refcite{AbdusSalam:2020rdj} for a discussion of the benefits of random versus grid scanning in this context, though both could perhaps be improved by quasi-random sampling). We optimize locally from each chosen point using a local optimization algorithm using a Nelder-Mead-type algorithm~\cite{rowan1990functional} in \code{nlopt}~\cite{johnson2014nlopt} and keep only unique minima identified by this process. We increment the temperature by $\Delta T$, but rather than repeating the global minimization, we optimize locally starting from near the previous minima. We must take care not to fall into the basin of attraction of a different phase when tracing. We could avoid that by reducing the step $\Delta T$. This would be at the expense of performance as more steps would be required, so we in fact start from the previous minima perturbed by
\begin{equation}
\Delta \phi = \frac{d\phi}{dT} \Delta T ,
\end{equation}  
and we continuously tune $\Delta T$ to be as big as possible while respecting the fact that the starting guess indeed lies close the local minima that is later found from that guess.  Of course, a minimum may disappear during tracing. This happens, for example, when a barrier separating minima is reduced as temperature changes until it disappears. At such points, tracing stops. Finally, after tracing minima upwards and downwards in temperature, we keep the unique sets of paths, which we call phases. There are alternative methods and simpler methods for only partially tracing the minima as the potential changes. The code \code{BSMPT} \cite{Basler:2018cwe,Basler:2020nrq}, for example, identifies a single transition from the origin in field space to non-zero field values.

\begin{figure}
\centering
\includegraphics[width=0.49\textwidth]{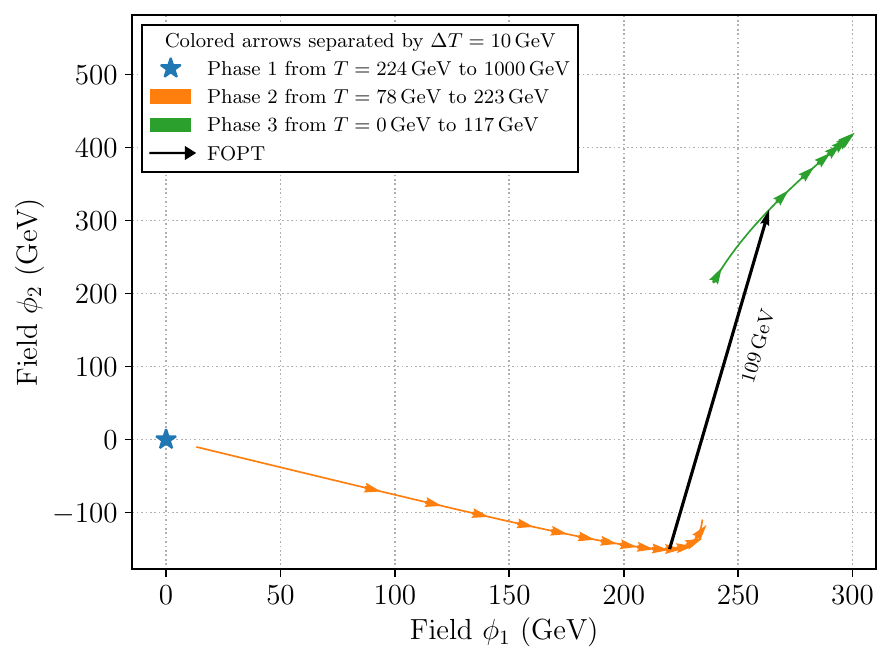}
\includegraphics[width=0.49\textwidth]{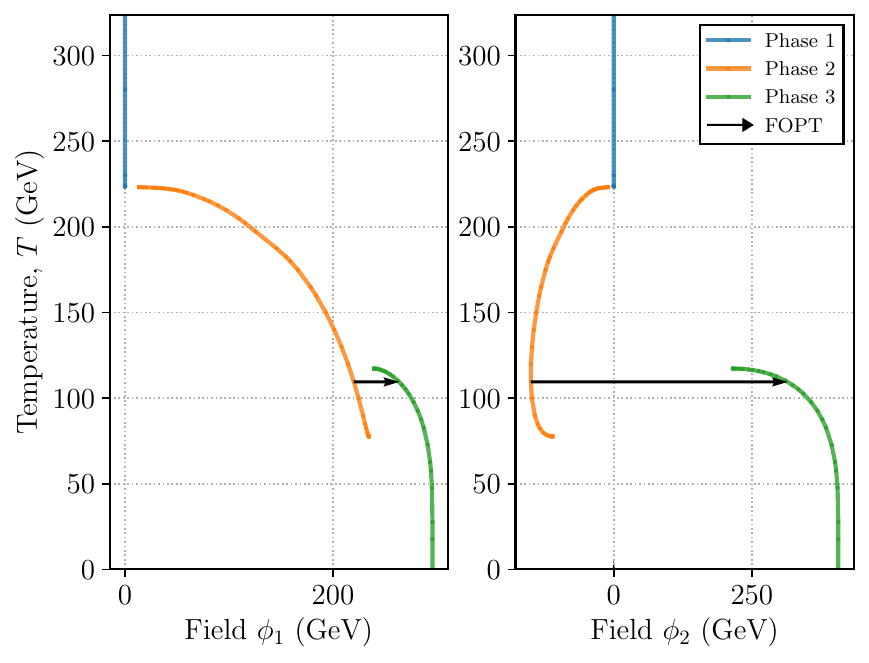}
\caption{The evolution with temperature of the minima of an effective potential with two scalar fields.}
\label{fig:tracing}
\end{figure}

To visualize the phenomenon, we consider a simplified model with two scalar fields, $\phi_1$ and $\phi_2$, that mix through a bilinear operator $\phi_1 \phi_2$ (see Eq.~(20) in \refcite{Athron:2020sbe}). We plot all minima as a function of temperature for this model in \cref{fig:tracing}. The left-hand-side shows the minima evolving parametrically with temperature; the right-hand-side shows the minima as a function of temperature. Three distinct phases are visible and are shown by blue, orange and green colors, and a first-order phase transition is indicated by a black arrow. We see that the fields lie at the origin at high temperature (blue), make a smooth second-order transition to non-zero field values at about $T \approx 200\gev$ (orange), and finally make a first-order transition at about $T \approx 100\gev$. The final phase (green) only appears below about $T \lesssim 120\gev$ and the intermediate phase (orange) disappears at about $T\approx 80\gev$. If there are many scalar fields, disconnectivity graphs could be used to visualize the structure of the potential at each temperature (see e.g., \refcite{wales2003energy}).

Tracing phases allows one to identify when transitions between the phases become possible. Specifically, the critical temperature $T_c$ --- defined as the temperature when the free energy densities of two phases are equal --- can be easily extracted from the phase structure. However, determining if and when a transition completes is considerably more complicated. We discuss this problem in \cref{sec:transition_rates,sec:transitionAnalysis}.

\section{Transition rates}\label{sec:transition_rates}

\begin{figure}
\centering
\includegraphics[width=0.9\linewidth]{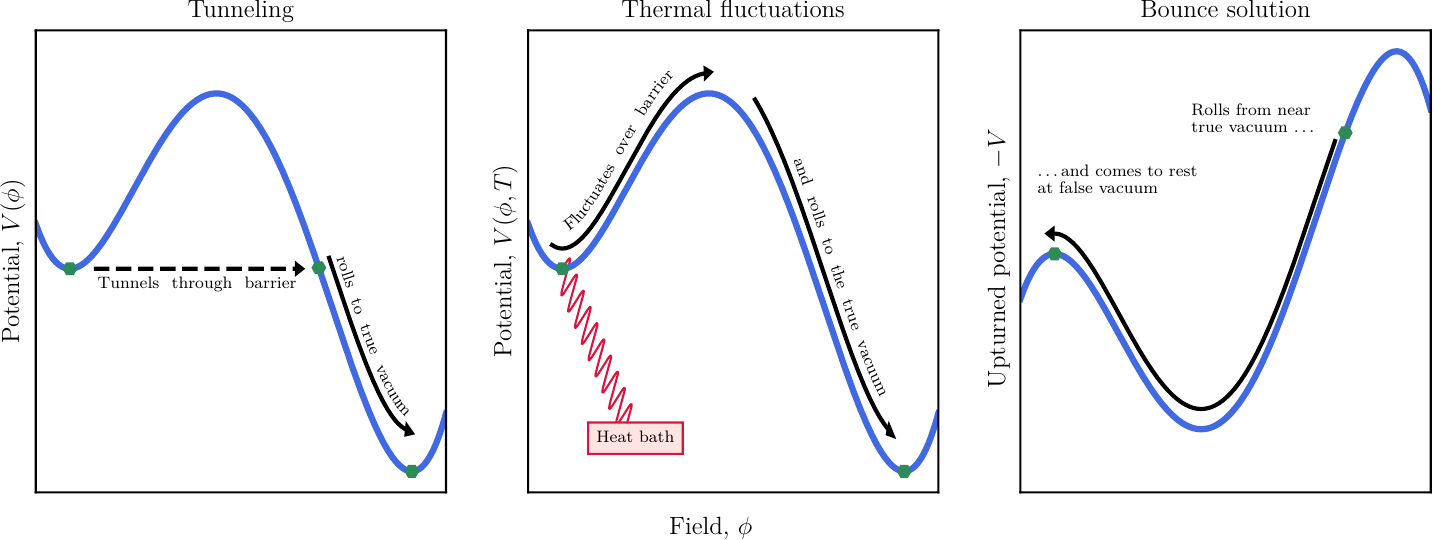}
\caption{The field may transition from the false vacuum by (left) tunneling through the barrier and then rolling into the true vacuum or (center) thermal fluctuations over the barrier. The transition probability in both cases involves the bounce solution (right), where we solve for the initial field position from which the field rolls down the upturned potential, along the barrier, and comes to a rest at the false vacuum.}
\label{fig:bounce}
\end{figure}

The tree-level potentials and effective potentials introduced in \cref{sec:effective_potential} should be bounded from below, but could have more than one minimum. The minima correspond to distinct vacua of the theory. In a quantum theory, only the lowest lying vacua need be stable, as the system may tunnel through any barriers between different vacua. Moreover, at finite temperature a thermal fluctuation could lift the system over a barrier. These possibilities are sketched in \cref{fig:bounce}. Transitions could occur anywhere in spacetime, leading to so-called bubbles, analogous to bubbles in boiling water. Away from the bubbles, the Universe remains in the false vacuum. As we shall discuss, bubbles that grow and that could dominate our Universe are impeded by sizeable barriers between states (and generally sizeable bounce actions), by the expansion of the Universe and by interactions with the surrounding plasma of particles, such that it cannot be taken for granted that a transition would complete and a numerical calculation is required.  In addition to spontaneous bubbles, inhomogeneous nucleation in cosmological phase transitions has been studied in \refcite{Affleck:1979px, Samuel:1991zs, Christiansen:1995ic, Shukla:2000dx, Ignatius:2000cz, Megevand:2003tg, Schwarz:2003du}, and with black holes and topological defects as preferred nucleation sites in \refcite{Moss:1984zf, Hiscock:1987hn, Arnold:1989cq, Berezin:1990qs, Samuel:1992wt, Hiscock:1995ma, Gregory:2013hja, Burda:2015isa, Burda:2015yfa, Tetradis:2016vqb, Canko:2017ebb, Mukaida:2017bgd, Hayashi:2020ocn, Shkerin:2021zbf, Strumia:2022jil}.

We first focus on computing the transition probability per unit time per unit volume. This was first described in \refcite{Kobzarev:1974cp, Coleman:1977py, Callan:1977pt} for zero-temperature quantum tunneling through the barrier and in \refcite{Linde:1977mm, Linde:1980tt, Linde:1981zj} at finite-temperature, though continues to the subject of fresh work (see e.g., the theses~\refcite{Citron:2013bje, Masoumi:2013nix, Lee:2014yud, hirvonen, Guada:2021lnb,Wickens:2021vzr} and \refcite{Devoto:2022qen} for a recent review). Schematically,
\begin{equation}
p(t) \equiv \Gamma / V = |A| e^{-B} \label{eq:p_0T}
\end{equation}
for the former and
\begin{equation}
p(t; T) \equiv \Gamma / V = |A(T)| e^{-B(T) / T} \label{eq:p_T}
\end{equation}
for the latter. There are in fact two contributions at finite-temperature to the factor $A(T)$ in \cref{eq:p_T}: tunneling through the barrier at finite temperature and thermal fluctuations over the barrier. Remarkably, an approximate quantitative description of tunneling, finite-temperature tunneling and thermal fluctuations boils down to a single non-linear ordinary differential equation called the bounce equation, which leads to the exponent $B$, which we discuss in \cref{sec:rates_qm,sec:rates_finite_t,sec:qft_bounce} in quantum mechanics, at finite-temperature and in quantum field theory (QFT), respectively.\footnote{Though see \refcite{Shen:2022xcm} for a recent discussion of lattice approaches to vacuum decay.} We continue by discussing the computational methods for the bounce action and the prefactor, $A$, in \cref{sec:compute_bounce,sec:prefactor}, respectively. Finally, we combine this probability with considerations about the age and expansion of the Universe in \cref{sec:transitionAnalysis}, showing how we may quantify the completion of a transition and compute whether a transition completes, and if so at what temperature.

\subsection{Rates in one-dimensional quantum mechanics}\label{sec:rates_qm}

We first sketch out the physical intuition behind \cref{eq:p_0T,eq:p_T} following \refcite{Callan:1977pt,Coleman:1978ae} (see \refcite{PhysRevD.19.667,Andreassen:2016cff,Blum:2023wnb,batini2023realtime} for alternative approaches and \refcite{KRAMERS1940284,Langer:1969bc,LANGER197461} for related methods in classical statistical physics). Our ultimate goal is to compute the probability per unit time per unit volume, $\Gamma / V$ of a transition from the false vacuum, \fv, to the true vacuum, \tv.

For simplicity, we first focus on a particle in a one-dimensional quantum mechanical potential starting at $t_i$ at the false vacuum at position $q_0$ and ending at $t_f$ at the same position $q_0$~\cite{Paranjape:2017}. The amplitude of this transition must be related to the path integral by
\begin{equation}\label{eq:path_integral}
\langle q_0 | e^{-i \hat H (t_f - t_i)}| q_0 \rangle = \int^{q(t_f) = q_0}_{q(t_i) = q_0} \, [\dint q] \, e^{i S[q]} ,
\end{equation}
where the action
\begin{equation}
S[q] = \int_{t_f}^{t_i} \left[\frac12 \left(\frac{\partial q}{\partial t}\right)^2 - V(q)\right] \dint t.
\end{equation}
We make a Wick rotation, $\tau = i t$. On the right-hand side in \cref{eq:path_integral}, we obtain the Euclidean action in the exponent
\begin{equation}\label{eq:euclidean_action}
i S[q] =  - \int_{\tau_i}^{\tau_f} \left[\frac12 \left(\frac{\partial q}{\partial \tau}\right)^2 + V(q)\right] \dint \tau \equiv - \SE[q].
\end{equation}
We tackle the integral approximately by the method of steepest descent. We expand the Euclidean action, $\SE \approx B + \delta \SE$, around the extremum $B$,
\begin{equation}\label{eq:path_one_bounce}
\int^{q(\tau_f) = q_f}_{q(\tau_i) = q_i} \, [\dint q] \, e^{- \SE[q]} \approx \left[\int^{q(\tau_f) = q_f}_{q(\tau_i) = q_i} \, [\dint q] \, e^{- \delta \SE[q]}\right] e^{-B} \equiv C e^{-B},
\end{equation}
where $B$ is the action for a non-trivial path that extremizes the Euclidean action subject to the boundary conditions of the integral and $C$ results from summing quadratic fluctuations, $\delta \SE$, about that path. Restoring $\hbar$ through $\SE \to \SE / \hbar$, the relative error in the method of steepest descent is $\order{\hbar}$; this is a semi-classical approximation in which $\hbar$ must be considered small. This path thus satisfies the Euler-Lagrange equation of motion,
\begin{equation}
\frac{\dint^2 q}{\dint \tau^2} = \frac{\partial V}{\partial q}.
\end{equation}
As we shall discuss, this resembles a classical mechanics problem in which the potential has been upturned, $V \to -V$. This solution to the Euclideanized equations of motion is an instanton (see e.g., \cite{marino_2015}) that we call the \emph{bounce}, $q_b$, since it begins and ends at $q_0$. The factor $C$ originates from second order variations of the action,
\begin{equation}\label{eq:second_order_variation}
\delta \SE = \frac12 \int \left[\delta q \left(- \frac{\dint^2}{\dint \tau^2} + V^{\prime\prime}(q_b)\right) \delta q\right] \dint \tau.
\end{equation}
The first-order variations vanish as we are expanding around an extremum. By viewing $\tau$ as an index, \cref{eq:second_order_variation} may be seen as a matrix expression of the form 
\begin{equation}
\frac 12 \sum_{ij} (x - x_0)_i  M_{ij}(x_0) (x - x_0)_j.
\end{equation}
Thus we may make a Gaussian integral over the second order variations, giving
\begin{equation}\label{eq:product_eigenvalues}
C = (\det M)^{-1/2} = \prod \frac{1}{\sqrt{\lambda}} ,
\end{equation}
where 
\begin{equation}\label{eq:matrix}
M = - \frac{\dint^2}{\dint \tau^2} + V^{\prime\prime}(q_b)
\end{equation}
and $\lambda$ are the eigenvalues of $M$ subject to the boundary condition that the fluctuations vanish at $t_i$ and $t_f$. This is similar to the simple case of a Laplace approximation for a finite-dimensional integral in which the product of the eigenvalues of the Hessian matrix would appear in the result.

The matrix $M$, however, is singular. As our problem is time translation invariant, we are free to place the center of the bounce solution anywhere in Euclidean time and the action wouldn't change. This corresponds to a zero eigenvalue in \cref{eq:matrix}. We may verify this by considering the function $\dot q_b$,
\begin{equation}
M \dot q_b = \left(- \frac{\dint^2}{\dint \tau^2} + V^{\prime\prime}(q_b) \right) \dot q_b =  \frac{\dint}{\dint \tau} \left(- \frac{\dint^2 q_b}{\dint \tau^2} + V^{\prime}(q_b) \right) = 0 ,
\end{equation}
This final factor in brackets vanishes as it is nothing but the equation of motion evaluated at the solution $q_b$. Rather than treating this mode through Gaussian integration, we directly integrate over the center of the bounce. This results in
\begin{equation}\label{eq:omit_zero_eigenvalues}
C = \Tau \, \sqrt{\frac{B}{2\pi}} \, (\detprime M)^{-1/2} ,
\end{equation}
where we place the center of the bounce anywhere between $\tau_i$ and $\tau_f$ resulting in a factor $\Tau \equiv \tau_f - \tau_i$; the second factor is a Jacobian from a change of variables to integrate over the center of the bounce; and $\detprime$ denotes the product of eigenvalues except the zero eigenvalue.

Finally, to fully compute the right-hand side in \cref{eq:path_integral}, we must account for more than one bounce. In \cref{eq:path_one_bounce}, we expanded around an extremum corresponding to a single bounce, but as bounces are localized in time the particle could approximately bounce $n$ times back and forth and the sum of these possibilities cannot be neglected. The action for $n$ bounces is approximately $B_n = n B$, since the action is dominated by time spent away from $q_0$. The prefactor for $n$ bounces $C_n$, though, requires care. The result is
\begin{equation}
C_n = \frac{1}{n!} \left(\sqrt{\frac{B}{2\pi}} \Tau\right)^n  (\det M_0)^{-1/2} \, \left(\frac{(\detprime M)^{-1/2} }{(\det M_0)^{-1/2}} \right)^n ,
\end{equation}
where 
\begin{equation}\label{eq:matrix_0}
M_0 = - \frac{\dint^2}{\dint \tau^2} + V^{\prime\prime}(q_0) 
\end{equation}
and we use the convention that $\SE[q_0] = 0$. Let us motivate this form. For $n$ bounces, each bounce may itself be translated resulting in $n$ zero eigenvalues. Thus we obtain the second factor. They must, however, be associated with a degeneracy factor $1 / n!$, which avoids double counting permutations of the order in which the bounces occurred. The third factor represents the persistence amplitude --- that the particle forever stays at rest without bouncing. The final factor adds $n$ bounces, localized in time, to that persistence amplitude and subtracts $n$ periods of persistence. Writing the prefactor for $n$ bounces as
\begin{equation}
C_n = (\det M_0)^{-1/2} \frac{(A \Tau)^n}{n!} ,
\end{equation}
where
\begin{equation}\label{eq:qm_prefactor}
A = \sqrt{\frac{B}{2\pi}} \, \frac{(\detprime M)^{-1/2} }{(\det M_0)^{-1/2}},
\end{equation}
the sum over such bounces thus exponentiates the right-hand side of \cref{eq:path_one_bounce}, such that we find
\begin{equation}
\int^{q(\tau_f) = q_f}_{q(\tau_i) = q_i} \, [\dint q] \, e^{- \SE[q]} = \sum_n (\det M_0)^{-1/2} \frac{(A e^{-B} \Tau)^n}{n!} = (\det M_0)^{-1/2} e^{Ae^{-B} \Tau }.
\end{equation}
This is the instanton gas approximation~\cite{Arnold:1987mh}. 

Let us now apply the Wick rotation to the left-hand side of \cref{eq:path_integral}. After expanding in eigenstates of the Hamiltonian, we obtain 
\begin{equation}\label{eq:eigenstates_path_integral}
\langle q_0 | e^{- \hat H \Tau}| q_0 \rangle = \sum_n e^{-E_n \Tau} \langle q_0 | n \rangle \langle n | q_0 \rangle ,
\end{equation}
Considering $\Tau \to \infty$, the dominant term involves the lowest energy level,
\begin{equation}
\sum_n e^{-E_n \Tau} \langle q_0 | n \rangle \langle n | q_0 \rangle \to e^{- E_0\Tau} |\psi_0(q_0)|^2 ,
\end{equation}
and thus we pick out the ground state. This happened, though, regardless of our choice of $q_0$. Putting our results together, we obtain
\begin{equation}\label{eq:one}
e^{- E_0 \Tau} |\psi_0(q_0)|^2 \approx (\det M_0)^{-1/2} e^{Ae^{-B} \Tau }.
\end{equation}
Finally, we must relate this result to the lifetime of the state. Unstable states are associated with complex eigenvalues of the Hamiltonian and decay as~(see e.g., \refcite{dau})
\begin{equation}
|\psi(t)|^2 \propto e^{2\Im E_0 t}. 
\end{equation}
That is, with decay rate 
\begin{equation}\label{eq:rate_from_imaginary_energy}
\Gamma = -2 \Im E_0.
\end{equation}
Thus we must compute the imaginary part of the energy of the ground state through \cref{eq:one}. 

Identifying the imaginary part of the energy, however, remains challenging, as there are pathologies associated with supposing that there are complex eigenvalues of the Hamiltonian and identifying an imaginary part of the \emph{real} Euclidean path integral~\cite{Andreassen:2016cvx}. Nevertheless, ultimately
\begin{equation}\label{eq:gamma_zero_temperature}
\Gamma = |A| e^{-B}.
\end{equation}
The trivial solution in which the particle rests at the false vacuum forever, moreover, cannot contribute to the imaginary part of the energy and so non-trivial bounces must be considered. This may be understood by noting that imaginary contributions originate from negative eigenvalues in \cref{eq:product_eigenvalues}. Coleman~\cite{Coleman:1987rm} proved that non-trivial bounces were associated with exactly one negative eigenvalue. To understand this, consider the fact for non-trivial solutions, in the zero-eigenvalue solution $\dot q_b$ the particle travels away from $q_0$ then turns around and comes back. It thus contains one mode and cannot be the ground-state.  This motivates our choice of $q_0$ --- picking non-trivial paths that begin and end at the false vacuum allows us to extract an imaginary part of the energy. The assumption that only a particular bounce and quadratic fluctuations about it contribute to the path integral is implicit in \cref{eq:gamma_zero_temperature}. If there are multiple solutions, they should be summed.

We close by trying to make sense of what we have done. It may be seen from the stationary phase approximation that dominant classical contributions to the path integral come from the classical equations of motion. There is, though, no classical analogue of tunneling, as a particle is forbidden from escaping the barrier by conservation of energy. We thus should not be completely surprised that the dominant contribution came from the \emph{Euclideanized} equations of motion in which the potential was upturned, creating a semi-classical analogue to tunneling and permitting us to find the leading semi-classical contribution.  

\subsection{Rates at finite temperature}\label{sec:rates_finite_t}

Let us motivate the finite-temperature expression \cref{eq:p_T} which describes thermally averaged tunneling or thermal fluctuations over the barrier. Consider the Euclidean path integral for a particle beginning at $q_0$ and ending at $q_0$ a Euclidean time $\beta$ later. By summing over possible states $q_0$, we obtain
\begin{equation}\label{eq:beta_path_integral}
\sum \langle q_0 | e^{-\beta \hat H }| q_0 \rangle =  \int\limits_{q(\tau) = q(\tau + \beta)} \!\!\!\!\!\!\!\! [\dint q] \, e^{- \SE[q]}.
\end{equation}
We are thus summing paths that are periodic in Euclidean time with period $\beta$. The left-hand side is nothing but $Z(\beta) \equiv \Tr e^{-\beta H}$, that is, the thermal partition function with temperature $T = 1 / \beta$. Similar to the case in QM, we may extract the imaginary part of the free energy from the thermal partition function since $Z(\beta) = e^{-\beta F(\beta)}$. 

Just as the imaginary part of the energy was related to the decay rate at zero temperature, the imaginary part of the free-energy is related to the decay rate at finite temperature. The finite-temperature decay rate is obtained by Boltzmann averaging the decay rate. If the temperature is small compared to the barrier height, averaging yields~\cite{Langer:1969bc,LANGER197461,Affleck:1980ac}
\begin{equation}\label{eq:width_finite_temperature_tunneling}
\Gamma = -2 \Im F(\beta).
\end{equation}
This may be seen as a thermally averaged tunneling rate \cref{eq:rate_from_imaginary_energy}. At temperatures above the barrier height, on the other hand, averaging results in~\cite{Affleck:1980ac}
\begin{equation}\label{eq:width_fluctuation}
\Gamma = - \frac{\sqrt{-\lambda_{-}} \beta}{2 \pi} \, 2 \Im F(\beta),
\end{equation}
where $\lambda_{-}$ is the negative eigenvalue of the bounce. Physically, it represents the exponential growth rate of the bubble because the negative eigenvalue corresponds to a dilatation of the bounce solution. Even this, though, neglects interactions between the escaping particle and the thermal plasma, which result in an additional damping factor that slows the exponential growth rate of the bubble~\cite{Berera:2019uyp}, and non-equilibrium considerations~\cite{Gleiser:1993hf}. The rates \cref{eq:width_finite_temperature_tunneling,eq:width_fluctuation} coincide at $T_0 = \sqrt{-\lambda_{-}} / (2 \pi)$. 

\begin{figure}
\centering
\includegraphics[width=0.6\linewidth]{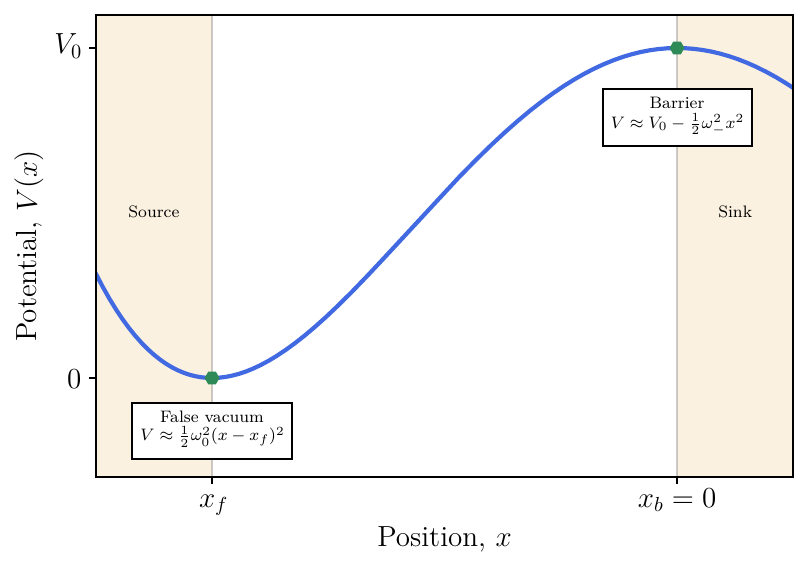}
\caption{A classical particle escapes the well by reaching the top of the barrier. In the computation, particles that reach the top disappear (the sink) and reappear again at the false vacuum (the source).}
\label{fig:escape}
\end{figure}

\Cref{eq:width_fluctuation} may be interpreted as particles fluctuating and escaping over the barrier. To see this, we now show that a remarkably similar result emerges from solving Kramers' escape problem~\cite{KRAMERS1940284,RevModPhys.62.251}. Consider the classical computation of the escape rate of the generic one-dimensional potential with a false vacuum and a barrier in \cref{fig:escape}. The result shouldn't depend on the potential beyond the barrier. In our approach to this problem that region plays the role of a sink and particles that reach it reappear at a source at the false vacuum. This enables us to compute the fluctuation rate over the barrier as the equilibrium rate at which particles go round this system. We approximate the rate by the probability that the particle reaches the barrier at $x = 0$ moving towards the sink multiplied by the expected rate at which it moves towards the sink, 
\begin{equation}\label{eq:gamma}
\Gamma \simeq \Pr(x = 0, p > 0) \cdot\left \langle \, p \, \middle| \, x = 0,  p > 0 \, \right\rangle =
\frac{\int e^{-\beta V(x)} \delta(x) \, \dint x
}{
\int e^{-\beta V(x)} \, \dint x
}
\frac{
\int_{p > 0} p e^{-\beta p^2 / 2} \, \dint p
}{
\int e^{-\beta p^2 / 2} \, \dint p
}
\simeq \frac{\omega_0}{2\pi} e^{-\beta V_0} ,
\end{equation}
where in the denominator we expanded the potential around the false vacuum to quadratic order, $\omega_0^2 \equiv V^{\prime\prime}(x_f)$. This assumes that equilibrium always holds and that nothing comes back over the barrier. On the other hand, the free energy may be expressed as
\begin{equation}\label{eq:beta_im_f}
\beta \Im F = \Im \log Z \simeq \frac{\Im Z}{Z} =
\frac{\Im \int e^{-\beta V(x)} \, \dint x
}{
\int e^{-\beta V(x)} \, \dint x
} 
\frac{
\int e^{-\beta p^2 / 2} \, \dint p
}{
\int e^{-\beta p^2 / 2} \, \dint p
}
\simeq
\frac{\omega_0}{\omega_{-}} e^{-\beta V_0},
\end{equation}
where  $\omega_-^2 \equiv -V^{\prime\prime}(x_b)$. We assumed that the dominant real contribution came from around the false vacuum and that the imaginary component came  from around the barrier, since, as in \cref{sec:rates_qm}, an imaginary component originates from the square root of a negative eigenvalue, corresponding to the convex region around the barrier. This allowed us to compute the numerator and denominator using Laplace approximations. Thus \cref{eq:gamma,eq:beta_im_f} are related by the factor ${\omega_{-} \beta} / {(2\pi)}$. A similar result may be obtained by modeling the thermal bath by stochastic forces perturbing the particle~\cite{Berera:2019uyp}.

Let us recapitulate. At $T \lesssim T_0$, \cref{eq:width_finite_temperature_tunneling} applies and may be interpreted as a thermally-averaged tunneling rate through the barrier, whereas at $T \gtrsim T_0$, \cref{eq:width_fluctuation} applies and may be interpreted as an escape rate over the barrier by thermal fluctuations. In other words, quantum tunneling and thermal fluctuations are limiting cases of a single quantum description~\cite{Affleck:1980ac}. There is a narrow regime at $T \simeq T_0$ in which neither formula precisely describes the rate. In both regimes \cref{eq:width_finite_temperature_tunneling,eq:width_fluctuation} the rate is proportional to the imaginary part of the free-energy.

Thus at finite-temperature we use an identical formalism as at zero temperature to extract the imaginary part from the path integral in \cref{eq:beta_path_integral} but consider only periods of Euclidean time $\beta$~\cite{Linde:1977mm, Linde:1980tt, Linde:1981zj}. The low-temperature $\beta \to \infty$ limit of tunneling reproduces the previous $T = 0$ tunneling results. In the high-temperature limit the periodicity of the bounce goes to zero such that it must be time-independent. In this limit the integration over a period $\beta$ in \cref{eq:euclidean_action} contributes a trivial factor $\beta$, such that
\begin{equation}\label{eq:trivial_beta_path_integral}
\SE = \beta S_3 = \frac{S_3}{T} ,
\end{equation}
where $S_3$ is the three-dimensional action and we search for $O(3)$ symmetric bounces. At low temperatures the periodicity becomes relevant and this breaks down, however. For tunneling, the Wick rotation facilitated a semi-classical description of a quantum phenomenon by upturning the potential. For the thermal tunneling, the Wick rotation facilitated a path integral description of thermal averaging. For fluctuations over the barrier, the Wick rotation together with \cref{eq:trivial_beta_path_integral} transforms $e^{i S}$ into a Boltzmann suppression factor $e^{-\beta \Delta F}$ where $\Delta F$ is the free energy required to escape over the barrier.

\subsection{The bounce in quantum field theory}\label{sec:qft_bounce}

Let us now move to the field theory setting, where a real multidimensional scalar field $\phi(\tau, \vec x)$ plays the role of position~\cite{Coleman:1977py}. The generalization of the one-dimensional quantum mechanical problem yields coupled ordinary differential equations as equations of motion,
\begin{equation}\label{eq:bounce_equation}
\grad^2 \phi_i = \frac{\partial V}{\partial \phi_i} ,
\end{equation}
where $\grad^2 = \frac{\partial^2}{\partial \tau^2} + \sum_i \frac{\partial^2}{\partial x_i^2}$ is the Laplace operator. The equations may be coupled through interaction terms in the potential such that the right-hand side depends in general on all the field coordinates. We must solve \cref{eq:bounce_equation} subject to the boundary conditions that the field begins and ends at rest at the false vacuum at $\tau = \pm\infty$. By convention, we take the moment at which the field crosses back and forth between the minima to be $\tau = 0$; other possibilities result in zero eigenvalues in \cref{eq:product_eigenvalues} and were accounted for by \cref{eq:omit_zero_eigenvalues}. Since the problem possesses a reflection symmetry $\tau \to -\tau$, the solutions from the false vacuum are the time-reversal of the solutions back to the false vacuum. 

We assume that the least action non-trivial bounce solution obeys an $O(d)$ symmetry around a center of the bounce. Thus by translation symmetry taking the center at $\tau = 0$ and $\vec x  = 0$ (other possible choices are ultimately accounted for in a manner similar to that in \cref{eq:omit_zero_eigenvalues}), the solution must be a function of
\begin{equation}
\rho = \sqrt{\tau^2 + |\vec{x}|^2}.
\end{equation}
This was proven in \refcite{Coleman:1977th} for $d > 2$ for a single field under certain regularity conditions for the potential. Progress towards a proof for multi-field cases is discussed in \refcite{Blum:2016ipp}. Thus bubbles nucleated through quantum tunneling are $O(4)$-symmetric, while bubbles nucleated through thermal fluctuations are $O(3)$-symmetric. 

Writing the Laplace operator in $d$-dimensional spherical coordinates and taking $\phi(\tau, \vec x) = \phi(\rho)$ leads to coupled bounce equations,
\begin{equation}\label{eq:rho_bounce_equation}
\frac{\partial^2 \phi_i}{\partial \rho^2} + \frac{d - 1}{\rho} \frac{\partial \phi_i}{\partial \rho} =  \frac{\partial V}{\partial \phi_i}. 
\end{equation}
The boundary conditions are that the particle begins at rest at $\rho = 0$, and asymptotically reaches the false vacuum,
\begin{equation}\label{eq:boundary_conditions}
\left.\frac{\partial \phi(\rho)}{\partial \rho}\right|_{\rho=0} = 0 \quad\text{and}\quad \lim_{\rho\to\infty}\phi(\rho) = \fv. 
\end{equation}
The former defines the center of the bounce. The latter follows from the fact that we must reach the false vacuum at $\tau = \pm \infty$ and the fact that the bounce action must be finite such that $\lim_{|\vec x| \to\infty} \phi(\tau, \vec x) = \fv$. This is consistent with our introductory remarks that the bounce is localized to a bubble and far away from it we remain in the false vacuum. The corresponding action may be written
\begin{equation}\label{eq:bounce_action}
\SE[\phi] = \mathcal{S}_{d-1} \int_0^\infty \rho^{d - 1} \left[\tfrac12\dot\phi^2 + V(\phi) \right] \dint \rho,  
\end{equation}
where $\mathcal{S}_{n}$ is the surface area of an $n$-sphere with unit radius, that is, the surface area of a unit sphere that is embedded in an $n+1$ dimensional space.

Since it is an extremum, we know that if we write a bounce solution as $\phi(a \rho)$, the action must be extremized by $a = 1$. By making a change of variables, $\sigma = a \rho$, we may write the stretched bounce action as~\cite{Derrick:1964ww}
\begin{equation}
\SE[\phi(a \rho)] = \frac{\mathcal{S}_{d-1}}{a^d} \int_0^\infty \sigma^{d - 1} \left[a^2 \frac12 \left(\frac{\dint \phi(\sigma)}{\dint \sigma}\right)^2 + V(\phi(\sigma)) \right] \dint \sigma .
\end{equation}
As we know $\delta \SE / \delta a|_{a=1} = 0$, the kinetic and potential terms in the bounce action must be related by
\begin{equation}\label{eq:action_relation}
\SV = \frac{2 - d}{d} \ST
\end{equation}
such that the bounce action may be written purely as kinetic or potential contributions,
\begin{equation}\label{eq:action_only_t_or_v}
\SE = \frac{2}{d} \ST = \frac{2}{2 - d} \SV. 
\end{equation}
Incidentally, by considering $\delta^2 \SE / \delta a^2|_{a=1}$, we may see that the dilatation $\phi(a \rho)$ corresponds to the negative eigenvalue. 

We may think of this as an unusual mechanics problem involving a particle in an upturned potential, where $\rho$ plays the role of time and $\phi$ plays the role of position. In this case, we seek a solution whereby a particle begins at rest near the global maximum, rolls down through the local minimum and comes to rest at the local maximum. The damping required to stop the particle shooting over the local maximum was provided by an unusual friction term that decreases with time as $1 / t$. This is illustrated in \cref{fig:bounce}. Although in this formulation energy is no longer manifestly conserved, there remains a symmetry $\rho \to -\rho$ such that were particle to start at $\rho = -\infty$, it would climb up to the starting position at $\rho = 0$ and then bounce back. The outstanding challenge, then, is to solve \cref{eq:rho_bounce_equation} subject to the boundary conditions in \cref{eq:boundary_conditions}. This amounts to finding the approximate coordinates $\phi_0 \equiv \phi(\rho = 0)$ from which the particle begins at rest and would subsequently roll to the false vacuum. This can be simple for a single scalar field, but challenging for even two or more. We discuss this numerical challenge in the next section.

\subsection{Computing the bounce}\label{sec:compute_bounce}

\begin{figure}[t]
\centering
\includegraphics[width=0.6\linewidth]{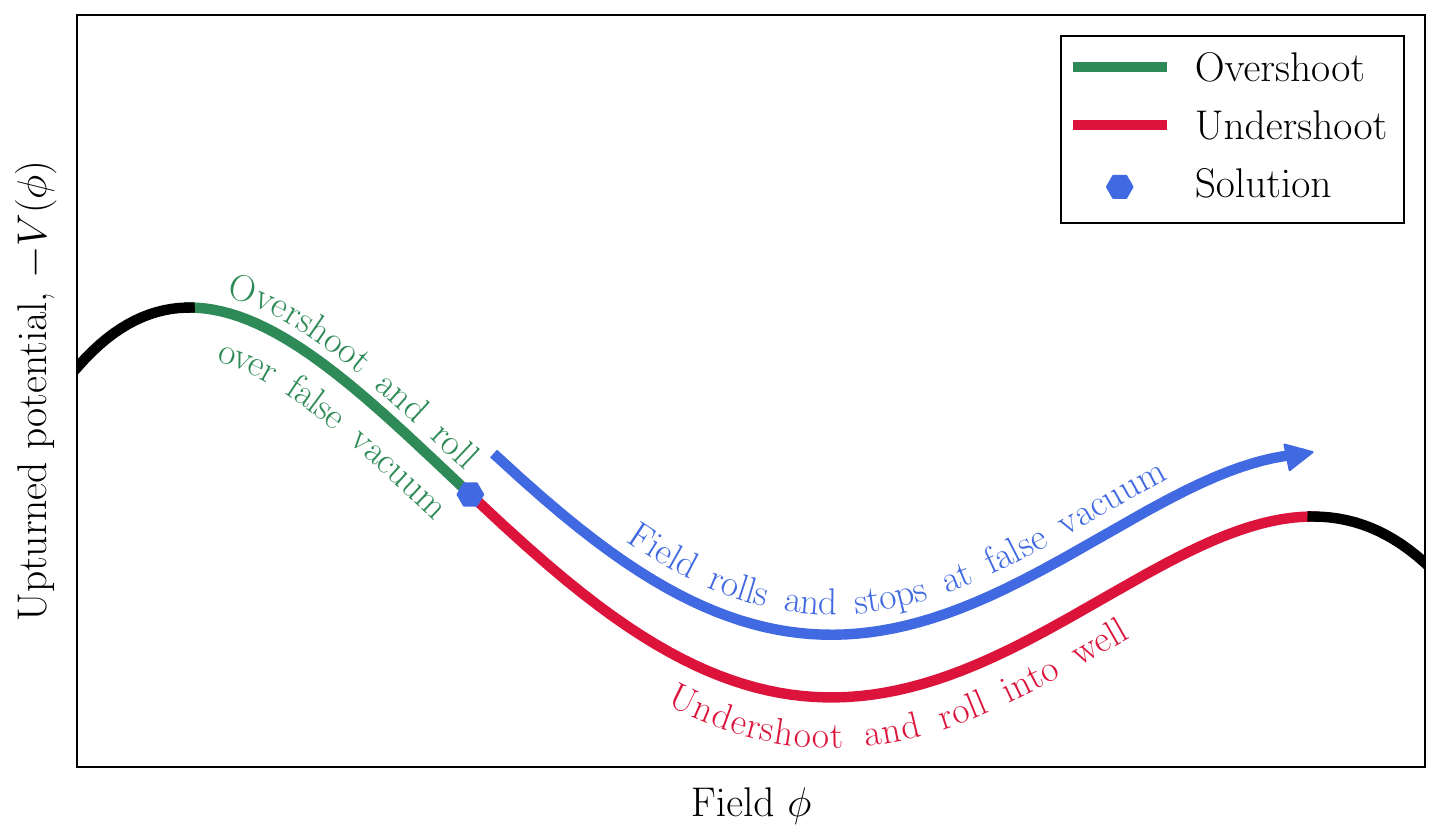}
\caption{The one-dimensional shooting method. The regions of the upturned potential are colored by the result of a shot from that position. At the boundary between overshots (green) and undershots (red), there exists a solution (blue) that asymptotically lands at the false vacuum.}
\label{fig:shoot}
\end{figure}

\begin{figure}[t]
\centering
\includegraphics[width=0.6\linewidth]{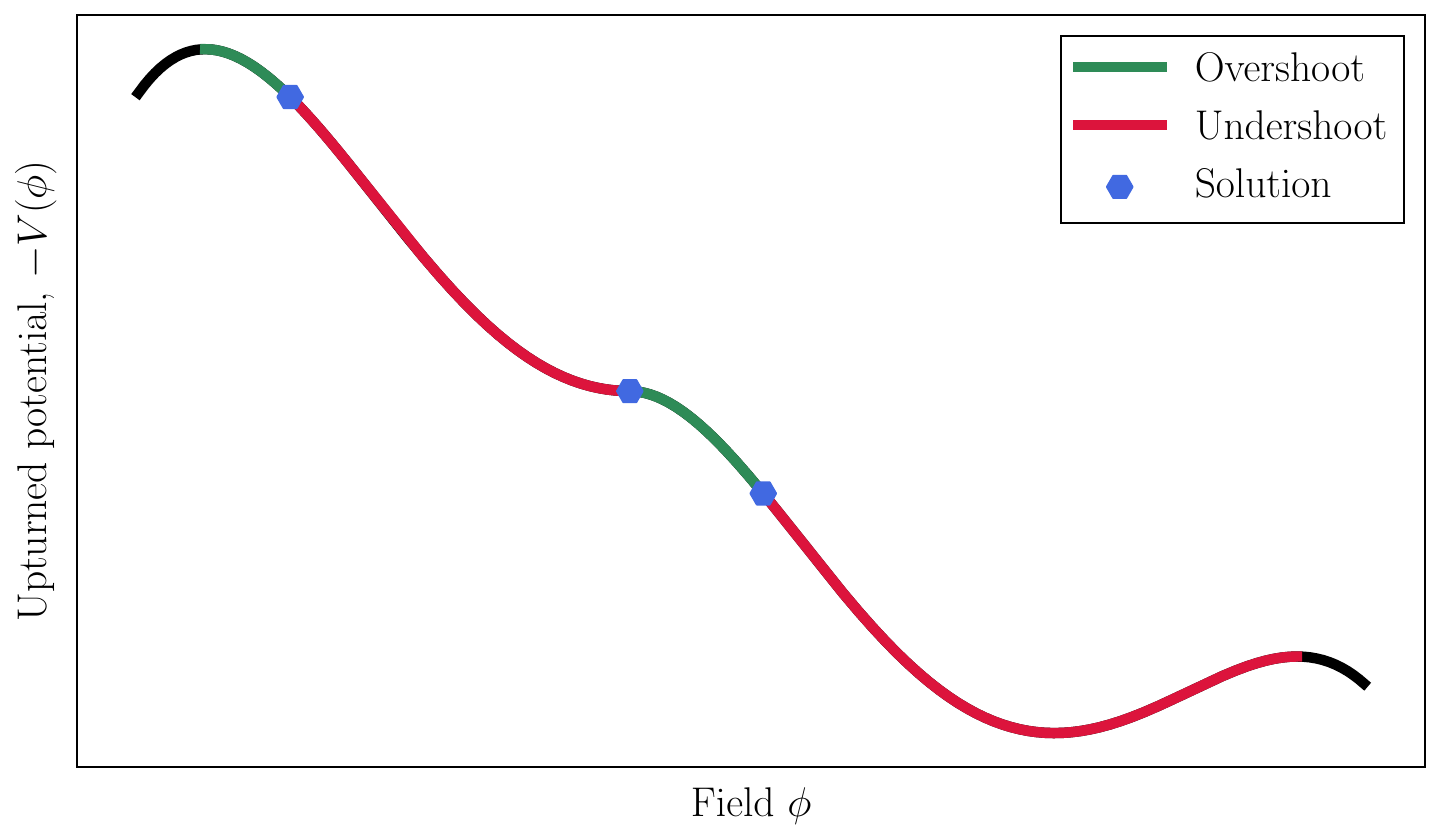}
\caption{Sketch of a one-dimensional potential with three bounce solutions at the boundaries between over and undershots. Based on figure 3 from \refcite{Masoumi:2016wot}.}
\label{fig:sols}
\end{figure}

Coleman~\cite{Coleman:1977py} proved that a bounce solution always exists for one scalar field with a true and false vacuum by an overshot/undershot argument. First, note that the friction term is dissipative,
\begin{equation}
\frac{\dint}{\dint \rho} \left(\frac12 \dot\phi^2 - V(\phi) \right) = - \frac{d-1}{\rho} \dot\phi^2 \le 0.
\end{equation}
Thus, as illustrated in \cref{fig:shoot}, for one-field it is easy to see by conservation of energy that there always exist starting positions from which the particle would fail to reach the false vacuum --- it would undershoot and roll back into the well. Similarly, by starting sufficiently close to the true vacuum, the particle can wait until friction which falls as $1 / t$ becomes negligible and roll over and past the false vacuum. By continuity, between these cases there must exist a bounce solution that asymptotically reaches the false vacuum. This only proves the existence of at least one solution for one field when true and false vacua exist. Dropping that condition and generally for more than one field a solution does not always exist~\cite{Mukhanov:2021kat}. The non-existence of a solution does not mean that the false vacuum is stable, just that the decay rate must be computed by other means~\cite{Mukhanov:2020pau, Mukhanov:2020wim, Espinosa:2019hbm, Espinosa:2021qeo}. Furthermore, even in one-dimension, the solution needn't be unique~\cite{Masoumi:2016wot} as there exist solutions at any boundaries between regions of undershots and overshots; see \cref{fig:sols} for an example.

\begin{figure}[t]
\centering
\includegraphics[width=0.9\linewidth]{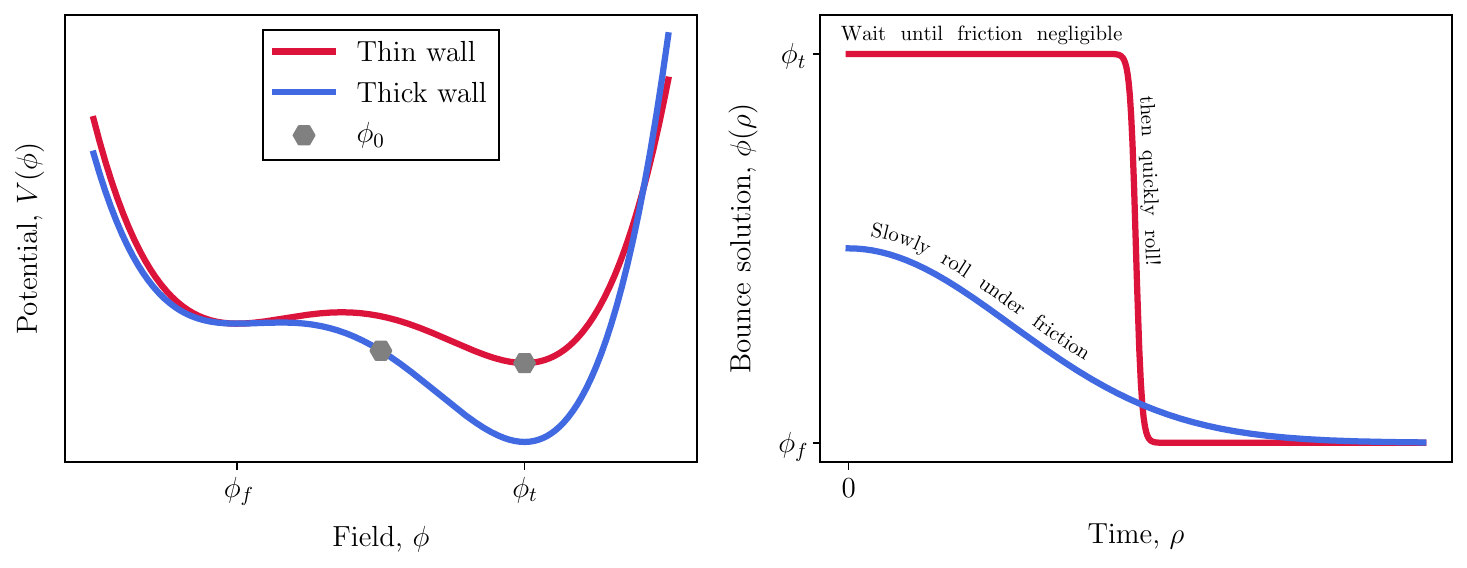}
\caption{Potentials (left) corresponding to thin-walled (red) and thick-walled (blue) bubbles (right).}
\label{fig:thin_thick}
\end{figure}

As shown in \cref{fig:thin_thick}, bounces are characterized by the width of the bubble wall. If the energy difference between the true and false vacuum is small, to avoid undershooting the particle cannot lose much energy to friction. The particle sits near the false vacuum until friction becomes negligible, then rapidly falls down the potential and towards the false vacuum. These cases are known as thin-walls; the wall being the region in Euclidean time in which the particle moves between the true and false vacuum. On the other hand, if the energy difference is substantial relative to the depth of the well, the particle may start closer to the false vacuum and roll slowly towards it; these cases are known as thick-walls. The bounce action diverges with thinness, as the particle sits for longer and longer at the true vacuum such that $\SV$ diverges.

There exist few exact analytic solutions for the field profile or action in one-dimension. Importantly, the thin-wall case was solved by Coleman~\cite{Coleman:1977py}. A one-dimensional thin-wall potential was written
\begin{equation}\label{eq:coleman_thin_wall_generic_example}
V(\phi) = V_+(\phi) + \frac\epsilon{2a} (\phi - a) ,
\end{equation}
where $V_+(\phi)$ is a symmetric potential with minima at $\phi = \pm a$ with $V_+(\pm a) = 0$. The second term breaks the symmetry and the small parameter $\epsilon$ controls thinness. To leading order, $\epsilon$ is the difference in energy density between the true and false vacua which lie at $\phi \simeq \pm a$. From the previous considerations, we expect that the field waits at the true vacuum at $\phi \simeq -a$ until friction is negligible, then quickly rolls to the false vacuum at $\phi \simeq a$. That is,
\begin{equation}
\phi(\rho) \simeq \begin{cases} 
          -a & \rho \ll R \\
          \phi_1(\rho - R) & \rho \approx R\\
          a & \rho \gg R \\
       \end{cases}
\end{equation}
where $R$ determines when the field rolls, and it rolls with profile $\phi_1(\rho - R)$ where $\phi_1$ is an odd function. We denote this profile by $\phi_1$ because as we shall see it is the bounce for a one-dimensional action. 

We may approximate the action by a piece for each of the three periods. The exterior of the bubble at $ \rho \gg R$ contributes nothing to the action, as the field is constant and the potential is zero. The interior of the bubble at $\rho \ll R$ contributes only through the potential $V(-a) = -\epsilon$, giving $-\mathcal{V}_d R^d \epsilon$, where $\mathcal{V}_n$ is the volume of an $n$-dimensional unit sphere. The bubble wall at $\rho \approx R$ contributes to the action through kinetic and potential energy, though may be approximated
\begin{align}\label{eq:S1_approx}
\SE ={}& \mathcal{S}_{d-1} \int_{\rho \approx R} \left[\frac1 2\dot\phi^2 + V(\phi) \right] \rho^{d-1}  \dint \rho\\
 \approx{}& \mathcal{S}_{d-1} R^{d-1} \int_{\rho \approx R} \left[\frac12 \dot\phi^2 + V_+(\phi) \right] \dint \rho\\ 
 \approx{}& \mathcal{S}_{d-1} R^{d-1} S_1 ,
\end{align}
where the one-dimensional action,
\begin{equation}\label{eq:1d_action}
S_1 = \int_{0}^{\infty} \left[\frac12 \dot\phi^2 + V_+(\phi) \right] \dint \rho
\end{equation}
depends on neither $R$ nor $\epsilon$. In the first approximation, we replace $\rho^{d-1} = R^{d-1}$ inside the integral, as it is approximately constant whereas the field velocity and potential are changing, and neglect the $\epsilon$-dependent term from the potential. In the second one, we change the integration limits to the whole bubble, as the regions in the interior and exterior now contribute nothing. The equation for motion for \cref{eq:1d_action} contains no friction term and so our ansatz $\phi_1(\rho - R)$ for the period $\rho \simeq R$ where friction is negligible is in fact the bounce for this one-dimensional action. 
Thus in total,
\begin{equation}
\SE \simeq \mathcal{S}_{d-1} R^{d - 1} S_1 - \mathcal{V}_d R^d \epsilon,
\end{equation}
where $R$ remains unknown.\footnote{There is a typo in the equivalent Eq.~(4.15) in \refcite{Coleman:1977py}: there should be a factor of $2$ in front of the second term.}
The terms may be interpreted as a competition between tension in the bubble wall  --- bubble surface area multiplied by surface energy density, $S_1$ --- and the outward pressure of the bubble interior --- the bubble volume multiplied by energy density difference between the true and false vacuum, $\epsilon$. By least action, we find the bubble radius
\begin{equation}
R = \frac{S_1}{\epsilon} \frac{d-1}{d} \frac{\mathcal{S}_{d-1}}{\mathcal{V}_d} = (d - 1) \frac{S_1}{\epsilon},
\end{equation}
where we used the recurrence relation $\mathcal{S}_{n-1} = n \mathcal{V}_n$ in the final equality, and so
\begin{equation}\label{eq:sn}
\SE = \frac{(d - 1)^{d-1} \mathcal{V}_d S_1^d}{\epsilon^{d-1}}
\end{equation}
such that action diverges with thinness, as anticipated. E.g., for $d = 4$ dimensions,
\begin{equation}\label{eq:R_dim_4}
R = 3 S_1 / \epsilon \quad\text{and}\quad \SE = 27 \pi^2 S_1^4 / (2 \epsilon^3) ,
\end{equation}
where we used $\mathcal{V}_4 = \pi^2 / 2$.

For a specific problem of the form \cref{eq:coleman_thin_wall_generic_example} such as 
\begin{equation}\label{eq:coleman_thin_wall_example}
V_+(\phi) = \frac{\lambda}{8} (\phi^2 - a^2)^2,
\end{equation}
it still remains to compute the one-dimensional action and the profile around $\rho \approx R$. The latter may be computed neglecting friction in the bounce equation (\cref{eq:rho_bounce_equation})  and the $\epsilon$-dependent parts of the potential, so depends on neither $\epsilon$ nor dimension. In the case \cref{eq:coleman_thin_wall_example}, \refcite{Coleman:1977py} gives the one-dimensional solution, 
\begin{equation}
S_1 = \frac{a^3 \sqrt{\lambda}}{3} \quad\text{and}\quad \phi_1(\rho - R) = a \tanh \left[a \sqrt{\lambda} (\rho - R) / 2 \right] ,
\end{equation}
and the solutions for $d = 4$ and $d=3$ are
\begin{equation}
S_4 = \frac{8 \pi^2 a^{12} \lambda^2}{3 \epsilon^3} \quad\text{and}\quad S_3 = \frac{128 \pi a^9 \lambda^{3/2}}{81 \epsilon^2}.
\end{equation}
There are other soluble cases. For example, a polynomial called the generalized Fubini potential~\cite{Fubini_1976,Lipatov:1976ny,Aravind:2014pva}, where
\begin{align}
V = \frac{4 u n^2 (n - 1)}{2 n + 1} \phi^{(2n + 1)/n} - 2 u v n^2 \phi^{(2n + 2) / n} , \\
\phi(\rho) = \frac{1}{\left(u + v \rho^2\right)^n} \quad\text{and}\quad \SE = \frac{n \pi^2}{4 n^2 - 1} \frac{1}{u v^{2n - 1}}
\end{align}
for $d = 4$ and with $u > 0$, $v > 0$ and $n > 1$.
Lastly, the simple logarithmic potential for $d = 4$~\cite{FerrazdeCamargo:1982sk},
\begin{equation}
V = \tfrac12 m^2 \phi^2 \left[1 - \ln\left(\frac{\phi^2}{\omega^2}\right)\right] \quad\text{and}\quad \phi(\rho) = \omega e^{-\tfrac12 m^2 \rho^2 + 2} \quad\text{and}\quad \SE = \frac{\pi^2 e^4}{2} \frac{w^2}{m^2}.
\end{equation}
There furthermore exist solutions for piece-wise potentials in which linear~\cite{DUNCAN1992109}, quadratic~\cite{Hamazaki:1995dy,Pastras:2011zr,Dutta:2011ej} and quartic~\cite{Lee:1985uv,Dutta:2011rc} potentials are joined. The one-dimensional case can in any case usually be solved numerically by bisecting between under and overshots.

There moreover exist relations between bounce solutions. The bounce action is invariant under $V \to V + \Delta$ and $\phi \to \phi + \Delta \phi$. Rescaling the potential $V \to \alpha^2 V$ the bounce equation \cref{eq:rho_bounce_equation} may be written
\begin{equation}
\alpha^2 \left[\frac{\partial^2 \psi_i(\rho)}{\partial (\alpha \rho)^2} + \frac{d - 1}{\alpha \rho} \frac{\partial \psi_i(\rho)}{\partial (\alpha \rho)}\right] =  \alpha^2 \frac{\partial V(\psi)}{\partial \psi_i} .
\end{equation}
Thus the bounce solution for the scaled potential is related to the solution to the original potential by $\psi(\rho) = \phi(\alpha\rho)$. Similarly, the action may be written as
\begin{equation}
\SV[\psi] = \alpha^2 \mathcal{S}_{d-1} \int \rho^{d - 1} V(\psi(\rho)) \, \dint \rho = \alpha^{2-d} \mathcal{S}_{d-1} \int (\alpha\rho)^{d - 1} V(\phi(\alpha \rho)) \, \dint (\alpha \rho) = \alpha^{2-d} \SV[\phi].
\end{equation}
By \cref{eq:action_only_t_or_v}, this means that the whole action scales as $\alpha^{2-d}$. Similarly, defining a new potential $V(\beta \psi)$, the bounce equation becomes
\begin{equation}
\frac{\partial^2 (\beta \psi_i(\rho))}{\partial (\beta\rho)^2} + \frac{d - 1}{\beta\rho} \frac{\partial (\beta\psi_i(\rho))}{\partial(\beta\rho)}  =  \frac{\partial V(\beta \psi)}{\partial (\beta \psi_i)}. 
\end{equation}
We see that the solution must be $\beta \psi(\rho) = \phi(\beta \rho)$ with action
\begin{equation}
\SV[\psi] = \mathcal{S}_{d-1} \int \rho^{d - 1} V(\phi(\beta\rho)) \, \dint \rho = \beta^{-d} \mathcal{S}_{d-1} \int (\beta\rho)^{d - 1} V(\phi(\beta\rho)) \, \dint (\beta\rho) = \beta^{-d} \SV[\phi].
\end{equation}
This is consistent with a change of units: under a change of units the potential changes to $u^4 V(\phi / u)$. Applying our two transformation rules we find, $S \to u^{4 - 2d} u^d S = u^{d - n} S$. This is exactly as we expect since $[S] = 4 - d$. The bounce solution changes to $u \phi(u^2 / u \rho) = u \phi(u\rho)$, which is just the original solution expressed in the new units.

In summary, so far we know i) that bounces always exist for one field, ii) a few explicit analytic solutions and iii) how bounce solutions are related through rescalings of the field and potential. Finding bounce solutions for general potentials, though, remains challenging, even for only one field and even numerically. First, the friction term blows up as $1/\rho$. Even if numerical singularities near $\rho = 0$ are avoided through e.g., changes of variable, evolving the field by the equations of motion during this period is slow due to friction. This can be overcome by Taylor expanding the potential to quadratic order and evolving using analytic solutions in this regime. 

Even for one field, thin-walled cases require extreme fine-tuning of the solution $\phi_0$. This may require working in logarithmic space and bisecting until we reach a solution.
For more than one field, we cannot simply bisect between undershots and overshots. Failed shots runaway to $|\phi| \to \infty$ in some direction (or possibly get stuck in wells), leaving us scant information about the location of a solution. Similarly, the evolution is unstable, as particles must be delicately balanced as they pass between maxima of the upturned potential. For this reason, tailored numerical methods were developed for the problem, many in publicly available software packages; see e.g., the codes
\code{AnyBubble}~\cite{Masoumi:2017trx}, 
\code{BubbleProfiler}~\cite{Athron:2019nbd},
\code{FindBounce}~\cite{Guada:2020xnz,Guada:2021lnb},
\code{CosmoTransitions}~\cite{Wainwright:2011kj},
and
\code{SimpleBounce}~\cite{Sato:2019axv,Sato:2019wpo}, and the \code{OptiBounce} algorithm~\cite{Bardsley:2021lmq}.
As discussed, this is not a minimization problem, as the bounce is a saddle point. One approach, though, is to turn it into a minimization problem by constructing an improved action~\cite{Kusenko:1995jv} or another function that is minimized by the bounce~\cite{Cline:1999wi,Espinosa:2018hue,Espinosa:2018voj,Espinosa:2018szu}. The problem is then placed on a lattice: we discretize the time coordinate and solve for the values of the fields at each time by minimization. We now review the path deformation algorithm, as it was proven to work well and because it is physically intuitive, being related to the motion of a particle moving through the potential on rails.

\subsubsection{Path deformation}

Path deformation was developed in the code \code{CosmoTransitions}~\cite{Wainwright:2011kj} and bears similarity to a method proposed in \refcite{Cline:1999wi}. First, the bounce equation is rewritten in intrinsic coordinates by parameterizing a path by $\phi(x)$ with $x = x(\rho)$ and $|\dint \phi / \dint x| = 1$. In such coordinates,
\begin{equation}
\dot \phi = \dot x \frac{\dint \phi}{\dint x} =  \dot x \hat e_t(x) \quad\text{and}\quad \ddot \phi = \ddot x \frac{\dint \phi}{\dint x} + \dot x^2 \frac{\dint^2 \phi}{\dint x^2} = \ddot x \hat e_t(x) + \dot x^2 \left|\frac{\dint^2 \phi}{\dint x^2}\right| \hat e_n(x) ,
\end{equation}
where we used $\dint \phi / \dint x \equiv \hat e_t(x)$, a normalized tangent to the path. On the other hand, $\dint^2 \phi / \dint x^2 \equiv |\dint^2 \phi / \dint x^2| \hat e_n(x)$ points in a normal direction to the path. With these coordinates, we may write the bounce equation of motion \cref{eq:rho_bounce_equation} as
\begin{equation}
\ddot x \hat e_t(x) + \dot x^2 \left|\frac{\dint^2 \phi}{\dint x^2}\right| \hat e_n(x) + \frac{d - 1}{\rho} \dot x \hat e_t(x) = \grad V(\phi(x)) ,
\end{equation}
where $\grad$ and the unit vectors are all in field-space. The equation thus splits into
\begin{align}
\ddot x + \frac{d - 1}{\rho} \dot x &= (\div \hat e_t(x)) V(\phi(x)) = \frac{\dint}{\dint x} V(\phi(x)) \label{eq:path_deformation_x}\\
\dot x^2 \left|\frac{\dint^2 \phi}{\dint x^2}\right| &= (\div \hat e_n(x)) V(\phi(x)) \label{eq:path_deformation_n}\\
0 &= (\div \hat e_b(x)) V(\phi(x)) , \label{eq:path_deformation_o}
\end{align}
where $\hat e_b$ represents any other binormal directions to the path (in general there are $n - 1$ normal directions for $n$ fields; the ones we denoted $\hat e_t(x)$ and $\hat e_n(x)$ form the time-dependent osculating plane). We thus may proceed as follows. We guess the path that the field takes (e.g., a straight line between the true and false vacua) and parameterize it as $\phi(x)$. We solve \cref{eq:path_deformation_x} for this path to find $x(\rho)$ using e.g., the shooting method; this supposes that the particle is just on rails rolling along the path. We then compute the normal forces along the path, $F_n(x)$, that would be required to keep the particle rolling on this track using \cref{eq:path_deformation_n}
\begin{align}
\dot x^2 \left|\frac{\dint^2 \phi}{\dint x^2}\right|  &= (\div \hat e_n(x)) V(\phi(x)) + F_n(x).
\end{align}
If the normal forces vanish, we know that our path satisfies \cref{eq:path_deformation_n}. Let us, then, try to deform the path so that the normal forces vanish. We deform it only in the normal direction,
\begin{equation}
\phi(x) \to \phi(x) + \Delta \phi_n(x) \hat e_n.
\end{equation}
Treating this a perturbation and attempting to make $F_n(x)$ vanish we find 
\begin{equation}\label{eq:delta_phi_normal}
\Delta \phi_n(x) = \frac{\dot x^2\left|\frac{\dint^2 \Delta \phi}{\dint x^2}\right| -  F_n(x)}{(\div \hat e_n(x))^2 V(\phi(x)) }.
\end{equation}
We may understand this as follows.  The change in the force provided by the potential, caused by the perturbation and found through a Taylor expansion,  must equal the change in acceleration caused by the perturbation and provide the normal forces to keep the particle on the rails. We don't solve this equation; instead in the path deformation algorithm we approximate the deformation by
\begin{equation}
\Delta \phi_n(x) = - f \frac{|\tv - \fv|}{\max_\phi |\grad V(\phi)|} F_n(x) ,
\end{equation}
where $f$ is a parameter of the algorithm. This neglects the first term in the numerator in \cref{eq:delta_phi_normal} and replaces the denominator by a constant relevant scale.

Thus we proceed iteratively; solving the speed along the path, then deforming the path, and repeating, until the normal forces vanish and we found a bounce solution.  We might wonder what happened to the equations of motion for the binormal directions in \cref{eq:path_deformation_o}; were they solved? So long as there is torsion and the tangent, normal and binormal directions change as we iterate, they are solved automatically. If this is a concern, we could extend the deformation method by deforming in binormal directions in a similar manner to \cref{eq:delta_phi_normal}
\begin{equation}\label{eq:delta_phi_binormal}
\Delta \phi_b(x) = \frac{- F_b(x)}{(\div \hat e_b(x))^2 V(\phi(x)) }.
\end{equation}
For example, if our initial guess for the path between the true and false vacua lies on the two-dimensional plane $z = 0$ of a three-field problem, without \cref{eq:delta_phi_binormal} there are only planar deformations on the $(x, y)$ plane and \cref{eq:path_deformation_o} won't be solved.

This is a fast and reliable method. The use of intrinsic coordinates means that the method scales well with the number of fields: regardless of the number of fields, we only solve the speed along the one-dimensional path, and deform the path in one normal direction in field space. The trouble caused by thin-walled cases is confined to the one-dimensional \cref{eq:path_deformation_x} where it can be carefully handled.

\subsection{Computing the prefactor}\label{sec:prefactor}

We now turn our consideration to the prefactors in \cref{eq:p_0T,eq:p_T}, which quantify the impact of one-loop corrections to the bounce. A precise computation of the prefactor has usually been considered to be of secondary importance to the exponent, since it is assumed that the latter determines the order of magnitude of the transition probability. \Refcite{Ekstedt:2023sqc}, however, argues that the prefactor generically takes an exponential form such that the relative importance of the prefactor is similar to that of any one-loop correction to a tree-level result. 

For zero-temperature tunneling and at leading order, the prefactor was found to be \cite{Callan:1977pt}
\begin{equation}\label{eq:prefactor_zero_temperature}
  A = \left(\frac{S_4[\bounce]}{2\pi} \right)^{\!\!2} \!\! \left(\frac{\detprime [-\laplacian + V''(\bounce)]}{\det[-\laplacian + V''(\fv)]} \right)^{\!\!-\half},
\end{equation}
where $S_4$ is the action of the bounce configuration $\bounce$ discussed in \cref{sec:qft_bounce}, $\fv$ is the field configuration of the false vacuum, and the prime denotes differentiation with respect to $\phi$ for the potential, and denotes the omission of zero eigenvalues from the functional determinant. Compared to \cref{eq:qm_prefactor}, the Jacobian factor is raised to the power four as this time there were four translation symmetries, and rather than just $\Tau$ in the treatment of zero eigenvalues in \cref{eq:omit_zero_eigenvalues}, we would find $\Tau \mathcal{V}$, and dividing by $\mathcal{V}$ lets us find a transition probability per unit volume.

Similarly, for finite-temperature tunneling the prefactor was found to be \cite{Linde:1981zj}
\begin{equation}
A(T) = T \left(\frac{S_3[\bounce(T)]}{2\pi T} \right)^{\!\!\frac{3}{2}} \!\! \left(\frac{\detprime[-\laplacian + V''(\bounce)]}{\det[-\laplacian + V''(\fv)]} \right)^{\!\!-\half}, \label{eq:nucleationRate-full-FT}
\end{equation}
where $S_3$ is the action of the three-dimensional bounce configuration and we work in three-dimensions as the time integration in the action was completed through \cref{eq:trivial_beta_path_integral}. There are thus three spatial translations and the Jacobian factor is raised to the power three. Rather than $\Tau \mathcal{V}$ in \cref{eq:omit_zero_eigenvalues}, however, we would find only $\mathcal{V}$, as we did not integrate over time translations. Since we are comparing with $\beta F(\beta)$ rather than $E \Tau$, however, we obtain an extra factor $T = 1 / \beta$.
Similarly, for thermal fluctuations over the barrier~\cite{Berera:2019uyp}
\begin{align}
A(T) &= \frac{\sqrt{-\lambda_{-}}}{2\pi} \left(\frac{S_3[\bounce(T)]}{2\pi T} \right)^{\!\!\frac{3}{2}} \!\! \left(\frac{\detprime[-\laplacian + V''(\bounce)]}{\det[-\laplacian + V''(\fv)]} \right)^{\!\!-\half} \\
&= \frac{1}{2\pi} \left(\frac{S_3[\bounce(T)]}{2\pi T} \right)^{\!\!\frac{3}{2}} \!\! \left(\frac{\detplus[-\laplacian + V''(\bounce)]}{\det[-\laplacian + V''(\fv)]} \right)^{\!\!-\half}
,
\end{align}
where $\detplus$ omits negative and zero eigenvalues, since by \cref{eq:width_fluctuation} we obtain an extra factor of $\sqrt{-\lambda_{-}} \beta / (2\pi)$. This extra factor may be interpreted as a dynamical factor describing the growth of supercritical bubbles; whereas \cref{eq:prefactor_zero_temperature} was an entropic statistical factor~\cite{Csernai:1992tj}, taking into account deformations of the bounce. This splitting into dynamical and statistical factors holds to all orders~\cite{Ekstedt:2022tqk}.

For a discussion of functional determinants in this context, see \refcite{Dunne:2007rt}. Whilst they are usually extremely difficult to compute, they have been computed, for example, in electroweak theory~\cite{Carson:1990jm}, the Standard Model~\cite{Isidori:2001bm} and in a quartic-quartic piecewise potential~\cite{Guada:2020ihz}. See also \refcite{Csernai:1992tj} for an improved estimation of the ratio of determinants without explicit computation.  The problem of finding the eigenvalues of \cref{eq:matrix} resembles that of solving the Schr\"{o}dinger equation.  There are few analytically solvable cases. For example, for the quartic potential in one-dimension,
\begin{equation}
V = -\frac12 q^2 + \frac{\lambda}{4} q^4
\end{equation}
with bounce solution
\begin{equation}
q(t) = \pm \sqrt{\frac{2}{\lambda}} \sech(t - t_0),
\end{equation}
we find
\begin{equation}\label{eq:poshl_teller}
M = \frac{\dint^2}{\dint t^2} + 1 - 6 \sech^2(t - t_0).
\end{equation}
Thus $M \psi(t) = \lambda \psi(t)$ is the same form as the Schr\"{o}dinger equation with a P\"{o}schl-Teller potential. The spectrum and thus product of the eigenvalues may be solved exactly. However, to find $\det M$ it is not necessary to compute a spectrum of eigenvalues, as we may exploit the Gel'fand-Yaglom theorem~\cite{Gelfand:1959nq}. Suppose we wish to find $\det M$ for
\begin{equation}
M \psi(t) = \lambda \psi(t) \quad\text{where}\quad \psi(0) = \psi(\beta) = 0.
\end{equation}
We instead may consider only
\begin{equation}
M \phi(t) = 0 \quad\text{where}\quad \phi(0) = 0 \quad\text{and}\quad \dot\phi(0) = 1.
\end{equation}
Having found $\phi(t)$, roughly speaking, the theorem tells us that
\begin{equation}
\det M = \phi(\beta).
\end{equation}
This remarkable result may be generalized to other boundary conditions and ratios of determinants. Using this result and expanding in partial waves in terms of hyperspherical harmonics, \refcite{Dunne:2005rt,Dunne:2006ct} showed that the required determinants may be written
\begin{equation}
\log \left(\frac{\detprime M}{\det M_0}\right) = \sum_{\ell = 0}^\infty g_\ell \log\left(\frac{\psi_\ell(\infty)}{\psi_{0\ell}(\infty)}\right) ,
\end{equation}
where $g_\ell$ is a degeneracy factor that depends on the number of dimensions,
\begin{equation}
g_\ell = \frac{(2\ell + d - 2) (\ell + d  - 3)}{\ell! (d - 2)!}.
\end{equation}
Unfortunately, this sum is divergent. \Refcite{Dunne:2005rt,Dunne:2006ct,Hur:2008yg} propose computing $\ell \lesssim 10$ numerically and regularizing the sum of terms $\ell \gtrsim 10$. Through approximations, the $\ell \gtrsim 10$ terms may be written analytically as integrals involving simple one-dimensional integrals that depend on the form of the potential. \Refcite{Ekstedt:2023sqc} introduces a computer program, \code{BubbleDet}, that implements a similar procedure that automatically computes and regularizes these sums.

In the absence of a proper calculation, it may be possible to approximate the functional determinants on dimensional grounds~\cite{Linde:1981zj}, though we do not know a simple way of checking the validity of such an approximation. At zero temperature, the relevant dimensional quantity is often taken to be the radius of a critical bubble, $R_0$, leading to
\begin{equation}
  A \simeq R_0^{-4} \left(\frac{S_4[\bounce(T)]}{2\pi} \right)^{\!\!2} ,  \label{eq:nucleationRate-ZT}
\end{equation}
and at finite temperature, 
\begin{equation}
  A(T) \simeq  T^4 \left(\frac{S_3[\bounce(T)]}{2\pi T} \right)^{\!\!\frac{3}{2}} \label{eq:nucleationRate-FT}
\end{equation}
as the relevant dimensional quantity is often taken to be the temperature. With such crude approximations, the prefactors for finite-temperature tunneling and thermal fluctuations are identical.

\subsection{Separation of scales}\label{sec:separation_of_scales}

There are inconsistencies in the popular approach that we sketch~\cite{Croon:2020cgk,Gould:2021ccf}. For example, if the vacuum changes due to temperature corrections, we are tempted to replace the tree-level potential by the real part of the effective potential and calculate the decay rate in the usual manner,
\begin{equation}
e^{i \int \dint^4 x (T + V)} \to e^{i \int \dint^4 x (T + \Re V_\text{eff})}.
\end{equation}
After all, the tree-level potential might not even possess more than one minimum, so we cannot begin the computation from it alone~\cite{Weinberg:1992ds}. However, there is no theoretical justification for this replacement. More alarmingly, by making it we double count the fluctuations (first to create the effective potential and second to consider fluctuations about the bounce), throw away a mysterious imaginary part of the effective potential as discussed in \cref{sec:imaginary_contributions} and introduce scale and gauge dependence as discussed in \cref{sec:ZeroTempDeltaV,sec:DeltaVT0GaugeDep}, respectively.

As anticipated in \cref{sec:3defftheory}, the solution is a separation of scales in an effective field theory (see e.g., \refcite{Strumia:1998nf} and \refcite{Gould:2021ccf} for a recent review). This shouldn't be surprising --- we already tackled the escape over the barrier using a separation of scales. In our description of the escape problem and in \cref{fig:bounce}, we focused on evolution of the scalar field (which plays the role of a macroscopic order parameter) and declined to model the microscopic degrees of freedom in the thermal bath, instead treating them as a stochastic force or assuming equilibrium. Effective field theory is thus the obvious tool to tackle this problem more generally. First, it avoids double counting as the IR and UV degrees of freedom appear separately. Second, it prevents hierarchies of scales breaking perturbation theory by resumming terms $\log \Lambda_\text{UV} / \Lambda_\text{IR}$. Roughly speaking, we introduce a scale $\Lambda_\text{UV}$ and compute an effective potential considering only UV fluctuations, $V_\text{UV}$. We make a saddle point approximation around a bounce solution of this UV effective potential. Finally, we consider fluctuations about that bounce solution. This incorporates IR fluctuations and leads to the prefactor in the decay rate, $A_\text{IR}$.

\section{First-order phase transitions} \label{sec:transitionAnalysis}

In the previous section, we demonstrated that the transition rate per unit time per unit volume was related to the imaginary part of the free energy, which could be extracted by considering a bounce. The bounces are not just a mechanism for extracting the transition rate, though; they are directly related to bubbles of true vacuum. In this section we explain how phase transitions proceed and complete through those bubbles.

\subsection{From bounces to bubbles}\label{sec:bounce_to_bubble}

In a semi-classical description of tunneling, a particle sits inside a well until it tunnels through a barrier quantum mechanically, reaches the other side at rest, and then propagates classically again from rest (see e.g.,~\refcite{Paranjape:2017}). We now describe that classical motion from rest in our four-dimensional problem following \refcite{Coleman:1978ae}. We call the classical motion of the field once it tunnels a bubble. The field is at rest at the center of the bounce, which is conventionally at $t = 0$. Thus the bubble materializes with shape given by the four-dimensional Euclidean bounce at $t=0$ and propagates classically. Fortuitously, since we already solved the Euclideanized equations of motion to find the bounce, we may find the classical propagation of the bubble by making an analytic continuation of the bounce solution,%
\footnote{See \refcite{Megevand:2023nin} for a generalization of this result, and a detailed analysis of the bubble wall evolution beyond the thin-wall approximation in a vacuum transition.}
\begin{equation}\label{eq:continue}
\phi(t, \vec x) = \phi(\sqrt{x^2 - t^2}).
\end{equation}
That is, the bounce solution in Euclidean space and the bubble in Minkowski space are related by an analytic continuation back to real time. The $O(4)$ symmetry becomes $O(3, 1)$, and so all Lorentz observers see the same bubble evolution.

To understand bubble evolution, consider a thin-wall bubble with the wall initially localized at $R_0$ at $t = 0$.
The solution \cref{eq:continue} requires $x^2 - t^2 \ge 0$ reflecting the fact that it holds only for the exterior of the bubble. Due to Lorentz the invariance of the bubble, the bubble wall is always constrained by $x^2 - t^2 = R_0^2$, so that  temporarily restoring factors of $c$, the bubble wall lies at 
\begin{equation}\label{eq:bubble_radius}
R \equiv |x| = \sqrt{c^2 t^2 + R_0^2}
\end{equation}
and so asymptotes to $|x| = c t$. That is, the bubble rapidly expands and asymptotically the bubble wall travels at the speed of light. The growth rate depends only on $R_0$ --- it roughly reaches the speed of light at time $R_0 / c$.\footnote{See \refcite{Ai:2022kqm} for a discussion of subtleties about the stability of accelerating bubbles.}
Thus the ``warning time'' between seeing a signal from the bubble wall and being hit by it is roughly $R_0 / c$. As $R_0$ is typically microscopic, this is no time at all. As we shall discuss in \cref{sec:friction}, however, this neglects friction which impedes the bubble wall velocity. See for example \refcite{Giblin:2013kea, Giblin:2014qia, Ellis:2019oqb} for the evolution of the bubble radius under the effects of friction from a coupled background plasma.

Finally, consider the energy carried by the bubble wall. The energy density of the wall for a critical bubble in the thin-wall limit was given by $S_1 = \epsilon R / 3$ by \cref{eq:R_dim_4}, where $R$ is the critical radius. To see that $S_1$ plays the role of energy density per unit area of the wall, consider the energy density around the bubble wall at $R$,
\begin{equation}
\mathcal{E} = \frac{1}{4\pi R^2} \int_{r \approx R} \left[\frac12 \dot\phi^2 + V(\phi) \right] 4\pi r^2 \, \dint r \approx S_1.
\end{equation}
The approximation follows the same reasoning as around \cref{eq:S1_approx}.

By a Lorentz transformation from the rest frame of the wall to the plasma frame, the moving wall carries an energy density $S_1 / \sqrt{1 - v^2}$, such that
\begin{equation}
E = 4 \pi |x|^2 S_1 / \sqrt{1 - v^2}.
\end{equation}
Using $v = d|x| / dt$, $|x| = \sqrt{t^2 + R_0^2}$ and assuming a critical bubble such that $R = R_0$, we obtain
\begin{equation}\label{eq:bubble_energy}
E = 4 \pi |x|^3 S_1 / R_0 = \frac{4}{3} \pi |x|^3 \epsilon.
\end{equation}
This is nothing but the volume of the bubble multiplied by the difference in energy density between the true and false vacuum. We thus see that all the energy released expands and accelerates the bubble wall --- there is no e.g., particle production and the expanding bubble leaves nothing behind but true vacuum. See e.g., \refcite{Ellis:2019oqb} for related discussions in the presence of friction impeding the bubble wall in the thin-wall and strong supercooling limits.

We can consider the relevance of gravity on bubble expansion following \refcite{Coleman:1980aw}. Whilst we might expect it to be irrelevant, we saw in \cref{eq:bubble_energy} that the energy carried by the bubble wall is proportional to the bubble volume. This means that the Schwarzschild radius (see e.g.,~\refcite{Carroll:1997ar}) of a bubble, $r_s$, could be comparable to or greater than the bubble radius, \cref{eq:bubble_radius}, for sufficiently large bubbles. This occurs when
\begin{equation}
r_s \equiv 2 G E \approx R,
\end{equation}
where the energy of the bubble, $E$, was given in \cref{eq:bubble_energy}. This usually occurs above about $1\,\text{km}$. Gravitational effects must be important for such bubbles. Indeed, the gravitational corrections impact the bounce action and appear as $R / r_s$; see e.g.\ \refcite{Samuel:1991zs, Salvio:2016mvj, Branchina:2016bws, Rajantie:2016hkj, Gialamas:2022gxv} for calculations. In the case in which transitions occurred in the past and we live in a stable or metastable ground state with vanishing vacuum energy, $V(\phi_t) = 0$, gravitational corrections increase the decay rate.

\subsection{Progress of the phase transition} \label{sec:jmak}

\subsubsection{Simplified transition scenario} \label{sec:simplifiedTransition}

A cosmological first-order phase transition (FOPT) may only be possible in a finite temperature interval in the history of the Universe, because the effective potential --- and therefore the phase structure --- changes with temperature. Within that interval, there are several important temperatures that mark significant milestones in the progress of a transition, which we discuss in \cref{sec:transitionMilestones}. To analyze a transition's progress, we make four simplifying assumptions:
\begin{enumerate}
	\item Only two phases exist during the progress of the transition.
	\item Both phases coexist across that entire temperature interval.
	\item The potential barrier between the phases vanishes at some finite temperature after the transition completes. That is, the phases are separated by a potential barrier throughout the transition, and the potential barrier does not persist at zero temperature.
	\item There is exactly one critical temperature. That is, the free energy density of one phase is always lower than that of the other phase below the critical temperature.
\end{enumerate}
We will relax these assumptions in \cref{sec:relaxAssumptions}. Note that the transition is guaranteed to complete under the assumption that the potential barrier vanishes at a finite temperature. This is because if the transition has not completed before the potential barrier vanishes, the disappearance of the potential barrier allows the field to roll down to the deeper phase, smoothly completing the transition through a second-order or cross-over transition \cite{Boyanovsky:2006bf, Bea:2021zol}.

We refer to the true vacuum at the end of the transition simply as the true vacuum, even though it does not correspond to the zero-temperature ground state of the potential. We suppose that at high enough temperature, there was a single vacuum at the origin.%
\footnote{We thereby employ the principle of symmetry restoration at high temperature \cite{Kirzhnits:1972iw,Kirzhnits:1972ut,Kirzhnits:1976ts,Weinberg:1974hy,Dolan:1973qd,Senaha:2020mop}. See e.g.\ \refcite{Biekotter:2021ysx, Carena:2021onl, Matsedonskyi:2021hti} for cases where the electroweak symmetry is not restored until well above the electroweak scale.} The highest temperature relevant to the transition is that at which the true vacuum appeared, at first as a point of inflection, $T_{\text{inf}}$. Below this temperature it became a minimum of the potential. For a FOPT, the true vacuum generically has a higher free energy density than the false vacuum at $T_{\text{inf}}$. As the Universe cools, the difference in free energy density between the two phases decreases until the critical temperature, $T_c$, is reached, which marks the temperature at which the two vacua have equal free energy density. Below this temperature, the true vacuum has a lower free energy density than the false vacuum, and a transition from the false vacuum to the true vacuum becomes possible.

\subsubsection{The JMAK equation}

The progress of a phase transition can be tracked by determining, at any point in time, the fraction of the Universe in the false and true vacuum.%
\footnote{See \refcite{Turner:1992tz} for alternative measures of the progress of the transition.}
In our simplified scenario, the entire Universe is in the false vacuum at the beginning of the transition, and the transition is complete when the entire Universe is in the true vacuum. The first appearance of the concept of false vacuum fraction is typically attributed to Guth, Tye and Weinberg from the late 1970s \cite{Guth:1979bh, Guth:1981uk}, with the false vacuum fraction $P_f$ given by
\begin{equation}
	P_f(t) = \exp(-\Vext(t)) , \label{eq:falseVacuumFraction}
\end{equation}
where $\Vext(t)$ is the fractional extended volume of true vacuum bubbles at time $t$. The meaning of extended volume will become clear when we derive this equation in the next section. The extended volume is divided by the total volume in which the transition occurs. Typically this total volume would be the entire Universe, and can be conveniently canceled out by noting that the nucleation rate is multiplied by the total volume \cite{Athron:2022mmm}. We use $\vol$ to denote a fractional volume in this section, and thus omit specifying the modifier `fractional' within the text. We note that the form of \cref{eq:falseVacuumFraction} has been known since the 1930s.%
\footnote{To our knowledge, this is mostly unknown or unacknowledged in research on cosmological phase transitions, but was recognized in \refcite{Kampfer:1988, Kampfer:1991, Yamaguchi:1994yt, Kampfer:2000gx} and Avrami's work was cited in \refcite{Csernai:1992as}.}
In other fields, this equation is known as the Johnson-Mehl-Avrami-Kolmogorov (JMAK) equation \cite{johnson1939reaction, Avrami1, Avrami2, Avrami3, kolmogorov1937statistical},%
\footnote{For an English translation of Kolmogorov's work, see \refcite{Shiryayev1992}, and for a review of the JMAK equation, see \refcite{fanfoni1998johnson}. This form was independently derived in \refcite{TF9454100365} a few years later.}
where much work has gone into verifying its correctness and applicability to a wide variety of phase transitions.

In the upcoming derivation of the JMAK equation, we endeavor to provide a clear picture of the correctness of the extended volume approach, because the concept of extended volume has caused much confusion in the past. Over the decades since the conception of the JMAK equation, various methods for deriving it have been used. Here we present one such derivation, but note that other methods lead to the same equation. We provide a very thorough derivation due to potential confusion arising from the omission of some steps. We follow the method of \refcite{Guth:1979bh, Guth:1981uk}, familiar to the field of cosmological phase transitions. We provide a novel alternative derivation of the JMAK equation in \cref{sec:altJMAK} that explicitly treats bubble nucleation as a stochastic process.

We start by emphasizing the equivalence of the false vacuum fraction $P_f$ and the probability of a point chosen at random being in the false vacuum, and indeed to the probability of a point chosen at random {\it not} being in the true vacuum. We first consider just bubbles of the true vacuum that have fractional volumes between $\vol_1$ and $\vol_2$ and define $P_f(\vol_1, \vol_2)$ to be the probability of a random point not being enveloped by at least one such bubble. Now consider an infinitesimally increased volume range $[\vol_1, \vol_2 + d\vol]$. The probability of a random point not being enveloped by at least one bubble within this volume range is
\begin{equation}
  P_f(\vol_1, \vol_2 + d\vol) = P_f(\vol_1, \vol_2) P_f(\vol_2, \vol_2 + d\vol) ,
  \label{Eq:PfV1V2pdV}
\end{equation}
since the two probabilities are independent. We may re-express this using the fact that $P_t(\vol_2, \vol_2+ d\vol)$, the probability of a random point
being inside true vacuum bubbles that have volumes in the range $[\vol_2, \vol_2 + d\vol]$, should be the total volume of bubbles with individual volumes of $\vol_2$, multiplied by the infinitesimal volume width $d\vol$,
\begin{equation}
	P_t(\vol_2, \vol_2 + d\vol) = n(\vol_2) \vol_2 d\vol , \label{eq:changeInTrueVacuumProbability}
\end{equation}
where $n(\vol)$ is the distribution for the number density of bubbles with volume $\vol$. We implicitly assume in \cref{eq:changeInTrueVacuumProbability} that bubbles of volume $\vol_2$ do not overlap each other. This assumption is valid if the fractional volume of bubbles is small and the number density of bubbles of volume $\vol_2$ is small. Using $P_f = 1 - P_t$ in \cref{eq:changeInTrueVacuumProbability} for a transition in isolation, we have
\begin{equation}
	P_f(\vol_2, \vol_2 + d\vol) = 1 -  n(\vol_2) \vol_2 d\vol . \label{eq:changeInFalseVacuumProbability}
\end{equation}
We can substitute \cref{eq:changeInFalseVacuumProbability} into \cref{Eq:PfV1V2pdV}, leading to the differential equation
\begin{equation}
	\recip{P_f(\vol_1, V_2)} \dv{P_f(\vol_1,\vol_2)}{\vol_2} = -n(\vol_2) \vol_2 .
\end{equation}
This can easily be solved by integrating over the volume, as done in the appendix of \refcite{Guth:1981uk}. However to put this in the context of the progress of the phase transition where bubbles can form at different times, we additionally write the volumes as functions $\vol(t',t)$ of the time they formed, $t'$, and the sample time $t$. We assume $\vol(t', t)$ is a monotonically decreasing function of the nucleation time $t'$. That is, bubbles nucleated earlier will have a larger volume, having grown over a duration of $t - t'$. We then integrate over the volumes of all bubbles that nucleated from the start of the transition, $t_0$, to some sample time $t$, with all bubbles growing up to time $t$, giving
\begin{align}
	\int_{\vol(t, t)}^{\vol(t_0, t)} d\vol' \, \dv{\vol'} \ln(P_f(\vol', \vol(t_0, t))) & = - \int_{\vol(t, t)}^{\vol(t_0, t)} d\vol' \, n(\vol') \vol' , \\[1ex]
	\left.\ln P_f(\vol', \vol(t_0, t)) \right|_{\vol'=\vol(t, t)}^{\vol'=\vol(t_0, t)} & = - \int_{\vol(t, t)}^{\vol(t_0, t)} d\vol' \, n(\vol') \vol'.
\end{align}
The lower integration bound $\vol(t, t)$ depends on the initial radius of bubbles that nucleate at time $t$. Noting that $P_f(\vol, \vol) = 1$, the false vacuum fraction at time $t$, $P_f(t)$, is given by
\begin{equation}
	P_f(t) = P_f(\vol(t, t), \vol(t_0, t)) = \exp(-\int_{\vol(t, t)}^{\vol(t_0, t)} d\vol' \, n(\vol') \vol') . \label{eq:falseVacuumFraction-volume}
\end{equation}
Alternatively, once \cref{eq:changeInFalseVacuumProbability} is established, one can immediately recognize a Volterra integral equation
\begin{equation}
	P_f(t) = \prod_{\vol(t_0)}^{\vol(t)} (1 - n(\vol) \vol d\vol) \equiv \lim_{\Delta \vol \rightarrow 0} \prod_{i=0}^{j} (1 - n(\vol_i) \vol_i \Delta \vol) , \label{eq:falseVacuumFraction-Volterra}
\end{equation}
noting that on the left, $\prod$ is a continuous \textit{product integral}, with the solution of \cref{eq:falseVacuumFraction-volume} \cite{ProductIntegration}. The index $j$ is such that $\vol_j = \vol(t_j) = \vol(t)$.

Before continuing, let us pause for a moment to discuss the true vacuum volume. The exponent in \cref{eq:falseVacuumFraction-volume} is not the fraction of the total volume in the true vacuum. Instead it is the `extended' volume of all real and phantom bubbles that we allow to nucleate at any point in space. Any overlapping region of bubbles is double-counted, bubbles are allowed to continue growing through the interface with other bubbles, and phantom bubbles are nucleated within other bubbles. This naturally gives an overestimate for the true vacuum volume, referred to as the extended volume. These issues do not invalidate \cref{eq:falseVacuumFraction-volume}, which captures the statistics correctly given a set of assumptions outlined in the next section \cite{PhysRevB.55.14071, PhysRevB.54.11845, ALEKSEECHKIN20113159}.%
\footnote{Although the necessity of phantoms in this approach is still debated \cite{PEREZCARDENAS2019100002, TOMELLINI2019119459, PEREZCARDENAS2019119458}.}
Alternative derivations that do not rely on an extended volume reproduce this result (see e.g.\ \refcite{cahn_1995, yu1995kinetics, yu1996kinetics, yu_lee_lai_1997, PhysRevB.54.11845} and \refcite{ALEKSEECHKIN20113159} for a review of different approaches). \Refcite{Megevand:2020klf} recently derived the JMAK equation for cosmological phase transitions using the differential critical-region method \cite{ALEKSEECHKIN20113159}.

The fact that the JMAK equation inherently accounts for impingement can be readily shown as follows \cite{Avrami2, PhysRevB.54.836, 10.1093/imamat/hxx012}. The fractional non-overlapping volume of the true vacuum is
\begin{equation}
	\vol_t(t) = P_t(t) . \label{eq:trueVacuumRatio}
\end{equation}
When the volume of the extended true vacuum increases, the volume of the true vacuum only increases if that vacuum conversion occurred in a region of false vacuum. Thus for a given $d\Vext$, the change in $\vol_t$ is $d\vol_t = 0$ with probability $P_t(t)$ and $d\vol_t = d\Vext$ with probability $1 - P_t(t)$. We then expect that on average
\begin{equation}
	d\vol_t = (1 - P_t(t)) d\Vext , \label{eq:actualAndExtendedVolumeRatio}
\end{equation}
because in order to convert a region of space to true vacuum through expansion or nucleation, that region must currently be in the false vacuum. The extended volume by definition has no such restriction, leading to a factor of $1 - P_t(t)$ between the volume increments $d\vol_t$ and $d\Vext$. Combining \cref{eq:trueVacuumRatio,eq:actualAndExtendedVolumeRatio}, we find
\begin{equation}
	d\vol_t = dP_t = (1 - P_t(t)) d\Vext , \label{eq:dV-impingement}
\end{equation}
which upon integrating yields
\begin{equation}
	P_f(t) = \exp(-\Vext(t)) . \label{eq:falseVacuumFraction-impingementDerivation}
\end{equation}
This demonstrates that the effects of phantom bubbles and impingement are correctly accounted for in the form of \cref{eq:falseVacuumFraction}.

Currently, the extended true vacuum volume in \cref{eq:falseVacuumFraction-volume} is expressed as a volume integral. It will be more convenient to integrate over time rather than volume. The false vacuum fraction can be expressed as
\begin{equation}
	P_f(t) = \exp(-\int_{t_0}^{t} dt' \, \pdv{\vol}{t'} n(\vol(t', t)) \vol(t', t)) , \label{eq:falseVacuumFraction-Jacobian}
\end{equation}
where again, $\vol(t', t)$ is the (fractional) volume of a bubble at $t$ that nucleated at $t'$. Although $t'$ is the time at which a bubble nucleated in $\vol(t', t)$, in the Jacobian $\pdv*{V}{t'}$ it should be interpreted as the time at which the volumes are evaluated. Thus, the Jacobian is positive as we would expect, because the volume of a bubble should increase over time. Note that with this interpretation of $t'$, the time integration bounds are opposite of the time inputs to the original volume integration bounds in \cref{eq:falseVacuumFraction-volume}. This prescription for the change of variables in the integral is used to more directly see the nucleation rate appear, as shown below \cref{eq:gammaFromN}.

To proceed, we define $N(\vol)$ as the total number of bubbles with volume $\vol' \leq \vol$;
\begin{equation}
	N(\vol) = \int_0^\vol d\vol' n(\vol') .
\end{equation}
The distribution of bubble volumes, $n(\vol)$, is given by
\begin{equation}
	n(\vol') = \left. \dv{N}{\vol} \right\vert_{\vol'} . \label{eq:bubbleNumberDistribution-volume}
\end{equation}
Additionally, we can define the nucleation rate $\Gamma(t)$ as the probability of nucleating a bubble of unspecified volume per unit time per unit volume. This can be expressed as
\begin{equation}
	\Gamma(t) = \dv{N}{t} = \dv{N}{\vol} \pdv{\vol}{t} , \label{eq:gammaFromN}
\end{equation}
with the nucleation time $t'$ of bubbles of volume $\vol$ left unspecified.
Here, $\pdv*{\vol}{t}$ is the change in an individual bubble's volume with respect to the time at which its volume is evaluated. This is equivalently $\pdv*{\vol}{t'}$ in \cref{eq:falseVacuumFraction-Jacobian} under the aforementioned prescription for the Jacobian. We can now recognize that the nucleation rate appears in \cref{eq:falseVacuumFraction-Jacobian}, which becomes
\begin{equation}
	P_f(t) = \exp(-\int_{t_0}^{t} dt' \, \Gamma(t') \vol(t', t)) . \label{eq:falseVacuumFraction-GammaV}
\end{equation}
Finally, to account for cosmic expansion, we must appropriately scale the nucleation rate to reflect the change in unit volume over time. This is achieved by a simple insertion of the ratio of scale factors into \cref{eq:falseVacuumFraction-GammaV}, leading to
\begin{equation}
	P_f(t) = \exp(-\int_{t_0}^{t} dt' \, \Gamma(t') \frac{a^3(t')}{a^3(t)} \vol(t', t)) . \label{eq:falseVacuumFraction-scaleFactors}
\end{equation}
Notice that the explicit form of $\vol(t', t)$ was not necessary. However, the nucleation rate and bubble volume must satisfy the assumptions outlined in the next section in order for this treatment of the false vacuum fraction to be valid.

\subsubsection{Assumptions of the JMAK equation}

Here we briefly discuss the assumptions central to the validity of the JMAK equation, and mention a few modifications that have been made that are potentially relevant to cosmological phase transitions. A relevant subset of assumptions, some of which can be relaxed with sufficient modification of the JMAK equation, is (see \refcite{kolmogorov1937statistical, ImpingementParam, BURBELKO2005429} for a more complete set of assumptions)
\begin{enumerate}
	\item The sample (in a cosmological context, the Universe) is initially homogeneous and presumably entirely in the false vacuum.
	\item Bubble nucleation is a random process, occurring at an equal rate throughout the sample.
	\item Bubble growth is not dependent on position in the sample.
	\item Bubbles are convex.
	\item Bubbles do not outgrow any previously nucleated bubbles. This implies that no part of a phantom bubble lies outside a real bubble.
	\item The initial radius of bubbles as they nucleate provides a negligible contribution to the total true vacuum volume.
	\item The number density of bubbles of each volume is small.
\end{enumerate}
Assumption 1 reflects the beginning of a phase transition with no seed nuclei. Assumptions 2 and 3 are directly related to the applicability of the underlying Poisson distribution of nucleation events. Assumptions 4 to 6 are necessary to prevent issues of overgrowth, where the non-overlapping extended volume exceeds the actual volume of true vacuum bubbles. It is crucial for the accuracy of the JMAK equation that the union of real and phantom bubbles matches the union of only real bubbles in spacetime. To understand this, consider again the proof that the JMAK equation correctly accounts for bubble impingement. In \cref{eq:dV-impingement} we assumed that $d\vol_t = d\Vext$ for conversion of any point in the false vacuum to the true vacuum. This no longer holds if some regions of the false vacuum are spuriously covered by the extended true vacuum volume. Vacuum conversion in these spuriously covered regions would yield an increase in $\Vext$ but not in $\vol_t$, and we would have $d\vol_t < (1 - P_t) d\Vext$. Hence, the JMAK equation incorrectly predicts the false vacuum fraction if any phantom bubble has some point outside a real bubble.

Assumptions 2 and 3 are evidently violated in a realistic phase transition, where phantom bubbles would not be realized. However, these assumptions and the introduction of phantom bubbles are merely tools to simplify the analysis. Other approaches avoid the use of phantoms; for instance instead considering the correlations between nucleation events, given that a true vacuum bubble excludes further nucleation events within its volume \cite{TOMELLINI200465, PhysRevE.85.021606, PhysRevE.95.022121}. Correlations (and consequently bubble clustering) have recently been considered for cosmological phase transitions \cite{Pirvu:2021roq, DeLuca:2021mlh}, suggesting that the JMAK equation may hold only approximately. Clustering has been studied in \refcite{VILLA20093714, FERNANDESIGNACIO2021777} and inhomogeneous nucleation in \refcite{PhysRevB.54.836, alekseechkin2001theory, RIOS20091199, tomellini2010kinetics}, albeit not in a cosmological context. 

\begin{figure}
  \centering
  \includegraphics[width=0.95\linewidth]{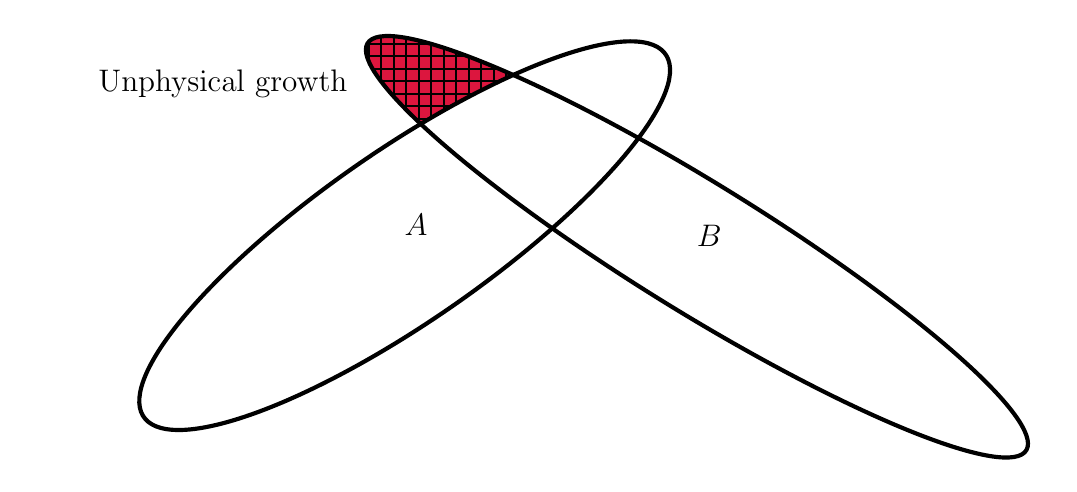}
  \caption{When bubble $B$ grows through bubble $A$, the hatched red region is unphysical. In this case, the total volume enclosed by the bubbles in the phase transition (the unhatched regions) is not accurately reproduced by considering the growth of each bubble in isolation. Based on a figure from \refcite{fanfoni1998johnson}.}
  \label{fig:screening}
\end{figure}

Assumptions 4 and 5 are naturally satisfied in a cosmological phase transition with spherical bubbles as shown in \cref{sec:transition_rates}, and monotonically increasing bubble wall velocities. Assumption 5 can be violated in non-cosmological phase transitions if bubble growth is diffusion-controlled (where the bubble wall velocity decreases with time) or bubbles are asymmetric and not oriented in parallel \cite{PhysRevB.55.14071, PhysRevE.85.021606, TOMELLINI200465, PhysRevE.95.022121, doi:10.1063/1.470052, pusztai1998monte, kooi2004monte}. For cosmological phase transitions, Assumption 5 could be violated due to reheating effects around bubble walls \cite{Megevand:2000da, Megevand:2003tg, Megevand:2017vtb, Cutting:2019zws}. Admittedly, such effects would also violate Assumptions 2 and 3. The consequence of violation of Assumptions 4 and 5 is that the extended volume includes points that could not be covered by true vacuum bubbles in an actual phase transition. An example of overgrowth is illustrated in \cref{fig:screening}. Assumption 6 is also necessary, because phantom bubbles with a finite initial size nucleated near the edge of another bubble would protrude through the wall of that pre-existing bubble. While bubbles do nucleate with a finite initial radius, this effect is typically negligible because the volume of a bubble rapidly dominates its initial volume. Further, bubbles nucleated at rest would be quickly overtaken by pre-existing bubbles expanding at their terminal velocity.

Assumption 7 allows independence of events to be assumed in obtaining \cref{eq:changeInTrueVacuumProbability}, because the probability of two bubbles of the same volume overlapping is negligible. Assumption 7 is satisfied if bubbles have a small fractional volume and there are few bubbles of each volume. The probability of overlap may become significant in models where a delta-function nucleation rate is appropriate, where the number density of bubbles of a specific volume can be high. Additionally, finite-volume effects need to be accounted for if the transforming medium is not much larger than the size of individual bubbles. This could occur during a simultaneous phase transition, where bubbles nucleate within bubbles of another metastable phase. A finite-volume spherical parent phase has been considered e.g.\ in \refcite{levine_narayan_kelton_1997, alekseechkin2001theory, alekseechkin2008kinetics}.

An alternative false vacuum fraction equation has seen limited use in cosmological phase transition literature; namely that of Csernai and Kapusta \cite{Csernai:1992as}, who attempt to mitigate some purported limitations of the model from Guth, Tye and Weinberg. Such limitations are not present, and the claims come from a misunderstanding of the implicit use of extended volume in \refcite{Guth:1979bh, Guth:1981uk}. Their equation is
\begin{equation}
	P_f(t) = 1 - \int_{t_0}^t dt' P_f(t) \Gamma(t') \vol(t', t) , \label{eq:falseVacuumFraction-CK}
\end{equation}
which accounts for nucleation only occurring in the false vacuum. However, impingement is not handled, leading to an overestimation of the true vacuum volume, as they observe. A brief discussion comparing \cref{eq:falseVacuumFraction-GammaV,eq:falseVacuumFraction-CK} is given in \refcite{Kampfer:1988, Ignatius:1993xb, JayOlson:2014vgy}, and the equivalence of \cref{eq:falseVacuumFraction-CK} to that of Ruckenstein and Ihm's equation \cite{Ruckenstein-Ihm} for the false vacuum fraction was noted in \refcite{Kampfer:1988, Kampfer:1991}.

Finally, we reiterate that the JMAK equation accounts for overlapping bubbles. In the presented derivation, this requires nucleation to occur even within existing bubbles. Incorrect attempts to rectify this counter-intuitive nucleation assumption --- by only allowing nucleation to occur in the false vacuum, with an otherwise similar derivation --- have been made and subsequently criticized \cite{ImpingementParam, ALEKSEECHKIN20113159}.

\subsection{Expected volume of a bubble} \label{sec:bubble_volume}

As seen in \cref{eq:falseVacuumFraction-GammaV}, the progress of a transition depends on the extended volume of true vacuum bubbles,
\begin{equation}
	V_t^{\text{ext}}(t) = \int_{t_0}^t dt' \Gamma(t') V(t', t). \label{eq:extendedTrueVacuumVolume}
\end{equation}
This is intuitively the combination of two factors: the rate at which bubbles are nucleated at time $t^\prime$ and the expected volume of those bubbles at a later time $t$. The nucleation rate $\Gamma$ is described in \cref{sec:transition_rates}. We can estimate the outstanding expected volume by \cite{Linde:1981zj, Megevand:2007sv, Athron:2022mmm}
\begin{equation}
	V(t', t) = \frac{4\pi}{3} \left[R(t',t)\right]^3, \label{eq:bubbleVolume}
\end{equation}
where
\begin{equation}
  R(t',t) = \frac{a(t)}{a(t')} R_0(t') + \int_{t'}^t dt'' v_w(t', t'') \frac{a(t)}{a(t'')}.
  \label{eq:bubbleRadius}
\end{equation}
\cref{eq:bubbleVolume} is the volume of a three-dimensional spherical bubble with an initial radius $R_0(t^\prime)$ that grows with a bubble wall velocity $v_w(t', t'')$. Both terms in the bubble radius \cref{eq:bubbleRadius} are multiplied by ratios of scale factors, accounting for the Hubble expansion of the Universe during the bubble's evolution. The bubble wall velocity is typically assumed to start from zero and increase due to the pressure difference between the phases. Eventually, the friction from the surrounding plasma may balance the driving pressure such that a terminal velocity is reached. Friction and the terminal velocity are discussed in \cref{sec:wallVelocity}. Estimates for the terminal velocity and the initial bubble radius in the strongly supercooled and thin-walled limits are described in \refcite{Ellis:2019oqb, Ellis:2020nnr, Lewicki:2022pdb}. \Refcite{Marques:1992ws} also provide a thorough discussion of the surface tension and critical bubble radius, similar to the aforementioned references, while \refcite{Megevand:2000zw} determine the initial bubble radius and wall thickness directly from the bubble equation of motion (see their appendix). See also \cref{sec:bounce_to_bubble}. Bubbles accelerating from rest to a time-dependent terminal velocity naturally gives a time dependence for the wall velocity of $v_w(t', t'')$.

A few approximations are often made to simplify \cref{eq:bubbleVolume}. The initial bubble radius, $R_0$, is often neglected in transitions well below the Planck scale (see e.g.\ \refcite{Megevand:2000zw, Megevand:2007sv} for justification). Bubbles are also assumed to quickly reach their terminal velocity, so the acceleration stage is ignored. Further, it is assumed that the terminal velocity remains constant over the course of a bubble's propagation.%
\footnote{The terminal velocity could change significantly throughout a transition in two obvious cases: if the transition is strongly supercooled and the pressure and friction vary significantly with temperature (see \cref{sec:wallVelocity}), or if reheating slows the expansion of bubbles (see \cref{sec:reheating}).}
With these approximations, \cref{eq:bubbleVolume} reduces to
\begin{equation}
	V(t', t) = \frac{4\pi}{3} \left[v_w \int_{t'}^t dt'' \frac{a(t)}{a(t'')} \right]^3 . \label{eq:bubbleVolume-const-vw}
\end{equation}
This can be simplified even further given particular approximations for the expansion of the Universe during the transition. In fact, neglecting cosmic expansion altogether, one finds
\begin{equation}
	V(t - t') = \frac{4\pi}{3} v_w^3 (t - t')^3 .
\end{equation}
This is a reasonable approximation for fast transitions, where the duration of the transition is much shorter than a Hubble time, but breaks down for strongly supercooled transitions. In general, to model the volume of a bubble we must understand the expansion of the Universe.

We have described the growth of an isolated bubble. In \cref{eq:extendedTrueVacuumVolume} we have considered a general wall velocity $v_w(t',t'')$ that can capture the effects on bubble growth from many phenomena. However, it is worth briefly mentioning a few of these effects. Once bubbles begin to collide, their growth into the false vacuum is facilitated by both the driving pressure between the phases, and the surface tension of the bubbles. The latter contribution refers to bubble coalescence, and may dominate shortly after the onset of percolation \cite{Witten:1984rs, Megevand:2003tg}. Once the false vacuum fraction decreases below the percolation threshold, regions of the false vacuum become isolated. These pockets (or droplets) of false vacuum shrink as the bubbles continue to grow \cite{Lu:2022paj}. If bubbles grow as subsonic deflagrations or hybrids, reheating occurs in the false vacuum and hinders the collapse of the reheated droplets of false vacuum \cite{Cutting:2019zws, Cutting:2022zgd}. In effect, the bubbles slow down once the false vacuum decreases sufficiently from unity. The reduction in wall velocity due to reheating has also been studied in the context of entropy production \cite{Heckler:1994uu, Megevand:2000da, Megevand:2003tg, Megevand:2017vtb}. The situation is considerably simpler if bubbles grow as supersonic detonations, in which case a constant wall velocity is expected to be a reasonable assumption.

\subsection{Impact of cosmic expansion on bubble volume and geometry} \label{sec:cosmicExpansion}

In \cref{sec:bubble_volume} we saw that the volume of true vacuum bubbles depends on the scale factor, $a(t)$. As shall be further discussed in the context of Einstein's equations in \cref{sec:GWobs}, the scale factor is the key quantity governing the expansion of the Universe. Thus, it is necessary to understand the evolution of the scale factor throughout a phase transition in order to predict the progress of the transition. The evolution of the scale factor can be tracked through the Hubble parameter 
\begin{equation}
	H(t) = \frac{\dot{a}(t)}{a(t)} , \label{eq:hubble}
\end{equation}
using Friedmann's equations (see \cref{eq:Friedmann1,eq:Friedmann2}). Assuming the metric in \cref{eq:FLRW-metric}, the Hubble parameter is given by
\begin{equation}
	H(t) = \sqrt{\frac{8\pi G}{3} \rhotot(t)} , \label{eq:Hubble}
\end{equation}
where $G = 6.7088 \! \times \! 10^{-39} \gev^{-2}$ is Newton's gravitational constant \cite{Wu:2019pbm}. The total energy density of the Universe, $\rhotot$, should be distinguished from the energy density of an arbitrary field configuration,
\begin{equation}
	\rho(\field, T) = V(\field, T) - T \pdv{V}{T} .
\end{equation}
Note that the energy density is most conveniently expressed in terms of temperature, following its definition based on the temperature-dependent effective potential. Thus, the time dependence of the temperature is often suppressed when stating the energy density. Additionally, we have assumed that the total energy density (and thus the Hubble parameter) is space independent. We discuss this further below.

The total energy density $\rhotot$ must appropriately reflect the changes in energy density throughout the course of a phase transition. Because the fraction of the Universe in a given field configuration $\field$ and at a given temperature $T$ changes over time, the energy density is clearly affected by the transition and the cooling of the Universe. There are two common approaches to estimating $\rhotot$. The first is to calculate the average energy density of the Universe by summing the contribution from each phase \cite{Megevand:2003tg, Megevand:2007sv, Azatov:2019png, Ellis:2019oqb},
\begin{equation}
	\rhotot(t) = \sum_i P_i(t) \rho_i(T_i(t)) - \rho_{\text{gs}} , \label{eq:energyDensity-average}
\end{equation}
where $P_i$ is the fraction of the Universe in phase $\field_i$, and we have introduced the shorthands
\begin{equation}
	\rho_i(T) \equiv \rho(\field_i(T), T) \quad\text{and}\quad \rho_{\text{gs}} \equiv V(\field_{\text{gs}}, 0)
\end{equation}
The zero-temperature ground state energy density is subtracted off to give a vanishing cosmological constant. The ground state could be either the current vacuum of the Universe or the global minimum of the zero-temperature potential, which are ideally one and the same. If only two phases coexist, \cref{eq:energyDensity-average} reduces to
\begin{equation}
	\rhotot(t) = P_f(t) \rho_f(T_f(t)) + P_t(t) \rho_t(T_t(t)) - \rho_{\text{gs}} , \label{eq:energyDensity-average-twoPhase}
\end{equation}
where the subscripts $f$ and $t$ denote the false and true vacua, or more generally the `from' and `to' phases in the transition.%
\footnote{The false vacuum temperature $T_f$ here is not to be confused with the completion temperature in \cref{sec:completion}.}
We assume that the Universe exists only in phases in \cref{eq:energyDensity-average}, not in perturbations about these phases. The average temperature of the false vacuum is known, provided the reheating in front of bubble walls (see \cref{sec:reheating}) is not significant. The average temperature of the true vacuum deep inside an isolated bubble can be determined by considering the hydrodynamics of the coupled field-fluid system (see \cref{sec:energyBudget}).

The second common approach to estimating $\rhotot$ is to apply energy conservation. The energy liberated by a true vacuum bubble builds up in the bubble wall and reheats the plasma in front of the bubble wall (except for supersonic detonations) or behind the bubble wall (except for subsonic deflagrations). One may then take \cite{Sato:1980yn, Heckler:1994uu, Megevand:2016lpr, Athron:2022mmm}
\begin{equation}
	\rhotot(t) = \rho_f(t) - \rho_{\text{gs}} . \label{eq:energyDensity-conservation}
\end{equation}
The energy liberated by vacuum conversion is converted to other forms (e.g.\ kinetic or thermal energy) which continue to contribute to $\rhotot$. The energy density decreases over time only due to cooling of the Universe.

The energy density is typically decomposed into several components that scale differently with temperature. The radiation component scales as $T^4$ and dominates at sufficiently high temperature. The exact form
\begin{equation}
	\rho_R(T) = \frac{\pi^2}{30} g_* T^4 \label{eq:rhoR}
\end{equation}
can be extracted from a high-temperature expansion of the effective potential, where $g_*$ is the effective number of relativistic degrees of freedom in the plasma (with $g_* = 106.75$ in the Standard Model).\footnote{Note that if the radiation energy density is to contain all $T^4$ terms in the effective potential, then daisy corrections would alter \cref{eq:rhoR}.}
The remaining contributions to $\rhotot$,
\begin{equation}
	\rho_V(T) = \rhotot(T) - \rho_R(T) ,
\end{equation}
are typically subdominant unless significant supercooling occurs. We denote these remaining contributions with a subscript $V$ for vacuum energy density, assuming that matter energy density is subdominant throughout the transition. See e.g.\ \refcite{Guo:2020grp, Ellis:2020nnr} for studies that also consider the matter contribution.

Interpreting $\rho_V$ as the vacuum energy density becomes more apparent when it is approximated. At sufficiently low temperatures, the radiation component can be ignored, leaving
\begin{equation}
	\rhotot(T) \approx \rho_V(T) \approx \rho_f(0) - \rho_\text{gs} = V(\phi_f(0), 0) - V(\phi_{\text{gs}}(0), 0) .
\end{equation}
In this case, the Universe is vacuum dominated. At high temperature, the Universe is instead radiation dominated, and we can take $\rhotot \approx \rho_R$. The energy density at an arbitrary temperature can be reasonably approximated by
\begin{equation}
	\rhotot(T) \approx \rho_R(T) + \rho_V(0) = \rho_R(T) + V(\phi_f(0), 0) - V(\phi_{\text{gs}}(0), 0) , \label{eq:rhotot-approxV}
\end{equation}
provided terms in the potential scaling as $T^4$ or $T^0$ are dominant. The approximation $\rhotot \approx \rho_R$ is common in `fast' transitions; those that complete before significant supercooling allows $\rho_V$ to be comparable to $\rho_R$. Additionally, fast transitions complete before the Universe expands noticeably since the onset of the transition, allowing for the approximation that the scale factor $a(t)$ is constant throughout the transition. For supercooled transitions, which are of interest due to stronger predicted gravitational wave (GW) signals (see \cref{Section:GWs-Sources}), the energy density is commonly approximated as in \cref{eq:rhotot-approxV} with the assumption $\phi_{\text{gs}} = \phi_t$ (i.e.\ the true vacuum is the ground state of the potential).

We still have not arrived at an explicit form for the scale factor (to be used e.g.\ in \cref{eq:bubbleVolume}). In fact, we only require the ratio of scale factors. By integrating $H(t) = \dot{a}(t)/a(t)$, one arrives at \cite{Athron:2022mmm}
\begin{equation}
	\frac{a(t_1)}{a(t_2)} = \exp(\int_{t_2}^{t_1} \! dt' \, H(t')) . \label{eq:scaleFactorRatio-time}
\end{equation}
We will see in the \cref{sec:timeTemperature} that in some cases the time-temperature Jacobian is $dT/dt = - T H(T)$, which offers a particularly simple ratio of scale factors:
\begin{equation}
	\frac{a(T_1)}{a(T_2)} = \frac{T_2}{T_1} . \label{eq:scaleFactorRatio-temperature}
\end{equation}
This convenient form, or equivalently 
\begin{equation}
	a(T) \propto \frac1T, \label{eq:scale_factor_adiabatic}
\end{equation}
is used throughout studies of phase transitions that neglect
reheating.  Note however that this expression relies on the assumption
(inherent in the bag model \cite{Chodos:1974je}) that the entropy varies with temperature
like $s\sim T^3$. While this is usually a valid assumption for the
period during a phase transition at high enough temperatures, it does
not hold in general because states drop out of thermal equilibrium and thus
change the effective degrees of freedom contributing to entropy.

We now briefly return to the discussion of spacetime geometry within bubbles. For simplicity we consider a vacuum (i.e.\ low-temperature) transition, such that the plasma can be ignored and the case in which transitions occurred in the past and we live in a stable or metastable ground state with vanishing vacuum energy. The transition from the false vacuum to the true vacuum alters the vacuum energy (i.e.\ the cosmological constant) from $\rho_V(\field_f) = V(\field_f)$ to $\rho_V(\field_t) = V(\field_t)$, where the latter is smaller by definition.  The cosmological constant and the spacetime geometry inside and outside a bubble are thus quite different.  It is usually assumed that gravitational effects do not break the $O(4)$ symmetry of the bounce solution and that the Euclidean metric obeys the $O(4)$ symmetry. In this case, the interior of the bubble obeys the ordinary Minkowski metric; similar to the shell theorem of Newtonian gravity, the gravity inside a spherical bubble with energy contained in the wall vanishes. Outside the bubble, the positive cosmological constant implies de Sitter space. This differs from the cosmological disaster in which our vacuum decays (see e.g.\ \refcite{Espinosa:2015qea}). In a phase transition at finite temperature, the energy density may not be uniform around bubbles. This suggests that the expansion rate of space may vary across the radial profile of a bubble, and the energy density in reheated droplets of false vacuum may deviate from the that deep within bubbles of true vacuum. This could impact the prediction of percolation and completion, which are important milestones discussed in \cref{sec:transitionMilestones}.

\subsection{Reheating} \label{sec:reheating}

By definition, the energy density of the true vacuum is lower than that of the false vacuum. Thus, energy is released during the phase transition. We discuss the distribution of this energy to different forms in \cref{sec:energyBudget}. For the purpose of this section, it is sufficient to note that some energy reheats the plasma.%
\footnote{\Refcite{Cai:2017tmh} mention another source of reheating that we do not discuss here: namely the field within the bubble relaxing from the exit point of the tunneling solution to the true phase configuration. This effect would be more significant in thick-walled bubbles; see \cref{fig:thin_thick}, where $\phi_0$ is the exit point.}
The rate and timing of this reheating can have important consequences for the evolution of the transition and for the corresponding GW signal. However, accurately modeling the reheating requires a hydrodynamic treatment which is beyond the scope of most studies of GWs from particle physics models.

First we consider the simplest case, where bubbles grow as supersonic detonations. The plasma outside the bubble remains undisturbed, so there is no reheating in the false vacuum. Instead, reheating occurs within the growing bubbles and in areas where bubbles have collided. The nucleation rate of bubbles is not affected because nucleation of bubbles only occurs in the false vacuum which is not reheated. Additionally, the expansion of a bubble is not affected until it collides with another bubble, where part of the wall then would propagate into the reheated true vacuum. However, this does not affect the union of all bubble volumes. Thus, the evolution of the false vacuum is not affected by reheating; the JMAK equation (see \cref{sec:jmak}) remains valid in this case. The reheating can be treated after the transition ends, where the entire Universe is in the reheated true vacuum.

The situation becomes more complicated if bubbles grow as subsonic deflagrations or hybrids, because reheating then occurs in front of the bubble walls. This results in a hydrodynamic obstruction \cite{Konstandin:2010dm} that affects both bubble nucleation and the growth of nearby bubbles in the reheated regions. A homogeneous nucleation rate equal to that far from bubble walls may become a poor approximation, particularly if the volume of shock fronts is significant. The growth of bubbles is hindered as they approach other shock fronts, because the pressure difference between phases typically decreases with temperature. The JMAK equation may then incorrectly model the progress of the transition in such cases. The reheating from slow bubble walls has been shown to significantly alter the temperature evolution and bubble wall velocity throughout the transition \cite{Heckler:1994uu, Megevand:2000da, Megevand:2017vtb}. This breaks two common assumptions used in simplifying the JMAK equation: a constant wall velocity, and a simple adiabatic time-temperature relation $\dv*{t}{T} = -T H(T)$. A recent approach to accounting for the thermal suppression of bubble nucleation is to determine an effective region where bubble nucleation is prevented, where the effective region is larger than the true vacuum volume \cite{Ajmi:2022nmq}. However, the bubble wall velocity was assumed to be constant throughout the transition. Reheating was intuitively found to reduce the bubble number density, and in turn increase the amplitude of the GWs. 

The local reheating around bubble walls eventually thermalizes.%
\footnote{This may not hold if the transition is completed by few large bubbles \cite{Guth:1982pn, Hawking:1982ga, Turner:1992tz}.} For slow bubble walls this reheating occurs during the transition, and in the limit of vanishing bubble wall velocity we can consider a uniform temperature throughout the Universe, regardless of the phase \cite{Heckler:1994uu}. In this case, the bubble nucleation rate is uniform but is still affected by the increased temperature (see \cref{eq:p_T}). This can result in a minimum in the bounce action $S(T(t))$, suggesting the nucleation rate is approximately Gaussian rather than exponential as is commonly assumed \cite{Megevand:2021llq}. \Refcite{Megevand:2017vtb} suggests that simultaneous nucleation is also a reasonable approximation in cases where the false vacuum reheats towards the critical temperature, because bubble nucleation is effectively switched off. We define the reheating temperature $\Treh$ to be the homogeneous temperature of the Universe \text{after} the transition completes. Determination of $\Treh$ is important for phenomenological predictions. For instance, $\Treh$ governs the extent of redshifting of GWs. Typically $\Treh$ is determined by applying conservation of energy, where the energy that goes into the phase transition is neglected since this should be negligible compared to the energy reheating the plasma. Here we list a few methods used to estimate $\Treh$.

One method is to enforce that the initial energy of the bubble $E_0 = \rho_f(\Ttr) V_0$ is conserved \cite{Heckler:1994uu}. \Refcite{Heckler:1994uu} consider the case of a particularly slow deflagration where the fluid kinetic energy in the shock can be ignored due to negligible fluid velocity. The reheat temperature can then be determined by solving
\begin{equation}
	\frac43 \pi \xi_{\text{sh}}^3 \rho_f(\Ttr) = \frac43 \pi \xi_w^3 \rho_t(\Treh) + 4\pi \! \int_{\xi_w}^{\xi_{\text{sh}}} \! d\xi \, \xi^2 \rho(\xi) . \label{eq:TrehHeckler}
\end{equation}
Here, $\xi = r/t$ is a position within the asymptotic fluid profile of the bubble (see \cref{sec:energyBudget}), with $r$ being the distance from the center of the bubble and $t$ the time since the nucleation of the bubble. Hence $\xi_{\text{sh}}$ is the position of the shock front, $\xi_w$ is the position of the bubble wall, $T_*$ is the temperature at which reheating takes place, and $r$ is the radius from the center of the bubble. The left-hand side of \cref{eq:TrehHeckler} is the false vacuum energy contained within a region of equal volume to the bubble. The right-hand side comprises of two parts: the true vacuum contained within the bubble up to the bubble wall, and the energy contained in the shock front. The energy density profile $\rho(\xi)$ within the shock front can be determined by solving the hydrodynamics of the bubble.

Assuming the reheating is instantaneous after the transition completes, $\Treh$ can be estimated using \cite{Athron:2022mmm}
\begin{equation}
	\rho(\field_f(T_f), T_f) = \rho(\field_t(\Treh), \Treh) ,
\end{equation}
where $T_f$ is the completion temperature (discussed in \cref{sec:completion}). The extent of reheating is then purely determined by the energy density difference between the phases. Alternatively, one can consider the dilution of energy density due to the expansion of space, compared to the time when reheating becomes significant, leading to \cite{Eichhorn:2020upj}
\begin{equation}
	a^3(\Treh) \rho_R(\Treh) = a^3(T_p) (\rho_R(T_p) + \Delta V) ,
\end{equation}
where $T_p$ is the percolation temperature (discussed in \cref{sec:percolation}). This states that the energy within a comoving volume is conserved, and assumes that the Universe is radiation dominated directly after reheating. \Refcite{Ellis:2019oqb} consider a case where the Universe can be matter dominated towards the end of the transition due to the oscillation of the scalar field about the true vacuum. The reheat temperature was then determined by the decay rate of the scalar field into radiation. A common approximation (see e.g.\ \refcite{Ellis:2018mja}) is
\begin{equation}
	\Treh = T_p \left(1 + \alpha \right)^{\frac14} ,
\label{Eq:Treh_from_Tp}
\end{equation}
where $\alpha$ is the transition strength parameter discussed in \cref{sec:kineticEnergy}. An accurate determination of the reheat temperature is necessary for accurate predictions of the GW signal. In the case of subsonic deflagrations and hybrids, a careful treatment of reheating throughout the transition is essential. However, the complications of reheating are often avoided by assuming bubbles grow as supersonic detonations, which is often reasonable in studies of strongly supercooled transitions \cite{Leitao:2015fmj, Megevand:2016lpr, Kobakhidze:2017mru, Cai:2017tmh, Ellis:2018mja, Ellis:2019oqb, Wang:2020jrd, Ellis:2020nnr, Athron:2022mmm}.

\subsection{Time and temperature} \label{sec:timeTemperature}

In \cref{eq:extendedTrueVacuumVolume} we express the nucleation rate as a function of time. However, in \cref{sec:transition_rates} we considered the nucleation rate to be a function of temperature. In \cref{sec:cosmicExpansion} we noted that the energy density (which enters the Hubble parameter) explicitly depends on temperature. Thus, we require a mapping between time and temperature in order to track the evolution of the transition. Generally, one assumes adiabatic expansion of the Universe, allowing for the application of conservation of entropy. From
\begin{equation}
	\dv{t} \left(s(t) a^3(t) \right) = 0 ,
\end{equation}
where $s$ is the entropy density, one can derive \cite{Athron:2022mmm}
\begin{equation}
	\dv{T}{t} = -3H(T) \frac{V'(\field_f(T), T)}{V''(\field_f(T), T)} , \label{eq:dTdt-general}
\end{equation}
with primes denoting partial derivatives with respect to temperature. The familiar form of
\begin{equation}
	\dv{T}{t} = -T H(T) \label{eq:dTdt-simple}
\end{equation}
is then obtained by assuming an equation of state that satisfies $V'/V'' = T/3$ in the false vacuum, such as the MIT bag equation of state \cite{Chodos:1974je}
\begin{equation}
	V(\field_f(T), T) = a T^4 + b , \label{eq:bagEoS-Vf}
\end{equation}
where $a$ and $b$ are temperature independent. The bag model describes hadrons as quark-gluon matter confined to a particular region of spacetime. There are efforts to improve this model for use in this cosmological context~\cite{Leitao:2014pda, Giese:2020rtr, Giese:2020znk, Wang:2020nzm, Wang:2022lyd, Wang:2023jto}. Then if radiation domination is assumed, the time-temperature relation is
\begin{equation}
	t = \sqrt{\frac{45}{\pi^3 G g_*}} \recip{4 T^2} , \label{eq:timeFunction}
\end{equation}
where $g_*$ and $G$ are as defined in \cref{sec:cosmicExpansion}. An analytic solution for $t(T)$ can be obtained for the approximation for $\rhotot$ given in \cref{eq:rhotot-approxV} (see e.g.\ \refcite{Einhorn:1980ik, Suhonen:1982ee, DeGrand:1984uq, Megevand:2003tg, Megevand:2016lpr}). For more complicated equations of state, \cref{eq:dTdt-general} can be numerically integrated.

The effect on the time-temperature relation due to reheating in the false vacuum was considered in \refcite{Heckler:1994uu} and later in \refcite{Megevand:2000da, Megevand:2003tg, Megevand:2017vtb}. It was argued that if the energy density liberated by true vacuum bubbles can be assumed to thermalize (as in the case of slow bubble walls), the time-temperature Jacobian receives an additional contribution \cite{Heckler:1994uu}. Ignoring for a moment the cosmic expansion, we can use energy conservation by equating the time derivative of \cref{eq:energyDensity-average-twoPhase} to zero. This equation can be solved for $\dv*{T}{t}$, and we can then add back \cref{eq:dTdt-simple} to account for cosmic expansion, leading to \cite{Heckler:1994uu}%
\footnote{\Refcite{Heckler:1994uu} use a simplified equation of state and neglect some $V''$ terms. Our \cref{eq:dTdt-reheating} matches their (51) only when $P_f \approx 1$ or $\rho_t'(T) - \rho_f'(T) \approx 0$.}
\begin{equation}
	\dv{T}{t} = \frac{\rho_f(T) - \rho_t(T)}{(\rho_t'(T) - \rho_f'(T)) P_f(t) - \rho_t'(T)} \dv{P_f}{t} - T H(T) . \label{eq:dTdt-reheating}
\end{equation}
The first term accounts for reheating while the last term accounts for the expansion of the Universe. \Refcite{Megevand:2017vtb} further consider entropy production and argue that it can be neglected for sufficiently slow bubble walls to good approximation.

If bubbles grow as supersonic detonations or if reheating can be otherwise be ignored, we may use \cref{eq:dTdt-simple} to simplify the transition analysis. The ratio of scale factors \cref{eq:scaleFactorRatio-time} reduces to a ratio of temperatures, as shown in \cref{eq:scaleFactorRatio-temperature}, and the volume of a bubble in \cref{eq:bubbleVolume-const-vw} becomes
\begin{equation}
	V(T', T) = \frac{4\pi}{3} \left[\frac{v_w}{T} \! \int_T^{T'} \!\! \frac{dT''}{H(T'')} \right]^3 ,
\end{equation}
where we have again neglected the initial bubble radius and assumed a constant wall velocity for simplicity. The false vacuum fraction \cref{eq:falseVacuumFraction-scaleFactors} can then be written as \cite{Athron:2022mmm}
\begin{equation}
	P_f(T) = \exp(-\frac{4\pi}{3} v_w^3 \int_T^{T_c} \! dT' \frac{\Gamma(T')}{T'^4 H(T')} \left(\int_T^{T'} \! \frac{dT''}{H(T'')} \right)^{\!\!3}) ,
\end{equation}
where $T_c$ is the critical temperature, which we take to correspond to $t_0$. However, if bubbles grow as subsonic deflagrations or hybrids, reheating must be accounted for. In this case, the false vacuum fraction and the time-temperature relation form a coupled system of equations that must be solved simultaneously.

\subsection{Transition milestones} \label{sec:transitionMilestones}

In this section we provide a discussion of the common milestones used in transition analyses and GW predictions. We will define these milestones in terms of time rather than temperature, because reheating effects cause temperature to be a non-monotonic function of time. That is, there are possibly multiple times at which the temperature has a particular value. The transition times can be converted to temperatures if $T(t)$ is known. Otherwise, one is presumably using temperature directly --- ideally in cases where reheating can be ignored.

\subsubsection{Nucleation}

The nucleation time $t_n$ is defined as the time at which there is on average one bubble nucleated per Hubble volume. An implicit equation for $t_n$ is
\begin{equation}
	N(t_n) = \frac{4\pi}{3} \int_{t_c}^{t_n} dt \, \frac{\Gamma(T(t)) P_f(t)}{H^3(t)} = 1 , \label{eq:tn}
\end{equation}
where $t_c$ is first time the temperature reaches the critical temperature $T_c$. We refer to the event of having at least one bubble per Hubble volume as unit nucleation. The factor of $4\pi/3$ is often omitted because one can consider a volume of the order $H^{-3}$, noting that the Hubble radius equals the inverse Hubble parameter. In a fast transition, where there is little supercooling, many small bubbles nucleate before a significant fraction of the Universe is converted to the true vacuum. In this case we can approximate $P_f(t \leq t_n) \approx 1$, leaving the more common form
\begin{equation}
	N(t_n) \approx \frac{4\pi}{3} \int_{t_c}^{t_n} dt \, \frac{\Gamma(T(t))}{H^3(t)} = 1 . \label{eq:tn-approx}
\end{equation}
Phantom bubbles are counted in \cref{eq:tn-approx}. However, they should provide a negligible contribution provided $P_t(t_n) \ll 1$.  This condition is satisfied under the typical assumption that the nucleation time marks the start of a transition, where $P_t$ is indeed small.

The nucleation time can be approximated further by estimating the number of bubbles of in a Hubble volume $H^{-3}$ nucleated in a Hubble time $H^{-1}$ and equating it to unity, through
\begin{equation}
	\frac{\Gamma(T(t_n))}{H^4(t_n)} = 1 .
\end{equation}
A simpler method of estimating the nucleation time is through constraints on the value of the bounce action, $S(T)$, which dominates the behavior of the nucleation rate $\Gamma(T)$. A popular action constraint is (see e.g.\ \refcite{Wainwright:2011kj})
\begin{equation}
	S(T(t_n)) \sim 140 , \label{eq:tn-140}
\end{equation}
which is valid only at the electroweak scale and during the radiation dominated era. See Appendix B of \refcite{Ignatius:1993xb} for an example of a more sophisticated heuristic for the nucleation time with clear assumptions for validity, and \refcite{Levi:2022bzt} for an equation for the nucleation temperature in a minimal Coleman-Weinberg model. The assumptions (such as a linear action around $t_n$ in the former heuristic) can break down for slow transitions, where supercooling can become significant. For effective potentials with a zero-temperature potential barrier between the phases, the action has a minimum at finite temperature. This minimum can be above 140, in which case the simple condition in \cref{eq:tn-140} would suggest there is no nucleation time in the transition. And, if the nucleation time were to consistently mark the beginning of a transition, then the transition must not occur.

However, it was recently found that it is possible for a transition to complete without having one bubble per Hubble volume \cite{Athron:2022mmm}. This scenario could arise for very strongly supercooled transitions where nucleation is either slow or cut off, and the transition must complete due to prolonged bubble growth. Bubbles can be in causal contact with other bubbles beyond their Hubble volume because the Hubble volume is smaller than the causal volume \cite{Davis:2003ad, Ellis:2015wdi}. \Refcite{Athron:2022mmm} further argue that the nucleation time does not mark an important milestone in a transition because GWs are generated later, and because unit nucleation and completion are independent events (i.e.\ the occurrence of one does not imply the occurrence of the other). In cases where the nucleation criterion \cref{eq:tn} is not satisfied before the nucleation rate $\Gamma$ is maximized, the nucleation time is sometimes replaced with the time at which the nucleation rate is maximized, $t_{\Gamma}$ (see e.g.\ \refcite{Cai:2017tmh, Ellis:2018mja}).

Still, the nucleation time provides a reasonable approximation for the time at which GWs are generated during a fast transition. This is unsurprising because the duration of a fast transition is small. The nucleation time is a poor indication of the time at which GWs are generated during a slow transition \cite{Megevand:2016lpr, Athron:2022mmm}. The nucleation condition \cref{eq:tn} may be satisfied well before bubbles or sound waves in the plasma collide. More useful is an event that corresponds directly to when GWs are generated, in both fast and slow transitions. This brings us to the event of bubble percolation.

\subsubsection{Percolation} \label{sec:percolation}

Consider a fast transition where the expansion of space can be neglected. As bubbles continue to nucleate and grow, collisions between these bubbles are inevitable. The transition begins with few isolated, small bubbles of true vacuum. Over time more bubbles nucleate and existing bubbles grow, until the bubbles start colliding and coalescing. GWs are generated during this collision stage of a phase transition, due to collisions of the bubbles and disruptions to the surrounding plasma. If one were to seek a characteristic time for GW production, as done when mapping transition properties to GW predictions, a natural time would be when bubbles were colliding. See \cref{sec:transitionTemperature} for further discussion.

\begin{figure}
  \centering
  \includegraphics[width=0.65\linewidth]{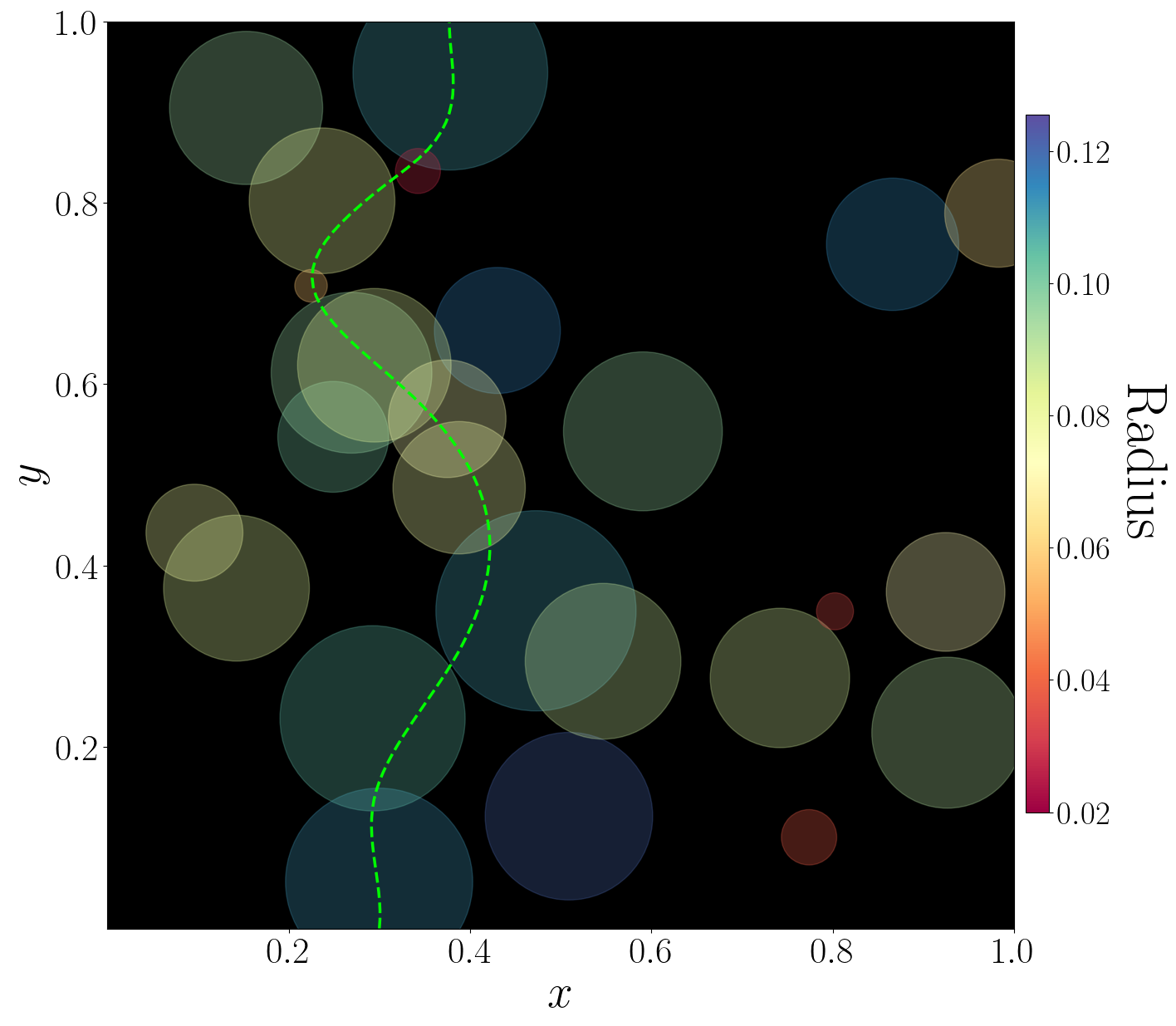}
  \caption{Two-dimensional slice of a simulation of bubble nucleation and growth in a three-dimensional unit cube. There is a path (green dashed line) of connected bubbles from one side of the cube to the other, meaning that the bubbles have percolated. Taken from \refcite{Athron:2022mmm}.}
  \label{fig:percolation}
\end{figure}

Conveniently, we can extract such a time from the evolution of the false vacuum fraction by considering bubble percolation. Bubbles are said to have percolated if there is a connected cluster of bubbles that spans the entire Universe, as depicted in \cref{fig:percolation}. A connected cluster requires collisions between bubbles, hence offering a source of GWs. The onset of percolation is when this sufficiently large connected cluster first forms, and is marked by the percolation time, $t_p$. Percolation theory \cite{Broadbent:1957rm, doi:10.1080/00018737100101261, hunt2014percolation} states that although nucleation is a stochastic process, a large connected cluster almost surely forms at a critical threshold of the true vacuum fraction: the percolation threshold. Percolation studies of uniformly nucleated spherical bubbles demonstrate the percolation threshold relevant to a cosmological phase transition is $P_t \approx 0.29$ \cite{doi:10.1063/1.1338506, LIN2018299, LI2020112815}. Thus, the percolation time is found by solving
\begin{equation}
	P_f(t_p) \approx 0.71 . \label{eq:tp}
\end{equation}
Sometimes the percolation time is instead defined through
\begin{equation}
	P_f(t_e) = e^{-1} , \label{eq:te}
\end{equation}
based on the requirement $\Vext(t_e) = 1$. We denote this alternative percolation time with a different subscript, $t_e$. Although \cref{eq:te} does not represent the onset of percolation, it is a conservative indication for whether percolation has occurred and also corresponds to a time when bubbles are colliding.

We now consider a strongly supercooled transition, where the Universe remains in the false vacuum long after it becomes metastable. The expansion of space must be taken into account. The growth of bubbles now competes with the expansion of the false vacuum, which introduces a caveat to the percolation condition \cref{eq:tp}. If the volume of the false vacuum monotonically increases up to $t_p$, then bubble percolation is not guaranteed at $t_p$ or even at later times. Although the true vacuum bubbles continue to grow over time, so does the space in the false vacuum between them.

The physical volume of the false vacuum is characterized by \cite{Turner:1992tz}
\begin{equation}
	\Vphys(t) = a^3(t) P_f(t) ,
\end{equation}
where $a(t)$ is again the scale factor. The time derivative,
\begin{equation}
	\dv{\Vphys}{t} = \Vphys(t) \left[\dv{t} \ln(P_f(t)) + 3 H(t) \right] ,
\end{equation}
then indicates whether the physical volume of the false vacuum fraction is decreasing at a particular time. Percolation is questionable if
\begin{equation}
	\dv{t} \ln(P_f(t)) \geq -3 H(t) \label{eq:dvphys-dt-cond}
\end{equation}
at $t_p$ \cite{Turner:1992tz, Ellis:2018mja, Athron:2022mmm}, because the physical volume of the false vacuum is not decreasing. In such cases, although the false vacuum fraction may be decreasing, the space between bubbles can be increasing, preventing the bubbles from meeting. To our knowledge, a percolation study has not been performed in expanding spacetime. The duration of the phase transition and the time at which GWs are produced could change significantly from the predictions in Minkowski spacetime.

\subsubsection{Completion} \label{sec:completion}

It is often important to determine whether a transition completes. To avoid cosmological constraints on phase transitions that leave finite regions of the Universe in the false vacuum, one can demand that there is vanishing probability of such a region existing in our causal volume. A simple completion criterion is that the false vacuum fraction becomes arbitrarily small.%
\footnote{See \refcite{Turner:1992tz} for a discussion of an alternative completion metric: the uncollided surface area of bubble walls, which is treated further in \refcite{Guo:2020grp, Megevand:2020klf}.}
Because \cref{eq:falseVacuumFraction} only approaches zero asymptotically, a completion threshold for the false vacuum fraction is often assumed. A completion or final time, $t_f$, can then be found by solving
\begin{equation}
	P_f(t_f) = \epsilon ,
\end{equation}
where $\epsilon \ll 1$. In some studies, $t_p$ and $t_e$ (defined in \cref{sec:percolation}) are treated as the completion time. This is equivalent to taking $\epsilon = 0.71$ and $\epsilon = 0.37$, respectively. The details of the end of a phase transition are difficult to model because the interaction between bubble walls affects their subsequent growth \cite{Turner:1992tz}, and because of the decay of false vacuum pockets surrounding coalescing bubbles \cite{Liu:2021svg, Cutting:2022zgd, Lu:2022paj, Kawana:2022lba}. Such complications are irrelevant if a precise determination of the end of a transition is not required. The completion time is very weakly dependent on the completion threshold $\epsilon$, except in strongly supercooled transitions where the asymptotic value of $P_f$ may be non-zero. 

The competition between regions of the false and true vacua also plays a role in determining whether a transition completes. Even with vanishing false vacuum fraction, finite regions of the false vacuum could exist and grow if $\Vphys$ increases during the transition. To be confident that no such regions of the false vacuum persist after the transition, one can impose more stringent constraints on the evolution of the physical volume \cite{Turner:1992tz, Ellis:2018mja, Athron:2022mmm}, such as \cref{eq:dvphys-dt-cond} which is the condition for the false vacuum to be decreasing at particular time $t$.  \Refcite{Athron:2022mmm} found that \cref{eq:dvphys-dt-cond} at the percolation temperature is more constraining than requiring $P_f < 0.01$ and several other conditions that may be considered, such as applying the same condition at the completion temperature. This was demonstrated both analytically for a simple model of simultaneous nucleation, and numerically for various scenarios from the scalar singlet model and a toy model of supercooling,

Requiring the phase transition completes in particular models can
compete with constraints from LIGO data \cite{Athron:2023aqe}, or
effectively rule out proposed explanations for data from pulsar timing
arrays \cite{Athron:2023mer}. However it is also possible that if
pockets of the false vacuum persist to low energies they can lead to
the formation of primordial black holes due to the total energy
density of the false vacuum remnants increasing \cite{Kodama:1982sf,
  Hall:1989hr, Liu:2021svg,Lewicki:2023ioy}.  The false vacuum
remnants can also form other macroscopic objects (see
e.g. \refcite{Kawana:2022lba}) such as quark nuggets
\cite{Witten:1984rs}, Q-balls\cite{Krylov:2013qe} and Fermi-balls
\cite{Hong:2020est}.

\subsection{Complicated phase histories} \label{sec:relaxAssumptions}

We now turn to complications that can arise when analyzing the phase history of a particle physics model. Here we relax the simplifying assumptions made in \cref{sec:simplifiedTransition}. We treat the relaxation of each assumption separately to streamline the discussion.

First consider a potential barrier that persists at zero temperature. The bounce action diverges at the critical temperature and is too large at zero temperature to allow for a low-temperature transition in many models (see e.g.\ \refcite{Megevand:2016lpr, Kobakhidze:2017mru, Cai:2017tmh}). Thus, if a phase transition is to have any chance to occur, the bounce action must have a minimum at some intermediate temperature, $T_S$. Remembering that reheating can increase the average temperature, the minimum action may be reached at more than one time, particularly if significant reheating occurs after the temperature first drops to $T_S$. The maximum nucleation rate roughly coincides with $T_S$ because the nucleation rate is exponentially suppressed by the bounce action \cite{Athron:2022mmm}. If the maximum nucleation rate is insufficient, too few bubbles may form for bubble percolation to occur, let alone for completion of the transition. The expansion of the false vacuum and finite lifetime of the Universe impose a lower bound on the maximum nucleation rate for a successful phase transition at the electroweak scale \cite{Turner:1992tz}.

Approximations of the nucleation rate are key to analytic treatments of a phase transition. The evolution of a phase transition is strongly characterized by the form of the nucleation rate. The nucleation rate is most commonly assumed to increase exponentially with time. An exponential nucleation rate is valid if a transition completes before $T_S$ is reached and if reheating is negligible during the transition. If the transition completes after $T_S$ is reached, then a Gaussian nucleation rate (centered about $T_S$) is appropriate \cite{Megevand:2016lpr}. If the nucleation rate is strongly suppressed after a period of rapid reheating, it can be approximated as a Gaussian or a delta function \cite{Megevand:2017vtb}. The delta function nucleation rate implies all bubbles nucleate simultaneously, which is convenient for both analytic studies and hydrodynamic simulations (for the latter, see e.g.\ \refcite{Hindmarsh:2015qta}). While these approximations for the nucleation rate are useful for analytic discussions of a phase transition, the nucleation rate --- and quantities dependent on the nucleation rate --- can still be determined numerically in each case.

All discussion of a phase transition has so far assumed only two phases of the Universe are involved at any moment in time. This has allowed us to label one phase as the metastable false vacuum and the other phase as the stable true vacuum. Even with only two relevant phases at a time, the true vacuum may only be temporarily stable, because it may be the false vacuum of a subsequent phase transition. The situation is further complicated when more than two phases are relevant at once. Scenarios exist where the false vacuum can decay into several distinct phases simultaneously with non-degenerate free energy densities. See \refcite{Croon:2018new, Morais:2018uou, Morais:2019fnm} for preliminary studies. For simplicity, consider three phases $\field_1$, $\field_2$ and $\field_3$ in descending order of free energy density, all in existence at the same time, and assume the entire Universe is initially in phase $\field_1$. Three transitions can happen simultaneously: $\Trans12$, $\Trans13$ and $\Trans23$, where $\Trans{i}{j}$ denotes a transition from phase $\field_i$ to phase $\field_j$. A competition exists between these transitions; $\Trans23$ will only occur if $\Trans12$ is not dominated by $\Trans13$. The existence of simultaneous transitions affects the progress of each transition. There is a competition over depletion of regions of $\field_1$ for transitions $\Trans12$ and $\Trans13$, and the existence of more than two minima along the direction of the bounce path could pose a problem for determining the nucleation rates (see e.g.~\refcite{Wainwright:2011kj}). The effects of overlapping and intersecting bubbles of different phases have not yet been rigorously studied in a cosmological context. The idea of percolation or completion marking the time of GW production becomes less reasonable if competing phases share similar fractions of the Universe.

In some cases there are multiple transitions (sometimes referred to as multistep transitions) that must be analyzed to determine whether the model can describe our current Universe, and to predict the total GW signal \cite{Land:1992sm, Zarikas:1995qb, Angelescu:2018dkk, Fabian:2020hny, Aoki:2021oez, Lewicki:2021pgr, Zhao:2022cnn, Benincasa:2022elt, Cao:2022ocg}. In fact, completion of a strongly supercooled electroweak phase transition may be assisted or triggered by another phase transition. If electroweak symmetry is unbroken at the QCD scale, a first-order chiral phase transition can occur and trigger electroweak symmetry breaking well below the electroweak scale \cite{Witten:1980ez, Iso:2017uuu, Arunasalam:2017ajm, vonHarling:2017yew, Baratella:2018pxi, Bodeker:2021mcj, Sagunski:2023ynd}.  Furthermore, first-order phase transitions can produce primordial black holes \cite{Hawking:1982ga,Kodama:1982sf} and other exotic macroscopic objects \cite{Kawana:2022lba,Witten:1984rs,Krylov:2013qe, Hong:2020est}.  These possibilities have been studied a lot in the recent literature \cite{Deng:2017uwc,Kawana:2021tde,Liu:2021svg,Jung:2021mku,Maeso:2021xvl,Hashino:2022tcs,Huang:2022him,Kawana:2022lba,Cline:2022xhx,Kawana:2022olo,Lewicki:2023ioy,Jinno:2023vnr}.

The phase structure of models with many scalar fields gaining VEVs can be arbitrarily complicated. The phase history may involve simultaneous transitions and multistep transitions, as described above. However, other possibilities exist that complicate the analysis of an individual transition. For instance, the free energy density of the true vacuum may exceed that of the false vacuum after the transition begins (thus reversing the transition) \cite{Rafelski:2015lva}, or a transition may be terminated by the disappearance of one of the phases or due to the false vacuum becoming unstable \cite{Einhorn:1980ik}. A thorough analysis of the phase history is required to understand whether a model predicts that the Universe reaches --- and remains in --- the current vacuum \cite{Balazs:2023kuk}. Further improvements to the analysis of these complicated scenarios is essential going forwards.

\section{Thermal parameters characterizing first-order phase transitions}\label{sec:thermal_parameters}

First-order phase transitions (FOPTs) can be described by a set of so-called thermal parameters, including the transition temperature, characteristic length scale, transition strength, and the bubble wall velocity (see \refcite{Caprini:2015zlo, Weir:2017wfa, Caprini:2019egz} for reviews). As well as characterizing the transition, these parameters can be fed into fits from lattice simulations or analytic models of gravitational waves (GWs), as discussed later in \cref{Section:GWs-Sources}. Thus the thermal parameters build a bridge between particle physics models and existing state-of-the-art GW computations.

\subsection{Transition temperature} \label{sec:transitionTemperature}

A major limitation of the thermal parameters is that they are evaluated at a single temperature: the transition temperature $\Ttr$. More realistically, GWs are produced throughout the phase transition. Meanwhile, the studies that provide the GW fits do not inherently provide a clear milestone associated with $\Ttr$. The correct transition temperature is not known in general; perhaps there is no single milestone that works in all cases. Typical choices for $\Ttr$ are the nucleation temperature, the percolation temperature, the completion temperature, and the reheating temperature.

The nucleation temperature, $T_n$, is a common choice because it is expected to indicate the start of the phase transition. For fast transitions, the nucleation temperature should not be far from the temperature at which bubbles are colliding. Further, it does not require determination of the false vacuum fraction, allowing for a simpler transition analysis. Its usage in strongly supercooled transitions should be avoided due to the hierarchy in temperatures that can arise (see e.g.\ \refcite{Megevand:2016lpr, Cai:2017tmh, Kobakhidze:2017mru, Ellis:2018mja, Wang:2020jrd, Athron:2022mmm} for polynomial potentials and e.g.\ \refcite{Ellis:2019oqb, Ellis:2020nnr} for conformal potentials), which can have a large impact on the thermal parameters (see e.g.\ \refcite{Leitao:2015fmj, Megevand:2016lpr}).

Motivated by the fact that GWs are sourced by collisions of bubbles and sound shells, and the subsequent turbulence, a more appropriate choice of $\Ttr$ is the percolation temperature, $T_p$. The onset of percolation involves the collision of bubbles by definition, so it corresponds to a time when GWs are produced, even in strongly supercooled transitions.%
\footnote{However, this is subject to the caveat of decreasing physical volume mentioned in \cref{sec:percolation}.}
Some authors (e.g.\ \refcite{Guo:2021qcq}) argue that the thermal parameters should instead be evaluated at the completion temperature, although they use the moment when $\Vext = 1$ to define completion. The reheating temperature is another relevant reference temperature. If the plasma reheats to a homogeneous temperature $\Treh$ after the transition completes, then the GWs should redshift from $\Treh$ rather than from $\Ttr$ \cite{Cai:2017tmh}. The characteristic length scale and the energy available for GW production are set by events that occur before global reheating takes effect, so they should not be evaluated at $\Treh$. In the case of slow deflagrations, the reheating can occur throughout the course of the phase transition such that there is no sudden reheating near the end \cite{Heckler:1994uu, Megevand:2017vtb}.

In the following subsections we discuss the other thermal parameters. We leave the transition temperature as a generic temperature, and do not fix it to a choice of milestone. Although the same symbol $\Ttr$ is used throughout, the relevant temperature in each case (and in the corresponding GW signals in \cref{Section:GWs-Sources}) may differ. Quantities with a subscript $*$ are evaluated at the transition temperature. The common approach in recent careful studies is to choose $\Ttr = T_p$ throughout. \Refcite{Wang:2020nzm} demonstrate in their Table 1 how the thermal parameters can vary when choosing $T_n$ versus $T_p$.

\subsection{Hydrodynamics and energy budget} \label{sec:energyBudget}

In this section we introduce key quantities for a hydrodynamic treatment of the phase transition, deferring much of the detail to the dedicated studies. A hydrodynamic treatment offers valuable information in the analysis of a phase transition and its aftereffects, such as the generation of GWs or baryon asymmetry. Important results are the energy budget and the aforementioned thermal parameters.

To begin, we state the thermodynamic quantities in terms of the scalar potential. All thermodynamic quantities depend on the temperature $T$ and the field configuration $\field$. We will suppress this dependence for convenience of notation. The free energy density is equal to the scalar potential: $\FED = V$, while the pressure is the negative of the free energy density: $p = -\FED$. The energy density $\rho$, the enthalpy density $w$, and the entropy density $s$, are respectively given by%
\footnote{These quantities (and \cref{eq:Tuv-plasma}) can be derived from the partition function \cite{Hindmarsh:2020hop}.}
\begin{align}
	\rho & = T \pdv{p}{T} - p = V - T \pdv{V}{T} , \label{eq:hydro-rho} \\
	w & = T \pdv{p}{T} = - T \pdv{V}{T} , \\
	s & = \pdv{p}{T} = -\pdv{V}{T} .
\end{align}

The energy-momentum tensor receives contributions \cite{Espinosa:2010hh} from the scalar field,%
\footnote{Note that we use the mostly negative metric signature $(+, -, -, -)$ here to match standard literature (e.g.\ \refcite{Espinosa:2010hh, Wang:2023jto}), in contrast to the mostly positive signature used in \cref{sec:GWobs}.}
\begin{equation}
	T_{\mu\nu}^{\field} = \partial_\mu \phi_i \partial_\nu \phi_i - g_{\mu\nu} \left[\half \partial_\rho \phi_i \partial^\rho \phi_i - V_0(\field) \right] , \label{eq:Tuv-scalar}
\end{equation}
and from the plasma,
\begin{equation}
	T_{\mu\nu}^{\text{plasma}} = \sum_j \! \int \! \frac{d^3 k}{(2\pi)^3 E_j} k_\mu k_\nu f_j(k, z) . \label{eq:Tuv-plasma}
\end{equation}
The sum in \cref{eq:Tuv-plasma} runs over the particle species that form the plasma, each of which have a corresponding distribution function $f_j$ that depends on the position $z$ in the bubble wall and the particle momentum $k$. We discuss the distribution function further in \cref{sec:friction}. Typically the plasma is assumed to be a perfect relativistic fluid \cite{Espinosa:2010hh}, simplifying \cref{eq:Tuv-plasma} to%
\footnote{See \refcite{Wang:2022txy} or Footnote 3 of \refcite{Cai:2020djd} for an explicit connection between \cref{eq:Tuv-plasma,eq:Tuv-plasma-perfect} without the assumption of local thermal equilibrium.}
\begin{equation}
	T_{\mu\nu}^{\text{plasma}} = w u_\mu u_\nu - g_{\mu\nu} p , \label{eq:Tuv-plasma-perfect}
\end{equation}
where $u^\mu = \gamma \left(1, \vecb{v} \right)$ is the fluid four-velocity in the reference frame of the bubble center and $\gamma = 1/\sqrt{1 - v^2}$ is the Lorentz factor. The combined energy-momentum tensor of the field-fluid system is
\begin{equation}\label{eq:combined_energy_momentum_tensor}
	T_{\mu\nu} = \partial_\mu \phi_i \partial_\nu \phi_i - \half g_{\mu\nu} \partial_\rho \phi_i \partial^\rho \phi_i + (\rho + p) u_\mu u_\nu - g_{\mu\nu} p ,
\end{equation}
where $V_0(\field)$ is absorbed into the definition of the pressure.

The total energy density is given by
\begin{equation}
	T_{00} = w \gamma^2 + \half \sum_i \left(\dot{\phi}_i^2 + (\grad{\phi_i})^2 \right) - p ,
\end{equation}
or, upon using $p = w - \rho$ and $\gamma^2 -1 = v^2\gamma^2$, equivalently
\begin{equation}
	T_{00} = w \gamma^2 v^2 + \half \sum_i \left(\dot{\phi}_i^2 + (\grad{\phi})^2 \right) + \rho .
\end{equation}
Using the trace anomaly
\begin{equation}
	\theta = \frac14 g^{\mu\nu} T_{\mu\nu}^{\text{plasma}} = \frac14 (\rho - 3p) , \label{eq:traceAnomaly}
\end{equation}
the last term can be further decomposed into \cite{Hindmarsh:2019phv, Cutting:2021tqt}
\begin{equation}
	T_{00} = w \gamma^2 v^2 + \half \sum_i \left(\dot{\phi}_i^2 + (\grad{\phi})^2 \right) + \frac34 w + \theta . \label{eq:energyBudget}
\end{equation}
This decomposition allows for identification of the energy budget. The fluid kinetic energy, $\rhokin$, is given by the first term, as seen through \cite{Leitao:2010yw}
\begin{equation}
	\rhokin = T_{00}(v) - T_{00}(0) = w \gamma^2 v^2 .
\end{equation}
The next two terms are evidently the kinetic and gradient energy of the scalar fields. These two contributions to the total energy density are typically negligible in transitions where the bubbles reach a terminal velocity due to friction (see \cref{sec:wallVelocity}). The liberated vacuum energy ($\rho_\theta = \theta$) is instead deposited into the surrounding fluid as thermal energy ($\rho_Q = \frac34 w$) or bulk fluid motion $\rhokin$.%
\footnote{This decomposition is motivated by recognizing that $\rho_R = (3/4) w$ in the bag model \cite{Chodos:1974je}. Thus, we refer to $(3/4) w$ as the thermal energy density. Let $\Delta \rho_x$ denote the difference in the energy contribution $\rho_x$ before and after the transition. Neglecting the scalar field contributions for simplicity, conservation of energy gives $\rhokin + \Delta \rho_Q + \Delta \rho_\theta = 0$, where we have assumed the fluid is at rest before the transition. Hence, the change in kinetic and thermal energy is sourced by the change in $\rho_\theta$, which must be the energy liberated in the transition \cite{Hindmarsh:2019phv, Cutting:2021tqt}.} 
See \refcite{Espinosa:2010hh, Hindmarsh:2019phv, Cutting:2021tqt} for further discussion of the energy budget.

Kinetic and gradient energy can contribute to GW production (see Section~3.2.3 of \refcite{Cutting:2021tqt}), whereas thermal energy reheats the plasma but does not source GWs.Thus, the fluid kinetic energy $\rhokin$ is usually the only component considered in phase transitions at finite temperature. However, energy from the scalar fields may be significant in strongly supercooled transitions with runaway walls \cite{Ellis:2019oqb, Ellis:2020nnr, Lewicki:2022pdb}.  

The source tensor for GWs is \cite{Hindmarsh:2013xza}
\begin{equation}
	\tau_{ij} = \tau_{ij}^\phi + \tau_{ij}^f , \label{eq:sourceTensor}
\end{equation}
where $\tau_{ij}^\phi = \partial_i \phi \partial_j \phi$ and $\tau_{ij}^f = w \gamma^2 v_i v_j$. In cases where the scalar contribution $\tau_{ij}^\phi$ is negligible, the total fluid kinetic energy of a bubble is an indicator of the energy available for sourcing GWs. This is discussed further in \cref{sec:kineticEnergy}.

The bubble wall reaches a terminal velocity if there is sufficient friction from the plasma to match the driving pressure (see \cref{sec:wallVelocity}). This steady state is reached quickly compared to the total duration of the phase transition \cite{Megevand:2020klf}. The plasma around the bubble wall acquires an asymptotic self-similar profile parameterized by $\xi = r/t$,%
\footnote{This fluid profile parameter is not to be confused with the gauge parameter used in quantum field theory.}
where $r$ is the distance from the bubble center and $t$ is the time since nucleation of the bubble \cite{LandauLifshitz-1D}. Isolated bubbles are assumed to be spherically symmetric, allowing one to consider the fluid profiles in the radial direction alone. The fluid profiles are determined using conservation of the energy-momentum tensor,
\begin{equation}
	\partial^\mu T_{\mu \nu} = 0 , 
\end{equation}
and integrating across the bubble wall. The bubble wall reaches a steady state once the terminal velocity is attained, and is treated as locally planar \cite{Steinhardt:1981ct}. Thus, taking the fluid velocity to be in the z-direction and considering the bubble wall's frame of reference, conservation of the energy-momentum tensor reduces to
\begin{equation}
	\partial^z T_{zz} = \partial^z T_{zt} = 0 . \label{eq:EMcons-planarSteady}
\end{equation}
This states that the flux of the $z$-component of momentum and the flux of energy are constant in the $z$-direction. Let the subscripts $-$ and $+$ respectively denote quantities evaluated just inside and just outside the bubble wall. Integrating \cref{eq:EMcons-planarSteady} across the bubble wall yields the matching equations
\begin{subequations} \label{eq:matchingConditions}
\begin{align}
	w_+ v_+^2 \gamma_+^2 + p_+ & = w_- v_-^2 \gamma_-^2 + p_- , \\
	w_+ v_+ \gamma_+^2 & = w_- v_- \gamma_-^2 .
\end{align}
\end{subequations}
Terms involving $\partial_z \phi$ vanish away from the wall (i.e.\ at $z_-$ and $z_+$) and have been dropped, conveniently leading to matching equations originally derived for a relativistic fluid in isolation~\cite{LandauLifshitz-relativistic}. The matching equations can also be applied across the shock front in a subsonic deflagration or hybrid, where the subscripts $+$ and $-$ would respectively denote quantities evaluated just in front of and just behind the shock front. Regardless of whether one considers the bubble wall or the shock front, the quantities are evaluated at their respective temperatures, $T_+$ and $T_-$. The fluid velocities just in front of and behind the bubble wall or shock front --- $v_+$ and $v_-$, respectively --- are then given by~\cite{LandauLifshitz-relativistic}
\begin{subequations}
\begin{align}
	v_+ & = \left[\frac{(p_- - p_+)(\rho_- + p_+)}{(\rho_- - \rho_+)(\rho_+ + p_-)} \right]^{\!\half} , \\
	v_- & = \left[\frac{(p_- - p_+)(\rho_+ + p_-)}{(\rho_- - \rho_+)(\rho_- + p_+)} \right]^{\!\half} ,
\end{align}
\end{subequations}
in the frame of reference of the bubble wall or shock front. The boundary conditions for $v$, $p$, $\rho$ and $T$ on each side of the bubble wall depend on the expansion mode of the bubble. We refer the reader to \refcite{Steinhardt:1981ct, Espinosa:2010hh, Leitao:2010yw, Giese:2020rtr, Giese:2020znk, Wang:2020nzm, Wang:2022lyd, Wang:2023jto} for calculations of the fluid profiles for the various expansion modes, and \refcite{Cai:2018teh} for an extension to slow phase transitions where the expansion of space cannot be neglected.

The fluid profiles can be solved by mapping to the bag equation of state~\cite{Espinosa:2010hh}, which is often used for its simplicity. \Refcite{Leitao:2014pda} developed the $\nu$-model to extend the bag equation of state. The $\nu$-model allows for a constant speed of sound differing from $c_s = 1/\sqrt{3}$ and illuminates the impact of the speed of sound on the energy budget. This model was also considered in \refcite{Giese:2020rtr, Giese:2020znk, Wang:2020nzm}. Two recent studies further improve on the equation of state, allowing for a non-constant speed of sound in each phase. \Refcite{Wang:2022lyd} consider a higher-order expansion of the finite-temperature potential. They find that higher-order terms provide negative contributions to the energy density and act to reduce the kinetic energy compared to the bag equation of state. \Refcite{Wang:2023jto} instead use the effective potential directly to compute the equation of state, not using any truncation or approximations beyond those used in the effective potential itself. They also solve the equations of motion across the bubble wall, rather than treating the bubble wall as a discontinuity. Their model-dependent method demonstrates that the fluid kinetic energy can be overestimated when using the bag equation of state, particularly for supersonic detonations and weak transitions.

Solving the fluid profiles gives the fluid velocity, enthalpy density and temperature around the bubble wall as functions of $\xi$. From there, the total fluid kinetic energy of an isolated bubble can be determined through
\begin{equation}
	\Ekin = 4\pi \int_0^\infty \! dr \, r^2 w(r) v^2(r) \gamma^2(v(r)) . \label{eq:Ekin-hydro}
\end{equation}
The integrand vanishes outside the fluid shell surrounding the bubble wall because $v(r)$ goes to zero. The hydrodynamic profiles are also useful for understanding local reheating, the terminal wall velocity, and other phenomenology that occurs near bubble walls. In the next section, we discuss how the kinetic energy is used in studies of GWs.

\subsection{Kinetic energy, RMS fluid velocity, and transition strength} \label{sec:kineticEnergy}

As described in \cref{sec:energyBudget} the energy available for GW
production from a phase transition can be sourced from energy in the
scalar field (bubble wall) and/or the fluid (see
\cref{eq:sourceTensor}).  The ratio of a source of energy to the total
energy density $\rho_\text{source}/\rhotot$ then has a strong
influence on any gravitational wave signal, with more energy corresponding
to larger signals.  In this section we will discuss how this can be
estimated for a given phase transition, focusing mostly on the case
where the energy goes into the fluid, though many aspects also apply to
the case where bubble collisions dominate.

As described in \cref{sec:energyBudget} if bubbles reach a terminal
velocity, energy that goes into the fluid dominates over the energy
available for collision sources (i.e.\ the energy remaining in the
scalar field).  Therefore the fluid kinetic energy represents the
energy available for GW production in transitions at finite
temperature when bubbles reach a terminal velocity. GW predictions then
depend on the kinetic energy fraction \cite{Hindmarsh:2019phv},
\begin{equation}
	K = \frac{\rhokinavg(\Ttr)}{\rhotot(\Ttr)} , \label{eq:kineticEnergyFraction}
\end{equation}
where $\rhokinavg$ is the volume-averaged kinetic energy density of the plasma, and the denominator is the average energy density of the plasma.  We take the latter to be $\rhotot(\Ttr) = \rho_f(\Ttr) - \rho_{\text{gs}}$ (i.e.\ the energy density in the false vacuum at the transition temperature $\Ttr$) using the same energy conservation argument used to arrive at \cref{eq:energyDensity-conservation}.%
\footnote{One must consistently subtract off the ground state energy
  density, $\rho_\text{gs}$, from the energy density and pressure when
  calculating important quantities such as the kinetic energy fraction
  and the Hubble parameter. Thus, it is more convenient to instead
  redefine the free energy density such that it vanishes in the ground
  state at zero temperature. We will assume such a redefinition in the
  remainder of this section.} %
Note that $K$ only represents the energy available for the production of
gravitational waves from the fluid.  Even if this is close to one most
of the energy will be transferred to heat during the progress of the
phase transition, leading to a suppression factor that is
incorporated in the gravitational wave fits.

We will discuss how the kinetic energy density appears in hydrodynamical simulations in \cref{sec:KE_hydrosims}. However, outside of lattice simulations, the usual approach of estimating the kinetic energy density is to consider a single, isolated bubble. Averaging the fluid kinetic energy around such a bubble, leads to \cite{Espinosa:2010hh}
\begin{equation}
	\rhokinavg = \frac{3}{\xi_w^3} \int_{0}^{\infty} \! d\xi \, \xi^2 w(\xi) v^2(\xi) \gamma^2(v(\xi)) . \label{eq:rhofl}
\end{equation}
where we have suppressed the implicit temperature dependence.  Note that in \cref{eq:rhofl}, the total kinetic energy \cref{eq:Ekin-hydro} is divided by the bubble volume up to the wall at $\xi_w$. This approach requires solving the hydrodynamic equations for the asymptotic fluid profile of the bubble, as discussed in \cref{sec:energyBudget}.

This has been done in \refcite{Espinosa:2010hh} in a rather model-independent fashion, but using the assumption of the bag equation of
state.  In this case the effective potential in the false vacuum can
be separated into the sum of constant vacuum energy density $\rho_V$
and a part which is proportional to $T^4$ (and the radiation energy density $\rho_R$), i.e.\ $V = \rho_V -\frac13 \rho_R = \epsilon - a T^4$. In the true vacuum $\rho_V=0$ is taken and
$\epsilon$ is the energy liberated by the phase transition.  As can be
understood from looking at the high- and low-temperature limits
discussed in \cref{sec:VT}
(\cref{eq:jb_high_t_exp,eq:jb_low_t_exp,Eq:jf_HT_exp,Eq:jf_LT_exp})
this will approximately hold when all particles have masses
$m_i \ll T$, where at leading order they only contribute to $\rho_R$, or
when $m_i\gg T$ where their contribution can be neglected due to the
Boltzmann suppression. However, it will break down when some masses are
close to the relevant temperature.

\subsubsection{Parameterizing the kinetic energy fraction}

It is useful to relate the kinetic energy fraction to the energy
released by the vacuum by introducing an efficiency coefficient that
represents how much of this energy is converted into the fluid kinetic
energy.  In the bag model the vacuum energy is given by the bag constant, $\epsilon$. Defining $\kappa = \rhokinavg / \epsilon$, so that $ K=\kappa \epsilon / \rhotot$, and further defining
$\alpha =\epsilon / \rho_R$ leads to \cite{Kamionkowski:1993fg}
\begin{equation} K = \frac{\kappa \alpha}{1+\alpha}.  \label{eq:Kparameterisation}   \end{equation}
If the bag equation of state holds one can compute $\epsilon$ to directly obtain $\alpha$ and find $\kappa$ using the fits
given in \refcite{Espinosa:2010hh}. The fits for $\kappa$ also require knowledge of
the bubble wall velocity and expansion mode (which are discussed in \cref{sec:wallVelocity}).

When the bag equation of state does not hold the energy density
does not separate so neatly into a constant vacuum contribution and a
$T^4$ radiation contribution.  However as discussed in
\cref{sec:energyBudget} the trace anomaly coincides with
$\epsilon$. Therefore $\theta_f - \theta_t$ may represent the energy
liberated from the vacuum, generalizing the bag constant $\epsilon$.
The total energy density can be written  $\rho_\text{tot} =
\theta_f + \frac34 w_f$, where $w_f$ is the enthalpy density in the false vacuum.  In this case the
efficiency coefficient is defined as
\begin{equation}
  \kappa = \frac{\rhokinavg}{\theta_f - \theta_t} ,
  \label{Eq:kappa_theta}
\end{equation}
and, as with the bag model equivalent, can be interpreted as the efficiency at which potential energy is converted to fluid kinetic energy.  The efficiency coefficient can then be estimated from fits provided in the appendix of \refcite{Espinosa:2010hh}, where it is parameterized by the terminal bubble wall velocity $\xi_w$ and $\alpha$. The latter is in turn defined by%
\footnote{\Refcite{Giese:2020znk} argue that the relevant temperature is actually the temperature in front of the wall, $T_+$, that enters the hydrodynamic matching equations, not $\Ttr$. While $T_+ = \Ttr$ for a supersonic detonation, equality does not hold for subsonic deflagrations or hybrids. Nevertheless, when using the pseudotrace \cref{eq:pseudotrace}, \refcite{Giese:2020znk} provide code in an appendix that numerically maps the transition strength parameter and the efficiency coefficient from $\Ttr$ to $T_+$.}
\begin{equation}
	\alpha = \frac{4(\theta_f(\Ttr) - \theta_t(\Ttr))}{3w_f(\Ttr)} , \label{eq:alpha-traceAnomaly}
\end{equation}
where $\theta$ is the trace anomaly defined in \cref{eq:traceAnomaly} and $w_f$ is the enthalpy density of the false vacuum. The latter generalizes the radiation energy density, with the relation $w_f = \frac43 \rho_R$ holding in the bag model. The enthalpy density (scaled by a factor $3/4$) now plays the role of the thermal energy as may be anticipated from its appearance alongside $\theta$ in the decomposition in \cref{eq:energyBudget} and text below it on the interpretation.

Using \cref{Eq:kappa_theta,eq:alpha-traceAnomaly} in \cref{eq:kineticEnergyFraction} we obtain \cite{Hindmarsh:2019phv}
\begin{equation}
	K = \frac{\kappa \alpha}{1 + \alpha + \delta} , \label{eq:Kparameterisation-delta}
\end{equation}
where
\begin{equation}
	\delta = \frac{4 \theta_t}{3 w_f}
\end{equation}
vanishes in the bag equation of state because $\theta_t = 0$.  This $\delta$ is usually neglected in the literature that uses the trace anomaly definition of the transition strength parameter $\alpha$ (\cref{eq:alpha-traceAnomaly}), such that the kinetic energy fraction is computed by combining \cref{eq:Kparameterisation,Eq:kappa_theta,eq:alpha-traceAnomaly}.  In fact usually $w_f$ is replaced by $\rho_R$, using the bag model relation  $w_f = \frac43 \rho_R$, and the expression is rewritten using the relations \cref{eq:hydro-rho} and $p = -V$
in \cref{eq:traceAnomaly} to obtain the form
\begin{equation}
	\alpha = \frac{\Delta(V - \frac14 T \partial_T V)|_{\Ttr}}{\rho_R(\Ttr)} , \label{eq:alpha-traceAnomaly-common}
\end{equation}
where $\Delta q = q_f - q_t$; that is, the difference in the quantity
between the false and true vacua. In \refcite{Athron:2023rfq} the
replacement of the enthalpy with $\rho_R$ was found to make little
difference in the scalar singlet model scenarios considered, and
similarly they found $\delta \ll \alpha +1$, such that neglecting it had less
than $1\%$ impact on the results.

Frequently in the literature the trace anomaly difference is replaced with the energy density difference or pressure difference. We can capture this by defining
\begin{equation}
	\alpha = \frac{4(q_f(\Ttr) - q_t(\Ttr))}{3w_f(\Ttr)} , \label{eq:transitionStrength}
\end{equation}
where $q$ can be the energy density ($\rho$), the trace anomaly ($\theta$), or the negative of the pressure ($-p$); i.e.\ the free energy.  The choice $q=\rho$ is commonly used in phase transition literature, along with the bag model relation to replace $w_f$ with $\rho_R$, to give
\begin{equation}
	\alpha = \frac{\rho_f(\Ttr) - \rho_t(\Ttr)}{\rho_R(\Ttr)} . \label{eq:latentHeat}
\end{equation}
The numerator is referred to as the \textit{latent heat} (density),%
\footnote{Latent heat density is actually the enthalpy density
  difference, which will only equal the energy density difference at
  the critical temperature. This was also noted in
  \refcite{Caprini:2019egz}.}  and the radiation energy density in the
denominator is evaluated using \cref{eq:rhoR}.  A similar expression
for $q=-p$ sometimes appears in the literature, where the numerator is
then replaced by $\Delta V$.

Using the energy density ($q=\rho$) in \cref{eq:transitionStrength} or \cref{eq:latentHeat} means that the $\rho_R$ is also included in the numerator. As a result, one may expect that this will overestimate the energy liberated from the vacuum and consequentially the kinetic energy fraction and resultant GW signals.  Meanwhile, using the pressure difference tends to underestimate the liberated energy.  \Refcite{Giese:2020rtr, Giese:2020znk} showed that choosing $q = - p$ or $q = \rho$ can lead to order of magnitude errors in the determination of $K$, while $q=\theta$ does considerably better.  Similarly \refcite{Athron:2023rfq} found scenarios with $\mathcal{O}(10)$ errors in the GW amplitude from such choices.  The difference vanishes in the limit $T\rightarrow 0$, so phase transitions at lower temperatures tend to be less sensitive.  Nonetheless we strongly recommend avoiding the choices $q=-p$ and $q=\rho$ in general. 





However even using $q=\theta$ in \cref{eq:transitionStrength} with \cref{eq:Kparameterisation} can lead to significant deviations in some cases \cite{Giese:2020rtr, Giese:2020znk}. This can  be improved upon by including the $\delta$ correction factor and introducing the pseudotrace \cite{Giese:2020rtr, Giese:2020znk}
\begin{equation}
	\thb(T) = \frac14 \! \left(\rho(T) - \frac{p(T)}{c_{s,t}^2(T)} \right) , \label{eq:pseudotrace}
\end{equation}
which generalizes the trace anomaly to models where the speed of sound differs from $c_s = 1/\sqrt{3}$. This provides another choice, $q = \thb$, for determining the transition strength. Now the correction factor $\delta$ in \cref{eq:Kparameterisation-delta} takes the form \cite{Athron:2023rfq}
\begin{equation}
	\delta = \frac{4}{3w_f} (\rhotot - \Delta \thb) - 1 .
\end{equation}

The kinetic energy fraction can be expressed as
\begin{equation}
	K = \frac{\thb_f(\Ttr) - \thb_t(\Ttr)}{\rho_f(\Ttr)} \kappa_{\thb}(\alpha_{\thb}, c_{s,f}^2, c_{s,t}^2) , \label{eq:Kpseudo}
\end{equation}
where $\alpha_{\thb}$ denotes the transition strength using the pseudotrace. This is equivalent to \cref{eq:Kparameterisation-delta} with the choice $q = \thb$ in \cref{eq:transitionStrength}. The appendices of \refcite{Giese:2020rtr,Giese:2020znk} include code for the determination of the efficiency coefficient $\kappa_{\thb}$. The inputs to $\kappa_{\thb}$ all implicitly depend on the temperature in front of the wall $T_+$, and the speed of sound in a given phase can be determined through
\begin{equation}
	c_{s,x}^2(T) = \left. \dv{p}{\rho} \right\vert_{\field_x(T)} = \left. \frac{\partial_T \FED}{T \partial_T^2 \FED} \right\vert_{\field_x(T)} ,
\end{equation}
with entropy held fixed \cite{LandauLifshitz-sound}. The efficiency coefficient is independent of $c_{s,f}$ for supersonic detonations. The derivation of \cref{eq:Kpseudo} assumes a weak transition such that $T_- \simeq T_+$, and the accuracy depends on how temperature independent the speed of sound is in the true vacuum.  We recommend using $q = \thb$ as it is the best parameterization approach currently available. Beyond that, one could integrate the hydrodynamic profiles around bubbles or perform a hydrodynamic simulation of the phase transition.

Irrespective of the method of determining the transition strength parameter, larger values lead to a stronger GW signal, as will be seen in \cref{Section:GWs-Sources}. Consequently, the transition strength $\alpha$ is synonymous with the strength of GW production. However, some approximations for the kinetic energy fraction assume $K \propto \alpha$, which would lead to a large overestimation in the GW amplitude for particularly strong transitions. The impact on GW predictions due to deviations from $c_s^2 = 1/3$ and more generally due to approximations in the determination of $K$ have recently been investigated in \refcite{Giese:2020znk, Wang:2020nzm, Wang:2021dwl, Tenkanen:2022tly,Athron:2023rfq}. We also reiterate that order of magnitude errors in the kinetic energy fraction (and consequently the GW spectrum) are possible when using common approximations such as \cref{eq:latentHeat}.

Finally we comment on the qualitative classification of transitions based on the transition strength parameter $\alpha$. Commonly the classification is that weak transitions have $\alpha \sim \mathcal{O}(0.01)$, intermediate transitions have $\alpha \sim \mathcal{O}(0.1)$, and strong transitions have $\alpha \gtrsim \mathcal{O}(1)$. Lattice simulations are currently unable to probe strong transitions or ultra-relativistic wall velocities due to the large scale hierarchies involved \cite{Hindmarsh:2015qta}. Thus, modeling of GW production is currently limited in these cases. Further work is required to understand when the turbulent regime begins and its effects on the GW signal, and to understand the effects of cosmic expansion in strongly supercooled transitions. GW predictions from models with large values for $\alpha$ are therefore extrapolated and subject to considerable --- and currently unquantifiable --- uncertainties.

\subsubsection{Kinetic energy from hydrodynamic simulations}
\label{sec:KE_hydrosims}
Hydrodynamic simulations determine the GW power from the fluid velocity field \cite{Hindmarsh:2019phv}. The GW power depends on the unequal time correlator of the fluid shear stress, which in turn depends on the volume-averaged fluid kinetic energy. However, $\rhokinavg$ is disguised in the form \cite{Hindmarsh:2013xza}
\begin{equation}
	\overbar{w} \Uf^2 = \recip{\vol} \int_\vol d^3 x \tau_{ii}^f , \label{eq:wUf}
\end{equation}
where $\tau_{ii}^f = \rhokin$, $\overbar{w}$ is the volume-averaged enthalpy density, and $\vol$ is the averaging volume. The right-hand side of \cref{eq:wUf} is equivalent to \cref{eq:rhofl} if $\vol$ is taken to be the bubble volume up to the wall. Assuming the asymptotic fluid profile is reached, the enthalpy-weighted mean square fluid four-velocity $\Uf^2$ from an isolated bubble is \cite{Hindmarsh:2019phv}
\begin{equation}
  \Uf^2 = \frac{\rhokinavg}{\overbar{w}} ,
  \label{Eq:Ufsq}
\end{equation}
where \cref{eq:rhofl} should be used for $\rhokinavg$ in \cref{Eq:Ufsq}. The kinetic energy is related to $\Uf$ through
\begin{equation}
	K = \Gamma \Uf^2 , \label{eq:kineticEnergyFraction-Uf}
\end{equation}
where $\Gamma = \overbar{w}(\Ttr)/\rho_f(\Ttr)$ is the mean adiabatic index of the plasma at the transition temperature, and we used energy conservation to estimate $\rho_f$ as the mean energy density. The adiabatic index is $\Gamma = 4/3$ in the bag equation of state.

Several approaches to estimating $\Uf$ have been compared in hydrodynamic simulations~\cite{Hindmarsh:2017gnf, Cutting:2019zws, Hindmarsh:2019phv}. Generally the prediction from the asymptotic fluid profile agrees reasonably well with the prediction from the fluid profile at the time of collision. However, the value for $\Uf$ extracted from the simulated fluid was recently found to be lower than expected from the isolated bubble profile \cite{Cutting:2019zws}. The effect is more noticeable for stronger transitions with lower wall velocities, mainly impacting subsonic deflagrations and presumably hybrids. The cause of the kinetic energy suppression is expected to be the reheating that occurs in front of the bubble walls (except for supersonic detonations), which increases the pressure in the false vacuum and hence delays its collapse. Additionally, bubbles may not reach their asymptotic fluid profiles by the end of the simulation. \Refcite{Guo:2021qcq, Gowling:2021gcy} account for this effect by using a GW suppression factor based on the ratio of the RMS fluid velocities predicted from the isolated bubble fluid profile and that extracted from simulations.

\subsection{Characteristic length scales and timescales} \label{sec:lengthscale}

The characteristic length scale, $\lenscale$, has an important impact on the resulting GW signal and the prospects of its detection. As will be shown in \cref{Section:GWs-Sources}, the characteristic length scale affects both the amplitude and peak frequency (i.e.\ the frequency at which the amplitude peaks) of the GW signal. The length scale informs the timescale of events such as the formation of shocks and the onset of turbulence. These timescales  directly affect the duration for which GWs are generated, and therefore the amplitude of the signal. Here we only briefly mention the timescales involved; see \refcite{Caprini:2019egz} and references therein for a more detailed discussion.

The GW amplitude depends on the lifetime of the source. For sound waves, the acoustic period ends due to the onset of turbulence, which happens at around the time shocks begin to form.\footnote{A survey of the lifetime of the sound wave source for various models was conducted in \refcite{Ellis:2020awk}.} The shock formation timescale can be estimated by 
\begin{equation}\label{eq:tau_shock}
\tau_{\text{sh}} \sim \frac{\lenscale}{\Ulong},
\end{equation}
where $\Ulong$ is the longitudinal component of the RMS fluid four-velocity. However, \refcite{Hindmarsh:2013xza,Hindmarsh:2015qta} determined that the maximum effective lifetime of the sound wave source of GWs was precisely the Hubble time, $H_*^{-1}$, even in the absence of turbulence. GW production ceases because the shear stress sourcing GWs is damped by the expansion of space and decorrelated. The effective lifetime of the source was subsequently often taken to actually be the Hubble time, since this is assumed in the numerical lattice simulations from which the GW fits are obtained (see \cref{Section:GWs-Sources}).  However this is only true if $H_*L_* > \Uf$.  \Refcite{Ellis:2018mja} showed that it is not fulfilled in realistic scenarios they investigated. To address this, \refcite{Ellis:2019oqb} introduced a suppression factor for the GW signal,\footnote{Contrary to the common definition of Hubble time being $H_0^{-1}$, we use $H_*^{-1}$ where $t_*$ is the transition time \cite{RoperPol:2019wvy}.}
\begin{equation}
  \Upsilon = \min(1, H_* \tau_{\text{sw}}).
  \label{Eq:sw_lifetime_min}
\end{equation}
However, \refcite{Guo:2020grp} found that the original findings of the maximum effective lifetime neglect effects from the expansion of space-time.  As a result  a careful calculation shows that \cref{Eq:sw_lifetime_min} was only an approximation to
\begin{equation}
  \Upsilon = 1 - \recip{\sqrt{1 + 2 H_*\tau_{\text{sw}} }} .
  \label{Eq:sw_lifetime_upsilon}
\end{equation}
The approximate \cref{Eq:sw_lifetime_min} matches this more precise form when $H_* \tau_{\text{sw}} \ll 1$ and when $H_* \tau_{\text{sw}} \gg 1$. \Cref{Eq:sw_lifetime_upsilon} assumes radiation domination and should be slightly altered in the case of matter domination~\cite{Guo:2020grp}. The suppression of the sound wave signal caused by expansion was seen in simulations that do not assume a flat background~\cite{Cai:2018teh} and found through analytic calculations to be an important effect for collisions as well as sound waves~\cite{Zhong:2021hgo}. Another possible mechanism for suppressing the sound wave source is through viscous damping; however, \refcite{Hindmarsh:2015qta} demonstrated that the timescale associated with viscous damping is considerably longer than a Hubble time.

There are two other timescales that are of great significance for GWs.  The first is the autocorrelation time, $\tau_{\text{c}}$, which ends when the sound wave source decorrelates.  This is expected to scale like the sound-crossing time
\begin{align}\label{eq:tau_c}
  \tau_{\text{c}} \sim \frac{\lenscale}{c_s}.
\end{align}
The other timescale is for the onset of turbulence,
\begin{align}\label{eq:tau_turbulence}
\tau_{\text{turb}} \sim \frac{\lenscale}{\Utrans},
 \end{align}
where $\Utrans$ is the transverse component of the RMS fluid
four-velocity.
We leave further discussion of these timescales to
\refcite{Caprini:2019egz}.

Because these timescales are only approximately estimated, they may contribute significantly to the uncertainty in GW predictions. The transition between the laminar flow and the turbulent regimes requires further study, particularly for strong phase transitions \cite{Cutting:2019zws}. Some of the timescales involve longitudinal and rotational components of the plasma four-velocity. These separate components cannot easily be predicted in non-lattice simulations, while the total four-velocity is more accessible. In light of the enhanced vorticity realized in a recent simulation of intermediate strength phase transitions, we currently lack a reasonable estimate of the splitting between longitudinal and rotational components for use in non-lattice studies. This compounds the already approximate nature of the timescales.

We see that the estimates for the timescales depend on a characteristic length scale. The length scale --- rather than the timescale --- is the quantity determined in non-lattice studies. Once $\lenscale$ is known, it can be used to estimate the relevant timescales. Several characteristic length scales see common use. In the remainder of this section, we enumerate them and discuss the motivation behind the various choices.

The most popular choice for $\lenscale$ is the mean bubble separation, $\bubsep$, which has been used in hydrodynamic simulations from which fits for the GW signals are obtained.%
\footnote{The usual notation for the mean bubble separation is $R_*$. We avoid this notation because by our convention it could be confused with any radius evaluated at the transition temperature.}
It can be determined from the bubble number density through
\begin{equation}
	\bubsep(t) = n(t)^{-\frac13} . \label{eq:meanBubbleSeparation}
\end{equation}
The bubble number density is given by \cite{Guth:1979bh}
\begin{equation}
	n(t) = \int_{t_c}^t dt' \Gamma(t') P_f(t') \frac{a^3(t')}{a^3(t)} , \label{eq:bubbleNumberDensity}
\end{equation}
and is evidently diluted by the expansion of space. The mean bubble separation is a directly controllable quantity in hydrodynamic simulations from which GW fits are extracted. Such simulations typically nucleate the bubbles simultaneously at a fixed separation and neglect the expansion of space, such that the mean bubble separation is constant throughout the simulation and effectively an input to the simulation (see e.g.\ \refcite{Hindmarsh:2015qta}). Naturally, the mean bubble separation appears as a parameter in the GW fits extracted from the hydrodynamic simulation data.

Other quantities have also been considered for use as a characteristic length scale in the literature. For example the mean bubble radius,
\begin{equation}
	\bubrad(t) = \recip{n(t)} \int_{t_c}^{t} dt' \Gamma(t') P_f(t') \frac{a^3(t')}{a^3(t)} R(t', t) , \label{eq:meanBubbleRadius}
\end{equation}
is also a candidate \cite{Megevand:2016lpr, Cai:2017tmh}, where the actual radius of the bubbles is given by \cref{eq:bubbleRadius} (usually with the initial bubble radius neglected).  The mean bubble separation and radius should be roughly of the same order of magnitude around the time of percolation and completion (as shown explicitly in \refcite{Leitao:2015fmj, Megevand:2016lpr, Megevand:2017vtb}). The mean bubble radius may be more directly connected to the stirring scale for turbulence in the plasma \cite{Kosowsky:2001xp, Caprini:2009yp}. However, the characteristic scale for turbulence is typically derived from the largest bubbles that collide, because these cause the largest scale eddies in the plasma. Eddies at smaller scales are further generated by the cascade of larger eddies.

Another candidate is the maximum of the energy-weighted distribution of bubble radii \cite{Huber:2007vva, Ellis:2018mja}. While small bubbles may be more numerous than large bubbles --- especially if the nucleation rate is exponential --- each large bubble contributes more energy to GW production. The distribution of bubble radii can be found by differentiating \cref{eq:bubbleNumberDensity} with respect to radius, leading to \cite{Turner:1992tz, Cai:2017tmh}
\begin{equation}
	\dv{n}{R}{}(t, R) = \frac{\Gamma(t') P_f(t')}{v_w(t')} \frac{a^4(t')}{a^4(t)} .
\end{equation}
Here, $t'(t, R)$ is the nucleation time of a bubble that has radius $R$ at time $t$. To good approximation, the energy of a bubble scales with its volume. Then the characteristic length scale can be approximated as the radius that maximizes the volume-weighted distribution of bubble radii, $\bubradweight$, by solving
\begin{equation}
	\dv{R} \! \left.\left(R^3 \dv{n}{R}{}(t_*, R) \right) \right\vert_{\bubradweight} \! = 0 . \label{eq:maximalEnergyRadius}
\end{equation}
\Refcite{Kobakhidze:2017mru} instead use the radius at which $dn/dR$ peaks as the characteristic length scale. This is motivated by the idea that the majority of bubbles colliding have this radius.%
\footnote{The distribution of bubble radii may be maximized at $R = 0$ in a fast transition with bubbles growing as supersonic detonations, even at the onset of percolation \cite{Megevand:2016lpr, Megevand:2017vtb}. On the other hand, the volume-weighted distribution was found to peak away from $R = 0$ at the onset of percolation in \refcite{Megevand:2016lpr, Megevand:2017vtb}.}

We currently recommend using $\lenscale = \bubsep$ because the hydrodynamic simulations that provide GW fits calculate $\bubsep$ rather than $\bubrad$ or $\bubradweight$.
However, it is very common in the literature to approximate the length scale based on the shape of the bounce action, $S$. The bounce action is Taylor expanded about the transition time $t_*$ in the expression for the nucleation rate, $\Gamma = A\exp(-S)$. A fast transition is assumed such that the Taylor expansion can be truncated at the linear term,
\begin{equation}
	S(t) \approx S(t_*) - \beta_* (t - t_*) ,
\end{equation}
where
\begin{equation}
	\beta_* = -\left. \dv{S}{t} \right\vert_{t_*} . \label{eq:beta}
\end{equation}
This gives an exponential nucleation rate,
\begin{equation}
  \Gamma(t) = \Gamma(t_*) \exp(\beta_* (t - t_*)).
  \label{Eq:exp_nuc_rate}
\end{equation}
An approximate relationship between $\beta_*$ and $\bubsep$ can be obtained when this exponential nucleation rate is assumed by additionally neglecting effects from the expansion of space (which should also be small for a fast transition). One finds \cite{Enqvist:1991xw, Megevand:2016lpr}
\begin{equation}
	\bubsep \approx (8\pi)^{\frac13} \frac{v_w}{\beta_*} , \label{eq:bubSepBeta}
\end{equation}
where $v_w$ is the bubble wall velocity,  and the factor of $(8\pi)^{\frac13}$ is sometimes omitted. Strictly speaking, this relation is obtained at the so-called $e$-folding time/temperature, where the false vacuum fraction $P_f = 1/e$, but it is often applied at the percolation or nucleation temperature. Clearly this linear approximation of the bounce action is only valid for fast transitions, where the differences between various transition times/temperatures will be small; however, $\dint S / \dint t$ could vary significantly.  \Refcite{Guo:2020grp} generalize this mapping to the case of an expanding Universe, which simply introduces a ratio of scale factors to the right-hand side of \cref{eq:bubSepBeta}. This additional factor should not be too large for fast transitions.

The mapping between the transition rate and the mean bubble separation
was slightly modified in \refcite{Caprini:2019egz}, where they instead
use
\begin{equation}
  \bubsep \approx (8\pi)^{\frac13} \frac{\max(v_w, c_{s,f})}{\beta_*} .
  \label{Eq:BubbleSepToBeta}
\end{equation}
This adjustment accounts for the collision of shocks preceding the bubble walls for bubbles expanding as subsonic deflagrations. Note that the shock front does not travel at exactly $c_{s,f}$ in general (see \cref{sec:bubbleExpansionModes}).

An exponential nucleation rate can be anticipated in many scenarios
because the FOPT is often caused by a barrier
that is induced by the temperature-dependent terms in the effective
potential (see \cref{sec:VT}).  Above the critical temperature the
nucleation rate should vanish, and below it the barrier inhibits
nucleation.  Assuming that below $T_c$ the barrier steadily decreases
as the Universe cools, until it dissolves at some lower temperature
$T_0$, the nucleation rate should monotonically increase until
$T_0$ is reached.\footnote{At $T_0$ if it has not already completed,
  the phase transition should happen instantly with no barrier to
  prevent the slide into the lower minimum.}

However, caution must be taken regarding the use of $\beta_*$ because
this picture is not always valid.  For example, if a large barrier
persists at low energies (or even $T=0$) then the transition will not
be fast and the barrier may inhibit nucleation,
leading to a maximum in the nucleation rate at some finite
temperature.  The linear coefficient in the Taylor
expansion, $\beta_*$, will vanish about the maximum, and at lower
temperatures $\beta_*$ will be negative. Keeping only the second
term in the expansion gives a Gaussian nucleation rate. Thus, for these
strongly supercooled transitions this linear approximation for $S$ must be avoided.
An alternative approximation for the mean bubble separation can then
also be made for a Gaussian nucleation rate with or without the
expansion of space \cite{Megevand:2016lpr}. Nonetheless, we advocate
avoiding both approximations and instead computing the mean
bubble separation directly using
\cref{eq:meanBubbleSeparation,eq:bubbleNumberDensity}. The assumption
of an exponential nucleation rate also breaks down if there is
substantial reheating before the transition is over
\cite{Heckler:1994uu, Megevand:2017vtb}. This can occur if the bubbles
expand as subsonic deflagrations or hybrids. A delta-function
nucleation rate was found to be a reasonable approximation for this
case in \refcite{Megevand:2017vtb}.

A quadratic correction term for $S$ was investigated in
\refcite{Jinno:2017ixd} and found to have an impact on GWs at the
$\mathcal{O}(10\%)$ level in several particle physics models. The
transition rate can also be used to estimate $\bubradweight$ (defined
in \cref{eq:maximalEnergyRadius}); see Appendix A of
\refcite{Huber:2007vva} and Appendix B.1 of \refcite{Levi:2022bzt} for
details. Approximations for the transition rate exist in simplified
particle physics models where the bounce action can be estimated (see
e.g.\ \refcite{Levi:2022bzt}).

A recent analysis \cite{Megevand:2021llq} suggests that the timescale rather than the length scale may be more relevant for GWs sourced by bubble collisions. However, the relevant timescale is the total duration of the phase transition, not the exponential decay constant $\beta_*$. The ratio of $\beta_*^{-1}$ and $t_* - t_c$ can be estimated as $\beta_*^{-1}/(t_* - t_c) \sim [8 \ln(M_P/v)]^{-1}$, where $M_P$ is the Planck mass and $v$ is the scale of the model \cite{Megevand:2017vtb}. This ratio is $\sim\!0.3\%$ at the electroweak scale, demonstrating that $\beta_*^{-1}$ is not a good indication of the duration of a phase transition.

Recent modeling \cite{Hindmarsh:2016lnk} and simulations \cite{Hindmarsh:2019phv} of the sound wave source of GWs have identified another relevant length scale. The sound shell model \cite{Hindmarsh:2016lnk} predicts a double-broken power law for the spectral shape of the GW signal (see \cref{Section:GWs-Sound}). Thus two length scales inform the spectral shape of the acoustic GWs. First is the mean bubble separation. Second is the thickness of fluid shells, which for subsonic deflagrations and supersonic detonations is approximated as
\begin{equation}
	\bubsep \Delta_w \approx \bubsep \abs{v_w - c_{s,f}} / v_w . \label{eq:fluidShellThickness}
\end{equation}
This approximation underestimates the thickness for subsonic deflagrations and hybrids, particularly for strong transitions. Although the thickness of the fluid shell could be computed from the asymptotic fluid profile of the bubble,%
\footnote{See \refcite{Espinosa:2010hh, Leitao:2010yw} for estimates of the fluid shell thickness based on the kinetic energy profile around the bubble wall.}
the precise relevance of this length scale has not been determined so an estimate is fitting. Further, the results of \refcite{Hindmarsh:2015qta} suggest that the integral scale derived from the velocity field is more fundamental than the mean bubble separation. However, the integral scale is not accessible in typical studies of GWs from particle physics models.

Deformations in the bubble wall arising from inhomogeneous reheating or hydrodynamic instabilities may introduce yet another length scale that influences the spectral shape of the GW signal \cite{Megevand:2021juo}. The corresponding peak is at a higher frequency than from collisions of bubbles and thin fluid shells.

The key takeaway from this discussion is that while the exact quantity defining the characteristic length scale is not yet known, simulations and theoretical estimates predict the spectral shape and amplitude of the GW signal should depend on some combination of $\bubsep$, $\bubrad$, $\bubradweight$ and $\Delta_w$, where the former three are likely to agree within an order of magnitude. The ambiguity in the choice of length scale is a source of uncertainty in GW predictions, as is the mapping between length and timescales.

\subsection{Bubble wall velocity} \label{sec:wallVelocity}

\subsubsection{Significance of the bubble wall velocity}

The bubble wall velocity is the most complicated thermal parameter to determine, often left as a free parameter or fixed arbitrarily in studies of GWs from cosmological phase transitions. Outside of lattice simulations, studies invariably consider the terminal bubble wall velocity; that is, bubbles have reached a steady state where the driving pressure equals the friction from the plasma. The wall velocity of an individual bubble is a time-dependent quantity, with bubble walls starting at rest and accelerating due to the pressure difference between the phases. Nevertheless, the acceleration stage is predicted to be much shorter than the lifetime of a bubble, particularly in a strongly supercooled transition where bubbles can grow for a prolonged duration before colliding. Under the assumption that the acceleration stage is negligible, we may proceed to consider only the terminal wall velocity.

There is a caveat to the above discussion: in some transitions the bubble walls can `run away', accelerating up to the time of collision. The runaway regime is encountered when the driving pressure always exceeds the friction. This can happen in transitions that occur at very low temperatures, where the plasma is sufficiently diluted. Such transitions are sometimes referred to as vacuum transitions. Bubble walls may also run away during transitions in dark (or other non-SM) sectors where the transition does not disturb the plasma significantly. Then the bubble walls are free to accelerate with negligible impact on and from the plasma. The energy budget in a transition with runaway bubble walls is different from the energy budget when bubbles reach a terminal wall velocity. The kinetic and gradient energy of the scalar field is no longer negligible, because the energy can build up in the bubble wall instead of being redistributed to the surrounding plasma. We will discuss whether bubble walls run away in \cref{sec:friction}, where we provide a detailed discussion of the literature on estimates of the friction.  Note that this topic in particular has a lot of activity and, as we will indicate, there are controversial and even incorrect results that could be misused. Therefore this literature should be read with care.   

The bubble wall velocity affects not only the energy budget, but the progress of the transition itself. Throughout \cref{sec:transitionAnalysis,sec:thermal_parameters} we have discussed how various results depended on whether reheating occurs in front of the bubble wall. Reheating occurs in front of the bubble walls if the bubbles grow as subsonic deflagrations or hybrids, but not supersonic detonations. The expansion mode is set by the bubble wall velocity and the equation of state. Thus, the dynamics of the phase transition depend on the expansion mode of the bubbles and in turn the bubble wall velocity. We now turn our attention to enumerating and discussing the possible expansion modes.

\subsubsection{Bubble expansion modes} \label{sec:bubbleExpansionModes}
\begin{figure}
	\centering
	\includegraphics[width=0.95\linewidth]{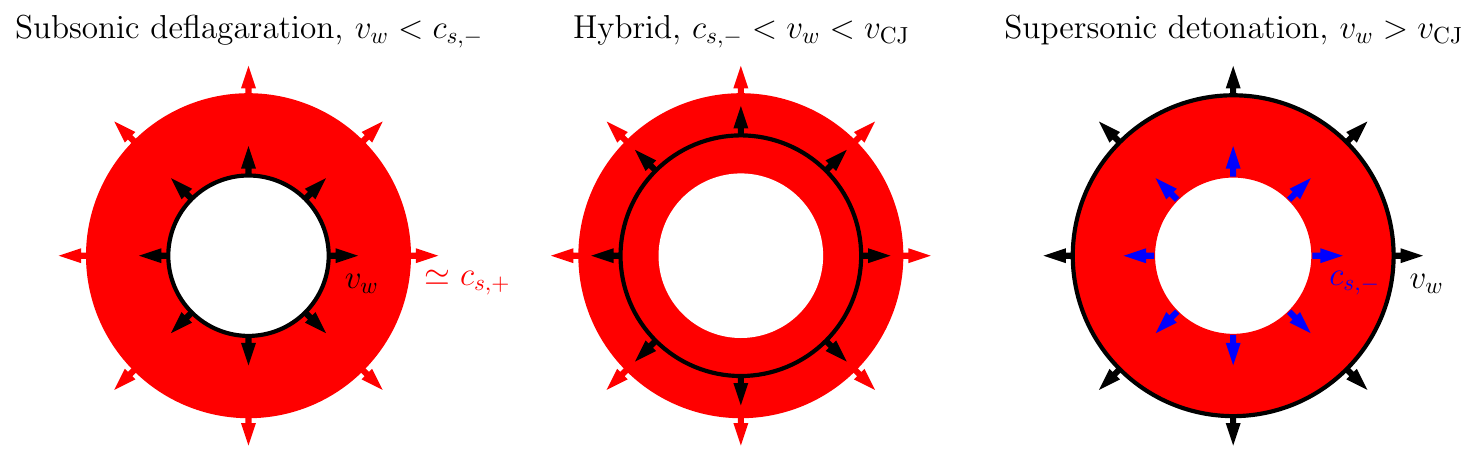}
	\caption{The expansion of a bubble wall (black) and the region of non-zero plasma velocity (red) in the frame of the bubble center, in four scenarios. The plasma velocity profile for supersonic deflagrations and subsonic detonations are indistinguishable in this illustration. These two expansion modes are collectively referred to as hybrid modes. Based on Fig.~3 in \refcite{Espinosa:2010hh}.}
	\label{fig:expansionModes}
\end{figure}

There are four expansion modes for bubbles in a cosmological phase transition: subsonic detonations, supersonic detonations, subsonic deflagrations, and supersonic deflagrations. In all expansion modes except supersonic detonations, there is a shock%
\footnote{The shock front corresponds to a discontinuity in thermodynamic quantities such as fluid velocity and temperature. The front sharpens over time into a shock \cite{Ignatius:1993qn}. The region between the bubble wall and the shock front is called a shock wave or sound shell.}
wave in front of the bubble wall, so the plasma in front of the bubble wall is disturbed. As a result, the plasma in front of the bubble wall is reheated, which has important implications on the evolution of the phase transition as previously discussed. In all expansion modes except subsonic deflagrations, there is a rarefaction behind the bubble wall, so the plasma just inside the bubble is not at rest with respect to the bubble center. We illustrate these expansion modes in \cref{fig:expansionModes}.

The distinction between detonations and deflagrations is based on the
relative velocities of the plasma just inside and just outside the
bubble wall.  Recall from \cref{sec:energyBudget} that we
  denote the velocity of the plasma just inside and outside the bubble wall (in the bubble wall frame of reference)
  as $v_-$ and $v_+$, respectively.  Detonations have $\abs{v_+} >
\abs{v_-}$, while deflagrations have $\abs{v_+} < \abs{v_-}$. A deflagration or detonation is weak
if the fluid velocities on both sides of the bubble wall, $v_-$ and
$v_+$, are both subsonic or both supersonic. Alternatively, a
deflagration or detonation is strong if only one of $v_-$ and $v_+$ is
supersonic, while the other is subsonic. The Chapman-Jouguet
condition, $v_- = c_{s,-}$, defines the border of the weak and strong
classification, where $c_{s,-}$ is the speed of sound just inside the
bubble. While strong detonations have been shown to be stable to
corrugations in the bubble wall \cite{Rezzolla:1996ey} (i.e.\ small
perturbations in the bubble wall surface), they are not possible in
cosmological phase transitions \cite{Steinhardt:1981ct,
  Laine:1993ey}. Strong deflagrations are also known to be an
impossible expansion mode for bubbles in a cosmological phase
transition \cite{Laine:1993ey}. This leaves weak and Chapman-Jouguet
deflagrations and detonations as the only possibilities. The stability
of deflagrations has been considered in the non-relativistic limit
\cite{Link:1992dm}, and for relativistic bubble wall velocities
\cite{Huet:1992ex}. Corrugations in the bubble wall grow exponentially
for sufficiently small wall velocities \cite{Huet:1992ex}. The
stability analysis was later improved in \refcite{Megevand:2013yua,
  Megevand:2014dua}, where it was found that the threshold velocity of
stability increases with supercooling, up to the sound speed. Thus,
for sufficiently strong supercooling, subsonic deflagrations are not a
stable expansion mode and the bubble wall accelerates to supersonic
velocities \cite{Kamionkowski:1992dc}.%
\footnote{See \refcite{Megevand:2014dua} for further discussion and references within for alternative possibilities for the evolution of an unstable wall.}
Meanwhile, \refcite{Ai:2023see} found that when in local equilibrium, any value of the transition strength parameter $\alpha$ that allows for detonations also allows for deflagrations. Because bubbles start from rest and would accelerate to the steady state deflagration solution first, it was argued that the bubbles would not expand as detonations. Such an argument may break down in the presence of hydrodynamic instabilities.

Supersonic deflagrations were first shown to be possible in \refcite{Kurki-Suonio:1995rrv}, and are commonly referred to as hybrids because they consist of both a rarefaction behind the wall and a shock wave in front of the wall. Another type of hybrid is the subsonic detonation mode, which is only possible when the speed of sound is smaller inside the bubble than it is outside the bubble (i.e.\ $c_{s,-} < c_{s,+}$). Subsonic detonations were first identified in \refcite{Leitao:2014pda}, and require going beyond the bag equation of state, which has $c_{s,-} = c_{s,+} = 1/\sqrt{3}$. Both hybrid modes satisfy the Chapman-Jouguet condition. Although many studies only consider supersonic deflagrations, the implications on the phase transition for both hybrid modes should be qualitatively similar. This is because the fluid velocity (in the plasma frame) is non-zero on both sides of the wall in both supersonic deflagrations and subsonic detonations \cite{Leitao:2014pda}.

Subsonic deflagrations were considered in \refcite{Gyulassy:1983rq} in $1+1$ dimensions, and shortly after in $3+1$ dimensions in \refcite{Kurki-Suonio:1984zeb}. While the qualitative features of the fluid profiles are similar in one and three spatial dimensions, they are quantitatively quite different. For instance, the fluid velocity is constant between the bubble wall and the shock front in a 1D subsonic deflagration. In a 3D subsonic deflagration, the fluid velocity instead peaks at the bubble wall and decays to a finite value at the shock front. In general, the shock wave is narrower and weaker as the number of spatial dimensions increases \cite{Kurki-Suonio:1984zeb}. This behavior is also nicely demonstrated in \refcite{Leitao:2010yw}. The shock front is typically located at $\xi_{\text{sh}} \simeq c_{s,+}$ for weak subsonic deflagrations \cite{Leitao:2014pda}, motivating the approximation \cref{eq:fluidShellThickness} for the thickness of the shock wave.

The Chapman-Jouguet detonation is the only possible expansion mode in chemical combustion \cite{LandauLifshitz-combustion}, and as a result originally \cite{Steinhardt:1981ct} this was assumed to also be the case in cosmological phase transitions. It was later found that the Chapman-Jouguet condition need not be satisfied in a cosmological phase transition \cite{Laine:1993ey}. Chapman-Jouguet detonations have been assumed in many studies of GWs from cosmological phase transitions. The corresponding bubble wall velocity is given by
\begin{equation}
	v_\text{CJ} = \frac{\sqrt{\alpha_+(2 + 3\alpha_+)} + 1}{\sqrt{3} (1 + \alpha_+)} \label{eq:vcj}
\end{equation}
in the bag equation of state,%
\footnote{Analytic equations for the Chapman-Jouguet velocity have also been determined in more general equations of state \cite{Leitao:2014pda, Giese:2020rtr}.}
where $\alpha_+$ is the ratio of the vacuum energy density to the radiation energy density (see e.g.\ \refcite{Steinhardt:1981ct, Espinosa:2010hh}). However, Chapman-Jouguet detonations are not hydrodynamically favored over weak detonations, so fixing $v_w = v_{\text{CJ}}$ is an arbitrary choice. In fact, $v_\text{CJ}$ is the smallest wall velocity corresponding to a weak supersonic detonation. If there is insufficient supercooling, the wall velocity may not reach $v_\text{CJ}$, ruling out the supersonic detonation mode altogether. If the wall velocity is below $c_{s,-}$, then a subsonic deflagration is the only possible expansion mode.

To summarize, we are left with six possible expansion modes: weak and Chapman-Jouguet subsonic deflagrations, Chapman-Jouguet supersonic deflagrations and subsonic detonations, and weak and Chapman-Jouguet supersonic detonations. The other six expansion modes are ruled out either due to having no consistent hydrodynamic solution, or due to being unstable to corrugations in the bubble wall. In general, the terminal bubble wall velocity, $v_w$, along with the equation of state, set the possible expansion modes for the bubble. However, the driving pressure and friction that govern the terminal bubble wall velocity are also affected by the expansion mode. Thus, the determination of the terminal bubble wall velocity and the expansion mode are inherently coupled problems.

\subsubsection{Friction on expanding bubbles} \label{sec:friction}

There are three unknown quantities (the bubble wall velocity, and the plasma temperature and velocity profiles around the bubble wall), while we only have two equations (\cref{eq:matchingConditions}) from the conservation of the energy-momentum tensor \cite{Turok:1992jp}. If the bubble wall velocity is fixed by hand, the hydrodynamic profiles around the bubble can be solved. However, determining the bubble wall velocity itself requires a third equation: the equation of motion of the scalar fields \cite{Turok:1992jp}. An alternative, approximate method that avoids the equations of motion is to assume local thermal equilibrium, which offers a simpler third equation; namely that of entropy conservation. We will discuss these methods and the various approximations often used in the first approach, as well as provide a commentary on developments in the literature.

Fundamentally, determining the terminal bubble wall velocity is simply a matter of finding when the driving pressure matches the friction from the plasma. That is, acceleration ceases when the total pressure,
\begin{equation}
	\ptot(t, v) = \pdr(t) - \pfr(t, v) , \label{eq:totalPressure}
\end{equation}
is zero, so the bubble expands at a constant rate. Importantly, the friction depends on the bubble wall velocity $v$. By fixing the time in \cref{eq:totalPressure}, say to the transition time $t_*$, and setting the total pressure to zero, one can obtain the \textit{instantaneous} terminal bubble wall velocity, $v_w$.%
\footnote{The terminal bubble wall velocity is time dependent. In fact, $\ptot$ (and consequently $v_w$) should also depend on the time since nucleation of the bubble because the hydrodynamic profile of the plasma around the bubble takes time to reach the asymptotic profile. Additionally, the bubble wall profile takes time to settle such that the exit point of the tunneling solution lies at the true vacuum.}
Thus, the terminal bubble wall velocity is usually defined such that $\ptot(t_*, v_w) = 0$. See \refcite{Liu:1992tn, Turok:1992jp, Dine:1992wr} for early works on the bubble wall velocity.

The driving pressure is simply the pressure difference between the phases,
\begin{equation}
	\pdr(t) = V(\field_f(t), T_+(t)) - V(\field_t(t), T_-(t)) . \label{eq:pdr}
\end{equation}
If one were to neglect hydrodynamics, then the total pressure would be zero when $\pfr(t) = V(\field_f(t), T(t)) - V(\field_t(t), T(t))$. The friction is unfortunately considerably more complicated. Intuitively, the friction arises because the passage of the bubble wall alters the equation of state of the plasma; and in the case of the electroweak transition, the mass of particles change upon entering the bubble. Sufficiently energetic particles may pass through the wall but transfer momentum to the wall, while other particles are reflected off the wall \cite{Bodeker:2009qy}. Additionally, particles may emit soft gauge bosons as they enter the wall, enhancing the friction \cite{Bodeker:2017cim}. While we focus on terminal bubble wall velocity studies in weakly coupled theories, the terminal bubble wall velocity in strongly coupled theories can be studied using holography \cite{Bea:2021zsu, Bigazzi:2021ucw} or hydrodynamics in the non-relativistic limit \cite{LiLi:2023dlc}.

The total pressure on the bubble wall can be determined by solving the equations of motion of the field-fluid system. The equation of motion for the scalar field $\phi_i$ is given by%
\footnote{In the WKB approximation, the operators for the quantum fields are thermally averaged over the ensemble of particles, leading to the form of \cref{eq:Tuv-plasma} and consequently \cref{eq:scalarEOM} \cite{Moore:1995si, Moore:1995ua}. The WKB approximation is discussed and justified in \refcite{Liu:1992tn, Moore:1995si}. Importantly, the wall thickness is larger than the inverse momenta of particles (characterized by $T^{-1}$).}
\begin{equation}
	\partial_\rho \partial^\rho \phi_i + \pdv{V_0}{\phi_i} + \sum_j \dv{m_j^2}{z} \! \int \! \frac{d^3 k}{(2\pi)^3 2E_j} f_j(k, z) = 0 , \label{eq:scalarEOM}
\end{equation}
where $f_j$ is the distribution function of particle species $j$, and we have assumed a Minkowski metric. More generally, the partial derivatives should be replaced with covariant derivatives. The total pressure can be obtained by integrating the second and third terms in \cref{eq:scalarEOM} across the bubble wall~\cite{Moore:1995si}:%
\footnote{To see this, one can identify force terms when multiplying the left-hand side of \cref{eq:scalarEOM} by $\partial^\mu \phi_i$, as done in \refcite{Moore:1995si}.}
\begin{equation}
	\ptot(t, v) = \int \! dz \dv{\phi_i}{z} \left[\pdv{V_0}{\phi_i} + \sum_j \dv{m_j^2}{z} \! \int \! \frac{d^3 k}{(2\pi)^3 2E_j} f_j(k, z) \right] . \label{eq:ptot-general}
\end{equation}
We assume the bubble wall expands in the $+z$ direction, with a positive $\ptot$ accelerating the expansion in the $+z$ direction. The time dependence of $\ptot$ enters through the evolution of the potential and consequently the equation of state, while the velocity dependence may be understood by considering its effect on the bubble expansion mode, and more subtly how the transport and interactions of particles within the bubble wall is affected by this velocity. The total pressure should vanish when the bubble wall reaches terminal velocity.

We decompose the distribution function as $f_j(k, z) = f_j^{\text{eq}}(k, z) + \delta f_j(k, z)$, where
\begin{equation}
	f_j^{\text{eq}}(k, z) = \left[\exp(\frac{E_j(k)}{T(z)}) - (-1)^{2s_j} \right]^{-1} \label{eq:feq}
\end{equation}
is the equilibrium distribution function in the plasma frame for a particle with spin $s_j$ and energy satisfying $E_j^2 = k^2 + m_j^2$. The equilibrium part of the last term in \cref{eq:ptot-general} can be absorbed into the tree-level potential term to yield an effective potential term~\cite{Konstandin:2014zta, Cai:2020djd}, leaving%
\footnote{\Refcite{Konstandin:2014zta} demonstrate that \cref{eq:ptot-decomposed} can also be obtained from Kadanoff-Baym equations.}
\begin{equation}
	\ptot(t, v) = \int \! dz \dv{\phi_i}{z} \left[\pdv{V}{\phi_i} + \sum_j \dv{m_j^2}{z} \! \int \! \frac{d^3 k}{(2\pi)^3 2E_j} \delta f_j(k, z) \right] . \label{eq:ptot-decomposed}
\end{equation}
The first term in \cref{eq:ptot-decomposed} term can be massaged by applying the chain rule to $\Delta V = \int \! dz \frac{dV}{dz}$,  yielding~\cite{Ai:2021kak}
\begin{equation}
	\int \! dz \dv{\phi_i}{z} \pdv{V}{\phi_i} = \Delta V - \int \! dz \dv{T}{z} \pdv{V}{T} , \label{eq:dVdphi-chain}
\end{equation}
where the operator $\Delta$ represents the difference in the quantity on each side of the wall. Substituting \cref{eq:dVdphi-chain} into \cref{eq:ptot-decomposed} gives
\begin{equation}
	\ptot(t, v) = \Delta V - \int \! dz \dv{T}{z} \pdv{V}{T} + \int \! dz \dv{\phi_i}{z} \sum_j \dv{m_j^2}{z} \! \int \! \frac{d^3 k}{(2\pi)^3 2E_j} \delta f_j(k, z) . \label{eq:ptot-chain}
\end{equation}
The first term in \cref{eq:ptot-chain} is precisely the driving force, $\pdr$ (\cref{eq:pdr}).
Thus, we can identify the friction as~\cite{Ai:2021kak}
\begin{equation}
	\pfr(t, v) = \int \! dz \dv{T}{z} \pdv{V}{T} - \int \! dz \dv{\phi_i}{z} \sum_a \dv{m_j^2}{\phi_i} \! \int \! \frac{d^3 k}{(2\pi)^3 2E_j} \delta f_j(k, z) . \label{eq:pfr}
\end{equation}
Usually only the second term in \cref{eq:pfr} is identified as friction. The first term is sometimes interpreted as a modification of the driving force (see e.g.\ \refcite{Espinosa:2010hh, Megevand:2013hwa}), and sometimes is mistakenly ignored by assuming $\text{d}T/\text{d}z = 0$. As a reminder, one can determine the (instantaneous) terminal bubble wall velocity by solving $\pdr(t) = \pfr(t, v_w)$ for $v_w$, such that the total pressure vanishes. However, we still need to know the form of $\dv*{T}{z}$ and $\delta f_j(k, z)$.  

Determining $v_w$ can be made simpler by assuming local thermal equilibrium such that $\delta f_j = 0$, provided the bubble wall is not ultra-relativistic. Because local thermal equilibrium implies entropy conservation \cite{Ai:2021kak, Ai:2023see}, we find an additional equation,
\begin{equation}
	s_+ \gamma_+ v_+ = s_- \gamma_- v_- ,
\end{equation}
to solve for the bubble wall velocity and the plasma velocity and temperature profiles around the bubble wall. Assuming local thermal equilibrium, one gets \cite{Balaji:2020yrx}
\begin{equation}
	\pfr(t, v) = \abs{\Delta \! \left\{(\gamma^2 - 1) T s \right\}} , \label{eq:pfr-lte-nonconst}
\end{equation}
contrary to the long-held expectation of vanishing friction in local thermal equilibrium \cite{Turok:1992jp}. \Refcite{Balaji:2020yrx} connect the reheating from conservation of entropy to the hydrodynamic obstruction demonstrated in \refcite{Konstandin:2010dm} for the subsonic deflagration mode. However, they find the friction also appears for the supersonic detonation mode. Particles lose entropy as they gain mass by entering the bubble wall, hence the expansion of the bubble lowers entropy. This is compensated by reheating of the plasma around the bubble wall.

\Cref{eq:pfr-lte-nonconst} is an improvement to the results of an earlier study \cite{BarrosoMancha:2020fay}%
\footnote{In the ultra-relativistic limit, \refcite{BarrosoMancha:2020fay} instead use the ballistic approximation (discussed below), and recover a saturated friction as found in \refcite{Bodeker:2009qy}.}
which found a similar form proportional to $\Delta s$ by (inconsistently) assuming a constant temperature and plasma velocity across the sound shell, and led to a (misleading) $\gamma^2$ scaling of the friction. Later, \refcite{Ai:2021kak} showed that treating temperature and velocity across the sound shell leads to zero friction in local thermal equilibrium. This can be seen directly in \cref{eq:pfr} by making the substitutions $\dv*{T}{z} = 0$ and $\delta f_j = 0$.
However, the temperature and fluid velocity are known to vary across the sound shell in $3+1$ and even $1+1$ dimensions~\cite{Leitao:2010yw}.
\Refcite{Ai:2021kak} found that $\pfr \propto \gamma_+^2 v_+ (v_+ - v_w)$ for subsonic deflagrations in the bag model. Further, they found that the friction decreases with $v_w$ for detonations, where $\pfr \propto \gamma_w^2 v_w (v_w - v_-)$. Importantly, $v_w - v_-$ decreases rapidly with $v_w$ for detonations. This reopens the possibility of runaway solutions in local thermal equilibrium, which previously seemed improbable given the $\gamma^2$ scaling of friction found in \refcite{BarrosoMancha:2020fay}. The analysis of \refcite{Ai:2021kak} was recently updated in \refcite{Ai:2023see} using a more general equation of state with different --- yet constant --- speeds of sound in each phase. They provide a fit for the terminal bubble wall velocity in local thermal equilibrium when $c_{s,f} = c_{s,t} = 1/\sqrt{3}$:
\begin{equation}
	v_w = \left(\abs{\frac{3\alpha + \Psi - 1}{2(2 - 3\Psi + \Psi^3)}}^{\frac{p}{2}} + \abs{v_{\text{CJ}} \! \left(1 - a \frac{(1 - \Psi)^b}{\alpha} \right)}^{\frac{p}{2}} \right)^{\!\!\recip{p}} , \label{eq:vwPred}
\end{equation}
Here, $\alpha$ is the transition strength parameter defined in \cref{eq:alpha-traceAnomaly}; $\Psi = w_t / w_f$ is the ratio of enthalpies; $v_{\text{CJ}}$ is the Chapman-Jouguet velocity given by \cref{eq:vcj}; and $a = 0.2233$, $b = 1.704$ and $p = -3.433$ are numerical factors for the fit. The temperature-dependent quantities $\alpha$, $\Psi$ and $v_{\text{CJ}}$ should be evaluated at the transition temperature, $T_*$. The fit is an interpolation of two regimes of the terminal bubble wall velocity, specifically a $p$-norm. The first term of \cref{eq:vwPred} corresponds to the case where $v_w \lesssim 0.5$, while the second term corresponds to the case where $v_w \to 1$.
A recent study \cite{Wang:2022txy} decomposed the total friction into several contributions: the bubble wall, the sound shell, and the shock front. The total friction from all three contributions was investigated numerically in \refcite{Wang:2022txy} and later analytically in the appendix of \refcite{LiLi:2023dlc}. It was demonstrated that the friction from the bubble wall alone reproduces \cref{eq:pfr-lte-nonconst} with $\Delta$ representing a difference across the bubble wall. Extrapolating $\Delta$ to be a difference between quantities evaluated at $\xi = 1$ and $\xi = 0$ does not reproduce the total friction from the bubble wall, sound shell, and shock front contributions.

The predicted value of $v_w$ from \eqref{eq:vwPred} is a rough upper bound because out-of-equilibrium effects would act to reduce $v_w$. In fact, this is a general expectation: the $v_w$ predictions from local thermal equilibrium studies should be larger than when out-of-equilibrium effects are taken into account. However, as we discuss below, the assumption of local thermal equilibrium is not reasonable when $\gamma \gg 1$, so it should not be used to analyze the runaway behavior of bubble walls. \Refcite{Laurent:2022jrs} found that allowing the top quark to be out of equilibrium enhances the friction, but only affects the resulting prediction of $v_w$ by a few percent on average. Yet, the difference in friction could change a supersonic detonation into a hybrid mode in some cases. A similar analysis was performed in \refcite{DeCurtis:2023hil}, where they reported a larger reduction in $v_w$ when including the out-of-equilibrium effects of the top quark than was found in \refcite{Laurent:2022jrs}. While local thermal equilibrium may only hold approximately and only in particular scenarios, the resulting predictions can provide an estimate of $v_w$, or at least an upper bound away from the ultra-relativistic regime. If the friction arising from local thermal equilibrium predicts a subsonic deflagration, then a more rigorous analysis should also predict a subsonic deflagration. The validity of the assumption of local thermal equilibrium is further discussed in \refcite{Ai:2023see}.

Going beyond the assumption of local thermal equilibrium, one must determine the deviations from thermal equilibrium, $\delta f_j$, for each particle species. The task is simplified by assuming light particle species (i.e.\ those with a small change in mass across the bubble wall) remain in thermal equilibrium, and consequently are distributed according to \cref{eq:feq}. Then only $\delta f_j$ for heavy particles such as the top quark and W and Z bosons need to be calculated.%
\footnote{Although the Higgs boson has a similar mass, its few degrees of freedom lessen its effect on the bubble wall \cite{Laurent:2020gpg}.}
Nevertheless, this remains a difficult problem. The distribution functions for heavy particle species can be determined by solving Boltzmann equations \cite{Moore:1995si}:
\begin{equation}
	d_t f_j \equiv \partial_t f_j + \dot{z} \partial_z f_j + \dot{k}_z \partial_{k_z} f_j = -C[f_j] , \label{eq:Boltzmann}
\end{equation}
where the partial time derivative vanishes in the wall frame in the steady state. The collision term $C[f_j]$ encapsulates the scattering processes affecting the particle species denoted by the subscript $j$. Typically a subset of possible scattering processes is considered, with light particles forming a thermal bath in local equilibrium and other scattering processes having little effect on the friction. The collision integrals are often computed in the leading-log approximation, where only the logarithmic terms in the squared scattering amplitude are calculated, and usually only for the $t$-channel processes as described in \refcite{Moore:1995si}. \Refcite{Hoche:2020ysm} resum large logarithms to all orders and consider the processes from all channels. Similarly, \refcite{Wang:2020zlf} improve the calculation of the collision integrals by considering the $s$-, $t$- and $u$-channel processes, but they also go beyond the leading-log approximation. Both \refcite{Hoche:2020ysm} and \refcite{Wang:2020zlf} found a reduction in the predicted terminal bubble wall velocity when improving the treatment of the collision integrals. Hence, the typical treatment in the leading-log approximation underestimates the friction.

Rather than solving the Boltzmann equation directly for each species of heavy particle, it is common to use an ansatz for the distribution functions.%
\footnote{Or, one can perform an $N$-body simulation of particles interacting with the bubble wall as done recently in \refcite{Lewicki:2022nba}.}
The fluid approximation \cite{Moore:1995si, Konstandin:2014zta} considers perturbations in the chemical potential, plasma velocity, and temperature, about the equilibrium values. The Boltzmann equation is truncated to first order in these perturbations and three moments of the Boltzmann equation are taken to solve for the perturbations. This procedure converts the integro-differential equation into a system of ordinary differential equations. Similarly, using a $\tanh$ profile ansatz for each scalar field further simplifies the problem \cite{Moore:1995si}. Each scalar field is given a separate bubble wall width and position \cite{Laurent:2022jrs}. Two moments of the equations of motion for each scalar field are taken, one for setting the total pressure to zero, and the other for making it so the wall does not stretch or compress. \Refcite{Friedlander:2020tnq} devised an iterative algorithm to modify the field profiles from an initial $\tanh$ ansatz. \Refcite{Laurent:2020gpg} adjusted the fluid ansatz to avoid a singularity in the friction when $v_w = c_{s,-}$, which arise in the original fluid ansatz from \refcite{Moore:1995si}. However, \refcite{Dorsch:2021nje} argued that the singularity in friction for $v_w = c_{s,-}$, the so-called `sonic boom', has a physical interpretation stemming from hydrodynamics. \Refcite{Laurent:2022jrs} again found friction to be a smooth function of the bubble wall velocity by going beyond the fluid approximation, instead computing the background perturbations of the plasma temperature and velocity nonlinearly and solving the Boltzmann equations using a spectral expansion. At around the same time, \refcite{DeCurtis:2022hlx} --- using no ansatz or imposed momentum dependence of the perturbations --- similarly found no peaks in the friction that otherwise appear when using the fluid ansatz. While not all computations of the friction explicitly assume an ansatz for the distribution functions, \refcite{Dorsch:2021nje} argue that choices made when computing the collision integrals are effectively equivalent to choosing an ansatz. In this case, the momentum dependence of the perturbations may not be consistent across the terms of the Boltzmann equation \cite{Dorsch:2021nje}.

Several studies have been performed in the ultra-relativistic limit to understand whether bubble walls can run away. That is, whether bubble walls accelerate continuously up to the time of collision, rather than reaching a terminal velocity. In fact, there is no steady state if bubble walls run away. In the ultra-relativistic limit, particle interactions within the bubble wall are negligible and the distribution functions remain unchanged when crossing the bubble wall \cite{Bodeker:2009qy}.%
\footnote{\Refcite{BarrosoMancha:2020fay} refer to the limit in which the bubble wall passes by faster than the thermalization timescale as the ballistic limit. The ballistic approximation similarly neglects the collision term in \cref{eq:Boltzmann}. Local thermal equilibrium is not a reasonable assumption in this limit.}
Thus, we can consider the friction coming from particles changing mass in equilibrium. The leading-order (LO) friction is given by \cite{Bodeker:2009qy, Gouttenoire:2021kjv}%
\footnote{\Refcite{GarciaGarcia:2022yqb} find additional $\gamma$-dependent contributions to the friction at leading order from gauge bosons (e.g.~dark photons) that change mass across the bubble wall and are massive on both sides of the bubble wall.}
\begin{equation}
	\pfrlo(t) = \sum_j g_j \! \int \! \frac{d^3 k}{(2\pi)^3} f_j(k) \times \Delta k_j^{\text{LO}}(t) ,
\end{equation}
where $g_j$ is the number of degrees of freedom and $\Delta k_j^{\text{LO}} \simeq \Delta m_j^2 / 2E_j$ is the change in momentum in the ultra-relativistic limit $E_j \gg \Delta m_j$. The friction can be calculated analytically in this limit, and is approximately given by
\begin{equation}
	\pfrlo(t) = \sum_j c_j g_j \frac{\Delta m_j^2 T^2}{24} ,
\end{equation}
where $c_j = 1$ for bosons and $c_j = 1/2$ for fermions. This leading-order contribution to the friction is independent of the bubble wall velocity because, in the wall frame, the thermal flux scales as $\gamma^1$ due to Lorentz contraction while the momentum transfer scales as $\gamma^{-1}$ \cite{Turok:1992jp, Bodeker:2009qy}. Then one can compare $\pdr(t_*)$ to $\pfrlo(t_*)$ to determine whether the bubble wall reaches a terminal velocity or continues to accelerate until it collides with another bubble. The latter case corresponds to the runaway regime and has important consequences on the energy budget of the phase transition as discussed in \cref{sec:energyBudget}. \Refcite{Megevand:2013hwa} argued that the condition $\pdr > \pfr$ for the runaway regime is necessary but not sufficient, because there can be cases where there is both a runaway and steady state solution. The steady state solution is stable if it is a detonation. Therefore, if a detonation solution is possible, then it would be favored over the runaway solution.

Next-to-leading-order corrections were considered in \refcite{Bodeker:2017cim}, where they found that the emission of a soft gauge boson enhances the friction to scale as $\gamma^1$. This phenomenon is a form of transition radiation that arises due to the change in the virtual cloud of particles when crossing the bubble wall. The analysis of \refcite{Bodeker:2017cim} was extended in \refcite{Hoche:2020ysm}, where they considered multiple emissions of soft gauge bosons (rather than just one emission) and resummed to all orders in the leading-log approximation. Controversially, they found the friction to scale as $\gamma^2$ and to be independent of the change in mass across the wall. This was disputed and criticized in a later study \cite{Gouttenoire:2021kjv} which recovered the $\gamma^1$ scaling found in \refcite{Bodeker:2017cim}, arguing that the computation in \refcite{Hoche:2020ysm} satisfies the Ward identity erroneously, because the presence of the bubble wall breaks four-momentum conservation. Additionally, \refcite{Azatov:2020ufh} noted that the $\gamma^2$ friction found in \refcite{Hoche:2020ysm} does not vanish in the limit where the heavy particles remain massless. \Refcite{Azatov:2020ufh} also found that particles much heavier than the scale of the phase transition can provide important contributions to the friction. These analyses \cite{Bodeker:2009qy, Bodeker:2017cim, Hoche:2020ysm, Azatov:2020ufh, Gouttenoire:2021kjv} assume the ultra-relativistic limit $\gamma \gg 1$, so an accurate determination of $v_w$ is not possible when $\gamma \sim 1$. However, they provide important insight into the conditions required for runaway bubble walls.

While some studies treat the terminal bubble wall velocity as a free parameter, a similar approach treats the friction as a free parameter. The phenomenological friction is parameterized by a dimensionless parameter $\tilde{\eta}$. By varying $\tilde{\eta}$, one is able to scan through a range of values for $v_w$. The benefit of this approach is that $\tilde{\eta}$ can be matched onto results from studies of the microphysics (i.e.\ calculations of the collision integrals) in a given particle physics model. The damping is expected to be proportional to $\text{d}\phi/\text{d}t$, leading to the Lorentz-invariant equivalent \cite{Heckler:1994uu, Megevand:2003tg}%
\footnote{While $\mathcal{K}$ depends on more than just the scalar field, we leave the parametric dependence as found in \refcite{Espinosa:2010hh}.}
\begin{equation}
	\mathcal{K}(\phi_i) = \eta u^\mu \partial_\mu \phi_i , \label{eq:friction-pheno}
\end{equation}
matching the form obtained earlier in \refcite{Ignatius:1993qn}, where $u^\mu$ is fluid four-velocity. The dimensionless parameter $\tilde{\eta}$ is defined through $\eta = \tilde{\eta} T$. The equation of motion \cref{eq:scalarEOM} can be written as \cite{Espinosa:2010hh}
\begin{equation}
	\partial_\rho \partial^\rho \phi_i + \pdv{V_0}{\phi_i} - \mathcal{K}(\phi_i) = 0 , \label{eq:scalarEOM-pheno}
\end{equation}
and we see that $\mathcal{K}$ models only the out-of-equilibrium contribution to the friction in \cref{eq:pfr}.

According to \cref{eq:friction-pheno}, the friction scales with $v_w \gamma(v_w)$.  After \refcite{Bodeker:2009qy} found that the friction saturates in the ultra-relativistic limit, some studies (e.g.\ \refcite{Espinosa:2010hh, Huber:2013kj}) adjusted $\mathcal{K}$ to scale with $v_w$ instead of $v_w \gamma(v_w)$. \Refcite{Megevand:2013hwa} noted such a modification does not allow one to consistently determine $\tilde{\eta}$ for both the non-relativistic and ultra-relativistic limits simultaneously. They proposed a phenomenological model for friction depending on two dimensionless parameters, both of which can be determined from microphysics calculations. However, these phenomenological friction models do not match the current expected scaling in the ultra-relativistic regime mentioned previously --- which, depending on the study, is either $\gamma^1$ or $\gamma^2$. \Refcite{Cai:2020djd} rectify this issue by
generalizing the $\gamma$ scaling as $\mathcal{K} \propto h(\gamma)$.
The terminal Lorentz factor can then be solved for in terms of $h(\gamma)$. Results for both $h(\gamma) = \gamma$ and $h(\gamma) = \gamma^2$ were demonstrated.

We now remark on the typical values of $v_w$ that have been found in recent studies. \Refcite{Laurent:2022jrs} found that the most likely terminal bubble wall velocities in the real scalar singlet model are either near the speed of sound or are ultra-relativistic with $\gamma \gtrsim 10$. This result is expected to hold for other models because the friction is strongly peaked near $v_{\text{CJ}}$ (see their Fig.~2, and Fig.~3 of \refcite{DeCurtis:2023hil}) for the following reasons. First, if the driving pressure is sufficient to overcome this peak and lead to a supersonic detonation, then the lower friction when $v_w > v_{\text{CJ}}$ is insufficient to slow the wall before reaching the ultra-relativistic regime. Second, the friction below the sound speed is much lower and grows roughly linearly with $v_w$, so very low values $v_w \lesssim 0.1$ are unlikely unless the driving pressure is weak. \Refcite{Ai:2023see} found solutions for $v_w$ arbitrarily close to zero in local thermal equilibrium. However, they note that these low wall velocities were obtained when the pressure difference was negligible, in which case bubble nucleation is highly inefficient. Thus, they argued that low wall velocities may not be realistic. Yet the inclusion of out-of-equilibrium effects can yield low wall velocities for larger driving pressures. For instance, \refcite{Wang:2020zlf} demonstrated a reduction in the prediction for $v_w$ in the Higgs sextic model by treating the scattering processes more rigorously. The benchmarks of \refcite{DeCurtis:2023hil} demonstrate that while $v_w$ may be around the sound speed in local thermal equilibrium, it is lowered (e.g.\ below $v_w \sim 0.4$) when out-of-equilibrium effects are taken into account. \Refcite{Friedlander:2020tnq} performed a scan over the parameter space of the real scalar singlet model, treating the out-of-equilibrium effects using the fluid approximation, and found bubble wall velocities as low as $v_w = 0.1$. Similarly, \refcite{Kozaczuk:2015owa} studied the real scalar singlet model using the fluid approximation, specifically targeting subsonic wall velocities in mildly FOPTs (with an order parameter $\phi_h(T_n)/T_n$ less than or near unity). They found a wide range of values for $v_w$ including below $0.1$, with the general trend that $v_w$ increases with the order parameter and decreases with the singlet mass. \Refcite{Jiang:2022btc} considered three benchmarks in a narrow region of the parameter space in the inert doublet model, and found $v_w \simeq 0.165$ for all benchmarks.

Using a phenomenological friction model, \refcite{Krajewski:2023clt} found that there is a velocity gap when increasing the friction. That is, as the friction is increased, the terminal bubble wall velocity changes discontinuously going from a supersonic detonation to a hybrid --- in some cases straight from a supersonic detonation to a subsonic deflagration. A range of terminal bubble wall velocities is ruled out, typically around the hybrid regime: $v_w \sim v_{\text{CJ}}$. This qualitatively matches the behavior found in \refcite{Laurent:2022jrs}, although it is a result from hydrodynamics alone because the friction is controlled through the parameter $\eta$ and does not peak near $v_w = v_{\text{CJ}}$. Additionally, semi-relativistic detonations were not disfavored in the results of \refcite{Krajewski:2023clt}.

An accurate determination of $v_w$ involves consistently solving the scalar field equations of motion and the Boltzmann equations. This is difficult without the use of an ansatz for the distribution functions or the bubble wall profile. There has been recent progress \cite{Laurent:2022jrs, DeCurtis:2023hil} in more efficient methods for evaluating the collision integrals. However, a fast and accessible method for determining $v_w$ accurately is not yet readily available to studies of GWs from FOPTs. To this end, \refcite{Laurent:2022jrs} detail a list of estimation methods that can be used with varying levels of sophistication. \Refcite{Ai:2023see} provides a Python code snippet to determine $v_w$ in local thermal equilibrium for models where the speed of sound in each phase can be different but are approximately temperature independent. At the very least, these recent results can help to restrict the range of expected values for $v_w$ in studies of GWs, preventing the need to scan over the whole range $v_w \in (0, 1)$.

\section{Gravitational waves from first-order phase transitions} \label{Section:GWs-Sources}

\subsection{Gravitational waves} \label{sec:GWobs}

Phase transitions in our cosmological history could result in gravitational waves (GWs) that are observable in ground- or space-based detectors. Let us briefly review GWs, including how they arise in general relativity and how they are detected. The central equations in general relativity are Einstein's equations~\cite{1916AnP...354..769E},\footnote{We adopt natural units where $\hbar = c  = 1$ throughout.}
\begin{equation}
  R_{\mu\nu} - \frac{1}{2} R g_{\mu\nu} = 8 \pi G T_{\mu\nu} ,
\label{eq:Einstein}
\end{equation} 
where $G = 6.7088 \! \times \! 10^{-39} \gev^{-2}$ is Newton's gravitational constant \cite{Wu:2019pbm}, $g_{\mu\nu}$ is the four-dimensional spacetime metric, the Ricci scalar curvature $R$ is the trace of the Ricci curvature tensor
\begin{equation}
  R_{\mu\nu} = \partial_\alpha \Gamma^\alpha_{\mu \nu} - \partial_\nu \Gamma^\alpha_{\mu \alpha} + 
               \Gamma^\alpha_{\sigma \alpha} \Gamma^\sigma_{\nu \mu} - \Gamma^\alpha_{\sigma \nu} \Gamma^\sigma_{\alpha \mu} ,
\label{eq:Ricci}
\end{equation} 
and the Christoffel symbols are defined as
\begin{equation}
  \Gamma^\alpha_{\mu \nu} = \frac{1}{2} g^{\alpha \lambda} 
                            \left( \partial_\mu g_{\nu \lambda} + \partial_\nu g_{\mu \lambda} - \partial_\lambda g_{\mu \nu} \right) .
\label{eq:Christoffel}
\end{equation}
In \cref{eq:Einstein}, $T_{\mu\nu}$ represents the energy-momentum tensor. For GWs sourced by a first-order phase transition (FOPT), there are contributions to the energy-momentum from scalar fields and from the thermal plasma, which is described by a perfect fluid; see \cref{eq:combined_energy_momentum_tensor}.

As \cref{eq:Ricci,eq:Christoffel} show, the left-hand side of \cref{eq:Einstein} is fully determined by the metric.  In this work we frequently use the standard Friedmann-Lemaitre-Robertson-Walker (FLRW) metric with zero Gaussian curvature:
\begin{equation}
	ds^2 = g_{\mu\nu} dx^\mu dx^\nu = - dt^2 + a^2(t) \delta_{ij} dx^i dx^j ,
\label{eq:FLRW-metric}
\end{equation}
where we are using the $(-,+,+,+)$ signature and the form follows from assumptions of homogeneity and isotropy. The scale factor $a(t)$ fully determines the metric and must be determined from Einstein's equations. With the FLRW metric (\cref{eq:FLRW-metric}) and the assumption of a perfect fluid filling the space, Einstein's equations yield Friedmann's two equations
\begin{align}
  {H}^2 &= \frac{8\pi G}{3}\rhotot ,
\label{eq:Friedmann1}\\
  \frac{\ddot{{a}}}{{a}} &= -\frac{4\pi G}{3}(\rhotot + 3 {p}),
\label{eq:Friedmann2}
\end{align}
where, as discussed previously, the Hubble parameter \cref{eq:hubble} gives us the rate of expansion of space and overdots denote a time derivative. 

Very soon after writing down his field equations, Einstein found solutions of the linearized version of these equations \cite{1916SPAW.......688E}.%
\footnote{See \refcite{Einstein:1916cc} for an English translation of \refcite{1916SPAW.......688E}.}
Assuming small perturbations $\epsilon h_{\mu\nu}$ with $\epsilon \ll 1$, over the background metric, $\eta_{\mu\nu}$, 
\begin{equation}
  g_{\mu\nu} = \eta_{\mu\nu} + \epsilon h_{\mu\nu} ,
\label{eq:MetricPerturbation}
\end{equation}  
he derived a wave equation, which in the Einstein gauge ($\partial_\alpha \partial^\alpha x^\mu = 0$) reads
\begin{equation}
  \partial_{\alpha} \partial^{\alpha} \left( h_{\mu \nu} - \frac{1}{2} \eta_{\mu \nu} h_{\beta \gamma} \eta^{\beta \gamma} \right) = - 16 \pi G T_{\mu\nu}.
\label{eq:LinearWave}
\end{equation} 
In the absence of sources ($T_{\mu\nu} = 0$) these equations imply wave equations for the metric perturbations
\begin{equation}
  \partial_{\alpha} \partial^{\alpha} h_{\mu \nu} = 0 .
\label{eq:LinearWaveNoSource}
\end{equation} 
This is nothing but a relativistic wave equation.
Because these are independent equations for the ten independent components of $h_{\mu \nu}$, they predict linear GWs.  If there were sources, far enough away from them these linearized waves are a good approximation of the waves that would emerge in the original non-linearized theory.

\begin{figure}[t]
  \centering
  \includegraphics[width=0.8\linewidth]{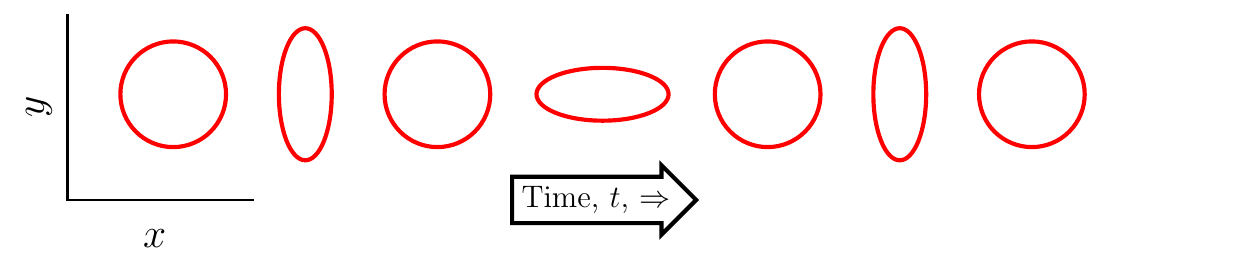}%
  \caption{Impact of a GW passing through the center of a ring of particles (along the $z$-axis) for $C_+ \neq 0$ and $C_\times = 0$. The coefficient for $C_\times\neq0$ stretches and compresses along diagonals in a similar manner. Based on \refcite{Carroll:1997ar}.}
  \label{fig:gw_impact}
\end{figure}

The solutions to \cref{eq:LinearWaveNoSource} are plane-waves, $h_{\mu \nu} = C_{\mu \nu}  e^{i k_\alpha x^\alpha}$ that satisfy $k^2 = 0$. That is, they are waves traveling at the speed of light. For a wave propagating in the $z$-direction, after exploiting gauge freedom, there are three free parameters in the plane wave solution: the frequency, and two polarizations: $C_+ \equiv C_{11} = - C_{22}$ and $C_{\times} \equiv C_{12} = C_{21}$. The other coefficients $C$ vanish. The impact of the $C_+$ coefficient is shown by how it would disturb a ring of particles on the $x$-$y$ plane in \cref{fig:gw_impact}. The fact that $C_{11} = - C_{22}$ means that the ring stretches on one axis whilst it compresses on the other. See e.g., \refcite{Carroll:1997ar,Spurio:2019xej,Bishop:2021rye} for further details. Importantly, note that spherically symmetric configurations cannot source GWs. This may be understood from the shell theorem: externally, a spherically symmetric configuration behaves as though its mass was concentrated at the center. For example, suppose a star explodes in a supernova or collapses to a black hole in a spherically symmetric manner. Externally, the gravitational field does not change and so there are no GWs.

After Einstein discovered the existence of linear GWs, however, the question remained whether GW solutions exist in the full theory of general relativity.  Twenty years after proposing GWs in the linearized theory, Einstein and Rosen found a gravitational plane-wave solution in the full theory \cite{EINSTEIN193743}.  Interestingly, they also found that the solution had singularities, and they deemed it unphysical.  In a subsequent paper Rosen argued that plane wave solutions do not exist, but only cylindrical GWs exist in the full theory \cite{rosen1937plane}.  It took a lot of work to show that plane waves can be defined in general relativity, that they can carry energy, and that such solutions do exist in the general theory.  That work began two decades later \cite{Pirani:1956wr, Bondi:1957dt, Bondi:1958aj, Penrose:1960eq, 1969eisp.book.....P}, rediscovered some mathematics from the 1920's \cite{Brinkmann1925}, and culminated in the papers of Trautman and Robinson \cite{Trautman:1958rph, Trautman:1958zdi, Robinson:1960zzb, Robinson:1962zz}. See \refcite{Cervantes-Cota:2016zjc,DiMauro:2021sah} for historical reviews.

Fast forwarding to 2015, in an amazing feat of experimental science, the Laser Interferometer Gravitational-Wave Observatory  (LIGO) detected the very GWs that were foreseen by Einstein about a century before and were predicted over five decades earlier~\cite{LIGOScientific:2016aoc}.  GW detectors are Michelson-interferometers with two orthogonal arms that are kilometers long. The interferometers are capable of registering minute changes in the arm lengths.  We can imagine the detector located within the circle in \cref{fig:gw_impact} with its arms in the $x$ and $y$ directions.  As the GW moves through the detector it morphs spacetime, dragging the arms with it and slightly changing their lengths.  Even for the strongest GWs originating from the nearest black hole coalescences, the arms, which are about a kilometer long, change length by less than the size of a proton.  The detection of such microscopic displacement of the mirrors, which are macroscopic objects, requires quantum optical measurements.  These measurements are made possible by the statistical fact that a measurement using many quanta provides a mean with a very small variance.  The next challenge is to isolate the mirrors from their surroundings to lower the background motion below the signal.  Extremely stable, and high energy, lasers are also a must for these experiments. To discriminate signals from remaining noise, it is essential to operate two or more independent detectors. See e.g., \refcite{Abbott:2016xvh} for a detailed discussion of the experimental setup and noise reduction techniques.

Following LIGO's discovery in 2015, Advanced LIGO and Advanced Virgo detected 90 compact-object merger events during their observing runs O(1) -- O(3)~\cite{LIGOScientific:2018mvr,LIGOScientific:2020ibl,LIGOScientific:2021usb,LIGOScientific:2021djp}, and in 2022 the KAGRA and GEO600 underground experiments reported the first joint observation~\cite{KAGRA:2022twx}. Perhaps not surprisingly, GW detection in the 2020s is becoming an industry.  We can divide the GW experiments into four major categories: present and planned ground-based detectors, planned space-based instruments, and pulsar timing observations.  
The ground-based detectors currently under operation are Advanced LIGO (aLIGO;  \cite{LIGOScientific:2014pky}), Advanced Virgo (AdV; \cite{VIRGO:2014yos}), GEO600 (GEO; \cite{Luck:2010rt, Affeldt:2014rza, Dooley:2015fpa} and KAGRA~\cite{KAGRA:2020tym, Somiya:2011np, Aso:2013eba}, which superseded TAMA \cite{Fujimoto:2000qbu, Ando:2002bv}, with armlengths of a few km and sensitivity to signals at frequencies of around Hz -- kHz.
Two future ground-based detectors are planned: the Cosmic Explorer (CE; \cite{Reitze:2019iox}) in the US with armlengths of 10\,km and the Einstein Telescope (ET; \cite{Punturo:2010zz, Sathyaprakash:2012jk, Maggiore:2019uih}) in Europe with armlengths of up to 40\,km.
Space-based detectors with huge baseline lengths of around $10^6\,\text{km}$ could come online around 2035 (see e.g., \refcite{Crowder:2005nr,Baibhav:2019rsa,Bailes:2021tot}) and plans include the Deci-hertz Interferometer GW Observatory (DECIGO; \cite{Kawamura:2011zz,Sato:2017dkf,Kawamura:2020pcg}), the Laser Interferometer Space Antenna (LISA;~\cite{Caprini:2019egz,Barausse:2020rsu}), TianQin~\cite{TianQin:2015yph,TianQin:2020hid} and Taiji~\cite{Hu:2017mde}. Farther in the future, LISA could be followed by the Advanced Laser Interferometer Antenna (ALIA;~\cite{Gong:2014mca}) or the Big Bang Observer (BBO; \cite{Harry_2006}). Roughly speaking, the peak sensitivity lies at frequencies around $c / L \sim \text{mHz}$; the inverse of the time it takes light to travel along the baseline.

\begin{figure}[t]
  \centering
  \includegraphics[width=0.8\linewidth]{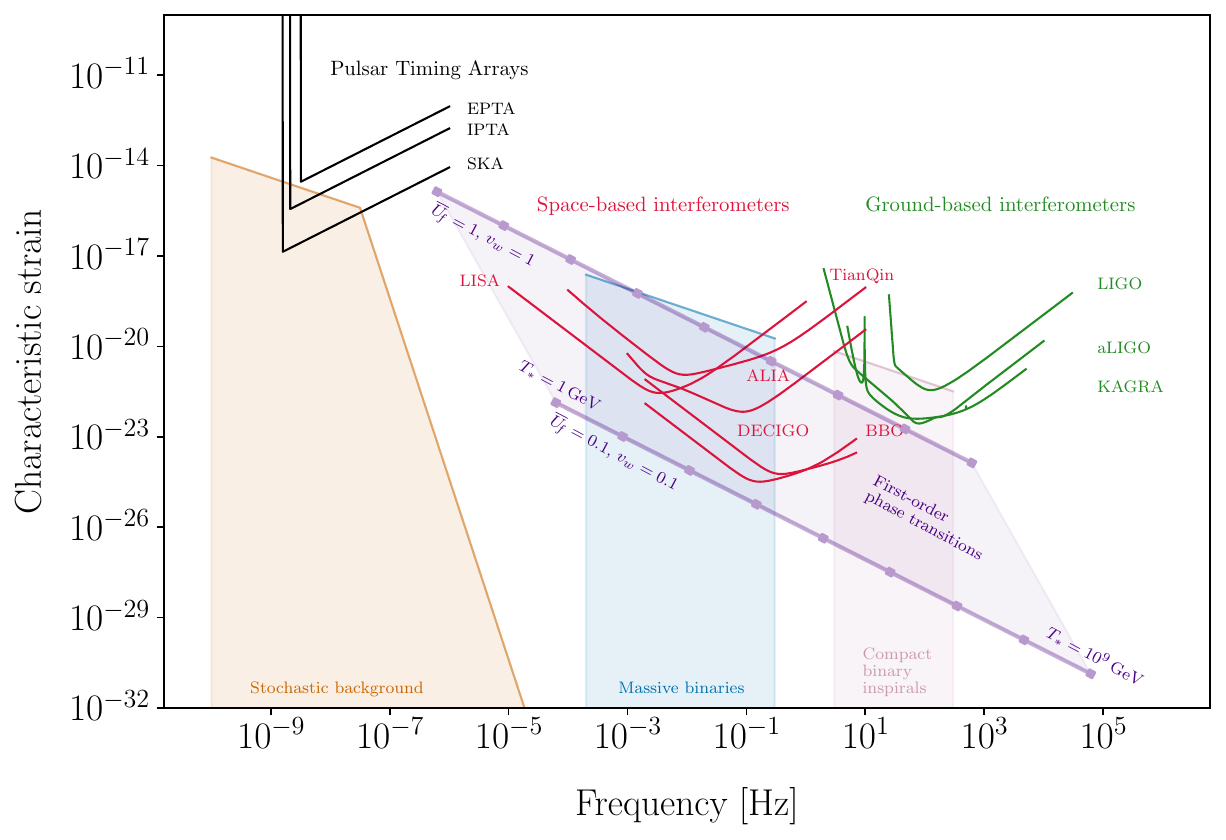}
  \caption{Sensitivity of current and future GW detectors. 
  The purple lines show the peak frequency and amplitude of the sound spectrum of GWs from a FOPT when the temperature was varied between $T_* = 1\gev$ and $10^9\gev$, for $\Uf = 0.1$ and $v_w=0.1$ (bottom line) and $\Uf = 1$ and $v_w = 1$ (top line). We fixed the other thermal parameters to $z_p=6.9$, $g_*=106.75$, $H / \beta=0.1$ and $\Gamma=4/3$.
   This plot was made using data from \href{https://github.com/robsci/GWplotter}{\code{GWplotter}}~\cite{Moore:2014lga}.
  }
  \label{fig:freq_strain}
\end{figure}

Pulsar timing array experiments (see e.g., \refcite{Verbiest:2021kmt}), such as 
the Chinese Pulsar Timing Array (CPTA; \cite{2016ASPC..502...19L}),
the European Pulsar Timing Array (EPTA; \cite{EPTA:2011kjn, Desvignes_2016}), 
the International Pulsar Timing Array (IPTA; \cite{Hobbs:2009yy, Manchester_2013}),
the North American Nanohertz Observatory (NANOGrav; \cite{Demorest:2009ex}),
the Parkes Pulsar Timing Array (PPTA; \cite{Manchester:2012za}), and 
the Square Kilometer Array (SKA; \cite{Janssen:2014dka}), are also sensitive to GWs, as GWs introduce correlated disturbances in the pulse arrival times. In particular, stochastic gravitational wave backgrounds (SGWB) lead to distinctive Hellings-Downs correlations~\cite{Hellings:1983fr}, enabling the discrimination of signal versus noise.
The experiments are sensitive to signals with a period similar to the length of the observation period, which corresponds to frequencies of about $1 / (10\,\text{years}) \sim \text{nHz}$. A graphical overview of the reach of some of these detectors is given in \cref{fig:freq_strain}.

Recently, CPTA \cite{Xu:2023wog}, EPTA \cite{EPTA:2023fyk}, NANOGrav \cite{NANOGrav:2023gor} and PPTA \cite{Reardon:2023gzh} reported the detection of a nano-Hertz GW signal. Within the uncertainties, this signal is consistent with a SGWB and possibly with GWs originating from a multitude of coalescing supermassive black hole binaries~\cite{NANOGrav:2023hfp, NANOGrav:2023tcn}. Measurements of the nano-Hertz signal are still in their infancy, and it is important to verify its source. Beyond cosmological phase transitions, black hole binaries and other astrophysical explanations, a nano-Hertz signal could be produced by cosmological inflation, scalar fluctuations, cosmic (super-)strings or domain walls \cite{NANOGrav:2023hvm}.

GW signals could be transient bursts, which are localized in time and originating from a specific event such as a binary merger; continuous and periodic, such as from a rotating neutron star; or stochastic, originating from a collection of independent sources and consequently isotropic, stationary and unpolarized~\cite{Maggiore:1999vm}.\footnote{The collection of e.g., binary mergers over the history of the Universe could lead to a stochastic background. See e.g., \refcite{Christensen:2018iqi} for a review of sources of stochastic backgrounds.} A stochastic background is thus characterized solely by the amplitude as a function of the frequency, $f$, which we refer to as a GW spectrum.  The amplitude may be quantified in several ways, the most widely used of which are the energy density (or energy density per logarithmic frequency) $\Omega_\text{GW}$, the spectral energy density $S_E$, the one-sided spectral density $S_h$, the characteristic strain $h_c$, and the frequency-domain strain $\tilde{h}$.  
The ratio $\rho_{\text{GW}}/\rhotot$ quantifies the energy density carried by the GWs normalized to the total energy density. Based on observational evidence for a flat Universe, the total energy density can be directly related to the critical density: $\rhotot = \rho_c = 3 H_0^2/(8 \pi G)$, where $H_0$ is the present value of the Hubble parameter and $G$ is Newton's gravitational constant.  The GW spectra can then be defined as the energy density per logarithmic frequency\footnote{
Note an alternative convention in the literature (see e.g.~\refcite{Caprini:2007xq}) in which $\Omega_\text{GW} \equiv {\rho_\text{GW}/}{\rhotot}$ and the  energy density per logarithmic frequency is written ${d \Omega_{\text{GW}}}/{d\ln f}$.}
\begin{equation}
  \label{Eq:GWsEngDenperLogf}
  \Omega_\text{GW}(f) \equiv \frac{1}{\rhotot}\frac{d \rho_{\text{GW}}}{d\ln f} .
\end{equation}  
The characteristic strain, the power spectral density, and the spectral energy density are various measures of the fractional change of the length of the macroscopic object used as a detector. The various amplitude measures are related by~\cite{Moore:2014lga}
\begin{equation}
  H_0^2 \Omega_\text{GW}(f) = \frac{8 \pi G}{3 c^2} f S_E(f) = \frac{2 \pi^2}{3} f^3 S_h(f) = \frac{2 \pi^2}{3} f^2 h_c^2(f) = \frac{8 \pi^2}{3} f^4 |\tilde{h}(f)|^2 ,
\end{equation}
where $c$ is the speed of light in vacuum.

\subsection{From first-order phase transitions to gravitational waves}\label{sec:from_fopt_to_gws}

We now consider the ways in which GWs can be produced by FOPTs. There are other possible sources of GWs, including known sources such as binary mergers and theoretical sources such as extensions to the inflationary model. See e.g.\ \refcite{Binetruy:2012ze, Caprini:2018mtu, Kuroyanagi:2018csn} for descriptions of many cosmological and astrophysical sources. These other sources act as background when searching for a GW signal from a FOPT. On top of these backgrounds, there exists a foreground (e.g.\ from white dwarf binaries) from which any signal must be distinguished \cite{Ruiter:2007xx, Yu:2010fq, Adams:2013qma, Colpi:2016fup}.

FOPTs proceed through the nucleation of bubbles and release latent heat. As discussed in \cref{sec:bounce_to_bubble}, the energy is carried by the bubble wall, which expands into the plasma of hot particles. Through the breaking of spherical symmetry in the bubble walls, gravitational radiation is released. Clearly, the collision of bubbles and their surrounding sound shells is a violent process with tremendous energy involved. The phenomenology depends on the expansion mode of bubbles discussed in \cref{sec:bubbleExpansionModes}. The bubbles lead to GWs from three sources. First, the collision of bubbles breaks their spherical symmetry. Second, waves of plasma accelerated by the bubble wall propagate along with the bubble wall. This can lead to GWs prior to the bubble wall collisions for both subsonic and supersonic deflagrations. Finally, shocks in the fluid may lead to turbulence, which also sources GWs.

GW spectrum calculations became increasingly sophisticated during the last decades since the initial predictions of GWs~\cite{Witten:1984rs, Hogan:1986qda}. For bubble collisions, numerical studies of the frequency distribution of GWs were pioneered in the early 1990s \cite{Kosowsky:1991ua, Kosowsky:1992rz, Kosowsky:1992vn, Kamionkowski:1993fg}. In general, because we are treating the surrounding plasma of particles as a fluid, the bubble wall dynamics can be understood through relativistic hydrodynamic simulations of the coupled scalar field and fluid systems. The energy-momentum tensor of the rapidly varying scalar field and that of the plasma morph space through linear perturbations in the metric. The energy flow from a scalar field to the plasma and spacetime during a FOPT can be deduced from the coupled equations of motion of the scalar and gravitational fields and that of the plasma.

\begin{figure}[t]
  \centering
  \includegraphics[width=0.55\linewidth]{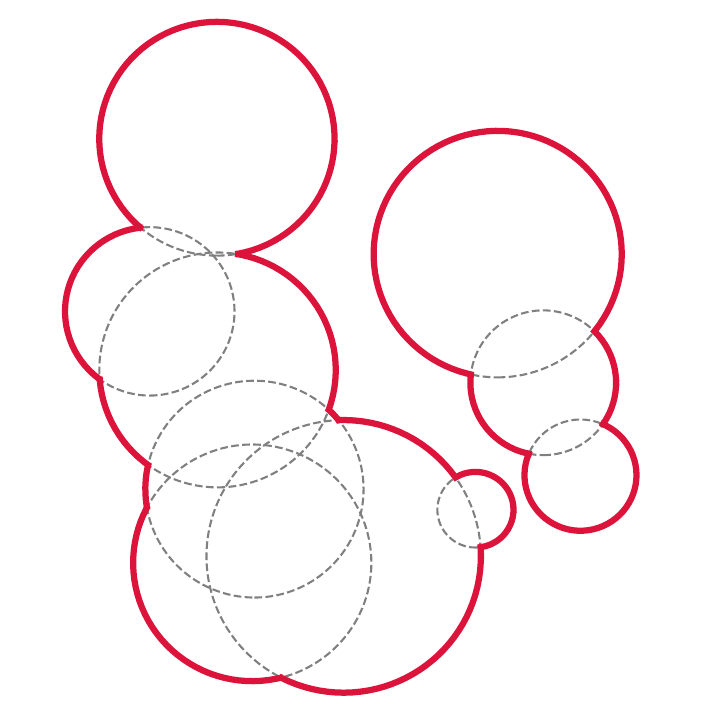}
  \caption{The envelope approximation for several spherical bubbles. The energy-momentum is all contained in envelope of bubble walls, and the energy-momentum in any intersecting region immediately vanishes.
  }
  \label{fig:envelope}
\end{figure}

Various approximations are introduced because it is challenging to solve the coupled equations even numerically. The energy-momentum tensor induced by the bubbles, for example, can be described in a simplified fashion by the so-called envelope approximation~\cite{Kosowsky:1992vn}.  This approximation assumes a highly simplified space and time dependence of the energy-momentum tensor.  One of the assumptions is that bubble walls are infinitesimally thin, consequently right after bubble nucleation the energy-momentum tensor vanishes everywhere in space except on infinitesimally thin, spherical shells.  The second assumption is that the energy-momentum tensor immediately vanishes in sections of the bubble wall contained inside other bubbles (after overlap), leaving only the envelope surrounding regions of the true vacuum. This is shown in \cref{fig:envelope}. Over the last decade, however, researchers began gradually relaxing these approximations \cite{Huber:2008hg,Child:2012qg,Hindmarsh:2013xza, Hindmarsh:2015qta, Weir:2016tov, Hindmarsh:2017gnf, Konstandin:2017sat, Cutting:2019zws, Gould:2021dpm}.
For example, \refcite{Weir:2016tov} explored the envelope approximation and \refcite{Gould:2021dpm} comprehensively studied the bubble wall thickness and velocity.  Finally, analytical derivations of the GW spectrum from bubble collisions were recently completed \cite{Jinno:2016vai, Zhong:2021hgo, Megevand:2021juo}. 

The approximate form of the GW spectrum can be found on simple dimensional grounds, following arguments similar to those in~\refcite{Kosowsky:1991ua}. We assume that the energy in the GWs must be proportional to Newton's constant, $G$. We assume, furthermore, that the energy depends on the available vacuum energy, $\kappa \rho_V$, where $\kappa$ denotes the fraction of vacuum energy available for GWs, the bubble wall velocity, $v_w$, and that the only other relevant dimensionful scale is the characteristic timescale $\beta^{-1}$ (not necessarily that derived from \cref{eq:beta}). On dimensional grounds, we must have that
\begin{equation}\label{eq:e_gw}
E_\text{GW} \sim G v_w^3 \kappa^2 \rho_V^2 \beta^{-5}.
\end{equation}
There may in fact be other relevant velocities, so there could be additional dimensionless factors formed using the bubble wall velocity, such  as $v_w / c$ or $v_w / c_s$. Even if there are other relevant time or length scales, so long as there are no other mass scales the $\rho_V^2$ dependence must remain.
The total liberated vacuum energy, on the other hand, is not proportional to $G$ and on similar dimensional grounds, we must have that
\begin{equation}
E_V \sim \rho_V v_w^3 \beta^{-3}.
\end{equation} 
Thus, the fraction of vacuum energy that is in GWs must go like
\begin{equation}
r \equiv \frac{E_\text{GW}}{E_V} \sim \frac{G \kappa^2 \rho_V^2 \beta^{-5}}{\rho_V \beta^{-3}} = \kappa^2 \left(\frac{\beta}{\sqrt{G \rho_V}}\right)^{\!\!-2} .
\end{equation}
We rewrite this using $H \sim \sqrt{G\rhotot}$ from \cref{eq:Hubble} and defining $\alpha = \rho_V/\rho_R$ such that $\rho_V / \rhotot = {\alpha}/(1 + \alpha)$, to give
\begin{equation}
r \sim \kappa^2 \left(\frac{\rho_V}{\rhotot}\right) \! \left(\frac{\beta}{\sqrt{G \rhotot}}\right)^{\!\!-2} \sim \kappa^2  \left(\frac{\alpha}{1 + \alpha}\right) \! \left(\frac{\beta}{H}\right)^{\!\!-2} .
\end{equation}
Using $r$, the normalized energy density of GWs may be written as 
\begin{equation}
\frac{\rho_\text{GW}}{\rhotot} \sim \frac{\rho_V}{\rhotot} \frac{E_\text{GW}}{E_V} \sim \kappa^2  \left(\frac{\alpha}{1 + \alpha}\right)^{\!2} \! \left(\frac{\beta}{H}\right)^{\!\!-2}.
\end{equation}
Finally, the spectrum may then be written as
\begin{equation}\label{eq:deduced}
\frac{1}{\rhotot} \frac{d\rho_\text{GW}}{d\ln f} \sim \kappa^2  \left(\frac{\alpha}{1 + \alpha}\right)^{\!2} \! \left(\frac{\beta}{H}\right)^{\!\!-2} \! g(f \beta^{-1}) ,
\end{equation}
where the function $g$ was forced to be a function of $f\beta^{-1}$ and we expect a peak frequency at around $\beta$ because $\beta^{-1}$ remains the only relevant timescale in the problem. 

From the perspective of energy conservation, the form \cref{eq:e_gw} seems surprising, since an energy is proportional to an energy squared. On dimensional grounds, the squared energy was necessary because the energy divided by Newton's constant contains two powers of mass, e.g., $V = G m_1 m_2 / r$. By energy conservation, we might expect that
\begin{equation}
E_\text{GW} \le \kappa E_V
\end{equation}
or equivalently that $r \le \kappa$. Roughly speaking, we thus expect
\begin{equation}
\kappa  \left(\frac{\alpha}{1 + \alpha}\right) \! \left(\frac{\beta}{H}\right)^{-2} \lesssim 1.
\end{equation}
By definition we have that $\kappa \le 1$ and that  ${\alpha} / (1 + \alpha) \le 1$. We might wonder, then, whether we require the lifetime is bounded by e.g., $(\beta / H)^{-1} \lesssim 1$? That is, whether the lifetime must be less than about a Hubble time? As we shall discuss, whilst there is nothing stopping a source of GWs lasting longer than a Hubble time, the effective lifetime over which a source may radiate GWs must be less than about a Hubble time; see the discussion of suppression factors in \cref{Eq:sw_lifetime_min,Eq:sw_lifetime_upsilon}. Furthermore, the peak frequency corresponds to a peak wavelength around $c \beta^{-1}$. The requirement  $(\beta / H)^{-1} \lesssim 1$ thus corresponds to a peak wavelength that is sub-horizon, $c \beta^{-1} \lesssim c H^{-1}$. This has been proposed as a constraint from causality that bounds the amplitude~\cite{Giblin:2014gra}. See e.g., \refcite{Caprini:2006jb} for further discussions of causality in the context of the GW spectrum.

To go beyond these simple arguments, we will turn to more detailed simulations and analysis of the spectrum from each source: bubble wall collisions, sound waves in the plasma, and long-term magnetohydrodynamic turbulence in the plasma.  We will describe the GW signal for each source, presenting explicit expressions for each contribution which arise from numerical fits to analytical calculations and simulations.  These fits arise because the GW signal for each source is fitted using a smoothly broken (or double-broken) power law, which consists of three pieces: the peak amplitude, the peak frequency, and the spectral shape.  For ease of use, the spectra are parameterized in terms of relevant thermal parameters, such as those discussed in \cref{sec:thermal_parameters}. More specifically, this usually involves a transition temperature $T_*$ (see \cref{sec:transitionTemperature}),%
\footnote{The transition temperature is often arbitrarily claimed to be the nucleation temperature. However, the background temperature of the plasma is held constant in these hydrodynamic simulations. Therefore, the transition temperature would more appropriately be one that corresponds to a time when bubbles and their sound shells are colliding, because those are the events that are occurring in the simulations that result in GW production.} a parameter for the transition strength $\alpha$ which is related to the energy released by the phase transition (\cref{sec:kineticEnergy}), a characteristic timescale that is often substituted by a characteristic length scale or $\beta_*^{-1}$ (\cref{sec:lengthscale}), and the bubble wall velocity (\cref{sec:wallVelocity}).  Notably, simulations and their approximate formulae are valid for a limited range of these thermal parameters.  Consequently, when using the resultant GW fits, one has to ensure that the approximations and assumptions are valid for the thermal parameters at hand. We will follow the standard convention in which the energy density spectrum is multiplied by $h^2$ \cite{Binetruy:2012ze}: 
\begin{equation}
  H_0 = 100 h \myunit{km s^{-1} Mpc^{-1}} ,
  \label{Eq:H0withLittleh}
\end{equation}
where $h = 0.674 \pm 0.005$ \cite{ParticleDataGroup:2020ssz} captures the uncertainty in the Hubble parameter.

However, before we can do that we must first account for the fact that the energy and frequency of GWs that would be seen today are changed significantly from their original energy and frequency.  A redshift between times $t_1$ and $t_2$ changes the frequency by a factor $a(t_1)/ a(t_2)$. The amplitude is related to the energy density divided by the critical density. An energy density redshifts by $(a(t_1) / a(t_2))^4$, whereas the critical density changes by a factor $(H(t_2) / H(t_1))^2$. The overall scaling is thus 
\begin{equation}\label{eq:redshift_amplitude}
\redshift \equiv  \left(\frac{a(t_1)}{a(t_2)}\right)^{\!\!4}
\left(\frac{H(t_1)}{H(t_2)}\right)^{\!\!2}.
\end{equation}

Standard expressions used in the literature for these redshifts can be obtained as follows \cite{Cai:2017tmh}.  First we assume adiabatic expansion, so that $s(t) a^3(t)$ is conserved, and use the fact that the entropy density scales with temperature as $s(T) \sim g_s(T) T^3$, where $g_s$ are the entropic degrees of freedom.  This gives   
\begin{align}
  \frac{a_1}{a_2} = \left(\frac{s_2}{s_1}\right)^{\!\!\frac13} =  \left(\frac{g_s(T_2) }{g_s(T_1) }\right)^{\!\!\frac13} \frac{T_2}{T_1} , \label{eq:scaleFactorRatio}
\end{align}
where we denote scale factors and temperatures at $t_1$ and $t_2$ with and an index 1 and 2 respectively.  If we assume that at $t_1$ all the effective degrees of freedom are entropic (i.e.\ that at this stage all states are in thermal equilibrium) so that $g_s(T_1) = g_\text{eff}(T_1)$, then we can use the Friedmann \cref{eq:Friedmann1} and the assumption of radiation domination to write 
\begin{align}
H_1^2 = \frac{8 \pi^3 G}{90}  g_s(T_1)T_1^4 . \label{eq:H-entropic}
\end{align}
We can use \cref{eq:H-entropic} to pull out a factor of $1/H_1$ in \cref{eq:scaleFactorRatio}, giving
\begin{align}\label{eq:a_ratio}
  \frac{a_1}{a_2}=  \sqrt{\frac{8\pi^3 G}{90}} \frac{T_2T_1}{H_1} g_s(T_2)^{\frac13} g_\text{eff}(T_1)^{\frac16},
\end{align}
for the redshift of a frequency. For the redshift of the amplitude,
\begin{align}
  \redshift(T_1,T_2) = 
  \left(\frac{g_s(T_2)}{g_s(T_1)}\right)^{\!\!\frac43} \!
  \left(\frac{T_2}{T_1}\right)^{\!\!4} \!
  \left(\frac{H_1}{H_2}\right)^{\!\!2} = 
  \frac{g_s(T_2)^{\frac43}}{g_s(T_1)^{\frac13}} \frac{8\pi^3G}{90} \frac{T^4_2}{H_2^2},
\end{align}
follows from combining \cref{eq:redshift_amplitude,eq:H-entropic,eq:a_ratio}.

We now specialize to the case where $T_2$ is the temperature today, $T_2 = T_0 =2.725 \,\text{K} = 3.3177\times10^{-5}\,\text{eV}$. At this temperature there are 
\begin{equation}
g_s(T_0) = 2+ \frac7{11} N_\text{eff}
\end{equation}
 degrees of freedom, where $N_\text{eff}=3.046$ is the effective number of neutrinos. With appropriate unit conversions and using $G= 6.674 \times 10^{-11}\,$m$^3$kg$^{-1}$s$^{-2}$, the frequencies today are given by \cite{Kamionkowski:1993fg}
\begin{align}
  f_0 = \frac{a_1}{a_0}f_1 &= 1.65 \times 10^{-5} \,\text{Hz} \, \left(\frac{g_\text{eff}(T_1)}{100}\right)^{\!\!\frac16} \! \left(\frac{T_1}{100\gev}\right) \! \frac{f_1}{H_1},
  \label{Eq:f_redshit_factor}
\end{align}
and using \cref{Eq:H0withLittleh} the amplitudes today are given by~\cite{Kamionkowski:1993fg}
\begin{align}
  \Omega_0 h^2 =  \redshift(T_1,T_0) \, \Omega_1 h^2 =  1.67\times 10^{-5}\left(\frac{100}{g_\text{eff}(T_1)}\right)^{\!\!\frac13}\Omega_1 .
  \label{Eq:Omega_redshift_factor}
\end{align}
We have written these final expression with explicit dependence on
the effective degrees of freedom and the temperature $T_1$ from which
we redshift, though the explicit dependence on this temperature canceled in \cref{Eq:Omega_redshift_factor}. We recommend using $T_1 =\Treh$ because redshifting should be taken from after reheating.\footnote{Note that in directly using \cref{Eq:f_redshit_factor} with $T_1 =\Treh$ we implicitly assume that the peak frequency at the time of the transition (i.e. at percolation) is not changed by reheating and we also assume any particle that fell out of thermal equilibrium as the Universe initially cooled from $\Treh$ to $T_p$ returns to thermal equilibrium when the universe reheats back to $\Treh$.} The reheating temperature $\Treh$ can be estimated from the percolation temperature using \cref{Eq:Treh_from_Tp}.

In obtaining these expressions we have assumed
radiation domination, adiabatic expansion and that all states are in
thermal equilibrium at the temperature $T_1$ from which we redshift. The assumption of radiation domination may break down. For example, if there is a subsequent strongly supercooled transition after the time of GW production from an earlier transition, a period of vacuum domination could affect the redshifting. The same goes for a simultaneous strongly supercooled phase transition, provided it continues for some time after GWs are produced. A period of vacuum or matter domination can affect the assumed form of the temperature dependence of the Hubble parameter, which in turn affects the redshift factor. See \refcite{Allahverdi:2020bys} for a review of other possible deviations from the standard assumption of radiation domination prior to Big Bang nucleosynthesis. For a treatment avoiding many of the assumptions in deriving \cref{Eq:f_redshit_factor,Eq:Omega_redshift_factor} --- that is appropriate for supercooled phase transitions and transitions where the degrees of freedom can vary significantly over relevant temperature ranges --- see \refcite{Athron:2023mer}.

\subsection{Bubble collisions} 
\label{Section:GWs-Coll}

As discussed in \cref{sec:GWobs}, a source of GWs must be spherically asymmetric. The most obvious source of anisotropies in the stress-energy tensor is when bubbles collide, breaking the spherical symmetry of the highly energetic walls.\footnote{This is often referred to as the scalar field contribution in the literature, see e.g.~\refcite{Caprini:2015zlo}, since it involves only the scalar field and not the plasma.} Analytically, this source is often described using the envelope approximation. This is typically combined with the thin-wall approximation where the bubble walls are taken to be infinitesimally thin. The envelope approximation, although originally developed for vacuum transitions (in the absence of a thermal plasma slowing the bubble walls), has been found to predict the GW signal from bubble collisions with a reasonable precision \cite{Weir:2016tov}, though for a recent treatment of GW spectra from thick-walled bubbles see \refcite{Cutting:2020nla}.

In 2016 an analytic derivation of the GW signal from colliding bubbles was performed in \refcite{Jinno:2016vai}.  Prior to that numerical simulation results of \refcite{Huber:2008hg} were often used \cite{Weir:2017wfa}. The analytic derivations make the envelope and thin-wall approximations mentioned above and assume an exponential nucleation rate as in \cref{Eq:exp_nuc_rate}. Motivated by ansatzes similar to \cref{eq:deduced}, they first write the GW spectrum after the phase transition ends as
\begin{equation}
  \Omega_{\text{coll}}(k) = \kappa_{\text{coll}}^2 \left(\frac{\alpha}{1+\alpha} \right)^{\!\!2} \! \left( \frac{\beta}{H}\right)^{\!\!-2} \! \Delta(k/\beta, v_w) ,
  \label{Eq:BubColl_GWs_PT_time}
\end{equation}
where $k$ is the wave number. The factor $\Delta(k/\beta,v)$ was defined as the departure of the full calculation from the ansatz, and may be written as an integral,
\begin{align}
    \Delta(k/\beta, v_w) &\equiv \frac{\Omega_{\text{coll}}(k)}{\kappa_{\text{coll}}^2 \left(\frac{\alpha}{1+\alpha} \right)^{\!2} \! \left(\frac{\beta}{H_*}\right)^{\!-2}}\\
    &= \frac{3}{4 \pi^2}\frac{\beta^2 k^3}{\kappa_{\text{coll}}^2\rho_0^2} \int^{t_\text{end}}_{t_\text{start}}dt_x \int^{t_\text{end}}_{t_\text{start}}dt_y \cos(k(t_x-t_y))\Pi(t_x,t_y,k) ,
    \label{Eq:Delta_Jinno}
\end{align}
where $\Pi$ is the source term for the GWs, which they compute from the energy-momentum tensor, and the integration limits are given by the lifetime of the source.
Although this is not explicitly written as a function of $\beta / k$ or $v_w$, based on the expected ansatz and dimensional analysis, when evaluated it is anticipated that it depends only on them.
\Refcite{Jinno:2016vai} express the GW spectrum that should be seen today as
\newcommand{\lrvph}{\vphantom{\frac{\beta}{H_*}}}%
  \begin{align}
    	h^2 \Omega_{\text{coll}} (f) & = \underbrace{1.67\times 10^{-5}
         \left(\lrvph\frac{g_*}{100} \right)^{\!\!-\recip{3}}
    }_{\textit{\textcolor{Red}{Redshift}}}
    \,\underbrace{
       \left(\lrvph\frac{\kappa_{\text{coll}} \alpha}{1 + \alpha} \right)^{\!\!2} \! \left(\lrvph\frac{\beta}{H_*} \right)^{\!\!-2} \! \Delta(f_{\text{coll}}, v_w)
    }_{\textit{\textcolor{Green}{Scaling}}}
    \,\underbrace{
        \vphantom{\left(\lrvph\right)}
        S_{\text{coll}}(f, v_w)
    }_{\textit{\textcolor{Blue}{Shape}}} , \label{GWCollPeakAmp} \\
	f_{\text{coll}} & = 
    \underbrace{
      1.65\times 10^{-5} \myunit{Hz} \left(\lrvph\frac{g_*}{100} \right)^{\!\!\recip{6}} \!
      \left(\lrvph\frac{\Ttr}{100 \myunit{GeV}} \right) \!
      \left(\lrvph\frac{f_*}{\beta} \right) \! \left(\lrvph\frac{\beta}{H_*} \right)
    }_\textit{\textcolor{Red}{Redshift}} ,
    \label{GWCollPeakFreq}
  \end{align}
where $f_* / \beta$ is the peak frequency at creation in Hubble units and may depend on the bubble wall velocity, and we have factorized $\Delta(f, v_w) \equiv \Delta(f_{\text{coll}}, v_w) S_{\text{coll}}(f, v_w)$ such that $S_{\text{coll}}(f_{\text{coll}}, v_w) = 1$. We underline the redshift factors in red and the scaling factors deduced on dimensional grounds in \cref{eq:deduced} in green. The last part of \cref{GWCollPeakAmp} underlined in blue, $S_{\text{coll}}$, gives the spectral shape of the GWs.

The main focus of \refcite{Jinno:2016vai} is on the integration needed to evaluate $\Delta$, which determines the shape $S_{\text{coll}}$ of the GW spectrum and the additional (bubble wall velocity dependent) scaling factor. Their analytical calculation ultimately yields double integral expressions that must be calculated numerically.  In this calculation they make an additional assumption by taking the integration limits in \cref{Eq:Delta_Jinno} to $\pm\infty$, which neglects additional effects from the duration of the phase transition.   They also present their results as fit formulae, including expressions for $\Delta_\text{peak}(v)$ to account for the case of subluminal bubble wall velocities.  These fits have been presented in a previous review \cite{Weir:2017wfa}, and can be found there or in the original paper. 

Recently, this work was extended in \refcite{Zhong:2021hgo} (see also \refcite{Megevand:2021juo}). While they
also assume the envelope approximation, thin walls and an exponential
nucleation rate, they include two important additional
effects. First, they include effects from the expansion of the Universe
in the integration to obtain $\Delta$. Second, they avoid taking
integration limits to infinity, which they find can have an impact on
the results.  The latter effect means that the factor $(\beta
/H_*)^{-2}$ no longer appears as a prefactor in their version of
\cref{GWCollPeakAmp} and instead the dependence on the duration is
incorporated into $\Delta$ and obtained through their integration.
These effects then appear in the analytical calculation of their
$\Delta$ which --- similarly to the previous work --- ultimately yields (in this case triple) integrals
that must be evaluated numerically.
They do not present fit formulae for these integrals and the
expressions are rather lengthy so we
do not reproduce them here.  However, these new effects can reduce the
signal by an order of magnitude, which suggests that using the updated
results is important if one needs to consider GWs from
bubble collisions.

There are studies of the scalar field contribution to GWs that go beyond the envelope approximation. The bulk flow model \cite{Jinno:2017fby, Konstandin:2017sat} allows the energy stored in the collided regions of bubble walls to decay over time, rather than immediately vanishing. \Refcite{Jinno:2019bxw} propose a trapping equation to more accurately model the dynamics of the collided regions of bubble walls. They discuss the cases in which the envelope approximation leads to reasonable predictions, and the cases in which the bulk flow model is more appropriate. A general approach that allows calculations of the GW spectrum from thin walls and also from fluid shells has also been developed \cite{Megevand:2021juo}.  

Insights regarding these models can also be obtained by comparing to results from a hybrid approach where full lattice simulations of two bubbles are performed to find the scaling of the gradient energy, and the full gravitational wave spectrum is then calculated from many-bubble simulations in the thin wall limit.  \Refcite{Lewicki:2020jiv} demonstrate fundamental differences in the spectral shape depending on whether the scalar field is real or the potential is $U(1)$ symmetric. They demonstrate (see their Fig.~4) that in both cases the high frequency scaling is predicted by the bulk flow model. However, the low frequency scaling for a real scalar field is well predicted by the envelope approximation, but for a $U(1)$-symmetric potential is instead better predicted by the bulk flow model.   This is because the energy dissipation in collided regions of bubble walls is slower when the bubbles have a different complex phase --- an effect ignored in the envelope approximation. This work was further extended in \refcite{Lewicki:2020azd} where they show that for a case where the $U(1)$ is a gauge symmetry the scaling is close to the real singlet case, and further fits can be found in \refcite{Lewicki:2022pdb}. The dependence on the shape of the potential was also explored in full simulations \cite{Cutting:2020nla}. In this case they varied a parameter controlling how thick the bubble wall is, finding only moderate variations in the peak amplitude (a factor of $\approx 1.3$) but potentially significant differences in the shape.

After collisions, a period of bubble coalescence \cite{Child:2012qg} and oscillation of the scalar field in collided regions \cite{Cutting:2018tjt} continues to source GWs. \Refcite{Child:2012qg} found an enhancement of the signal beyond the envelope approximation due to the period of bubble coalescence, but \refcite{Cutting:2018tjt} later found that this effect is always subdominant to the collision source.

Irrespective of these issues and the other assumptions, a crucial
factor in determining the GW spectra is the efficiency factor
$\kappa_{\text{coll}}$.  This describes how much of the available
energy goes into the bubble collisions and cannot be determined in the
analytical calculations of \refcite{Jinno:2016vai,Zhong:2021hgo}.  Due
to the next-to-leading-order friction effects discussed in
\cref{sec:friction}, $\kappa_{\text{coll}}$ can be approximated by
zero in most cases \cite{Cai:2018teh}.  The fraction of energy
released during the phase transition into the kinetic and gradient
energy of the bubble wall reduces as the bubble wall approaches its
terminal velocity (in the non-runaway case), while most of the energy
is transferred to the surrounding plasma before collision
\cite{Ellis:2019oqb}.  This is demonstrated nicely in a concrete model
in \refcite{Alanne:2019bsm}.  Analytic estimates of the efficiency
coefficient $\kappa_{\text{coll}}$ have been made in
\refcite{Ellis:2019oqb,Cai:2020djd,Lewicki:2022pdb}.  However, because
$\kappa_{\text{coll}} \lll 1$ in the non-runaway case, the collision
component of the GW spectrum is typically subdominant across the whole
frequency spectrum and it may be neglected entirely.  This is a common
approximation in recent literature on phase transitions at finite
temperature.

An exception to this is when the bubble wall runs away. This could
happen for transitions at temperatures so low that the plasma can be
ignored and thus there is negligible friction to oppose bubble wall
acceleration. For these transitions $\kappa_{\text{coll}}$ is close to
unity and the collision contribution would be effectively the only
source of GWs (except for oscillons \cite{Hawking:1982ga,
  Kosowsky:1992vn, Johnson:2011wt, Cutting:2018tjt}). For phase
transitions at high temperatures we expect that the transition occurs
in a surrounding plasma and we must consider GW production from bulk
motion of the plasma; that is, from sound waves and turbulence.

\subsection{Sound waves} 
\label{Section:GWs-Sound}

In general terms, sound waves correspond to relativistic bulk motion of the plasma, in many cases with significant energy \cite{Hogan:1986qda}.  As this energy directly enters into the energy-momentum tensor, in turn it directly contributes to the spacetime curvature creating GWs.  In the specific case of FOPTs the expanding bubble walls, carrying substantial energy density and pressure, transmit this energy to the surrounding plasma.  During and after bubble collisions these kinetic energy fluctuations of the plasma appear as sound waves.

Sound waves arising from FOPTs potentially contributing to GWs were proposed almost four decades ago \cite{Witten:1984rs, Hogan:1986qda, Kamionkowski:1993fg}.  Their contribution to GWs, however, has only been investigated in detail in the past decade~\cite{Hindmarsh:2013xza, Giblin:2014qia, Hindmarsh:2015qta, Hindmarsh:2016lnk, Hindmarsh:2017gnf, Cutting:2019zws, Hindmarsh:2019phv}.  Sound waves are predicted to be the dominant source of GWs in transitions with a strong coupling between the bubble wall and the plasma, as is typically the case for a first-order electroweak phase transition. The GW signal from sound waves dominates over bubble collisions if the bubble walls reach a terminal velocity rather than running away~\cite{Hindmarsh:2013xza}. While the overlapping regions of bubble walls are neglected in the envelope approximation, in reality these regions become freely propagating pulses of fluid kinetic energy.  The sound wave contribution gains a significant boost in signal since it persists long after the transition completes \cite{Hindmarsh:2015qta, Hindmarsh:2016lnk}.

There are multiple relevant timescales for GWs sourced from the sound waves, including the lifetime of the source, $\tau_\text{sw}$, and the autocorrelation time $\tau_c$: the time period for the anisotropic stresses from the bulk fluid motion to decorrelate.  In general, the GW amplitude at the time of their production can be written as \cite{Hindmarsh:2015qta, Hindmarsh:2017gnf, Caprini:2009yp, Caprini:2019egz}
\begin{equation}
  \Omega_\text{sw} \sim K^2  \left( H_* \tau_\text{sw} \right)  \left( H_* \tau_c \right)  S_\text{sw}(k),
  \label{eq:omega_gw_general}
\end{equation}
where $K$ is the kinetic energy fraction of the fluid. The kinetic energy fraction is obtained from the fluid velocity in hydrodynamic simulations via $K = \Gamma \Uf^2$, where $\Gamma$ is the mean adiabatic index and $\Uf^2$ is the enthalpy-weighted root-mean-square of the spatial part of the fluid four-velocity; see \cref{sec:kineticEnergy} for further discussion of these quantities.
Finally the last factor in \cref{eq:omega_gw_general},
  $S_{\text{sw}}(k)$, sets the shape of the spectrum which is
  determined by the original source. If the wave number is larger than the inverse autocorrelation time, $k > \tau_c^{-1}$, then $H_* \tau_c$ should be replaced by $H_*/k$.
As anticipated, this is similar to the form derived on dimensional grounds in \cref{eq:deduced}. Although there are now two timescales, the energy squared dependence remains since there are no other relevant mass scales.

The form of \cref{eq:omega_gw_general} must be modified for sources that are particularly long-lasting~\cite{Caprini:2019egz}, as can be seen from the discussion around \cref{Eq:sw_lifetime_min} in \cref{sec:lengthscale}. For transitions with $H_* \tau_\text{sw} > 1$, decorrelation is expected before a Hubble time and \cref{eq:omega_gw_general} must be modified to
\begin{equation}
  \Omega_\text{sw} \sim K^2  \left( H_* \tau_c \right)  S_\text{sw}(k).
  \label{eq:omega_gw_long_lasting}
\end{equation}
This is because even if a source is active longer than a Hubble time, the effective duration of GW production is of the order of a Hubble time. For short-lasting sound waves, the lifetime may be expressed as
\begin{equation}
  \tau_{\text{sw}} \sim \tau_{\text{sh}} \sim  \frac{L_*}{\Uf} = L_* \sqrt{\frac{\Gamma}{K}} ,
  \label{Eq:taush_to_L_and_K}
\end{equation}
where we assumed the source is cut off by the onset of shocks and that $\Ulong \approx \Uf$.
For both short- and long-lasting waves, the autocorrelation time may be approximated by the sound-crossing time, $\tau_c \sim L_*  / c_s$. With these approximations, for short-lasting waves \cref{eq:omega_gw_general} becomes
\begin{equation}
  \Omega_{\text{sw}} \sim K^{3/2}  \frac{(H_* L_*)^2}{c_s}  S_{\text{sw}}(k), \label{eq:sw-scaling-short}
\end{equation}
whereas for long-lasting waves \cref{eq:omega_gw_long_lasting} becomes
\begin{equation}
  \Omega_{\text{sw}} \sim K^2  \frac{H_* L_*}{c_s}  S_{\text{sw}}(k). \label{eq:sw-scaling-long}
\end{equation}
In the case of short-lasting waves, the enthalpy $\bar w$ in the adiabatic index $\Gamma$ provided a factor containing dimension mass, avoiding the $\rho^2$ arguments from dimensional analysis. In obtaining \cref{eq:sw-scaling-short,eq:sw-scaling-long}, our treatment of $\tau_\text{sw}$ corresponds to using \cref{Eq:sw_lifetime_min} as a suppression factor.  However, as discussed just below \cref{Eq:sw_lifetime_min}, a more careful treatment leads to the suppression factor in \cref{Eq:sw_lifetime_upsilon}.  Therefore, we only expect these relations to hold in the specific limits described there, and in what follows we will use \cref{Eq:sw_lifetime_upsilon}.

For more detailed modeling, early treatments attempted to apply the envelope approximation to the fluid (or sound) shells surrounding bubble walls \cite{Huber:2008hg}. However, it is known that the sound shells are not infinitesimally thin, instead scaling with the size of the bubble (see \cref{eq:fluidShellThickness}). The sound shell model \cite{Hindmarsh:2016lnk, Hindmarsh:2019phv} was proposed to take these effects into account, which expresses the velocity field of the fluid as a linear superposition of sound shells. GWs are produced at the boundary between two shells, where the pressure abruptly changes.  In the original version of the sound shell model, the sound shells are assumed to freely propagate, no longer driven by a scalar field. However \refcite{Cai:2023guc} have very recently performed preliminary modeling of the scalar-driven propagation of the sound shells during the early stages of a collision. Due to linearity, the sound shell model provides a computationally efficient way to calculate the GW signal. The dominant source of shear stress comes from the fluid, $\tau_{ij} = \gamma^2 w v_i v_j + \partial_i \phi \partial_j \phi \simeq \gamma^2 w v_i v_j$ (see \cref{eq:sourceTensor}). This is not true if the bubbles run away, in which case the sound shell model is not applicable~\cite{Hindmarsh:2019phv}. Although the sound shell model predicts a spectrum characterized solely by the peak amplitude and frequency~\cite{Curtin:2022ovx}, extracting more information about all the thermal parameters may be possible with more careful modeling~\cite{Gowling:2021gcy}.  

Turning to simulations, we take the corresponding GW
signal from recent large scale hydrodynamic simulations
\cite{Hindmarsh:2017gnf}.  Combining these with effects discussed in \refcite{Caprini:2019egz} and \refcite{Guo:2020grp} gives the following expressions for the signal and
peak frequency:
\renewcommand{\lrvph}{\vphantom{\frac{c}{\max(v_w, c_s)}}}%
\begin{align}
    h^2 \Omega_{\text{sw}}(f) & = 
    2.59 \times 10^{-6}
    \underbrace{
         \! \left(\lrvph\frac{g_*}{100} \right)^{\!\!-\recip{3}}
    }_{\textit{\textcolor{Red}{Redshift}}}
    \,\underbrace{
        \! \left(\lrvph\frac{\kappa_\text{sw} \alpha}{1+\alpha} \right)^{\!\!2} \left(\lrvph\frac{\max(v_w, c_{s,f})}{c}\right) \!
        \left(\frac{\beta}{H_*} \right)^{\!\!-1} \! \Upsilon(\tau_\text{sw})
    }_{\textit{\textcolor{Green}{Scaling}}}
    \,\underbrace{
        \vphantom{\left(\lrvph\right)}
        S_\text{sw}(f) 
    }_{\textit{\textcolor{Blue}{Shape}}} ,
  \label{GWSoundPeakAmp} \\
    f_{\text{sw}} & = 8.9 \times 10^{-6} \myunit{Hz} \,
	    \underbrace{
         \! \left(\lrvph\frac{g_*}{100} \right)^{\!\!\recip{6}} \! \left(\lrvph\frac{\Ttr}{100 \myunit{GeV}} \right)
    }_{\textit{\textcolor{Red}{Redshift}}}
    \,\underbrace{
        \left(\frac{c}{\max(v_w, c_{s,f})}\right) \! \left(\lrvph\frac{\beta}{H_*} \right) \! \left(\lrvph\frac{z_p}{10} \right)
    }_{\textit{\textcolor{Green}{Scaling}}},
\label{GWSoundPeakFreq}
\end{align}
where we have reinstated the factors of $c$ in \cref{GWSoundPeakAmp,GWSoundPeakFreq}
to make the dimensions clear.  The numerical prefactors absorb
contributions from redshift factors
(\cref{Eq:Omega_redshift_factor,Eq:f_redshit_factor}), substitutions
and the fit. We have included corrections to the numerical prefactor
based on the erratum of \refcite{Hindmarsh:2017gnf}. This disagrees
with the numerical prefactor given in \refcite{Weir:2017wfa} prior to
the publication of the erratum and these numerical errors have also
propagated into some literature after the erratum was published. The peak amplitude and frequency from these fits are shown in \cref{fig:freq_strain} for temperatures between $T_* = 1\gev$ and $10^9\gev$ for two choices of $\Uf$ and $v_w$. We see that future space-based experiments could be sensitive to transitions at around $1\gev \lesssim T_* \lesssim 10^6 \gev$, though we must emphasize the broadband nature of the signal.

The general form of the amplitude comes from
\cref{eq:sw-scaling-long} after making a few substitutions so that
we can write it in terms of the thermal parameters.  We relate the
mean bubble separation to $\beta$ by \cref{Eq:BubbleSepToBeta} and use
the relations $K = \Gamma \Uf^2 = \kappa_\text{sw} \alpha /
(1+\alpha)$ (see \cref{sec:kineticEnergy} for details).%
\footnote{However, the substitutions involving $\alpha$ and $\beta$ introduce further approximation. More fundamental quantities such as $K$ and $\lenscale$ should be used wherever possible.}
The factor $\kappa_\text{sw}$ is the efficiency of producing bulk fluid motion
from the vacuum energy, i.e.\ the fraction of energy available for GWs
sourced from sound waves. The efficiency factor can be determined via fits provided in the appendix of \refcite{Espinosa:2010hh} in the bag model \cite{Chodos:1974je}, or via code provided in the appendix of \refcite{Giese:2020znk} for a more general equation of state.
\Refcite{Ellis:2019oqb, Ellis:2020nnr} include an additional efficiency factor $1 - \kappa_{\text{coll}}$ to account for energy used in collisions.

We include a lifetime suppression factor $\Upsilon(\tau_\text{sw})$ from \refcite{Guo:2020grp}, which we define
in \cref{Eq:sw_lifetime_upsilon}, to account for both cases where $H_*
\tau_\text{sw} > 1$ and $H_* \tau_\text{sw} < 1$.  The factor $\tau_\text{sw}$ in $\Upsilon$ can be substituted using the relation given in \cref{Eq:taush_to_L_and_K}. 

Motivated by simulations in~\refcite{Hindmarsh:2015qta}, the spectral shape can be approximated by the ansatz~\cite{Caprini:2015zlo} 
\begin{equation}
	S_{\text{sw}}(f) = \left(\frac{f}{f_{\text{sw}}} \right)^{\!\!3} \! \left(\frac{7}{4 + 3(f/f_{\text{sw}})^2} \right)^{\!\!\frac{7}{2}},
\end{equation}
which takes a maximum $S_{\text{sw}}(f_{\text{sw}}) = 1$ at the peak frequency. This is a smoothly broken power law with low- and high-frequency regimes with powers $3$ and $-4$, respectively.

The sound shell model, on the other hand, predicts a  smoothly double broken power law~\cite{Hindmarsh:2019phv}. The three regimes can be understood as originating from the boundaries between two distinct characteristic scales: the mean bubble separation and the thickness of sound shells. The sound shell model with freely propagating sound shells predicts powers $9$, $1$ and $-4$ for the three regions.\footnote{\Refcite{Hindmarsh:2019phv} corrects an error in the earlier \refcite{Hindmarsh:2016lnk} that led to a low-frequency power of $5$ instead of $9$.}  In particular note that the power $9$ is much steeper than seen in simulations (see the discussion in section 6 of \refcite{Hindmarsh:2019phv} ).  Therefore this original version of the sound shell model predicts a higher growth rate at low frequencies than has been seen in the simulations. However as mentioned above very recently the sound shell model has been extended to include forced propagation driven by the uncollided envelope of the bubble walls and including this can reproduce the $k^3$ scaling at low $k$ seen in simulations \cite{Cai:2023guc}.

There are a few major caveats to using the fits from simulations. The large separation in length scales between the bubble wall width and the bubble radius does not allow for simulation of realistic phase transitions due to a significant lack in dynamic range of the lattice simulations \cite{Hindmarsh:2015qta}. However, a Higgsless approach has recently been developed \cite{Jinno:2020eqg, Jinno:2022mie} that removes the smallest length scale: the bubble wall width. The separation of length scales is then between the sound shell thickness and the bubble radius. The Higgsless approach evolves only the fluid, and not the scalar fields. The scalar fields only enter as boundary conditions through the space-dependent bag constant in the bag equation of state, which is used to determine the fluid profile. Hydrodynamic simulations typically simplify bubble nucleation by initializing the simulation with all bubbles already nucleated. This prevents a hierarchy in bubble radii that can appear in more realistic nucleation scenarios, such as an exponential nucleation rate. Both scalar only simulations (discussed in \cref{Section:GWs-Coll}) and the Higgsless approach are able to consider exponential nucleation because they do not have to evolve the more expensive coupled field-fluid system. Additionally, no simulations have probed $\alpha \gtrsim 1$ or the ultra-relativistic regime for bubble walls coupled to a plasma; however, there is recent and ongoing work to treat this regime analytically~\cite{Jinno:2019jhi}.
It is expected that strong transitions will shorten the duration of the acoustic regime, $\tau_{\text{sw}}$, and lead to an early onset of turbulence \cite{Caprini:2015zlo, Hindmarsh:2015qta, Hindmarsh:2017gnf, Ellis:2018mja}. Indeed, the amplitude may be dampened by the effective lifetime of the source, which is accounted for by the $\Upsilon$ factor. A corresponding factor $1 - \Upsilon$ in the contribution to the GW spectrum from turbulence is discussed in \cref{Section:GWs-Turb}.  While earlier simulations \cite{Hindmarsh:2013xza, Hindmarsh:2015qta, Hindmarsh:2017gnf} considered only $\alpha = \mathcal{O}(10^{-2} \text{--} 10^{-1})$, a recent large-scale simulation \cite{Cutting:2019zws} focusing on sound waves has considered strengths up to $\alpha = 0.67$ (defined using the trace anomaly as in \cref{eq:alpha-traceAnomaly}). It was found that significant rotational modes appear as the transition strength is increased. This is contrary to the prediction of \refcite{Hindmarsh:2015qta,Hindmarsh:2017gnf},  where they found the transverse component of $\Uf$ was at most $5-10\%$, and is also contrary to the sound shell model, which assumes the fluid motion is predominantly linear acoustic waves. The resultant loss in efficiency, as well as the formation of reheated droplets in front of the wall for subsonic deflagrations and hybrids, can suppress the GW signal by a factor of $\mathcal{O}(10^{-3})$ for subsonic deflagrations in strong transitions.\footnote{Briefly, reheating occurs in front of the bubble wall unless it expands as a supersonic detonation. The evolution of regions of reheated plasma in the false vacuum was studied in \refcite{Cutting:2022zgd}. See \cref{sec:reheating,sec:wallVelocity} for further discussions of the effects of reheating in front of the bubble wall.}

Finally, \refcite{Ellis:2018mja} showed that in concrete models the sound wave source does not last for a Hubble time. A later and more detailed survey of different types of models~\cite{Ellis:2020awk} showed that in the absence of supercooling the effective lifetime of the sound wave source can be only a fraction of a Hubble time because turbulence sets in before a Hubble time. As a result, the simulations which do not yet include turbulence likely fail to yield accurate GW predictions.  On the analytic side, the assumption of linear flow breaks down in the turbulence regime, which dominates long after the transition completes.  This naturally brings us to consider the turbulence contribution.

\subsection{Turbulence} \label{Section:GWs-Turb}

Turbulence may develop in the plasma after the collisions between bubbles. Massive amounts of energy are injected at particular length scales. Additionally, the Reynolds number of the plasma is enormous (possibly $10^{13}$~\cite{Caprini:2009yp}) at those scales. Indeed, shocks --- discontinuities in the properties of the fluid --- are anticipated to lead to turbulence at some stage after the timescale for the onset of shocks; see \cref{eq:tau_shock}. Roughly speaking, the shock timescale is inversely proportional to the strength of the transition. Turbulence is characterized by irregular eddy motions and typically modeled by Kolmogorov's theory~\cite{1991RSPSA.434....9K}. In this theory, energy from large eddies cascades down into smaller, statistically stationary, homogeneous and  isotropic eddies.\footnote{Though see \refcite{Kalaydzhyan:2014wca,Galtier:2017mve} for discussions of inverse cascades.} Eddies dissipate their energy through viscous heating once the Kolmogorov scale is reached because viscosity becomes relevant \cite{Mazumdar:2018dfl, Leitao:2012tx}. Because viscosity is neglected above the Kolmogorov scale and the eddies are self-similar, the only relevant scale is the rate at which energy flows through the length scales, and Kolmogorov's theory predicts the energy spectrum by dimensional analysis alone. The expansion and collision of bubbles acts to stir the plasma and thus the largest eddies are of order of the bubble radius --- this is the stirring scale.  However, there is ambiguity in the stirring scale of the turbulence from bubbles, because the stirring may be dominated by a few large bubbles or from a larger number of small bubbles \cite{Leitao:2015fmj} (see also the discussion around \cref{eq:meanBubbleRadius}). The plasma is only stirred for the timescale of the transition, which we take to also mark the stirring time. If the stirring time is shorter than the time for largest eddies to circulate, the circulation time $\tau_\text{turb}$, then stationary, homogeneous and isotropic turbulence will not develop because the plasma is not being continually stirred. On top of that, the fact that we are modeling a relativistic plasma means that this classical model of turbulence in an incompressible fluid may need some corrections~\cite{Kosowsky:2001xp}. For example, turbulent motion cannot be correlated across causal horizons~\cite{Caprini:2009yp} and so cannot be homogeneous and isotropic on horizon scales.

The turbulent motions in the fluid can result in GWs and, because turbulence transfers energy from bulk motion on large scales to small scales, impact the GWs from sound waves even if the GWs from turbulence are negligible. The fits common in the current literature from \refcite{Caprini:2009yp} are comparatively old, but provide important corrections to previous results (see \refcite[Section 2.2.2]{Binetruy:2012ze} for a review). The fits are \cite{Weir:2017wfa}
\begin{align}
	h^2 \Omega_{\text{turb}}(f) & = 3.35 \times 10^{-4}
    \underbrace{
          \left(\lrvph\frac{g_*}{100} \right)^{\!\!-\recip{3}}
    }_{\textit{\textcolor{Red}{Redshift}}}
    \,\underbrace{
         \! \left(\lrvph\frac{\beta}{H_*} \right)^{\!\!-1}
         \! \left(\lrvph\frac{\kappa_{\text{turb}} \alpha}{1 + \alpha} \right)^{\!\!\frac{3}{2}} 
         \!\left(\lrvph\frac{\max(v_w, c_{s,f})}{c}\right)
    }_{\textit{\textcolor{Green}{Scaling}}}
    \,\underbrace{
        \vphantom{\left(\lrvph\right)}
        S_{\text{turb}}(f)
    }_{\textit{\textcolor{Blue}{Shape}}}, \\
	f_{\text{turb}} & =  2.7\times 10^{-5} \myunit{Hz} 
    \underbrace{
         \left(\lrvph\frac{g_*}{100} \right)^{\!\!\recip{6}} \!  \left(\lrvph\frac{\Ttr}{100 \myunit{GeV}} \right)
    }_{\textit{\textcolor{Red}{Redshift}}}
    \,\underbrace{
         \left(\lrvph\frac{c}{\max(v_w, c_{s,f})}\right) \! \left(\lrvph\frac{\beta}{H_*} \right)
    }_{\textit{\textcolor{Green}{Scaling}}},
\end{align}
where we have reinstated factors of $c$, and again the characteristic length scale is modified as in \cref{Eq:BubbleSepToBeta}. As for sound waves the numerical prefactor in the amplitude combines factors from the redshifting, fitting and substitutions. The form for the amplitude may be deduced from the general expression for a long-lasting source in \cref{eq:omega_gw_long_lasting} by setting the autocorrelation time to the eddy circulation time for the largest scales (see \cref{eq:tau_turbulence}),
\begin{equation}
\tau_{c} \sim \tau_{\text{turb}}  \sim \frac{\lenscale}{\Utrans} \sim \lenscale \sqrt{\frac{\Gamma}{K}} ,
\end{equation}
leading to the $K^{3/2}$ dependence. On the other hand, if the turbulence lasts less than a Hubble time, all timescales are similar, $\tau_{\text{sw}} \sim \tau_c \sim \tau_{\text{turb}}$, leading to only a $K$ dependence.  The dimensional analysis arguments in \cref{eq:deduced} fail because the enthalpy of the sound waves becomes relevant, just as it did in the case of sound waves. There are attempts to improve these results by numerical simulations that relax assumptions of Kolmogorov turbulence and include the expansion of the Universe and magnetic fields~\cite{RoperPol:2018sap,Niksa:2018ofa,Sharma:2022ysf,Brandenburg:2021bvg,Auclair:2022jod}. The spectral shape follows analytically by combining Kolmogorov theory with causality, leading to~\cite{Caprini:2009yp}
\begin{equation}
	S_{\text{turb}}(f) = \frac{(f/f_{\text{turb}})^3}{(1 + f/f_{\text{turb}})^{\frac{11}{3}} (1 + 8\pi f/h_*)},
\end{equation}
where,
\begin{equation}
  h_* = 1.65\times10^{-5}\,\textrm{Hz}\left(\frac{g_*}{100} \right)^{\frac16} \left(\frac{T_*}{100 \textrm{GeV}}\right) 
\end{equation}
The peak frequency for turbulence is roughly three times larger than for sound waves, but according to these fits scales the same with the thermal parameters.

For turbulence to result in GWs, it must set in before a Hubble time. There are several reasons why turbulence is connected to the Hubble time. First, redshift by cosmic expansion damps turbulence. This means that the circulation time should be less than a Hubble time.
Second, only eddies with circulation times less than a Hubble time lead to an impact on GWs, because slower eddies are effectively constant~\cite{Niksa:2018ofa}. 
Lastly, and most importantly, as discussed in \cref{sec:lengthscale}, decorrelation and damping by redshift cause sources to have an effective lifetime of a Hubble time~\cite{Hindmarsh:2015qta,Hindmarsh:2017gnf}.
Turbulence may have set in before the end of the simulations in \refcite{Cutting:2019zws}, but not for earlier simulations of weak transitions (e.g.\ \refcite{Hindmarsh:2015qta, Hindmarsh:2017gnf}). However, none of these simulations run up to a Hubble time. Turbulence can be challenging to simulate, because it sets in relatively slowly in weak transitions, thus requiring longer simulations. Strong transitions, on the other hand, lead to hierarchies between the bubble wall thickness and the bubble radii that are challenging to simulate.

Because the plasma is fully ionized and because of the substantial magnetic Reynolds number of the plasma, turbulence must be treated along with the magnetic fields, and thus leads to magnetohydrodynamic (MHD) turbulence. So-called seed magnetic fields are amplified by the turbulent flow of charged particles, and energy is transferred between the fluid and the magnetic field until equipartition is reached. Although the GWs from magnetic fields may be subdominant compared to ordinary turbulence~\cite{Kosowsky:2001xp}, the energy decay of the turbulence may depend on whether primordial magnetic fields exist at the scale of the transition \cite{Kahniashvili:2012uj, Hindmarsh:2017gnf}.
See \refcite{Niksa:2018ofa, RoperPol:2019wvy, Brandenburg:2021tmp, Brandenburg:2021bvg} for simulations of MHD turbulence.

The efficiency factor, $\kappa_{\text{turb}}$, which gives the fraction of vacuum energy converted into turbulence, remains poorly understood and is arguably the greatest source of uncertainty in predictions for turbulence. Many studies take this to be a few percent of the sound wave efficiency, $\kappa_{\text{turb}} = \epsilon \kappa_{\text{sw}}$, where $\epsilon = 5\text{--}10\%$. This choice, however, is based on simulation results that do not probe the turbulence regime \cite{Caprini:2015zlo}. For instance, \refcite{Caprini:2015zlo} uses $\epsilon = 0.05$, \refcite{Alves:2018jsw, Azatov:2019png} use $\epsilon = 0.1$, and \refcite{FitzAxen:2018vdt} uses $\epsilon = 1$ for strong transitions. 
\Refcite{Ellis:2019oqb,Alanne:2019bsm}, on the other hand, suppose that the acoustic period breaks down after a time $\tau_{\text{sw}}$ but could last up to a Hubble time, $H_*^{-1}$ if not interrupted by turbulence~\cite{Hindmarsh:2017gnf}.  They thus estimate the fraction of energy available for turbulence by the fraction of the possible acoustic period disturbed by turbulence, $\epsilon = (1 - \min(H_* \tau_{\text{sw}}, 1))^{2/3}$, where the power appears because this suppression is really applied as a linear factor to the amplitude. This is always an upper bound, because energy may instead be converted to thermal energy. This may need to be further generalized to take account of the difference between \cref{Eq:sw_lifetime_min,Eq:sw_lifetime_upsilon}.

\subsection{Detectability}
\label{sec:Detectability}
The GWs from a phase transition would be a stochastic background in GW experiments --- being isotropic, stationary and unpolarized. Thus, the predicted signal $s$ would be difficult to distinguish from background noise, $b$, and possibly other irreducible sources of stochastic background and astrophysical foregrounds. Because the noise in two distant detectors should be uncorrelated, two or more detectors may be used to disentangle a signal from noise. Writing the detection at experiment $i$ as a function of time by  $S_i(t) = s_i(t) + b_i(t)$, the cross-correlation between the measurements at two detectors would be
\begin{equation}
S_{12} \equiv \int \dint t \dint t^\prime \, S_1(t) S_2(t^\prime) \, Q(t - t^\prime)
\end{equation}
for a filter $Q$, e.g.\ $Q = \delta(t - t^\prime)$, though other filters may be optimal. This correlation is expected to vanish in the absence of a signal and be positive otherwise, with the only contribution coming from genuine stochastic backgrounds at both detectors.  Thus, the cross-correlation may be used as a discriminator to detect the presence of signals in noisy data. On the other hand, the cross-correlation itself would be noisy, with some correlation seen between the detectors by chance even in the absence of a signal. Thus, the detectability of a signal scales as
\begin{equation}\label{eq:ratio}
\frac{\text{Expected $S_{12}$ if signal}}{\text{Noise in $S_{12}$ in absence of signal}} \equiv \frac{\text{E}[S_{12}]}{N} .
\end{equation}
This reasoning is used in the most common measure for detectability, the so-called signal-to-noise ratio (SNR) \cite{Allen:1997ad}
\begin{equation}
\text{SNR} \equiv \sqrt{\frac{\text{E}[S_{12}]}{N}} .
\end{equation}
The SNR is the square root of \cref{eq:ratio}, with the square root motivated by the fact that the numerator and denominator in \cref{eq:ratio} are quadratic in the signal already. Signals with $\text{SNR} \gtrsim 10$ are usually considered to be detectable~\cite{Caprini:2015zlo}, in that they are anticipated to lead to compelling evidence for a GW signal in a Bayesian framework or detection of a GW signal at an acceptable false alarm rate in a frequentist framework~\cite{Romano:2016dpx}.

The SNR can be calculated for a particular experiment and an expected GW spectrum through
\begin{equation}
\text{SNR} = \sqrt{\Tau \! \int \! \dint f \left(\frac{\Omega_\text{GW}(f)}{\Omega_\text{sens}(f)}\right)^{\!\!2}}.
\end{equation}
As is typical in statistics-limited settings, the detectability scales with the square root of the detection time, $\Tau$. The properties of the detectors are encoded in the function $\Omega_\text{sens}(f)$. The SNR measure takes into account the spectral shape and broadband nature of stochastic sources because it integrates over the whole spectrum. The noise energy density power spectrum,  $\Omega_\text{sens}(f)$, may be connected to the noise strain by
\begin{equation}
\Omega_\text{sens}(f) = \frac{4\pi^2}{3H_0^2} f^3 S_n(f) .
\end{equation}
The strain may be written in terms of the noise power spectral density, $P_n$,
\begin{equation}
S_n(f) = \frac{P_n(f)}{R(f)} ,
\end{equation}
where $R$ is the polarization- and sky-averaged response function of the detector. Explicit forms of the noise power spectral density and the detector response function can be published by experiments. For example, for the planned LISA experiment these functions may be approximately written as~\cite{Caprini:2019pxz}
\begin{align}
	P_n(f) &\simeq 16 \sin[2](\frac{2\pi fL}{c}) \left\{P_\text{OMS}(f) + \left[3 + \cos(\frac{4\pi fL}{c}) \right] P_\text{acc}(f) \right\} ,\\
	P_\text{oms}(f) & \simeq 9 \times 10^{-26} \, \text{s} \times \left(\frac{f}{\text{Hz}}\right)^{\!\!2} \left[1 + \left(\frac{2\times 10^{-3}\, \myunit{Hz}}{f} \right)^{\!\!4} \right], \\
	P_\text{acc}(f) & \simeq 3 \times 10^{-33} \, \text{s} \times \left(\frac{\text{Hz}}{f}\right)^{\!\!2}  \left[1 + \left(\frac{4\times 10^{-4} \,\myunit{Hz}}{f} \right)^{\!\!2} \right] \! \left[1 + \left(\frac{f}{8\times 10^{-3}\, \myunit{Hz}} \right)^{\!\!4} \right],\\
	R(f) &\simeq 16 \sin[2](\frac{2\pi fL}{c}) \! \left(\frac{3}{10 + 6(2\pi fL/c)^2}\right) \! \left(\frac{2\pi fL}{c} \right)^{\!\!2},
\end{align}
where $L = 2.5 \times 10^9\,\text{m}$ is the arm length for LISA and the noise was divided into the optical metrology system noise (oms) and single test mass acceleration noise (acc).

\begin{figure}[t]
  \centering
  \includegraphics[width=0.8\linewidth]{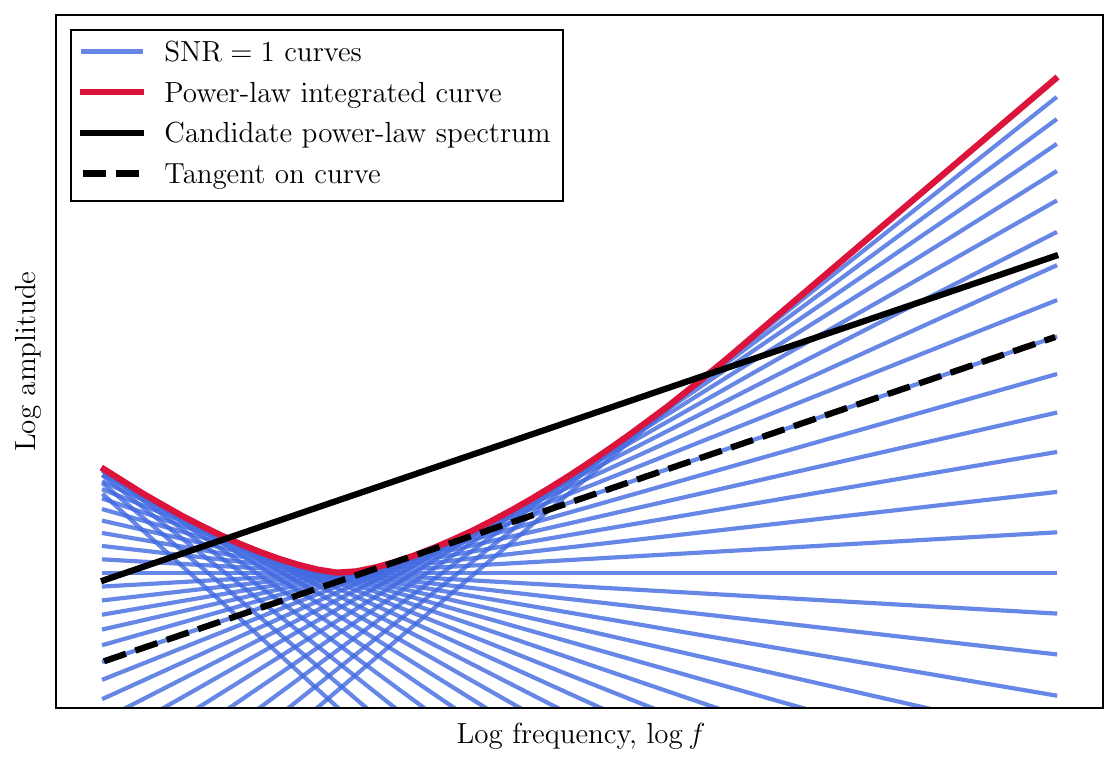}
  \caption{Illustration of a power-law integrated curve. The $\text{SNR} = 1$ lines (blue) for a specific experiment form a curve (red line). If a power-law GW spectrum lies above that curve (e.g., the black line), it must be detectable at $\text{SNR} > 1$ as it is above and tangent to an $\text{SNR} = 1$ curve (here, dashed black line).}
  \label{fig:power_curves}
\end{figure} 

Rather than computing the SNR, we can visualize detectability using a sensitivity curve or a power-law integrated curve~\cite{Thrane:2013oya}. The sensitivity curve shows whether a signal at a specific frequency would result in $\text{SNR} > 1$, indicating some sensitivity to a signal at that frequency. Power-law integrated curves, on the other hand, take into account the broadband nature of a power-law stochastic background. The shape of a power-law background is defined by the power $p$,
\begin{equation}
\Omega(f) \propto f^p .
\end{equation}
To understand the power-law integrated curve, first note that for each $p$ we may find the amplitude that would result in $\text{SNR} = 1$. The envelope of this collection of $\text{SNR} = 1$ lines forms the power-law integrated curve. To understand the utility of this curve, consider a power-law spectrum that intersects the power-law integrated curve somewhere in the spectrum. By the mean value theorem, it must be parallel to the power-law integrated curve at some point in the region in which it lies above it. Since it is parallel to and above a tangent on the power-law integrated curve, it must have $\text{SNR} > 1$, and so the experiment is sensitive to such a signal. This is shown in \cref{fig:power_curves}. This idea can be extended to other spectral shapes. Indeed, \refcite{Schmitz:2020syl} describes constructing similar curves based on a generic spectral shape typical of GWs from a FOPT.

\subsection{Predictions from physics beyond the Standard Model}\label{sec:bsm}
Although GW predictions from FOPTs in theories of beyond the standard
model (BSM) physics had been considered long before the first
observations of GWs by LIGO, this discovery led to an explosion of
papers making such predictions.  To avoid constraints from collider
physics many SM extensions introduce new physics above the EW
scale. Phase transitions that take place at scales between $100$ GeV
and $10^7$ GeV may be detectable by the second generation GW detectors
\cite{Grojean:2006bp}.\footnote{However, phase transitions from light
  new physics are also possible. We discuss the case of the axion in
  \cref{sec:axions}.}

Since it is very common for extensions of the Standard Model (SM) to
either introduce new symmetries that are broken or modify the
electroweak phase transition (EWPT), there is a huge array of models for which
GW signatures could be generated. For example, GW predictions have been
presented for many different models that modify the Higgs sector of
the SM. This includes the scalar singlet extensions of the SM
\cite{Ashoorioon:2009nf,Jinno:2015doa,Huang:2016cjm,Hashino:2016xoj,Balazs:2016tbi,Vaskonen:2016yiu,Kang:2017mkl,Huang:2018aja, Ellis:2018mja,Alves:2018oct,Hashino:2018wee,Beniwal:2017eik,Beniwal:2018hyi,Alves:2018jsw,Gould:2019qek,Alves:2019igs,Alanne:2019bsm,Zhou:2019uzq,Bian:2019bsn,Alves:2020bpi,Xie:2020wzn,Abdussalam:2020ssl,Liu:2021jyc,Chiang:2019oms,Zhou:2020ojf,Chen:2019ebq,Freitas:2021yng,Chao:2017vrq,Guo:2021qcq,Ellis:2022lft}, $O(N)$ scalar singlet extensions
\cite{Kakizaki:2015wua,Hashino:2016rvx,Hashino:2018wee}, two-Higgs
doublet models
\cite{Dorsch:2016nrg,Wang:2019pet,Zhou:2020irf,Goncalves:2021egx,Aoki:2021oez,Biekotter:2022kgf}
including the inert Higgs doublet variant \cite{Benincasa:2022elt} and
two-Higgs doublet plus singlet extensions
\cite{Morais:2019fnm,Han:2020ekm,Zhang:2021alu,Biekotter:2021ysx},
triplet Higgs extensions \cite{Chala:2018opy} including the
Georgi-Machacek (GM) model \cite{Bian:2019bsn}, the minimal scotogenic
model \cite{Borah:2020wut}, the Zee-Babu model \cite{Phong:2021lea}, scalar leptoquark extensions \cite{Fu:2022eun} 
and the minimal and next-to-minimal supersymmetric extensions of the
SM~\cite{Apreda:2001us,Apreda:2001tj,Huber:2007vva,Leitao:2012tx,Kozaczuk:2014kva,
  Huber:2015znp,Bian:2017wfv,Chatterjee:2022pxf} including also split SUSY versions
\cite{Demidov:2017lzf,Fornal:2021ovz}, additional right-handed neutrinos \cite{Borah:2023zsb} and an extension with extra
triplets \cite{Garcia-Pepin:2016hvs}. In addition to specific models,
a more model-independent approach has sometimes been taken by adding
higher dimensional operators to the SM and treating it as an EFT
\cite{Huber:2007vva,Delaunay:2007wb,Leitao:2015fmj,Kobakhidze:2015xlz,Huang:2016odd,Kobakhidze:2016mch,Cai:2017tmh,Kobakhidze:2017mru,Chala:2018ari, Ellis:2018mja,Ellis:2019oqb,Wang:2020jrd,Ellis:2020nnr,Croon:2020cgk,Lewicki:2021pgr, Di:2020ivg, Hashino:2022ghd}. Other models extend the symmetries of the SM, giving rise to GW
signatures from the breaking of the new symmetry.  Examples of this
include gauged $U(1)_{B-L}$ extensions of the SM
\cite{Jinno:2016knw,Chao:2017ilw,Okada:2018xdh,Marzo:2018nov,Ellis:2019oqb,Bian:2019szo,Hasegawa:2019amx,Ellis:2020nnr}
including supersymmetric variants \cite{Haba:2019qol,Dong:2021cxn},
Majoron models
\cite{Addazi:2017oge,Imtiaz:2018dfn,Addazi:2019dqt,DiBari:2021dri},
left-right symmetric models
\cite{Li:2020eun,Brdar:2019fur,Graf:2021xku}, Pati-Salam extensions
\cite{Croon:2018kqn,Huang:2020bbe}, a $U(1)_X$ extension from the
breaking of an $SO(10)$ GUT \cite{Okada:2020vvb}, gauged lepton number
\cite{Madge:2018gfl}, 331 models \cite{Huang:2017laj}, variants of
technicolor \cite{Jarvinen:2009mh,Chen:2017cyc,Miura:2018dsy},
composite Higgs models \cite{Aziz:2013fga,Chala:2016ykx,Bruggisser:2018mrt,Bian:2019kmg,Xie:2020bkl},
mirror Higgs parity \cite{Dunsky:2019upk}, a radion phase transition
in the theories with warped extra dimensions
\cite{Randall:2006py,Megias:2018sxv}, $R$-symmetry breaking in models of
SUSY breaking \cite{Craig:2020jfv}, dark photon models
\cite{Zhou:2021cfu,Borah:2021ocu}, other dark sector models
\cite{Schwaller:2015tja,Addazi:2016fbj,Arunasalam:2017ajm,Addazi:2017gpt,Baldes:2018emh,Schwaller:2015tja,Jaeckel:2016jlh,Aoki:2017aws,Aoki:2019mlt,Halverson:2020xpg,Fairbairn:2019xog,Huang:2020crf,Kang:2021epo,Reichert:2021cvs,Bigazzi:2020avc,Kierkla:2022odc} with some steps towards model discrimination \cite{Croon:2018erz}
and a model of asymmetric dark matter \cite{Baldes:2017rcu}.  There
has also been a recent attempt to formulate a model-independent
description of radiatively broken symmetries with strong supercooling
\cite{Salvio:2023qgb}.

The methods used vary a lot because the community interested in GW
signatures from BSM physics is growing rapidly, and the community is gradually maturing
as new entrants become more familiar with the topic in the wake
of the LIGO discovery.  Treatments of the GW signal have varied from
simply plotting the predicted power spectrum versus frequency and
comparing it to sensitivity curves, to computing the signal-to-noise
ratio as described in \cref{sec:Detectability} and to using actual
LIGO data to extract constraints \cite{Romero:2021kby}.  Similarly, the
predicted signals have varied in rigor and sophistication for the
calculation of the effective potential and phase transition properties
that enter fit formulae for the GW spectra.  For example, the thermal
parameters vary from just using the critical temperature, or heuristic-based estimates of the nucleation temperature for the transition
temperature, to detailed tracking of the false vacuum fraction to
obtain the percolation temperature
\cite{Leitao:2015fmj,Cai:2017tmh,Kobakhidze:2017mru,Ellis:2018mja,Beniwal:2018hyi,Ellis:2019oqb,Wang:2020jrd,Ellis:2020nnr,Freitas:2021yng,Kierkla:2022odc}.   On
the effective potential side, analyses vary from just including
tree-level plus leading-order terms from the high-temperature
expansion, to treatments constructing a three-dimensional effective
field theory (3dEFT) and even then using non-perturbative lattice Monte Carlo
simulations of the phase transition
\cite{Kainulainen:2019kyp,Niemi:2020hto,Gould:2021dzl}. Finally the
effects from the finite duration of the sound wave source have also
been taken into account in recent papers, for example
\refcite{Ellis:2019oqb,Alanne:2019bsm,Romero:2021kby,Arcadi:2022lpp,Ellis:2022lft,Ghosh:2022fzp,Liu:2023sey,Borah:2023zsb}.

The literature is already too vast to review and summarize thoroughly,
and the actual predictions are very sensitive to the specific
treatment used.  Detailed comparisons of the impact of various effects
are often lacking though some uncertainties have been examined
\cite{Croon:2020cgk}. Instead in the following we briefly survey three 
popular and representative classes of models.

\subsubsection{Scalar singlet extensions}

We first consider scalar singlet extensions of the SM, which are by far the most studied class of models for gravitational wave signatures, and arguably the simplest extensions of the SM. In these models, the EWPT is modified by extending the SM Higgs sector by a single real or complex scalar singlet. For example,
\begin{equation}
V = V_\text{SM} + m_S^2 S^2 + \kappa_S S^3 + \lambda_S S^4 + \kappa_{HS} S |H|^2 + \lambda_{HS} S^2 |H|^2.
\end{equation}
A discrete $\mathbb{Z}_2$ symmetry is often imposed, eliminating the cubic terms. The interaction between the Higgs and the singlet impacts the EWPT, potentially making it strongly first order~\cite{Profumo:2007wc,Espinosa:2011ax,Profumo:2014opa}, and causes the Higgs and singlet to mix after EWSB. The mixing angle is constrained by LHC measurements indicating that the observed Higgs is SM-like.

This simple setup with just one real scalar singlet may yield observable GWs because the singlet may undergo or assist a strong FOPT~\cite{Ashoorioon:2009nf,Jinno:2015doa,Huang:2016cjm,Hashino:2016xoj,Balazs:2016tbi,Vaskonen:2016yiu,Kang:2017mkl,Ellis:2018mja,Huang:2018aja,Alves:2018oct,Hashino:2018wee,Beniwal:2018hyi,Alves:2018jsw,Gould:2019qek,Alves:2019igs,Alanne:2019bsm,Zhou:2019uzq,Bian:2019bsn,Alves:2020bpi,Xie:2020wzn,Liu:2021jyc,Guo:2021qcq,Ellis:2022lft}. Similarly, the potential GW phenomenology in the SM Higgs sector by a complex scalar singlet has been considered~\cite{Chiang:2019oms,Zhou:2020ojf,Chen:2019ebq,Chen:2020wvu,Freitas:2021yng}, while \refcite{Chao:2017vrq} considers two real singlets where one is scalar singlet dark matter and the other mixes with the SM Higgs. The GW signals could be a unique window to study this model as the singlet could be hard to detect at the LHC~\cite{Barger:2007im,Curtin:2014jma,Costa:2015llh,Han:2016gyy}.  On the other hand the physics could instead be connected to BSM phenomenology (see e.g., \refcite{GAMBIT:2017gge} for constraints), including baryogenesis~\cite{Vaskonen:2016yiu}, dark matter~\cite{McDonald:1993ex,Cline:2013gha}, vacuum stability~\cite{Balazs:2016tbi} and leave imprints at future muon colliders~\cite{Liu:2021jyc}.

Even such a simple extension has quite a rich set of possible
scenarios and possible cosmological phase histories to consider. For
example, even restricting only to a single real scalar singlet, the
model can be studied with and without $\mathbb{Z}_2$ or $\mathbb{Z}_3$ symmetries, and
has scenarios with a single step phase transition that breaks
electroweak symmetry and generates a VEV in the singlet direction at
the same time and two-step transitions where the singlet develops a
VEV $v_s$ first, and then in a subsequent step electroweak symmetry is
broken while $v_s=0$ is restored.  Furthermore, it can admit FOPTs
at scales far above the electroweak scale
(e.g.\ $10^7\gev$) as well as close to the electroweak scale.

Many of the different approaches we have discussed throughout the
review are implemented in the literature for this model, and there are
almost as many approaches as there are papers.  For example, the
effective potential has been calculated simply using a
high-temperature
expansion~\cite{Vaskonen:2016yiu,Huang:2018aja,Zhou:2019uzq,Liu:2021jyc,Alves:2018jsw,Alves:2018jsw,Bian:2019bsn},
using the full one-loop finite-temperature potential
\cite{Alanne:2019bsm} alone, or augmented by daisy resummation using
the Parwani
approach~\cite{Ashoorioon:2009nf,Beniwal:2017eik,Beniwal:2018hyi} and
Arnold-Espinosa
approach~\cite{Xie:2020wzn,Jinno:2015doa,Hashino:2016xoj,Balazs:2016tbi},
or lastly by matching to a 3dEFT~\cite{Gould:2019qek}. For the thermal
parameters, while there are exceptions using the percolation
temperature already mentioned earlier
\cite{Beniwal:2018hyi,Freitas:2021yng,Ellis:2022lft}, it is very
common to use the nucleation temperature as the transition temperature,
often calculated through simple heuristics such as
requiring $S(T_n)/T_n = 140$. Similarly the $\alpha$ parameter is most
often calculated using the latent heat definition given in
\cref{eq:latentHeat}, but a significant number use the expression
derived from the trace anomaly
\cite{Vaskonen:2016yiu,Alanne:2019bsm,Chao:2017vrq,
  Freitas:2021yng,Ellis:2022lft} (see
\cref{eq:alpha-traceAnomaly-common}).  The sources for gravitational
wave predictions also vary. Some consider runaway bubble walls and use
the collision contribution, while others consider non-runaway bubble
walls and include sound waves and turbulence as sources, or just sound
waves. Others consider all three contributions, with various estimates
for the $\kappa_i$ efficiency factors.  For testing the gravitational
wave predictions, there are many papers using the signal-to-noise
ratio when estimating the detectability of signatures at future
gravitational wave experiments, though there have also been papers
that just compare GW signals directly to sensitivity curves.  Finally, 
we note that \refcite{Guo:2021qcq} considers a variety of approaches,
grouping them under low, moderate and high diligence.
Comparing the GW predictions of low and moderate diligence approaches to that of the high diligence approach provides a rough estimate of the uncertainty involved in simplified treatments. However, many improvements are made going from moderate diligence to high diligence, so we currently lack an understanding of the individual impact of many of the approximations and improvements.

\subsubsection{Gauged $B-L$ extensions}

Instead of modifying the electroweak symmetry, BSM theories may
also extend the SM with new symmetries that can be broken to
generate GWs. The most popular example of this seems to be
$U(1)_{B-L}$ extensions, which gauge the quantum number $B-L$.  This
symmetry is motivated by the fact that it is a simple way to create an
anomaly-free gauge extension of the SM, and because it is occurs in left-right symmetric
models and in Pati-Salam grand unification models,  where the
$U(1)_{B-L}$ symmetry can survive to the TeV scale.

For example, in the classically conformal $B - L$ model introduced in
\refcite{Iso:2009ss,Iso:2009nw,Oda:2017kwl} the quarks and leptons
carry $B - L$ charges $1/3$ and $-1$, respectively. The SM is
extended by a right-handed neutrino, required by anomaly cancellation,
and by a complex scalar SM singlet $\Phi$ with Yukawa couplings
generating a Majorana mass for the right-handed neutrino and thus
carrying $B - L = 2$.  The tree-level scalar potential contains no
dimensional parameters,
\begin{equation}
V = \lambda_H |H|^4 + \lambda_{\Phi} |\Phi|^4 + \lambda_{\Phi H} |\Phi|^2 |H|^2,
\end{equation}
as a classical conformal symmetry is assumed. Of course, this symmetry
is broken at the quantum level because the Higgs and SM singlet fields
obtain VEVs through the Coleman-Weinberg mechanism. This breaks EW and
$U(1)_{B-L}$ symmetries and generates masses for ordinary SM particles
and the right-handed neutrinos. As in scalar singlet extensions, the
Higgs and SM fields mix. Unlike that model though, there is kinetic
mixing between the SM and $B-L$ neutral bosons. In addition to
providing a simple way to create an anomaly-free $U(1)$ extension, this
setup is also motivated because the right-handed neutrinos may play a
role in leptogenesis.

GWs from $B-L$ extensions of the SM were studied in
\refcite{Jinno:2016knw,Chao:2017ilw,Okada:2018xdh,Marzo:2018nov,Ellis:2019oqb,Bian:2019szo,Hasegawa:2019amx,Haba:2019qol,Dong:2021cxn}. GW
spectra have been predicted for the classical conformally invariant
model described above, in scenarios where the $U(1)_{B-L}$ is broken at
the TeV scale \cite{Jinno:2016knw,Marzo:2018nov}.  For breaking at higher scales, predictions for this minimal
version \cite{Hasegawa:2019amx}  and also for a non-minimal \cite{Okada:2018xdh} version have been made. \Refcite{Chao:2017ilw} found that when
the minimal $U(1)_{B-L}$ model is broken at the TeV scale it gives GW
signatures that are too weak, but that with an extended Higgs sector
they can be detectable. GWs from supersymmetric
extensions of this model have also been considered in
\refcite{Haba:2019qol,Dong:2021cxn}.

\subsubsection{Axions and axion-like particles}
\label{sec:axions}
Finally we consider something quite different to the kind of SM extensions we have focused on in this review. Specifically we consider axions and axion-like particles (ALPs) --- pseudo-scalar particles that could be connected to solutions to the strong CP problem~\cite{Peccei:1977hh,Weinberg:1977ma,Wilczek:1977pj,Kim:1979if} and dark matter~\cite{Preskill:1982cy,Abbott:1982af,Dine:1982ah}, and may arise naturally in fundamental theories including string theory~\cite{Svrcek:2006yi,Arvanitaki:2009fg,Cicoli:2012sz}. We start from the QCD axion. In QCD a topological term that breaks CP is permitted,
\begin{equation}
{\cal L}^{ \text {QCD}} _{\theta}  = \frac{\alpha_S}{32\pi} \theta_\text{QCD} G_{\mu\nu} \tilde G^{\mu\nu} ,
\label{Eq:theta}
\end{equation}
where $\alpha_S$ is the strong fine-structure constant, $G_{\mu\nu}$ is the gluon field strength tensor and $\tilde G^{\mu\nu}=\epsilon^{\mu\nu\alpha\beta}G_{\alpha\beta}/2$ is its dual. Although \cref{Eq:theta} is a total derivative, the parameter $\theta_\text{QCD}$ impacts observables including dipole moments through non-perturbative effects. From experimental measurements we thus know that $\theta_\text{QCD} \lesssim 10^{-10}$~\cite{Baker:2006ts}. The Peccei-Quinn (PQ) mechanism explains why it is so small~\cite{Peccei:1977hh}. With the addition of a pseudo-scalar axion, the low-energy Lagrangian becomes
\begin{equation}
{\cal L}^{ \text {QCD}} _{\theta}  = \frac{\alpha_S}{32\pi} \left(\frac{a}{f_a} + \theta_\text{QCD}\right) G_{\mu\nu} \tilde G^{\mu\nu},
\end{equation}
where $f_a$ is the axion decay constant. The QCD dynamics result in a VEV for the axion field that satisfies
\begin{equation}
 \left\langle\frac{a}{f_a} + \theta_\text{QCD}\right\rangle =0,
\end{equation}
thus dynamically setting an effective $\theta_\text{QCD}$ parameter to zero. The axion may be embedded in a UV theory if, for example, it is a pseudo-Goldstone boson associated with the spontaneous breakdown of a $U(1)_\text{PQ}$ global symmetry. There are two popular realizations of this idea in which the PQ symmetry is broken by a complex scalar that obtains a VEV: Dine-Fischler-Srednicki-Zhitnitsky (DFSZ; \cite{Zhitnitsky:1980tq,Dine:1981rt}) and Kim-Shifman-Vainshtein-Zakharov (KSVZ;~\cite{Kim:1979if,Shifman:1979if}). On the other hand, many different theories of BSM physics lead to similar pseudo-scalar fields in the low-energy effective theory. These more general pseudo-scalars are usually called ALPs and might not be connected to the strong CP problem. Various types of axions and ALPs have been discussed in particle physics and cosmology~(see recent examples in \refcite{Kim:2008hd,Sikivie:2009qn,Marsh:2015xka,Ballesteros:2016euj,DiLuzio:2020wdo,Choi:2020rgn} and references therein).

If the axion results from PQ symmetry breaking, the transition could result in GWs.  Although in weakly coupled theories the PQ transition is typically second order, FOPTs often occur in strongly coupled theories. To produce a detectable signal, the PQ transition should take place after inflation. If the PQ symmetry was ever restored after inflation, axion strings could produce an additional contribution to the stochastic GW background~\cite{Gorghetto:2021fsn}. A dark axion can also give rise to GW signals \cite{Bigazzi:2020avc}. The phase transition could be supercooled in the QCD axion model~\cite{VonHarling:2019rgb}, which would lower the transition temperature and thus enhance the GWs signals~\cite{DelleRose:2019pgi}. The GW signals resulting from a PQ or $U(1)$ breaking phase transition may in fact already lie within the sensitivity of LIGO and could potentially be probed by future GW experiments~\cite{Dev:2019njv,Ghoshal:2020vud,Ringwald:2020vei}. 

\section{Closing remarks}\label{sec:summary}

Gravitational waves (GWs) present us with a very exciting opportunity
to learn about fundamental physics. The techniques to go from a
particle physics model, described by a Lagrangian, to predictions for
observable GWs from vacuum decays that induce first-order phase
transitions (FOPTs) are still being regularly improved. The chain of
steps involves building an effective potential, identifying the
relevant first-order phase transitions, determining the properties of
the phase transitions, tracking the bubble dynamics and the progress
of the transitions, and lastly understanding the hydrodynamic
effects that lead to a stochastic background of GWs. In this review we
have presented a pedagogical and up-to-date summary of the methods for
doing this, with a focus on perturbative methods that are familiar to
particle physicists, and attempted to point out the limitations and
subtleties along the way. We focus on physics beyond the Standard
Model because we are interested in first-order phase transitions.  We
see many open questions and sources of uncertainty in GW predictions
and hope that our review serves as a starting point for further
developments that would allow us to fully exploit GW observations in
the future. In particular, there are several areas where further
research is needed to improve our understanding and predictions of
GWs.

Cosmological phase transitions that can give rise to GWs
are transitions between minima of the effective
potential. Calculations of the effective potential are therefore
central to the predictions of the phase transition properties and any
GW signals that may be induced.  However, significant
challenges are posed by the gauge dependence and
inclusion of finite-temperature effects.  While the effective
potential {\it should} be gauge dependent and is not an observable,
this can be potentially problematic for calculations of phase
transitions and requires careful treatment. Methods for handling the
gauge dependence exist, but combining them with precision calculations
of the effective potential can pose problems and there remains some
dispute within the community as to the right approach for handling
this.

One can compute finite-temperature corrections perturbatively, but they
transform the phase structure over the course of the phase transition
provoking questions of whether they are really perturbative.  More
precisely, the validity of the perturbative treatment can be spoiled
when the temperature is much bigger than the field. As we have reviewed,
daisy resummation and three-dimensional effective field theories (3dEFTs)
provide methods for resumming these corrections.  Recent studies
suggest that the 3dEFT approach
(sometimes called dimensional reduction) may provide more precise
predictions for the GW spectra and this method can
also simplify the application of certain gauge-independent
treatments. Nonetheless, further understanding of the uncertainties
and when this more complicated approach is really necessary would help
significantly. Computational tools have very recently been made
available for automating the 3dEFT
approach, which may significantly reduce the computational cost and
make this approach more accessible in studies of physics beyond the
Standard Model (SM).

Once the shape of the effective potential is understood and the phases
and possible FOPTs have been mapped out, one needs
to determine the decay rate of the false vacuum for the relevant
transitions and check if the possible phase transition actually
completes.  Whilst the path-deformation technique for computing the
false vacuum decay rate is well known, there remain outstanding
theoretical puzzles about why it works. More practically, computing
functional determinants remains challenging in most models and the
bounce equation can be hard and slow to solve reliably, especially for
multiple coupled scalar fields. That said, there are established
techniques in publicly available software packages for this.

With the vacuum decay rate at hand, one can then analyze the phase transition by tracking the evolution of the false vacuum fraction. This allows one to determine when milestone events (such as bubble percolation) occur, and also determine thermal parameters relevant for GW predictions. Approximations and assumptions creep into analyses of phase transitions in many places. For instance, one may use a simplified treatment of the equation of state (e.g.~the bag model), neglect the expansion of space, or neglect reheating. Furthermore,  the onset of bubble nucleation is frequently used as the time of GW production and is used to indicate completion of the phase transition. Many of these approximations break down in strongly supercooled phase transitions. Thus, predictions of strongly supercooled phase transitions are highly uncertain in simple analyses. The onset of percolation is the recommended milestone in these cases, yet simulations have not yet directly confirmed how appropriate it is.

Bubble percolation and completion of the phase transition are important milestones in a FOPT. The latter milestone is desired for a phase transition that leads to our current Universe. The percolation of bubbles necessarily requires them to have collided, suggesting that GWs are produced around the time of percolation. However, the condition for bubble percolation is derived from simulations in static spacetime. One can extend the usual percolation analysis to expanding space by checking that the physical volume of the false vacuum decreases over time. Yet such a determination does not give an updated estimate of when percolation or completion occurs in expanding space. Simultaneous phase transitions complicate matters further. In such cases, it is no longer enough to consider a false and true vacuum; one must consider a larger set of phases and model the time evolution of the fractions of the Universe encompassed by each phase.

The thermal parameters relevant for GW predictions can be determined once the FOPT has been analyzed. The terminal bubble wall velocity, $v_w$, is the most difficult thermal parameter to determine. Non-lattice studies of FOPTs (i.e.\ those that treat bubble nucleation and growth statistically) often treat $v_w$ as a free parameter or set it by hand, whereas hydrodynamic (lattice) simulations assume a constant $v_w$ set by a phenomenological friction parameter. Reheating that occurs when bubbles do not expand as supersonic detonations can induce significant time dependence in $v_w$, but reheating is usually ignored in non-lattice studies. This could pose issues for interpreting the results of hydrodynamic simulations because the GW fits are stated in terms of a constant $v_w$. Thus, there is currently an ambiguity in how to use the existing GW fits in cases where reheating effects would have a significant impact on the GW signal.

Recently, the assumption of local thermal equilibrium has offered the possibility of a much simpler and cheaper method for estimating $v_w$. This method is accessible to non-lattice studies that use a hydrodynamic treatment for bubble expansion. However, one must assume that bubbles reach their asymptotic fluid profiles, and recent hydrodynamic simulations suggest this may not happen before GWs are produced. The validity and precision of predictions in local thermal equilibrium are still being investigated. Additionally, more rigorous treatments of out-of-equilibrium effects on friction are being developed every year, offering important updates to predictions of the runaway regime.

In addition to the thermal parameters, there are several timescales that affect both the amplitude and peak frequency of the GW signal. The timescales used in GW fits are only roughly estimated using combinations of characteristic length scales and relevant velocities. Additionally, the transition from laminar flow (assumed in the sound shell model) to the turbulent regime is not well understood. Thus, the end time of the sound wave source and the start time of the turbulence source are largely unknown. Turbulence in the plasma could have a significant impact on the total GW signal but remains difficult to probe in hydrodynamic simulations.

We now we turn to open issues in hydrodynamic simulations. Strong FOPTs are expected to generate a stronger GW signal than weak FOPTs. However, the predicted strength from particle physics models can exceed the largest strengths probed by hydrodynamic simulations. In this case, the GW fits must be extrapolated. Extrapolation is fraught with danger because new phenomena become apparent or more significant as the transition strength is increased. Beyond the gap in hydrodynamic simulations for strong FOPTs, there is also a lack of data for slow phase transitions where the expansion of space becomes significant, and for phase transitions with ultra-relativistic bubble walls.

Another limitation of current hydrodynamic simulations is the simplified treatment of the effective potential. The equation of state is taken to be that of the bag model rather than derived from a realistic effective potential, the temperature of the background plasma is held constant, and bubbles are often assumed to nucleate simultaneously. The latter limitation is lifted in a recent Higgsless simulation, where a more realistic exponential nucleation scenario is considered. The background temperature of the plasma should decrease over time as the Universe expands, which becomes especially relevant in slow phase transitions. If the bubble nucleation rate is suppressed and reaches a maximum at some finite temperature, simultaneous and even exponential nucleation models are no longer appropriate. A useful investigation bridging the gap between lattice and non-lattice studies would be to include an analysis of the false vacuum fraction and the corresponding derived quantities alongside a hydrodynamic simulation. Then one could explore the time --- and equivalently temperature --- dependence of thermal parameters rather than evaluating them at a single `transition' temperature. Such an analysis could also inform a more appropriate choice of the transition temperature.

In addition to using the bag equation of state, hydrodynamic simulations effectively take the terminal bubble wall velocity $v_w$ and the transition strength $\alpha$ as input. Applying fits derived from the bag model can introduce errors in that the hydrodynamics could be starkly different in a different equation of state. The bubble expansion mode may be different, which could affect reheating around the bubble and the consequent evolution of the phase transition. A given combination of $\alpha$ and $v_w$ may not be realizable in both equations of state. Although the thermal parameters are sometimes viewed as independent quantities, they are in fact interconnected. In a realistic nucleation scenario, the characteristic length scales will be affected by the expansion of bubbles, which is in turn affected by the equation of state through the corresponding hydrodynamics. Thus, the GW fits from hydrodynamic simulations using the bag model may not be directly applicable to a particle physics model with an equation of state that deviates significantly from the bag model.

While we have highlighted many open issues and many limitations of current methods, we note that our understanding of GWs from FOPTs improves every year. Some current issues stem from computational limitations, while others can be addressed by further theoretical analyses. We hope this summary serves to focus future research efforts in the prediction of GWs from FOPTs. Additionally, we hope the review as a whole can provide a good starting point for a deeper understanding of many concepts relevant for the analysis of FOPTs and the prediction of GW signals.

At the time of writing, we are almost a decade on from the
revolutionary discovery of GWs. By now our collective
mastery of sophisticated experiments, data analysis, relevant
astrophysical processes and theory means that GW
detections are a regular occurrence. This is the era of ground-based
detectors. Whilst this is extraordinary, the best may be yet to
come. The era of space-based detectors, hopefully commencing in the
2030s, might enable us to see signatures of new physics that would
otherwise be completely inaccessible, because future space-based detectors could be able to detect signals from FOPTs.
If, as we hope, a stochastic GW background is discovered in the future, it cannot emanate from known SM physics, and the theoretical techniques surveyed here would be essential for uncovering what lies beyond the SM.

\section*{Acknowledgments}

We thank our collaborators Graham White and Yang Zhang for many
helpful discussions and insights gained during projects related to
this topic. PA and CB also thank Tom\'as E. Gonzalo and Matthew Pearce
for related discussion while working on another gravitational waves
project.  AF thanks Oliver Gould and Joonas Hirvonen for invaluable
discussions and insights on transition rates. LM would like to thank
Ariel M\'egevand for useful discussions about the energy density of
the Universe, Mark Hindmarsh for clarifying the identification of
thermal energy density in the energy budget, and Eric Thrane for a
discussion about the detectability of gravitational waves.
The work of PA, AF and
LW are supported by the National Natural Science Foundation of China
(NNSFC) under grants No.~11950410509, 12150610460 and 12275134. PA was
also supported by the ARC Future fellowship FT160100274 and Discovery
project DP180102209 grants in the early parts of this project. LM is
supported by an Australian Government Research Training Program (RTP)
Scholarship and a Monash Graduate Excellence Scholarship (MGES).  The 
work of CB is also supported by the Australian Research Council 
Discovery Project grant DP210101636. 

\appendix

\section{Alternative derivation of the JMAK equation}
\label{sec:altJMAK}
\newcommand{\given}{\,|\,}
\newcommand{\total}{\mathcal{V}_T}

Here we provide an alternative derivation of the JMAK equation to that in \cref{sec:jmak}. This derivation uses a statistical model such that the assumptions and statistical process of nucleation are explicit. We begin by assuming that at time $t$ there are $n$ bubbles with volumes $V_1, V_2, \ldots V_n$, and that the bubbles are independently and uniformly distributed within the total volume $\total$. Ignoring boundary effects, regardless of our location, the probability of being inside the false vacuum at time $t$ is
\begin{equation}\label{eq:cond}
	P_f(t \given V_1, V_2, \ldots V_n) = \prod_{i=1}^n (1 - V_i / \total)
\end{equation}
That is, it is the probability of not being enveloped by any bubble of true vacuum. Note that since we assume that the bubbles are independently distributed this includes the possibility of bubbles overlapping or one being entirely inside another. Thus \cref{eq:cond} differs from the result that one would obtain if the bubbles were arranged such that none of them overlapped, $1 -\sum_i V_i/\total$. 
We don't know the number of bubbles or their volumes at time $t$, though. 

Let us model the bubble nucleation process by assuming that the total number of bubbles at time $t$, follows a Poisson distribution with mean $N$, and that the volumes at time $t$ independently follow the distribution $f(V)$. Lastly, we assume that the times at which bubbles are born independently follow the distribution $g(t)$. We don't, however, assume independence between bubble volume and the time at which a bubble was born. The Poisson distribution is a natural choice for the number of bubbles nucleated as we assume bubbles are nucleated one at a time, independently and that previously nucleated bubbles don't impact the rate of nucleation.\footnote{Note that the rate of events needn't be constant; denote it by $\lambda(t)$. These conditions define an inhomogeneous Poisson process in which the number of events at time $t$ follows a Poisson with mean $\int \lambda(t) \,\dint t$.}

Since we are only interested in the probability of not being enveloped by a bubble of the true vacuum (or equivalently the false vacuum fraction), we don't need to know the individual bubble volumes or the number of them. Therefore we can average over these, weighting them by their probability distributions, the Poisson distribution and $f(V)$. More formally, in the language of statistics, this means we can now marginalize the unknown number of bubbles and their volumes in \cref{eq:cond},
\begin{equation}
  P_f(t) = \sum_{n=0}^\infty \text{Po}(n | N) \left\{ \int \left[\prod_{i=1}^n \dint V_i \, f(V_i)\right] P_f(t \given V_1, V_2, \ldots V_n)\right\}
\end{equation}
This is an application of the law of total probability that $p(x) = \int p(x \given y) p(y) \, \dint y$. The result now becomes a mathematical exercise and follows quickly,
\begin{align}
	P_f(t) ={}& \sum_{n=0}^\infty \text{Po}(n | N)  \prod_{i=1}^n \int_0^\infty (1 - V_i / \total) f(V_i) \, \dint V_i\\
	={}& \sum_{n=0}^\infty \frac{e^{-N} N^n}{n!} \left(1 - \int_0^\infty V / \total f(V) \, \dint V\right)^n\\
	={}& e^{-N / \total \int_0^\infty V f(V) \, \dint V} \\
  ={}& e^{-N / \total \langle V \rangle} 
\end{align}
where in the second line we relabel the integration variables $V_i$, making it transparent that we just have $n$ factors of the same integral and also use $\int_0^\infty f(V) dV = 1$, i.e.\ the fact that the probabilities sum to unity.  In the third line we use the fact that $e^{\lambda} = \sum_{n=0}^\infty \lambda^n / n!$. Finally in the last line we define $\langle V \rangle \equiv \int_0^\infty f(V) V \dint V$ as the expected volume of a bubble at time $t$. This makes it clear that the extended volume in the usual JMAK result is nothing but the expected number of bubbles multiplied by the expected volume of a bubble and normalized to the total volume. The result works even if $N$ and $\total$ both diverge so long as the density of bubbles $N / \total$ remains finite. 

We may re-write our result in the usual form of the JMAK equation. The average volume at time $t$ may itself be written as an average over the average volumes of bubbles born before $t$,
\begin{equation}
\langle V \rangle = \int_0^t \langle V(t^\prime) \rangle \, g(t^\prime) \, \dint t^\prime
\end{equation}
where $g(t^\prime)$ is the probability density that a bubble was born at time $t^\prime$, 
and $\langle V(t^\prime) \rangle$ is the expected volume at time $t$ of a bubble born at time $t^\prime$. Thus we may write
\begin{equation}
\frac{N \langle V \rangle}{\total} = \int_0^t \langle V(t^\prime) \rangle \,  \frac{N g(t^\prime)}{\total} \, \dint t^\prime
\end{equation}
and identify the factors $V(t^\prime) \equiv \langle V(t^\prime) \rangle$ as the expected volume at time $t$ of a bubble born at time $t^\prime$, and $\Gamma(t^\prime) \equiv N g(t^\prime) / \total$ as the expected rate of bubble nucleation at time $t^\prime$ per unit volume.

Having developed this machinery, we may identify an even quicker proof of the JMAK result. Consider a point in the false vacuum. We assume that at any moment zero or one bubbles could be born that would later envelop that point. The probability of a point being enveloped by time $t$ by a bubble of volume $V$ that was born at between $t^\prime$ and $t^\prime + \dint t$ would be
\begin{equation}
\dint\lambda(t^\prime) =  \frac{V}{\total} \, \total \Gamma(t^\prime) \, \dint t^\prime,
\end{equation}
that is, the probability that the bubble envelops the point multiplied by the probability a bubble was born anywhere at time $t$. The factor $\total \Gamma(t) = \int \Gamma(t) \, \dint V$ integrates the probability of a bubble per unit time per unit volume with respect to volume to obtain the probability per unit time.
Averaging over the unknown bubble volume and simplifying, we obtain
\begin{equation}
\dint\lambda(t^\prime) =  \langle V(t^\prime) \rangle \Gamma(t^\prime) \, \dint t^\prime,
\end{equation}
where $\langle V(t^\prime) \rangle$ is the average volume at time $t$ of a bubble born at $t^\prime$. We may now identify this as an inhomogeneous Poisson process with time-dependent rate $\dint \lambda(t^\prime) / \dint t^\prime$. The number of events by time $t$ follows a Poisson distribution with mean $\lambda = \int_0^t \dint \lambda(t^\prime) / \dint t^\prime \, \dint t^\prime$. Thus the probability that a point is not enveloped by any bubbles by time $t$ (that is, zero events by time $t$) is simply,
\begin{equation}
P_f(t) = e^{-\lambda} = e^{-\int_0^t \langle V(t^\prime) \rangle \Gamma(t^\prime) \, \dint t^\prime}
\end{equation}
This again verifies the JMAK equation, lets us interpret the exponent as the expected number of bubbles that envelope a point and readily generalizes to the probability that a point is enveloped by one or more bubbles.

\section{Ultraviolet-divergent sum over Matsubara frequencies}\label{sec:FTVStrick}

To handle the infinite sum in \cref{Eq:1L_Scalar_T}, we use
the following trick from \refcite{Dolan:1973qd}.  We define
\begin{align}
    \label{Eq:v_omega}
    v(\omega)=\sum_{n=-\infty}^{\infty} \log(\omega_n^2+\omega^2),
\end{align}
where for scalars we have $\omega_n = n\pi/\beta$. We wish to extract the finite, $\omega$-dependent part of this sum. Taking the derivative of this yields
\begin{align}
   \label{Eq:dvomega}
   \frac{\partial v}{\partial\omega}&=
   \sum_{n=-\infty}^{\infty}\frac{2\omega}{\omega_n^2+\omega^2} \\
   &= \frac{2\omega}{\omega_0^2 + \omega^2} + 2  \sum_{n=1}^{\infty} \frac{2\omega}{\omega_n^2+\omega^2} ,
\end{align}
where in the last step we have split the sum up into sums over the
positive, negative and zero Matsubara frequencies and used the fact that
$\omega_n = \omega_{-n}$ to combine the first two of these, and the
first term will be simplified further using $\omega_0=0$.  The
remaining sum can be transformed using a clever identity:
\begin{equation}\label{eq:clever_identity}
\sum_{n=1}^{\infty}\frac{2y}{y^2 +n^2} = - \frac{1}{y}+\pi\left(1 +  \frac{2 e^{-2\pi y}}{1 - e^{-2\pi y} }\right).
\end{equation}
By integrating \cref{Eq:dvomega} again with respect to $\omega$, we will finally extract the finite, $\omega$-dependent part of \cref{Eq:v_omega}.

To understand \cref{eq:clever_identity}, we start with the infinite product series for $\sin(\pi z)$ that is obtained from the Weierstrass factorization theorem,
\begin{align}
  \sin(\pi z) = \pi z \, \prod_{n=0}^{\infty}\left[1 -\left(\frac{z}{n}\right)^{\!\!2} \right].
\end{align}
We re-express this as an identity for $\textrm{sinc}(\pi z) \equiv \sin(\pi z)/(\pi z)$, then on both sides of the equation take the natural logarithm and differentiate. After simple rearrangement this yields the identity
\begin{align}
 \pi \cot(\pi z) = \frac{1}{z} + \sum_{n=1}^{\infty}\frac{2z}{n^2 -z^2} .
\end{align}
Now choose $z=-iy$, and use
\begin{align}
  \cot(-i\pi y) = i\frac{1 + \exp{-2\pi y} }{ 1 - \exp{-2\pi y} }
\end{align}
to rearrange to obtain our identity \cref{eq:clever_identity},\footnote{One could instead use $\text{coth}(z) = i \cot(iz)$ here to obtain the specific identity quoted in Eq.~(3.11) of \refcite{Dolan:1973qd} and then re-express $\text{coth}(z)$.}
\begin{align}
  f(y) &:=  \sum_{n=1}^{\infty}\frac{2y}{y^2 +n^2}\\
  &= -\frac{1}{y}+\pi \frac{1 + \exp{-2\pi y} }{ 1 - \exp{-2\pi y} }\\
                                      &= - \frac{1}{y}+\pi\left(1 +  \frac{2 e^{-2\pi y}}{1 - e^{-2\pi y} }\right) .
\end{align}

Having justified the identity \cref{eq:clever_identity}, we use it in \cref{Eq:dvomega} with $y= \beta\omega /(2\pi)$ to give the result
\begin{align}
  \frac{\partial v}{\partial\omega}&=  2 \beta \left[\frac12 + \pi\frac{ \exp{-\beta\omega} }{ 1 -\exp{-\beta\omega} }\right].
\end{align}
Integrating over $\omega$ in this equation, we can then isolate the $\omega$-dependent term in \cref{Eq:v_omega} as
\begin{equation}
    \label{vefin}
    v(\omega)=2\beta\left[\frac{w}{2}+\frac{1}{\beta}\log\left(
    1-e^{-\beta\omega}\right) \right]+\ldots,
\end{equation}
where the ellipsis represent terms that are independent of $\omega$.

There is a similar infinite sum that needs to be handled in \cref{Eq:V1L_fer_T} for the contribution to the effective potential from fermions.  \Cref{Eq:v_omega} is the same except now the sum is over the fermion Matsubara frequencies, which are odd integer multiples of $\pi /\beta$, $\omega_n = (2n+1)\pi/\beta$.  Differentiating and using $\omega_n = \omega_{-n}$ now gives
\begin{align}
   \label{Eq:dvdomegaF}
   \frac{\partial v}{\partial\omega} &=  2  \sum_{n=1}^{\infty} \frac{2\omega}{\omega_n^2+\omega^2} \\
                                     &=  \sum_{n=1}^{\infty}\beta/\pi\frac{4\omega \beta/\pi}{(2n+1)^2 +\omega^2\beta^2/\pi^2} ,
\end{align}
where in the second line we have substituted for $\omega_n$ and rearranged in a way that will be convenient for us.
From the fact that the sum over odd positive integers and the sum over even positive integers can be combined to give the sum over all integers, we have
\begin{align}
   \sum_{n=1}^{\infty}\frac{2y}{y^2 +(2n+1)^2} &= f(y) - \sum_{n=1}^{\infty}\frac{2y}{y^2 +4n^2}  \\
   &= f(y) - \frac12 f(y/2)\\
   & = \frac{\pi}{2} - \frac{\pi}{1 + \exp{\pi y}} ,
\end{align}
and applying this with $y=\omega\beta/ \pi$ we get
\begin{equation}
   \label{Eq:der_ve_fer}
    \frac{\partial v}{\partial\omega}= 2\beta\left(\frac{1}{2} - \frac{1}{1 + \exp{\omega\beta}}\right) .
\end{equation}
This integrates to
\begin{align}
  v(\omega)& =  -\beta \omega + 2 \log [1 + \exp{\beta \omega}] + \ldots\\
           & = \beta \omega + 2 \log[1 + \exp{-\beta\omega}] + \ldots ,
  \end{align}
where again the ellipses represent terms that are independent of $\omega$ and in the second line we have rewritten the expression using $-2\beta\omega = 2\log[\exp{-2\beta\omega}]$.

\section{Convexity of the effective potential}\label{App:Convex}

Taking a step back, we can understand the arguments about the convexity of the potential in \cref{sec:imaginary_contributions} classically (here we extend an argument made in \refcite{Weinberg:1987vp}). We interpret $\phi_{\text{cl}}$ as an ordinary average value of a random variable, 
\begin{equation}\label{eq:constraint}
\phi_{\text{cl}} = \int p(\phi) \phi \,d\phi ,
\end{equation}
over a distribution $p(\phi)$, and interpret $\langle \psi | H | \psi \rangle$ as the average of a potential energy function $V(\phi)$,
\begin{equation}\label{eq:average_energy}
\langle \psi | H | \psi \rangle = \int p(\phi) V(\phi) \,d\phi.
\end{equation}
Continuing the classical interpretation, we now obtain $V_\text{eff}(\phi_{\text{cl}})$ as the minimum average potential energy \cref{eq:average_energy} for any $p(\phi)$ subject to the constraint \cref{eq:constraint} on $p(\phi)$,
\begin{equation}\label{eq:problem}
V_\text{eff}(\phi_{\text{cl}}) = \min_p \left[\int p(\phi) V(\phi) \,d\phi \right] \quad\text{subject to}\quad \int p(\phi) \phi \,d\phi = \phi_{\text{cl}}.
\end{equation}
To tackle this problem, consider the convex hull of $V(\phi)$, $V_\text{hull}(\phi)$. By Jensen's inequality, any average of the convex hull of the potential evaluated at a set of points must be greater than the value of the convex hull at the average of those points,
\begin{equation}
\int p(\phi) V_\text{hull}(\phi) \, d\phi \ge V_\text{hull}\left(\int p(\phi) \phi \, d\phi\right) .
\end{equation}
Then because $V(\phi) \ge V_\text{hull}(\phi)$, we have
\begin{equation}
\int p(\phi) V(\phi) \, d\phi \ge V_\text{hull}\left(\int p(\phi) \phi \, d\phi\right) .
\end{equation}
Applying our constraint \cref{eq:constraint}, we obtain
\begin{equation}
\int p(\phi) V(\phi) \, d\phi \ge V_\text{hull}(\phi_{\text{cl}}) .
\end{equation}
Thus if we find $p(\phi)$ that achieves the lower bound and satisfies \cref{eq:constraint}, we know it is the desired solution that minimizes \cref{eq:problem} and that $V_\text{eff}(\phi) = V_\text{hull}(\phi_{\text{cl}})$. We may quickly verify the existence of a solution by considering \cref{fig:VeffConvexHull}: in the convex region in which the convex hull equals the potential, $p(\phi) = \delta(\phi - \phi_{\text{cl}})$. In the non-convex region between the true and false minima, we write $p(\phi) = \alpha \delta(\phi - \phi_t) + (1 - \alpha) \delta(\phi - \phi_f)$; that is, as a mixture of lumps at the true and false minima, where by \cref{eq:constraint} $\alpha  \phi_t + (1 - \alpha) \phi_f = \phi_{\text{cl}}$. Thus, in this region
\begin{equation}
V_\text{eff}(\phi_{\text{cl}}) = \int p(\phi) V(\phi) \,d\phi = \alpha  V(\phi_t) + (1 - \alpha) V(\phi_f) ,
\end{equation}
which is a straight line between the minima; that is, it is the convex hull of $V(\phi)$. We arrived at the convex hull because we permitted the mixture of lumps at the true and false minima. If we restricted ourselves to a single, localized values of $\phi$, we would have found the potential rather than its convex hull.

\bibliography{all}

\providecommand{\href}[2]{#2}\begingroup\raggedright\begin{thebibliography}{100}

\bibitem{LIGOScientific:2016aoc}
{\scshape LIGO Scientific, Virgo} collaboration, \emph{{Observation of
  Gravitational Waves from a Binary Black Hole Merger}},
  \href{https://doi.org/10.1103/PhysRevLett.116.061102}{\emph{Phys. Rev. Lett.}
  {\bfseries 116} (2016) 061102}
  [\href{https://arxiv.org/abs/1602.03837}{{\ttfamily 1602.03837}}].

\bibitem{Meszaros:2019xej}
P.~M\'esz\'aros, D.B.~Fox, C.~Hanna and K.~Murase, \emph{{Multi-Messenger
  Astrophysics}}, \href{https://doi.org/10.1038/s42254-019-0101-z}{\emph{Nature
  Rev. Phys.} {\bfseries 1} (2019) 585}
  [\href{https://arxiv.org/abs/1906.10212}{{\ttfamily 1906.10212}}].

\bibitem{LIGO-data}
\emph{{LIGO/Virgo Data}},
  \href{https://www.ligo.caltech.edu/page/ligo-data}{https://www.ligo.caltech.edu/page/ligo-data},
  Accessed: 2022 March 1.

\bibitem{Langacker:1980js}
P.~Langacker, \emph{{Grand Unified Theories and Proton Decay}},
  \href{https://doi.org/10.1016/0370-1573(81)90059-4}{\emph{Phys. Rept.}
  {\bfseries 72} (1981) 185}.

\bibitem{Mazumdar:2018dfl}
A.~Mazumdar and G.~White, \emph{{Review of cosmic phase transitions: their
  significance and experimental signatures}},
  \href{https://doi.org/10.1088/1361-6633/ab1f55}{\emph{Rept. Prog. Phys.}
  {\bfseries 82} (2019) 076901}
  [\href{https://arxiv.org/abs/1811.01948}{{\ttfamily 1811.01948}}].

\bibitem{Cai:2017cbj}
R.-G.~Cai, Z.~Cao, Z.-K.~Guo, S.-J.~Wang and T.~Yang, \emph{{The
  Gravitational-Wave Physics}},
  \href{https://doi.org/10.1093/nsr/nwx029}{\emph{Natl. Sci. Rev.} {\bfseries
  4} (2017) 687} [\href{https://arxiv.org/abs/1703.00187}{{\ttfamily
  1703.00187}}].

\bibitem{Bian:2021ini}
L.~Bian et~al., \emph{{The Gravitational-Wave Physics II: Progress}},
  \href{https://doi.org/10.1007/s11433-021-1781-x}{\emph{Sci. China Phys. Mech.
  Astron.} {\bfseries 64} (2021) 120401}
  [\href{https://arxiv.org/abs/2106.10235}{{\ttfamily 2106.10235}}].

\bibitem{Binetruy:2012ze}
P.~Binetruy, A.~Bohe, C.~Caprini and J.-F.~Dufaux, \emph{{Cosmological
  Backgrounds of Gravitational Waves and eLISA/NGO: Phase Transitions, Cosmic
  Strings and Other Sources}},
  \href{https://doi.org/10.1088/1475-7516/2012/06/027}{\emph{JCAP} {\bfseries
  06} (2012) 027} [\href{https://arxiv.org/abs/1201.0983}{{\ttfamily
  1201.0983}}].

\bibitem{Caprini:2015zlo}
C.~Caprini et~al., \emph{{Science with the space-based interferometer eLISA.
  II: Gravitational waves from cosmological phase transitions}},
  \href{https://doi.org/10.1088/1475-7516/2016/04/001}{\emph{JCAP} {\bfseries
  04} (2016) 001} [\href{https://arxiv.org/abs/1512.06239}{{\ttfamily
  1512.06239}}].

\bibitem{Caprini:2019egz}
C.~Caprini et~al., \emph{{Detecting gravitational waves from cosmological phase
  transitions with LISA: an update}},
  \href{https://doi.org/10.1088/1475-7516/2020/03/024}{\emph{JCAP} {\bfseries
  03} (2020) 024} [\href{https://arxiv.org/abs/1910.13125}{{\ttfamily
  1910.13125}}].

\bibitem{Hindmarsh:2020hop}
M.B.~Hindmarsh, M.~L\"uben, J.~Lumma and M.~Pauly, \emph{{Phase transitions in
  the early universe}},
  \href{https://doi.org/10.21468/SciPostPhysLectNotes.24}{\emph{SciPost Phys.
  Lect. Notes} {\bfseries 24} (2021) 1}
  [\href{https://arxiv.org/abs/2008.09136}{{\ttfamily 2008.09136}}].

\bibitem{Cohen:1993nk}
A.G.~Cohen, D.B.~Kaplan and A.E.~Nelson, \emph{{Progress in electroweak
  baryogenesis}},
  \href{https://doi.org/10.1146/annurev.ns.43.120193.000331}{\emph{Ann. Rev.
  Nucl. Part. Sci.} {\bfseries 43} (1993) 27}
  [\href{https://arxiv.org/abs/hep-ph/9302210}{{\ttfamily hep-ph/9302210}}].

\bibitem{Trodden:1998ym}
M.~Trodden, \emph{{Electroweak baryogenesis}},
  \href{https://doi.org/10.1103/RevModPhys.71.1463}{\emph{Rev. Mod. Phys.}
  {\bfseries 71} (1999) 1463}
  [\href{https://arxiv.org/abs/hep-ph/9803479}{{\ttfamily hep-ph/9803479}}].

\bibitem{Cline:2006ts}
J.M.~Cline, \emph{{Baryogenesis}},  in \emph{{Les Houches Summer School -
  Session 86: Particle Physics and Cosmology: The Fabric of Spacetime}}, 9,
  2006 [\href{https://arxiv.org/abs/hep-ph/0609145}{{\ttfamily
  hep-ph/0609145}}].

\bibitem{Morrissey:2012db}
D.E.~Morrissey and M.J.~Ramsey-Musolf, \emph{{Electroweak baryogenesis}},
  \href{https://doi.org/10.1088/1367-2630/14/12/125003}{\emph{New J. Phys.}
  {\bfseries 14} (2012) 125003}
  [\href{https://arxiv.org/abs/1206.2942}{{\ttfamily 1206.2942}}].

\bibitem{White:2016nbo}
G.A.~White, \emph{{A Pedagogical Introduction to Electroweak Baryogenesis}},
  {IOP} Publishing (2016),
  \href{https://doi.org/10.1088/978-1-6817-4457-5}{10.1088/978-1-6817-4457-5}.

\bibitem{Kosowsky:1991ua}
A.~Kosowsky, M.S.~Turner and R.~Watkins, \emph{{Gravitational radiation from
  colliding vacuum bubbles}},
  \href{https://doi.org/10.1103/PhysRevD.45.4514}{\emph{Phys. Rev. D}
  {\bfseries 45} (1992) 4514}.

\bibitem{Heisenberg:1936nmg}
W.~Heisenberg and H.~Euler, \emph{{Consequences of Dirac's theory of
  positrons}}, \href{https://doi.org/10.1007/BF01343663}{\emph{Z. Phys.}
  {\bfseries 98} (1936) 714}
  [\href{https://arxiv.org/abs/physics/0605038}{{\ttfamily physics/0605038}}].

\bibitem{Schwinger:1951nm}
J.S.~Schwinger, \emph{{On gauge invariance and vacuum polarization}},
  \href{https://doi.org/10.1103/PhysRev.82.664}{\emph{Phys. Rev.} {\bfseries
  82} (1951) 664}.

\bibitem{Jona-Lasinio:1964zvf}
G.~Jona-Lasinio, \emph{{Relativistic field theories with symmetry breaking
  solutions}}, \href{https://doi.org/10.1007/BF02750573}{\emph{Nuovo Cim.}
  {\bfseries 34} (1964) 1790}.

\bibitem{Goldstone:1962es}
J.~Goldstone, A.~Salam and S.~Weinberg, \emph{{Broken Symmetries}},
  \href{https://doi.org/10.1103/PhysRev.127.965}{\emph{Phys. Rev.} {\bfseries
  127} (1962) 965}.

\bibitem{Coleman:1973jx}
S.R.~Coleman and E.J.~Weinberg, \emph{{Radiative Corrections as the Origin of
  Spontaneous Symmetry Breaking}},
  \href{https://doi.org/10.1103/PhysRevD.7.1888}{\emph{Phys. Rev. D} {\bfseries
  7} (1973) 1888}.

\bibitem{Jackiw:1974cv}
R.~Jackiw, \emph{{Functional evaluation of the effective potential}},
  \href{https://doi.org/10.1103/PhysRevD.9.1686}{\emph{Phys. Rev. D} {\bfseries
  9} (1974) 1686}.

\bibitem{Brandenberger:1984cz}
R.H.~Brandenberger, \emph{{Quantum Field Theory Methods and Inflationary
  Universe Models}}, \href{https://doi.org/10.1103/RevModPhys.57.1}{\emph{Rev.
  Mod. Phys.} {\bfseries 57} (1985) 1}.

\bibitem{Sher:1988mj}
M.~Sher, \emph{{Electroweak Higgs Potentials and Vacuum Stability}},
  \href{https://doi.org/10.1016/0370-1573(89)90061-6}{\emph{Phys. Rept.}
  {\bfseries 179} (1989) 273}.

\bibitem{Quiros:1999jp}
M.~Quiros, \emph{{Finite temperature field theory and phase transitions}},  in
  \emph{{ICTP Summer School in High-Energy Physics and Cosmology}},
  pp.~187--259, 1, 1999 [\href{https://arxiv.org/abs/hep-ph/9901312}{{\ttfamily
  hep-ph/9901312}}].

\bibitem{Espinosa:2011ax}
J.R.~Espinosa, T.~Konstandin and F.~Riva, \emph{{Strong Electroweak Phase
  Transitions in the Standard Model with a Singlet}},
  \href{https://doi.org/10.1016/j.nuclphysb.2011.09.010}{\emph{Nucl. Phys. B}
  {\bfseries 854} (2012) 592}
  [\href{https://arxiv.org/abs/1107.5441}{{\ttfamily 1107.5441}}].

\bibitem{Lee:1974fj}
S.Y.~Lee and A.M.~Sciaccaluga, \emph{{Evaluation of Higher Order Effective
  Potentials with Dimensional Regularization}},
  \href{https://doi.org/10.1016/0550-3213(75)90341-7}{\emph{Nucl. Phys. B}
  {\bfseries 96} (1975) 435}.

\bibitem{Dolan:1974gu}
L.~Dolan and R.~Jackiw, \emph{{Gauge Invariant Signal for Gauge Symmetry
  Breaking}}, \href{https://doi.org/10.1103/PhysRevD.9.2904}{\emph{Phys. Rev.
  D} {\bfseries 9} (1974) 2904}.

\bibitem{Coleman:1974jh}
S.R.~Coleman, R.~Jackiw and H.D.~Politzer, \emph{{Spontaneous Symmetry Breaking
  in the O(N) Model for Large N*}},
  \href{https://doi.org/10.1103/PhysRevD.10.2491}{\emph{Phys. Rev. D}
  {\bfseries 10} (1974) 2491}.

\bibitem{Schnitzer:1974ue}
H.J.~Schnitzer, \emph{{The Hartree Approximation in Relativistic Field
  Theory}}, \href{https://doi.org/10.1103/PhysRevD.10.2042}{\emph{Phys. Rev. D}
  {\bfseries 10} (1974) 2042}.

\bibitem{Peskin:1995ev}
M.E.~Peskin and D.V.~Schroeder, \emph{{An Introduction to quantum field
  theory}}, Addison-Wesley, Reading, USA (1995).

\bibitem{Camargo-Molina:2016moz}
J.E.~Camargo-Molina, A.P.~Morais, R.~Pasechnik, M.O.P.~Sampaio and J.~Wess\'en,
  \emph{{All one-loop scalar vertices in the effective potential approach}},
  \href{https://doi.org/10.1007/JHEP08(2016)073}{\emph{JHEP} {\bfseries 08}
  (2016) 073} [\href{https://arxiv.org/abs/1606.07069}{{\ttfamily
  1606.07069}}].

\bibitem{Leibbrandt:1975dj}
G.~Leibbrandt, \emph{{Introduction to the Technique of Dimensional
  Regularization}}, \href{https://doi.org/10.1103/RevModPhys.47.849}{\emph{Rev.
  Mod. Phys.} {\bfseries 47} (1975) 849}.

\bibitem{Jack:1994rk}
I.~Jack, D.R.T.~Jones, S.P.~Martin, M.T.~Vaughn and Y.~Yamada,
  \emph{{Decoupling of the epsilon scalar mass in softly broken
  supersymmetry}}, \href{https://doi.org/10.1103/PhysRevD.50.R5481}{\emph{Phys.
  Rev. D} {\bfseries 50} (1994) R5481}
  [\href{https://arxiv.org/abs/hep-ph/9407291}{{\ttfamily hep-ph/9407291}}].

\bibitem{Martin:2001vx}
S.P.~Martin, \emph{{Two Loop Effective Potential for a General Renormalizable
  Theory and Softly Broken Supersymmetry}},
  \href{https://doi.org/10.1103/PhysRevD.65.116003}{\emph{Phys. Rev. D}
  {\bfseries 65} (2002) 116003}
  [\href{https://arxiv.org/abs/hep-ph/0111209}{{\ttfamily hep-ph/0111209}}].

\bibitem{Siegel:1979wq}
W.~Siegel, \emph{{Supersymmetric Dimensional Regularization via Dimensional
  Reduction}}, \href{https://doi.org/10.1016/0370-2693(79)90282-X}{\emph{Phys.
  Lett. B} {\bfseries 84} (1979) 193}.

\bibitem{Capper:1979ns}
D.M.~Capper, D.R.T.~Jones and P.~van Nieuwenhuizen, \emph{{Regularization by
  Dimensional Reduction of Supersymmetric and Nonsupersymmetric Gauge
  Theories}}, \href{https://doi.org/10.1016/0550-3213(80)90244-8}{\emph{Nucl.
  Phys. B} {\bfseries 167} (1980) 479}.

\bibitem{Anderson:1991zb}
G.W.~Anderson and L.J.~Hall, \emph{{The Electroweak phase transition and
  baryogenesis}}, \href{https://doi.org/10.1103/PhysRevD.45.2685}{\emph{Phys.
  Rev. D} {\bfseries 45} (1992) 2685}.

\bibitem{Basler:2018cwe}
P.~Basler and M.~M\"uhlleitner, \emph{{BSMPT (Beyond the Standard Model Phase
  Transitions): A tool for the electroweak phase transition in extended Higgs
  sectors}}, \href{https://doi.org/10.1016/j.cpc.2018.11.006}{\emph{Comput.
  Phys. Commun.} {\bfseries 237} (2019) 62}
  [\href{https://arxiv.org/abs/1803.02846}{{\ttfamily 1803.02846}}].

\bibitem{Xie:2020wzn}
K.-P.~Xie, \emph{{Lepton-mediated electroweak baryogenesis, gravitational waves
  and the $4\tau$ final state at the collider}},
  \href{https://doi.org/10.1007/JHEP02(2021)090}{\emph{JHEP} {\bfseries 02}
  (2021) 090} [\href{https://arxiv.org/abs/2011.04821}{{\ttfamily
  2011.04821}}].

\bibitem{Fujikawa:1972fe}
K.~Fujikawa, B.W.~Lee and A.I.~Sanda, \emph{{Generalized Renormalizable Gauge
  Formulation of Spontaneously Broken Gauge Theories}},
  \href{https://doi.org/10.1103/PhysRevD.6.2923}{\emph{Phys. Rev. D} {\bfseries
  6} (1972) 2923}.

\bibitem{Yao:1973am}
Y.-P.~Yao, \emph{{Quantization and gauge freedom in a theory with spontaneously
  broken symmetry}}, \href{https://doi.org/10.1103/PhysRevD.7.1647}{\emph{Phys.
  Rev. D} {\bfseries 7} (1973) 1647}.

\bibitem{Patel:2011th}
H.H.~Patel and M.J.~Ramsey-Musolf, \emph{{Baryon Washout, Electroweak Phase
  Transition, and Perturbation Theory}},
  \href{https://doi.org/10.1007/JHEP07(2011)029}{\emph{JHEP} {\bfseries 07}
  (2011) 029} [\href{https://arxiv.org/abs/1101.4665}{{\ttfamily 1101.4665}}].

\bibitem{DiLuzio:2014bua}
L.~Di~Luzio and L.~Mihaila, \emph{{On the gauge dependence of the Standard
  Model vacuum instability scale}},
  \href{https://doi.org/10.1007/JHEP06(2014)079}{\emph{JHEP} {\bfseries 06}
  (2014) 079} [\href{https://arxiv.org/abs/1404.7450}{{\ttfamily 1404.7450}}].

\bibitem{Arnold:1992fb}
P.B.~Arnold, \emph{{Phase transition temperatures at next-to-leading order}},
  \href{https://doi.org/10.1103/PhysRevD.46.2628}{\emph{Phys. Rev. D}
  {\bfseries 46} (1992) 2628}
  [\href{https://arxiv.org/abs/hep-ph/9204228}{{\ttfamily hep-ph/9204228}}].

\bibitem{Laine:1994bf}
M.~Laine, \emph{{The Two loop effective potential of the 3-d SU(2) Higgs model
  in a general covariant gauge}},
  \href{https://doi.org/10.1016/0370-2693(94)91409-5}{\emph{Phys. Lett. B}
  {\bfseries 335} (1994) 173}
  [\href{https://arxiv.org/abs/hep-ph/9406268}{{\ttfamily hep-ph/9406268}}].

\bibitem{Laine:1994zq}
M.~Laine, \emph{{Gauge dependence of the high temperature two loop effective
  potential for the Higgs field}},
  \href{https://doi.org/10.1103/PhysRevD.51.4525}{\emph{Phys. Rev. D}
  {\bfseries 51} (1995) 4525}
  [\href{https://arxiv.org/abs/hep-ph/9411252}{{\ttfamily hep-ph/9411252}}].

\bibitem{Andreassen:2013hpa}
A.J.~Andreassen, \emph{{Gauge Dependence of the Quantum Field Theory Effective
  Potential}},  Master's thesis, Norwegian U. Sci. Tech., 2013.

\bibitem{Andreassen:2014eha}
A.~Andreassen, W.~Frost and M.D.~Schwartz, \emph{{Consistent Use of Effective
  Potentials}}, \href{https://doi.org/10.1103/PhysRevD.91.016009}{\emph{Phys.
  Rev. D} {\bfseries 91} (2015) 016009}
  [\href{https://arxiv.org/abs/1408.0287}{{\ttfamily 1408.0287}}].

\bibitem{Kastening:1993zn}
B.M.~Kastening, \emph{{A New gauge for computing effective potentials in
  spontaneously broken gauge theories}},
  \href{https://doi.org/10.1103/PhysRevD.51.265}{\emph{Phys. Rev. D} {\bfseries
  51} (1995) 265} [\href{https://arxiv.org/abs/hep-ph/9307220}{{\ttfamily
  hep-ph/9307220}}].

\bibitem{Papaefstathiou:2020iag}
A.~Papaefstathiou and G.~White, \emph{{The electro-weak phase transition at
  colliders: confronting theoretical uncertainties and complementary
  channels}}, \href{https://doi.org/10.1007/JHEP05(2021)099}{\emph{JHEP}
  {\bfseries 05} (2021) 099}
  [\href{https://arxiv.org/abs/2010.00597}{{\ttfamily 2010.00597}}].

\bibitem{Martin:2018emo}
S.P.~Martin and H.H.~Patel, \emph{{Two-loop effective potential for generalized
  gauge fixing}}, \href{https://doi.org/10.1103/PhysRevD.98.076008}{\emph{Phys.
  Rev. D} {\bfseries 98} (2018) 076008}
  [\href{https://arxiv.org/abs/1808.07615}{{\ttfamily 1808.07615}}].

\bibitem{Kripfganz:1995jx}
J.~Kripfganz, A.~Laser and M.G.~Schmidt, \emph{{The High temperature two loop
  effective potential of the electroweak theory in a general 't Hooft
  background gauge}},
  \href{https://doi.org/10.1016/0370-2693(95)00382-U}{\emph{Phys. Lett. B}
  {\bfseries 351} (1995) 266}
  [\href{https://arxiv.org/abs/hep-ph/9501317}{{\ttfamily hep-ph/9501317}}].

\bibitem{Athron:2022jyi}
P.~Athron, C.~Balazs, A.~Fowlie, L.~Morris, G.~White and Y.~Zhang, \emph{{How
  arbitrary are perturbative calculations of the electroweak phase
  transition?}}, \href{https://doi.org/10.1007/JHEP01(2023)050}{\emph{JHEP}
  {\bfseries 01} (2023) 050}
  [\href{https://arxiv.org/abs/2208.01319}{{\ttfamily 2208.01319}}].

\bibitem{Ford:1992pn}
C.~Ford, I.~Jack and D.R.T.~Jones, \emph{{The Standard model effective
  potential at two loops}},
  \href{https://doi.org/10.1016/0550-3213(92)90165-8}{\emph{Nucl. Phys. B}
  {\bfseries 387} (1992) 373}
  [\href{https://arxiv.org/abs/hep-ph/0111190}{{\ttfamily hep-ph/0111190}}],
  [Erratum:
  \href{https://doi.org/10.1016/S0550-3213(97)00532-4}{\textit{Nucl.~Phys.~B}
  \textbf{04} (1997) 551}].

\bibitem{Martin:2013gka}
S.P.~Martin, \emph{{Three-Loop Standard Model Effective Potential at Leading
  Order in Strong and Top Yukawa Couplings}},
  \href{https://doi.org/10.1103/PhysRevD.89.013003}{\emph{Phys. Rev. D}
  {\bfseries 89} (2014) 013003}
  [\href{https://arxiv.org/abs/1310.7553}{{\ttfamily 1310.7553}}].

\bibitem{Martin:2017lqn}
S.P.~Martin, \emph{{Effective potential at three loops}},
  \href{https://doi.org/10.1103/PhysRevD.96.096005}{\emph{Phys. Rev. D}
  {\bfseries 96} (2017) 096005}
  [\href{https://arxiv.org/abs/1709.02397}{{\ttfamily 1709.02397}}].

\bibitem{Martin:2015eia}
S.P.~Martin, \emph{{Four-Loop Standard Model Effective Potential at Leading
  Order in QCD}}, \href{https://doi.org/10.1103/PhysRevD.92.054029}{\emph{Phys.
  Rev. D} {\bfseries 92} (2015) 054029}
  [\href{https://arxiv.org/abs/1508.00912}{{\ttfamily 1508.00912}}].

\bibitem{Kirzhnits:1972iw}
D.A.~Kirzhnits, \emph{{Weinberg model in the hot universe}}, {\emph{JETP Lett.}
  {\bfseries 15} (1972) 529}.

\bibitem{Kirzhnits:1976ts}
D.A.~Kirzhnits and A.D.~Linde, \emph{{Symmetry Behavior in Gauge Theories}},
  \href{https://doi.org/10.1016/0003-4916(76)90279-7}{\emph{Annals Phys.}
  {\bfseries 101} (1976) 195}.

\bibitem{Weinberg:1974hy}
S.~Weinberg, \emph{{Gauge and Global Symmetries at High Temperature}},
  \href{https://doi.org/10.1103/PhysRevD.9.3357}{\emph{Phys. Rev. D} {\bfseries
  9} (1974) 3357}.

\bibitem{Dolan:1973qd}
L.~Dolan and R.~Jackiw, \emph{{Symmetry Behavior at Finite Temperature}},
  \href{https://doi.org/10.1103/PhysRevD.9.3320}{\emph{Phys. Rev. D} {\bfseries
  9} (1974) 3320}.

\bibitem{Matsubara:1955ws}
T.~Matsubara, \emph{{A New approach to quantum statistical mechanics}},
  \href{https://doi.org/10.1143/PTP.14.351}{\emph{Prog. Theor. Phys.}
  {\bfseries 14} (1955) 351}.

\bibitem{Kubo:1957mj}
R.~Kubo, \emph{{Statistical mechanical theory of irreversible processes. 1.
  General theory and simple applications in magnetic and conduction problems}},
  \href{https://doi.org/10.1143/JPSJ.12.570}{\emph{J. Phys. Soc. Jap.}
  {\bfseries 12} (1957) 570}.

\bibitem{Martin:1959jp}
P.C.~Martin and J.S.~Schwinger, \emph{{Theory of many particle systems. 1.}},
  \href{https://doi.org/10.1103/PhysRev.115.1342}{\emph{Phys. Rev.} {\bfseries
  115} (1959) 1342}.

\bibitem{Schwinger:1960qe}
J.S.~Schwinger, \emph{{Brownian motion of a quantum oscillator}},
  \href{https://doi.org/10.1063/1.1703727}{\emph{J. Math. Phys.} {\bfseries 2}
  (1961) 407}.

\bibitem{Keldysh:1964ud}
L.V.~Keldysh, \emph{{Diagram technique for nonequilibrium processes}},
  {\emph{Zh. Eksp. Teor. Fiz.} {\bfseries 47} (1964) 1515}.

\bibitem{Bernard:1974bq}
C.W.~Bernard, \emph{{Feynman Rules for Gauge Theories at Finite Temperature}},
  \href{https://doi.org/10.1103/PhysRevD.9.3312}{\emph{Phys. Rev. D} {\bfseries
  9} (1974) 3312}.

\bibitem{Landsman:1986uw}
N.P.~Landsman and C.G.~van Weert, \emph{{Real and Imaginary Time Field Theory
  at Finite Temperature and Density}},
  \href{https://doi.org/10.1016/0370-1573(87)90121-9}{\emph{Phys. Rept.}
  {\bfseries 145} (1987) 141}.

\bibitem{Quiros:1994dr}
M.~Quiros, \emph{{Field theory at finite temperature and phase transitions}},
  {\emph{Helv. Phys. Acta} {\bfseries 67} (1994) 451}.

\bibitem{Landshoff:1998ku}
P.V.~Landshoff, \emph{{Introduction to equilibrium thermal field theory}},  in
  \emph{{9th Jorge Andre Swieca Summer School: Particles and Fields}}, 8, 1998
  [\href{https://arxiv.org/abs/hep-ph/9808362}{{\ttfamily hep-ph/9808362}}].

\bibitem{Zinn-Justin:2000ecv}
J.~Zinn-Justin, \emph{{Quantum field theory at finite temperature: An
  Introduction}},  \href{https://arxiv.org/abs/hep-ph/0005272}{{\ttfamily
  hep-ph/0005272}}.

\bibitem{Lombardo:2000rs}
M.P.~Lombardo, \emph{{Finite temperature field theory and phase transitions}},
  {\emph{ICTP Lect. Notes Ser.} {\bfseries 4} (2001) 115}
  [\href{https://arxiv.org/abs/hep-ph/0103141}{{\ttfamily hep-ph/0103141}}].

\bibitem{Das:2000ft}
A.K.~Das, \emph{{Topics in finite temperature field theory}},
  \href{https://arxiv.org/abs/hep-ph/0004125}{{\ttfamily hep-ph/0004125}}.

\bibitem{Laine:2016hma}
M.~Laine and A.~Vuorinen, \emph{{Basics of Thermal Field Theory}}, vol.~925,
  Springer (2016),
  \href{https://doi.org/10.1007/978-3-319-31933-9}{10.1007/978-3-319-31933-9},
  [\href{https://arxiv.org/abs/1701.01554}{{\ttfamily 1701.01554}}].

\bibitem{Kapusta:1989tk}
J.I.~Kapusta, \emph{{Finite Temperature Field Theory}}, Cambridge Monographs on
  Mathematical Physics, Cambridge University Press, Cambridge (1989).

\bibitem{Bellac:2011kqa}
M.L.~Bellac, \emph{{Thermal Field Theory}}, Cambridge Monographs on
  Mathematical Physics, Cambridge University Press (3, 2011),
  \href{https://doi.org/10.1017/CBO9780511721700}{10.1017/CBO9780511721700}.

\bibitem{Kapusta:2006pm}
J.I.~Kapusta and C.~Gale, \emph{{Finite-temperature field theory: Principles
  and applications}}, Cambridge Monographs on Mathematical Physics, Cambridge
  University Press (2011),
  \href{https://doi.org/10.1017/CBO9780511535130}{10.1017/CBO9780511535130}.

\bibitem{Bakshi:1962dv}
P.M.~Bakshi and K.T.~Mahanthappa, \emph{{Expectation value formalism in quantum
  field theory. 1.}}, \href{https://doi.org/10.1063/1.1703883}{\emph{J. Math.
  Phys.} {\bfseries 4} (1963) 1}.

\bibitem{Bakshi:1963bn}
P.M.~Bakshi and K.T.~Mahanthappa, \emph{{Expectation value formalism in quantum
  field theory. 2.}}, \href{https://doi.org/10.1063/1.1703879}{\emph{J. Math.
  Phys.} {\bfseries 4} (1963) 12}.

\bibitem{Matsumoto:1982ry}
H.~Matsumoto, Y.~Nakano, H.~Umezawa, F.~Mancini and M.~Marinaro, \emph{{Thermo
  Field Dynamics in Interaction Representation}},
  \href{https://doi.org/10.1143/PTP.70.599}{\emph{Prog. Theor. Phys.}
  {\bfseries 70} (1983) 599}.

\bibitem{Matsumoto:1984au}
H.~Matsumoto, Y.~Nakano and H.~Umezawa, \emph{{An equivalence class of quantum
  field theories at finite temperature}},
  \href{https://doi.org/10.1063/1.526023}{\emph{J. Math. Phys.} {\bfseries 25}
  (1984) 3076}.

\bibitem{Curtin:2016urg}
D.~Curtin, P.~Meade and H.~Ramani, \emph{{Thermal Resummation and Phase
  Transitions}},
  \href{https://doi.org/10.1140/epjc/s10052-018-6268-0}{\emph{Eur. Phys. J. C}
  {\bfseries 78} (2018) 787}
  [\href{https://arxiv.org/abs/1612.00466}{{\ttfamily 1612.00466}}].

\bibitem{Fowlie:2018eiu}
A.~Fowlie, \emph{{A fast C++ implementation of thermal functions}},
  \href{https://doi.org/10.1016/j.cpc.2018.02.015}{\emph{Comput. Phys. Commun.}
  {\bfseries 228} (2018) 264}
  [\href{https://arxiv.org/abs/1802.02720}{{\ttfamily 1802.02720}}].

\bibitem{Cline:1996mga}
J.M.~Cline and P.-A.~Lemieux, \emph{{Electroweak phase transition in two Higgs
  doublet models}}, \href{https://doi.org/10.1103/PhysRevD.55.3873}{\emph{Phys.
  Rev. D} {\bfseries 55} (1997) 3873}
  [\href{https://arxiv.org/abs/hep-ph/9609240}{{\ttfamily hep-ph/9609240}}].

\bibitem{Wainwright:2011kj}
C.L.~Wainwright, \emph{{CosmoTransitions: Computing Cosmological Phase
  Transition Temperatures and Bubble Profiles with Multiple Fields}},
  \href{https://doi.org/10.1016/j.cpc.2012.04.004}{\emph{Comput. Phys. Commun.}
  {\bfseries 183} (2012) 2006}
  [\href{https://arxiv.org/abs/1109.4189}{{\ttfamily 1109.4189}}].

\bibitem{Athron:2020sbe}
P.~Athron, C.~Bal\'azs, A.~Fowlie and Y.~Zhang, \emph{{PhaseTracer: tracing
  cosmological phases and calculating transition properties}},
  \href{https://doi.org/10.1140/epjc/s10052-020-8035-2}{\emph{Eur. Phys. J. C}
  {\bfseries 80} (2020) 567}
  [\href{https://arxiv.org/abs/2003.02859}{{\ttfamily 2003.02859}}].

\bibitem{Symanzik:1969ek}
K.~Symanzik, \emph{{Renormalizable models with simple symmetry breaking. 1.
  Symmetry breaking by a source term}},
  \href{https://doi.org/10.1007/BF01645494}{\emph{Commun. Math. Phys.}
  {\bfseries 16} (1970) 48}.

\bibitem{Iliopoulos:1974ur}
J.~Iliopoulos, C.~Itzykson and A.~Martin, \emph{{Functional Methods and
  Perturbation Theory}},
  \href{https://doi.org/10.1103/RevModPhys.47.165}{\emph{Rev. Mod. Phys.}
  {\bfseries 47} (1975) 165}.

\bibitem{Fujimoto:1982tc}
Y.~Fujimoto, L.~O'Raifeartaigh and G.~Parravicini, \emph{{Effective Potential
  for Nonconvex Potentials}},
  \href{https://doi.org/10.1016/0550-3213(83)90305-X}{\emph{Nucl. Phys. B}
  {\bfseries 212} (1983) 268}.

\bibitem{Weinberg:1987vp}
E.J.~Weinberg and A.-q.~Wu, \emph{{Understanding complex perturbative effective
  potentials}}, \href{https://doi.org/10.1103/PhysRevD.36.2474}{\emph{Phys.
  Rev. D} {\bfseries 36} (1987) 2474}.

\bibitem{Dannenberg:1987fw}
A.~Dannenberg, \emph{{Dysfunctional Methods and the Effective Potential}},
  \href{https://doi.org/10.1016/0370-2693(88)90862-3}{\emph{Phys. Lett. B}
  {\bfseries 202} (1988) 110}.

\bibitem{Langer:1967ax}
J.S.~Langer, \emph{{Theory of the condensation point}},
  \href{https://doi.org/10.1016/0003-4916(67)90200-X}{\emph{Annals Phys.}
  {\bfseries 41} (1967) 108}.

\bibitem{Langer:1969bc}
J.S.~Langer, \emph{{Statistical theory of the decay of metastable states}},
  \href{https://doi.org/10.1016/0003-4916(69)90153-5}{\emph{Annals Phys.}
  {\bfseries 54} (1969) 258}.

\bibitem{Plascencia:2015pga}
A.D.~Plascencia and C.~Tamarit, \emph{{Convexity, gauge-dependence and
  tunneling rates}}, \href{https://doi.org/10.1007/JHEP10(2016)099}{\emph{JHEP}
  {\bfseries 10} (2016) 099}
  [\href{https://arxiv.org/abs/1510.07613}{{\ttfamily 1510.07613}}].

\bibitem{Martin:2014bca}
S.P.~Martin, \emph{{Taming the Goldstone contributions to the effective
  potential}}, \href{https://doi.org/10.1103/PhysRevD.90.016013}{\emph{Phys.
  Rev. D} {\bfseries 90} (2014) 016013}
  [\href{https://arxiv.org/abs/1406.2355}{{\ttfamily 1406.2355}}].

\bibitem{Devoto:2022qen}
F.~Devoto, S.~Devoto, L.~Di~Luzio and G.~Ridolfi, \emph{{False vacuum decay: an
  introductory review}},
  \href{https://doi.org/10.1088/1361-6471/ac7f24}{\emph{J. Phys. G} {\bfseries
  49} (2022) 103001} [\href{https://arxiv.org/abs/2205.03140}{{\ttfamily
  2205.03140}}].

\bibitem{Delamotte:2002vw}
B.~Delamotte, \emph{{A Hint of renormalization}},
  \href{https://doi.org/10.1119/1.1624112}{\emph{Am. J. Phys.} {\bfseries 72}
  (2004) 170} [\href{https://arxiv.org/abs/hep-th/0212049}{{\ttfamily
  hep-th/0212049}}].

\bibitem{Kastening:1991gv}
B.M.~Kastening, \emph{{Renormalization group improvement of the effective
  potential in massive $\phi^4$ theory}},
  \href{https://doi.org/10.1016/0370-2693(92)90021-U}{\emph{Phys. Lett. B}
  {\bfseries 283} (1992) 287}.

\bibitem{Coleman:1977py}
S.R.~Coleman, \emph{{The Fate of the False Vacuum. 1. Semiclassical Theory}},
  \href{https://doi.org/10.1103/PhysRevD.16.1248}{\emph{Phys. Rev. D}
  {\bfseries 15} (1977) 2929} [Erratum:
  \href{https://doi.org/10.1103/PhysRevD.16.1248}{\textit{Phys.~Rev.~D}~\textbf{16}
  (1977) 1248}].

\bibitem{Isidori:2001bm}
G.~Isidori, G.~Ridolfi and A.~Strumia, \emph{{On the metastability of the
  standard model vacuum}},
  \href{https://doi.org/10.1016/S0550-3213(01)00302-9}{\emph{Nucl. Phys. B}
  {\bfseries 609} (2001) 387}
  [\href{https://arxiv.org/abs/hep-ph/0104016}{{\ttfamily hep-ph/0104016}}].

\bibitem{Buttazzo:2013uya}
D.~Buttazzo, G.~Degrassi, P.P.~Giardino, G.F.~Giudice, F.~Sala, A.~Salvio
  et~al., \emph{{Investigating the near-criticality of the Higgs boson}},
  \href{https://doi.org/10.1007/JHEP12(2013)089}{\emph{JHEP} {\bfseries 12}
  (2013) 089} [\href{https://arxiv.org/abs/1307.3536}{{\ttfamily 1307.3536}}].

\bibitem{Espinosa:2015qea}
J.R.~Espinosa, G.F.~Giudice, E.~Morgante, A.~Riotto, L.~Senatore, A.~Strumia
  et~al., \emph{{The cosmological Higgstory of the vacuum instability}},
  \href{https://doi.org/10.1007/JHEP09(2015)174}{\emph{JHEP} {\bfseries 09}
  (2015) 174} [\href{https://arxiv.org/abs/1505.04825}{{\ttfamily
  1505.04825}}].

\bibitem{Andreassen:2017rzq}
A.~Andreassen, W.~Frost and M.D.~Schwartz, \emph{{Scale Invariant Instantons
  and the Complete Lifetime of the Standard Model}},
  \href{https://doi.org/10.1103/PhysRevD.97.056006}{\emph{Phys. Rev. D}
  {\bfseries 97} (2018) 056006}
  [\href{https://arxiv.org/abs/1707.08124}{{\ttfamily 1707.08124}}].

\bibitem{Chigusa:2017dux}
S.~Chigusa, T.~Moroi and Y.~Shoji, \emph{{State-of-the-Art Calculation of the
  Decay Rate of Electroweak Vacuum in the Standard Model}},
  \href{https://doi.org/10.1103/PhysRevLett.119.211801}{\emph{Phys. Rev. Lett.}
  {\bfseries 119} (2017) 211801}
  [\href{https://arxiv.org/abs/1707.09301}{{\ttfamily 1707.09301}}].

\bibitem{Croon:2020cgk}
D.~Croon, O.~Gould, P.~Schicho, T.V.I.~Tenkanen and G.~White,
  \emph{{Theoretical uncertainties for cosmological first-order phase
  transitions}}, \href{https://doi.org/10.1007/JHEP04(2021)055}{\emph{JHEP}
  {\bfseries 04} (2021) 055}
  [\href{https://arxiv.org/abs/2009.10080}{{\ttfamily 2009.10080}}].

\bibitem{Ford:1992mv}
C.~Ford, D.R.T.~Jones, P.W.~Stephenson and M.B.~Einhorn, \emph{{The Effective
  potential and the renormalization group}},
  \href{https://doi.org/10.1016/0550-3213(93)90206-5}{\emph{Nucl. Phys. B}
  {\bfseries 395} (1993) 17}
  [\href{https://arxiv.org/abs/hep-lat/9210033}{{\ttfamily hep-lat/9210033}}].

\bibitem{Elias-Miro:2014pca}
J.~Elias-Miro, J.R.~Espinosa and T.~Konstandin, \emph{{Taming Infrared
  Divergences in the Effective Potential}},
  \href{https://doi.org/10.1007/JHEP08(2014)034}{\emph{JHEP} {\bfseries 08}
  (2014) 034} [\href{https://arxiv.org/abs/1406.2652}{{\ttfamily 1406.2652}}].

\bibitem{Nielsen:1975fs}
N.K.~Nielsen, \emph{{On the Gauge Dependence of Spontaneous Symmetry Breaking
  in Gauge Theories}},
  \href{https://doi.org/10.1016/0550-3213(75)90301-6}{\emph{Nucl. Phys. B}
  {\bfseries 101} (1975) 173}.

\bibitem{Aitchison:1983ns}
I.J.R.~Aitchison and C.M.~Fraser, \emph{{Gauge Invariance and the Effective
  Potential}}, \href{https://doi.org/10.1016/0003-4916(84)90209-4}{\emph{Annals
  Phys.} {\bfseries 156} (1984) 1}.

\bibitem{Loinaz:1997td}
W.~Loinaz and R.S.~Willey, \emph{{Gauge dependence of lower bounds on the Higgs
  mass derived from electroweak vacuum stability constraints}},
  \href{https://doi.org/10.1103/PhysRevD.56.7416}{\emph{Phys. Rev. D}
  {\bfseries 56} (1997) 7416}
  [\href{https://arxiv.org/abs/hep-ph/9702321}{{\ttfamily hep-ph/9702321}}].

\bibitem{Espinosa:2016uaw}
J.R.~Espinosa, M.~Garny and T.~Konstandin, \emph{{Interplay of Infrared
  Divergences and Gauge-Dependence of the Effective Potential}},
  \href{https://doi.org/10.1103/PhysRevD.94.055026}{\emph{Phys. Rev. D}
  {\bfseries 94} (2016) 055026}
  [\href{https://arxiv.org/abs/1607.08432}{{\ttfamily 1607.08432}}].

\bibitem{Casas:1994us}
J.A.~Casas, J.R.~Espinosa, M.~Quiros and A.~Riotto, \emph{{The Lightest Higgs
  boson mass in the minimal supersymmetric standard model}},
  \href{https://doi.org/10.1016/0550-3213(94)00508-C}{\emph{Nucl. Phys. B}
  {\bfseries 436} (1995) 3}
  [\href{https://arxiv.org/abs/hep-ph/9407389}{{\ttfamily hep-ph/9407389}}],
  [Erratum:
  \href{https://doi.org/10.1016/0550-3213(95)00057-Y}{\textit{Nucl.~Phys.~B}
  \textbf{439} (1995) 466}].

\bibitem{Espinosa:2017aew}
J.R.~Espinosa and T.~Konstandin, \emph{{Resummation of Goldstone Infrared
  Divergences: A Proof to All Orders}},
  \href{https://doi.org/10.1103/PhysRevD.97.056020}{\emph{Phys. Rev. D}
  {\bfseries 97} (2018) 056020}
  [\href{https://arxiv.org/abs/1712.08068}{{\ttfamily 1712.08068}}].

\bibitem{Pilaftsis:2015bbs}
A.~Pilaftsis and D.~Teresi, \emph{{Symmetry-Improved 2PI Approach to the
  Goldstone-Boson IR Problem of the SM Effective Potential}},
  \href{https://doi.org/10.1016/j.nuclphysb.2016.03.018}{\emph{Nucl. Phys. B}
  {\bfseries 906} (2016) 381}
  [\href{https://arxiv.org/abs/1511.05347}{{\ttfamily 1511.05347}}].

\bibitem{Marko:2016wtw}
G.~Mark\'o, U.~Reinosa and Z.~Sz\'ep, \emph{{Loss of solution in the symmetry
  improved \ensuremath{\Phi}-derivable expansion scheme}},
  \href{https://doi.org/10.1016/j.nuclphysb.2016.09.022}{\emph{Nucl. Phys. B}
  {\bfseries 913} (2016) 405}
  [\href{https://arxiv.org/abs/1604.04193}{{\ttfamily 1604.04193}}].

\bibitem{Pilaftsis:2017enx}
A.~Pilaftsis and D.~Teresi, \emph{{Exact RG Invariance and Symmetry Improved
  2PI Effective Potential}},
  \href{https://doi.org/10.1016/j.nuclphysb.2017.04.015}{\emph{Nucl. Phys. B}
  {\bfseries 920} (2017) 298}
  [\href{https://arxiv.org/abs/1703.02079}{{\ttfamily 1703.02079}}].

\bibitem{Kumar:2016ltb}
N.~Kumar and S.P.~Martin, \emph{{Resummation of Goldstone boson contributions
  to the MSSM effective potential}},
  \href{https://doi.org/10.1103/PhysRevD.94.014013}{\emph{Phys. Rev. D}
  {\bfseries 94} (2016) 014013}
  [\href{https://arxiv.org/abs/1605.02059}{{\ttfamily 1605.02059}}].

\bibitem{Braathen:2016cqe}
J.~Braathen and M.D.~Goodsell, \emph{{Avoiding the Goldstone Boson Catastrophe
  in general renormalisable field theories at two loops}},
  \href{https://doi.org/10.1007/JHEP12(2016)056}{\emph{JHEP} {\bfseries 12}
  (2016) 056} [\href{https://arxiv.org/abs/1609.06977}{{\ttfamily
  1609.06977}}].

\bibitem{Braathen:2017izn}
J.~Braathen, M.D.~Goodsell and F.~Staub, \emph{{Supersymmetric and
  non-supersymmetric models without catastrophic Goldstone bosons}},
  \href{https://doi.org/10.1140/epjc/s10052-017-5303-x}{\emph{Eur. Phys. J. C}
  {\bfseries 77} (2017) 757}
  [\href{https://arxiv.org/abs/1706.05372}{{\ttfamily 1706.05372}}].

\bibitem{Kirzhnits:1974as}
D.A.~Kirzhnits and A.D.~Linde, \emph{{A Relativistic phase transition}},
  {\emph{Zh. Eksp. Teor. Fiz.} {\bfseries 67} (1974) 1263}.

\bibitem{Linde:1978px}
A.D.~Linde, \emph{{Phase Transitions in Gauge Theories and Cosmology}},
  \href{https://doi.org/10.1088/0034-4885/42/3/001}{\emph{Rept. Prog. Phys.}
  {\bfseries 42} (1979) 389}.

\bibitem{Linde:1980ts}
A.D.~Linde, \emph{{Infrared Problem in Thermodynamics of the Yang-Mills Gas}},
  \href{https://doi.org/10.1016/0370-2693(80)90769-8}{\emph{Phys. Lett. B}
  {\bfseries 96} (1980) 289}.

\bibitem{Gross:1980br}
D.J.~Gross, R.D.~Pisarski and L.G.~Yaffe, \emph{{QCD and Instantons at Finite
  Temperature}}, \href{https://doi.org/10.1103/RevModPhys.53.43}{\emph{Rev.
  Mod. Phys.} {\bfseries 53} (1981) 43}.

\bibitem{Takahashi:1985vx}
K.~Takahashi, \emph{{Perturbative calculations at finite temperatures}},
  \href{https://doi.org/10.1007/BF01551804}{\emph{Z. Phys. C} {\bfseries 26}
  (1985) 601}.

\bibitem{Fendley:1987ef}
P.~Fendley, \emph{{The Effective Potential and the Coupling Constant at High
  Temperature}},
  \href{https://doi.org/10.1016/0370-2693(87)90599-5}{\emph{Phys. Lett. B}
  {\bfseries 196} (1987) 175}.

\bibitem{Pisarski:1988vd}
R.D.~Pisarski, \emph{{Scattering Amplitudes in Hot Gauge Theories}},
  \href{https://doi.org/10.1103/PhysRevLett.63.1129}{\emph{Phys. Rev. Lett.}
  {\bfseries 63} (1989) 1129}.

\bibitem{Braaten:1989mz}
E.~Braaten and R.D.~Pisarski, \emph{{Soft Amplitudes in Hot Gauge Theories: A
  General Analysis}},
  \href{https://doi.org/10.1016/0550-3213(90)90508-B}{\emph{Nucl. Phys. B}
  {\bfseries 337} (1990) 569}.

\bibitem{Carrington:1991hz}
M.E.~Carrington, \emph{{The Effective potential at finite temperature in the
  Standard Model}}, \href{https://doi.org/10.1103/PhysRevD.45.2933}{\emph{Phys.
  Rev. D} {\bfseries 45} (1992) 2933}.

\bibitem{Parwani:1991gq}
R.R.~Parwani, \emph{{Resummation in a hot scalar field theory}},
  \href{https://doi.org/10.1103/PhysRevD.45.4695}{\emph{Phys. Rev. D}
  {\bfseries 45} (1992) 4695}
  [\href{https://arxiv.org/abs/hep-ph/9204216}{{\ttfamily hep-ph/9204216}}],
  [Erratum:
  \href{https://doi.org/10.1103/PhysRevD.48.5965.2}{\textit{Phys.~Rev.~D}
  \textbf{48} (1993) 5965}].

\bibitem{Arnold:1992rz}
P.B.~Arnold and O.~Espinosa, \emph{{The Effective potential and first order
  phase transitions: Beyond leading-order}},
  \href{https://doi.org/10.1103/PhysRevD.47.3546}{\emph{Phys. Rev. D}
  {\bfseries 47} (1993) 3546}
  [\href{https://arxiv.org/abs/hep-ph/9212235}{{\ttfamily hep-ph/9212235}}],
  [Erratum:
  \href{https://doi.org/10.1103/physrevd.50.6662.2}{\textit{Phys.~Rev.~D}
  \textbf{50} (1994) 6662}].

\bibitem{Espinosa:1992kf}
J.R.~Espinosa, M.~Quiros and F.~Zwirner, \emph{{On the nature of the
  electroweak phase transition}},
  \href{https://doi.org/10.1016/0370-2693(93)90450-V}{\emph{Phys. Lett. B}
  {\bfseries 314} (1993) 206}
  [\href{https://arxiv.org/abs/hep-ph/9212248}{{\ttfamily hep-ph/9212248}}].

\bibitem{Lofgren:2023sep}
J.~L\"ofgren, \emph{{Stop comparing resummation methods}},
  \href{https://arxiv.org/abs/2301.05197}{{\ttfamily 2301.05197}}.

\bibitem{Brahm:1991nh}
D.E.~Brahm and S.D.H.~Hsu, \emph{{Infrared divergences and the electroweak
  phase transition}}, .

\bibitem{Brahm:1991rr}
D.E.~Brahm and S.D.H.~Hsu, \emph{{Infrared divergences, finite temperature
  effective actions and the electroweak phase transition}}, .

\bibitem{Espinosa:1992gq}
J.R.~Espinosa, M.~Quiros and F.~Zwirner, \emph{{On the phase transition in the
  scalar theory}},
  \href{https://doi.org/10.1016/0370-2693(92)90129-R}{\emph{Phys. Lett. B}
  {\bfseries 291} (1992) 115}
  [\href{https://arxiv.org/abs/hep-ph/9206227}{{\ttfamily hep-ph/9206227}}].

\bibitem{Quiros:1992ez}
M.~Quiros, \emph{{On daisy and superdaisy resummation of the effective
  potential at finite temperature}},  in \emph{{4th Hellenic School on
  Elementary Particle Physics}}, pp.~502--511, 9, 1992
  [\href{https://arxiv.org/abs/hep-ph/9304284}{{\ttfamily hep-ph/9304284}}].

\bibitem{Dine:1992vs}
M.~Dine, R.G.~Leigh, P.~Huet, A.D.~Linde and D.A.~Linde, \emph{{Comments on the
  electroweak phase transition}},
  \href{https://doi.org/10.1016/0370-2693(92)90026-Z}{\emph{Phys. Lett. B}
  {\bfseries 283} (1992) 319}
  [\href{https://arxiv.org/abs/hep-ph/9203201}{{\ttfamily hep-ph/9203201}}].

\bibitem{Dine:1992wr}
M.~Dine, R.G.~Leigh, P.Y.~Huet, A.D.~Linde and D.A.~Linde, \emph{{Towards the
  theory of the electroweak phase transition}},
  \href{https://doi.org/10.1103/PhysRevD.46.550}{\emph{Phys. Rev. D} {\bfseries
  46} (1992) 550} [\href{https://arxiv.org/abs/hep-ph/9203203}{{\ttfamily
  hep-ph/9203203}}].

\bibitem{Boyd:1992xn}
C.G.~Boyd, D.E.~Brahm and S.D.H.~Hsu, \emph{{Corrections to the electroweak
  effective action at finite temperature}},
  \href{https://doi.org/10.1103/PhysRevD.48.4952}{\emph{Phys. Rev. D}
  {\bfseries 48} (1993) 4952}
  [\href{https://arxiv.org/abs/hep-ph/9206235}{{\ttfamily hep-ph/9206235}}].

\bibitem{Laine:2017hdk}
M.~Laine, M.~Meyer and G.~Nardini, \emph{{Thermal phase transition with full
  2-loop effective potential}},
  \href{https://doi.org/10.1016/j.nuclphysb.2017.04.023}{\emph{Nucl. Phys. B}
  {\bfseries 920} (2017) 565}
  [\href{https://arxiv.org/abs/1702.07479}{{\ttfamily 1702.07479}}].

\bibitem{Boyd:1993tz}
C.G.~Boyd, D.E.~Brahm and S.D.H.~Hsu, \emph{{Resummation methods at finite
  temperature: The Tadpole way}},
  \href{https://doi.org/10.1103/PhysRevD.48.4963}{\emph{Phys. Rev. D}
  {\bfseries 48} (1993) 4963}
  [\href{https://arxiv.org/abs/hep-ph/9304254}{{\ttfamily hep-ph/9304254}}].

\bibitem{Curtin:2022ovx}
D.~Curtin, J.~Roy and G.~White, \emph{{Gravitational waves and tadpole
  resummation: Efficient and easy convergence of finite temperature QFT}},
  \href{https://arxiv.org/abs/2211.08218}{{\ttfamily 2211.08218}}.

\bibitem{Senaha:2020mop}
E.~Senaha, \emph{{Symmetry Restoration and Breaking at Finite Temperature: An
  Introductory Review}},
  \href{https://doi.org/10.3390/sym12050733}{\emph{Symmetry} {\bfseries 12}
  (2020) 733}.

\bibitem{Kajantie:1995dw}
K.~Kajantie, M.~Laine, K.~Rummukainen and M.E.~Shaposhnikov, \emph{{Generic
  rules for high temperature dimensional reduction and their application to the
  standard model}},
  \href{https://doi.org/10.1016/0550-3213(95)00549-8}{\emph{Nucl. Phys. B}
  {\bfseries 458} (1996) 90}
  [\href{https://arxiv.org/abs/hep-ph/9508379}{{\ttfamily hep-ph/9508379}}].

\bibitem{Ginsparg:1980ef}
P.H.~Ginsparg, \emph{{First Order and Second Order Phase Transitions in Gauge
  Theories at Finite Temperature}},
  \href{https://doi.org/10.1016/0550-3213(80)90418-6}{\emph{Nucl. Phys. B}
  {\bfseries 170} (1980) 388}.

\bibitem{Appelquist:1981vg}
T.~Appelquist and R.D.~Pisarski, \emph{{High-Temperature Yang-Mills Theories
  and Three-Dimensional Quantum Chromodynamics}},
  \href{https://doi.org/10.1103/PhysRevD.23.2305}{\emph{Phys. Rev. D}
  {\bfseries 23} (1981) 2305}.

\bibitem{Nadkarni:1982kb}
S.~Nadkarni, \emph{{Dimensional Reduction in Hot QCD}},
  \href{https://doi.org/10.1103/PhysRevD.27.917}{\emph{Phys. Rev. D} {\bfseries
  27} (1983) 917}.

\bibitem{Farakos:1994kx}
K.~Farakos, K.~Kajantie, K.~Rummukainen and M.E.~Shaposhnikov, \emph{{3-D
  physics and the electroweak phase transition: Perturbation theory}},
  \href{https://doi.org/10.1016/0550-3213(94)90173-2}{\emph{Nucl. Phys. B}
  {\bfseries 425} (1994) 67}
  [\href{https://arxiv.org/abs/hep-ph/9404201}{{\ttfamily hep-ph/9404201}}].

\bibitem{Braaten:1995cm}
E.~Braaten and A.~Nieto, \emph{{Effective field theory approach to high
  temperature thermodynamics}},
  \href{https://doi.org/10.1103/PhysRevD.51.6990}{\emph{Phys. Rev. D}
  {\bfseries 51} (1995) 6990}
  [\href{https://arxiv.org/abs/hep-ph/9501375}{{\ttfamily hep-ph/9501375}}].

\bibitem{Ekstedt:2022bff}
A.~Ekstedt, P.~Schicho and T.V.I.~Tenkanen, \emph{{DRalgo: A package for
  effective field theory approach for thermal phase transitions}},
  \href{https://doi.org/10.1016/j.cpc.2023.108725}{\emph{Comput. Phys. Commun.}
  {\bfseries 288} (2023) 108725}
  [\href{https://arxiv.org/abs/2205.08815}{{\ttfamily 2205.08815}}].

\bibitem{Schicho:2021gca}
P.M.~Schicho, T.V.I.~Tenkanen and J.~\"Osterman, \emph{{Robust approach to
  thermal resummation: Standard Model meets a singlet}},
  \href{https://doi.org/10.1007/JHEP06(2021)130}{\emph{JHEP} {\bfseries 06}
  (2021) 130} [\href{https://arxiv.org/abs/2102.11145}{{\ttfamily
  2102.11145}}].

\bibitem{Kainulainen:2019kyp}
K.~Kainulainen, V.~Keus, L.~Niemi, K.~Rummukainen, T.V.I.~Tenkanen and
  V.~Vaskonen, \emph{{On the validity of perturbative studies of the
  electroweak phase transition in the Two Higgs Doublet model}},
  \href{https://doi.org/10.1007/JHEP06(2019)075}{\emph{JHEP} {\bfseries 06}
  (2019) 075} [\href{https://arxiv.org/abs/1904.01329}{{\ttfamily
  1904.01329}}].

\bibitem{Niemi:2020hto}
L.~Niemi, M.J.~Ramsey-Musolf, T.V.I.~Tenkanen and D.J.~Weir,
  \emph{{Thermodynamics of a Two-Step Electroweak Phase Transition}},
  \href{https://doi.org/10.1103/PhysRevLett.126.171802}{\emph{Phys. Rev. Lett.}
  {\bfseries 126} (2021) 171802}
  [\href{https://arxiv.org/abs/2005.11332}{{\ttfamily 2005.11332}}].

\bibitem{Gould:2021dzl}
O.~Gould, \emph{{Real scalar phase transitions: a nonperturbative analysis}},
  \href{https://doi.org/10.1007/JHEP04(2021)057}{\emph{JHEP} {\bfseries 04}
  (2021) 057} [\href{https://arxiv.org/abs/2101.05528}{{\ttfamily
  2101.05528}}].

\bibitem{Andersen:2017ika}
J.O.~Andersen, T.~Gorda, A.~Helset, L.~Niemi, T.V.I.~Tenkanen, A.~Tranberg
  et~al., \emph{{Nonperturbative Analysis of the Electroweak Phase Transition
  in the Two Higgs Doublet Model}},
  \href{https://doi.org/10.1103/PhysRevLett.121.191802}{\emph{Phys. Rev. Lett.}
  {\bfseries 121} (2018) 191802}
  [\href{https://arxiv.org/abs/1711.09849}{{\ttfamily 1711.09849}}].

\bibitem{Gould:2019qek}
O.~Gould, J.~Kozaczuk, L.~Niemi, M.J.~Ramsey-Musolf, T.V.I.~Tenkanen and
  D.J.~Weir, \emph{{Nonperturbative analysis of the gravitational waves from a
  first-order electroweak phase transition}},
  \href{https://doi.org/10.1103/PhysRevD.100.115024}{\emph{Phys. Rev. D}
  {\bfseries 100} (2019) 115024}
  [\href{https://arxiv.org/abs/1903.11604}{{\ttfamily 1903.11604}}].

\bibitem{Gould:2022ran}
O.~Gould, S.~G\"uyer and K.~Rummukainen, \emph{{First-order electroweak phase
  transitions: A nonperturbative update}},
  \href{https://doi.org/10.1103/PhysRevD.106.114507}{\emph{Phys. Rev. D}
  {\bfseries 106} (2022) 114507}
  [\href{https://arxiv.org/abs/2205.07238}{{\ttfamily 2205.07238}}].

\bibitem{Losada:1996ju}
M.~Losada, \emph{{High temperature dimensional reduction of the MSSM and other
  multiscalar models}},
  \href{https://doi.org/10.1103/PhysRevD.56.2893}{\emph{Phys. Rev. D}
  {\bfseries 56} (1997) 2893}
  [\href{https://arxiv.org/abs/hep-ph/9605266}{{\ttfamily hep-ph/9605266}}].

\bibitem{Farrar:1996cp}
G.R.~Farrar and M.~Losada, \emph{{SUSY and the electroweak phase transition}},
  \href{https://doi.org/10.1016/S0370-2693(97)00663-1}{\emph{Phys. Lett. B}
  {\bfseries 406} (1997) 60}
  [\href{https://arxiv.org/abs/hep-ph/9612346}{{\ttfamily hep-ph/9612346}}].

\bibitem{Cline:1997bm}
J.M.~Cline and K.~Kainulainen, \emph{{Supersymmetric electroweak phase
  transition: Dimensional reduction versus effective potential}},
  \href{https://doi.org/10.1016/S0550-3213(97)00570-1}{\emph{Nucl. Phys. B}
  {\bfseries 510} (1998) 88}
  [\href{https://arxiv.org/abs/hep-ph/9705201}{{\ttfamily hep-ph/9705201}}].

\bibitem{Bodeker:1996pc}
D.~Bodeker, P.~John, M.~Laine and M.G.~Schmidt, \emph{{The Two loop MSSM finite
  temperature effective potential with stop condensation}},
  \href{https://doi.org/10.1016/S0550-3213(97)00252-6}{\emph{Nucl. Phys. B}
  {\bfseries 497} (1997) 387}
  [\href{https://arxiv.org/abs/hep-ph/9612364}{{\ttfamily hep-ph/9612364}}].

\bibitem{Andersen:1998br}
J.O.~Andersen, \emph{{Dimensional reduction of the two Higgs doublet model at
  high temperature}}, \href{https://doi.org/10.1007/s100520050655}{\emph{Eur.
  Phys. J. C} {\bfseries 11} (1999) 563}
  [\href{https://arxiv.org/abs/hep-ph/9804280}{{\ttfamily hep-ph/9804280}}].

\bibitem{Gorda:2018hvi}
T.~Gorda, A.~Helset, L.~Niemi, T.V.I.~Tenkanen and D.J.~Weir,
  \emph{{Three-dimensional effective theories for the two Higgs doublet model
  at high temperature}},
  \href{https://doi.org/10.1007/JHEP02(2019)081}{\emph{JHEP} {\bfseries 02}
  (2019) 081} [\href{https://arxiv.org/abs/1802.05056}{{\ttfamily
  1802.05056}}].

\bibitem{Brauner:2016fla}
T.~Brauner, T.V.I.~Tenkanen, A.~Tranberg, A.~Vuorinen and D.J.~Weir,
  \emph{{Dimensional reduction of the Standard Model coupled to a new singlet
  scalar field}}, \href{https://doi.org/10.1007/JHEP03(2017)007}{\emph{JHEP}
  {\bfseries 03} (2017) 007}
  [\href{https://arxiv.org/abs/1609.06230}{{\ttfamily 1609.06230}}].

\bibitem{Niemi:2021qvp}
L.~Niemi, P.~Schicho and T.V.I.~Tenkanen, \emph{{Singlet-assisted electroweak
  phase transition at two loops}},
  \href{https://doi.org/10.1103/PhysRevD.103.115035}{\emph{Phys. Rev. D}
  {\bfseries 103} (2021) 115035}
  [\href{https://arxiv.org/abs/2103.07467}{{\ttfamily 2103.07467}}].

\bibitem{Niemi:2018asa}
L.~Niemi, H.H.~Patel, M.J.~Ramsey-Musolf, T.V.I.~Tenkanen and D.J.~Weir,
  \emph{{Electroweak phase transition in the real triplet extension of the SM:
  Dimensional reduction}},
  \href{https://doi.org/10.1103/PhysRevD.100.035002}{\emph{Phys. Rev. D}
  {\bfseries 100} (2019) 035002}
  [\href{https://arxiv.org/abs/1802.10500}{{\ttfamily 1802.10500}}].

\bibitem{Kang:1974yj}
J.S.~Kang, \emph{{Gauge Invariance of the Scalar-Vector Mass Ratio in the
  Coleman-Weinberg Model}},
  \href{https://doi.org/10.1103/PhysRevD.10.3455}{\emph{Phys. Rev. D}
  {\bfseries 10} (1974) 3455}.

\bibitem{Fischler:1974ue}
W.~Fischler and R.~Brout, \emph{{Gauge Invariance in Spontaneously Broken
  Symmetry}}, \href{https://doi.org/10.1103/PhysRevD.11.905}{\emph{Phys. Rev.
  D} {\bfseries 11} (1975) 905}.

\bibitem{Frere:1974ia}
J.M.~Frere and P.~Nicoletopoulos, \emph{{Gauge Invariant Content of the
  Effective Potential}},
  \href{https://doi.org/10.1103/PhysRevD.11.2332}{\emph{Phys. Rev. D}
  {\bfseries 11} (1975) 2332}.

\bibitem{Fukuda:1975di}
R.~Fukuda and T.~Kugo, \emph{{Gauge Invariance in the Effective Action and
  Potential}}, \href{https://doi.org/10.1103/PhysRevD.13.3469}{\emph{Phys. Rev.
  D} {\bfseries 13} (1976) 3469}.

\bibitem{Kobes:1990dc}
R.~Kobes, G.~Kunstatter and A.~Rebhan, \emph{{Gauge dependence identities and
  their application at finite temperature}},
  \href{https://doi.org/10.1016/0550-3213(91)90300-M}{\emph{Nucl. Phys. B}
  {\bfseries 355} (1991) 1}.

\bibitem{DelCima:1999gg}
O.M.~Del~Cima, D.H.T.~Franco and O.~Piguet, \emph{{Gauge independence of the
  effective potential revisited}},
  \href{https://doi.org/10.1016/S0550-3213(99)00226-6}{\emph{Nucl. Phys. B}
  {\bfseries 551} (1999) 813}
  [\href{https://arxiv.org/abs/hep-th/9902084}{{\ttfamily hep-th/9902084}}].

\bibitem{DelCima:1999dr}
O.M.~Del~Cima, \emph{{Probing the Nielsen identities}},
  \href{https://doi.org/10.1016/S0370-2693(99)00554-7}{\emph{Phys. Lett. B}
  {\bfseries 457} (1999) 307}
  [\href{https://arxiv.org/abs/hep-th/9903004}{{\ttfamily hep-th/9903004}}].

\bibitem{Alexander:2008hd}
L.P.~Alexander and A.~Pilaftsis, \emph{{The One-Loop Effective Potential in
  Non-Linear Gauges}},
  \href{https://doi.org/10.1088/0954-3899/36/4/045006}{\emph{J. Phys. G}
  {\bfseries 36} (2009) 045006}
  [\href{https://arxiv.org/abs/0809.1580}{{\ttfamily 0809.1580}}].

\bibitem{Nielsen:2014spa}
N.K.~Nielsen, \emph{{Removing the gauge parameter dependence of the effective
  potential by a field redefinition}},
  \href{https://doi.org/10.1103/PhysRevD.90.036008}{\emph{Phys. Rev. D}
  {\bfseries 90} (2014) 036008}
  [\href{https://arxiv.org/abs/1406.0788}{{\ttfamily 1406.0788}}].

\bibitem{Ekstedt:2018ftj}
A.~Ekstedt and J.~L\"ofgren, \emph{{On the relationship between gauge
  dependence and IR divergences in the $\hbar$-expansion of the effective
  potential}}, \href{https://doi.org/10.1007/JHEP01(2019)226}{\emph{JHEP}
  {\bfseries 01} (2019) 226}
  [\href{https://arxiv.org/abs/1810.01416}{{\ttfamily 1810.01416}}].

\bibitem{Ekstedt:2020abj}
A.~Ekstedt and J.~L\"ofgren, \emph{{A Critical Look at the Electroweak Phase
  Transition}}, \href{https://doi.org/10.1007/JHEP12(2020)136}{\emph{JHEP}
  {\bfseries 12} (2020) 136}
  [\href{https://arxiv.org/abs/2006.12614}{{\ttfamily 2006.12614}}].

\bibitem{Wainwright:2011qy}
C.~Wainwright, S.~Profumo and M.J.~Ramsey-Musolf, \emph{{Gravity Waves from a
  Cosmological Phase Transition: Gauge Artifacts and Daisy Resummations}},
  \href{https://doi.org/10.1103/PhysRevD.84.023521}{\emph{Phys. Rev. D}
  {\bfseries 84} (2011) 023521}
  [\href{https://arxiv.org/abs/1104.5487}{{\ttfamily 1104.5487}}].

\bibitem{Metaxas:1995ab}
D.~Metaxas and E.J.~Weinberg, \emph{{Gauge independence of the bubble
  nucleation rate in theories with radiative symmetry breaking}},
  \href{https://doi.org/10.1103/PhysRevD.53.836}{\emph{Phys. Rev. D} {\bfseries
  53} (1996) 836} [\href{https://arxiv.org/abs/hep-ph/9507381}{{\ttfamily
  hep-ph/9507381}}].

\bibitem{Garny:2012cg}
M.~Garny and T.~Konstandin, \emph{{On the gauge dependence of vacuum
  transitions at finite temperature}},
  \href{https://doi.org/10.1007/JHEP07(2012)189}{\emph{JHEP} {\bfseries 07}
  (2012) 189} [\href{https://arxiv.org/abs/1205.3392}{{\ttfamily 1205.3392}}].

\bibitem{Endo:2017gal}
M.~Endo, T.~Moroi, M.M.~Nojiri and Y.~Shoji, \emph{{On the Gauge Invariance of
  the Decay Rate of False Vacuum}},
  \href{https://doi.org/10.1016/j.physletb.2017.05.057}{\emph{Phys. Lett. B}
  {\bfseries 771} (2017) 281}
  [\href{https://arxiv.org/abs/1703.09304}{{\ttfamily 1703.09304}}].

\bibitem{Endo:2017tsz}
M.~Endo, T.~Moroi, M.M.~Nojiri and Y.~Shoji, \emph{{False Vacuum Decay in Gauge
  Theory}}, \href{https://doi.org/10.1007/JHEP11(2017)074}{\emph{JHEP}
  {\bfseries 11} (2017) 074}
  [\href{https://arxiv.org/abs/1704.03492}{{\ttfamily 1704.03492}}].

\bibitem{Lofgren:2020zzn}
J.~L\"ofgren, \emph{{The Powers of Perturbation Theory : Loops and Gauge
  Invariance in Particle Physics}}, Ph.D. thesis, Uppsala U., 2020.
\newblock
  \href{https://inspirehep.net/files/f61d73c5bc8f3b96a3266e26ed17288f}{https://inspirehep.net/files/f61d73c5bc8f3b96a3266e26ed17288f}.

\bibitem{Arunasalam:2021zrs}
S.~Arunasalam and M.J.~Ramsey-Musolf, \emph{{Tunneling potentials for the
  tunneling action: gauge invariance}},
  \href{https://doi.org/10.1007/JHEP08(2022)138}{\emph{JHEP} {\bfseries 08}
  (2022) 138} [\href{https://arxiv.org/abs/2105.07588}{{\ttfamily
  2105.07588}}].

\bibitem{Hirvonen:2021zej}
J.~Hirvonen, J.~L\"ofgren, M.J.~Ramsey-Musolf, P.~Schicho and T.V.I.~Tenkanen,
  \emph{{Computing the gauge-invariant bubble nucleation rate in finite
  temperature effective field theory}},
  \href{https://doi.org/10.1007/JHEP07(2022)135}{\emph{JHEP} {\bfseries 07}
  (2022) 135} [\href{https://arxiv.org/abs/2112.08912}{{\ttfamily
  2112.08912}}].

\bibitem{Lofgren:2021ogg}
J.~L\"ofgren, M.J.~Ramsey-Musolf, P.~Schicho and T.V.I.~Tenkanen,
  \emph{{Nucleation at Finite Temperature: A Gauge-Invariant Perturbative
  Framework}},
  \href{https://doi.org/10.1103/PhysRevLett.130.251801}{\emph{Phys. Rev. Lett.}
  {\bfseries 130} (2023) 251801}
  [\href{https://arxiv.org/abs/2112.05472}{{\ttfamily 2112.05472}}].

\bibitem{Croon:2021vtc}
D.~Croon, E.~Hall and H.~Murayama, \emph{{Non-perturbative methods for false
  vacuum decay}},  \href{https://arxiv.org/abs/2104.10687}{{\ttfamily
  2104.10687}}.

\bibitem{Farakos:1994xh}
K.~Farakos, K.~Kajantie, K.~Rummukainen and M.E.~Shaposhnikov, \emph{{3-d
  physics and the electroweak phase transition: A Framework for lattice Monte
  Carlo analysis}},
  \href{https://doi.org/10.1016/0550-3213(95)80129-4}{\emph{Nucl. Phys. B}
  {\bfseries 442} (1995) 317}
  [\href{https://arxiv.org/abs/hep-lat/9412091}{{\ttfamily hep-lat/9412091}}].

\bibitem{Schicho:2022wty}
P.~Schicho, T.V.I.~Tenkanen and G.~White, \emph{{Combining thermal resummation
  and gauge invariance for electroweak phase transition}},
  \href{https://doi.org/10.1007/JHEP11(2022)047}{\emph{JHEP} {\bfseries 11}
  (2022) 047} [\href{https://arxiv.org/abs/2203.04284}{{\ttfamily
  2203.04284}}].

\bibitem{Kirzhnits:1972ut}
D.A.~Kirzhnits and A.D.~Linde, \emph{{Macroscopic Consequences of the Weinberg
  Model}}, \href{https://doi.org/10.1016/0370-2693(72)90109-8}{\emph{Phys.
  Lett. B} {\bfseries 42} (1972) 471}.

\bibitem{Profumo:2007wc}
S.~Profumo, M.J.~Ramsey-Musolf and G.~Shaughnessy, \emph{{Singlet Higgs
  phenomenology and the electroweak phase transition}},
  \href{https://doi.org/10.1088/1126-6708/2007/08/010}{\emph{JHEP} {\bfseries
  08} (2007) 010} [\href{https://arxiv.org/abs/0705.2425}{{\ttfamily
  0705.2425}}].

\bibitem{Barger:2008jx}
V.~Barger, P.~Langacker, M.~McCaskey, M.~Ramsey-Musolf and G.~Shaughnessy,
  \emph{{Complex Singlet Extension of the Standard Model}},
  \href{https://doi.org/10.1103/PhysRevD.79.015018}{\emph{Phys. Rev. D}
  {\bfseries 79} (2009) 015018}
  [\href{https://arxiv.org/abs/0811.0393}{{\ttfamily 0811.0393}}].

\bibitem{Vaskonen:2016yiu}
V.~Vaskonen, \emph{{Electroweak baryogenesis and gravitational waves from a
  real scalar singlet}},
  \href{https://doi.org/10.1103/PhysRevD.95.123515}{\emph{Phys. Rev. D}
  {\bfseries 95} (2017) 123515}
  [\href{https://arxiv.org/abs/1611.02073}{{\ttfamily 1611.02073}}].

\bibitem{Chiang:2017nmu}
C.-W.~Chiang, M.J.~Ramsey-Musolf and E.~Senaha, \emph{{Standard Model with a
  Complex Scalar Singlet: Cosmological Implications and Theoretical
  Considerations}},
  \href{https://doi.org/10.1103/PhysRevD.97.015005}{\emph{Phys. Rev. D}
  {\bfseries 97} (2018) 015005}
  [\href{https://arxiv.org/abs/1707.09960}{{\ttfamily 1707.09960}}].

\bibitem{Ghorbani:2020xqv}
P.~Ghorbani, \emph{{Vacuum structure and electroweak phase transition in
  singlet scalar dark matter}},
  \href{https://doi.org/10.1016/j.dark.2021.100861}{\emph{Phys. Dark Univ.}
  {\bfseries 33} (2021) 100861}
  [\href{https://arxiv.org/abs/2010.15708}{{\ttfamily 2010.15708}}].

\bibitem{Maniatis:2006jd}
M.~Maniatis, A.~von Manteuffel and O.~Nachtmann, \emph{{Determining the global
  minimum of Higgs potentials via Groebner bases: Applied to the NMSSM}},
  \href{https://doi.org/10.1140/epjc/s10052-006-0186-2}{\emph{Eur. Phys. J. C}
  {\bfseries 49} (2007) 1067}
  [\href{https://arxiv.org/abs/hep-ph/0608314}{{\ttfamily hep-ph/0608314}}].

\bibitem{Maniatis:2012ex}
M.~Maniatis and D.~Mehta, \emph{{Minimizing Higgs Potentials via Numerical
  Polynomial Homotopy Continuation}},
  \href{https://doi.org/10.1140/epjp/i2012-12091-1}{\emph{Eur. Phys. J. Plus}
  {\bfseries 127} (2012) 91} [\href{https://arxiv.org/abs/1203.0409}{{\ttfamily
  1203.0409}}].

\bibitem{Camargo-Molina:2013qva}
J.E.~Camargo-Molina, B.~O'Leary, W.~Porod and F.~Staub,
  \emph{{$\mathbf{Vevacious}$: A Tool For Finding The Global Minima Of One-Loop
  Effective Potentials With Many Scalars}},
  \href{https://doi.org/10.1140/epjc/s10052-013-2588-2}{\emph{Eur. Phys. J. C}
  {\bfseries 73} (2013) 2588}
  [\href{https://arxiv.org/abs/1307.1477}{{\ttfamily 1307.1477}}].

\bibitem{Mehta:2011xs}
D.~Mehta, \emph{{Finding All the Stationary Points of a Potential Energy
  Landscape via Numerical Polynomial Homotopy Continuation Method}},
  \href{https://doi.org/10.1103/PhysRevE.84.025702}{\emph{Phys. Rev. E}
  {\bfseries 84} (2011) 025702}
  [\href{https://arxiv.org/abs/1104.5497}{{\ttfamily 1104.5497}}].

\bibitem{Mehta:2012qr}
D.~Mehta, J.D.~Hauenstein and M.~Kastner, \emph{{Energy landscape analysis of
  the two-dimensional nearest-neighbor $\phi^4$ model}},
  \href{https://doi.org/10.1103/PhysRevE.85.061103}{\emph{Phys. Rev. E}
  {\bfseries 85} (2012) 061103}
  [\href{https://arxiv.org/abs/1202.3320}{{\ttfamily 1202.3320}}].

\bibitem{Mehta:2011wj}
D.~Mehta, \emph{{Numerical Polynomial Homotopy Continuation Method and String
  Vacua}}, \href{https://doi.org/10.1155/2011/263937}{\emph{Adv. High Energy
  Phys.} {\bfseries 2011} (2011) 263937}
  [\href{https://arxiv.org/abs/1108.1201}{{\ttfamily 1108.1201}}].

\bibitem{AbdusSalam:2020rdj}
S.S.~AbdusSalam et~al., \emph{{Simple and statistically sound recommendations
  for analysing physical theories}},
  \href{https://doi.org/10.1088/1361-6633/ac60ac}{\emph{Rept. Prog. Phys.}
  {\bfseries 85} (2022) 052201}
  [\href{https://arxiv.org/abs/2012.09874}{{\ttfamily 2012.09874}}].

\bibitem{rowan1990functional}
T.H.~Rowan, \emph{Functional stability analysis of numerical algorithms}, The
  University of Texas at Austin (1990).

\bibitem{johnson2014nlopt}
S.G.~Johnson, \emph{The nlopt nonlinear-optimization package},  2014.

\bibitem{Basler:2020nrq}
P.~Basler, M.~M\"uhlleitner and J.~M\"uller, \emph{{BSMPT v2 a tool for the
  electroweak phase transition and the baryon asymmetry of the universe in
  extended Higgs Sectors}},
  \href{https://doi.org/10.1016/j.cpc.2021.108124}{\emph{Comput. Phys. Commun.}
  {\bfseries 269} (2021) 108124}
  [\href{https://arxiv.org/abs/2007.01725}{{\ttfamily 2007.01725}}].

\bibitem{wales2003energy}
D.~Wales et~al., \emph{Energy landscapes: Applications to clusters,
  biomolecules and glasses}, Cambridge University Press (2003).

\bibitem{Affleck:1979px}
I.K.~Affleck and F.~De~Luccia, \emph{{Induced vacuum decay}},
  \href{https://doi.org/10.1103/PhysRevD.20.3168}{\emph{Phys. Rev. D}
  {\bfseries 20} (1979) 3168}.

\bibitem{Samuel:1991zs}
D.A.~Samuel, \emph{{Numerical Analysis of Bubble Nucleation Processes for First
  - Order Phase Transitions within Quantum Fields}},  other thesis, Montana
  State University, 12, 1991,
  \href{http://wwwlib.umi.com/dissertations/fullcit?p9216657}{http://wwwlib.umi.com/dissertations/fullcit?p9216657}.

\bibitem{Christiansen:1995ic}
M.B.~Christiansen and J.~Madsen, \emph{{Large nucleation distances from
  impurities in the cosmological quark-hadron transition}},
  \href{https://doi.org/10.1103/PhysRevD.53.5446}{\emph{Phys. Rev. D}
  {\bfseries 53} (1996) 5446}
  [\href{https://arxiv.org/abs/astro-ph/9602071}{{\ttfamily
  astro-ph/9602071}}].

\bibitem{Shukla:2000dx}
P.~Shukla, A.K.~Mohanty, S.K.~Gupta and M.~Gleiser, \emph{{Inhomogeneous
  nucleation in quark hadron phase transition}},
  \href{https://doi.org/10.1103/PhysRevC.62.054904}{\emph{Phys. Rev. C}
  {\bfseries 62} (2000) 054904}
  [\href{https://arxiv.org/abs/hep-ph/0006071}{{\ttfamily hep-ph/0006071}}].

\bibitem{Ignatius:2000cz}
J.~Ignatius and D.J.~Schwarz, \emph{{The QCD phase transition in the
  inhomogeneous universe}},
  \href{https://doi.org/10.1103/PhysRevLett.86.2216}{\emph{Phys. Rev. Lett.}
  {\bfseries 86} (2001) 2216}
  [\href{https://arxiv.org/abs/hep-ph/0004259}{{\ttfamily hep-ph/0004259}}].

\bibitem{Megevand:2003tg}
A.~Megevand, \emph{{First order cosmological phase transitions in the radiation
  dominated era}},
  \href{https://doi.org/10.1103/PhysRevD.69.103521}{\emph{Phys. Rev. D}
  {\bfseries 69} (2004) 103521}
  [\href{https://arxiv.org/abs/hep-ph/0312305}{{\ttfamily hep-ph/0312305}}].

\bibitem{Schwarz:2003du}
D.J.~Schwarz, \emph{{The first second of the universe}},
  \href{https://doi.org/10.1002/andp.200310010}{\emph{Annalen Phys.} {\bfseries
  12} (2003) 220} [\href{https://arxiv.org/abs/astro-ph/0303574}{{\ttfamily
  astro-ph/0303574}}].

\bibitem{Moss:1984zf}
I.G.~Moss, \emph{{Black hole bubbles}},
  \href{https://doi.org/10.1103/PhysRevD.32.1333}{\emph{Phys. Rev. D}
  {\bfseries 32} (1985) 1333}.

\bibitem{Hiscock:1987hn}
W.A.~Hiscock, \emph{{Can black holes nucleate vacuum phase transitions?}},
  \href{https://doi.org/10.1103/PhysRevD.35.1161}{\emph{Phys. Rev. D}
  {\bfseries 35} (1987) 1161}.

\bibitem{Arnold:1989cq}
P.B.~Arnold, \emph{{Gravity and false vacuum decay rates: $O(3)$ solutions}},
  \href{https://doi.org/10.1016/0550-3213(90)90243-7}{\emph{Nucl. Phys. B}
  {\bfseries 346} (1990) 160}.

\bibitem{Berezin:1990qs}
V.A.~Berezin, V.A.~Kuzmin and I.I.~Tkachev, \emph{{Black holes initiate false
  vacuum decay}}, \href{https://doi.org/10.1103/PhysRevD.43.R3112}{\emph{Phys.
  Rev. D} {\bfseries 43} (1991) 3112}.

\bibitem{Samuel:1992wt}
D.A.~Samuel and W.A.~Hiscock, \emph{{Gravitationally compact objects as
  nucleation sites for first order vacuum phase transitions}},
  \href{https://doi.org/10.1103/PhysRevD.45.4411}{\emph{Phys. Rev. D}
  {\bfseries 45} (1992) 4411}.

\bibitem{Hiscock:1995ma}
W.A.~Hiscock, \emph{{Nucleation of vacuum phase transitions by topological
  defects}}, \href{https://doi.org/10.1016/0370-2693(95)01345-8}{\emph{Phys.
  Lett. B} {\bfseries 366} (1996) 77}
  [\href{https://arxiv.org/abs/gr-qc/9510003}{{\ttfamily gr-qc/9510003}}].

\bibitem{Gregory:2013hja}
R.~Gregory, I.G.~Moss and B.~Withers, \emph{{Black holes as bubble nucleation
  sites}}, \href{https://doi.org/10.1007/JHEP03(2014)081}{\emph{JHEP}
  {\bfseries 03} (2014) 081} [\href{https://arxiv.org/abs/1401.0017}{{\ttfamily
  1401.0017}}].

\bibitem{Burda:2015isa}
P.~Burda, R.~Gregory and I.~Moss, \emph{{Gravity and the stability of the Higgs
  vacuum}}, \href{https://doi.org/10.1103/PhysRevLett.115.071303}{\emph{Phys.
  Rev. Lett.} {\bfseries 115} (2015) 071303}
  [\href{https://arxiv.org/abs/1501.04937}{{\ttfamily 1501.04937}}].

\bibitem{Burda:2015yfa}
P.~Burda, R.~Gregory and I.~Moss, \emph{{Vacuum metastability with black
  holes}}, \href{https://doi.org/10.1007/JHEP08(2015)114}{\emph{JHEP}
  {\bfseries 08} (2015) 114}
  [\href{https://arxiv.org/abs/1503.07331}{{\ttfamily 1503.07331}}].

\bibitem{Tetradis:2016vqb}
N.~Tetradis, \emph{{Black holes and Higgs stability}},
  \href{https://doi.org/10.1088/1475-7516/2016/09/036}{\emph{JCAP} {\bfseries
  09} (2016) 036} [\href{https://arxiv.org/abs/1606.04018}{{\ttfamily
  1606.04018}}].

\bibitem{Canko:2017ebb}
D.~Canko, I.~Gialamas, G.~Jelic-Cizmek, A.~Riotto and N.~Tetradis, \emph{{On
  the Catalysis of the Electroweak Vacuum Decay by Black Holes at High
  Temperature}},
  \href{https://doi.org/10.1140/epjc/s10052-018-5808-y}{\emph{Eur. Phys. J. C}
  {\bfseries 78} (2018) 328}
  [\href{https://arxiv.org/abs/1706.01364}{{\ttfamily 1706.01364}}].

\bibitem{Mukaida:2017bgd}
K.~Mukaida and M.~Yamada, \emph{{False Vacuum Decay Catalyzed by Black Holes}},
  \href{https://doi.org/10.1103/PhysRevD.96.103514}{\emph{Phys. Rev. D}
  {\bfseries 96} (2017) 103514}
  [\href{https://arxiv.org/abs/1706.04523}{{\ttfamily 1706.04523}}].

\bibitem{Hayashi:2020ocn}
T.~Hayashi, K.~Kamada, N.~Oshita and J.~Yokoyama, \emph{{On catalyzed vacuum
  decay around a radiating black hole and the crisis of the electroweak
  vacuum}}, \href{https://doi.org/10.1007/JHEP08(2020)088}{\emph{JHEP}
  {\bfseries 08} (2020) 088}
  [\href{https://arxiv.org/abs/2005.12808}{{\ttfamily 2005.12808}}].

\bibitem{Shkerin:2021zbf}
A.~Shkerin and S.~Sibiryakov, \emph{{Black hole induced false vacuum decay from
  first principles}},
  \href{https://doi.org/10.1007/JHEP11(2021)197}{\emph{JHEP} {\bfseries 11}
  (2021) 197} [\href{https://arxiv.org/abs/2105.09331}{{\ttfamily
  2105.09331}}].

\bibitem{Strumia:2022jil}
A.~Strumia, \emph{{Black holes don\textquoteright{}t source fast Higgs vacuum
  decay}}, \href{https://doi.org/10.1007/JHEP03(2023)039}{\emph{JHEP}
  {\bfseries 03} (2023) 039}
  [\href{https://arxiv.org/abs/2209.05504}{{\ttfamily 2209.05504}}].

\bibitem{Kobzarev:1974cp}
I.Y.~Kobzarev, L.B.~Okun and M.B.~Voloshin, \emph{{Bubbles in Metastable
  Vacuum}}, {\emph{Yad. Fiz.} {\bfseries 20} (1974) 1229}
  \href{{https://inis.iaea.org/search/search.aspx?orig\_q=RN:7276129}}{{https://inis.iaea.org/search/search.aspx?orig\_q=RN:7276129}}.

\bibitem{Callan:1977pt}
C.G.~Callan, Jr. and S.R.~Coleman, \emph{{The Fate of the False Vacuum. 2.
  First Quantum Corrections}},
  \href{https://doi.org/10.1103/PhysRevD.16.1762}{\emph{Phys. Rev. D}
  {\bfseries 16} (1977) 1762}.

\bibitem{Linde:1977mm}
A.D.~Linde, \emph{{On the Vacuum Instability and the Higgs Meson Mass}},
  \href{https://doi.org/10.1016/0370-2693(77)90664-5}{\emph{Phys. Lett. B}
  {\bfseries 70} (1977) 306}.

\bibitem{Linde:1980tt}
A.D.~Linde, \emph{{Fate of the False Vacuum at Finite Temperature: Theory and
  Applications}},
  \href{https://doi.org/10.1016/0370-2693(81)90281-1}{\emph{Phys. Lett. B}
  {\bfseries 100} (1981) 37}.

\bibitem{Linde:1981zj}
A.D.~Linde, \emph{{Decay of the False Vacuum at Finite Temperature}},
  \href{https://doi.org/10.1016/0550-3213(83)90072-X}{\emph{Nucl. Phys. B}
  {\bfseries 216} (1983) 421} [Erratum:
  \href{https://doi.org/10.1016/0550-3213(83)90293-6}{\textit{Nucl. Phys. B}
  \textbf{223} (1983) 544}].

\bibitem{Citron:2013bje}
M.~Citron, \emph{{Spontaneous and induced false vacuum decay}},  Master's
  thesis, Imperial College London, 2013,
  \href{https://inspirehep.net/files/a4ecfe3a3f4b9eed3ec001d4aa5cd6f4}{https://inspirehep.net/files/a4ecfe3a3f4b9eed3ec001d4aa5cd6f4}.

\bibitem{Masoumi:2013nix}
A.~Masoumi, \emph{{Topics in vacuum decay}}, Ph.D. thesis, Columbia U., 2013.
\newblock \href{https://arxiv.org/abs/1505.06397}{{\ttfamily 1505.06397}}.
\newblock [\href{https://arxiv.org/abs/1505.06397}{{\ttfamily 1505.06397}}].

\bibitem{Lee:2014yud}
H.J.~Lee, \emph{{Negative Modes in Vacuum Decay}}, Ph.D. thesis, Columbia U.,
  2014.
\newblock \href{https://doi.org/10.7916/D84X55Z3}{10.7916/D84X55Z3}.

\bibitem{hirvonen}
J.~Hirvonen, \emph{Nucleation rate in a radiatively induced first-order phase
  transition},  Master's thesis, University of Helsinki, 2020,
  \href{https://helda.helsinki.fi/handle/10138/318464}{https://helda.helsinki.fi/handle/10138/318464}.

\bibitem{Guada:2021lnb}
V.~Guada, \emph{{False vacuum decay with multiple scalar fields}}, Ph.D.
  thesis, Ljubljana U., 2021.
\newblock
  \href{https://inspirehep.net/files/ed7319cd5a150fdd6fd5026c113b1b60}{https://inspirehep.net/files/ed7319cd5a150fdd6fd5026c113b1b60}.

\bibitem{Wickens:2021vzr}
A.~Wickens, \emph{{Vacuum decay, gravitational waves and magnetic fields in the
  early universe}}, Ph.D. thesis, King's Coll. London, 2021.
\newblock
  \href{https://inspirehep.net/files/72146bfa2e39a7afd9263f99a27fc3aa}{https://inspirehep.net/files/72146bfa2e39a7afd9263f99a27fc3aa}.

\bibitem{Shen:2022xcm}
J.~Shen, P.~Draper and A.X.~El-Khadra, \emph{{Vacuum decay and Euclidean
  lattice Monte~Carlo}},
  \href{https://doi.org/10.1103/PhysRevD.107.094506}{\emph{Phys. Rev. D}
  {\bfseries 107} (2023) 094506}
  [\href{https://arxiv.org/abs/2210.05925}{{\ttfamily 2210.05925}}].

\bibitem{Coleman:1978ae}
S.R.~Coleman, \emph{{The Uses of Instantons}}, {\emph{Subnucl. Ser.} {\bfseries
  15} (1979) 805}
  \href{https://lib-extopc.kek.jp/preprints/PDF/1978/7805/7805043.pdf}{https://lib-extopc.kek.jp/preprints/PDF/1978/7805/7805043.pdf}.

\bibitem{PhysRevD.19.667}
C.L.~Hammer, J.E.~Shrauner and B.~De~Facio, \emph{Alternate derivation of
  vacuum tunneling}, \href{https://doi.org/10.1103/PhysRevD.19.667}{\emph{Phys.
  Rev. D} {\bfseries 19} (1979) 667}.

\bibitem{Andreassen:2016cff}
A.~Andreassen, D.~Farhi, W.~Frost and M.D.~Schwartz, \emph{{Direct Approach to
  Quantum Tunneling}},
  \href{https://doi.org/10.1103/PhysRevLett.117.231601}{\emph{Phys. Rev. Lett.}
  {\bfseries 117} (2016) 231601}
  [\href{https://arxiv.org/abs/1602.01102}{{\ttfamily 1602.01102}}].

\bibitem{Blum:2023wnb}
K.~Blum and O.~Rosner, \emph{{Unraveling the bounce: a real time perspective on
  tunneling}},  \href{https://arxiv.org/abs/2309.07585}{{\ttfamily
  2309.07585}}.

\bibitem{batini2023realtime}
L.~Batini, A.~Chatrchyan and J.~Berges, \emph{Real-time dynamics of false
  vacuum decay},  \href{https://arxiv.org/abs/2310.04206}{{\ttfamily
  2310.04206}}.

\bibitem{KRAMERS1940284}
H.~Kramers, \emph{Brownian motion in a field of force and the diffusion model
  of chemical reactions},
  \href{https://doi.org/10.1016/S0031-8914(40)90098-2}{\emph{Physica}
  {\bfseries 7} (1940) 284}.

\bibitem{LANGER197461}
J.S.~Langer, \emph{Metastable states},
  \href{https://doi.org/10.1016/0031-8914(74)90226-2}{\emph{Physica} {\bfseries
  73} (1974) 61}.

\bibitem{Paranjape:2017}
M.~Paranjape, \emph{The Theory and Applications of Instanton Calculations},
  Cambridge University Press (Nov., 2017),
  \href{https://doi.org/10.1017/9781316658741}{10.1017/9781316658741}.

\bibitem{marino_2015}
M.~Mariño, \emph{Instantons and Large $N$: An Introduction to Non-Perturbative
  Methods in Quantum Field Theory}, Cambridge University Press (2015),
  \href{https://doi.org/10.1017/CBO9781107705968}{10.1017/CBO9781107705968}.

\bibitem{Arnold:1987mh}
P.B.~Arnold and L.D.~McLerran, \emph{{Sphalerons, Small Fluctuations and Baryon
  Number Violation in Electroweak Theory}},
  \href{https://doi.org/10.1103/PhysRevD.36.581}{\emph{Phys. Rev. D} {\bfseries
  36} (1987) 581}.

\bibitem{dau}
L.D.~Landau and E.~Lifshitz, \emph{Quantum Mechanics --- Nonrelativistic
  Theory}, vol.~III of \emph{Course of theoretical physics}, Pergamon Press,
  3rd~ed. (1977).

\bibitem{Andreassen:2016cvx}
A.~Andreassen, D.~Farhi, W.~Frost and M.D.~Schwartz, \emph{{Precision decay
  rate calculations in quantum field theory}},
  \href{https://doi.org/10.1103/PhysRevD.95.085011}{\emph{Phys. Rev. D}
  {\bfseries 95} (2017) 085011}
  [\href{https://arxiv.org/abs/1604.06090}{{\ttfamily 1604.06090}}].

\bibitem{Coleman:1987rm}
S.R.~Coleman, \emph{{Quantum Tunneling and Negative Eigenvalues}},
  \href{https://doi.org/10.1016/0550-3213(88)90308-2}{\emph{Nucl. Phys. B}
  {\bfseries 298} (1988) 178}.

\bibitem{Affleck:1980ac}
I.~Affleck, \emph{{Quantum Statistical Metastability}},
  \href{https://doi.org/10.1103/PhysRevLett.46.388}{\emph{Phys. Rev. Lett.}
  {\bfseries 46} (1981) 388}.

\bibitem{Berera:2019uyp}
A.~Berera, J.~Mabillard, B.W.~Mintz and R.O.~Ramos, \emph{{Formulating the
  Kramers problem in field theory}},
  \href{https://doi.org/10.1103/PhysRevD.100.076005}{\emph{Phys. Rev. D}
  {\bfseries 100} (2019) 076005}
  [\href{https://arxiv.org/abs/1906.08684}{{\ttfamily 1906.08684}}].

\bibitem{Gleiser:1993hf}
M.~Gleiser, G.C.~Marques and R.O.~Ramos, \emph{{On the evaluation of thermal
  corrections to false vacuum decay rates}},
  \href{https://doi.org/10.1103/PhysRevD.48.1571}{\emph{Phys. Rev. D}
  {\bfseries 48} (1993) 1571}
  [\href{https://arxiv.org/abs/hep-ph/9304234}{{\ttfamily hep-ph/9304234}}].

\bibitem{RevModPhys.62.251}
P.~H\"anggi, P.~Talkner and M.~Borkovec, \emph{Reaction-rate theory: fifty
  years after kramers},
  \href{https://doi.org/10.1103/RevModPhys.62.251}{\emph{Rev. Mod. Phys.}
  {\bfseries 62} (1990) 251}.

\bibitem{Coleman:1977th}
S.R.~Coleman, V.~Glaser and A.~Martin, \emph{{Action Minima Among Solutions to
  a Class of Euclidean Scalar Field Equations}},
  \href{https://doi.org/10.1007/BF01609421}{\emph{Commun. Math. Phys.}
  {\bfseries 58} (1978) 211}.

\bibitem{Blum:2016ipp}
K.~Blum, M.~Honda, R.~Sato, M.~Takimoto and K.~Tobioka, \emph{{O($N$)
  Invariance of the Multi-Field Bounce}},
  \href{https://doi.org/10.1007/JHEP05(2017)109}{\emph{JHEP} {\bfseries 05}
  (2017) 109} [\href{https://arxiv.org/abs/1611.04570}{{\ttfamily
  1611.04570}}], [Erratum:
  \href{https://doi.org/10.1007/JHEP06(2017)060}{\textit{JHEP} \textbf{06}
  (2017) 060}].

\bibitem{Derrick:1964ww}
G.H.~Derrick, \emph{{Comments on nonlinear wave equations as models for
  elementary particles}}, \href{https://doi.org/10.1063/1.1704233}{\emph{J.
  Math. Phys.} {\bfseries 5} (1964) 1252}.

\bibitem{Mukhanov:2021kat}
V.F.~Mukhanov and A.S.~Sorin, \emph{{On the existence of the Coleman
  instantons}},
  \href{https://doi.org/10.1088/1475-7516/2021/10/066}{\emph{JCAP} {\bfseries
  10} (2021) 066} [\href{https://arxiv.org/abs/2104.12661}{{\ttfamily
  2104.12661}}].

\bibitem{Mukhanov:2020pau}
V.F.~Mukhanov, E.~Rabinovici and A.S.~Sorin, \emph{{Quantum Fluctuations and
  New Instantons I: Linear Unbounded Potential}},
  \href{https://doi.org/10.1002/prop.202000100}{\emph{Fortsch. Phys.}
  {\bfseries 69} (2021) 2000100}
  [\href{https://arxiv.org/abs/2009.12445}{{\ttfamily 2009.12445}}].

\bibitem{Mukhanov:2020wim}
V.~Mukhanov, E.~Rabinovici and A.~Sorin, \emph{{Quantum Fluctuations and New
  Instantons II: Quartic Unbounded Potential}},
  \href{https://doi.org/10.1002/prop.202000101}{\emph{Fortsch. Phys.}
  {\bfseries 69} (2021) 2000101}
  [\href{https://arxiv.org/abs/2009.12444}{{\ttfamily 2009.12444}}].

\bibitem{Espinosa:2019hbm}
J.R.~Espinosa, \emph{{Tunneling without Bounce}},
  \href{https://doi.org/10.1103/PhysRevD.100.105002}{\emph{Phys. Rev. D}
  {\bfseries 100} (2019) 105002}
  [\href{https://arxiv.org/abs/1908.01730}{{\ttfamily 1908.01730}}].

\bibitem{Espinosa:2021qeo}
J.R.~Espinosa and J.~Huertas, \emph{{Pseudo-bounces vs. new instantons}},
  \href{https://doi.org/10.1088/1475-7516/2021/12/029}{\emph{JCAP} {\bfseries
  12} (2021) 029} [\href{https://arxiv.org/abs/2106.04541}{{\ttfamily
  2106.04541}}].

\bibitem{Masoumi:2016wot}
A.~Masoumi, K.D.~Olum and B.~Shlaer, \emph{{Efficient numerical solution to
  vacuum decay with many fields}},
  \href{https://doi.org/10.1088/1475-7516/2017/01/051}{\emph{JCAP} {\bfseries
  01} (2017) 051} [\href{https://arxiv.org/abs/1610.06594}{{\ttfamily
  1610.06594}}].

\bibitem{Fubini_1976}
S.~Fubini, \emph{A new approach to conformal invariant field theories},
  \href{https://doi.org/10.1007/bf02785664}{\emph{Il Nuovo Cimento A}
  {\bfseries 34} (1976) 521}.

\bibitem{Lipatov:1976ny}
L.N.~Lipatov, \emph{{Divergence of the Perturbation Theory Series and the
  Quasiclassical Theory}}, {\emph{Sov. Phys. JETP} {\bfseries 45} (1977) 216}
  \href{{{http://jetp.ras.ru/cgi-bin/dn/e\_045\_02\_0216.pdf}}}{{{http://jetp.ras.ru/cgi-bin/dn/e\_045\_02\_0216.pdf}}}.

\bibitem{Aravind:2014pva}
A.~Aravind, B.S.~DiNunno, D.~Lorshbough and S.~Paban, \emph{{Analyzing
  multifield tunneling with exact bounce solutions}},
  \href{https://doi.org/10.1103/PhysRevD.91.025026}{\emph{Phys. Rev. D}
  {\bfseries 91} (2015) 025026}
  [\href{https://arxiv.org/abs/1412.3160}{{\ttfamily 1412.3160}}].

\bibitem{FerrazdeCamargo:1982sk}
A.~Ferraz~de Camargo, R.C.~Shellard and G.C.~Marques, \emph{{Vacuum Decay in a
  Soluble Model}}, \href{https://doi.org/10.1103/PhysRevD.29.1147}{\emph{Phys.
  Rev. D} {\bfseries 29} (1984) 1147}.

\bibitem{DUNCAN1992109}
M.~Duncan and L.G.~Jensen, \emph{Exact tunnelling solutions in scalar field
  theory},
  \href{https://doi.org/https://doi.org/10.1016/0370-2693(92)90128-Q}{\emph{Physics
  Letters B} {\bfseries 291} (1992) 109}.

\bibitem{Hamazaki:1995dy}
T.~Hamazaki, M.~Sasaki, T.~Tanaka and K.~Yamamoto, \emph{{Selfexcitation of the
  tunneling scalar field in false vacuum decay}},
  \href{https://doi.org/10.1103/PhysRevD.53.2045}{\emph{Phys. Rev. D}
  {\bfseries 53} (1996) 2045}
  [\href{https://arxiv.org/abs/gr-qc/9507006}{{\ttfamily gr-qc/9507006}}].

\bibitem{Pastras:2011zr}
G.~Pastras, \emph{{Exact Tunneling Solutions in Minkowski Spacetime and a
  Candidate for Dark Energy}},
  \href{https://doi.org/10.1007/JHEP08(2013)075}{\emph{JHEP} {\bfseries 08}
  (2013) 075} [\href{https://arxiv.org/abs/1102.4567}{{\ttfamily 1102.4567}}].

\bibitem{Dutta:2011ej}
K.~Dutta, P.M.~Vaudrevange and A.~Westphal, \emph{{An Exact Tunneling Solution
  in a Simple Realistic Landscape}},
  \href{https://doi.org/10.1088/0264-9381/29/6/065011}{\emph{Class. Quant.
  Grav.} {\bfseries 29} (2012) 065011}
  [\href{https://arxiv.org/abs/1102.4742}{{\ttfamily 1102.4742}}].

\bibitem{Lee:1985uv}
K.-M.~Lee and E.J.~Weinberg, \emph{{TUNNELING WITHOUT BARRIERS}},
  \href{https://doi.org/10.1016/0550-3213(86)90150-1}{\emph{Nucl. Phys. B}
  {\bfseries 267} (1986) 181}.

\bibitem{Dutta:2011rc}
K.~Dutta, C.~Hector, P.M.~Vaudrevange and A.~Westphal, \emph{{More Exact
  Tunneling Solutions in Scalar Field Theory}},
  \href{https://doi.org/10.1016/j.physletb.2012.01.026}{\emph{Phys. Lett. B}
  {\bfseries 708} (2012) 309}
  [\href{https://arxiv.org/abs/1110.2380}{{\ttfamily 1110.2380}}].

\bibitem{Masoumi:2017trx}
A.~Masoumi, K.D.~Olum and J.M.~Wachter, \emph{{Approximating tunneling rates in
  multi-dimensional field spaces}},
  \href{https://doi.org/10.1088/1475-7516/2017/10/022}{\emph{JCAP} {\bfseries
  10} (2017) 022} [\href{https://arxiv.org/abs/1702.00356}{{\ttfamily
  1702.00356}}].

\bibitem{Athron:2019nbd}
P.~Athron, C.~Bal\'azs, M.~Bardsley, A.~Fowlie, D.~Harries and G.~White,
  \emph{{BubbleProfiler: finding the field profile and action for cosmological
  phase transitions}},
  \href{https://doi.org/10.1016/j.cpc.2019.05.017}{\emph{Comput. Phys. Commun.}
  {\bfseries 244} (2019) 448}
  [\href{https://arxiv.org/abs/1901.03714}{{\ttfamily 1901.03714}}].

\bibitem{Guada:2020xnz}
V.~Guada, M.~Nemev\v{s}ek and M.~Pintar, \emph{{FindBounce: Package for
  multi-field bounce actions}},
  \href{https://doi.org/10.1016/j.cpc.2020.107480}{\emph{Comput. Phys. Commun.}
  {\bfseries 256} (2020) 107480}
  [\href{https://arxiv.org/abs/2002.00881}{{\ttfamily 2002.00881}}].

\bibitem{Sato:2019axv}
R.~Sato, \emph{{Simple Gradient Flow Equation for the Bounce Solution}},
  \href{https://doi.org/10.1103/PhysRevD.101.016012}{\emph{Phys. Rev. D}
  {\bfseries 101} (2020) 016012}
  [\href{https://arxiv.org/abs/1907.02417}{{\ttfamily 1907.02417}}].

\bibitem{Sato:2019wpo}
R.~Sato, \emph{{SimpleBounce : a simple package for the false vacuum decay}},
  \href{https://doi.org/10.1016/j.cpc.2020.107566}{\emph{Comput. Phys. Commun.}
  {\bfseries 258} (2021) 107566}
  [\href{https://arxiv.org/abs/1908.10868}{{\ttfamily 1908.10868}}].

\bibitem{Bardsley:2021lmq}
M.~Bardsley, \emph{{An optimisation based algorithm for finding the nucleation
  temperature of cosmological phase transitions}},
  \href{https://doi.org/10.1016/j.cpc.2021.108252}{\emph{Comput. Phys. Commun.}
  {\bfseries 273} (2022) 108252}
  [\href{https://arxiv.org/abs/2103.01985}{{\ttfamily 2103.01985}}].

\bibitem{Kusenko:1995jv}
A.~Kusenko, \emph{{Improved action method for analyzing tunneling in quantum
  field theory}},
  \href{https://doi.org/10.1016/0370-2693(95)00994-V}{\emph{Phys. Lett. B}
  {\bfseries 358} (1995) 51}
  [\href{https://arxiv.org/abs/hep-ph/9504418}{{\ttfamily hep-ph/9504418}}].

\bibitem{Cline:1999wi}
J.M.~Cline, G.D.~Moore and G.~Servant, \emph{{Was the electroweak phase
  transition preceded by a color broken phase?}},
  \href{https://doi.org/10.1103/PhysRevD.60.105035}{\emph{Phys. Rev. D}
  {\bfseries 60} (1999) 105035}
  [\href{https://arxiv.org/abs/hep-ph/9902220}{{\ttfamily hep-ph/9902220}}].

\bibitem{Espinosa:2018hue}
J.R.~Espinosa, \emph{{A Fresh Look at the Calculation of Tunneling Actions}},
  \href{https://doi.org/10.1088/1475-7516/2018/07/036}{\emph{JCAP} {\bfseries
  07} (2018) 036} [\href{https://arxiv.org/abs/1805.03680}{{\ttfamily
  1805.03680}}].

\bibitem{Espinosa:2018voj}
J.R.~Espinosa, \emph{{Fresh look at the calculation of tunneling actions
  including gravitational effects}},
  \href{https://doi.org/10.1103/PhysRevD.100.104007}{\emph{Phys. Rev. D}
  {\bfseries 100} (2019) 104007}
  [\href{https://arxiv.org/abs/1808.00420}{{\ttfamily 1808.00420}}].

\bibitem{Espinosa:2018szu}
J.R.~Espinosa and T.~Konstandin, \emph{{A Fresh Look at the Calculation of
  Tunneling Actions in Multi-Field Potentials}},
  \href{https://doi.org/10.1088/1475-7516/2019/01/051}{\emph{JCAP} {\bfseries
  01} (2019) 051} [\href{https://arxiv.org/abs/1811.09185}{{\ttfamily
  1811.09185}}].

\bibitem{Ekstedt:2023sqc}
A.~Ekstedt, O.~Gould and J.~Hirvonen, \emph{{BubbleDet: A Python package to
  compute functional determinants for bubble nucleation}},
  \href{https://arxiv.org/abs/2308.15652}{{\ttfamily 2308.15652}}.

\bibitem{Csernai:1992tj}
L.P.~Csernai and J.I.~Kapusta, \emph{{Nucleation of relativistic first order
  phase transitions}},
  \href{https://doi.org/10.1103/PhysRevD.46.1379}{\emph{Phys. Rev. D}
  {\bfseries 46} (1992) 1379}.

\bibitem{Ekstedt:2022tqk}
A.~Ekstedt, \emph{{Bubble nucleation to all orders}},
  \href{https://doi.org/10.1007/JHEP08(2022)115}{\emph{JHEP} {\bfseries 08}
  (2022) 115} [\href{https://arxiv.org/abs/2201.07331}{{\ttfamily
  2201.07331}}].

\bibitem{Dunne:2007rt}
G.V.~Dunne, \emph{{Functional determinants in quantum field theory}},
  \href{https://doi.org/10.1088/1751-8113/41/30/304006}{\emph{J. Phys. A}
  {\bfseries 41} (2008) 304006}
  [\href{https://arxiv.org/abs/0711.1178}{{\ttfamily 0711.1178}}].

\bibitem{Carson:1990jm}
L.~Carson, X.~Li, L.D.~McLerran and R.-T.~Wang, \emph{{Exact Computation of the
  Small Fluctuation Determinant Around a Sphaleron}},
  \href{https://doi.org/10.1103/PhysRevD.42.2127}{\emph{Phys. Rev. D}
  {\bfseries 42} (1990) 2127}.

\bibitem{Guada:2020ihz}
V.~Guada and M.~Nemev\v{s}ek, \emph{{Exact one-loop false vacuum decay rate}},
  \href{https://doi.org/10.1103/PhysRevD.102.125017}{\emph{Phys. Rev. D}
  {\bfseries 102} (2020) 125017}
  [\href{https://arxiv.org/abs/2009.01535}{{\ttfamily 2009.01535}}].

\bibitem{Gelfand:1959nq}
I.M.~Gelfand and A.M.~Yaglom, \emph{{Integration in functional spaces and it
  applications in quantum physics}},
  \href{https://doi.org/10.1063/1.1703636}{\emph{J. Math. Phys.} {\bfseries 1}
  (1960) 48}.

\bibitem{Dunne:2005rt}
G.V.~Dunne and H.~Min, \emph{{Beyond the thin-wall approximation: Precise
  numerical computation of prefactors in false vacuum decay}},
  \href{https://doi.org/10.1103/PhysRevD.72.125004}{\emph{Phys. Rev. D}
  {\bfseries 72} (2005) 125004}
  [\href{https://arxiv.org/abs/hep-th/0511156}{{\ttfamily hep-th/0511156}}].

\bibitem{Dunne:2006ct}
G.V.~Dunne and K.~Kirsten, \emph{{Functional determinants for radial
  operators}}, \href{https://doi.org/10.1088/0305-4470/39/38/017}{\emph{J.
  Phys. A} {\bfseries 39} (2006) 11915}
  [\href{https://arxiv.org/abs/hep-th/0607066}{{\ttfamily hep-th/0607066}}].

\bibitem{Hur:2008yg}
J.~Hur and H.~Min, \emph{{A Fast Way to Compute Functional Determinants of
  Radially Symmetric Partial Differential Operators in General Dimensions}},
  \href{https://doi.org/10.1103/PhysRevD.77.125033}{\emph{Phys. Rev. D}
  {\bfseries 77} (2008) 125033}
  [\href{https://arxiv.org/abs/0805.0079}{{\ttfamily 0805.0079}}].

\bibitem{Gould:2021ccf}
O.~Gould and J.~Hirvonen, \emph{{Effective field theory approach to thermal
  bubble nucleation}},
  \href{https://doi.org/10.1103/PhysRevD.104.096015}{\emph{Phys. Rev. D}
  {\bfseries 104} (2021) 096015}
  [\href{https://arxiv.org/abs/2108.04377}{{\ttfamily 2108.04377}}].

\bibitem{Weinberg:1992ds}
E.J.~Weinberg, \emph{{Vacuum decay in theories with symmetry breaking by
  radiative corrections}},
  \href{https://doi.org/10.1103/PhysRevD.47.4614}{\emph{Phys. Rev. D}
  {\bfseries 47} (1993) 4614}
  [\href{https://arxiv.org/abs/hep-ph/9211314}{{\ttfamily hep-ph/9211314}}].

\bibitem{Strumia:1998nf}
A.~Strumia and N.~Tetradis, \emph{{A Consistent calculation of bubble
  nucleation rates}},
  \href{https://doi.org/10.1016/S0550-3213(98)00804-9}{\emph{Nucl. Phys. B}
  {\bfseries 542} (1999) 719}
  [\href{https://arxiv.org/abs/hep-ph/9806453}{{\ttfamily hep-ph/9806453}}].

\bibitem{Megevand:2023nin}
A.~M\'egevand and F.A.~Membiela, \emph{{Thin and thick bubble walls. Part I.
  Vacuum phase transitions}},
  \href{https://doi.org/10.1088/1475-7516/2023/06/007}{\emph{JCAP} {\bfseries
  06} (2023) 007} [\href{https://arxiv.org/abs/2302.13349}{{\ttfamily
  2302.13349}}].

\bibitem{Ai:2022kqm}
W.-Y.~Ai, J.S.~Cruz, B.~Garbrecht and C.~Tamarit, \emph{{Instability of bubble
  expansion at zero temperature}},
  \href{https://doi.org/10.1103/PhysRevD.107.036014}{\emph{Phys. Rev. D}
  {\bfseries 107} (2023) 036014}
  [\href{https://arxiv.org/abs/2209.00639}{{\ttfamily 2209.00639}}].

\bibitem{Giblin:2013kea}
J.T.~Giblin, Jr. and J.B.~Mertens, \emph{{Vacuum Bubbles in the Presence of a
  Relativistic Fluid}},
  \href{https://doi.org/10.1007/JHEP12(2013)042}{\emph{JHEP} {\bfseries 12}
  (2013) 042} [\href{https://arxiv.org/abs/1310.2948}{{\ttfamily 1310.2948}}].

\bibitem{Giblin:2014qia}
J.T.~Giblin and J.B.~Mertens, \emph{{Gravitional radiation from first-order
  phase transitions in the presence of a fluid}},
  \href{https://doi.org/10.1103/PhysRevD.90.023532}{\emph{Phys. Rev. D}
  {\bfseries 90} (2014) 023532}
  [\href{https://arxiv.org/abs/1405.4005}{{\ttfamily 1405.4005}}].

\bibitem{Ellis:2019oqb}
J.~Ellis, M.~Lewicki, J.M.~No and V.~Vaskonen, \emph{{Gravitational wave energy
  budget in strongly supercooled phase transitions}},
  \href{https://doi.org/10.1088/1475-7516/2019/06/024}{\emph{JCAP} {\bfseries
  06} (2019) 024} [\href{https://arxiv.org/abs/1903.09642}{{\ttfamily
  1903.09642}}].

\bibitem{Coleman:1980aw}
S.R.~Coleman and F.~De~Luccia, \emph{{Gravitational Effects on and of Vacuum
  Decay}}, \href{https://doi.org/10.1103/PhysRevD.21.3305}{\emph{Phys. Rev. D}
  {\bfseries 21} (1980) 3305}.

\bibitem{Carroll:1997ar}
S.M.~Carroll, \emph{{Lecture notes on general relativity}},
  \href{https://arxiv.org/abs/gr-qc/9712019}{{\ttfamily gr-qc/9712019}}.

\bibitem{Salvio:2016mvj}
A.~Salvio, A.~Strumia, N.~Tetradis and A.~Urbano, \emph{{On gravitational and
  thermal corrections to vacuum decay}},
  \href{https://doi.org/10.1007/JHEP09(2016)054}{\emph{JHEP} {\bfseries 09}
  (2016) 054} [\href{https://arxiv.org/abs/1608.02555}{{\ttfamily
  1608.02555}}].

\bibitem{Branchina:2016bws}
V.~Branchina, E.~Messina and D.~Zappala, \emph{{Impact of Gravity on Vacuum
  Stability}}, \href{https://doi.org/10.1209/0295-5075/116/21001}{\emph{EPL}
  {\bfseries 116} (2016) 21001}
  [\href{https://arxiv.org/abs/1601.06963}{{\ttfamily 1601.06963}}].

\bibitem{Rajantie:2016hkj}
A.~Rajantie and S.~Stopyra, \emph{{Standard Model vacuum decay with gravity}},
  \href{https://doi.org/10.1103/PhysRevD.95.025008}{\emph{Phys. Rev. D}
  {\bfseries 95} (2017) 025008}
  [\href{https://arxiv.org/abs/1606.00849}{{\ttfamily 1606.00849}}].

\bibitem{Gialamas:2022gxv}
I.D.~Gialamas, A.~Karam and T.D.~Pappas, \emph{{Gravitational corrections to
  electroweak vacuum decay: metric vs. Palatini}},
  \href{https://doi.org/10.1016/j.physletb.2023.137885}{\emph{Phys. Lett. B}
  {\bfseries 840} (2023) 137885}
  [\href{https://arxiv.org/abs/2212.03052}{{\ttfamily 2212.03052}}].

\bibitem{Boyanovsky:2006bf}
D.~Boyanovsky, H.J.~de~Vega and D.J.~Schwarz, \emph{{Phase transitions in the
  early and the present universe}},
  \href{https://doi.org/10.1146/annurev.nucl.56.080805.140539}{\emph{Ann. Rev.
  Nucl. Part. Sci.} {\bfseries 56} (2006) 441}
  [\href{https://arxiv.org/abs/hep-ph/0602002}{{\ttfamily hep-ph/0602002}}].

\bibitem{Bea:2021zol}
Y.~Bea, J.~Casalderrey-Solana, T.~Giannakopoulos, A.~Jansen, S.~Krippendorf,
  D.~Mateos et~al., \emph{{Spinodal Gravitational Waves}},
  \href{https://arxiv.org/abs/2112.15478}{{\ttfamily 2112.15478}}.

\bibitem{Biekotter:2021ysx}
T.~Biek\"otter, S.~Heinemeyer, J.M.~No, M.O.~Olea and G.~Weiglein, \emph{{Fate
  of electroweak symmetry in the early Universe: Non-restoration and trapped
  vacua in the N2HDM}},
  \href{https://doi.org/10.1088/1475-7516/2021/06/018}{\emph{JCAP} {\bfseries
  06} (2021) 018} [\href{https://arxiv.org/abs/2103.12707}{{\ttfamily
  2103.12707}}].

\bibitem{Carena:2021onl}
M.~Carena, C.~Krause, Z.~Liu and Y.~Wang, \emph{{New approach to electroweak
  symmetry nonrestoration}},
  \href{https://doi.org/10.1103/PhysRevD.104.055016}{\emph{Phys. Rev. D}
  {\bfseries 104} (2021) 055016}
  [\href{https://arxiv.org/abs/2104.00638}{{\ttfamily 2104.00638}}].

\bibitem{Matsedonskyi:2021hti}
O.~Matsedonskyi, J.~Unwin and Q.~Wang, \emph{{Electroweak symmetry
  non-restoration from dark matter}},
  \href{https://doi.org/10.1007/JHEP12(2021)167}{\emph{JHEP} {\bfseries 12}
  (2021) 167} [\href{https://arxiv.org/abs/2107.07560}{{\ttfamily
  2107.07560}}].

\bibitem{Turner:1992tz}
M.S.~Turner, E.J.~Weinberg and L.M.~Widrow, \emph{{Bubble nucleation in first
  order inflation and other cosmological phase transitions}},
  \href{https://doi.org/10.1103/PhysRevD.46.2384}{\emph{Phys. Rev. D}
  {\bfseries 46} (1992) 2384}.

\bibitem{Guth:1979bh}
A.H.~Guth and S.H.H.~Tye, \emph{{Phase Transitions and Magnetic Monopole
  Production in the Very Early Universe}},
  \href{https://doi.org/10.1103/PhysRevLett.44.631}{\emph{Phys. Rev. Lett.}
  {\bfseries 44} (1980) 631} [Erratum:
  \href{https://doi.org/10.1103/PhysRevLett.44.963.2}{\textit{Phys.~Rev.~Lett.}
  \textbf{44} (1980) 963}].

\bibitem{Guth:1981uk}
A.H.~Guth and E.J.~Weinberg, \emph{{Cosmological Consequences of a First Order
  Phase Transition in the SU(5) Grand Unified Model}},
  \href{https://doi.org/10.1103/PhysRevD.23.876}{\emph{Phys. Rev. D} {\bfseries
  23} (1981) 876}.

\bibitem{Athron:2022mmm}
P.~Athron, C.~Bal\'azs and L.~Morris, \emph{{Supercool subtleties of
  cosmological phase transitions}},
  \href{https://doi.org/10.1088/1475-7516/2023/03/006}{\emph{JCAP} {\bfseries
  03} (2023) 006} [\href{https://arxiv.org/abs/2212.07559}{{\ttfamily
  2212.07559}}].

\bibitem{Kampfer:1988}
B.~Kämpfer, \emph{Phenomenological models of cosmic phase transitions ii.
  application of the classical nucleation theory},
  \href{https://doi.org/https://doi.org/10.1002/asna.2113090602}{\emph{Astronomische
  Nachrichten} {\bfseries 309} (1988) 347}
  [\href{https://arxiv.org/abs/https://onlinelibrary.wiley.com/doi/pdf/10.1002/asna.2113090602}{{\ttfamily
  https://onlinelibrary.wiley.com/doi/pdf/10.1002/asna.2113090602}}].

\bibitem{Kampfer:1991}
B.~Kämpfer, B.~Lukács and G.~Paál, \emph{Phenomenology of phase transitions
  in the early universe}, {\emph{Fiz. Ehlem. Chastits At. Yadra, Vol. 22, No.
  1, p. 63 - 139} (1991) }.

\bibitem{Yamaguchi:1994yt}
A.~Yamaguchi and A.~Sugamoto, \emph{{Electroweak baryogenesis and the phase
  transition dynamics}},
  \href{https://doi.org/10.1142/S0217732394002446}{\emph{Mod. Phys. Lett. A}
  {\bfseries 9} (1994) 2599}
  [\href{https://arxiv.org/abs/hep-ph/9408201}{{\ttfamily hep-ph/9408201}}].

\bibitem{Kampfer:2000gx}
B.~Kampfer, \emph{{Cosmic phase transitions}},
  \href{https://doi.org/10.1002/1521-3889(200009)9:8<605::AID-ANDP605>3.0.CO;2-6}{\emph{Annalen
  Phys.} {\bfseries 9} (2000) 605}
  [\href{https://arxiv.org/abs/astro-ph/0004403}{{\ttfamily
  astro-ph/0004403}}].

\bibitem{Csernai:1992as}
L.P.~Csernai and J.I.~Kapusta, \emph{{Dynamics of the QCD phase transition}},
  \href{https://doi.org/10.1103/PhysRevLett.69.737}{\emph{Phys. Rev. Lett.}
  {\bfseries 69} (1992) 737}.

\bibitem{johnson1939reaction}
W.A.~Johnson, \emph{Reaction kinetics in processes of nucleation and growth},
  {\emph{Am. Inst. Min. Metal. Petro. Eng.} {\bfseries 135} (1939) 416}.

\bibitem{Avrami1}
M.~Avrami, \emph{{Kinetics of Phase Change. I General Theory}},
  \href{https://doi.org/10.1063/1.1750380}{\emph{The Journal of Chemical
  Physics} {\bfseries 7} (1939) 1103}.

\bibitem{Avrami2}
M.~Avrami, \emph{{Kinetics of Phase Change. II Transformation-Time Relations
  for Random Distribution of Nuclei}},
  \href{https://doi.org/10.1063/1.1750631}{\emph{The Journal of Chemical
  Physics} {\bfseries 8} (1940) 212}.

\bibitem{Avrami3}
M.~Avrami, \emph{{Granulation, Phase Change, and Microstructure Kinetics of
  Phase Change. III}}, \href{https://doi.org/10.1063/1.1750872}{\emph{The
  Journal of Chemical Physics} {\bfseries 9} (1941) 177}.

\bibitem{kolmogorov1937statistical}
A.~Kolmogorov, \emph{On the statistical theory of metal crystallization},
  {\emph{Izv. Akad. Nauk SSSR, Ser. Math} {\bfseries 1} (1937) 335}.

\bibitem{Shiryayev1992}
A.N.~Shiryayev, \emph{On the statistical theory of metal crystallization},  in
  \emph{Selected Works of A. N. Kolmogorov: Volume II Probability Theory and
  Mathematical Statistics}, A.N.~Shiryayev, ed., (Dordrecht), pp.~188--192,
  Springer Netherlands (1992),
  \href{https://doi.org/10.1007/978-94-011-2260-3_22}{DOI}.

\bibitem{fanfoni1998johnson}
M.~Fanfoni and M.~Tomellini, \emph{{The Johnson-Mehl-Avrami-Kolmogorov model: a
  brief review}}, \href{https://doi.org/10.1007/BF03185527}{\emph{Il Nuovo
  Cimento D} {\bfseries 20} (1998) 1171}.

\bibitem{TF9454100365}
U.R.~Evans, \emph{The laws of expanding circles and spheres in relation to the
  lateral growth of surface films and the grain-size of metals},
  \href{https://doi.org/10.1039/TF9454100365}{\emph{Trans. Faraday Soc.}
  {\bfseries 41} (1945) 365}.

\bibitem{ProductIntegration}
A.~Slavík, \emph{Product integration, its history and applications},
  Matfyzpress (01, 2007),
  \href{https://doi.org/10.13140/2.1.2056.4488}{10.13140/2.1.2056.4488}.

\bibitem{PhysRevB.55.14071}
M.~Tomellini and M.~Fanfoni, \emph{{Why phantom nuclei must be considered in
  the Johnson-Mehl-Avrami-Kolmogoroff kinetics}},
  \href{https://doi.org/10.1103/PhysRevB.55.14071}{\emph{Phys. Rev. B}
  {\bfseries 55} (1997) 14071}.

\bibitem{PhysRevB.54.11845}
C.D.~Van~Siclen, \emph{Random nucleation and growth kinetics},
  \href{https://doi.org/10.1103/PhysRevB.54.11845}{\emph{Phys. Rev. B}
  {\bfseries 54} (1996) 11845}.

\bibitem{ALEKSEECHKIN20113159}
N.V.~Alekseechkin, \emph{Extension of the kolmogorov–johnson–mehl–avrami
  theory to growth laws of diffusion type},
  \href{https://doi.org/https://doi.org/10.1016/j.jnoncrysol.2011.05.007}{\emph{Journal
  of Non-Crystalline Solids} {\bfseries 357} (2011) 3159}.

\bibitem{PEREZCARDENAS2019100002}
F.C.~Pérez-Cárdenas, \emph{The irrelevance of phantom nuclei in
  crystallization kinetics: An integral equation approach},
  \href{https://doi.org/https://doi.org/10.1016/j.nocx.2018.100002}{\emph{Journal
  of Non-Crystalline Solids: X} {\bfseries 1} (2019) 100002}.

\bibitem{TOMELLINI2019119459}
M.~Tomellini and M.~Fanfoni, \emph{{Comment on “The irrelevance of phantom
  nuclei in crystallization kinetics: An integral equation approach”}},
  \href{https://doi.org/https://doi.org/10.1016/j.jnoncrysol.2019.05.035}{\emph{Journal
  of Non-Crystalline Solids} {\bfseries 520} (2019) 119459}.

\bibitem{PEREZCARDENAS2019119458}
F.C.~Pérez-Cárdenas, \emph{Response to comment on “the irrelevance of
  phantom nuclei in crystallization kinetics: An integral equation
  approach”},
  \href{https://doi.org/https://doi.org/10.1016/j.jnoncrysol.2019.05.034}{\emph{Journal
  of Non-Crystalline Solids} {\bfseries 520} (2019) 119458}.

\bibitem{cahn_1995}
J.W.~Cahn, \emph{The time cone method for nucleation and growth kinetics on a
  finite domain}, \href{https://doi.org/10.1557/PROC-398-425}{\emph{MRS
  Proceedings} {\bfseries 398} (1995) 425}.

\bibitem{yu1995kinetics}
G.~Yu and J.~Lai, \emph{Kinetics of transformation with nucleation and growth
  mechanism: Fundamentals of derivation and one-dimensional model},
  \href{https://doi.org/10.1063/1.360599}{\emph{Journal of applied physics}
  {\bfseries 78} (1995) 5965}.

\bibitem{yu1996kinetics}
G.~Yu and J.~Lai, \emph{Kinetics of transformation with nucleation and growth
  mechanism: Two-and three-dimensional models},
  \href{https://doi.org/10.1063/1.361376}{\emph{Journal of applied physics}
  {\bfseries 79} (1996) 3504}.

\bibitem{yu_lee_lai_1997}
G.~Yu, S.T.~Lee and J.K.L.~Lai, \emph{Derivation of the kinetics of phase
  transformations with nucleation and growth mechanism},
  \href{https://doi.org/10.1557/PROC-481-97}{\emph{MRS Proceedings} {\bfseries
  481} (1997) 97}.

\bibitem{Megevand:2020klf}
A.~M\'egevand and F.A.~Membiela, \emph{{Bubble wall correlations in
  cosmological phase transitions}},
  \href{https://doi.org/10.1103/PhysRevD.102.103514}{\emph{Phys. Rev. D}
  {\bfseries 102} (2020) 103514}
  [\href{https://arxiv.org/abs/2008.01873}{{\ttfamily 2008.01873}}].

\bibitem{PhysRevB.54.836}
V.~Sessa, M.~Fanfoni and M.~Tomellini, \emph{{Validity of Avrami's kinetics for
  random and nonrandom distributions of germs}},
  \href{https://doi.org/10.1103/PhysRevB.54.836}{\emph{Phys. Rev. B} {\bfseries
  54} (1996) 836}.

\bibitem{10.1093/imamat/hxx012}
D.~Hömberg, F.S.~Patacchini, K.~Sakamoto and J.~Zimmer, \emph{{A revisited
  Johnson–Mehl–Avrami–Kolmogorov model and the evolution of grain-size
  distributions in steel}},
  \href{https://doi.org/10.1093/imamat/hxx012}{\emph{IMA Journal of Applied
  Mathematics} {\bfseries 82} (2017) 763}.

\bibitem{ImpingementParam}
M.~Starink, \emph{On the meaning of the impingement parameter in kinetic
  equations for nucleation and growth reactions},
  \href{https://doi.org/10.1023/A:1017974517877}{\emph{Journal of Materials
  Science} {\bfseries 36} (2001) 4433}.

\bibitem{BURBELKO2005429}
A.~Burbelko, E.~Fraś and W.~Kapturkiewicz, \emph{{About Kolmogorov's
  statistical theory of phase transformation}},
  \href{https://doi.org/https://doi.org/10.1016/j.msea.2005.08.161}{\emph{Materials
  Science and Engineering: A} {\bfseries 413-414} (2005) 429} International
  Conference on Advances in Solidification Processes.

\bibitem{TOMELLINI200465}
M.~Tomellini and M.~Fanfoni, \emph{{Eliminating overgrowth effects in
  Kolmogorov–Johnson–Mehl–Avrami model through the correlation among
  actual nuclei}},
  \href{https://doi.org/https://doi.org/10.1016/j.physa.2003.09.066}{\emph{Physica
  A: Statistical Mechanics and its Applications} {\bfseries 333} (2004) 65}.

\bibitem{PhysRevE.85.021606}
M.~Tomellini and M.~Fanfoni, \emph{{Beyond the constraints underlying
  Kolmogorov-Johnson-Mehl-Avrami theory related to the growth laws}},
  \href{https://doi.org/10.1103/PhysRevE.85.021606}{\emph{Phys. Rev. E}
  {\bfseries 85} (2012) 021606}.

\bibitem{PhysRevE.95.022121}
J.M.~Rickman and K.~Barmak, \emph{Kinetics of first-order phase transitions
  with correlated nuclei},
  \href{https://doi.org/10.1103/PhysRevE.95.022121}{\emph{Phys. Rev. E}
  {\bfseries 95} (2017) 022121}.

\bibitem{Pirvu:2021roq}
D.~Pirvu, J.~Braden and M.C.~Johnson, \emph{{Bubble clustering in cosmological
  first order phase transitions}},
  \href{https://doi.org/10.1103/PhysRevD.105.043510}{\emph{Phys. Rev. D}
  {\bfseries 105} (2022) 043510}
  [\href{https://arxiv.org/abs/2109.04496}{{\ttfamily 2109.04496}}].

\bibitem{DeLuca:2021mlh}
V.~De~Luca, G.~Franciolini and A.~Riotto, \emph{{Bubble correlation in
  first-order phase transitions}},
  \href{https://doi.org/10.1103/PhysRevD.104.123539}{\emph{Phys. Rev. D}
  {\bfseries 104} (2021) 123539}
  [\href{https://arxiv.org/abs/2110.04229}{{\ttfamily 2110.04229}}].

\bibitem{VILLA20093714}
E.~Villa and P.R.~Rios, \emph{Transformation kinetics for nucleus clusters},
  \href{https://doi.org/https://doi.org/10.1016/j.actamat.2009.04.014}{\emph{Acta
  Materialia} {\bfseries 57} (2009) 3714}.

\bibitem{FERNANDESIGNACIO2021777}
N.~{Fernandes Ignácio}, M.~{Silva Fernandes}, D.~{Magalhães Baía},
  A.G.~{Conceição dos Santos}, F.~{da Silva Siqueira}, W.~{Luiz da Silva
  Assis} et~al., \emph{A study on the effect of the number of clusters at the
  phase transformation kinetics},
  \href{https://doi.org/https://doi.org/10.1016/j.jmrt.2021.08.040}{\emph{Journal
  of Materials Research and Technology} {\bfseries 15} (2021) 777}.

\bibitem{alekseechkin2001theory}
N.~Alekseechkin, \emph{On the theory of phase transformations with
  position-dependent nucleation rates},
  \href{https://doi.org/10.1134/1.1332151}{\emph{Journal of Physics: Condensed
  Matter} {\bfseries 13} (2001) 3083}.

\bibitem{RIOS20091199}
P.~Rios and E.~Villa, \emph{Transformation kinetics for inhomogeneous
  nucleation},
  \href{https://doi.org/https://doi.org/10.1016/j.actamat.2008.11.003}{\emph{Acta
  Materialia} {\bfseries 57} (2009) 1199}.

\bibitem{tomellini2010kinetics}
M.~Tomellini, \emph{On the kinetics of nucleation and growth reactions in
  inhomogeneous systems},
  \href{https://doi.org/10.1007/s10853-009-3992-8}{\emph{Journal of materials
  science} {\bfseries 45} (2010) 733}.

\bibitem{doi:10.1063/1.470052}
D.P.~Birnie and M.C.~Weinberg, \emph{Kinetics of transformation for anisotropic
  particles including shielding effects},
  \href{https://doi.org/10.1063/1.470052}{\emph{The Journal of Chemical
  Physics} {\bfseries 103} (1995) 3742}.

\bibitem{pusztai1998monte}
T.~Pusztai and L.~Gr{\'a}n{\'a}sy, \emph{Monte carlo simulation of first-order
  phase transformations with mutual blocking of anisotropically growing
  particles up to all relevant orders},
  \href{https://doi.org/10.1103/PhysRevB.57.14110}{\emph{Physical Review B}
  {\bfseries 57} (1998) 14110}.

\bibitem{kooi2004monte}
B.~Kooi, \emph{{Monte Carlo simulations of phase transformations caused by
  nucleation and subsequent anisotropic growth: extension of the
  Johnson-Mehl-Avrami-Kolmogorov theory}},
  \href{https://doi.org/10.1103/PhysRevB.70.224108}{\emph{Physical Review B}
  {\bfseries 70} (2004) 224108}.

\bibitem{Megevand:2000da}
A.~Megevand, \emph{{Effect of reheating on electroweak baryogenesis}},
  \href{https://doi.org/10.1103/PhysRevD.64.027303}{\emph{Phys. Rev. D}
  {\bfseries 64} (2001) 027303}
  [\href{https://arxiv.org/abs/hep-ph/0011019}{{\ttfamily hep-ph/0011019}}].

\bibitem{Megevand:2017vtb}
A.~M\'egevand and S.~Ram\'\i{}rez, \emph{{Bubble nucleation and growth in slow
  cosmological phase transitions}},
  \href{https://doi.org/10.1016/j.nuclphysb.2018.01.012}{\emph{Nucl. Phys. B}
  {\bfseries 928} (2018) 38}
  [\href{https://arxiv.org/abs/1710.06279}{{\ttfamily 1710.06279}}].

\bibitem{Cutting:2019zws}
D.~Cutting, M.~Hindmarsh and D.J.~Weir, \emph{{Vorticity, kinetic energy, and
  suppressed gravitational wave production in strong first order phase
  transitions}},
  \href{https://doi.org/10.1103/PhysRevLett.125.021302}{\emph{Phys. Rev. Lett.}
  {\bfseries 125} (2020) 021302}
  [\href{https://arxiv.org/abs/1906.00480}{{\ttfamily 1906.00480}}].

\bibitem{levine_narayan_kelton_1997}
L.E.~Levine, K.L.~Narayan and K.F.~Kelton, \emph{{Finite size corrections for
  the Johnson–Mehl–Avrami–Kolmogorov equation}},
  \href{https://doi.org/10.1557/JMR.1997.0020}{\emph{Journal of Materials
  Research} {\bfseries 12} (1997) 124–132}.

\bibitem{alekseechkin2008kinetics}
N.~Alekseechkin, \emph{{On the kinetics of phase transformation of small
  particles in Kolmogorov's model}}, {\emph{Condensed Matter Physics} (2008) }
  \href{https://www.icmp.lviv.ua/journal/zbirnyk.56/002/art02.pdf}{https://www.icmp.lviv.ua/journal/zbirnyk.56/002/art02.pdf}.

\bibitem{Ignatius:1993xb}
J.~Ignatius, \emph{{Cosmological phase transitions}}, Ph.D. thesis, University
  of Helsinki, 10, 1993.
\newblock \href{https://arxiv.org/abs/hep-ph/9312293}{{\ttfamily
  hep-ph/9312293}}.
\newblock [\href{https://arxiv.org/abs/hep-ph/9312293}{{\ttfamily
  hep-ph/9312293}}].

\bibitem{JayOlson:2014vgy}
S.~Jay~Olson, \emph{{Homogeneous cosmology with aggressively expanding
  civilizations}},
  \href{https://doi.org/10.1088/0264-9381/32/21/215025}{\emph{Class. Quant.
  Grav.} {\bfseries 32} (2015) 215025}
  [\href{https://arxiv.org/abs/1411.4359}{{\ttfamily 1411.4359}}].

\bibitem{Ruckenstein-Ihm}
E.~Ruckenstein and S.K.~Ihm, \emph{Kinetics of glass formation},
  \href{https://doi.org/10.1039/F19767200764}{\emph{J. Chem. Soc.{,} Faraday
  Trans. 1} {\bfseries 72} (1976) 764}.

\bibitem{Megevand:2007sv}
A.~Megevand and A.D.~Sanchez, \emph{{Supercooling and phase coexistence in
  cosmological phase transitions}},
  \href{https://doi.org/10.1103/PhysRevD.77.063519}{\emph{Phys. Rev. D}
  {\bfseries 77} (2008) 063519}
  [\href{https://arxiv.org/abs/0712.1031}{{\ttfamily 0712.1031}}].

\bibitem{Ellis:2020nnr}
J.~Ellis, M.~Lewicki and V.~Vaskonen, \emph{{Updated predictions for
  gravitational waves produced in a strongly supercooled phase transition}},
  \href{https://doi.org/10.1088/1475-7516/2020/11/020}{\emph{JCAP} {\bfseries
  11} (2020) 020} [\href{https://arxiv.org/abs/2007.15586}{{\ttfamily
  2007.15586}}].

\bibitem{Lewicki:2022pdb}
M.~Lewicki and V.~Vaskonen, \emph{{Gravitational waves from bubble collisions
  and fluid motion in strongly supercooled phase transitions}},
  \href{https://doi.org/10.1140/epjc/s10052-023-11241-3}{\emph{Eur. Phys. J. C}
  {\bfseries 83} (2023) 109}
  [\href{https://arxiv.org/abs/2208.11697}{{\ttfamily 2208.11697}}].

\bibitem{Marques:1992ws}
G.C.~Marques and R.O.~Ramos, \emph{{Phase transitions and formation of bubbles
  in the early universe}},
  \href{https://doi.org/10.1103/PhysRevD.45.4400}{\emph{Phys. Rev. D}
  {\bfseries 45} (1992) 4400}.

\bibitem{Megevand:2000zw}
A.~Megevand, \emph{{Development of the electroweak phase transition and
  baryogenesis}}, \href{https://doi.org/10.1142/S0218271800000724}{\emph{Int.
  J. Mod. Phys. D} {\bfseries 9} (2000) 733}
  [\href{https://arxiv.org/abs/hep-ph/0006177}{{\ttfamily hep-ph/0006177}}].

\bibitem{Witten:1984rs}
E.~Witten, \emph{{Cosmic Separation of Phases}},
  \href{https://doi.org/10.1103/PhysRevD.30.272}{\emph{Phys. Rev. D} {\bfseries
  30} (1984) 272}.

\bibitem{Lu:2022paj}
P.~Lu, K.~Kawana and K.-P.~Xie, \emph{{Old phase remnants in first-order phase
  transitions}}, \href{https://doi.org/10.1103/PhysRevD.105.123503}{\emph{Phys.
  Rev. D} {\bfseries 105} (2022) 123503}
  [\href{https://arxiv.org/abs/2202.03439}{{\ttfamily 2202.03439}}].

\bibitem{Cutting:2022zgd}
D.~Cutting, E.~Vilhonen and D.J.~Weir, \emph{{Droplet collapse during strongly
  supercooled transitions}},
  \href{https://doi.org/10.1103/PhysRevD.106.103524}{\emph{Phys. Rev. D}
  {\bfseries 106} (2022) 103524}
  [\href{https://arxiv.org/abs/2204.03396}{{\ttfamily 2204.03396}}].

\bibitem{Heckler:1994uu}
A.F.~Heckler, \emph{{The Effects of electroweak phase transition dynamics on
  baryogenesis and primordial nucleosynthesis}},
  \href{https://doi.org/10.1103/PhysRevD.51.405}{\emph{Phys. Rev. D} {\bfseries
  51} (1995) 405} [\href{https://arxiv.org/abs/astro-ph/9407064}{{\ttfamily
  astro-ph/9407064}}].

\bibitem{Wu:2019pbm}
J.~Wu, Q.~Li, J.~Liu, C.~Xue, S.~Y~ang, C.~Shao et~al., \emph{{Progress in
  Precise Measurements of the Gravitational Constan t}},
  \href{https://doi.org/10.1002/andp.201900013}{\emph{Annalen Phys.} {\bfseries
  531} (2019) 1900013}.

\bibitem{Azatov:2019png}
A.~Azatov, D.~Barducci and F.~Sgarlata, \emph{{Gravitational traces of broken
  gauge symmetries}},
  \href{https://doi.org/10.1088/1475-7516/2020/07/027}{\emph{JCAP} {\bfseries
  07} (2020) 027} [\href{https://arxiv.org/abs/1910.01124}{{\ttfamily
  1910.01124}}].

\bibitem{Sato:1980yn}
K.~Sato, \emph{{First Order Phase Transition of a Vacuum and Expansion of the
  Universe}}, \href{https://doi.org/10.1093/mnras/195.3.467}{\emph{Mon. Not.
  Roy. Astron. Soc.} {\bfseries 195} (1981) 467}.

\bibitem{Megevand:2016lpr}
A.~Megevand and S.~Ramirez, \emph{{Bubble nucleation and growth in very strong
  cosmological phase transitions}},
  \href{https://doi.org/10.1016/j.nuclphysb.2017.03.009}{\emph{Nucl. Phys. B}
  {\bfseries 919} (2017) 74}
  [\href{https://arxiv.org/abs/1611.05853}{{\ttfamily 1611.05853}}].

\bibitem{Guo:2020grp}
H.-K.~Guo, K.~Sinha, D.~Vagie and G.~White, \emph{{Phase Transitions in an
  Expanding Universe: Stochastic Gravitational Waves in Standard and
  Non-Standard Histories}},
  \href{https://doi.org/10.1088/1475-7516/2021/01/001}{\emph{JCAP} {\bfseries
  01} (2021) 001} [\href{https://arxiv.org/abs/2007.08537}{{\ttfamily
  2007.08537}}].

\bibitem{Chodos:1974je}
A.~Chodos, R.L.~Jaffe, K.~Johnson, C.B.~Thorn and V.F.~Weisskopf, \emph{{A New
  Extended Model of Hadrons}},
  \href{https://doi.org/10.1103/PhysRevD.9.3471}{\emph{Phys. Rev. D} {\bfseries
  9} (1974) 3471}.

\bibitem{Cai:2017tmh}
R.-G.~Cai, M.~Sasaki and S.-J.~Wang, \emph{{The gravitational waves from the
  first-order phase transition with a dimension-six operator}},
  \href{https://doi.org/10.1088/1475-7516/2017/08/004}{\emph{JCAP} {\bfseries
  08} (2017) 004} [\href{https://arxiv.org/abs/1707.03001}{{\ttfamily
  1707.03001}}].

\bibitem{Konstandin:2010dm}
T.~Konstandin and J.M.~No, \emph{{Hydrodynamic obstruction to bubble
  expansion}}, \href{https://doi.org/10.1088/1475-7516/2011/02/008}{\emph{JCAP}
  {\bfseries 02} (2011) 008} [\href{https://arxiv.org/abs/1011.3735}{{\ttfamily
  1011.3735}}].

\bibitem{Ajmi:2022nmq}
M.A.~Ajmi and M.~Hindmarsh, \emph{{Thermal suppression of bubble nucleation at
  first-order phase transitions in the early Universe}},
  \href{https://doi.org/10.1103/PhysRevD.106.023505}{\emph{Phys. Rev. D}
  {\bfseries 106} (2022) 023505}
  [\href{https://arxiv.org/abs/2205.04097}{{\ttfamily 2205.04097}}].

\bibitem{Guth:1982pn}
A.H.~Guth and E.J.~Weinberg, \emph{{Could the Universe Have Recovered from a
  Slow First Order Phase Transition?}},
  \href{https://doi.org/10.1016/0550-3213(83)90307-3}{\emph{Nucl. Phys. B}
  {\bfseries 212} (1983) 321}.

\bibitem{Hawking:1982ga}
S.W.~Hawking, I.G.~Moss and J.M.~Stewart, \emph{{Bubble Collisions in the Very
  Early Universe}}, \href{https://doi.org/10.1103/PhysRevD.26.2681}{\emph{Phys.
  Rev. D} {\bfseries 26} (1982) 2681}.

\bibitem{Megevand:2021llq}
A.~Megevand and F.A.~Membiela, \emph{{Model-independent features of
  gravitational waves from bubble collisions}},
  \href{https://doi.org/10.1103/PhysRevD.104.123532}{\emph{Phys. Rev. D}
  {\bfseries 104} (2021) 123532}
  [\href{https://arxiv.org/abs/2108.07034}{{\ttfamily 2108.07034}}].

\bibitem{Eichhorn:2020upj}
A.~Eichhorn, J.~Lumma, J.M.~Pawlowski, M.~Reichert and M.~Yamada,
  \emph{{Universal gravitational-wave signatures from heavy new physics in the
  electroweak sector}},
  \href{https://doi.org/10.1088/1475-7516/2021/05/006}{\emph{JCAP} {\bfseries
  05} (2021) 006} [\href{https://arxiv.org/abs/2010.00017}{{\ttfamily
  2010.00017}}].

\bibitem{Ellis:2018mja}
J.~Ellis, M.~Lewicki and J.M.~No, \emph{{On the Maximal Strength of a
  First-Order Electroweak Phase Transition and its Gravitational Wave Signal}},
  \href{https://doi.org/10.1088/1475-7516/2019/04/003}{\emph{JCAP} {\bfseries
  04} (2019) 003} [\href{https://arxiv.org/abs/1809.08242}{{\ttfamily
  1809.08242}}].

\bibitem{Leitao:2015fmj}
L.~Leitao and A.~Megevand, \emph{{Gravitational waves from a very strong
  electroweak phase transition}},
  \href{https://doi.org/10.1088/1475-7516/2016/05/037}{\emph{JCAP} {\bfseries
  05} (2016) 037} [\href{https://arxiv.org/abs/1512.08962}{{\ttfamily
  1512.08962}}].

\bibitem{Kobakhidze:2017mru}
A.~Kobakhidze, C.~Lagger, A.~Manning and J.~Yue, \emph{{Gravitational waves
  from a supercooled electroweak phase transition and their detection with
  pulsar timing arrays}},
  \href{https://doi.org/10.1140/epjc/s10052-017-5132-y}{\emph{Eur. Phys. J. C}
  {\bfseries 77} (2017) 570}
  [\href{https://arxiv.org/abs/1703.06552}{{\ttfamily 1703.06552}}].

\bibitem{Wang:2020jrd}
X.~Wang, F.P.~Huang and X.~Zhang, \emph{{Phase transition dynamics and
  gravitational wave spectra of strong first-order phase transition in
  supercooled universe}},
  \href{https://doi.org/10.1088/1475-7516/2020/05/045}{\emph{JCAP} {\bfseries
  05} (2020) 045} [\href{https://arxiv.org/abs/2003.08892}{{\ttfamily
  2003.08892}}].

\bibitem{Leitao:2014pda}
L.~Leitao and A.~Megevand, \emph{{Hydrodynamics of phase transition fronts and
  the speed of sound in the plasma}},
  \href{https://doi.org/10.1016/j.nuclphysb.2014.12.008}{\emph{Nucl. Phys. B}
  {\bfseries 891} (2015) 159}
  [\href{https://arxiv.org/abs/1410.3875}{{\ttfamily 1410.3875}}].

\bibitem{Giese:2020rtr}
F.~Giese, T.~Konstandin and J.~van~de Vis, \emph{{Model-independent energy
  budget of cosmological first-order phase transitions\textemdash{}A sound
  argument to go beyond the bag model}},
  \href{https://doi.org/10.1088/1475-7516/2020/07/057}{\emph{JCAP} {\bfseries
  07} (2020) 057} [\href{https://arxiv.org/abs/2004.06995}{{\ttfamily
  2004.06995}}].

\bibitem{Giese:2020znk}
F.~Giese, T.~Konstandin, K.~Schmitz and J.~Van De~Vis, \emph{{Model-independent
  energy budget for LISA}},
  \href{https://doi.org/10.1088/1475-7516/2021/01/072}{\emph{JCAP} {\bfseries
  01} (2021) 072} [\href{https://arxiv.org/abs/2010.09744}{{\ttfamily
  2010.09744}}].

\bibitem{Wang:2020nzm}
X.~Wang, F.P.~Huang and X.~Zhang, \emph{{Energy budget and the gravitational
  wave spectra beyond the bag model}},
  \href{https://doi.org/10.1103/PhysRevD.103.103520}{\emph{Phys. Rev. D}
  {\bfseries 103} (2021) 103520}
  [\href{https://arxiv.org/abs/2010.13770}{{\ttfamily 2010.13770}}].

\bibitem{Wang:2022lyd}
S.-J.~Wang and Z.-Y.~Yuwen, \emph{{The energy budget of cosmological
  first-order phase transitions beyond the bag equation of state}},
  \href{https://doi.org/10.1088/1475-7516/2022/10/047}{\emph{JCAP} {\bfseries
  10} (2022) 047} [\href{https://arxiv.org/abs/2206.01148}{{\ttfamily
  2206.01148}}].

\bibitem{Wang:2023jto}
X.~Wang, C.~Tian and F.P.~Huang, \emph{{Model-dependent analysis method for
  energy budget of the cosmological first-order phase transition}},
  \href{https://doi.org/10.1088/1475-7516/2023/07/006}{\emph{JCAP} {\bfseries
  07} (2023) 006} [\href{https://arxiv.org/abs/2301.12328}{{\ttfamily
  2301.12328}}].

\bibitem{Einhorn:1980ik}
M.B.~Einhorn and K.~Sato, \emph{{Monopole Production in the Very Early Universe
  in a First Order Phase Transition}},
  \href{https://doi.org/10.1016/0550-3213(81)90057-2}{\emph{Nucl. Phys. B}
  {\bfseries 180} (1981) 385}.

\bibitem{Suhonen:1982ee}
E.~Suhonen, \emph{{The Quark - Hadron Phase Transition in the Early Universe}},
  \href{https://doi.org/10.1016/0370-2693(82)90248-9}{\emph{Phys. Lett. B}
  {\bfseries 119} (1982) 81}.

\bibitem{DeGrand:1984uq}
T.A.~DeGrand and K.~Kajantie, \emph{{Supercooling, Entropy Production and
  Bubble Kinetics in the Quark - Hadron Phase Transition in the Early
  Universe}}, \href{https://doi.org/10.1016/0370-2693(84)90115-1}{\emph{Phys.
  Lett. B} {\bfseries 147} (1984) 273}.

\bibitem{Levi:2022bzt}
N.~Levi, T.~Opferkuch and D.~Redigolo, \emph{{The supercooling window at weak
  and strong coupling}},
  \href{https://doi.org/10.1007/JHEP02(2023)125}{\emph{JHEP} {\bfseries 02}
  (2023) 125} [\href{https://arxiv.org/abs/2212.08085}{{\ttfamily
  2212.08085}}].

\bibitem{Davis:2003ad}
T.M.~Davis and C.H.~Lineweaver, \emph{{Expanding confusion: common
  misconceptions of cosmological horizons and the superluminal expansion of the
  universe}}, \href{https://doi.org/10.1071/AS03040}{\emph{Publ. Astron. Soc.
  Austral.} {\bfseries 21} (2004) 97}
  [\href{https://arxiv.org/abs/astro-ph/0310808}{{\ttfamily
  astro-ph/0310808}}].

\bibitem{Ellis:2015wdi}
G.F.R.~Ellis and J.-P.~Uzan, \emph{{Causal structures in inflation}},
  \href{https://doi.org/10.1016/j.crhy.2015.07.005}{\emph{Comptes Rendus
  Physique} {\bfseries 16} (2015) 928}
  [\href{https://arxiv.org/abs/1612.01084}{{\ttfamily 1612.01084}}].

\bibitem{Broadbent:1957rm}
S.R.~Broadbent and J.M.~Hammersley, \emph{{Percolation processes. 1. Crystals
  and mazes}}, \href{https://doi.org/10.1017/S0305004100032680}{\emph{Proc.
  Cambridge Phil. Soc.} {\bfseries 53} (1957) 629}.

\bibitem{doi:10.1080/00018737100101261}
V.K.S.~Shante and S.~Kirkpatrick, \emph{An introduction to percolation theory},
  \href{https://doi.org/10.1080/00018737100101261}{\emph{Advances in Physics}
  {\bfseries 20} (1971) 325}
  [\href{https://arxiv.org/abs/https://doi.org/10.1080/00018737100101261}{{\ttfamily
  https://doi.org/10.1080/00018737100101261}}].

\bibitem{hunt2014percolation}
A.~Hunt, R.~Ewing and B.~Ghanbarian, \emph{Percolation theory for flow in
  porous media}, vol.~880, Springer (2014).

\bibitem{doi:10.1063/1.1338506}
C.D.~Lorenz and R.M.~Ziff, \emph{{Precise determination of the critical
  percolation threshold for the three-dimensional “Swiss cheese” model
  using a growth algorithm}},
  \href{https://doi.org/10.1063/1.1338506}{\emph{The Journal of Chemical
  Physics} {\bfseries 114} (2001) 3659}
  [\href{https://arxiv.org/abs/https://doi.org/10.1063/1.1338506}{{\ttfamily
  https://doi.org/10.1063/1.1338506}}].

\bibitem{LIN2018299}
J.~Lin and H.~Chen, \emph{Continuum percolation of porous media via random
  packing of overlapping cube-like particles},
  \href{https://doi.org/https://doi.org/10.1016/j.taml.2018.05.007}{\emph{Theoretical
  and Applied Mechanics Letters} {\bfseries 8} (2018) 299}.

\bibitem{LI2020112815}
M.~Li, H.~Chen and J.~Lin, \emph{Numerical study for the percolation threshold
  and transport properties of porous composites comprising
  non-centrosymmetrical superovoidal pores},
  \href{https://doi.org/https://doi.org/10.1016/j.cma.2019.112815}{\emph{Computer
  Methods in Applied Mechanics and Engineering} {\bfseries 361} (2020) 112815}.

\bibitem{Liu:2021svg}
J.~Liu, L.~Bian, R.-G.~Cai, Z.-K.~Guo and S.-J.~Wang, \emph{{Primordial black
  hole production during first-order phase transitions}},
  \href{https://doi.org/10.1103/PhysRevD.105.L021303}{\emph{Phys. Rev. D}
  {\bfseries 105} (2022) L021303}
  [\href{https://arxiv.org/abs/2106.05637}{{\ttfamily 2106.05637}}].

\bibitem{Kawana:2022lba}
K.~Kawana, P.~Lu and K.-P.~Xie, \emph{{First-order phase transition and fate of
  false vacuum remnants}},
  \href{https://doi.org/10.1088/1475-7516/2022/10/030}{\emph{JCAP} {\bfseries
  10} (2022) 030} [\href{https://arxiv.org/abs/2206.09923}{{\ttfamily
  2206.09923}}].

\bibitem{Athron:2023aqe}
P.~Athron, C.~Bal\'azs, T.E.~Gonzalo and M.~Pearce, \emph{{Falsifying
  Pati-Salam models with LIGO}},
  \href{https://arxiv.org/abs/2307.02544}{{\ttfamily 2307.02544}}.

\bibitem{Athron:2023mer}
P.~Athron, A.~Fowlie, C.-T.~Lu, L.~Morris, L.~Wu, Y.~Wu et~al., \emph{{Can
  supercooled phase transitions explain the gravitational wave background
  observed by pulsar timing arrays?}},
  \href{https://arxiv.org/abs/2306.17239}{{\ttfamily 2306.17239}}.

\bibitem{Kodama:1982sf}
H.~Kodama, M.~Sasaki and K.~Sato, \emph{{Abundance of Primordial Holes Produced
  by Cosmological First Order Phase Transition}},
  \href{https://doi.org/10.1143/PTP.68.1979}{\emph{Prog. Theor. Phys.}
  {\bfseries 68} (1982) 1979}.

\bibitem{Hall:1989hr}
L.J.~Hall and S.~Hsu, \emph{{Cosmological Production of Black Holes}},
  \href{https://doi.org/10.1103/PhysRevLett.64.2848}{\emph{Phys. Rev. Lett.}
  {\bfseries 64} (1990) 2848}.

\bibitem{Lewicki:2023ioy}
M.~Lewicki, P.~Toczek and V.~Vaskonen, \emph{{Primordial black holes from
  strong first-order phase transitions}},
  \href{https://doi.org/10.1007/JHEP09(2023)092}{\emph{JHEP} {\bfseries 09}
  (2023) 092} [\href{https://arxiv.org/abs/2305.04924}{{\ttfamily
  2305.04924}}].

\bibitem{Krylov:2013qe}
E.~Krylov, A.~Levin and V.~Rubakov, \emph{{Cosmological phase transition,
  baryon asymmetry and dark matter Q-balls}},
  \href{https://doi.org/10.1103/PhysRevD.87.083528}{\emph{Phys. Rev. D}
  {\bfseries 87} (2013) 083528}
  [\href{https://arxiv.org/abs/1301.0354}{{\ttfamily 1301.0354}}].

\bibitem{Hong:2020est}
J.-P.~Hong, S.~Jung and K.-P.~Xie, \emph{{Fermi-ball dark matter from a
  first-order phase transition}},
  \href{https://doi.org/10.1103/PhysRevD.102.075028}{\emph{Phys. Rev. D}
  {\bfseries 102} (2020) 075028}
  [\href{https://arxiv.org/abs/2008.04430}{{\ttfamily 2008.04430}}].

\bibitem{Hindmarsh:2015qta}
M.~Hindmarsh, S.J.~Huber, K.~Rummukainen and D.J.~Weir, \emph{{Numerical
  simulations of acoustically generated gravitational waves at a first order
  phase transition}},
  \href{https://doi.org/10.1103/PhysRevD.92.123009}{\emph{Phys. Rev. D}
  {\bfseries 92} (2015) 123009}
  [\href{https://arxiv.org/abs/1504.03291}{{\ttfamily 1504.03291}}].

\bibitem{Croon:2018new}
D.~Croon and G.~White, \emph{{Exotic Gravitational Wave Signatures from
  Simultaneous Phase Transitions}},
  \href{https://doi.org/10.1007/JHEP05(2018)210}{\emph{JHEP} {\bfseries 05}
  (2018) 210} [\href{https://arxiv.org/abs/1803.05438}{{\ttfamily
  1803.05438}}].

\bibitem{Morais:2018uou}
A.P.~Morais, R.~Pasechnik and T.~Vieu, \emph{{Multi-peaked signatures of
  primordial gravitational waves from multi-step electroweak phase
  transition}}, \href{https://doi.org/10.22323/1.364.0054}{\emph{PoS}
  {\bfseries EPS-HEP2019} (2020) 054}
  [\href{https://arxiv.org/abs/1802.10109}{{\ttfamily 1802.10109}}].

\bibitem{Morais:2019fnm}
A.P.~Morais and R.~Pasechnik, \emph{{Probing multi-step electroweak phase
  transition with multi-peaked primordial gravitational waves spectra}},
  \href{https://doi.org/10.1088/1475-7516/2020/04/036}{\emph{JCAP} {\bfseries
  04} (2020) 036} [\href{https://arxiv.org/abs/1910.00717}{{\ttfamily
  1910.00717}}].

\bibitem{Land:1992sm}
D.~Land and E.D.~Carlson, \emph{{Two stage phase transition in two Higgs
  models}}, \href{https://doi.org/10.1016/0370-2693(92)90616-C}{\emph{Phys.
  Lett. B} {\bfseries 292} (1992) 107}
  [\href{https://arxiv.org/abs/hep-ph/9208227}{{\ttfamily hep-ph/9208227}}].

\bibitem{Zarikas:1995qb}
V.~Zarikas, \emph{{The Phase transition of the two Higgs extension of the
  standard model}},
  \href{https://doi.org/10.1016/0370-2693(96)00701-0}{\emph{Phys. Lett. B}
  {\bfseries 384} (1996) 180}
  [\href{https://arxiv.org/abs/hep-ph/9509338}{{\ttfamily hep-ph/9509338}}].

\bibitem{Angelescu:2018dkk}
A.~Angelescu and P.~Huang, \emph{{Multistep Strongly First Order Phase
  Transitions from New Fermions at the TeV Scale}},
  \href{https://doi.org/10.1103/PhysRevD.99.055023}{\emph{Phys. Rev. D}
  {\bfseries 99} (2019) 055023}
  [\href{https://arxiv.org/abs/1812.08293}{{\ttfamily 1812.08293}}].

\bibitem{Fabian:2020hny}
S.~Fabian, F.~Goertz and Y.~Jiang, \emph{{Dark matter and nature of electroweak
  phase transition with an inert doublet}},
  \href{https://doi.org/10.1088/1475-7516/2021/09/011}{\emph{JCAP} {\bfseries
  09} (2021) 011} [\href{https://arxiv.org/abs/2012.12847}{{\ttfamily
  2012.12847}}].

\bibitem{Aoki:2021oez}
M.~Aoki, T.~Komatsu and H.~Shibuya, \emph{{Possibility of a multi-step
  electroweak phase transition in the two-Higgs doublet models}},
  \href{https://doi.org/10.1093/ptep/ptac068}{\emph{PTEP} {\bfseries 2022}
  (2022) 063B05} [\href{https://arxiv.org/abs/2106.03439}{{\ttfamily
  2106.03439}}].

\bibitem{Lewicki:2021pgr}
M.~Lewicki, M.~Merchand and M.~Zych, \emph{{Electroweak bubble wall expansion:
  gravitational waves and baryogenesis in Standard Model-like thermal plasma}},
  \href{https://doi.org/10.1007/JHEP02(2022)017}{\emph{JHEP} {\bfseries 02}
  (2022) 017} [\href{https://arxiv.org/abs/2111.02393}{{\ttfamily
  2111.02393}}].

\bibitem{Zhao:2022cnn}
Z.~Zhao, Y.~Di, L.~Bian and R.-G.~Cai, \emph{{Probing the electroweak symmetry
  breaking history with Gravitational waves}},
  \href{https://arxiv.org/abs/2204.04427}{{\ttfamily 2204.04427}}.

\bibitem{Benincasa:2022elt}
N.~Benincasa, L.~Delle~Rose, K.~Kannike and L.~Marzola, \emph{{Multi-step phase
  transitions and gravitational waves in the inert doublet model}},
  \href{https://doi.org/10.1088/1475-7516/2022/12/025}{\emph{JCAP} {\bfseries
  12} (2022) 025} [\href{https://arxiv.org/abs/2205.06669}{{\ttfamily
  2205.06669}}].

\bibitem{Cao:2022ocg}
Q.-H.~Cao, K.~Hashino, X.-X.~Li and J.-H.~Yue, \emph{{Multi-step phase
  transition and gravitational wave from general $\mathbb{Z}_2$ scalar
  extensions}},  \href{https://arxiv.org/abs/2212.07756}{{\ttfamily
  2212.07756}}.

\bibitem{Witten:1980ez}
E.~Witten, \emph{{Cosmological Consequences of a Light Higgs Boson}},
  \href{https://doi.org/10.1016/0550-3213(81)90182-6}{\emph{Nucl. Phys. B}
  {\bfseries 177} (1981) 477}.

\bibitem{Iso:2017uuu}
S.~Iso, P.D.~Serpico and K.~Shimada, \emph{{QCD-Electroweak First-Order Phase
  Transition in a Supercooled Universe}},
  \href{https://doi.org/10.1103/PhysRevLett.119.141301}{\emph{Phys. Rev. Lett.}
  {\bfseries 119} (2017) 141301}
  [\href{https://arxiv.org/abs/1704.04955}{{\ttfamily 1704.04955}}].

\bibitem{Arunasalam:2017ajm}
S.~Arunasalam, A.~Kobakhidze, C.~Lagger, S.~Liang and A.~Zhou, \emph{{Low
  temperature electroweak phase transition in the Standard Model with hidden
  scale invariance}},
  \href{https://doi.org/10.1016/j.physletb.2017.11.017}{\emph{Phys. Lett. B}
  {\bfseries 776} (2018) 48}
  [\href{https://arxiv.org/abs/1709.10322}{{\ttfamily 1709.10322}}].

\bibitem{vonHarling:2017yew}
B.~von Harling and G.~Servant, \emph{{QCD-induced Electroweak Phase
  Transition}}, \href{https://doi.org/10.1007/JHEP01(2018)159}{\emph{JHEP}
  {\bfseries 01} (2018) 159}
  [\href{https://arxiv.org/abs/1711.11554}{{\ttfamily 1711.11554}}].

\bibitem{Baratella:2018pxi}
P.~Baratella, A.~Pomarol and F.~Rompineve, \emph{{The Supercooled Universe}},
  \href{https://doi.org/10.1007/JHEP03(2019)100}{\emph{JHEP} {\bfseries 03}
  (2019) 100} [\href{https://arxiv.org/abs/1812.06996}{{\ttfamily
  1812.06996}}].

\bibitem{Bodeker:2021mcj}
D.~B\"odeker, \emph{{Remarks on the QCD-electroweak phase transition in a
  supercooled universe}},
  \href{https://doi.org/10.1103/PhysRevD.104.L111501}{\emph{Phys. Rev. D}
  {\bfseries 104} (2021) L111501}
  [\href{https://arxiv.org/abs/2108.11966}{{\ttfamily 2108.11966}}].

\bibitem{Sagunski:2023ynd}
L.~Sagunski, P.~Schicho and D.~Schmitt, \emph{{Supercool exit: Gravitational
  waves from QCD-triggered conformal symmetry breaking}},
  \href{https://doi.org/10.1103/PhysRevD.107.123512}{\emph{Phys. Rev. D}
  {\bfseries 107} (2023) 123512}
  [\href{https://arxiv.org/abs/2303.02450}{{\ttfamily 2303.02450}}].

\bibitem{Deng:2017uwc}
H.~Deng and A.~Vilenkin, \emph{{Primordial black hole formation by vacuum
  bubbles}}, \href{https://doi.org/10.1088/1475-7516/2017/12/044}{\emph{JCAP}
  {\bfseries 12} (2017) 044}
  [\href{https://arxiv.org/abs/1710.02865}{{\ttfamily 1710.02865}}].

\bibitem{Kawana:2021tde}
K.~Kawana and K.-P.~Xie, \emph{{Primordial black holes from a cosmic phase
  transition: The collapse of Fermi-balls}},
  \href{https://doi.org/10.1016/j.physletb.2021.136791}{\emph{Phys. Lett. B}
  {\bfseries 824} (2022) 136791}
  [\href{https://arxiv.org/abs/2106.00111}{{\ttfamily 2106.00111}}].

\bibitem{Jung:2021mku}
T.H.~Jung and T.~Okui, \emph{{Primordial black holes from bubble collisions
  during a first-order phase transition}},
  \href{https://arxiv.org/abs/2110.04271}{{\ttfamily 2110.04271}}.

\bibitem{Maeso:2021xvl}
D.N.~Maeso, L.~Marzola, M.~Raidal, V.~Vaskonen and H.~Veerm\"ae,
  \emph{{Primordial black holes from spectator field bubbles}},
  \href{https://doi.org/10.1088/1475-7516/2022/02/017}{\emph{JCAP} {\bfseries
  02} (2022) 017} [\href{https://arxiv.org/abs/2112.01505}{{\ttfamily
  2112.01505}}].

\bibitem{Hashino:2022tcs}
K.~Hashino, S.~Kanemura, T.~Takahashi and M.~Tanaka, \emph{{Probing first-order
  electroweak phase transition via primordial black holes in the effective
  field theory}},
  \href{https://doi.org/10.1016/j.physletb.2023.137688}{\emph{Phys. Lett. B}
  {\bfseries 838} (2023) 137688}
  [\href{https://arxiv.org/abs/2211.16225}{{\ttfamily 2211.16225}}].

\bibitem{Huang:2022him}
P.~Huang and K.-P.~Xie, \emph{{Primordial black holes from an electroweak phase
  transition}}, \href{https://doi.org/10.1103/PhysRevD.105.115033}{\emph{Phys.
  Rev. D} {\bfseries 105} (2022) 115033}
  [\href{https://arxiv.org/abs/2201.07243}{{\ttfamily 2201.07243}}].

\bibitem{Cline:2022xhx}
J.M.~Cline, B.~Laurent, S.~Raby and J.-S.~Roux, \emph{{PeV-scale leptogenesis,
  gravitational waves, and black holes from a SUSY-breaking phase transition}},
  \href{https://doi.org/10.1103/PhysRevD.107.095011}{\emph{Phys. Rev. D}
  {\bfseries 107} (2023) 095011}
  [\href{https://arxiv.org/abs/2211.00422}{{\ttfamily 2211.00422}}].

\bibitem{Kawana:2022olo}
K.~Kawana, T.~Kim and P.~Lu, \emph{{PBH Formation from Overdensities in Delayed
  Vacuum Transitions}},  \href{https://arxiv.org/abs/2212.14037}{{\ttfamily
  2212.14037}}.

\bibitem{Jinno:2023vnr}
R.~Jinno, J.~Kume and M.~Yamada, \emph{{Super-slow phase transition catalyzed
  by BHs and the birth of baby BHs}},
  \href{https://arxiv.org/abs/2310.06901}{{\ttfamily 2310.06901}}.

\bibitem{Rafelski:2015lva}
J.~Rafelski and J.~Birrell, \emph{{Dynamical Emergence of the Universe into the
  False Vacuum}},
  \href{https://doi.org/10.1088/1475-7516/2015/11/035}{\emph{JCAP} {\bfseries
  11} (2015) 035} [\href{https://arxiv.org/abs/1510.05001}{{\ttfamily
  1510.05001}}].

\bibitem{Balazs:2023kuk}
C.~Bal\'azs, Y.~Xiao, J.M.~Yang and Y.~Zhang, \emph{{New vacuum stability limit
  from cosmological history}},
  \href{https://arxiv.org/abs/2301.09283}{{\ttfamily 2301.09283}}.

\bibitem{Weir:2017wfa}
D.J.~Weir, \emph{{Gravitational waves from a first order electroweak phase
  transition: a brief review}},
  \href{https://doi.org/10.1098/rsta.2017.0126}{\emph{Phil. Trans. Roy. Soc.
  Lond. A} {\bfseries 376} (2018) 20170126}
  [\href{https://arxiv.org/abs/1705.01783}{{\ttfamily 1705.01783}}].

\bibitem{Guo:2021qcq}
H.-K.~Guo, K.~Sinha, D.~Vagie and G.~White, \emph{{The benefits of diligence:
  how precise are predicted gravitational wave spectra in models with phase
  transitions?}}, \href{https://doi.org/10.1007/JHEP06(2021)164}{\emph{JHEP}
  {\bfseries 06} (2021) 164}
  [\href{https://arxiv.org/abs/2103.06933}{{\ttfamily 2103.06933}}].

\bibitem{Espinosa:2010hh}
J.R.~Espinosa, T.~Konstandin, J.M.~No and G.~Servant, \emph{{Energy Budget of
  Cosmological First-order Phase Transitions}},
  \href{https://doi.org/10.1088/1475-7516/2010/06/028}{\emph{JCAP} {\bfseries
  06} (2010) 028} [\href{https://arxiv.org/abs/1004.4187}{{\ttfamily
  1004.4187}}].

\bibitem{Wang:2022txy}
S.-J.~Wang and Z.-Y.~Yuwen, \emph{{Hydrodynamic backreaction force of
  cosmological bubble expansion}},
  \href{https://doi.org/10.1103/PhysRevD.107.023501}{\emph{Phys. Rev. D}
  {\bfseries 107} (2023) 023501}
  [\href{https://arxiv.org/abs/2205.02492}{{\ttfamily 2205.02492}}].

\bibitem{Cai:2020djd}
R.-G.~Cai and S.-J.~Wang, \emph{{Effective picture of bubble expansion}},
  \href{https://doi.org/10.1088/1475-7516/2021/03/096}{\emph{JCAP} {\bfseries
  03} (2021) 096} [\href{https://arxiv.org/abs/2011.11451}{{\ttfamily
  2011.11451}}].

\bibitem{Hindmarsh:2019phv}
M.~Hindmarsh and M.~Hijazi, \emph{{Gravitational waves from first order
  cosmological phase transitions in the Sound Shell Model}},
  \href{https://doi.org/10.1088/1475-7516/2019/12/062}{\emph{JCAP} {\bfseries
  12} (2019) 062} [\href{https://arxiv.org/abs/1909.10040}{{\ttfamily
  1909.10040}}].

\bibitem{Cutting:2021tqt}
D.~Cutting, \emph{{Simulations of early universe phase transitions and
  gravitational waves}}, Ph.D. thesis, University of Sussex, Sussex U., 2021.
\newblock
  \href{http://sro.sussex.ac.uk/id/eprint/96437/}{http://sro.sussex.ac.uk/id/eprint/96437/}.

\bibitem{Leitao:2010yw}
L.~Leitao and A.~Megevand, \emph{{Spherical and non-spherical bubbles in
  cosmological phase transitions}},
  \href{https://doi.org/10.1016/j.nuclphysb.2010.11.012}{\emph{Nucl. Phys. B}
  {\bfseries 844} (2011) 450}
  [\href{https://arxiv.org/abs/1010.2134}{{\ttfamily 1010.2134}}].

\bibitem{Hindmarsh:2013xza}
M.~Hindmarsh, S.J.~Huber, K.~Rummukainen and D.J.~Weir, \emph{{Gravitational
  waves from the sound of a first order phase transition}},
  \href{https://doi.org/10.1103/PhysRevLett.112.041301}{\emph{Phys. Rev. Lett.}
  {\bfseries 112} (2014) 041301}
  [\href{https://arxiv.org/abs/1304.2433}{{\ttfamily 1304.2433}}].

\bibitem{LandauLifshitz-1D}
L.~Landau and E.~Lifshitz, \emph{{CHAPTER X --- ONE-DIMENSIONAL GAS FLOW}},  in
  \emph{Fluid Mechanics (Second Edition)}, L.~Landau and E.~Lifshitz, eds.,
  pp.~361--413, Pergamon (1987),
  \href{https://doi.org/https://doi.org/10.1016/B978-0-08-033933-7.50018-0}{DOI}.

\bibitem{Steinhardt:1981ct}
P.J.~Steinhardt, \emph{{Relativistic Detonation Waves and Bubble Growth in
  False Vacuum Decay}},
  \href{https://doi.org/10.1103/PhysRevD.25.2074}{\emph{Phys. Rev. D}
  {\bfseries 25} (1982) 2074}.

\bibitem{LandauLifshitz-relativistic}
L.~Landau and E.~Lifshitz, \emph{{CHAPTER XV - RELATIVISTIC FLUID DYNAMICS}},
  in \emph{Fluid Mechanics (Second Edition)}, L.~Landau and E.~Lifshitz, eds.,
  pp.~505--514, Pergamon (1987),
  \href{https://doi.org/https://doi.org/10.1016/B978-0-08-033933-7.50023-4}{DOI}.

\bibitem{Cai:2018teh}
R.-G.~Cai and S.-J.~Wang, \emph{{Energy budget of cosmological first-order
  phase transition in FLRW background}},
  \href{https://doi.org/10.1007/s11433-018-9216-7}{\emph{Sci. China Phys. Mech.
  Astron.} {\bfseries 61} (2018) 080411}
  [\href{https://arxiv.org/abs/1803.03002}{{\ttfamily 1803.03002}}].

\bibitem{Kamionkowski:1993fg}
M.~Kamionkowski, A.~Kosowsky and M.S.~Turner, \emph{{Gravitational radiation
  from first order phase transitions}},
  \href{https://doi.org/10.1103/PhysRevD.49.2837}{\emph{Phys. Rev. D}
  {\bfseries 49} (1994) 2837}
  [\href{https://arxiv.org/abs/astro-ph/9310044}{{\ttfamily
  astro-ph/9310044}}].

\bibitem{Athron:2023rfq}
P.~Athron, L.~Morris and Z.~Xu, \emph{{How robust are gravitational wave
  predictions from cosmological phase transitions?}},
  \href{https://arxiv.org/abs/2309.05474}{{\ttfamily 2309.05474}}.

\bibitem{LandauLifshitz-sound}
L.~Landau and E.~Lifshitz, \emph{{Chapter VIII --- Sound}},  in \emph{Fluid
  Mechanics (Second Edition)}, L.~Landau and E.~Lifshitz, eds., pp.~251--312,
  Pergamon (1987),
  \href{https://doi.org/https://doi.org/10.1016/B978-0-08-033933-7.50016-7}{DOI}.

\bibitem{Wang:2021dwl}
X.~Wang, F.P.~Huang and Y.~Li, \emph{{Sound velocity effects on the phase
  transition gravitational wave spectrum in the sound shell model}},
  \href{https://doi.org/10.1103/PhysRevD.105.103513}{\emph{Phys. Rev. D}
  {\bfseries 105} (2022) 103513}
  [\href{https://arxiv.org/abs/2112.14650}{{\ttfamily 2112.14650}}].

\bibitem{Tenkanen:2022tly}
T.V.I.~Tenkanen and J.~van~de Vis, \emph{{Speed of sound in cosmological phase
  transitions and effect on gravitational waves}},
  \href{https://doi.org/10.1007/JHEP08(2022)302}{\emph{JHEP} {\bfseries 08}
  (2022) 302} [\href{https://arxiv.org/abs/2206.01130}{{\ttfamily
  2206.01130}}].

\bibitem{Hindmarsh:2017gnf}
M.~Hindmarsh, S.J.~Huber, K.~Rummukainen and D.J.~Weir, \emph{{Shape of the
  acoustic gravitational wave power spectrum from a first order phase
  transition}}, \href{https://doi.org/10.1103/PhysRevD.96.103520}{\emph{Phys.
  Rev. D} {\bfseries 96} (2017) 103520}
  [\href{https://arxiv.org/abs/1704.05871}{{\ttfamily 1704.05871}}], [Erratum:
  \href{https://doi.org/10.1103/PhysRevD.101.089902}{\textit{Phys.~Rev.~D}
  \textbf{101} (2020) 089902}].

\bibitem{Gowling:2021gcy}
C.~Gowling and M.~Hindmarsh, \emph{{Observational prospects for phase
  transitions at LISA: Fisher matrix analysis}},
  \href{https://doi.org/10.1088/1475-7516/2021/10/039}{\emph{JCAP} {\bfseries
  10} (2021) 039} [\href{https://arxiv.org/abs/2106.05984}{{\ttfamily
  2106.05984}}].

\bibitem{Ellis:2020awk}
J.~Ellis, M.~Lewicki and J.M.~No, \emph{{Gravitational waves from first-order
  cosmological phase transitions: lifetime of the sound wave source}},
  \href{https://doi.org/10.1088/1475-7516/2020/07/050}{\emph{JCAP} {\bfseries
  07} (2020) 050} [\href{https://arxiv.org/abs/2003.07360}{{\ttfamily
  2003.07360}}].

\bibitem{RoperPol:2019wvy}
A.~Roper~Pol, S.~Mandal, A.~Brandenburg, T.~Kahniashvili and A.~Kosowsky,
  \emph{{Numerical simulations of gravitational waves from early-universe
  turbulence}}, \href{https://doi.org/10.1103/PhysRevD.102.083512}{\emph{Phys.
  Rev. D} {\bfseries 102} (2020) 083512}
  [\href{https://arxiv.org/abs/1903.08585}{{\ttfamily 1903.08585}}].

\bibitem{Zhong:2021hgo}
H.~Zhong, B.~Gong and T.~Qiu, \emph{{Gravitational waves from bubble collisions
  in FLRW spacetime}},  \href{https://arxiv.org/abs/2107.01845}{{\ttfamily
  2107.01845}}.

\bibitem{Kosowsky:2001xp}
A.~Kosowsky, A.~Mack and T.~Kahniashvili, \emph{{Gravitational radiation from
  cosmological turbulence}},
  \href{https://doi.org/10.1103/PhysRevD.66.024030}{\emph{Phys. Rev. D}
  {\bfseries 66} (2002) 024030}
  [\href{https://arxiv.org/abs/astro-ph/0111483}{{\ttfamily
  astro-ph/0111483}}].

\bibitem{Caprini:2009yp}
C.~Caprini, R.~Durrer and G.~Servant, \emph{{The stochastic gravitational wave
  background from turbulence and magnetic fields generated by a first-order
  phase transition}},
  \href{https://doi.org/10.1088/1475-7516/2009/12/024}{\emph{JCAP} {\bfseries
  12} (2009) 024} [\href{https://arxiv.org/abs/0909.0622}{{\ttfamily
  0909.0622}}].

\bibitem{Huber:2007vva}
S.J.~Huber and T.~Konstandin, \emph{{Production of gravitational waves in the
  nMSSM}}, \href{https://doi.org/10.1088/1475-7516/2008/05/017}{\emph{JCAP}
  {\bfseries 05} (2008) 017} [\href{https://arxiv.org/abs/0709.2091}{{\ttfamily
  0709.2091}}].

\bibitem{Enqvist:1991xw}
K.~Enqvist, J.~Ignatius, K.~Kajantie and K.~Rummukainen, \emph{{Nucleation and
  bubble growth in a first order cosmological electroweak phase transition}},
  \href{https://doi.org/10.1103/PhysRevD.45.3415}{\emph{Phys. Rev. D}
  {\bfseries 45} (1992) 3415}.

\bibitem{Jinno:2017ixd}
R.~Jinno, S.~Lee, H.~Seong and M.~Takimoto, \emph{{Gravitational waves from
  first-order phase transitions: Towards model separation by bubble nucleation
  rate}}, \href{https://doi.org/10.1088/1475-7516/2017/11/050}{\emph{JCAP}
  {\bfseries 11} (2017) 050}
  [\href{https://arxiv.org/abs/1708.01253}{{\ttfamily 1708.01253}}].

\bibitem{Hindmarsh:2016lnk}
M.~Hindmarsh, \emph{{Sound shell model for acoustic gravitational wave
  production at a first-order phase transition in the early Universe}},
  \href{https://doi.org/10.1103/PhysRevLett.120.071301}{\emph{Phys. Rev. Lett.}
  {\bfseries 120} (2018) 071301}
  [\href{https://arxiv.org/abs/1608.04735}{{\ttfamily 1608.04735}}].

\bibitem{Megevand:2021juo}
A.~Megevand and F.A.~Membiela, \emph{{Gravitational waves from bubble walls}},
  \href{https://doi.org/10.1088/1475-7516/2021/10/073}{\emph{JCAP} {\bfseries
  10} (2021) 073} [\href{https://arxiv.org/abs/2108.05510}{{\ttfamily
  2108.05510}}].

\bibitem{Ignatius:1993qn}
J.~Ignatius, K.~Kajantie, H.~Kurki-Suonio and M.~Laine, \emph{{The growth of
  bubbles in cosmological phase transitions}},
  \href{https://doi.org/10.1103/PhysRevD.49.3854}{\emph{Phys. Rev. D}
  {\bfseries 49} (1994) 3854}
  [\href{https://arxiv.org/abs/astro-ph/9309059}{{\ttfamily
  astro-ph/9309059}}].

\bibitem{Rezzolla:1996ey}
L.~Rezzolla, \emph{{Stability of cosmological detonation front}},
  \href{https://doi.org/10.1103/PhysRevD.54.1345}{\emph{Phys. Rev. D}
  {\bfseries 54} (1996) 1345}
  [\href{https://arxiv.org/abs/astro-ph/9605033}{{\ttfamily
  astro-ph/9605033}}].

\bibitem{Laine:1993ey}
M.~Laine, \emph{{Bubble growth as a detonation}},
  \href{https://doi.org/10.1103/PhysRevD.49.3847}{\emph{Phys. Rev. D}
  {\bfseries 49} (1994) 3847}
  [\href{https://arxiv.org/abs/hep-ph/9309242}{{\ttfamily hep-ph/9309242}}].

\bibitem{Link:1992dm}
B.~Link, \emph{{Deflagration instability in the quark - hadron phase
  transition}}, \href{https://doi.org/10.1103/PhysRevLett.68.2425}{\emph{Phys.
  Rev. Lett.} {\bfseries 68} (1992) 2425}.

\bibitem{Huet:1992ex}
P.Y.~Huet, K.~Kajantie, R.G.~Leigh, B.-H.~Liu and L.D.~McLerran,
  \emph{{Hydrodynamic stability analysis of burning bubbles in electroweak
  theory and in QCD}},
  \href{https://doi.org/10.1103/PhysRevD.48.2477}{\emph{Phys. Rev. D}
  {\bfseries 48} (1993) 2477}
  [\href{https://arxiv.org/abs/hep-ph/9212224}{{\ttfamily hep-ph/9212224}}].

\bibitem{Megevand:2013yua}
A.~Megevand and F.A.~Membiela, \emph{{Stability of cosmological deflagration
  fronts}}, \href{https://doi.org/10.1103/PhysRevD.89.103507}{\emph{Phys. Rev.
  D} {\bfseries 89} (2014) 103507}
  [\href{https://arxiv.org/abs/1311.2453}{{\ttfamily 1311.2453}}].

\bibitem{Megevand:2014dua}
A.~Megevand, F.A.~Membiela and A.D.~Sanchez, \emph{{Lower bound on the
  electroweak wall velocity from hydrodynamic instability}},
  \href{https://doi.org/10.1088/1475-7516/2015/03/051}{\emph{JCAP} {\bfseries
  03} (2015) 051} [\href{https://arxiv.org/abs/1412.8064}{{\ttfamily
  1412.8064}}].

\bibitem{Kamionkowski:1992dc}
M.~Kamionkowski and K.~Freese, \emph{{Instability and subsequent evolution of
  electroweak bubbles}},
  \href{https://doi.org/10.1103/PhysRevLett.69.2743}{\emph{Phys. Rev. Lett.}
  {\bfseries 69} (1992) 2743}
  [\href{https://arxiv.org/abs/hep-ph/9208202}{{\ttfamily hep-ph/9208202}}].

\bibitem{Ai:2023see}
W.-Y.~Ai, B.~Laurent and J.~van~de Vis, \emph{{Model-independent bubble wall
  velocities in local thermal equilibrium}},
  \href{https://doi.org/10.1088/1475-7516/2023/07/002}{\emph{JCAP} {\bfseries
  07} (2023) 002} [\href{https://arxiv.org/abs/2303.10171}{{\ttfamily
  2303.10171}}].

\bibitem{Kurki-Suonio:1995rrv}
H.~Kurki-Suonio and M.~Laine, \emph{{Supersonic deflagrations in cosmological
  phase transitions}},
  \href{https://doi.org/10.1103/PhysRevD.51.5431}{\emph{Phys. Rev. D}
  {\bfseries 51} (1995) 5431}
  [\href{https://arxiv.org/abs/hep-ph/9501216}{{\ttfamily hep-ph/9501216}}].

\bibitem{Gyulassy:1983rq}
M.~Gyulassy, K.~Kajantie, H.~Kurki-Suonio and L.D.~McLerran,
  \emph{{Deflagrations and Detonations as a Mechanism of Hadron Bubble Growth
  in Supercooled Quark Gluon Plasma}},
  \href{https://doi.org/10.1016/0550-3213(84)90004-X}{\emph{Nucl. Phys. B}
  {\bfseries 237} (1984) 477}.

\bibitem{Kurki-Suonio:1984zeb}
H.~Kurki-Suonio, \emph{{Deflagration Bubbles in the Quark - Hadron Phase
  Transition}}, \href{https://doi.org/10.1016/0550-3213(85)90135-X}{\emph{Nucl.
  Phys. B} {\bfseries 255} (1985) 231}.

\bibitem{LandauLifshitz-combustion}
L.~Landau and E.~Lifshitz, \emph{{Chapter XIV --- Fluid Dynamics of
  Combustion}},  in \emph{Fluid Mechanics (Second Edition)}, L.~Landau and
  E.~Lifshitz, eds., pp.~484--504, Pergamon (1987),
  \href{https://doi.org/https://doi.org/10.1016/B978-0-08-033933-7.50022-2}{DOI}.

\bibitem{Turok:1992jp}
N.~Turok, \emph{{Electroweak bubbles: Nucleation and growth}},
  \href{https://doi.org/10.1103/PhysRevLett.68.1803}{\emph{Phys. Rev. Lett.}
  {\bfseries 68} (1992) 1803}.

\bibitem{Liu:1992tn}
B.-H.~Liu, L.D.~McLerran and N.~Turok, \emph{{Bubble nucleation and growth at a
  baryon number producing electroweak phase transition}},
  \href{https://doi.org/10.1103/PhysRevD.46.2668}{\emph{Phys. Rev. D}
  {\bfseries 46} (1992) 2668}.

\bibitem{Bodeker:2009qy}
D.~Bodeker and G.D.~Moore, \emph{{Can electroweak bubble walls run away?}},
  \href{https://doi.org/10.1088/1475-7516/2009/05/009}{\emph{JCAP} {\bfseries
  05} (2009) 009} [\href{https://arxiv.org/abs/0903.4099}{{\ttfamily
  0903.4099}}].

\bibitem{Bodeker:2017cim}
D.~Bodeker and G.D.~Moore, \emph{{Electroweak Bubble Wall Speed Limit}},
  \href{https://doi.org/10.1088/1475-7516/2017/05/025}{\emph{JCAP} {\bfseries
  05} (2017) 025} [\href{https://arxiv.org/abs/1703.08215}{{\ttfamily
  1703.08215}}].

\bibitem{Bea:2021zsu}
Y.~Bea, J.~Casalderrey-Solana, T.~Giannakopoulos, D.~Mateos,
  M.~Sanchez-Garitaonandia and M.~Zilh\~ao, \emph{{Bubble wall velocity from
  holography}}, \href{https://doi.org/10.1103/PhysRevD.104.L121903}{\emph{Phys.
  Rev. D} {\bfseries 104} (2021) L121903}
  [\href{https://arxiv.org/abs/2104.05708}{{\ttfamily 2104.05708}}].

\bibitem{Bigazzi:2021ucw}
F.~Bigazzi, A.~Caddeo, T.~Canneti and A.L.~Cotrone, \emph{{Bubble wall velocity
  at strong coupling}},
  \href{https://doi.org/10.1007/JHEP08(2021)090}{\emph{JHEP} {\bfseries 08}
  (2021) 090} [\href{https://arxiv.org/abs/2104.12817}{{\ttfamily
  2104.12817}}].

\bibitem{LiLi:2023dlc}
L.~Li, S.-J.~Wang and Z.-Y.~Yuwen, \emph{{Bubble expansion at strong
  coupling}},  \href{https://arxiv.org/abs/2302.10042}{{\ttfamily 2302.10042}}.

\bibitem{Moore:1995si}
G.D.~Moore and T.~Prokopec, \emph{{How fast can the wall move? A Study of the
  electroweak phase transition dynamics}},
  \href{https://doi.org/10.1103/PhysRevD.52.7182}{\emph{Phys. Rev. D}
  {\bfseries 52} (1995) 7182}
  [\href{https://arxiv.org/abs/hep-ph/9506475}{{\ttfamily hep-ph/9506475}}].

\bibitem{Moore:1995ua}
G.D.~Moore and T.~Prokopec, \emph{{Bubble wall velocity in a first order
  electroweak phase transition}},
  \href{https://doi.org/10.1103/PhysRevLett.75.777}{\emph{Phys. Rev. Lett.}
  {\bfseries 75} (1995) 777}
  [\href{https://arxiv.org/abs/hep-ph/9503296}{{\ttfamily hep-ph/9503296}}].

\bibitem{Konstandin:2014zta}
T.~Konstandin, G.~Nardini and I.~Rues, \emph{{From Boltzmann equations to
  steady wall velocities}},
  \href{https://doi.org/10.1088/1475-7516/2014/09/028}{\emph{JCAP} {\bfseries
  09} (2014) 028} [\href{https://arxiv.org/abs/1407.3132}{{\ttfamily
  1407.3132}}].

\bibitem{Ai:2021kak}
W.-Y.~Ai, B.~Garbrecht and C.~Tamarit, \emph{{Bubble wall velocities in local
  equilibrium}},
  \href{https://doi.org/10.1088/1475-7516/2022/03/015}{\emph{JCAP} {\bfseries
  03} (2022) 015} [\href{https://arxiv.org/abs/2109.13710}{{\ttfamily
  2109.13710}}].

\bibitem{Megevand:2013hwa}
A.~M\'egevand, \emph{{Friction forces on phase transition fronts}},
  \href{https://doi.org/10.1088/1475-7516/2013/07/045}{\emph{JCAP} {\bfseries
  07} (2013) 045} [\href{https://arxiv.org/abs/1303.4233}{{\ttfamily
  1303.4233}}].

\bibitem{Balaji:2020yrx}
S.~Balaji, M.~Spannowsky and C.~Tamarit, \emph{{Cosmological bubble friction in
  local equilibrium}},
  \href{https://doi.org/10.1088/1475-7516/2021/03/051}{\emph{JCAP} {\bfseries
  03} (2021) 051} [\href{https://arxiv.org/abs/2010.08013}{{\ttfamily
  2010.08013}}].

\bibitem{BarrosoMancha:2020fay}
M.~Barroso~Mancha, T.~Prokopec and B.~Swiezewska, \emph{{Field-theoretic
  derivation of bubble-wall force}},
  \href{https://doi.org/10.1007/JHEP01(2021)070}{\emph{JHEP} {\bfseries 01}
  (2021) 070} [\href{https://arxiv.org/abs/2005.10875}{{\ttfamily
  2005.10875}}].

\bibitem{Laurent:2022jrs}
B.~Laurent and J.M.~Cline, \emph{{First principles determination of bubble wall
  velocity}}, \href{https://doi.org/10.1103/PhysRevD.106.023501}{\emph{Phys.
  Rev. D} {\bfseries 106} (2022) 023501}
  [\href{https://arxiv.org/abs/2204.13120}{{\ttfamily 2204.13120}}].

\bibitem{DeCurtis:2023hil}
S.~De~Curtis, L.~Delle~Rose, A.~Guiggiani, A.~Gil~Muyor and G.~Panico,
  \emph{{Collision integrals for cosmological phase transitions}},
  \href{https://doi.org/10.1007/JHEP05(2023)194}{\emph{JHEP} {\bfseries 05}
  (2023) 194} [\href{https://arxiv.org/abs/2303.05846}{{\ttfamily
  2303.05846}}].

\bibitem{Laurent:2020gpg}
B.~Laurent and J.M.~Cline, \emph{{Fluid equations for fast-moving electroweak
  bubble walls}},
  \href{https://doi.org/10.1103/PhysRevD.102.063516}{\emph{Phys. Rev. D}
  {\bfseries 102} (2020) 063516}
  [\href{https://arxiv.org/abs/2007.10935}{{\ttfamily 2007.10935}}].

\bibitem{Hoche:2020ysm}
S.~H\"oche, J.~Kozaczuk, A.J.~Long, J.~Turner and Y.~Wang, \emph{{Towards an
  all-orders calculation of the electroweak bubble wall velocity}},
  \href{https://doi.org/10.1088/1475-7516/2021/03/009}{\emph{JCAP} {\bfseries
  03} (2021) 009} [\href{https://arxiv.org/abs/2007.10343}{{\ttfamily
  2007.10343}}].

\bibitem{Wang:2020zlf}
X.~Wang, F.P.~Huang and X.~Zhang, \emph{{Bubble wall velocity beyond
  leading-log approximation in electroweak phase transition}},
  \href{https://arxiv.org/abs/2011.12903}{{\ttfamily 2011.12903}}.

\bibitem{Lewicki:2022nba}
M.~Lewicki, V.~Vaskonen and H.~Veerm\"ae, \emph{{Bubble dynamics in fluids with
  N-body simulations}},
  \href{https://doi.org/10.1103/PhysRevD.106.103501}{\emph{Phys. Rev. D}
  {\bfseries 106} (2022) 103501}
  [\href{https://arxiv.org/abs/2205.05667}{{\ttfamily 2205.05667}}].

\bibitem{Friedlander:2020tnq}
A.~Friedlander, I.~Banta, J.M.~Cline and D.~Tucker-Smith, \emph{{Wall speed and
  shape in singlet-assisted strong electroweak phase transitions}},
  \href{https://doi.org/10.1103/PhysRevD.103.055020}{\emph{Phys. Rev. D}
  {\bfseries 103} (2021) 055020}
  [\href{https://arxiv.org/abs/2009.14295}{{\ttfamily 2009.14295}}].

\bibitem{Dorsch:2021nje}
G.C.~Dorsch, S.J.~Huber and T.~Konstandin, \emph{{A sonic boom in bubble wall
  friction}}, \href{https://doi.org/10.1088/1475-7516/2022/04/010}{\emph{JCAP}
  {\bfseries 04} (2022) 010}
  [\href{https://arxiv.org/abs/2112.12548}{{\ttfamily 2112.12548}}].

\bibitem{DeCurtis:2022hlx}
S.~De~Curtis, L.D.~Rose, A.~Guiggiani, A.G.~Muyor and G.~Panico, \emph{{Bubble
  wall dynamics at the electroweak phase transition}},
  \href{https://doi.org/10.1007/JHEP03(2022)163}{\emph{JHEP} {\bfseries 03}
  (2022) 163} [\href{https://arxiv.org/abs/2201.08220}{{\ttfamily
  2201.08220}}].

\bibitem{Gouttenoire:2021kjv}
Y.~Gouttenoire, R.~Jinno and F.~Sala, \emph{{Friction pressure on relativistic
  bubble walls}}, \href{https://doi.org/10.1007/JHEP05(2022)004}{\emph{JHEP}
  {\bfseries 05} (2022) 004}
  [\href{https://arxiv.org/abs/2112.07686}{{\ttfamily 2112.07686}}].

\bibitem{GarciaGarcia:2022yqb}
I.~Garcia~Garcia, G.~Koszegi and R.~Petrossian-Byrne, \emph{{Reflections on
  bubble walls}}, \href{https://doi.org/10.1007/JHEP09(2023)013}{\emph{JHEP}
  {\bfseries 09} (2023) 013}
  [\href{https://arxiv.org/abs/2212.10572}{{\ttfamily 2212.10572}}].

\bibitem{Azatov:2020ufh}
A.~Azatov and M.~Vanvlasselaer, \emph{{Bubble wall velocity: heavy physics
  effects}}, \href{https://doi.org/10.1088/1475-7516/2021/01/058}{\emph{JCAP}
  {\bfseries 01} (2021) 058}
  [\href{https://arxiv.org/abs/2010.02590}{{\ttfamily 2010.02590}}].

\bibitem{Huber:2013kj}
S.J.~Huber and M.~Sopena, \emph{{An efficient approach to electroweak bubble
  velocities}},  \href{https://arxiv.org/abs/1302.1044}{{\ttfamily 1302.1044}}.

\bibitem{Kozaczuk:2015owa}
J.~Kozaczuk, \emph{{Bubble Expansion and the Viability of Singlet-Driven
  Electroweak Baryogenesis}},
  \href{https://doi.org/10.1007/JHEP10(2015)135}{\emph{JHEP} {\bfseries 10}
  (2015) 135} [\href{https://arxiv.org/abs/1506.04741}{{\ttfamily
  1506.04741}}].

\bibitem{Jiang:2022btc}
S.~Jiang, F.P.~Huang and X.~Wang, \emph{{Bubble wall velocity during
  electroweak phase transition in the inert doublet model}},
  \href{https://doi.org/10.1103/PhysRevD.107.095005}{\emph{Phys. Rev. D}
  {\bfseries 107} (2023) 095005}
  [\href{https://arxiv.org/abs/2211.13142}{{\ttfamily 2211.13142}}].

\bibitem{Krajewski:2023clt}
T.~Krajewski, M.~Lewicki and M.~Zych, \emph{{Hydrodynamical constraints on
  bubble wall velocity}},  \href{https://arxiv.org/abs/2303.18216}{{\ttfamily
  2303.18216}}.

\bibitem{1916AnP...354..769E}
A.~{Einstein}, \emph{{Die Grundlage der allgemeinen Relativit{\"a}tstheorie}},
  \href{https://doi.org/10.1002/andp.19163540702}{\emph{Annalen der Physik}
  {\bfseries 354} (1916) 769}.

\bibitem{1916SPAW.......688E}
A.~{Einstein}, \emph{{N{\"a}herungsweise Integration der Feldgleichungen der
  Gravitation}},
  \href{https://doi.org/10.1007/978-3-662-57411-9_13}{\emph{Sitzungsberichte
  der K{\"o}niglich Preu{\ss}ischen Akademie der Wissenschaften (Berlin} (1916)
  688}.

\bibitem{Einstein:1916cc}
A.~Einstein, \emph{{Approximative Integration of the Field Equations of
  Gravitation}}, {\emph{Sitzungsber. Preuss. Akad. Wiss. Berlin (Math. Phys. )}
  {\bfseries 1916} (1916) 688}.

\bibitem{Spurio:2019xej}
M.~Spurio, \emph{{An introduction to astrophysical observables in gravitational
  wave detections}},  \href{https://arxiv.org/abs/1906.03643}{{\ttfamily
  1906.03643}}.

\bibitem{Bishop:2021rye}
N.T.~Bishop, \emph{Introduction to gravitational wave astronomy},  in
  \emph{Handbook of Gravitational Wave Astronomy}, pp.~1--31, Springer
  Singapore (2021), \href{https://doi.org/10.1007/978-981-15-4702-7_1-1}{DOI}
  [\href{https://arxiv.org/abs/2103.07675}{{\ttfamily 2103.07675}}].

\bibitem{EINSTEIN193743}
A.~Einstein and N.~Rosen, \emph{On gravitational waves},
  \href{https://doi.org/https://doi.org/10.1016/S0016-0032(37)90583-0}{\emph{Journal
  of the Franklin Institute} {\bfseries 223} (1937) 43}.

\bibitem{rosen1937plane}
N.~Rosen, \emph{Plane polarized waves in the general theory of relativity},
  {\emph{Phys. Z. Sowjetunion} {\bfseries 12} (1937) 366}.

\bibitem{Pirani:1956wr}
F.A.E.~Pirani, \emph{{Invariant formulation of gravitational radiation
  theory}}, \href{https://doi.org/10.1103/PhysRev.105.1089}{\emph{Phys. Rev.}
  {\bfseries 105} (1957) 1089}.

\bibitem{Bondi:1957dt}
H.~Bondi, \emph{{Plane gravitational waves in general relativity}},
  \href{https://doi.org/10.1038/1791072a0}{\emph{Nature} {\bfseries 179} (1957)
  1072}.

\bibitem{Bondi:1958aj}
H.~Bondi, F.A.E.~Pirani and I.~Robinson, \emph{{Gravitational waves in general
  relativity. 3. Exact plane waves}},
  \href{https://doi.org/10.1098/rspa.1959.0124}{\emph{Proc. Roy. Soc. Lond. A}
  {\bfseries 251} (1959) 519}.

\bibitem{Penrose:1960eq}
R.~Penrose, \emph{{A Spinor approach to general relativity}},
  \href{https://doi.org/10.1016/0003-4916(60)90021-X}{\emph{Annals Phys.}
  {\bfseries 10} (1960) 171}.

\bibitem{1969eisp.book.....P}
A.Z.~{Petrov}, \emph{{Einstein spaces}}, Pergamon Press, Oxford, UK (1969).

\bibitem{Brinkmann1925}
H.~Brinkmann, \emph{Einstein spaces which are mapped conformally on each
  other}, {\emph{Mathematische Annalen} {\bfseries 94} (1925) 119}
  \href{http://eudml.org/doc/159101}{http://eudml.org/doc/159101}.

\bibitem{Trautman:1958rph}
A.~Trautman, \emph{{Boundary Conditions at Infinity for Physical Theories}},
  {\emph{Bull. Acad. Pol. Sci. Ser. Sci. Math. Astron. Phys.} {\bfseries 6}
  (1958) 403} [\href{https://arxiv.org/abs/1604.03144}{{\ttfamily
  1604.03144}}].

\bibitem{Trautman:1958zdi}
A.~Trautman, \emph{{Radiation and Boundary Conditions in the Theory of
  Gravitation}}, {\emph{Bull. Acad. Pol. Sci. Ser. Sci. Math. Astron. Phys.}
  {\bfseries 6} (1958) 407} [\href{https://arxiv.org/abs/1604.03145}{{\ttfamily
  1604.03145}}].

\bibitem{Robinson:1960zzb}
I.~Robinson and A.~Trautman, \emph{{Spherical Gravitational Waves}},
  \href{https://doi.org/10.1103/PhysRevLett.4.431}{\emph{Phys. Rev. Lett.}
  {\bfseries 4} (1960) 431}.

\bibitem{Robinson:1962zz}
I.~Robinson and A.~Trautman, \emph{{Some spherical gravitational waves in
  general relativity}},
  \href{https://doi.org/10.1098/rspa.1962.0036}{\emph{Proc. Roy. Soc. Lond. A}
  {\bfseries 265} (1962) 463}.

\bibitem{Cervantes-Cota:2016zjc}
J.L.~Cervantes-Cota, S.~Galindo-Uribarri and G.-F..~Smoot, \emph{{A Brief
  History of Gravitational Waves}},
  \href{https://doi.org/10.3390/universe2030022}{\emph{Universe} {\bfseries 2}
  (2016) 22} [\href{https://arxiv.org/abs/1609.09400}{{\ttfamily 1609.09400}}].

\bibitem{DiMauro:2021sah}
M.~Di~Mauro, S.~Esposito and A.~Naddeo, \emph{{Towards detecting gravitational
  waves: A contribution by Richard Feynman}},  in \emph{{16th Marcel Grossmann
  Meeting on~Recent Developments in Theoretical and Experimental General
  Relativity, Astrophysics and Relativistic Field Theories}}, 10, 2021,
  \href{https://doi.org/10.1142/9789811269776_0297}{DOI}
  [\href{https://arxiv.org/abs/2111.00330}{{\ttfamily 2111.00330}}].

\bibitem{Abbott:2016xvh}
B.P.~Abbott et~al., \emph{{Sensitivity of the Advanced LIGO detectors at the
  beginning of gravitational wave astronomy}},
  \href{https://doi.org/10.1103/PhysRevD.93.112004}{\emph{Phys. Rev. D}
  {\bfseries 93} (2016) 112004}
  [\href{https://arxiv.org/abs/1604.00439}{{\ttfamily 1604.00439}}], [Addendum:
  \href{https://doi.org/10.1103/PhysRevD.97.059901}{\textit{Phys. Rev. D}
  \textbf{97} (2018) 059901}].

\bibitem{LIGOScientific:2018mvr}
{\scshape LIGO Scientific, Virgo} collaboration, \emph{{GWTC-1: A
  Gravitational-Wave Transient Catalog of Compact Binary Mergers Observed by
  LIGO and Virgo during the First and Second Observing Runs}},
  \href{https://doi.org/10.1103/PhysRevX.9.031040}{\emph{Phys. Rev. X}
  {\bfseries 9} (2019) 031040}
  [\href{https://arxiv.org/abs/1811.12907}{{\ttfamily 1811.12907}}].

\bibitem{LIGOScientific:2020ibl}
{\scshape LIGO Scientific, Virgo} collaboration, \emph{{GWTC-2: Compact Binary
  Coalescences Observed by LIGO and Virgo During the First Half of the Third
  Observing Run}},
  \href{https://doi.org/10.1103/PhysRevX.11.021053}{\emph{Phys. Rev. X}
  {\bfseries 11} (2021) 021053}
  [\href{https://arxiv.org/abs/2010.14527}{{\ttfamily 2010.14527}}].

\bibitem{LIGOScientific:2021usb}
{\scshape LIGO Scientific, VIRGO} collaboration, \emph{{GWTC-2.1: Deep Extended
  Catalog of Compact Binary Coalescences Observed by LIGO and Virgo During the
  First Half of the Third Observing Run}},
  \href{https://arxiv.org/abs/2108.01045}{{\ttfamily 2108.01045}}.

\bibitem{LIGOScientific:2021djp}
{\scshape LIGO Scientific, VIRGO, KAGRA} collaboration, \emph{{GWTC-3: Compact
  Binary Coalescences Observed by LIGO and Virgo During the Second Part of the
  Third Observing Run}},  \href{https://arxiv.org/abs/2111.03606}{{\ttfamily
  2111.03606}}.

\bibitem{KAGRA:2022twx}
{\scshape KAGRA, VIRGO, LIGO Scientific} collaboration, \emph{{First joint
  observation by the underground gravitational-wave detector KAGRA with GEO
  600}}, \href{https://doi.org/10.1093/ptep/ptac073}{\emph{PTEP} {\bfseries
  2022} (2022) 063F01} [\href{https://arxiv.org/abs/2203.01270}{{\ttfamily
  2203.01270}}].

\bibitem{LIGOScientific:2014pky}
{\scshape LIGO Scientific} collaboration, \emph{{Advanced LIGO}},
  \href{https://doi.org/10.1088/0264-9381/32/7/074001}{\emph{Class. Quant.
  Grav.} {\bfseries 32} (2015) 074001}
  [\href{https://arxiv.org/abs/1411.4547}{{\ttfamily 1411.4547}}].

\bibitem{VIRGO:2014yos}
{\scshape VIRGO} collaboration, \emph{{Advanced Virgo: a second-generation
  interferometric gravitational wave detector}},
  \href{https://doi.org/10.1088/0264-9381/32/2/024001}{\emph{Class. Quant.
  Grav.} {\bfseries 32} (2015) 024001}
  [\href{https://arxiv.org/abs/1408.3978}{{\ttfamily 1408.3978}}].

\bibitem{Luck:2010rt}
H.~Luck et~al., \emph{{The upgrade of GEO600}},
  \href{https://doi.org/10.1088/1742-6596/228/1/012012}{\emph{J. Phys. Conf.
  Ser.} {\bfseries 228} (2010) 012012}
  [\href{https://arxiv.org/abs/1004.0339}{{\ttfamily 1004.0339}}].

\bibitem{Affeldt:2014rza}
C.~Affeldt et~al., \emph{{Advanced techniques in GEO 600}},
  \href{https://doi.org/10.1088/0264-9381/31/22/224002}{\emph{Class. Quant.
  Grav.} {\bfseries 31} (2014) 224002}.

\bibitem{Dooley:2015fpa}
K.L.~Dooley et~al., \emph{{GEO 600 and the GEO-HF upgrade program: successes
  and challenges}},
  \href{https://doi.org/10.1088/0264-9381/33/7/075009}{\emph{Class. Quant.
  Grav.} {\bfseries 33} (2016) 075009}
  [\href{https://arxiv.org/abs/1510.00317}{{\ttfamily 1510.00317}}].

\bibitem{KAGRA:2020tym}
{\scshape KAGRA} collaboration, \emph{{Overview of KAGRA: Detector design and
  construction history}},
  \href{https://doi.org/10.1093/ptep/ptaa125}{\emph{PTEP} {\bfseries 2021}
  (2021) 05A101} [\href{https://arxiv.org/abs/2005.05574}{{\ttfamily
  2005.05574}}].

\bibitem{Somiya:2011np}
{\scshape KAGRA} collaboration, \emph{{Detector configuration of KAGRA: The
  Japanese cryogenic gravitational-wave detector}},
  \href{https://doi.org/10.1088/0264-9381/29/12/124007}{\emph{Class. Quant.
  Grav.} {\bfseries 29} (2012) 124007}
  [\href{https://arxiv.org/abs/1111.7185}{{\ttfamily 1111.7185}}].

\bibitem{Aso:2013eba}
{\scshape KAGRA} collaboration, \emph{{Interferometer design of the KAGRA
  gravitational wave detector}},
  \href{https://doi.org/10.1103/PhysRevD.88.043007}{\emph{Phys. Rev. D}
  {\bfseries 88} (2013) 043007}
  [\href{https://arxiv.org/abs/1306.6747}{{\ttfamily 1306.6747}}].

\bibitem{Fujimoto:2000qbu}
{\scshape TAMA} collaboration, \emph{{TAMA ground-based interferometer for the
  detection of gravitational waves}},
  \href{https://doi.org/10.1016/S0273-1177(99)00979-5}{\emph{Adv. Space Res.}
  {\bfseries 25} (2000) 1161}.

\bibitem{Ando:2002bv}
{\scshape TAMA} collaboration, \emph{{Current status of TAMA}},
  \href{https://doi.org/10.1088/0264-9381/19/7/324}{\emph{Class. Quant. Grav.}
  {\bfseries 19} (2002) 1409}.

\bibitem{Reitze:2019iox}
D.~Reitze et~al., \emph{{Cosmic Explorer: The U.S. Contribution to
  Gravitational-Wave Astronomy beyond LIGO}}, {\emph{Bull. Am. Astron. Soc.}
  {\bfseries 51} (2019) 035}
  [\href{https://arxiv.org/abs/1907.04833}{{\ttfamily 1907.04833}}].

\bibitem{Punturo:2010zz}
M.~Punturo et~al., \emph{{The Einstein Telescope: A third-generation
  gravitational wave observatory}},
  \href{https://doi.org/10.1088/0264-9381/27/19/194002}{\emph{Class. Quant.
  Grav.} {\bfseries 27} (2010) 194002}.

\bibitem{Sathyaprakash:2012jk}
B.~Sathyaprakash et~al., \emph{{Scientific Objectives of Einstein Telescope}},
  \href{https://doi.org/10.1088/0264-9381/29/12/124013}{\emph{Class. Quant.
  Grav.} {\bfseries 29} (2012) 124013}
  [\href{https://arxiv.org/abs/1206.0331}{{\ttfamily 1206.0331}}], [Erratum:
  \href{https://doi.org/10.1088/0264-9381/30/7/079501}{\textit{Class. Quant.
  Grav.} \textbf{30} (2013) 079501}].

\bibitem{Maggiore:2019uih}
M.~Maggiore et~al., \emph{{Science Case for the Einstein Telescope}},
  \href{https://doi.org/10.1088/1475-7516/2020/03/050}{\emph{JCAP} {\bfseries
  03} (2020) 050} [\href{https://arxiv.org/abs/1912.02622}{{\ttfamily
  1912.02622}}].

\bibitem{Crowder:2005nr}
J.~Crowder and N.J.~Cornish, \emph{{Beyond LISA: Exploring future gravitational
  wave missions}},
  \href{https://doi.org/10.1103/PhysRevD.72.083005}{\emph{Phys. Rev. D}
  {\bfseries 72} (2005) 083005}
  [\href{https://arxiv.org/abs/gr-qc/0506015}{{\ttfamily gr-qc/0506015}}].

\bibitem{Baibhav:2019rsa}
V.~Baibhav et~al., \emph{{Probing the nature of black holes: Deep in the mHz
  gravitational-wave sky}},
  \href{https://doi.org/10.1007/s10686-021-09741-9}{\emph{Exper. Astron.}
  {\bfseries 51} (2021) 1385}
  [\href{https://arxiv.org/abs/1908.11390}{{\ttfamily 1908.11390}}].

\bibitem{Bailes:2021tot}
M.~Bailes et~al., \emph{{Gravitational-wave physics and astronomy in the 2020s
  and 2030s}}, \href{https://doi.org/10.1038/s42254-021-00303-8}{\emph{Nature
  Rev. Phys.} {\bfseries 3} (2021) 344}.

\bibitem{Kawamura:2011zz}
S.~Kawamura et~al., \emph{{The Japanese space gravitational wave antenna:
  DECIGO}}, \href{https://doi.org/10.1088/0264-9381/28/9/094011}{\emph{Class.
  Quant. Grav.} {\bfseries 28} (2011) 094011}.

\bibitem{Sato:2017dkf}
S.~Sato et~al., \emph{{The status of DECIGO}},
  \href{https://doi.org/10.1088/1742-6596/840/1/012010}{\emph{J. Phys. Conf.
  Ser.} {\bfseries 840} (2017) 012010}.

\bibitem{Kawamura:2020pcg}
S.~Kawamura et~al., \emph{{Current status of space gravitational wave antenna
  DECIGO and B-DECIGO}},
  \href{https://doi.org/10.1093/ptep/ptab019}{\emph{PTEP} {\bfseries 2021}
  (2021) 05A105} [\href{https://arxiv.org/abs/2006.13545}{{\ttfamily
  2006.13545}}].

\bibitem{Barausse:2020rsu}
E.~Barausse et~al., \emph{{Prospects for Fundamental Physics with LISA}},
  \href{https://doi.org/10.1007/s10714-020-02691-1}{\emph{Gen. Rel. Grav.}
  {\bfseries 52} (2020) 81} [\href{https://arxiv.org/abs/2001.09793}{{\ttfamily
  2001.09793}}].

\bibitem{TianQin:2015yph}
{\scshape TianQin} collaboration, \emph{{TianQin: a space-borne gravitational
  wave detector}},
  \href{https://doi.org/10.1088/0264-9381/33/3/035010}{\emph{Class. Quant.
  Grav.} {\bfseries 33} (2016) 035010}
  [\href{https://arxiv.org/abs/1512.02076}{{\ttfamily 1512.02076}}].

\bibitem{TianQin:2020hid}
{\scshape TianQin} collaboration, \emph{{The TianQin project: current progress
  on science and technology}},
  \href{https://doi.org/10.1093/ptep/ptaa114}{\emph{PTEP} {\bfseries 2021}
  (2021) 05A107} [\href{https://arxiv.org/abs/2008.10332}{{\ttfamily
  2008.10332}}].

\bibitem{Hu:2017mde}
W.-R.~Hu and Y.-L.~Wu, \emph{{The Taiji Program in Space for gravitational wave
  physics and the nature of gravity}},
  \href{https://doi.org/10.1093/nsr/nwx116}{\emph{Natl. Sci. Rev.} {\bfseries
  4} (2017) 685}.

\bibitem{Gong:2014mca}
X.~Gong et~al., \emph{{Descope of the ALIA mission}},
  \href{https://doi.org/10.1088/1742-6596/610/1/012011}{\emph{J. Phys. Conf.
  Ser.} {\bfseries 610} (2015) 012011}
  [\href{https://arxiv.org/abs/1410.7296}{{\ttfamily 1410.7296}}].

\bibitem{Harry_2006}
G.M.~Harry, P.~Fritschel, D.A.~Shaddock, W.~Folkner and E.S.~Phinney,
  \emph{Laser interferometry for the big bang observer},
  \href{https://doi.org/10.1088/0264-9381/23/15/008}{\emph{Classical and
  Quantum Gravity} {\bfseries 23} (2006) 4887}.

\bibitem{Verbiest:2021kmt}
J.P.W.~Verbiest, S.~Os{\l}owski and S.~Burke-Spolaor, \emph{{Pulsar Timing
  Array Experiments}},  in \emph{Handbook of Gravitational Wave Astronomy},
  pp.~157--198, Springer (2022),
  \href{https://doi.org/10.1007/978-981-15-4702-7_4-1}{DOI}
  [\href{https://arxiv.org/abs/2101.10081}{{\ttfamily 2101.10081}}].

\bibitem{2016ASPC..502...19L}
K.J.~{Lee}, \emph{{Prospects of Gravitational Wave Detection Using Pulsar
  Timing Array for Chinese Future Telescopes}},  in \emph{Frontiers in Radio
  Astronomy and FAST Early Sciences Symposium 2015}, L.~{Qain} and D.~{Li},
  eds., vol.~502 of \emph{Astronomical Society of the Pacific Conference
  Series}, p.~19, Feb., 2016.

\bibitem{EPTA:2011kjn}
{\scshape EPTA} collaboration, \emph{{Placing limits on the stochastic
  gravitational-wave background using European Pulsar Timing Array data}},
  \href{https://doi.org/10.1111/j.1365-2966.2011.18613.x}{\emph{Mon. Not. Roy.
  Astron. Soc.} {\bfseries 414} (2011) 3117}
  [\href{https://arxiv.org/abs/1103.0576}{{\ttfamily 1103.0576}}], [Erratum:
  Mon.Not.Roy.Astron.Soc. 425, 1597 (2012)].

\bibitem{Desvignes_2016}
G.~Desvignes, R.N.~Caballero, L.~Lentati, J.P.W.~Verbiest, D.J.~Champion,
  B.W.~Stappers et~al., \emph{High-precision timing of 42 millisecond pulsars
  with the european pulsar timing array},
  \href{https://doi.org/10.1093/mnras/stw483}{\emph{Monthly Notices of the
  Royal Astronomical Society} {\bfseries 458} (2016) 3341}.

\bibitem{Hobbs:2009yy}
G.~Hobbs et~al., \emph{{The international pulsar timing array project: using
  pulsars as a gravitational wave detector}},
  \href{https://doi.org/10.1088/0264-9381/27/8/084013}{\emph{Class. Quant.
  Grav.} {\bfseries 27} (2010) 084013}
  [\href{https://arxiv.org/abs/0911.5206}{{\ttfamily 0911.5206}}].

\bibitem{Manchester_2013}
R.N.~Manchester, \emph{The international pulsar timing array},
  \href{https://doi.org/10.1088/0264-9381/30/22/224010}{\emph{Classical and
  Quantum Gravity} {\bfseries 30} (2013) 224010}.

\bibitem{Demorest:2009ex}
{\scshape NANOGrav} collaboration, \emph{{Gravitational Wave Astronomy Using
  Pulsars: Massive Black Hole Mergers \& the Early Universe}},
  \href{https://arxiv.org/abs/0902.2968}{{\ttfamily 0902.2968}}.

\bibitem{Manchester:2012za}
R.N.~Manchester et~al., \emph{{The Parkes Pulsar Timing Array Project}},
  \href{https://doi.org/10.1017/pasa.2012.017}{\emph{Publ. Astron. Soc.
  Austral.} {\bfseries 30} (2013) 17}
  [\href{https://arxiv.org/abs/1210.6130}{{\ttfamily 1210.6130}}].

\bibitem{Janssen:2014dka}
G.~Janssen et~al., \emph{{Gravitational wave astronomy with the SKA}},
  \href{https://doi.org/10.22323/1.215.0037}{\emph{PoS} {\bfseries AASKA14}
  (2015) 037} [\href{https://arxiv.org/abs/1501.00127}{{\ttfamily
  1501.00127}}].

\bibitem{Hellings:1983fr}
R.W.~Hellings and G.S.~Downs, \emph{{Upper limits on the isotropic
  gravitational radiation background from pulsar timing analysis}},
  \href{https://doi.org/10.1086/183954}{\emph{Astrophys. J. Lett.} {\bfseries
  265} (1983) L39}.

\bibitem{Xu:2023wog}
H.~Xu et~al., \emph{{Searching for the Nano-Hertz Stochastic Gravitational Wave
  Background with the Chinese Pulsar Timing Array Data Release I}},
  \href{https://doi.org/10.1088/1674-4527/acdfa5}{\emph{Res. Astron.
  Astrophys.} {\bfseries 23} (2023) 075024}
  [\href{https://arxiv.org/abs/2306.16216}{{\ttfamily 2306.16216}}].

\bibitem{EPTA:2023fyk}
{\scshape EPTA} collaboration, \emph{{The second data release from the European
  Pulsar Timing Array III. Search for gravitational wave signals}},
  \href{https://arxiv.org/abs/2306.16214}{{\ttfamily 2306.16214}}.

\bibitem{NANOGrav:2023gor}
{\scshape NANOGrav} collaboration, \emph{{The NANOGrav 15 yr Data Set: Evidence
  for a Gravitational-wave Background}},
  \href{https://doi.org/10.3847/2041-8213/acdac6}{\emph{Astrophys. J. Lett.}
  {\bfseries 951} (2023) L8}
  [\href{https://arxiv.org/abs/2306.16213}{{\ttfamily 2306.16213}}].

\bibitem{Reardon:2023gzh}
D.J.~Reardon et~al., \emph{{Search for an Isotropic Gravitational-wave
  Background with the Parkes Pulsar Timing Array}},
  \href{https://doi.org/10.3847/2041-8213/acdd02}{\emph{Astrophys. J. Lett.}
  {\bfseries 951} (2023) L6}
  [\href{https://arxiv.org/abs/2306.16215}{{\ttfamily 2306.16215}}].

\bibitem{NANOGrav:2023hfp}
{\scshape NANOGrav} collaboration, \emph{{The NANOGrav 15 yr Data Set:
  Constraints on Supermassive Black Hole Binaries from the Gravitational-wave
  Background}},
  \href{https://doi.org/10.3847/2041-8213/ace18b}{\emph{Astrophys. J. Lett.}
  {\bfseries 952} (2023) L37}
  [\href{https://arxiv.org/abs/2306.16220}{{\ttfamily 2306.16220}}].

\bibitem{NANOGrav:2023tcn}
{\scshape NANOGrav} collaboration, \emph{{The NANOGrav 15-year Data Set: Search
  for Anisotropy in the Gravitational-Wave Background}},
  \href{https://arxiv.org/abs/2306.16221}{{\ttfamily 2306.16221}}.

\bibitem{NANOGrav:2023hvm}
{\scshape NANOGrav} collaboration, \emph{{The NANOGrav 15 yr Data Set: Search
  for Signals from New Physics}},
  \href{https://doi.org/10.3847/2041-8213/acdc91}{\emph{Astrophys. J. Lett.}
  {\bfseries 951} (2023) L11}
  [\href{https://arxiv.org/abs/2306.16219}{{\ttfamily 2306.16219}}].

\bibitem{Moore:2014lga}
C.J.~Moore, R.H.~Cole and C.P.L.~Berry, \emph{{Gravitational-wave sensitivity
  curves}}, \href{https://doi.org/10.1088/0264-9381/32/1/015014}{\emph{Class.
  Quant. Grav.} {\bfseries 32} (2015) 015014}
  [\href{https://arxiv.org/abs/1408.0740}{{\ttfamily 1408.0740}}].

\bibitem{Maggiore:1999vm}
M.~Maggiore, \emph{{Gravitational wave experiments and early universe
  cosmology}}, \href{https://doi.org/10.1016/S0370-1573(99)00102-7}{\emph{Phys.
  Rept.} {\bfseries 331} (2000) 283}
  [\href{https://arxiv.org/abs/gr-qc/9909001}{{\ttfamily gr-qc/9909001}}].

\bibitem{Christensen:2018iqi}
N.~Christensen, \emph{{Stochastic Gravitational Wave Backgrounds}},
  \href{https://doi.org/10.1088/1361-6633/aae6b5}{\emph{Rept. Prog. Phys.}
  {\bfseries 82} (2019) 016903}
  [\href{https://arxiv.org/abs/1811.08797}{{\ttfamily 1811.08797}}].

\bibitem{Caprini:2007xq}
C.~Caprini, R.~Durrer and G.~Servant, \emph{{Gravitational wave generation from
  bubble collisions in first-order phase transitions: An analytic approach}},
  \href{https://doi.org/10.1103/PhysRevD.77.124015}{\emph{Phys. Rev. D}
  {\bfseries 77} (2008) 124015}
  [\href{https://arxiv.org/abs/0711.2593}{{\ttfamily 0711.2593}}].

\bibitem{Caprini:2018mtu}
C.~Caprini and D.G.~Figueroa, \emph{{Cosmological Backgrounds of Gravitational
  Waves}}, \href{https://doi.org/10.1088/1361-6382/aac608}{\emph{Class. Quant.
  Grav.} {\bfseries 35} (2018) 163001}
  [\href{https://arxiv.org/abs/1801.04268}{{\ttfamily 1801.04268}}].

\bibitem{Kuroyanagi:2018csn}
S.~Kuroyanagi, T.~Chiba and T.~Takahashi, \emph{{Probing the Universe through
  the Stochastic Gravitational Wave Background}},
  \href{https://doi.org/10.1088/1475-7516/2018/11/038}{\emph{JCAP} {\bfseries
  11} (2018) 038} [\href{https://arxiv.org/abs/1807.00786}{{\ttfamily
  1807.00786}}].

\bibitem{Ruiter:2007xx}
A.J.~Ruiter, K.~Belczynski, M.~Benacquista, S.L.~Larson and G.~Williams,
  \emph{{The LISA Gravitational Wave Foreground: A Study of Double White
  Dwarfs}},
  \href{https://doi.org/10.1088/0004-637X/717/2/1006}{\emph{Astrophys. J.}
  {\bfseries 717} (2010) 1006}
  [\href{https://arxiv.org/abs/0705.3272}{{\ttfamily 0705.3272}}].

\bibitem{Yu:2010fq}
S.~Yu and C.S.~Jeffery, \emph{{The gravitational wave signal from diverse
  populations of double white dwarf binaries in the Galaxy}},
  \href{https://doi.org/10.1051/0004-6361/201014827}{\emph{Astron. Astrophys.}
  {\bfseries 521} (2010) A85}
  [\href{https://arxiv.org/abs/1007.4267}{{\ttfamily 1007.4267}}].

\bibitem{Adams:2013qma}
M.R.~Adams and N.J.~Cornish, \emph{{Detecting a Stochastic Gravitational Wave
  Background in the presence of a Galactic Foreground and Instrument Noise}},
  \href{https://doi.org/10.1103/PhysRevD.89.022001}{\emph{Phys. Rev. D}
  {\bfseries 89} (2014) 022001}
  [\href{https://arxiv.org/abs/1307.4116}{{\ttfamily 1307.4116}}].

\bibitem{Colpi:2016fup}
M.~Colpi and A.~Sesana, \emph{{Gravitational Wave Sources in the Era of
  Multi-Band Gravitational Wave Astronomy}},  in \emph{{An Overview of
  Gravitational Waves}: {Theory, Sources and Detection}}, G.~Auger and
  E.~Plagnol, eds., pp.~43--140 (2017),
  \href{https://doi.org/10.1142/9789813141766_0002}{DOI}
  [\href{https://arxiv.org/abs/1610.05309}{{\ttfamily 1610.05309}}].

\bibitem{Hogan:1986qda}
C.J.~Hogan, \emph{{Gravitational radiation from cosmological phase
  transitions}}, \href{https://doi.org/10.1093/mnras/218.4.629}{\emph{Mon. Not.
  Roy. Astron. Soc.} {\bfseries 218} (1986) 629}.

\bibitem{Kosowsky:1992rz}
A.~Kosowsky, M.S.~Turner and R.~Watkins, \emph{{Gravitational waves from first
  order cosmological phase transitions}},
  \href{https://doi.org/10.1103/PhysRevLett.69.2026}{\emph{Phys. Rev. Lett.}
  {\bfseries 69} (1992) 2026}.

\bibitem{Kosowsky:1992vn}
A.~Kosowsky and M.S.~Turner, \emph{{Gravitational radiation from colliding
  vacuum bubbles: envelope approximation to many bubble collisions}},
  \href{https://doi.org/10.1103/PhysRevD.47.4372}{\emph{Phys. Rev. D}
  {\bfseries 47} (1993) 4372}
  [\href{https://arxiv.org/abs/astro-ph/9211004}{{\ttfamily
  astro-ph/9211004}}].

\bibitem{Huber:2008hg}
S.J.~Huber and T.~Konstandin, \emph{{Gravitational Wave Production by
  Collisions: More Bubbles}},
  \href{https://doi.org/10.1088/1475-7516/2008/09/022}{\emph{JCAP} {\bfseries
  09} (2008) 022} [\href{https://arxiv.org/abs/0806.1828}{{\ttfamily
  0806.1828}}].

\bibitem{Child:2012qg}
H.L.~Child and J.T.~Giblin, Jr., \emph{{Gravitational Radiation from
  First-Order Phase Transitions}},
  \href{https://doi.org/10.1088/1475-7516/2012/10/001}{\emph{JCAP} {\bfseries
  10} (2012) 001} [\href{https://arxiv.org/abs/1207.6408}{{\ttfamily
  1207.6408}}].

\bibitem{Weir:2016tov}
D.J.~Weir, \emph{{Revisiting the envelope approximation: gravitational waves
  from bubble collisions}},
  \href{https://doi.org/10.1103/PhysRevD.93.124037}{\emph{Phys. Rev. D}
  {\bfseries 93} (2016) 124037}
  [\href{https://arxiv.org/abs/1604.08429}{{\ttfamily 1604.08429}}].

\bibitem{Konstandin:2017sat}
T.~Konstandin, \emph{{Gravitational radiation from a bulk flow model}},
  \href{https://doi.org/10.1088/1475-7516/2018/03/047}{\emph{JCAP} {\bfseries
  03} (2018) 047} [\href{https://arxiv.org/abs/1712.06869}{{\ttfamily
  1712.06869}}].

\bibitem{Gould:2021dpm}
O.~Gould, S.~Sukuvaara and D.J.~Weir, \emph{{Vacuum bubble collisions: From
  microphysics to gravitational waves}},
  \href{https://doi.org/10.1103/PhysRevD.104.075039}{\emph{Phys. Rev. D}
  {\bfseries 104} (2021) 075039}
  [\href{https://arxiv.org/abs/2107.05657}{{\ttfamily 2107.05657}}].

\bibitem{Jinno:2016vai}
R.~Jinno and M.~Takimoto, \emph{{Gravitational waves from bubble collisions: An
  analytic derivation}},
  \href{https://doi.org/10.1103/PhysRevD.95.024009}{\emph{Phys. Rev. D}
  {\bfseries 95} (2017) 024009}
  [\href{https://arxiv.org/abs/1605.01403}{{\ttfamily 1605.01403}}].

\bibitem{Giblin:2014gra}
J.T.~Giblin and E.~Thrane, \emph{{Estimates of maximum energy density of
  cosmological gravitational-wave backgrounds}},
  \href{https://doi.org/10.1103/PhysRevD.90.107502}{\emph{Phys. Rev. D}
  {\bfseries 90} (2014) 107502}
  [\href{https://arxiv.org/abs/1410.4779}{{\ttfamily 1410.4779}}].

\bibitem{Caprini:2006jb}
C.~Caprini and R.~Durrer, \emph{{Gravitational waves from stochastic
  relativistic sources: Primordial turbulence and magnetic fields}},
  \href{https://doi.org/10.1103/PhysRevD.74.063521}{\emph{Phys. Rev. D}
  {\bfseries 74} (2006) 063521}
  [\href{https://arxiv.org/abs/astro-ph/0603476}{{\ttfamily
  astro-ph/0603476}}].

\bibitem{ParticleDataGroup:2020ssz}
{\scshape Particle Data Group} collaboration, \emph{{Review of Particle
  Physics}}, \href{https://doi.org/10.1093/ptep/ptaa104}{\emph{PTEP} {\bfseries
  2020} (2020) 083C01}.

\bibitem{Allahverdi:2020bys}
R.~Allahverdi et~al., \emph{{The First Three Seconds: a Review of Possible
  Expansion Histories of the Early Universe}},
  \href{https://arxiv.org/abs/2006.16182}{{\ttfamily 2006.16182}}.

\bibitem{Cutting:2020nla}
D.~Cutting, E.G.~Escartin, M.~Hindmarsh and D.J.~Weir, \emph{{Gravitational
  waves from vacuum first order phase transitions II: from thin to thick
  walls}}, \href{https://doi.org/10.1103/PhysRevD.103.023531}{\emph{Phys. Rev.
  D} {\bfseries 103} (2021) 023531}
  [\href{https://arxiv.org/abs/2005.13537}{{\ttfamily 2005.13537}}].

\bibitem{Jinno:2017fby}
R.~Jinno and M.~Takimoto, \emph{{Gravitational waves from bubble dynamics:
  Beyond the Envelope}},
  \href{https://doi.org/10.1088/1475-7516/2019/01/060}{\emph{JCAP} {\bfseries
  01} (2019) 060} [\href{https://arxiv.org/abs/1707.03111}{{\ttfamily
  1707.03111}}].

\bibitem{Jinno:2019bxw}
R.~Jinno, T.~Konstandin and M.~Takimoto, \emph{{Relativistic bubble
  collisions\textemdash{}a closer look}},
  \href{https://doi.org/10.1088/1475-7516/2019/09/035}{\emph{JCAP} {\bfseries
  09} (2019) 035} [\href{https://arxiv.org/abs/1906.02588}{{\ttfamily
  1906.02588}}].

\bibitem{Lewicki:2020jiv}
M.~Lewicki and V.~Vaskonen, \emph{{Gravitational wave spectra from strongly
  supercooled phase transitions}},
  \href{https://doi.org/10.1140/epjc/s10052-020-08589-1}{\emph{Eur. Phys. J. C}
  {\bfseries 80} (2020) 1003}
  [\href{https://arxiv.org/abs/2007.04967}{{\ttfamily 2007.04967}}].

\bibitem{Lewicki:2020azd}
M.~Lewicki and V.~Vaskonen, \emph{{Gravitational waves from colliding vacuum
  bubbles in gauge theories}},
  \href{https://doi.org/10.1140/epjc/s10052-021-09232-3}{\emph{Eur. Phys. J. C}
  {\bfseries 81} (2021) 437}
  [\href{https://arxiv.org/abs/2012.07826}{{\ttfamily 2012.07826}}], [Erratum:
  Eur.Phys.J.C 81, 1077 (2021)].

\bibitem{Cutting:2018tjt}
D.~Cutting, M.~Hindmarsh and D.J.~Weir, \emph{{Gravitational waves from vacuum
  first-order phase transitions: from the envelope to the lattice}},
  \href{https://doi.org/10.1103/PhysRevD.97.123513}{\emph{Phys. Rev. D}
  {\bfseries 97} (2018) 123513}
  [\href{https://arxiv.org/abs/1802.05712}{{\ttfamily 1802.05712}}].

\bibitem{Alanne:2019bsm}
T.~Alanne, T.~Hugle, M.~Platscher and K.~Schmitz, \emph{{A fresh look at the
  gravitational-wave signal from cosmological phase transitions}},
  \href{https://doi.org/10.1007/JHEP03(2020)004}{\emph{JHEP} {\bfseries 03}
  (2020) 004} [\href{https://arxiv.org/abs/1909.11356}{{\ttfamily
  1909.11356}}].

\bibitem{Johnson:2011wt}
M.C.~Johnson, H.V.~Peiris and L.~Lehner, \emph{{Determining the outcome of
  cosmic bubble collisions in full General Relativity}},
  \href{https://doi.org/10.1103/PhysRevD.85.083516}{\emph{Phys. Rev. D}
  {\bfseries 85} (2012) 083516}
  [\href{https://arxiv.org/abs/1112.4487}{{\ttfamily 1112.4487}}].

\bibitem{Cai:2023guc}
R.-G.~Cai, S.-J.~Wang and Z.-Y.~Yuwen, \emph{{Hydrodynamic sound shell model}},
  \href{https://doi.org/10.1103/PhysRevD.108.L021502}{\emph{Phys. Rev. D}
  {\bfseries 108} (2023) L021502}
  [\href{https://arxiv.org/abs/2305.00074}{{\ttfamily 2305.00074}}].

\bibitem{Jinno:2020eqg}
R.~Jinno, T.~Konstandin and H.~Rubira, \emph{{A hybrid simulation of
  gravitational wave production in first-order phase transitions}},
  \href{https://doi.org/10.1088/1475-7516/2021/04/014}{\emph{JCAP} {\bfseries
  04} (2021) 014} [\href{https://arxiv.org/abs/2010.00971}{{\ttfamily
  2010.00971}}].

\bibitem{Jinno:2022mie}
R.~Jinno, T.~Konstandin, H.~Rubira and I.~Stomberg, \emph{{Higgsless
  simulations of cosmological phase transitions and gravitational waves}},
  \href{https://doi.org/10.1088/1475-7516/2023/02/011}{\emph{JCAP} {\bfseries
  02} (2023) 011} [\href{https://arxiv.org/abs/2209.04369}{{\ttfamily
  2209.04369}}].

\bibitem{Jinno:2019jhi}
R.~Jinno, H.~Seong, M.~Takimoto and C.M.~Um, \emph{{Gravitational waves from
  first-order phase transitions: Ultra-supercooled transitions and the fate of
  relativistic shocks}},
  \href{https://doi.org/10.1088/1475-7516/2019/10/033}{\emph{JCAP} {\bfseries
  10} (2019) 033} [\href{https://arxiv.org/abs/1905.00899}{{\ttfamily
  1905.00899}}].

\bibitem{1991RSPSA.434....9K}
A.N.~{Kolmogorov}, \emph{{The Local Structure of Turbulence in Incompressible
  Viscous Fluid for Very Large Reynolds Numbers}},
  \href{https://doi.org/10.1098/rspa.1991.0075}{\emph{Proceedings of the Royal
  Society of London Series A} {\bfseries 434} (1991) 9}.

\bibitem{Kalaydzhyan:2014wca}
T.~Kalaydzhyan and E.~Shuryak, \emph{{Gravity waves generated by sounds from
  big bang phase transitions}},
  \href{https://doi.org/10.1103/PhysRevD.91.083502}{\emph{Phys. Rev. D}
  {\bfseries 91} (2015) 083502}
  [\href{https://arxiv.org/abs/1412.5147}{{\ttfamily 1412.5147}}].

\bibitem{Galtier:2017mve}
S.~Galtier and S.V.~Nazarenko, \emph{{Turbulence of Weak Gravitational Waves in
  the Early Universe}},
  \href{https://doi.org/10.1103/PhysRevLett.119.221101}{\emph{Phys. Rev. Lett.}
  {\bfseries 119} (2017) 221101}
  [\href{https://arxiv.org/abs/1703.09069}{{\ttfamily 1703.09069}}].

\bibitem{Leitao:2012tx}
L.~Leitao, A.~Megevand and A.D.~Sanchez, \emph{{Gravitational waves from the
  electroweak phase transition}},
  \href{https://doi.org/10.1088/1475-7516/2012/10/024}{\emph{JCAP} {\bfseries
  10} (2012) 024} [\href{https://arxiv.org/abs/1205.3070}{{\ttfamily
  1205.3070}}].

\bibitem{RoperPol:2018sap}
A.~Roper~Pol, A.~Brandenburg, T.~Kahniashvili, A.~Kosowsky and S.~Mandal,
  \emph{{The timestep constraint in solving the gravitational wave equations
  sourced by hydromagnetic turbulence}},
  \href{https://doi.org/10.1080/03091929.2019.1653460}{\emph{Geophys.
  Astrophys. Fluid Dynamics} {\bfseries 114} (2020) 130}
  [\href{https://arxiv.org/abs/1807.05479}{{\ttfamily 1807.05479}}].

\bibitem{Niksa:2018ofa}
P.~Niksa, M.~Schlederer and G.~Sigl, \emph{{Gravitational Waves produced by
  Compressible MHD Turbulence from Cosmological Phase Transitions}},
  \href{https://doi.org/10.1088/1361-6382/aac89c}{\emph{Class. Quant. Grav.}
  {\bfseries 35} (2018) 144001}
  [\href{https://arxiv.org/abs/1803.02271}{{\ttfamily 1803.02271}}].

\bibitem{Sharma:2022ysf}
R.~Sharma and A.~Brandenburg, \emph{{Low frequency tail of gravitational wave
  spectra from hydromagnetic turbulence}},
  \href{https://doi.org/10.1103/PhysRevD.106.103536}{\emph{Phys. Rev. D}
  {\bfseries 106} (2022) 103536}
  [\href{https://arxiv.org/abs/2206.00055}{{\ttfamily 2206.00055}}].

\bibitem{Brandenburg:2021bvg}
A.~Brandenburg, G.~Gogoberidze, T.~Kahniashvili, S.~Mandal, A.~Roper~Pol and
  N.~Shenoy, \emph{{The scalar, vector, and tensor modes in gravitational wave
  turbulence simulations}},
  \href{https://doi.org/10.1088/1361-6382/ac011c}{\emph{Class. Quant. Grav.}
  {\bfseries 38} (2021) 145002}
  [\href{https://arxiv.org/abs/2103.01140}{{\ttfamily 2103.01140}}].

\bibitem{Auclair:2022jod}
P.~Auclair, C.~Caprini, D.~Cutting, M.~Hindmarsh, K.~Rummukainen, D.A.~Steer
  et~al., \emph{{Generation of gravitational waves from freely decaying
  turbulence}},
  \href{https://doi.org/10.1088/1475-7516/2022/09/029}{\emph{JCAP} {\bfseries
  09} (2022) 029} [\href{https://arxiv.org/abs/2205.02588}{{\ttfamily
  2205.02588}}].

\bibitem{Kahniashvili:2012uj}
T.~Kahniashvili, A.G.~Tevzadze, A.~Brandenburg and A.~Neronov, \emph{{Evolution
  of Primordial Magnetic Fields from Phase Transitions}},
  \href{https://doi.org/10.1103/PhysRevD.87.083007}{\emph{Phys. Rev. D}
  {\bfseries 87} (2013) 083007}
  [\href{https://arxiv.org/abs/1212.0596}{{\ttfamily 1212.0596}}].

\bibitem{Brandenburg:2021tmp}
A.~Brandenburg, E.~Clarke, Y.~He and T.~Kahniashvili, \emph{{Can we observe the
  QCD phase transition-generated gravitational waves through pulsar timing
  arrays?}}, \href{https://doi.org/10.1103/PhysRevD.104.043513}{\emph{Phys.
  Rev. D} {\bfseries 104} (2021) 043513}
  [\href{https://arxiv.org/abs/2102.12428}{{\ttfamily 2102.12428}}].

\bibitem{Alves:2018jsw}
A.~Alves, T.~Ghosh, H.-K.~Guo, K.~Sinha and D.~Vagie, \emph{{Collider and
  Gravitational Wave Complementarity in Exploring the Singlet Extension of the
  Standard Model}}, \href{https://doi.org/10.1007/JHEP04(2019)052}{\emph{JHEP}
  {\bfseries 04} (2019) 052}
  [\href{https://arxiv.org/abs/1812.09333}{{\ttfamily 1812.09333}}].

\bibitem{FitzAxen:2018vdt}
M.~Fitz~Axen, S.~Banagiri, A.~Matas, C.~Caprini and V.~Mandic,
  \emph{{Multiwavelength observations of cosmological phase transitions using
  LISA and Cosmic Explorer}},
  \href{https://doi.org/10.1103/PhysRevD.98.103508}{\emph{Phys. Rev. D}
  {\bfseries 98} (2018) 103508}
  [\href{https://arxiv.org/abs/1806.02500}{{\ttfamily 1806.02500}}].

\bibitem{Allen:1997ad}
B.~Allen and J.D.~Romano, \emph{{Detecting a stochastic background of
  gravitational radiation: Signal processing strategies and sensitivities}},
  \href{https://doi.org/10.1103/PhysRevD.59.102001}{\emph{Phys. Rev. D}
  {\bfseries 59} (1999) 102001}
  [\href{https://arxiv.org/abs/gr-qc/9710117}{{\ttfamily gr-qc/9710117}}].

\bibitem{Romano:2016dpx}
J.D.~Romano and N.J.~Cornish, \emph{{Detection methods for stochastic
  gravitational-wave backgrounds: a unified treatment}},
  \href{https://doi.org/10.1007/s41114-017-0004-1}{\emph{Living Rev. Rel.}
  {\bfseries 20} (2017) 2} [\href{https://arxiv.org/abs/1608.06889}{{\ttfamily
  1608.06889}}].

\bibitem{Caprini:2019pxz}
C.~Caprini, D.G.~Figueroa, R.~Flauger, G.~Nardini, M.~Peloso, M.~Pieroni
  et~al., \emph{{Reconstructing the spectral shape of a stochastic
  gravitational wave background with LISA}},
  \href{https://doi.org/10.1088/1475-7516/2019/11/017}{\emph{JCAP} {\bfseries
  11} (2019) 017} [\href{https://arxiv.org/abs/1906.09244}{{\ttfamily
  1906.09244}}].

\bibitem{Thrane:2013oya}
E.~Thrane and J.D.~Romano, \emph{{Sensitivity curves for searches for
  gravitational-wave backgrounds}},
  \href{https://doi.org/10.1103/PhysRevD.88.124032}{\emph{Phys. Rev. D}
  {\bfseries 88} (2013) 124032}
  [\href{https://arxiv.org/abs/1310.5300}{{\ttfamily 1310.5300}}].

\bibitem{Schmitz:2020syl}
K.~Schmitz, \emph{{New Sensitivity Curves for Gravitational-Wave Signals from
  Cosmological Phase Transitions}},
  \href{https://doi.org/10.1007/JHEP01(2021)097}{\emph{JHEP} {\bfseries 01}
  (2021) 097} [\href{https://arxiv.org/abs/2002.04615}{{\ttfamily
  2002.04615}}].

\bibitem{Grojean:2006bp}
C.~Grojean and G.~Servant, \emph{{Gravitational Waves from Phase Transitions at
  the Electroweak Scale and Beyond}},
  \href{https://doi.org/10.1103/PhysRevD.75.043507}{\emph{Phys. Rev. D}
  {\bfseries 75} (2007) 043507}
  [\href{https://arxiv.org/abs/hep-ph/0607107}{{\ttfamily hep-ph/0607107}}].

\bibitem{Ashoorioon:2009nf}
A.~Ashoorioon and T.~Konstandin, \emph{{Strong electroweak phase transitions
  without collider traces}},
  \href{https://doi.org/10.1088/1126-6708/2009/07/086}{\emph{JHEP} {\bfseries
  07} (2009) 086} [\href{https://arxiv.org/abs/0904.0353}{{\ttfamily
  0904.0353}}].

\bibitem{Jinno:2015doa}
R.~Jinno, K.~Nakayama and M.~Takimoto, \emph{{Gravitational waves from the
  first order phase transition of the Higgs field at high energy scales}},
  \href{https://doi.org/10.1103/PhysRevD.93.045024}{\emph{Phys. Rev. D}
  {\bfseries 93} (2016) 045024}
  [\href{https://arxiv.org/abs/1510.02697}{{\ttfamily 1510.02697}}].

\bibitem{Huang:2016cjm}
P.~Huang, A.J.~Long and L.-T.~Wang, \emph{{Probing the Electroweak Phase
  Transition with Higgs Factories and Gravitational Waves}},
  \href{https://doi.org/10.1103/PhysRevD.94.075008}{\emph{Phys. Rev. D}
  {\bfseries 94} (2016) 075008}
  [\href{https://arxiv.org/abs/1608.06619}{{\ttfamily 1608.06619}}].

\bibitem{Hashino:2016xoj}
K.~Hashino, M.~Kakizaki, S.~Kanemura, P.~Ko and T.~Matsui, \emph{{Gravitational
  waves and Higgs boson couplings for exploring first order phase transition in
  the model with a singlet scalar field}},
  \href{https://doi.org/10.1016/j.physletb.2016.12.052}{\emph{Phys. Lett. B}
  {\bfseries 766} (2017) 49}
  [\href{https://arxiv.org/abs/1609.00297}{{\ttfamily 1609.00297}}].

\bibitem{Balazs:2016tbi}
C.~Balazs, A.~Fowlie, A.~Mazumdar and G.~White, \emph{{Gravitational waves at
  aLIGO and vacuum stability with a scalar singlet extension of the Standard
  Model}}, \href{https://doi.org/10.1103/PhysRevD.95.043505}{\emph{Phys. Rev.
  D} {\bfseries 95} (2017) 043505}
  [\href{https://arxiv.org/abs/1611.01617}{{\ttfamily 1611.01617}}].

\bibitem{Kang:2017mkl}
Z.~Kang, P.~Ko and T.~Matsui, \emph{{Strong first order EWPT $\&$ strong
  gravitational waves in Z$_{3}$-symmetric singlet scalar extension}},
  \href{https://doi.org/10.1007/JHEP02(2018)115}{\emph{JHEP} {\bfseries 02}
  (2018) 115} [\href{https://arxiv.org/abs/1706.09721}{{\ttfamily
  1706.09721}}].

\bibitem{Huang:2018aja}
F.P.~Huang, Z.~Qian and M.~Zhang, \emph{{Exploring dynamical CP violation
  induced baryogenesis by gravitational waves and colliders}},
  \href{https://doi.org/10.1103/PhysRevD.98.015014}{\emph{Phys. Rev. D}
  {\bfseries 98} (2018) 015014}
  [\href{https://arxiv.org/abs/1804.06813}{{\ttfamily 1804.06813}}].

\bibitem{Alves:2018oct}
A.~Alves, T.~Ghosh, H.-K.~Guo and K.~Sinha, \emph{{Resonant Di-Higgs Production
  at Gravitational Wave Benchmarks: A Collider Study using Machine Learning}},
  \href{https://doi.org/10.1007/JHEP12(2018)070}{\emph{JHEP} {\bfseries 12}
  (2018) 070} [\href{https://arxiv.org/abs/1808.08974}{{\ttfamily
  1808.08974}}].

\bibitem{Hashino:2018wee}
K.~Hashino, R.~Jinno, M.~Kakizaki, S.~Kanemura, T.~Takahashi and M.~Takimoto,
  \emph{{Selecting models of first-order phase transitions using the synergy
  between collider and gravitational-wave experiments}},
  \href{https://doi.org/10.1103/PhysRevD.99.075011}{\emph{Phys. Rev. D}
  {\bfseries 99} (2019) 075011}
  [\href{https://arxiv.org/abs/1809.04994}{{\ttfamily 1809.04994}}].

\bibitem{Beniwal:2017eik}
A.~Beniwal, M.~Lewicki, J.D.~Wells, M.~White and A.G.~Williams,
  \emph{{Gravitational wave, collider and dark matter signals from a scalar
  singlet electroweak baryogenesis}},
  \href{https://doi.org/10.1007/JHEP08(2017)108}{\emph{JHEP} {\bfseries 08}
  (2017) 108} [\href{https://arxiv.org/abs/1702.06124}{{\ttfamily
  1702.06124}}].

\bibitem{Beniwal:2018hyi}
A.~Beniwal, M.~Lewicki, M.~White and A.G.~Williams, \emph{{Gravitational waves
  and electroweak baryogenesis in a global study of the extended scalar singlet
  model}}, \href{https://doi.org/10.1007/JHEP02(2019)183}{\emph{JHEP}
  {\bfseries 02} (2019) 183}
  [\href{https://arxiv.org/abs/1810.02380}{{\ttfamily 1810.02380}}].

\bibitem{Alves:2019igs}
A.~Alves, D.~Gon\c{c}alves, T.~Ghosh, H.-K.~Guo and K.~Sinha, \emph{{Di-Higgs
  Production in the $4b$ Channel and Gravitational Wave Complementarity}},
  \href{https://doi.org/10.1007/JHEP03(2020)053}{\emph{JHEP} {\bfseries 03}
  (2020) 053} [\href{https://arxiv.org/abs/1909.05268}{{\ttfamily
  1909.05268}}].

\bibitem{Zhou:2019uzq}
R.~Zhou, L.~Bian and H.-K.~Guo, \emph{{Connecting the electroweak sphaleron
  with gravitational waves}},
  \href{https://doi.org/10.1103/PhysRevD.101.091903}{\emph{Phys. Rev. D}
  {\bfseries 101} (2020) 091903}
  [\href{https://arxiv.org/abs/1910.00234}{{\ttfamily 1910.00234}}].

\bibitem{Bian:2019bsn}
L.~Bian, H.-K.~Guo, Y.~Wu and R.~Zhou, \emph{{Gravitational wave and collider
  searches for electroweak symmetry breaking patterns}},
  \href{https://doi.org/10.1103/PhysRevD.101.035011}{\emph{Phys. Rev. D}
  {\bfseries 101} (2020) 035011}
  [\href{https://arxiv.org/abs/1906.11664}{{\ttfamily 1906.11664}}].

\bibitem{Alves:2020bpi}
A.~Alves, D.~Gon\c{c}alves, T.~Ghosh, H.-K.~Guo and K.~Sinha, \emph{{Di-Higgs
  Blind Spots in Gravitational Wave Signals}},
  \href{https://doi.org/10.1016/j.physletb.2021.136377}{\emph{Phys. Lett. B}
  {\bfseries 818} (2021) 136377}
  [\href{https://arxiv.org/abs/2007.15654}{{\ttfamily 2007.15654}}].

\bibitem{Abdussalam:2020ssl}
S.~Abdussalam, M.J.~Kazemi and L.~Kalhor, \emph{{Upper limit on first-order
  electroweak phase transition strength}},
  \href{https://doi.org/10.1142/S0217751X21920032}{\emph{Int. J. Mod. Phys. A}
  {\bfseries 36} (2021) 2150024}
  [\href{https://arxiv.org/abs/2001.05973}{{\ttfamily 2001.05973}}].

\bibitem{Liu:2021jyc}
W.~Liu and K.-P.~Xie, \emph{{Probing electroweak phase transition with
  multi-TeV muon colliders and gravitational waves}},
  \href{https://doi.org/10.1007/JHEP04(2021)015}{\emph{JHEP} {\bfseries 04}
  (2021) 015} [\href{https://arxiv.org/abs/2101.10469}{{\ttfamily
  2101.10469}}].

\bibitem{Chiang:2019oms}
C.-W.~Chiang and B.-Q.~Lu, \emph{{First-order electroweak phase transition in a
  complex singlet model with $\mathbb{Z}_3$ symmetry}},
  \href{https://doi.org/10.1007/JHEP07(2020)082}{\emph{JHEP} {\bfseries 07}
  (2020) 082} [\href{https://arxiv.org/abs/1912.12634}{{\ttfamily
  1912.12634}}].

\bibitem{Zhou:2020ojf}
R.~Zhou, J.~Yang and L.~Bian, \emph{{Gravitational Waves from first-order phase
  transition and domain wall}},
  \href{https://doi.org/10.1007/JHEP04(2020)071}{\emph{JHEP} {\bfseries 04}
  (2020) 071} [\href{https://arxiv.org/abs/2001.04741}{{\ttfamily
  2001.04741}}].

\bibitem{Chen:2019ebq}
N.~Chen, T.~Li, Y.~Wu and L.~Bian, \emph{{Complementarity of the future $e^+
  e^-$ colliders and gravitational waves in the probe of complex singlet
  extension to the standard model}},
  \href{https://doi.org/10.1103/PhysRevD.101.075047}{\emph{Phys. Rev. D}
  {\bfseries 101} (2020) 075047}
  [\href{https://arxiv.org/abs/1911.05579}{{\ttfamily 1911.05579}}].

\bibitem{Freitas:2021yng}
F.F.~Freitas, G.~Louren\c{c}o, A.P.~Morais, A.~Nunes, J.a.~Ol\'\i{}via,
  R.~Pasechnik et~al., \emph{{Impact of SM parameters and of the vacua of the
  Higgs potential in gravitational waves detection}},
  \href{https://doi.org/10.1088/1475-7516/2022/03/046}{\emph{JCAP} {\bfseries
  03} (2022) 046} [\href{https://arxiv.org/abs/2108.12810}{{\ttfamily
  2108.12810}}].

\bibitem{Chao:2017vrq}
W.~Chao, H.-K.~Guo and J.~Shu, \emph{{Gravitational Wave Signals of Electroweak
  Phase Transition Triggered by Dark Matter}},
  \href{https://doi.org/10.1088/1475-7516/2017/09/009}{\emph{JCAP} {\bfseries
  09} (2017) 009} [\href{https://arxiv.org/abs/1702.02698}{{\ttfamily
  1702.02698}}].

\bibitem{Ellis:2022lft}
J.~Ellis, M.~Lewicki, M.~Merchand, J.M.~No and M.~Zych, \emph{{The scalar
  singlet extension of the Standard Model: gravitational waves versus
  baryogenesis}}, \href{https://doi.org/10.1007/JHEP01(2023)093}{\emph{JHEP}
  {\bfseries 01} (2023) 093}
  [\href{https://arxiv.org/abs/2210.16305}{{\ttfamily 2210.16305}}].

\bibitem{Kakizaki:2015wua}
M.~Kakizaki, S.~Kanemura and T.~Matsui, \emph{{Gravitational waves as a probe
  of extended scalar sectors with the first order electroweak phase
  transition}}, \href{https://doi.org/10.1103/PhysRevD.92.115007}{\emph{Phys.
  Rev. D} {\bfseries 92} (2015) 115007}
  [\href{https://arxiv.org/abs/1509.08394}{{\ttfamily 1509.08394}}].

\bibitem{Hashino:2016rvx}
K.~Hashino, M.~Kakizaki, S.~Kanemura and T.~Matsui, \emph{{Synergy between
  measurements of gravitational waves and the triple-Higgs coupling in probing
  the first-order electroweak phase transition}},
  \href{https://doi.org/10.1103/PhysRevD.94.015005}{\emph{Phys. Rev. D}
  {\bfseries 94} (2016) 015005}
  [\href{https://arxiv.org/abs/1604.02069}{{\ttfamily 1604.02069}}].

\bibitem{Dorsch:2016nrg}
G.C.~Dorsch, S.J.~Huber, T.~Konstandin and J.M.~No, \emph{{A Second Higgs
  Doublet in the Early Universe: Baryogenesis and Gravitational Waves}},
  \href{https://doi.org/10.1088/1475-7516/2017/05/052}{\emph{JCAP} {\bfseries
  05} (2017) 052} [\href{https://arxiv.org/abs/1611.05874}{{\ttfamily
  1611.05874}}].

\bibitem{Wang:2019pet}
X.~Wang, F.P.~Huang and X.~Zhang, \emph{{Gravitational wave and collider
  signals in complex two-Higgs doublet model with dynamical CP-violation at
  finite temperature}},
  \href{https://doi.org/10.1103/PhysRevD.101.015015}{\emph{Phys. Rev. D}
  {\bfseries 101} (2020) 015015}
  [\href{https://arxiv.org/abs/1909.02978}{{\ttfamily 1909.02978}}].

\bibitem{Zhou:2020irf}
R.~Zhou and L.~Bian, \emph{{Gravitational wave and electroweak baryogenesis
  with two Higgs doublet models}},
  \href{https://doi.org/10.1016/j.physletb.2022.137105}{\emph{Phys. Lett. B}
  {\bfseries 829} (2022) 137105}
  [\href{https://arxiv.org/abs/2001.01237}{{\ttfamily 2001.01237}}].

\bibitem{Goncalves:2021egx}
D.~Gon\c{c}alves, A.~Kaladharan and Y.~Wu, \emph{{Electroweak phase transition
  in the 2HDM: Collider and gravitational wave complementarity}},
  \href{https://doi.org/10.1103/PhysRevD.105.095041}{\emph{Phys. Rev. D}
  {\bfseries 105} (2022) 095041}
  [\href{https://arxiv.org/abs/2108.05356}{{\ttfamily 2108.05356}}].

\bibitem{Biekotter:2022kgf}
T.~Biek\"otter, S.~Heinemeyer, J.M.~No, M.O.~Olea-Romacho and G.~Weiglein,
  \emph{{The trap in the early Universe: impact on the interplay between
  gravitational waves and LHC physics in the 2HDM}},
  \href{https://doi.org/10.1088/1475-7516/2023/03/031}{\emph{JCAP} {\bfseries
  03} (2023) 031} [\href{https://arxiv.org/abs/2208.14466}{{\ttfamily
  2208.14466}}].

\bibitem{Han:2020ekm}
X.-F.~Han, L.~Wang and Y.~Zhang, \emph{{Dark matter, electroweak phase
  transition, and gravitational waves in the type II two-Higgs-doublet model
  with a singlet scalar field}},
  \href{https://doi.org/10.1103/PhysRevD.103.035012}{\emph{Phys. Rev. D}
  {\bfseries 103} (2021) 035012}
  [\href{https://arxiv.org/abs/2010.03730}{{\ttfamily 2010.03730}}].

\bibitem{Zhang:2021alu}
Z.~Zhang, C.~Cai, X.-M.~Jiang, Y.-L.~Tang, Z.-H.~Yu and H.-H.~Zhang,
  \emph{{Phase transition gravitational waves from pseudo-Nambu-Goldstone dark
  matter and two Higgs doublets}},
  \href{https://doi.org/10.1007/JHEP05(2021)160}{\emph{JHEP} {\bfseries 05}
  (2021) 160} [\href{https://arxiv.org/abs/2102.01588}{{\ttfamily
  2102.01588}}].

\bibitem{Chala:2018opy}
M.~Chala, M.~Ramos and M.~Spannowsky, \emph{{Gravitational wave and collider
  probes of a triplet Higgs sector with a low cutoff}},
  \href{https://doi.org/10.1140/epjc/s10052-019-6655-1}{\emph{Eur. Phys. J. C}
  {\bfseries 79} (2019) 156}
  [\href{https://arxiv.org/abs/1812.01901}{{\ttfamily 1812.01901}}].

\bibitem{Borah:2020wut}
D.~Borah, A.~Dasgupta, K.~Fujikura, S.K.~Kang and D.~Mahanta, \emph{{Observable
  Gravitational Waves in Minimal Scotogenic Model}},
  \href{https://doi.org/10.1088/1475-7516/2020/08/046}{\emph{JCAP} {\bfseries
  08} (2020) 046} [\href{https://arxiv.org/abs/2003.02276}{{\ttfamily
  2003.02276}}].

\bibitem{Phong:2021lea}
V.Q.~Phong, N.C.~Thao and H.N.~Long, \emph{{Baryogenesis and gravitational
  waves in the Zee\textendash{}Babu model}},
  \href{https://doi.org/10.1140/epjc/s10052-022-10961-2}{\emph{Eur. Phys. J. C}
  {\bfseries 82} (2022) 1005}
  [\href{https://arxiv.org/abs/2107.13823}{{\ttfamily 2107.13823}}].

\bibitem{Fu:2022eun}
B.~Fu and S.F.~King, \emph{{Gravitational wave signals from leptoquark-induced
  first-order electroweak phase transitions}},
  \href{https://doi.org/10.1088/1475-7516/2023/05/055}{\emph{JCAP} {\bfseries
  05} (2023) 055} [\href{https://arxiv.org/abs/2209.14605}{{\ttfamily
  2209.14605}}].

\bibitem{Apreda:2001us}
R.~Apreda, M.~Maggiore, A.~Nicolis and A.~Riotto, \emph{{Gravitational waves
  from electroweak phase transitions}},
  \href{https://doi.org/10.1016/S0550-3213(02)00264-X}{\emph{Nucl. Phys. B}
  {\bfseries 631} (2002) 342}
  [\href{https://arxiv.org/abs/gr-qc/0107033}{{\ttfamily gr-qc/0107033}}].

\bibitem{Apreda:2001tj}
R.~Apreda, M.~Maggiore, A.~Nicolis and A.~Riotto, \emph{{Supersymmetric phase
  transitions and gravitational waves at LISA}},
  \href{https://doi.org/10.1088/0264-9381/18/23/101}{\emph{Class. Quant. Grav.}
  {\bfseries 18} (2001) L155}
  [\href{https://arxiv.org/abs/hep-ph/0102140}{{\ttfamily hep-ph/0102140}}].

\bibitem{Kozaczuk:2014kva}
J.~Kozaczuk, S.~Profumo, L.S.~Haskins and C.L.~Wainwright, \emph{{Cosmological
  Phase Transitions and their Properties in the NMSSM}},
  \href{https://doi.org/10.1007/JHEP01(2015)144}{\emph{JHEP} {\bfseries 01}
  (2015) 144} [\href{https://arxiv.org/abs/1407.4134}{{\ttfamily 1407.4134}}].

\bibitem{Huber:2015znp}
S.J.~Huber, T.~Konstandin, G.~Nardini and I.~Rues, \emph{{Detectable
  Gravitational Waves from Very Strong Phase Transitions in the General
  NMSSM}}, \href{https://doi.org/10.1088/1475-7516/2016/03/036}{\emph{JCAP}
  {\bfseries 03} (2016) 036}
  [\href{https://arxiv.org/abs/1512.06357}{{\ttfamily 1512.06357}}].

\bibitem{Bian:2017wfv}
L.~Bian, H.-K.~Guo and J.~Shu, \emph{{Gravitational Waves, baryon asymmetry of
  the universe and electric dipole moment in the CP-violating NMSSM}},
  \href{https://doi.org/10.1088/1674-1137/42/9/093106}{\emph{Chin. Phys. C}
  {\bfseries 42} (2018) 093106}
  [\href{https://arxiv.org/abs/1704.02488}{{\ttfamily 1704.02488}}], [Erratum:
  Chin.Phys.C 43, 129101 (2019)].

\bibitem{Chatterjee:2022pxf}
A.~Chatterjee, A.~Datta and S.~Roy, \emph{{Electroweak phase transition in the
  Z$_{3}$-invariant NMSSM: Implications of LHC and Dark matter searches and
  prospects of detecting the gravitational waves}},
  \href{https://doi.org/10.1007/JHEP06(2022)108}{\emph{JHEP} {\bfseries 06}
  (2022) 108} [\href{https://arxiv.org/abs/2202.12476}{{\ttfamily
  2202.12476}}].

\bibitem{Demidov:2017lzf}
S.V.~Demidov, D.S.~Gorbunov and D.V.~Kirpichnikov, \emph{{Gravitational waves
  from phase transition in split NMSSM}},
  \href{https://doi.org/10.1016/j.physletb.2018.02.007}{\emph{Phys. Lett. B}
  {\bfseries 779} (2018) 191}
  [\href{https://arxiv.org/abs/1712.00087}{{\ttfamily 1712.00087}}].

\bibitem{Fornal:2021ovz}
B.~Fornal, B.~Shams Es~Haghi, J.-H.~Yu and Y.~Zhao, \emph{{Gravitational waves
  from minisplit SUSY}},
  \href{https://doi.org/10.1103/PhysRevD.104.115005}{\emph{Phys. Rev. D}
  {\bfseries 104} (2021) 115005}
  [\href{https://arxiv.org/abs/2104.00747}{{\ttfamily 2104.00747}}].

\bibitem{Borah:2023zsb}
P.~Borah, P.~Ghosh, S.~Roy and A.K.~Saha, \emph{{Electroweak phase transition
  in a right-handed neutrino superfield extended NMSSM}},
  \href{https://doi.org/10.1007/JHEP08(2023)029}{\emph{JHEP} {\bfseries 08}
  (2023) 029} [\href{https://arxiv.org/abs/2301.05061}{{\ttfamily
  2301.05061}}].

\bibitem{Garcia-Pepin:2016hvs}
M.~Garcia-Pepin and M.~Quiros, \emph{{Strong electroweak phase transition from
  Supersymmetric Custodial Triplets}},
  \href{https://doi.org/10.1007/JHEP05(2016)177}{\emph{JHEP} {\bfseries 05}
  (2016) 177} [\href{https://arxiv.org/abs/1602.01351}{{\ttfamily
  1602.01351}}].

\bibitem{Delaunay:2007wb}
C.~Delaunay, C.~Grojean and J.D.~Wells, \emph{{Dynamics of Non-renormalizable
  Electroweak Symmetry Breaking}},
  \href{https://doi.org/10.1088/1126-6708/2008/04/029}{\emph{JHEP} {\bfseries
  04} (2008) 029} [\href{https://arxiv.org/abs/0711.2511}{{\ttfamily
  0711.2511}}].

\bibitem{Kobakhidze:2015xlz}
A.~Kobakhidze, L.~Wu and J.~Yue, \emph{{Electroweak Baryogenesis with Anomalous
  Higgs Couplings}}, \href{https://doi.org/10.1007/JHEP04(2016)011}{\emph{JHEP}
  {\bfseries 04} (2016) 011}
  [\href{https://arxiv.org/abs/1512.08922}{{\ttfamily 1512.08922}}].

\bibitem{Huang:2016odd}
F.P.~Huang, Y.~Wan, D.-G.~Wang, Y.-F.~Cai and X.~Zhang, \emph{{Hearing the
  echoes of electroweak baryogenesis with gravitational wave detectors}},
  \href{https://doi.org/10.1103/PhysRevD.94.041702}{\emph{Phys. Rev. D}
  {\bfseries 94} (2016) 041702}
  [\href{https://arxiv.org/abs/1601.01640}{{\ttfamily 1601.01640}}].

\bibitem{Kobakhidze:2016mch}
A.~Kobakhidze, A.~Manning and J.~Yue, \emph{{Gravitational waves from the phase
  transition of a nonlinearly realized electroweak gauge symmetry}},
  \href{https://doi.org/10.1142/S0218271817501140}{\emph{Int. J. Mod. Phys. D}
  {\bfseries 26} (2017) 1750114}
  [\href{https://arxiv.org/abs/1607.00883}{{\ttfamily 1607.00883}}].

\bibitem{Chala:2018ari}
M.~Chala, C.~Krause and G.~Nardini, \emph{{Signals of the electroweak phase
  transition at colliders and gravitational wave observatories}},
  \href{https://doi.org/10.1007/JHEP07(2018)062}{\emph{JHEP} {\bfseries 07}
  (2018) 062} [\href{https://arxiv.org/abs/1802.02168}{{\ttfamily
  1802.02168}}].

\bibitem{Di:2020ivg}
Y.~Di, J.~Wang, R.~Zhou, L.~Bian, R.-G.~Cai and J.~Liu, \emph{{Magnetic Field
  and Gravitational Waves from the First-Order Phase Transition}},
  \href{https://doi.org/10.1103/PhysRevLett.126.251102}{\emph{Phys. Rev. Lett.}
  {\bfseries 126} (2021) 251102}
  [\href{https://arxiv.org/abs/2012.15625}{{\ttfamily 2012.15625}}].

\bibitem{Hashino:2022ghd}
K.~Hashino and D.~Ueda, \emph{{SMEFT effects on the gravitational wave spectrum
  from an electroweak phase transition}},
  \href{https://doi.org/10.1103/PhysRevD.107.095022}{\emph{Phys. Rev. D}
  {\bfseries 107} (2023) 095022}
  [\href{https://arxiv.org/abs/2210.11241}{{\ttfamily 2210.11241}}].

\bibitem{Jinno:2016knw}
R.~Jinno and M.~Takimoto, \emph{{Probing a classically conformal B-L model with
  gravitational waves}},
  \href{https://doi.org/10.1103/PhysRevD.95.015020}{\emph{Phys. Rev. D}
  {\bfseries 95} (2017) 015020}
  [\href{https://arxiv.org/abs/1604.05035}{{\ttfamily 1604.05035}}].

\bibitem{Chao:2017ilw}
W.~Chao, W.-F.~Cui, H.-K.~Guo and J.~Shu, \emph{{Gravitational wave imprint of
  new symmetry breaking}},
  \href{https://doi.org/10.1088/1674-1137/abb4cb}{\emph{Chin. Phys. C}
  {\bfseries 44} (2020) 123102}
  [\href{https://arxiv.org/abs/1707.09759}{{\ttfamily 1707.09759}}].

\bibitem{Okada:2018xdh}
N.~Okada and O.~Seto, \emph{{Probing the seesaw scale with gravitational
  waves}}, \href{https://doi.org/10.1103/PhysRevD.98.063532}{\emph{Phys. Rev.
  D} {\bfseries 98} (2018) 063532}
  [\href{https://arxiv.org/abs/1807.00336}{{\ttfamily 1807.00336}}].

\bibitem{Marzo:2018nov}
C.~Marzo, L.~Marzola and V.~Vaskonen, \emph{{Phase transition and vacuum
  stability in the classically conformal B\textendash{}L model}},
  \href{https://doi.org/10.1140/epjc/s10052-019-7076-x}{\emph{Eur. Phys. J. C}
  {\bfseries 79} (2019) 601}
  [\href{https://arxiv.org/abs/1811.11169}{{\ttfamily 1811.11169}}].

\bibitem{Bian:2019szo}
L.~Bian, W.~Cheng, H.-K.~Guo and Y.~Zhang, \emph{{Cosmological implications of
  a B \ensuremath{-} L charged hidden scalar: leptogenesis and gravitational
  waves}}, \href{https://doi.org/10.1088/1674-1137/ac1e09}{\emph{Chin. Phys. C}
  {\bfseries 45} (2021) 113104}
  [\href{https://arxiv.org/abs/1907.13589}{{\ttfamily 1907.13589}}].

\bibitem{Hasegawa:2019amx}
T.~Hasegawa, N.~Okada and O.~Seto, \emph{{Gravitational waves from the minimal
  gauged $U(1)_{B-L}$ model}},
  \href{https://doi.org/10.1103/PhysRevD.99.095039}{\emph{Phys. Rev. D}
  {\bfseries 99} (2019) 095039}
  [\href{https://arxiv.org/abs/1904.03020}{{\ttfamily 1904.03020}}].

\bibitem{Haba:2019qol}
N.~Haba and T.~Yamada, \emph{{Gravitational waves from phase transition in
  minimal SUSY $U(1)_{B-L}$ model}},
  \href{https://doi.org/10.1103/PhysRevD.101.075027}{\emph{Phys. Rev. D}
  {\bfseries 101} (2020) 075027}
  [\href{https://arxiv.org/abs/1911.01292}{{\ttfamily 1911.01292}}].

\bibitem{Dong:2021cxn}
X.-X.~Dong, T.-F.~Feng, H.-B.~Zhang, S.-M.~Zhao and J.-L.~Yang,
  \emph{{Gravitational waves from the phase transition in the B-LSSM}},
  \href{https://doi.org/10.1007/JHEP12(2021)052}{\emph{JHEP} {\bfseries 12}
  (2021) 052} [\href{https://arxiv.org/abs/2106.11084}{{\ttfamily
  2106.11084}}].

\bibitem{Addazi:2017oge}
A.~Addazi and A.~Marciano, \emph{{Limiting majoron self-interactions from
  gravitational wave experiments}},
  \href{https://doi.org/10.1088/1674-1137/42/2/023105}{\emph{Chin. Phys. C}
  {\bfseries 42} (2018) 023105}
  [\href{https://arxiv.org/abs/1705.08346}{{\ttfamily 1705.08346}}].

\bibitem{Imtiaz:2018dfn}
B.~Imtiaz, Y.-F.~Cai and Y.~Wan, \emph{{Two-field cosmological phase
  transitions and gravitational waves in the singlet Majoron model}},
  \href{https://doi.org/10.1140/epjc/s10052-019-6532-y}{\emph{Eur. Phys. J. C}
  {\bfseries 79} (2019) 25} [\href{https://arxiv.org/abs/1804.05835}{{\ttfamily
  1804.05835}}].

\bibitem{Addazi:2019dqt}
A.~Addazi, A.~Marcian\`o, A.P.~Morais, R.~Pasechnik, R.~Srivastava and
  J.W.F.~Valle, \emph{{Gravitational footprints of massive neutrinos and lepton
  number breaking}},
  \href{https://doi.org/10.1016/j.physletb.2020.135577}{\emph{Phys. Lett. B}
  {\bfseries 807} (2020) 135577}
  [\href{https://arxiv.org/abs/1909.09740}{{\ttfamily 1909.09740}}].

\bibitem{DiBari:2021dri}
P.~Di~Bari, D.~Marfatia and Y.-L.~Zhou, \emph{{Gravitational waves from
  first-order phase transitions in Majoron models of neutrino mass}},
  \href{https://doi.org/10.1007/JHEP10(2021)193}{\emph{JHEP} {\bfseries 10}
  (2021) 193} [\href{https://arxiv.org/abs/2106.00025}{{\ttfamily
  2106.00025}}].

\bibitem{Li:2020eun}
M.~Li, Q.-S.~Yan, Y.~Zhang and Z.~Zhao, \emph{{Prospects of gravitational waves
  in the minimal left-right symmetric model}},
  \href{https://doi.org/10.1007/JHEP03(2021)267}{\emph{JHEP} {\bfseries 03}
  (2021) 267} [\href{https://arxiv.org/abs/2012.13686}{{\ttfamily
  2012.13686}}].

\bibitem{Brdar:2019fur}
V.~Brdar, L.~Graf, A.J.~Helmboldt and X.-J.~Xu, \emph{{Gravitational Waves as a
  Probe of Left-Right Symmetry Breaking}},
  \href{https://doi.org/10.1088/1475-7516/2019/12/027}{\emph{JCAP} {\bfseries
  12} (2019) 027} [\href{https://arxiv.org/abs/1909.02018}{{\ttfamily
  1909.02018}}].

\bibitem{Graf:2021xku}
L.~Gr\'af, S.~Jana, A.~Kaladharan and S.~Saad, \emph{{Gravitational wave
  imprints of left-right symmetric model with minimal Higgs sector}},
  \href{https://doi.org/10.1088/1475-7516/2022/05/003}{\emph{JCAP} {\bfseries
  05} (2022) 003} [\href{https://arxiv.org/abs/2112.12041}{{\ttfamily
  2112.12041}}].

\bibitem{Croon:2018kqn}
D.~Croon, T.E.~Gonzalo and G.~White, \emph{{Gravitational Waves from a
  Pati-Salam Phase Transition}},
  \href{https://doi.org/10.1007/JHEP02(2019)083}{\emph{JHEP} {\bfseries 02}
  (2019) 083} [\href{https://arxiv.org/abs/1812.02747}{{\ttfamily
  1812.02747}}].

\bibitem{Huang:2020bbe}
W.-C.~Huang, F.~Sannino and Z.-W.~Wang, \emph{{Gravitational Waves from
  Pati-Salam Dynamics}},
  \href{https://doi.org/10.1103/PhysRevD.102.095025}{\emph{Phys. Rev. D}
  {\bfseries 102} (2020) 095025}
  [\href{https://arxiv.org/abs/2004.02332}{{\ttfamily 2004.02332}}].

\bibitem{Okada:2020vvb}
N.~Okada, O.~Seto and H.~Uchida, \emph{{Gravitational waves from breaking of an
  extra $U(1)$ in $SO(10)$ grand unification}},
  \href{https://doi.org/10.1093/ptep/ptab003}{\emph{PTEP} {\bfseries 2021}
  (2021) 033B01} [\href{https://arxiv.org/abs/2006.01406}{{\ttfamily
  2006.01406}}].

\bibitem{Madge:2018gfl}
E.~Madge and P.~Schwaller, \emph{{Leptophilic dark matter from gauged lepton
  number: Phenomenology and gravitational wave signatures}},
  \href{https://doi.org/10.1007/JHEP02(2019)048}{\emph{JHEP} {\bfseries 02}
  (2019) 048} [\href{https://arxiv.org/abs/1809.09110}{{\ttfamily
  1809.09110}}].

\bibitem{Huang:2017laj}
F.P.~Huang and X.~Zhang, \emph{{Probing the gauge symmetry breaking of the
  early universe in 3-3-1 models and beyond by gravitational waves}},
  \href{https://doi.org/10.1016/j.physletb.2018.11.024}{\emph{Phys. Lett. B}
  {\bfseries 788} (2019) 288}
  [\href{https://arxiv.org/abs/1701.04338}{{\ttfamily 1701.04338}}].

\bibitem{Jarvinen:2009mh}
M.~Jarvinen, C.~Kouvaris and F.~Sannino, \emph{{Gravitational Techniwaves}},
  \href{https://doi.org/10.1103/PhysRevD.81.064027}{\emph{Phys. Rev. D}
  {\bfseries 81} (2010) 064027}
  [\href{https://arxiv.org/abs/0911.4096}{{\ttfamily 0911.4096}}].

\bibitem{Chen:2017cyc}
Y.~Chen, M.~Huang and Q.-S.~Yan, \emph{{Gravitation waves from QCD and
  electroweak phase transitions}},
  \href{https://doi.org/10.1007/JHEP05(2018)178}{\emph{JHEP} {\bfseries 05}
  (2018) 178} [\href{https://arxiv.org/abs/1712.03470}{{\ttfamily
  1712.03470}}].

\bibitem{Miura:2018dsy}
K.~Miura, H.~Ohki, S.~Otani and K.~Yamawaki, \emph{{Gravitational Waves from
  Walking Technicolor}},
  \href{https://doi.org/10.1007/JHEP10(2019)194}{\emph{JHEP} {\bfseries 10}
  (2019) 194} [\href{https://arxiv.org/abs/1811.05670}{{\ttfamily
  1811.05670}}].

\bibitem{Aziz:2013fga}
S.~Aziz and B.~Ghosh, \emph{{Phenomenology of electroweak bubbles and
  gravitational waves in the littlest Higgs model with $T$ parity}},
  \href{https://doi.org/10.1103/PhysRevD.89.013004}{\emph{Phys. Rev. D}
  {\bfseries 89} (2014) 013004}
  [\href{https://arxiv.org/abs/1304.2997}{{\ttfamily 1304.2997}}].

\bibitem{Chala:2016ykx}
M.~Chala, G.~Nardini and I.~Sobolev, \emph{{Unified explanation for dark matter
  and electroweak baryogenesis with direct detection and gravitational wave
  signatures}}, \href{https://doi.org/10.1103/PhysRevD.94.055006}{\emph{Phys.
  Rev. D} {\bfseries 94} (2016) 055006}
  [\href{https://arxiv.org/abs/1605.08663}{{\ttfamily 1605.08663}}].

\bibitem{Bruggisser:2018mrt}
S.~Bruggisser, B.~Von~Harling, O.~Matsedonskyi and G.~Servant,
  \emph{{Electroweak Phase Transition and Baryogenesis in Composite Higgs
  Models}}, \href{https://doi.org/10.1007/JHEP12(2018)099}{\emph{JHEP}
  {\bfseries 12} (2018) 099}
  [\href{https://arxiv.org/abs/1804.07314}{{\ttfamily 1804.07314}}].

\bibitem{Bian:2019kmg}
L.~Bian, Y.~Wu and K.-P.~Xie, \emph{{Electroweak phase transition with
  composite Higgs models: calculability, gravitational waves and collider
  searches}}, \href{https://doi.org/10.1007/JHEP12(2019)028}{\emph{JHEP}
  {\bfseries 12} (2019) 028}
  [\href{https://arxiv.org/abs/1909.02014}{{\ttfamily 1909.02014}}].

\bibitem{Xie:2020bkl}
K.-P.~Xie, L.~Bian and Y.~Wu, \emph{{Electroweak baryogenesis and gravitational
  waves in a composite Higgs model with high dimensional fermion
  representations}}, \href{https://doi.org/10.1007/JHEP12(2020)047}{\emph{JHEP}
  {\bfseries 12} (2020) 047}
  [\href{https://arxiv.org/abs/2005.13552}{{\ttfamily 2005.13552}}].

\bibitem{Dunsky:2019upk}
D.~Dunsky, L.J.~Hall and K.~Harigaya, \emph{{Dark Matter, Dark Radiation and
  Gravitational Waves from Mirror Higgs Parity}},
  \href{https://doi.org/10.1007/JHEP02(2020)078}{\emph{JHEP} {\bfseries 02}
  (2020) 078} [\href{https://arxiv.org/abs/1908.02756}{{\ttfamily
  1908.02756}}].

\bibitem{Randall:2006py}
L.~Randall and G.~Servant, \emph{{Gravitational waves from warped spacetime}},
  \href{https://doi.org/10.1088/1126-6708/2007/05/054}{\emph{JHEP} {\bfseries
  05} (2007) 054} [\href{https://arxiv.org/abs/hep-ph/0607158}{{\ttfamily
  hep-ph/0607158}}].

\bibitem{Megias:2018sxv}
E.~Meg\'\i{}as, G.~Nardini and M.~Quir\'os, \emph{{Cosmological Phase
  Transitions in Warped Space: Gravitational Waves and Collider Signatures}},
  \href{https://doi.org/10.1007/JHEP09(2018)095}{\emph{JHEP} {\bfseries 09}
  (2018) 095} [\href{https://arxiv.org/abs/1806.04877}{{\ttfamily
  1806.04877}}].

\bibitem{Craig:2020jfv}
N.~Craig, N.~Levi, A.~Mariotti and D.~Redigolo, \emph{{Ripples in Spacetime
  from Broken Supersymmetry}},
  \href{https://doi.org/10.1007/JHEP02(2021)184}{\emph{JHEP} {\bfseries 21}
  (2020) 184} [\href{https://arxiv.org/abs/2011.13949}{{\ttfamily
  2011.13949}}].

\bibitem{Zhou:2021cfu}
R.~Zhou, L.~Bian and J.~Shu, \emph{{Probing new physics for $(g-2)_\mu$ and
  gravitational waves}},  \href{https://arxiv.org/abs/2104.03519}{{\ttfamily
  2104.03519}}.

\bibitem{Borah:2021ocu}
D.~Borah, A.~Dasgupta and S.K.~Kang, \emph{{Gravitational waves from a dark
  U(1)D phase transition in light of NANOGrav 12.5~yr data}},
  \href{https://doi.org/10.1103/PhysRevD.104.063501}{\emph{Phys. Rev. D}
  {\bfseries 104} (2021) 063501}
  [\href{https://arxiv.org/abs/2105.01007}{{\ttfamily 2105.01007}}].

\bibitem{Schwaller:2015tja}
P.~Schwaller, \emph{{Gravitational Waves from a Dark Phase Transition}},
  \href{https://doi.org/10.1103/PhysRevLett.115.181101}{\emph{Phys. Rev. Lett.}
  {\bfseries 115} (2015) 181101}
  [\href{https://arxiv.org/abs/1504.07263}{{\ttfamily 1504.07263}}].

\bibitem{Addazi:2016fbj}
A.~Addazi, \emph{{Limiting First Order Phase Transitions in Dark Gauge Sectors
  from Gravitational Waves experiments}},
  \href{https://doi.org/10.1142/S0217732317500493}{\emph{Mod. Phys. Lett. A}
  {\bfseries 32} (2017) 1750049}
  [\href{https://arxiv.org/abs/1607.08057}{{\ttfamily 1607.08057}}].

\bibitem{Addazi:2017gpt}
A.~Addazi and A.~Marciano, \emph{{Gravitational waves from dark first order
  phase transitions and dark photons}},
  \href{https://doi.org/10.1088/1674-1137/42/2/023107}{\emph{Chin. Phys. C}
  {\bfseries 42} (2018) 023107}
  [\href{https://arxiv.org/abs/1703.03248}{{\ttfamily 1703.03248}}].

\bibitem{Baldes:2018emh}
I.~Baldes and C.~Garcia-Cely, \emph{{Strong gravitational radiation from a
  simple dark matter model}},
  \href{https://doi.org/10.1007/JHEP05(2019)190}{\emph{JHEP} {\bfseries 05}
  (2019) 190} [\href{https://arxiv.org/abs/1809.01198}{{\ttfamily
  1809.01198}}].

\bibitem{Jaeckel:2016jlh}
J.~Jaeckel, V.V.~Khoze and M.~Spannowsky, \emph{{Hearing the signal of dark
  sectors with gravitational wave detectors}},
  \href{https://doi.org/10.1103/PhysRevD.94.103519}{\emph{Phys. Rev. D}
  {\bfseries 94} (2016) 103519}
  [\href{https://arxiv.org/abs/1602.03901}{{\ttfamily 1602.03901}}].

\bibitem{Aoki:2017aws}
M.~Aoki, H.~Goto and J.~Kubo, \emph{{Gravitational Waves from Hidden QCD Phase
  Transition}}, \href{https://doi.org/10.1103/PhysRevD.96.075045}{\emph{Phys.
  Rev. D} {\bfseries 96} (2017) 075045}
  [\href{https://arxiv.org/abs/1709.07572}{{\ttfamily 1709.07572}}].

\bibitem{Aoki:2019mlt}
M.~Aoki and J.~Kubo, \emph{{Gravitational waves from chiral phase transition in
  a conformally extended standard model}},
  \href{https://doi.org/10.1088/1475-7516/2020/04/001}{\emph{JCAP} {\bfseries
  04} (2020) 001} [\href{https://arxiv.org/abs/1910.05025}{{\ttfamily
  1910.05025}}].

\bibitem{Halverson:2020xpg}
J.~Halverson, C.~Long, A.~Maiti, B.~Nelson and G.~Salinas, \emph{{Gravitational
  waves from dark Yang-Mills sectors}},
  \href{https://doi.org/10.1007/JHEP05(2021)154}{\emph{JHEP} {\bfseries 05}
  (2021) 154} [\href{https://arxiv.org/abs/2012.04071}{{\ttfamily
  2012.04071}}].

\bibitem{Fairbairn:2019xog}
M.~Fairbairn, E.~Hardy and A.~Wickens, \emph{{Hearing without seeing:
  gravitational waves from hot and cold hidden sectors}},
  \href{https://doi.org/10.1007/JHEP07(2019)044}{\emph{JHEP} {\bfseries 07}
  (2019) 044} [\href{https://arxiv.org/abs/1901.11038}{{\ttfamily
  1901.11038}}].

\bibitem{Huang:2020crf}
W.-C.~Huang, M.~Reichert, F.~Sannino and Z.-W.~Wang, \emph{{Testing the dark
  SU(N) Yang-Mills theory confined landscape: From the lattice to gravitational
  waves}}, \href{https://doi.org/10.1103/PhysRevD.104.035005}{\emph{Phys. Rev.
  D} {\bfseries 104} (2021) 035005}
  [\href{https://arxiv.org/abs/2012.11614}{{\ttfamily 2012.11614}}].

\bibitem{Kang:2021epo}
Z.~Kang, S.~Matsuzaki and J.~Zhu, \emph{{Dark confinement-deconfinement phase
  transition: a roadmap from Polyakov loop models to gravitational waves}},
  \href{https://doi.org/10.1007/JHEP09(2021)060}{\emph{JHEP} {\bfseries 09}
  (2021) 060} [\href{https://arxiv.org/abs/2101.03795}{{\ttfamily
  2101.03795}}].

\bibitem{Reichert:2021cvs}
M.~Reichert, F.~Sannino, Z.-W.~Wang and C.~Zhang, \emph{{Dark confinement and
  chiral phase transitions: gravitational waves vs matter representations}},
  \href{https://doi.org/10.1007/JHEP01(2022)003}{\emph{JHEP} {\bfseries 01}
  (2022) 003} [\href{https://arxiv.org/abs/2109.11552}{{\ttfamily
  2109.11552}}].

\bibitem{Bigazzi:2020avc}
F.~Bigazzi, A.~Caddeo, A.L.~Cotrone and A.~Paredes, \emph{{Dark Holograms and
  Gravitational Waves}},
  \href{https://doi.org/10.1007/JHEP04(2021)094}{\emph{JHEP} {\bfseries 04}
  (2021) 094} [\href{https://arxiv.org/abs/2011.08757}{{\ttfamily
  2011.08757}}].

\bibitem{Kierkla:2022odc}
M.~Kierkla, A.~Karam and B.~Swiezewska, \emph{{Conformal model for
  gravitational waves and dark matter: a status update}},
  \href{https://doi.org/10.1007/JHEP03(2023)007}{\emph{JHEP} {\bfseries 03}
  (2023) 007} [\href{https://arxiv.org/abs/2210.07075}{{\ttfamily
  2210.07075}}].

\bibitem{Croon:2018erz}
D.~Croon, V.~Sanz and G.~White, \emph{{Model Discrimination in Gravitational
  Wave spectra from Dark Phase Transitions}},
  \href{https://doi.org/10.1007/JHEP08(2018)203}{\emph{JHEP} {\bfseries 08}
  (2018) 203} [\href{https://arxiv.org/abs/1806.02332}{{\ttfamily
  1806.02332}}].

\bibitem{Baldes:2017rcu}
I.~Baldes, \emph{{Gravitational waves from the asymmetric-dark-matter
  generating phase transition}},
  \href{https://doi.org/10.1088/1475-7516/2017/05/028}{\emph{JCAP} {\bfseries
  05} (2017) 028} [\href{https://arxiv.org/abs/1702.02117}{{\ttfamily
  1702.02117}}].

\bibitem{Salvio:2023qgb}
A.~Salvio, \emph{{Model-independent radiative symmetry breaking and
  gravitational waves}},
  \href{https://doi.org/10.1088/1475-7516/2023/04/051}{\emph{JCAP} {\bfseries
  04} (2023) 051} [\href{https://arxiv.org/abs/2302.10212}{{\ttfamily
  2302.10212}}].

\bibitem{Romero:2021kby}
A.~Romero, K.~Martinovic, T.A.~Callister, H.-K.~Guo, M.~Mart\'\i{}nez,
  M.~Sakellariadou et~al., \emph{{Implications for First-Order Cosmological
  Phase Transitions from the Third LIGO-Virgo Observing Run}},
  \href{https://doi.org/10.1103/PhysRevLett.126.151301}{\emph{Phys. Rev. Lett.}
  {\bfseries 126} (2021) 151301}
  [\href{https://arxiv.org/abs/2102.01714}{{\ttfamily 2102.01714}}].

\bibitem{Arcadi:2022lpp}
G.~Arcadi, N.~Benincasa, A.~Djouadi and K.~Kannike,
  \emph{{Two-Higgs-doublet-plus-pseudoscalar model: Collider, dark matter, and
  gravitational wave signals}},
  \href{https://doi.org/10.1103/PhysRevD.108.055010}{\emph{Phys. Rev. D}
  {\bfseries 108} (2023) 055010}
  [\href{https://arxiv.org/abs/2212.14788}{{\ttfamily 2212.14788}}].

\bibitem{Ghosh:2022fzp}
P.~Ghosh, T.~Ghosh and S.~Roy, \emph{{Interplay among gravitational waves, dark
  matter and collider signals in the singlet scalar extended type-II seesaw
  model}},  \href{https://arxiv.org/abs/2211.15640}{{\ttfamily 2211.15640}}.

\bibitem{Liu:2023sey}
S.~Liu and L.~Wang, \emph{{Spontaneous CP violation electroweak baryogenesis
  and gravitational wave through multistep phase transitions}},
  \href{https://doi.org/10.1103/PhysRevD.107.115008}{\emph{Phys. Rev. D}
  {\bfseries 107} (2023) 115008}
  [\href{https://arxiv.org/abs/2302.04639}{{\ttfamily 2302.04639}}].

\bibitem{Profumo:2014opa}
S.~Profumo, M.J.~Ramsey-Musolf, C.L.~Wainwright and P.~Winslow,
  \emph{{Singlet-catalyzed electroweak phase transitions and precision Higgs
  boson studies}},
  \href{https://doi.org/10.1103/PhysRevD.91.035018}{\emph{Phys. Rev. D}
  {\bfseries 91} (2015) 035018}
  [\href{https://arxiv.org/abs/1407.5342}{{\ttfamily 1407.5342}}].

\bibitem{Chen:2020wvu}
N.~Chen, T.~Li and Y.~Wu, \emph{{The gravitational waves from the collapsing
  domain walls in the complex singlet model}},
  \href{https://doi.org/10.1007/JHEP08(2020)117}{\emph{JHEP} {\bfseries 08}
  (2020) 117} [\href{https://arxiv.org/abs/2004.10148}{{\ttfamily
  2004.10148}}].

\bibitem{Barger:2007im}
V.~Barger, P.~Langacker, M.~McCaskey, M.J.~Ramsey-Musolf and G.~Shaughnessy,
  \emph{{LHC Phenomenology of an Extended Standard Model with a Real Scalar
  Singlet}}, \href{https://doi.org/10.1103/PhysRevD.77.035005}{\emph{Phys. Rev.
  D} {\bfseries 77} (2008) 035005}
  [\href{https://arxiv.org/abs/0706.4311}{{\ttfamily 0706.4311}}].

\bibitem{Curtin:2014jma}
D.~Curtin, P.~Meade and C.-T.~Yu, \emph{{Testing Electroweak Baryogenesis with
  Future Colliders}},
  \href{https://doi.org/10.1007/JHEP11(2014)127}{\emph{JHEP} {\bfseries 11}
  (2014) 127} [\href{https://arxiv.org/abs/1409.0005}{{\ttfamily 1409.0005}}].

\bibitem{Costa:2015llh}
R.~Costa, M.~M\"uhlleitner, M.O.P.~Sampaio and R.~Santos, \emph{{Singlet
  Extensions of the Standard Model at LHC Run 2: Benchmarks and Comparison with
  the NMSSM}}, \href{https://doi.org/10.1007/JHEP06(2016)034}{\emph{JHEP}
  {\bfseries 06} (2016) 034}
  [\href{https://arxiv.org/abs/1512.05355}{{\ttfamily 1512.05355}}].

\bibitem{Han:2016gyy}
H.~Han, J.M.~Yang, Y.~Zhang and S.~Zheng, \emph{{Collider Signatures of
  Higgs-portal Scalar Dark Matter}},
  \href{https://doi.org/10.1016/j.physletb.2016.03.010}{\emph{Phys. Lett. B}
  {\bfseries 756} (2016) 109}
  [\href{https://arxiv.org/abs/1601.06232}{{\ttfamily 1601.06232}}].

\bibitem{GAMBIT:2017gge}
{\scshape GAMBIT} collaboration, \emph{{Status of the scalar singlet dark
  matter model}},
  \href{https://doi.org/10.1140/epjc/s10052-017-5113-1}{\emph{Eur. Phys. J. C}
  {\bfseries 77} (2017) 568}
  [\href{https://arxiv.org/abs/1705.07931}{{\ttfamily 1705.07931}}].

\bibitem{McDonald:1993ex}
J.~McDonald, \emph{{Gauge singlet scalars as cold dark matter}},
  \href{https://doi.org/10.1103/PhysRevD.50.3637}{\emph{Phys. Rev. D}
  {\bfseries 50} (1994) 3637}
  [\href{https://arxiv.org/abs/hep-ph/0702143}{{\ttfamily hep-ph/0702143}}].

\bibitem{Cline:2013gha}
J.M.~Cline, K.~Kainulainen, P.~Scott and C.~Weniger, \emph{{Update on scalar
  singlet dark matter}},
  \href{https://doi.org/10.1103/PhysRevD.88.055025}{\emph{Phys. Rev. D}
  {\bfseries 88} (2013) 055025}
  [\href{https://arxiv.org/abs/1306.4710}{{\ttfamily 1306.4710}}], [Erratum:
  \href{https://doi.org/10.1103/PhysRevD.92.039906}{\textit{Phys.~Rev.~D}
  \textbf{92} (2015) 039906}].

\bibitem{Iso:2009ss}
S.~Iso, N.~Okada and Y.~Orikasa, \emph{{Classically conformal $B - L$ extended
  Standard Model}},
  \href{https://doi.org/10.1016/j.physletb.2009.04.046}{\emph{Phys. Lett. B}
  {\bfseries 676} (2009) 81} [\href{https://arxiv.org/abs/0902.4050}{{\ttfamily
  0902.4050}}].

\bibitem{Iso:2009nw}
S.~Iso, N.~Okada and Y.~Orikasa, \emph{{The minimal $B-L$ model naturally
  realized at TeV scale}},
  \href{https://doi.org/10.1103/PhysRevD.80.115007}{\emph{Phys. Rev. D}
  {\bfseries 80} (2009) 115007}
  [\href{https://arxiv.org/abs/0909.0128}{{\ttfamily 0909.0128}}].

\bibitem{Oda:2017kwl}
S.~Oda, N.~Okada and D.-s.~Takahashi, \emph{{Right-handed neutrino dark matter
  in the classically conformal U(1)' extended standard model}},
  \href{https://doi.org/10.1103/PhysRevD.96.095032}{\emph{Phys. Rev. D}
  {\bfseries 96} (2017) 095032}
  [\href{https://arxiv.org/abs/1704.05023}{{\ttfamily 1704.05023}}].

\bibitem{Peccei:1977hh}
R.D.~Peccei and H.R.~Quinn, \emph{{CP Conservation in the Presence of
  Instantons}}, \href{https://doi.org/10.1103/PhysRevLett.38.1440}{\emph{Phys.
  Rev. Lett.} {\bfseries 38} (1977) 1440}.

\bibitem{Weinberg:1977ma}
S.~Weinberg, \emph{{A New Light Boson?}},
  \href{https://doi.org/10.1103/PhysRevLett.40.223}{\emph{Phys. Rev. Lett.}
  {\bfseries 40} (1978) 223}.

\bibitem{Wilczek:1977pj}
F.~Wilczek, \emph{{Problem of Strong $P$ and $T$ Invariance in the Presence of
  Instantons}}, \href{https://doi.org/10.1103/PhysRevLett.40.279}{\emph{Phys.
  Rev. Lett.} {\bfseries 40} (1978) 279}.

\bibitem{Kim:1979if}
J.E.~Kim, \emph{{Weak Interaction Singlet and Strong CP Invariance}},
  \href{https://doi.org/10.1103/PhysRevLett.43.103}{\emph{Phys. Rev. Lett.}
  {\bfseries 43} (1979) 103}.

\bibitem{Preskill:1982cy}
J.~Preskill, M.B.~Wise and F.~Wilczek, \emph{{Cosmology of the Invisible
  Axion}}, \href{https://doi.org/10.1016/0370-2693(83)90637-8}{\emph{Phys.
  Lett. B} {\bfseries 120} (1983) 127}.

\bibitem{Abbott:1982af}
L.F.~Abbott and P.~Sikivie, \emph{{A Cosmological Bound on the Invisible
  Axion}}, \href{https://doi.org/10.1016/0370-2693(83)90638-X}{\emph{Phys.
  Lett. B} {\bfseries 120} (1983) 133}.

\bibitem{Dine:1982ah}
M.~Dine and W.~Fischler, \emph{{The Not So Harmless Axion}},
  \href{https://doi.org/10.1016/0370-2693(83)90639-1}{\emph{Phys. Lett. B}
  {\bfseries 120} (1983) 137}.

\bibitem{Svrcek:2006yi}
P.~Svrcek and E.~Witten, \emph{{Axions In String Theory}},
  \href{https://doi.org/10.1088/1126-6708/2006/06/051}{\emph{JHEP} {\bfseries
  06} (2006) 051} [\href{https://arxiv.org/abs/hep-th/0605206}{{\ttfamily
  hep-th/0605206}}].

\bibitem{Arvanitaki:2009fg}
A.~Arvanitaki, S.~Dimopoulos, S.~Dubovsky, N.~Kaloper and J.~March-Russell,
  \emph{{String Axiverse}},
  \href{https://doi.org/10.1103/PhysRevD.81.123530}{\emph{Phys. Rev. D}
  {\bfseries 81} (2010) 123530}
  [\href{https://arxiv.org/abs/0905.4720}{{\ttfamily 0905.4720}}].

\bibitem{Cicoli:2012sz}
M.~Cicoli, M.~Goodsell and A.~Ringwald, \emph{{The type IIB string axiverse and
  its low-energy phenomenology}},
  \href{https://doi.org/10.1007/JHEP10(2012)146}{\emph{JHEP} {\bfseries 10}
  (2012) 146} [\href{https://arxiv.org/abs/1206.0819}{{\ttfamily 1206.0819}}].

\bibitem{Baker:2006ts}
C.A.~Baker et~al., \emph{{An Improved experimental limit on the electric dipole
  moment of the neutron}},
  \href{https://doi.org/10.1103/PhysRevLett.97.131801}{\emph{Phys. Rev. Lett.}
  {\bfseries 97} (2006) 131801}
  [\href{https://arxiv.org/abs/hep-ex/0602020}{{\ttfamily hep-ex/0602020}}].

\bibitem{Zhitnitsky:1980tq}
A.R.~Zhitnitsky, \emph{{On Possible Suppression of the Axion Hadron
  Interactions. (In Russian)}}, {\emph{Sov. J. Nucl. Phys.} {\bfseries 31}
  (1980) 260}.

\bibitem{Dine:1981rt}
M.~Dine, W.~Fischler and M.~Srednicki, \emph{{A Simple Solution to the Strong
  CP Problem with a Harmless Axion}},
  \href{https://doi.org/10.1016/0370-2693(81)90590-6}{\emph{Phys. Lett. B}
  {\bfseries 104} (1981) 199}.

\bibitem{Shifman:1979if}
M.A.~Shifman, A.I.~Vainshtein and V.I.~Zakharov, \emph{{Can Confinement Ensure
  Natural CP Invariance of Strong Interactions?}},
  \href{https://doi.org/10.1016/0550-3213(80)90209-6}{\emph{Nucl. Phys. B}
  {\bfseries 166} (1980) 493}.

\bibitem{Kim:2008hd}
J.E.~Kim and G.~Carosi, \emph{{Axions and the Strong CP Problem}},
  \href{https://doi.org/10.1103/RevModPhys.82.557}{\emph{Rev. Mod. Phys.}
  {\bfseries 82} (2010) 557} [\href{https://arxiv.org/abs/0807.3125}{{\ttfamily
  0807.3125}}], [Erratum:
  \href{https://doi.org/10.1103/RevModPhys.91.049902}{\textit{Rev.~Mod.~Phys.}
  \textbf{91} (2019) 049902}].

\bibitem{Sikivie:2009qn}
P.~Sikivie and Q.~Yang, \emph{{Bose-Einstein Condensation of Dark Matter
  Axions}}, \href{https://doi.org/10.1103/PhysRevLett.103.111301}{\emph{Phys.
  Rev. Lett.} {\bfseries 103} (2009) 111301}
  [\href{https://arxiv.org/abs/0901.1106}{{\ttfamily 0901.1106}}].

\bibitem{Marsh:2015xka}
D.J.E.~Marsh, \emph{{Axion Cosmology}},
  \href{https://doi.org/10.1016/j.physrep.2016.06.005}{\emph{Phys. Rept.}
  {\bfseries 643} (2016) 1} [\href{https://arxiv.org/abs/1510.07633}{{\ttfamily
  1510.07633}}].

\bibitem{Ballesteros:2016euj}
G.~Ballesteros, J.~Redondo, A.~Ringwald and C.~Tamarit, \emph{{Unifying
  inflation with the axion, dark matter, baryogenesis and the seesaw
  mechanism}},
  \href{https://doi.org/10.1103/PhysRevLett.118.071802}{\emph{Phys. Rev. Lett.}
  {\bfseries 118} (2017) 071802}
  [\href{https://arxiv.org/abs/1608.05414}{{\ttfamily 1608.05414}}].

\bibitem{DiLuzio:2020wdo}
L.~Di~Luzio, M.~Giannotti, E.~Nardi and L.~Visinelli, \emph{{The landscape of
  QCD axion models}},
  \href{https://doi.org/10.1016/j.physrep.2020.06.002}{\emph{Phys. Rept.}
  {\bfseries 870} (2020) 1} [\href{https://arxiv.org/abs/2003.01100}{{\ttfamily
  2003.01100}}].

\bibitem{Choi:2020rgn}
K.~Choi, S.H.~Im and C.~Sub~Shin, \emph{{Recent Progress in the Physics of
  Axions and Axion-Like Particles}},
  \href{https://doi.org/10.1146/annurev-nucl-120720-031147}{\emph{Ann. Rev.
  Nucl. Part. Sci.} {\bfseries 71} (2021) 225}
  [\href{https://arxiv.org/abs/2012.05029}{{\ttfamily 2012.05029}}].

\bibitem{Gorghetto:2021fsn}
M.~Gorghetto, E.~Hardy and H.~Nicolaescu, \emph{{Observing invisible axions
  with gravitational waves}},
  \href{https://doi.org/10.1088/1475-7516/2021/06/034}{\emph{JCAP} {\bfseries
  06} (2021) 034} [\href{https://arxiv.org/abs/2101.11007}{{\ttfamily
  2101.11007}}].

\bibitem{VonHarling:2019rgb}
B.~Von~Harling, A.~Pomarol, O.~Pujol\`as and F.~Rompineve, \emph{{Peccei-Quinn
  Phase Transition at LIGO}},
  \href{https://doi.org/10.1007/JHEP04(2020)195}{\emph{JHEP} {\bfseries 04}
  (2020) 195} [\href{https://arxiv.org/abs/1912.07587}{{\ttfamily
  1912.07587}}].

\bibitem{DelleRose:2019pgi}
L.~Delle~Rose, G.~Panico, M.~Redi and A.~Tesi, \emph{{Gravitational Waves from
  Supercool Axions}},
  \href{https://doi.org/10.1007/JHEP04(2020)025}{\emph{JHEP} {\bfseries 04}
  (2020) 025} [\href{https://arxiv.org/abs/1912.06139}{{\ttfamily
  1912.06139}}].

\bibitem{Dev:2019njv}
P.S.B.~Dev, F.~Ferrer, Y.~Zhang and Y.~Zhang, \emph{{Gravitational Waves from
  First-Order Phase Transition in a Simple Axion-Like Particle Model}},
  \href{https://doi.org/10.1088/1475-7516/2019/11/006}{\emph{JCAP} {\bfseries
  11} (2019) 006} [\href{https://arxiv.org/abs/1905.00891}{{\ttfamily
  1905.00891}}].

\bibitem{Ghoshal:2020vud}
A.~Ghoshal and A.~Salvio, \emph{{Gravitational waves from fundamental axion
  dynamics}}, \href{https://doi.org/10.1007/JHEP12(2020)049}{\emph{JHEP}
  {\bfseries 12} (2020) 049}
  [\href{https://arxiv.org/abs/2007.00005}{{\ttfamily 2007.00005}}].

\bibitem{Ringwald:2020vei}
A.~Ringwald, K.~Saikawa and C.~Tamarit, \emph{{Primordial gravitational waves
  in a minimal model of particle physics and cosmology}},
  \href{https://doi.org/10.1088/1475-7516/2021/02/046}{\emph{JCAP} {\bfseries
  02} (2021) 046} [\href{https://arxiv.org/abs/2009.02050}{{\ttfamily
  2009.02050}}].

\end{thebibliography}\endgroup
\bibliographystyle{JHEP}

\end{document}